\documentclass[a4paper,12pt]{report}
\usepackage{amssymb}
\usepackage{amsfonts}
\usepackage{amsmath}
\usepackage{geometry}
\usepackage{indentfirst}
\usepackage{graphicx}
\usepackage{hyperref}
\usepackage[english]{babel}
\usepackage{slashed}
\usepackage{fancyhdr}
\usepackage[latin1]{inputenc}
\usepackage{ragged2e}
\usepackage{subfigure}

\fancypagestyle{plain}{
    \fancyhead{}      }

\begin{document}

\ \thispagestyle{empty}

\begin{center}
{\LARGE Centro Brasileiro de Pesquisas Físicas-CBPF}

{\LARGE COSMO}

\bigskip

\bigskip

\bigskip

\bigskip

\bigskip

\bigskip

\bigskip

\bigskip

\bigskip

\bigskip

\bigskip

\bigskip

\bigskip

\bigskip

\bigskip

\bigskip

\bigskip

\bigskip

\bigskip

\bigskip

\bigskip

\bigskip

{\large The Odd 2D Bubbles, 4D Triangles, and Einstein and Weyl Anomalies in
2D Gravitational Fermionic amplitudes: The Role of Breaking Integration
Linearity for Anomalies}

\bigskip

\bigskip

\bigskip

\bigskip

\bigskip

\bigskip

\bigskip

\bigskip

\bigskip

\bigskip

{\large Luciana Ebani}

\bigskip

\bigskip

\bigskip

\bigskip

\bigskip

\bigskip

\bigskip

\bigskip

\bigskip

\bigskip
\end{center}

\begin{flushright}
\bigskip 

\bigskip 

Advisor: Prof. Dr. José Abdalla Helayel Neto-CBPF

\bigskip

Doctoral Thesis
\end{flushright}

\begin{flushleft}
\bigskip

\bigskip

\bigskip

\bigskip

\bigskip

\bigskip

\bigskip

\bigskip

\bigskip

\bigskip
\end{flushleft}

\begin{center}
\bigskip

\bigskip \bigskip

Rio de Janeiro

2023

\newpage

\vfill\thispagestyle{empty}

\bigskip
\end{center}

\begin{flushright}
\hfill \vfill"Do \textit{mainstream} não se exige nada, mas da crítica razoá%
vel se

quer que se mostre até a existência dos átomos com a~qual~se~escreve."

(José Fernando~Thuorst)
\end{flushright}

\begin{center}
\vspace*{1cm}

\baselineskip=18pt
\end{center}

\baselineskip=18pt

\chapter*{Acknowledgments}

There are many people that I am grateful for being part of this journey,
especially to...

\begin{itemize}
\item My mother for teaching me to be strong and resilient. For helping you
go through this phase of uncertainty and lack of scholarship without
questioning my choice to continue. My father, who even though he is no
longer here, taught me from an early age that study and dedication would get
me where I wanted to go. To my brothers Patrícia, Rodrigo, and Ana and my
lifelong friend André, for always being with me, smiling in the victories,
and taking me through the falls (which were not few). To my nephews Alice
and Léo for existing and complementing my life in a unique way. To my cousin
Alexandre for being my family and my haven in Rio.

\item My two workmates: José, for being my tireless companion on this
journey and never letting me forget who I am and where I wanted to go. To
Thalis for bringing me wisdom, friendship, and dedication. From room 1012
that we shared at INPE-UFSM, early mornings at CCNE... UFLA, USP... to the
corridors of Diracstan, more than a decade passed... a PhD made us stronger.
If one had given up halfway through... maybe this thesis would never exist.

\item Professor Orimar, for our many discussions (and those we didn't have),
which left me with an immense desire to explore the hidden paths of quantum
field theories. And also for having introduced me and strengthened my
relationship with Professor Helayel, who is an example for everyone who
knows him as a teacher and as a human being. To Professor Tião, for his
great contribution to the discussions that fostered this thesis.

\item Miguel, for the best years I had in Rio and for being so generous to
me. To Guilherme, Erich, Pablo, Jade, Ivana, and many other friends I made
at CBPF... between "muretas" and discussions, the days were much better for
having you around. To Luiza for being one of my favorite people in the world
and Flores for our "reunion" for life.

\item My brother-in-law Rodrigo and friends from Estância Velha, Carol,
Lari, Fe, and Chuca, for bringing me back to reality when I needed it.

\item The APG José Leite Lopes. To Bete, Ricardo (in memoriam), and Cláudia
for always being willing to help me with the CBPF bureaucracies.

\item All the professors, especially those at UFSM, UFLA and CBPF who helped
me to become the physicist I am today.

\item CNPQ, for the financial support from this trip called doctorate;
without that, it would never be possible.

\item Dilma and Lula, who allowed me to study and get to a quality public
university. And finally, to the Brazilian people who once again dreamed of
better days (which will come!).\thispagestyle{empty}
\end{itemize}

\chapter*{Abstract}

We investigated Relations Among Green Functions defined in the context of an
alternative strategy for coping with the divergences, also called the
Implicit Regularization Method (IREG). This procedure does not use specific
rules for the context being investigated: the mathematical content
(divergent and finite) will remain intact until the calculations end. The
divergent part will be organized through standardized objects free of
physical quantities. In contrast, the finite part is projected in a class of
well-behaved functions that carry all the amplitudes' physical content. That
relations arise in fermionic amplitudes in even space-time dimensions, where
anomalous tensors connect to finite amplitudes as in the bubbles and
triangles in two and four dimensions. Those tensors depend on surface terms,
whose non-zero values arise from finite amplitudes as requirements of
consistency with the linearity of integration and uniqueness. Maintaining
these terms implies breaking momentum-space homogeneity and, in a later
step, the Ward identities. Meanwhile, eliminating them allows more than one
mathematical expression for the same amplitude. That is a consequence of
choices related to the involved Dirac traces. Independently of divergences,
it is impossible to satisfy all symmetry implications by simultaneously
requiring vanishing surface terms and linearity. Then we approach the 1-loop
level fermionic correction for the propagation of the graviton in a
space-time $D=1+1$ through the action of a Weyl fermion in curved
space-time. In this context, gravitational anomalies arise, and the
amplitudes investigated have the highest degree of divergence quadratic.
That imposes a substantial algebraic effort; however, the conclusions are in
agreement with the non-gravitational amplitudes. At the end of the
calculations, we show how it is possible to fix the value of the divergent
part through the relations imposed for amplitudes.

Keywords: Anomalies, Gravitational Anomalies, Divergences, Implicit
Regularization.\thispagestyle{empty}

\chapter*{Resumo}

Investigamos Relações entre Funções de Green definidas no contexto de uma
estratégia alternativa para lidar com as divergências, também conhecida como
Método de Regularização Implícita (IREG). Este procedimento não utiliza
regras específicas para o contexto que está sendo investigado: o conteúdo
matemático (divergente e finito) permanecerá intacto até o final dos cá%
lculos. A parte divergente será organizada através de objetos padronizados
livres de grandezas físicas. Em contraste, a parte finita é projetada em uma
classe de funções bem comportadas que carregam todo o conteúdo físico das
amplitudes. Essas relações surgem em amplitudes fermiônicas em dimensões espa%
ço-temporais pares, onde tensores anômalos se conectam a amplitudes finitas
como nas bolhas e triângulos em duas e quatro dimensões. Esses tensores
dependem de termos de superfície, cujos valores diferentes de zero surgem de
amplitudes finitas como requisitos de consistência com a linearidade de
integração e unicidade. Manter esses termos implica quebrar a homogeneidade
do espaço-momento e, em uma etapa posterior, as Identidades de Ward.
Entretanto, eliminá-los permite mais de uma expressão matemática para a
mesma amplitude. Isso é consequência de escolhas relacionadas aos traços de
Dirac envolvidos. Independentemente das divergências, é impossível
satisfazer todas as implicações de simetria exigindo simultaneamente termos
de superfície nulos e linearidade. Em seguida, abordamos a correção fermiô%
nica ao nível 1-loop para a propagação do gráviton em um espaço-tempo $D=1+1$
através da ação de um férmion de Weyl em um espaço-tempo curvo. Nesse
contexto, surgem as anomalias gravitacionais, sendo que as amplitudes
investigadas apresentam o maior grau de divergência quadrática. Isso impõe
um esforço algébrico substancial; no entanto, as conclusões estão de acordo
com as amplitudes sem acoplamento derivativo. Ao final dos cálculos,
mostramos como é possível fixar o valor da parte divergente através das relaç%
ões impostas para as amplitudes.

Palavras-chave: Anomalias, Anomalias Gravitacionais, Divergências, Regulariza%
ção Implícita.

\bigskip \newpage\thispagestyle{empty}

\begin{center}
{\LARGE Acronyms and Abbreviations}
\end{center}

\bigskip

{\large QFT - Quantum Field Theory }

{\large 2D - Two Dimensions}

{\large 4D - Four Dimensions}

{\large RAGFs - Relations Among Green Functions}

{\large WIs - Ward Identities}

{\large IREG - Implicit Regularization}

{\large DR - Dimensional Regularization}

{\large LHS - Left-Hand Side}

{\large RHS - Right-Hand Side}

{\large S - Scalar}

{\large P - Pseudoscalar}

{\large A- Axial}

{\large V - Vector}

\thispagestyle{empty}%
\tableofcontents%
\thispagestyle{empty}%
\listoffigures%
\listoftables%

\chapter{Introduction}

\setcounter{page}{1}Since their inception, anomalies have played an
important role in Quantum Field Theories (QFTs). The authors \cite%
{Fukuda1949, Steinberger1949, Schwinger1951, Rosenberg1963} first met the
subject in the forties and fifties. Then, it was rediscovered in two
dimensions ($2D$) by Johnson \cite{Johnson1963}; through the
non-conservation of the axial current in the two-point functions. And in
four dimensions ($4D$) in the context of the\textit{\ ABJ anomaly of the
triangle's graph} \cite{Adler1969, Bardeen1969, Jackiw1969}. In this case,
it manifests when two vector currents couple to an axial current via a
fermionic propagator loop. The anomalous term (i.e., not expected from the
canonical equations) in the divergence of the axial current that violates
the PCAC (partial conservation of the axial current) would be responsible
for the decay rate of some mesons, including the electromagnetic decay of
the neutral pion, $\pi ^{0}\rightarrow \gamma \gamma $, observed
experimentally. Later, many studies considered perturbative and
non-perturbative approaches to investigate these phenomena. Among them, the
Fujikawa interpretation of the path-integral measure \cite{Bastianelli2006},
heat kernel \cite{Vassilevich2003}, and cohomological methods \cite%
{Bertlmann1996}.

It is well-known that anomalies prevent the quantum counterparts of Noether
currents from satisfying their classical conservation laws, which break Ward
Identities (WI). Meanwhile, these constraints are necessary to ensure the
perturbative renormalizability of gauge models. That also applies to
theories with spontaneous symmetry breaking as the Standard Model \cite%
{BenLee1972, Gross1972}. The anomaly cancellation mechanism corroborates
with the number of quark generations that ultimately implies the prediction
of the top quark, for example, see the book \cite{Bertlmann1996}, and the
maintenance of the renormalizability of the standard model ensures internal
consistency of the theory.

Similarly, there are anomalies present when fermionic fields couple to
gravitational fields. Delbourgo and Salam in \cite{Salam1972} and Kimura in 
\cite{Kimura1969} established that in the physical dimension, $D=1+3$, two
gravitons contribute to the axial anomaly from a triangle diagram. Two
energy-momentum tensors couple to an axial current via a fermionic
propagator loop. This anomaly would indicate \cite{Gaume1983} the
impossibility of obtaining a gauge theory in a gravitational context unless
there is an anomaly cancellation mechanism.

Alvarez-Gaumé and Witten also show in \cite{Gaume1983} that the violation of
the diffeomorphism invariance (Einstein anomalies) at $D=4k+2$ occurs in
"purely gravitational" anomalies, without gauge coupling, in curved
spacetime for Weyl fermions with spin $1/2$ or $3/2$ coupled to the
gravitational field via energy-momentum tensor. When there is a violation of
the conformal symmetry as we have the Weyl anomaly (or trace anomaly).
Capper and Duff in \cite{DuffCapper1974, Duff1994} studied such anomalies in
the graviton propagation by interaction with photons and Weyl fermions at
the 1-loop level, and more recently, the contribution of the Pontryagin
density to the Weyl anomalies has been revisited by Bonara et al., \cite%
{Bonora2014}, \cite{Bonora2015}, and \cite{Bonoraetal:2017}. Furthermore,
for gravitation, we have Lorentz anomalies: They signify an antisymmetric
part in the energy-momentum tensor, in even dimensions, in particular $2D$,
they can be traded by the Einstein anomalies \cite{Bertlmann1996} using the
local Bardeen-Zummino polynomial \cite{BZumino1984}. The same polynomial
transforms the consistency into the covariant form for anomalies.

Among the places where anomalies manifest, we have the perturbative scenario
for correlators of axial and vector currents that are divergent odd tensors.
Some of them $AV^{n}$ amplitudes in $d=2n$ dimensions, which cannot satisfy
all WIs, (see \cite{JohnsonJackiw1969}). These are $\left( n+1\right) $%
th-rank tensors of odd-parity and functions of $n$ momenta variables.
Consequently, they have a set of low-energy theorems obtained through
momenta contractions. In one loop, they contain Dirac traces having two more
gamma matrices than the number of dimensions. These traces are linear
combinations of monomials in Levi-Civita tensor and metric, displaying
equivalent expressions that differ regarding index arrangement, signs, and
the number of monomials. In addition, the power counting of the integrals
indicates the presence of surface terms, making these structures depend on
the graph's momenta routing (outside the amplitude $AV$ in $d=2$). Since
perturbative solutions admit arbitrary choices for routings and Dirac
traces, the final results show many possibilities.

This last proposition is inseparable from the fact that divergences are the
rule to get model predictions of QFT in perturbation theory. Regularization
methods are adopted to obtain information about the amplitudes' kinematic
dependence and symmetry consequences. Some examples of these techniques are
Cut-off, Pauli-Villars, Analytic Regularization, Dimensional Regularization
(DR)\textrm{\ }\cite{Bollini1972, tHooft1972},\textrm{\ }High Covariant
Regularization \cite{Slavnov1972, Bakeyev1996}, Differential Renormalization
(\cite{AguilaVictoria1998}). However, these regularization methods can
compromise the theory's predictive power by modifying amplitudes and making
the divergent structures finite. Beyond its limits of applicability in
theories involving the chiral matrix, manipulations not guaranteed to the
original expressions take effect as shifts in the integration variable%
\footnote{%
Take the DR as an example; it eliminates surface terms as a condition to
achieve symmetry preservation.}. Furthermore, new methods to deal with
multi-loop calculations aiming for algorithmic implementation of precision
numerical predictions \cite{Pittau2012}, \cite{TodGnendiger2017}. The
prescription also may prescribe rules, not inherent to Feynman's ones, for
which properties of the algebras are valid or not \cite{Breitenlohner1977,
Jegerlehner2001, Tsai2011a, Tsai2011b, RFerrari2017, Bruque2018}.

On the other hand, tensor Feynman integrals exhibiting diverging power
counting have surface terms. For the linearly diverging ones, a shift in the
integration variable requires compensation through non-zero surface terms 
\cite{Treiman1985}, \cite{ChengLi1984}, and \cite{Bertlmann1996}. They
cannot be free-shifted and need arbitrary labels for internal momenta.
Energy-momentum conservation sets differences in the routings as functions
of the physical momenta; however, internal momenta are arbitrary (by
themselves and their sums) and may assume non-covariant expressions \cite%
{Sterman1993}. Since non-zero surface terms imply the breaking of
translational symmetry in the momentum space and this operation is needed to
prove WIs, other symmetries violations also occur. By exploring tensor
properties, we investigate symmetry maintenance and its relation with the
mathematical content of the diagrams. That materializes into a discussion
about the linearity of integration and choices for perturbative solutions
related to their uniqueness\footnote{%
To uniqueness, which needs a particular definition to work its consequences,
we provide it along the thesis.}.

For one of our purposes, we use a general model coupling spin-$1/2$ fermions
(through their bilinear and without derivatives, eventually with fermions of
distinct masses) with boson fields of even and odd parity (spins $0$ and $1$%
). The $n$-vertex polygon graphs of spin-$1/2$ internal propagators are one
part of the analysis, specifically the $2D$-$AV$ and $VA$ bubbles, $4D$-$AVV$%
, $VAV$, $VVA$, and $AAA$. In the e-print \cite{Preprint}, the extension to
the $6D$-$AVVV$ box is also explored with the same conclusions. In two
dimensions, the $AV$-$VA$ amplitudes worked with arbitrary masses; the
author has the publication \cite{Ebani2018}.

The amplitudes are obtained within a procedure to handle divergent and
finite integrals introduced in the Ph.D. thesis of O.A. Battistel \cite%
{PhdBattistel1999}. Several investigations applied this strategy in $2D$, $%
4D $, $6D$, and $5D.$ This method has no limit of applicability; without
specific rules to the context being investigated. We can use it for theories
in even and odd dimensions simultaneously, in addition to careful
investigation into chiral theories \cite{Battistel2012, Battistel2018} \cite%
{Battistel2002a, Battistel2002b} \cite{Fonseca2014} \cite{Fonseca2013} \cite%
{Battistel2014}. Other investigations use the name Implicit Regularization
(IREG), having a similar approach \cite{Viglioni2016, Vieira2016,
Ferreira2012, Porto2018}.

This procedure uses a general identity to isolate divergences that do not
interfere with Feynman's rules. Since we do not evaluate divergent integrals
explicitly, amplitudes are not modified at any stage of calculations. Also,
we use arbitrary routings for the momenta of internal lines. In this
strategy, we devise a notational scheme to systematize finite integrals and
their divergent parts based on previous works on the subject \cite%
{Battistel2006}, \cite{Dallabona2012}, and \cite{SunYi2012}. Three relevant
ingredients to our discussion are irreducible divergent objects, tensor
surface terms, and finite functions. The only assumption is linearity
applies to the Feynman integrals, which manifests through Relations Among
Green Functions (RAGFs). This aspect is one of the main points of this
investigation.

In this way, having studied, in the last instance, chiral anomalies in two
and four dimensions, we proceed to see how the conclusions extend for the
two-dimensional gravitational anomalies \cite{Gaume1983, Langouche1984,
Leutwyler1986, Hwang1987, Berger1990}. To that end, we explore couplings
with currents involving derivatives in the fermion field. The physical
scenario is described by a model from a Weyl fermion coupled to a background
gravitational field using the same model as the references \cite%
{Bertlmann2001a, Bertlmann2001b}. In an expansion around the Minkowski
metric, the matter field induces corrections through the two-point function
of its (linearized) stress tensor. Taking advantage of the strategy, we
write all the expressions similar to the case without derivative coupling,
which point to many similarities for the elements in the root of symmetry
violations.

By carrying intact the divergent content, until the end of all computations,
our stance on the perturbative amplitudes enables a detailed view of the
elements that yield different results. It also clarifies the connection
among the surface terms in amplitudes with ambiguities of routings, traces,
and symmetry violations. Any interpretation of divergences that sets surface
terms as zero for even amplitudes makes their results symmetric concerning
the symmetries related to momenta contractions but not metric contractions.
Nevertheless, these prescriptions break integration linearity for odd
amplitudes since equal integrands give rise to different integrals. Hence,
an uncountable number of tensors follows from the same expression.

On the other hand, by adopting the value of the surface term that preserves
linearity, all manipulation on the traces provides one and only one tensor
of the routing variables. Therefore the physical interpretation requires
arbitrary parameters to fix the symmetries. The freedom allows us to improve
the known and desired content of the results (for non-derivative couplings).
However, the consequence is that even amplitudes will more often violate
their WIs if they ask universality to play a role.

We organized the work as follows. In Chapter (\ref{ModlDef}), we have the
general model, definitions, and a preliminary discussion. Chapter (\ref{IREG}%
) discusses the strategy to handle the amplitudes, where we define
irreducible objects, tensor surface terms, and finite parts. The compilation
of the effects of traces and surface terms in $2D$ appears in the Chapters (%
\ref{2Dim2Pt}; \ref{2masses}) through complete and independent computation
of all the quantities related to RAGFs. The consequences of the results
preserving linearity or saving translational symmetry are presented and
interpreted in light of low-energy theorems. Chapter (\ref{4Dim3Pt}) deals
with all odd triangles in $4D$, their RAGFs, and the concept of uniqueness.
The Sections (\ref{LE4D}) and (\ref{LED4DSTS}) deal with general properties
of low-energy theorems and offer a proposition that connects linearity,
low-energy behavior of finite amplitudes and surface terms. Chapter (\ref%
{modlDef} and \ref{GAno}) extend the propositions to a gravitational
scenario. In the last Chapter (\ref{finalremarks}), we discuss some points
implied by the investigation for other scenarios.

\chapter{Notation, Definitions, Model and Preliminaries}

\label{ModlDef}

Feynman rules, vertices, and propagators employed in this investigation come
from a model where fermionic currents couple to bosonic fields of even and
odd parity $\{\Phi \left( x\right) ,V_{\mu }\left( x\right) $ $,\Pi \left(
x\right) ,A_{\mu }\left( x\right) \}$ through the general interacting action%
\begin{equation}
\mathcal{S}_{I}=\int \mathrm{d}^{2n}x\left[ e_{S}S\left( x\right) \Phi
\left( x\right) +e_{\Pi }P\left( x\right) \Pi \left( x\right) +e_{V}J^{\mu
}\left( x\right) V_{\mu }\left( x\right) +e_{A}J_{\ast }^{\mu }\left(
x\right) A_{\mu }\left( x\right) \right] .  \label{Action}
\end{equation}%
The currents $\{S,P,J_{\mu },J_{\ast \mu }\}$ are bilinears in the fermionic
fields $J_{i;ab}\left( x\right) =\left( \bar{\psi}_{a}\Gamma _{i}\psi
_{b}\right) \left( x\right) $. They deliver the vertices proportional%
\footnote{%
The proportionality comes from the coupling constants $\left\{ e_{S},e_{\Pi
},e,e_{A}\right\} ,$ taken as the unit for our purposes.} to%
\begin{equation}
\Gamma _{i}\in \left( S,P,V,A\right) =(1,\gamma _{\ast },\gamma _{\mu
},\gamma _{\ast }\gamma _{\mu }),  \label{SetofVertexes}
\end{equation}%
where $\gamma _{\mu }$ are the generators of the Clifford algebra of Dirac
matrices satisfying $\{\gamma ^{\mu _{1}},\gamma ^{\mu _{2}}\}=2g^{\mu
_{1}\mu _{2}}$. The chiral matrix, which is the algebra's highest-weight
element, satisfies $\{\gamma _{\ast },\gamma ^{\mu _{k}}\}=0$ and assumes
the explicit form 
\begin{equation}
\gamma _{\ast }=i^{n-1}\gamma _{0}\gamma _{1}\cdots \gamma _{2n-1}=\frac{%
i^{n-1}}{\left( 2n\right) !}\varepsilon _{\nu _{1}\cdots \nu _{2n}}\gamma
^{\nu _{1}\cdots \nu _{2n}}.  \label{gammastar}
\end{equation}%
We often adopt a merging notation to products of matrices $\gamma ^{\nu
_{1}\cdots \nu _{2n}}=\gamma ^{\nu _{1}}\gamma ^{\nu _{2}}\cdots \gamma
^{\nu _{2n}}$, adapting to Lorentz indexes $\mu _{1}\mu _{2}\cdots \mu
_{s}=\mu _{12\cdots s}$ when convenient. The behavior under the permutation
of the indexes is determined by the objects: $g_{\mu _{1}\mu _{2}}=g_{\mu
_{12}}=g_{\mu _{21}}$ or $\varepsilon _{\mu _{1}\mu _{2}\cdots \mu
_{2n}}=\varepsilon _{\mu _{12\cdots 2n}}=-\varepsilon _{\mu _{21}\cdots \mu
_{2n}}$. For the $2n$-dimensional, follow the normalization $\varepsilon
^{0123\cdots 2n-1}=1$.

The algebra elements are the antisymmetrized products of gamma matrices%
\begin{equation}
\gamma _{\left[ \mu _{1}\cdots \mu _{r}\right] }=\frac{1}{r!}\sum_{\pi \in
S_{r}}\mathrm{sign}\left( \pi \right) \gamma _{\mu _{\pi \left( 1\right)
}\cdots \mu _{\pi \left( r\right) }}.
\end{equation}%
They satisfy general identities as seen in the appendix of the reference 
\cite{deWit1986}: 
\begin{equation}
\gamma _{\ast }\gamma _{\left[ \mu _{1}\cdots \mu _{r}\right] }=\frac{%
i^{n-1+r\left( r+1\right) }}{\left( 2n-r\right) !}\varepsilon _{\mu
_{1}\cdots \mu _{r}}^{\hspace{0.95cm}\nu _{r+1}\cdots \nu _{2n}}\gamma _{%
\left[ \nu _{r+1}\cdots \nu _{2n}\right] }.  \label{Chiral-Id}
\end{equation}%
These identities are needed when taking traces with the chiral matrix. For
products of tensors, we adopted the antisymmetrization notation 
\begin{equation}
A_{[\alpha _{1}\cdots \alpha _{r}}B_{\alpha _{r+1}\cdots \alpha _{s}]}=\frac{%
1}{s!}\sum_{\pi \in S_{s}}\mathrm{sign}(\pi )A_{\alpha _{\pi \left( 1\right)
}\cdots \alpha _{\pi \left( r\right) }}B_{\alpha _{\pi \left( r+1\right)
}\cdots \alpha _{\pi \left( s\right) }},
\end{equation}%
where the normalization factor does not interfere with the used identities.

The spinorial Feynman propagators come from the standard kinetic term of
Dirac fermions%
\begin{equation}
S_{F}\left( K_{i}\right) =\frac{1}{(\slashed{K}_{i}-m_{i}+i0^{+})}=\frac{(\slashed%
{K}_{i}+m_{i})}{D_{i}},  \label{Prop}
\end{equation}%
where $D_{i}=K_{i}^{2}-m_{i}^{2}$ with $K_{i}=k+k_{i}$ and $m_{i}$
corresponding the mass of the $i$-particle. The momentum $k$ is the
unrestricted loop momentum while $k_{i}$ are routings that keep track of the
flux of external momenta through the graph, see \cite{Sterman1993}\footnote{%
Consult section (4.1) for a comment on the arbitrariness of these routings.}%
. They cannot be written as a function of the kinematical data in divergent
integrals. In our approach, they codify conditions of the satisfaction of
symmetries or lack thereof. Nonetheless, their differences relate to
external momenta through the definition%
\begin{equation}
p_{ij}=k_{i}-k_{j},  \label{pij}
\end{equation}%
using momenta conservation in the vertices of the diagram in figure (\ref%
{diag1}). 
\begin{figure}[tbph]
\begin{equation*}
T^{\Gamma _{1}\Gamma _{2}\cdots \Gamma _{n_{1}}}=%
\begin{array}{c}
\includegraphics[scale=0.8]{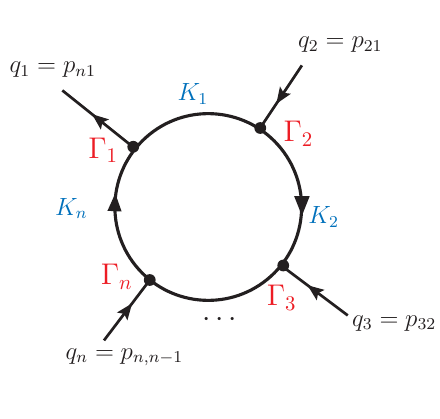}%
\end{array}%
\end{equation*}%
\caption{General diagram for the one-loop amplitudes of this work. }
\label{diag1}
\end{figure}

The integrand of these amplitudes follows from Feynman rules%
\begin{equation}
t^{\Gamma _{1}\Gamma _{2}\cdots \Gamma _{n_{1}}}\left( k_{1},\cdots
,k_{n_{1}}\right) =\text{\textrm{tr}}[\Gamma _{1}S_{F}(K_{1})\Gamma
_{2}S_{F}(K_{2})\cdots \Gamma _{n_{1}}S_{F}(K_{n_{1}})].  \label{t}
\end{equation}%
That is a well-defined function of the external momenta and sums
undetermined by momentum conservation%
\begin{equation}
P_{ij}=k_{i}+k_{j}.  \label{Pij}
\end{equation}%
Often we adopt the simplification $S\left( i\right) \equiv S_{F}(K_{i})$,
where the numerical index $i$ represents all parameters of the corresponding
line. The total amplitude comes from integration in the loop momenta%
\begin{equation}
T^{\Gamma _{1}\Gamma _{2}\cdots \Gamma _{s}}\left( 1,\cdots ,s\right) =\int 
\frac{\mathrm{d}^{2n}k}{(2\pi )^{2n}}t^{\Gamma _{1}\Gamma _{2}\cdots \Gamma
_{s}}\left( 1,\cdots ,s\right) .  \label{T}
\end{equation}%
When replacing the specific vertex operators $\Gamma _{i}$ from (\ref%
{SetofVertexes}), the notation accompanies the Lorentz indexes in order with
the operators. In addition, we set aside the minus signs for closed loops.

\section{Relation Among Green Functions (RAGF)}

As a part of the investigation, we establish identities among Green
functions that display Lorentz indices of vector and axial currents. These
are commonly called Relations Among Green Functions (RAGFs) and have been
used in investigations in the IREG scenario \cite{Battistel2012}\cite%
{Battistel2002a}\cite{Battistel2014}. They can be considered conditions on
the linearity of integration even before WIs are asked to play some role in
perturbation amplitudes.

Let us take the amplitude $AV^{r-1}$ to introduce these relations since they
are part of our analysis,%
\begin{equation}
t_{\mu _{1}\mu _{2}\cdots \mu _{r}}^{AV\cdots V}=\text{\textrm{tr}}[\gamma
_{\ast }\gamma _{\mu _{1}}S\left( 1\right) \gamma _{\mu _{2}}S\left(
2\right) \cdots \gamma _{\mu _{r}}S\left( r\right) ].
\end{equation}%
When contracted with $p_{21}^{\mu _{2}}$ in the vector vertex $\gamma _{\mu
_{2}}$, we remove one propagator using $K_{i}=k+k_{i}$ and $S^{-1}\left(
i\right) =\slashed{K}_{i}-m$ through the standard manipulation 
\begin{equation}
\slashed{p}_{ab}=\slashed{K}_{a}-\slashed{K}_{b}=S^{-1}\left( a\right) -S^{-1}\left(
b\right) +\left( m_{a}-m_{b}\right)
\end{equation}%
This result leads to the vector RAGF, a difference between two amplitudes
built out of the same rules%
\begin{equation}
p_{21}^{\mu _{2}}t_{\mu _{1}\mu _{2}\cdots \mu _{r}}^{AV\cdots V}=[t_{\mu
_{1}\hat{\mu}_{2}\cdots \mu _{r}}^{AV\cdots V}(1,\hat{2},\cdots ,r)-t_{\mu
_{1}\hat{\mu}_{2}\cdots \mu _{r}}^{AV\cdots V}(\hat{1},2,\cdots ,r)]+\left(
m_{2}-m_{1}\right) t_{\mu _{1}\hat{\mu}_{2}\cdots \mu _{r}}^{ASV\cdots V}.
\label{RAGF1}
\end{equation}%
The "hats" mean the omission of the propagator corresponding to that routing
and the vertices corresponding to the Lorentz indexes. In other words, the
RHS contains lower-point functions that are in general more singular under
integration (but not always). Now, observe the contraction of the axial
vertex with $p_{r1}^{\mu _{1}}$ 
\begin{eqnarray}
p_{r1}^{\mu _{1}}t_{\mu _{12}\cdots \mu _{r}}^{AV\cdots V} &=&\text{\textrm{%
tr}}[S\left( r\right) \gamma _{\ast }S^{-1}\left( r\right) S\left( 1\right)
\gamma _{\mu _{2}}S\left( 2\right) \cdots \gamma _{\mu _{r-1}}S\left(
r-1\right) \gamma _{\mu _{r}}] \\
&&-\text{\textrm{tr}}[\gamma _{\ast }\gamma _{\mu _{2}}S\left( 2\right)
\cdots \gamma _{\mu _{r}}S\left( r\right) ].  \notag
\end{eqnarray}%
Using the commutation of the chiral and Dirac matrices that implies in the
identity 
\begin{equation}
S\left( r\right) \gamma _{\ast }S^{-1}\left( r\right) =\left( -\gamma _{\ast
}-2mS\left( r\right) \gamma _{\ast }\right) ,
\end{equation}%
leading to the axial RAGF%
\begin{eqnarray}
p_{r1}^{\mu _{1}}t_{\mu _{12}\cdots \mu _{r}}^{AV\cdots V} &=&[t_{\mu _{r}%
\hat{\mu}_{1}\mu _{2}\cdots \mu _{r-1}}^{AV\cdots V}(1,2,\cdots ,\hat{r})-t_{%
\hat{\mu}_{1}\mu _{2}\cdots \mu _{r}}^{AV\cdots V}(\hat{1},2,\cdots ,r)]
\label{RAGF2} \\
&&-\left( m_{r}+m_{1}\right) t_{\mu _{2}\cdots \mu _{r}}^{PV\cdots V}. 
\notag
\end{eqnarray}

After integration, the relations achieved above become%
\begin{eqnarray}
p_{r1}^{\mu _{1}}T_{\mu _{12}\cdots \mu _{r}}^{AV\cdots V} &=&[T_{\mu _{r}%
\hat{\mu}_{1}\cdots \mu _{r-1}}^{AV\cdots V}(1,2,\cdots ,\hat{r})-T_{\hat{\mu%
}_{1}\cdots \mu _{r}}^{AV\cdots V}(\hat{1},2,\cdots ,r)] \\
&&-\left( m_{r}+m_{1}\right) T_{\mu _{2}\cdots \mu _{r}}^{PV\cdots V}  \notag
\\
p_{21}^{\mu _{2}}T_{\mu _{12}\cdots \mu _{r}}^{AV\cdots V} &=&[T_{\mu _{1}%
\hat{\mu}_{2}\cdots \mu _{r}}^{AV\cdots V}(1,\hat{2},\cdots ,r)-T_{\mu _{1}%
\hat{\mu}_{2}\cdots \mu _{r}}^{AV\cdots V}(\hat{1},2,\cdots ,r)] \\
&&+\left( m_{2}-m_{1}\right) T_{\mu _{1}\hat{\mu}_{2}\cdots \mu
_{r}}^{ASV\cdots V}.  \notag
\end{eqnarray}%
These equations embody assumptions of linearity of integration in
perturbative computations; however, this characteristic is not guaranteed
for divergent amplitudes. We expose this scenario through complete
calculations of amplitudes and their relations. Although these equations are
a structural property of the operations, they are not a priori linked to the
particularities of the model and its symmetries. However, after summing up
all contributions from the crossed diagrams (if applicable), the properties
for the total sum of lower-point Green functions coming from the momenta
contraction should make the expression correspond to the WIs.

The WIs are equations satisfied by Green functions as a consequence of
continuous symmetries of the action. They are valid in perturbative
approximations built on Feynman rules unless they are inevitably anomalous.
They arise from the joint application of the algebra of quantized currents
and equations of motion to these currents: $\partial _{\mu }J^{\mu }=0$ and $%
\partial _{\mu }J_{\ast }^{\mu }=-2miP$. Their expressions in the position
space for axial and vector WIs are%
\begin{eqnarray}
\partial _{\mu _{1}}^{x_{1}}\left\langle J_{\ast }^{\mu _{1}}\left(
x_{1}\right) J_{\mu _{2}}\left( x_{2}\right) \cdots J_{\mu _{r}}\left(
x_{r}\right) \right\rangle &=&-2mi\left\langle P\left( x_{1}\right) J_{\mu
_{2}}\left( x_{2}\right) \cdots J_{\mu _{r}}\left( x_{r}\right)
\right\rangle ,  \label{AWI} \\
\partial _{\mu _{2}}^{x_{2}}\left\langle J_{\ast \mu _{1}}\left(
x_{1}\right) J^{\mu _{2}}\left( x_{2}\right) \cdots J_{\mu _{r}}\left(
x_{r}\right) \right\rangle &=&0,  \label{VWI}
\end{eqnarray}%
where $\left\langle \cdots \right\rangle =\left\langle 0\left\vert T\left[
\cdots \right] \right\vert 0\right\rangle $ is an abbreviation for the time
ordering of the currents. In our notation for perturbative amplitudes, we
would have analogous equations%
\begin{equation}
q_{1}^{\mu _{1}}T_{\mu _{12}\cdots \mu _{r}}^{A\rightarrow V\cdots
V}=-2mT_{\mu _{2}\cdots \mu _{r}}^{P\rightarrow V\cdots V};\quad q_{2}^{\mu
_{2}}T_{\mu _{12}\cdots \mu _{r}}^{A\rightarrow V\cdots V}=0;\cdots \quad
q_{r}^{\mu _{r}}T_{\mu _{12}\cdots \mu _{r}}^{A\rightarrow V\cdots V}=0.
\end{equation}%
The arrow means the mentioned sum of contributions. The connection involving
RAGFs and WIs is straightforward, so that violations of RAGFs imply
violations of WIs. This way, maintaining all WIs depends on satisfying all
RAGFs while having translational invariance in the momentum space. We show
how this requirement is impossible for a class of amplitudes as those
introduced in the sequence. These objects share similar tensor structures,
contain diverging surface terms, and produce the same consequences regards
anomalies in their specific dimensions. All of them are divergent odd
tensors: they have logarithmic power counting in 2D and linear power
counting in 4D.

\begin{itemize}
\item The $2D$ Bubbles: $T_{\mu _{12}}^{AV};$ $T_{\mu _{12}}^{VA};$

\item The $4D$ Triangles: $T_{\mu _{123}}^{AVV};$ $T_{\mu _{123}}^{VAV};$ $%
T_{\mu _{123}}^{VVA};$ $T_{\mu _{123}}^{AAA};$
\end{itemize}

In the second part starting in the Chapter (\ref{modlDef}), we explore the
consequences in a gravitational scenario, we will also consider the
perturbative amplitudes with derivative coupling in 2D (defined in the same
chapter). They have linear and quadratic power counting and appear in
associated with the study of Einstein and Weyl anomalies.

\begin{itemize}
\item The Gravitational Amplitudes Even: $T_{\mu _{12;}\alpha _{1}}^{VV};$ $%
T_{\mu _{12;}\alpha _{12}}^{VV};T_{\mu _{12;}\alpha _{1}}^{AA};$ $T_{\mu
_{12;}\alpha _{12}}^{AA};$

\item The Gravitational Amplitudes Odd: $T_{\mu _{12;}\alpha _{1}}^{AV};$ $%
T_{\mu _{12;}\alpha _{12}}^{AV};$
\end{itemize}

To compute these amplitudes, we have to take the Dirac traces. After that,
any amplitude is expressed as linear combinations of bare Feynman integrals
following the definition\footnote{%
We simplify the dependence of the functions on their arguments $f\left(
k_{1},k_{2},\cdots \right) =f\left( 1,2,\cdots \right) $, omitting them if
it is clear.},\footnote{%
Changing from a reference routing $k_{j}$ to another $k_{i}$ is a matter of
recognizing the definition of $p_{ij}$ in (\ref{pij}) and writing $%
K_{i}=K_{j}+p_{ij}$.}%
\begin{equation}
\bar{J}_{n_{2}}^{\left( 2n\right) \mu _{1}\mu _{2}\cdots \mu _{n_{1}}}\left(
1,2,\cdots ,n_{2}\right) =\int \frac{\mathrm{d}^{2n}k}{\left( 2\pi \right)
^{2n}}\frac{K_{i}^{\mu _{1}}\cdots K_{i}^{\mu _{n_{1}}}}{D_{1}D_{2}\cdots
D_{n_{2}}}.  \label{JInt}
\end{equation}%
These integrals have power counting $\omega =2n+n_{1}-2n_{2}$, where $n_{1}$
is the tensor rank and $n_{2}$ is the number of denominators. A set of five
types of integrals arise within each amplitude, which is the subject of
subsection (\ref{FinFcts}). But first, we develop a procedure to deal with
divergent quantities in the sequence.

\chapter{Procedure to Handle the Divergences and the Finite Integrals}

\label{IREG}Before presenting the strategy to solve the divergent
amplitudes, let us digress into the divergent-integrals issue in QFT. It is
well-known that the products of propagators that are not regular
distribution are ill-defined in general. A good example is the equation%
\begin{equation}
\int \frac{\mathrm{d}^{4}k}{\left( 2\pi \right) ^{4}}\mathrm{tr}[S_{F}\left(
k\right) S_{F}\left( k-p\right) ]=\int \mathrm{d}^{4}x\mathrm{tr}[\hat{S}%
_{F}\left( x\right) \hat{S}_{F}\left( -x\right) ]\mathrm{e}^{ip\cdot x}.
\end{equation}%
The LHS displays a divergent convolution of two Feynman propagators in
momentum space. The RHS is the Fourier transform of a product of propagators
in position space. So both sides do not define distributions because when
the point-wise product of distributions does not exist, the convolution
product of their Fourier transform does not also.

These short-distance UV singularities manifest in divergences of loop
momentum integrals. Their origins trace back to multiplications of
distributions by discontinuous step function in the chronological ordering
of operators in the interaction picture. That leads, through the Wick
theorem, to the Feynman rules; see \cite{Scharf2014, Scharf2010}, originally
in Epstein and Glaser \cite{EpsteinGlaser1973}. Although the undefined
Feynman diagrams can be circumvented by carefully studying the splitting of
distributions with causal support in the setting of causal perturbation
theory \cite{Aste1997,Aste2003,Aste2008} (where no divergent integral
appears at all), we work with Feynman rules in the context of
regularizations.

We use the systematic procedure known as Implicit Regularization (IREG) to
handle the divergences. Its development dates back to the late 1990s in the
Ph.D. thesis of O.A. Battistel \cite{PhdBattistel1999}, having its first
investigations in the references \cite{Battistel1997,BattistelNemes1999}.
Its objective is to keep the connection at all times with the expression of
the "bare" Feynman rules while removing physical parameters (i.e., routings
and masses) from divergent integrals and putting them in strictly finite
integrals. The divergent ones do not suffer any modification besides an
organization through surface terms and irreducible scalar integrals.

This objective is realized by noticing that all Feynman integrals depend on
the propagators-like structures $D_{i}=[\left( k+k_{i}\right) ^{2}-m^{2}]$
defined in equation (\ref{Prop}). Thus, by introducing a parameter $\lambda
^{2}$, it is possible to construct an identity to separate quantities
depending on physical parameters%
\begin{equation}
\frac{1}{D_{i}}=\frac{1}{D_{\lambda }+A_{i}}=\frac{1}{D_{\lambda }}\frac{1}{%
[1-(-A_{i}/D_{\lambda })]},  \label{DecompDi}
\end{equation}%
where $D_{\lambda }=(k^{2}-\lambda ^{2})$ and $A_{i}=2k\cdot
k_{i}+(k_{i}^{2}+\lambda ^{2}-m^{2})$. Now, we use the sum of the geometric
progression of order $N$ and ratio $(-A_{i}/D_{\lambda })$ to write%
\begin{equation}
\frac{1}{[1-(-A_{i}/D_{\lambda })]}=\sum_{r=0}^{N}(-A_{i}/D_{\lambda
})^{r}+(-A_{i}/D_{\lambda })^{N+1}\frac{1}{[1-(-A_{i}/D_{\lambda })]}.
\label{IdentGeom}
\end{equation}%
Immediately it is possible to determine the asymptotic behavior at infinity
of the powers $(-A_{i}/D_{\lambda })^{r}$ as $\left\Vert k\right\Vert ^{-r}$%
. Observe that those terms in the summation sign depend on the routings only
in the numerator through a polynomial.

With the help of equations (\ref{IdentGeom}) and (\ref{DecompDi}), we get%
\begin{equation}
\frac{1}{D_{i}}=\sum_{r=0}^{N}\left( -1\right) ^{r}\frac{A_{i}^{r}}{%
D_{\lambda }^{r+1}}+\left( -1\right) ^{N+1}\frac{A_{i}^{N+1}}{D_{\lambda
}^{N+1}D_{i}}.  \label{id}
\end{equation}%
As this identity is valid for arbitrary $N$, choosing $N$ as equal to or
greater than the power counting is possible. The integration of the last
term is finite under these circumstances, exhibiting dependence on the
external momenta $p_{ij}=k_{i}-k_{j}$ when treating a product of
propagators. The parameters $\lambda ^{2}$ generate a connection between
divergent and finite parts of integrals. That implies specific behavior to
the divergent scalar integrals that is straightforwardly satisfied. We adopt
the mass of the propagator $\lambda ^{2}=m^{2}$ as the scale\footnote{%
The identity is independent of the parameter $\lambda ^{2}$, which is clear
when taking the derivative with this parameter.}.

To modularize the analysis, we organize divergences without modifications in
the first subsection. After that, we introduce the finite functions
necessary to express the amplitudes. Lastly, we introduce integrals
pertinent to this work, discussing some examples.

\section{Divergent Terms\label{DivTerms}}

After applying the identity (\ref{id}), we express the Feynman integrals
through surface terms, irreducible divergent objects, and finite functions.
Divergent terms follow the structure of the summation part of the identity
and appear as a set of pure integration-momentum integrals%
\begin{equation}
\int \frac{\mathrm{d}^{2n}k}{\left( 2\pi \right) ^{2n}}\frac{1}{D_{\lambda
}^{a}},\quad \int \frac{\mathrm{d}^{2n}k}{\left( 2\pi \right) ^{2n}}\frac{%
k_{\mu _{1}}k_{\mu _{2}}}{D_{\lambda }^{a+1}},\cdots \quad \int \frac{%
\mathrm{d}^{2n}k}{\left( 2\pi \right) ^{2n}}\frac{k_{\mu _{1}}k_{\mu
_{2}}\cdots k_{\mu _{2b-1}}k_{\mu _{2b}}}{D_{\lambda }^{a+b}},
\label{IntsDiv}
\end{equation}%
with $n\geq a$. Since they have the same power counting, combining them into
surface terms is always possible%
\begin{equation}
-\frac{\partial }{\partial k^{\mu _{1}}}\frac{k_{\mu _{2}}\cdots k_{\mu
_{2n}}}{D_{\lambda }^{a}}=2a\frac{k_{\mu _{1}}k_{\mu _{2}}\cdots k_{\mu
_{2n}}}{D_{\lambda }^{a+1}}-g_{\mu _{1}\mu _{2}}\frac{k^{\mu _{3}}\cdots
k^{\mu _{2n}}}{D_{\lambda }^{a}}-\text{permutations.}
\end{equation}

Observing the equation above, note that a surface term combines into
lower-order surface terms. That produces a chain of associations, leading to
scalar integrals that encode the divergent content of the original
expression. They preserve the possibility or not of shifting the integration
variable, which means we are trading the freedom of the operation of
translation in the momentum space for the arbitrary choice of the routings
in these perturbative corrections. These surface terms are always present
for linear and higher divergent or logarithmic-divergent tensor integrals.
Although their coefficients depend on ambiguous momenta (\ref{Pij}) in the
first case, only external momenta (\ref{pij}) appear in the second.

We define combinations that arise for this investigation for the abelian
chiral anomalies as follows%
\begin{equation}
\Delta _{\left( n+1\right) ;\mu _{1}\mu _{2}}^{\left( 2n\right) }(\lambda
^{2})=\int \frac{\mathrm{d}^{2n}k}{\left( 2\pi \right) ^{2n}}\left( \frac{%
2nk_{\mu _{1}}k_{\mu _{2}}}{D_{\lambda }^{n+1}}-g_{\mu _{1}\mu _{2}}\frac{1}{%
D_{\lambda }^{n}}\right) =-\int \frac{\mathrm{d}^{2n}k}{\left( 2\pi \right)
^{2n}}\frac{\partial }{\partial k^{\mu _{1}}}\frac{k_{\mu _{2}}}{D_{\lambda
}^{n}},  \label{delta2n}
\end{equation}%
where the superscript $n=1,2$ indicates respectively two and four
dimensions. The corresponding irreducible scalar comes from the definition%
\begin{equation}
I_{\log }^{\left( 2n\right) }\left( \lambda ^{2}\right) =\int \frac{\mathrm{d%
}^{2n}k}{\left( 2\pi \right) ^{2n}}\frac{1}{D_{\lambda }^{n}}.  \label{Ilog}
\end{equation}

The separation highlights diverging structures and organizes them without
performing any analytic operation. Moreover, it makes evident that the
divergent content is a local polynomial in the ambiguous and physical
momenta obtained without expansions or limits.

For the gravitational case, the integrals show superior power counting; the
iterative use of this systematization from the first tensor term allows to
recombine of all the tensor integrals in terms of surface plus scalar
integrals, whose coefficients are symmetrical combinations of the metric
tensor,%
\begin{equation}
\Delta _{2\mu _{12}}^{\left( 2\right) }=\int \frac{\mathrm{d}^{2}k}{\left(
2\pi \right) ^{2}}\left[ \frac{2k_{\mu _{12}}}{D_{\lambda }^{2}}-g_{\mu
_{12}}\frac{1}{D_{\lambda }}\right] =-\int \frac{\mathrm{d}^{2}k}{\left(
2\pi \right) ^{2}}\frac{\partial }{\partial k^{\mu _{1}}}\frac{k_{\mu _{2}}}{%
D_{\lambda }}.  \label{delta2}
\end{equation}%
The 4th-rank surface term%
\begin{equation}
\square _{3\mu _{1234}}^{\left( 2\right) }=\int \frac{\mathrm{d}^{2}k}{%
\left( 2\pi \right) ^{2}}\left[ \frac{8k_{\mu _{1234}}}{D_{\lambda }^{3}}-%
\frac{g_{(\mu _{12}}k_{\mu _{12})}}{D_{\lambda }^{2}}\right] =-\frac{1}{2}%
\sum_{i=1}^{4}\int \frac{\mathrm{d}^{2}k}{\left( 2\pi \right) ^{2}}\frac{%
\partial }{\partial k^{\mu _{i}}}\frac{k_{\mu _{1}\cdots \hat{\mu}_{i}\cdots
\mu _{4}}}{D_{\lambda }^{2}}
\end{equation}%
and the longest one, the 6th-rank surface term%
\begin{equation}
\Sigma _{4\mu _{123456}}^{\left( 2\right) }=\int \frac{\mathrm{d}^{2}k}{%
\left( 2\pi \right) ^{2}}\left[ \frac{48k_{\mu _{123456}}}{D_{\lambda }^{4}}-%
\frac{8}{3}\frac{g_{(\mu _{12}}k_{\mu _{1234})}}{D_{\lambda }^{3}}\right] =-%
\frac{4}{3}\sum_{i=1}^{6}\int \frac{\mathrm{d}^{2}k}{\left( 2\pi \right) ^{2}%
}\frac{\partial }{\partial k^{\mu _{i}}}\frac{k_{\mu _{1}\cdots \hat{\mu}%
_{i}\cdots \mu _{6}}}{D_{\lambda }^{3}}.
\end{equation}

For the symmetrization of indices, we use%
\begin{equation}
A_{(\alpha _{1}\cdots \alpha _{r}}B_{\alpha _{r+1}\cdots \alpha
_{s})}=\sum_{\pi \in S_{s}^{non}}A_{\alpha _{\pi \left( 1\right) }\cdots
\alpha _{\pi \left( r\right) }}B_{\alpha _{\pi \left( r+1\right) }\cdots
\alpha _{\pi \left( s\right) }}.
\end{equation}%
In our notation, $S_{s}^{non}\subset S_{s}$ is a subgroup of the permutation
group of $s$ elements that does not count terms that are already symmetric.
It means the total sum has all terms that make the tensor completely
antisymmetric without repetition of terms with a coefficient equal to the
unit. We are using the convention of condensing the indices $k_{\mu
_{1}}\cdots k_{\mu _{n}}=k_{\mu _{1}\cdots _{n}}$ and the same for vector $k$%
. These surface terms, therefore, have the character of being explicitly
completely symmetric, a handy property in computations. Beyond the
logarithmic objects defined above also appear quadratically divergent
integrals organized in the objects:%
\begin{eqnarray}
\Delta _{1\mu _{12}}^{\left( 2\right) } &=&\int \frac{\mathrm{d}^{2}k}{%
\left( 2\pi \right) ^{2}}\left[ \frac{2k_{\mu _{12}}}{D_{\lambda }}-g_{\mu
_{1}\mu _{2}}\log \frac{(k^{2}-m^{2})}{k^{2}}\right]  \label{delta1} \\
\square _{2\mu _{1234}}^{\left( 2\right) } &=&\int \frac{\mathrm{d}^{2}k}{%
\left( 2\pi \right) ^{2}}\left[ \frac{4k_{\mu _{1234}}}{D_{\lambda }^{2}}-%
\frac{g_{(\mu _{12}}k_{\mu _{34})}}{D_{\lambda }}\right] .
\end{eqnarray}%
And the quadratic scalar%
\begin{equation}
I_{\mathrm{quad}}^{\left( 2\right) }=\int \frac{\mathrm{d}^{2}k}{\left( 2\pi
\right) ^{2}}\log \frac{\left( k^{2}-m^{2}\right) }{k^{2}}.  \label{Iquad}
\end{equation}

\textbf{Important note}: the complete symmetrization of the indices that
appear as the product of the metrics can cause the expressions for the
surface terms to have dozens of terms. For the sake of clarity, let us
define the combinations,%
\begin{eqnarray}
W_{4\mu _{123456}} &=&\Sigma _{4\mu _{123456}}^{\left( 2\right) }+\frac{1}{3}%
g_{(\mu _{12}}\square _{3\mu _{3456})}^{\left( 2\right) }+\frac{1}{3}g_{(\mu
_{12}}g_{\mu _{34}}\Delta _{2\mu _{56})}^{\left( 2\right) }  \label{W4} \\
W_{3\mu _{1234}} &=&\square _{3\mu _{1234}}^{\left( 2\right) }+\frac{1}{2}%
g_{(\mu _{12}}\Delta _{2\mu _{34})}^{\left( 2\right) }  \label{W3} \\
W_{2\mu _{1234}} &=&\square _{2\mu _{1234}}^{\left( 2\right) }+\frac{1}{2}%
g_{(\mu _{12}}\Delta _{1\mu _{34})}^{\left( 2\right) }.
\end{eqnarray}%
The first row has sixty-one terms, while the second and third rows have
seven terms. They allow us to write the integrals often present in the
separation of divergent terms as%
\begin{eqnarray}
\int \frac{\mathrm{d}^{2}k}{\left( 2\pi \right) ^{2}}\frac{48k_{\mu
_{123456}}}{D_{\lambda }^{4}} &=&W_{4\mu _{123456}}+g_{(\mu _{12}}g_{\mu
_{34}}g_{\mu _{56})}I_{\text{\textrm{log}}}^{\left( 2\right) }
\label{IntDiv} \\
\int \frac{\mathrm{d}^{2}k}{\left( 2\pi \right) ^{2}}\frac{8k_{\mu _{1234}}}{%
D_{\lambda }^{3}} &=&W_{3\mu _{1234}}+g_{(\mu _{12}}g_{\mu _{34})}I_{\text{%
\textrm{log}}}^{\left( 2\right) }  \notag \\
\int \frac{\mathrm{d}^{2}k}{\left( 2\pi \right) ^{2}}\frac{4k_{\mu _{1234}}}{%
D_{\lambda }^{2}} &=&W_{2\mu _{1234}}+g_{(\mu _{12}}g_{\mu _{34})}I_{\text{%
\textrm{quad}}}^{\left( 2\right) }  \notag \\
\int \frac{\mathrm{d}^{2}k}{\left( 2\pi \right) ^{2}}\frac{2k_{\mu _{12}}}{%
D_{\lambda }^{2}} &=&\Delta _{2\mu _{12}}^{\left( 2\right) }+g_{\mu _{12}}I_{%
\text{\textrm{log}}}^{\left( 2\right) }  \notag \\
\int \frac{\mathrm{d}^{2}k}{\left( 2\pi \right) ^{2}}\frac{2k_{\mu _{12}}}{%
D_{\lambda }} &=&\Delta _{1\mu _{12}}^{\left( 2\right) }+g_{\mu _{12}}I_{%
\text{\textrm{quad}}}^{\left( 2\right) }.  \notag
\end{eqnarray}

For the trace of $W_{4\mu _{123456}}$ and $W_{3\mu _{1234}}$, we begin with 
\begin{equation}
W_{4\rho \mu _{1234}}^{\rho }=\Sigma _{4\rho \mu _{1234}}^{\rho }+\frac{10}{3%
}\square _{3\mu _{1234}}+\frac{1}{3}g_{(\mu _{12}}\square _{3\mu _{34})\rho
}^{\rho }+\frac{8}{3}g_{(\mu _{12}}\Delta _{2\mu _{34})}+\frac{1}{3}g_{(\mu
_{12}}g_{\mu _{34})}\Delta _{2\rho }^{\rho }
\end{equation}%
\begin{equation}
W_{3\rho \mu _{12}}^{\rho }=\square _{3\rho \mu _{12}}^{\rho }+3\Delta
_{2\mu _{12}}+\frac{1}{2}g_{\mu _{12}}\Delta _{2\rho }^{\rho }.
\end{equation}%
They arise from a simple combinatorial analysis: For $g_{(\mu _{12}}\square
_{3\mu _{3456})}$ there are fifteen terms where only in one of the indices $%
\mu _{56}$ appears in the metric and six terms where both indices appear in $%
\square _{3\mu _{3456}}$, the remaining ones have the indices $\mu _{5}$ or $%
\mu _{6}$ in the metric and the other in the surface term. In the first and
last set of permutations, we get a factor of ten for $\square _{3\mu
_{1234}},$ and the other six generate a complete symmetric combination of
the trace and metric, namely%
\begin{equation}
g^{\mu _{56}}g_{(\mu _{12}}\square _{3\mu _{3456})}=10\square _{3\mu
_{1234}}+g_{(\mu _{12}}\square _{3\mu _{34})\rho }^{\rho }.
\end{equation}%
As for the term $g_{(\mu _{12}}g_{\mu _{34}}\Delta _{2\mu _{56})}$, they are
forty-five terms, in eighteen of them the $\mu _{56}$ indices are in the
metric and twenty-four the metric and the surface term share them. These
terms generate a factor of eight multiplied by the symmetric combinations of 
$g_{(\mu _{12}}\Delta _{2\mu _{34})}$, the remaining three yield the total
result 
\begin{equation}
g^{\mu _{56}}g_{(\mu _{12}}g_{\mu _{34}}\Delta _{2\mu _{56})}=8g_{(\mu
_{12}}\Delta _{2\mu _{34})}+g_{(\mu _{12}}g_{\mu _{34})}\Delta _{2\rho
}^{\rho },
\end{equation}%
where $\Delta _{2\rho }^{\rho }$ is the trace of the divergent object.

As a last observation, two essential combinations appear in the verification
process of RAGF, resulting from traces with the metric. It is possible to
immediately express the features of $W$-tensors defined above in the
following ways%
\begin{equation}
2W_{3\rho \mu _{12}}^{\rho }-8\Delta _{2\mu _{12}}^{\left( 2\right)
}=[2(\square _{3\rho \mu _{12}}^{\left( 2\right) \rho }-\Delta _{2\mu
_{12}}^{\left( 2\right) })-g_{\mu _{12}}\Delta _{2\rho }^{\left( 2\right)
\rho }]+2g_{\mu _{12}}\Delta _{2\rho }^{\left( 2\right) \rho }  \label{trW3}
\end{equation}%
\begin{eqnarray}
3W_{4\rho \mu _{1234}}^{\rho }-18W_{3\mu _{1234}} &=&[3\Sigma _{4\rho \mu
_{1234}}^{\left( 2\right) \rho }-8\square _{3\mu _{1234}}^{\left( 2\right)
}-g_{(\mu _{12}}g_{\mu _{34})}\Delta _{2\rho }^{\left( 2\right) \rho }]
\label{trW4} \\
&&+g_{(\mu _{12}}[\square _{3\rho \mu _{34})}^{\left( 2\right) \rho }-\Delta
_{2\mu _{34})}^{\left( 2\right) }-\frac{1}{2}g_{\mu _{34})}\Delta _{2\rho
}^{\left( 2\right) \rho }]+3g_{(\mu _{12}}g_{\mu _{34})}\Delta _{2\rho
}^{\left( 2\right) \rho }.  \notag
\end{eqnarray}%
Its determination follows from the combinatorial analysis of the terms
symmetrized in their definitions. The term $g_{(\mu _{2}\alpha _{2}}g_{\nu
_{12})}\Delta _{2\rho }^{\rho }$ inside the parentheses is equal to $%
2g_{(\mu _{2}\alpha _{2}}g_{\nu _{12})}\Delta _{2\rho }^{\rho }$ due to
metric degeneracy. The term%
\begin{equation}
g_{(\mu _{12}}[\square _{3\rho \mu _{34})}^{\left( 2\right) \rho }-\Delta
_{2\mu _{34})}^{\left( 2\right) }-\frac{1}{2}g_{\mu _{34})}\Delta _{2\rho
}^{\left( 2\right) \rho }]
\end{equation}%
represents the six permutations for it to be completely symmetric. When one
splits it into three terms, the last one is symmetric with just three terms
of the type $g_{\mu _{12}}g_{\mu _{34}}$. Hence we get a factor of one
instead of a half, which is identical to the combination we have begun. This
arrangement makes the expression similar to the one shown for the trace of $%
W_{3}$.

These relations were exposed here because the expansion on the basic surface
terms becomes excessively long and unnecessary. The surface terms in the
leading integrals (highest rank-tensor) do not need expansion. The RAGF
conditions of satisfaction only require these terms to be ranked by their
indices and the number of contractions, as we will see in the Chapter on
gravitational two-point functions.

\section{Finite Functions\label{FinFcts}}

\subsection{Two Dimensions}

After separating the finite part, we solve the integrals through techniques
of perturbative calculations and project their results into a family of
functions. Two-point basic functions assume the form%
\begin{eqnarray}
Z_{n_{1}}^{\left( -1\right) } &=&\int_{0}^{1}\mathrm{d}x\frac{x^{n_{1}}}{Q};
\label{Zn(-1)} \\
Z_{n_{1}}^{\left( 0\right) } &=&\int_{0}^{1}\mathrm{d}xx^{n_{1}}\log \frac{Q%
}{-\lambda ^{2}},  \label{Zn(0)}
\end{eqnarray}%
with $n_{i}\in \mathbb{N}$, and the $Q$ is a polynomial given by 
\begin{equation}
Q\left( q^{2},m_{2},m_{1}\right) =q^{2}x\left( 1-x\right) +\left(
m_{1}^{2}-m_{2}^{2}\right) x-m_{1}^{2}.
\end{equation}%
An important point that will be explored is $q^{2}=0$ for equal masses $%
m_{1}=m_{2}$, where%
\begin{equation}
Z_{n_{1}}^{\left( -1\right) }\left( 0\right) =-\frac{1}{m^{2}\left(
n_{1}+1\right) };\text{ }\quad Z_{n_{1}}^{\left( 0\right) }\left( 0\right)
=0.  \label{LetZ2D}
\end{equation}%
And the combination between $Z_{1}^{\left( -1\right) }$ and $Z_{0}^{\left(
-1\right) }$ given by 
\begin{equation}
\left[ \left( m_{1}^{2}-m_{2}^{2}\right) Z_{1}^{\left( -1\right)
}-m_{1}^{2}Z_{0}^{\left( -1\right) }\right] _{q^{2}=0}=\int_{0}^{1}\mathrm{d}%
x\frac{\left( m_{1}^{2}-m_{2}^{2}\right) x-m_{1}^{2}}{Q\left(
0,m_{2},m_{1}\right) }=1;  \label{LetZ2M}
\end{equation}%
It has a nice limit that will appear in investigating the $AV$ of different
masses.

\textbf{Reductions:} $Z_{n_{1}}^{\left( k\right) }$ in both parameters and
the ones required for this work are%
\begin{eqnarray}
Z_{0}^{\left( 0\right) } &=&\log \frac{m_{2}^{2}}{\lambda ^{2}}%
+2q^{2}Z_{2}^{\left( -1\right) }-\left( q^{2}+m_{1}^{2}-m_{2}^{2}\right)
Z_{1}^{\left( -1\right) } \\
2q^{2}Z_{1}^{\left( -1\right) } &=&\left( q^{2}+m_{1}^{2}-m_{2}^{2}\right)
Z_{0}^{\left( -1\right) }+\log \frac{m_{1}^{2}}{m_{2}^{2}}  \label{Z1Z0} \\
q^{2}Z_{n_{1}+2}^{\left( -1\right) } &=&\left(
q^{2}+m_{1}^{2}-m_{2}^{2}\right) Z_{n_{1}+1}^{\left( -1\right)
}-m_{1}^{2}Z_{n_{1}}^{\left( -1\right) }-\frac{1}{\left( n_{1}+1\right) },
\end{eqnarray}%
with $n_{1}\geq 0$. In the gravitational setting (where only equal masses
integrals will are explored), we have the function%
\begin{equation}
Z_{0}^{\left( 1\right) }=\int_{0}^{1}\mathrm{d}xQ\log \frac{Q}{-\lambda ^{2}}%
.
\end{equation}%
Adopting $m_{1}=m_{2}=\lambda $, the reductions needed for that scenario are%
\begin{equation}
Z_{0}^{\left( 1\right) }-m^{2}=2q^{2}Z_{2}^{\left( 0\right)
}-q^{2}Z_{1}^{\left( 0\right) }
\end{equation}%
\begin{equation}
2Z_{1}^{\left( 0\right) }=Z_{0}^{\left( 0\right) }
\end{equation}%
\begin{equation}
(n_{1}+3)q^{2}Z_{n_{1}+2}^{(0)}=(n_{1}+2)q^{2}Z_{n_{1}+1}^{(0)}-(n_{1}+1)m^{2}Z_{n_{1}}^{(0)}-%
\frac{(n_{1}+1)}{(n_{1}+2)(n_{1}+3)}q^{2},
\end{equation}%
with $n_{1}\geq 0$.

\subsection{Four Dimensions\label{FinFcts4d}}

For the three-point amplitudes\footnote{%
These polynomials can be written in terms of Symanzik polynomials
constructed using the spanning trees and two-forests of the graph.}, we have
the polynomial 
\begin{eqnarray}
Q\left( p,q,m_{2},m_{3},m_{1}\right) &=&p^{2}x_{1}\left( 1-x_{1}\right)
+q^{2}x_{2}\left( 1-x_{2}\right) -2\left( p\cdot q\right) x_{1}x_{2} \\
&&+\left( m_{1}^{2}-m_{2}^{2}\right) x_{1}+\left( m_{1}^{2}-m_{3}^{2}\right)
x_{2}-m_{1}^{2}.  \notag
\end{eqnarray}%
And the corresponding basic functions,%
\begin{eqnarray}
Z_{n_{1}n_{2}}^{\left( -1\right) } &=&\int_{0}^{1}\mathrm{d}%
x_{1}\int_{0}^{1-x_{1}}\mathrm{d}x_{2}\frac{x_{1}^{n_{1}}x_{2}^{n_{2}}}{Q}
\label{Znm(-1)} \\
Z_{n_{1}n_{2}}^{\left( 0\right) } &=&\int_{0}^{1}\mathrm{d}%
x_{1}\int_{0}^{1-x_{1}}\mathrm{d}x_{2}x_{1}^{n_{1}}x_{2}^{n_{2}}\log \frac{Q%
}{-\lambda ^{2}}.
\end{eqnarray}%
At the point where all bilinears are zero, and for equal masses $%
m_{1}=m_{2}=m_{3}$, they satisfy%
\begin{equation}
Z_{n_{1}n_{2}}^{\left( -1\right) }\left( 0\right) =-\frac{n_{1}!n_{2}!}{m^{2}%
\left[ \left( n_{1}+n_{2}+2\right) !\right] };\quad Z_{n_{1}n_{2}}^{\left(
0\right) }\left( 0\right) =0.  \label{LetZ4D}
\end{equation}

Writing the parameters in terms of derivatives of the polynomials and using
partial integration follows relations among these functions. More precisely,
they are reductions of involved parameter powers $n_{1}+n_{2}$ for equation (%
\ref{Znm(-1)}). They were approached in the
papers \cite{Battistel2006}\cite{Dallabona2012}\cite{SunYi2012}. This
resource is necessary for the operations performed throughout this
investigation.

Let us start by making the derivative of the $Q$ polynomial for equal masses
concerning the parameter $x_{i}$ and multiplying by $1/Q$; we construct the
result%
\begin{eqnarray}
x_{1}^{n_{1}}x_{2}^{n_{2}}\frac{\partial }{\partial x_{1}}\log Q &=&-2\left[
p^{2}x_{1}^{n_{1}+1}x_{2}^{n_{2}}+\left( p\cdot q\right)
x_{1}^{n_{1}}x_{2}^{n_{2}+1}\right] \frac{1}{Q}+p^{2}\frac{%
x_{1}^{n_{1}}x_{2}^{n_{2}}}{Q} \\
x_{1}^{n_{1}}x_{2}^{n_{2}}\frac{\partial }{\partial x_{2}}\log Q &=&-2\left[
q^{2}x_{1}^{n_{1}}x_{2}^{n_{2}+1}+\left( p\cdot q\right)
x_{1}^{n_{1}+1}x_{2}^{n_{2}}\right] \frac{1}{Q}+q^{2}\frac{%
x_{1}^{n_{1}}x_{2}^{n_{2}}}{Q}.
\end{eqnarray}%
When integrating $\int_{0}^{1-x_{1}}\mathrm{d}x_{2}$, in some cases, we need
to commute the integral and a derivative. The upper limit of the integral is
not a constant; in that situation, we applied it to the Leibnitz formula 
\begin{equation}
\int_{a\left( x\right) }^{b\left( x\right) }\mathrm{d}z\frac{\partial }{%
\partial x}F\left( x,z\right) =\frac{\partial }{\partial x}\int_{a\left(
x\right) }^{b\left( x\right) }\mathrm{d}zF\left( x,z\right) -\left[ F\left(
x,b\left( x\right) \right) \frac{\partial b\left( x\right) }{\partial x}%
-F\left( x,a\left( x\right) \right) \frac{\partial a\left( x\right) }{%
\partial x}\right] .
\end{equation}%
For our purposes $b^{\prime }\left( x\right) =-1$ and $a^{\prime }\left(
x\right) =0$, hence%
\begin{equation}
\int_{0}^{b\left( x\right) }\mathrm{d}z\frac{\partial }{\partial x}F\left(
x,z\right) =\frac{\partial }{\partial x}\int_{0}^{b\left( x\right) }\mathrm{d%
}zF\left( x,z\right) +F\left( x,b\left( x\right) \right) .
\end{equation}%
The limits of integration will bring a binomial expansion as well%
\begin{equation}
\left( 1-x_{1}\right) ^{n_{2}}=\sum_{s=0}^{n_{2}}\left( -1\right) ^{s}\binom{%
n_{2}}{s}x_{1}^{s}.
\end{equation}

Through the application of these elements, it is derived the formulae%
\begin{eqnarray}
2[p^{2}Z_{n_{1}+1;n_{2}}^{\left( -1\right) }+\left( p\cdot q\right)
Z_{n_{1};n_{2}+1}^{\left( -1\right) }] &=&p^{2}Z_{n_{1};n_{2}}^{\left(
-1\right) }+\left( 1-\delta _{n_{1}0}\right) n_{1}Z_{n_{1}-1,n_{2}}^{\left(
0\right) } \\
&&+\delta _{n_{1}0}Z_{n_{2}}^{\left( 0\right) }\left( q\right)
-\sum_{s=0}^{n_{2}}\left( -1\right) ^{s}\binom{n_{2}}{s}Z_{n_{1}+s}^{\left(
0\right) }\left( q-p\right)  \notag
\end{eqnarray}%
\begin{eqnarray}
2[q^{2}Z_{n_{1};n_{2}+1}^{\left( -1\right) }+\left( p\cdot q\right)
Z_{n_{1}+1;n_{2}}^{\left( -1\right) }] &=&q^{2}Z_{n_{1};n_{2}}^{\left(
-1\right) }+\left( 1-\delta _{n_{2}0}\right) n_{2}Z_{n_{1};n_{2}-1}^{\left(
0\right) } \\
&&+\delta _{n_{2}0}Z_{n_{1}}^{\left( 0\right) }\left( p\right)
-\sum_{s=0}^{n_{2}}\left( -1\right) ^{s}\binom{n_{2}}{s}Z_{n_{1}+s}^{\left(
0\right) }\left( q-p\right) .  \notag
\end{eqnarray}%
They represent a reduction in $n_{i}$ from a situation of $%
n_{1}+n_{2}+1\rightarrow n_{1}+n_{2}$ appearing in the RAGFs and WI
verifications. It is also necessary to use another reduction 
\begin{equation}
2Z_{00}^{\left( 0\right) }=\left[ p^{2}Z_{10}^{\left( -1\right)
}+q^{2}Z_{01}^{\left( -1\right) }\right] -2m^{2}Z_{00}^{\left( -1\right)
}-1+2Z_{1}^{\left( 0\right) }\left( q-p\right) .
\end{equation}%
That comes from the previous ones and the use of 
\begin{equation}
\frac{1}{2}=-p^{2}Z_{20}^{\left( -1\right) }-q^{2}Z_{02}^{\left( -1\right)
}-p^{2}Z_{10}^{\left( -1\right) }-q^{2}Z_{01}^{\left( -1\right) }-2\left(
p\cdot q\right) Z_{11}^{\left( -1\right) }-m^{2}Z_{00}^{\left( -1\right) }
\end{equation}%
from integrating the identity $\frac{Q}{Q}=1$. This set of mathematical
results is enough to develop any computation concerning the finite parts in
this thesis.

\section{Basis of Feynman Integrals\label{BasisFI}}

At the end of Section (\ref{ModlDef}), we introduced a set of $\left(
n+1\right) $-point amplitudes in $2n$ dimensions. In the same context,
equation (\ref{JInt}) presented a general definition for integrals that
appear after taking Dirac traces. We describe in a nutshell those that arise
within the amplitudes. At two dimensions, the needed integrals are defined by%
\begin{equation}
\left[ \bar{J}_{1}^{\left( 2\right) }\left( k_{i}\right) ;\bar{J}%
_{1}^{\left( 2\right) \mu _{1}}\left( k_{i}\right) ;\bar{J}_{1}^{\left(
2\right) \mu _{12}}\left( k_{i}\right) ;\bar{J}_{1}^{\left( 2\right) \mu
_{123}}\left( k_{i}\right) \right] =\int \frac{\mathrm{d}^{2}k}{\left( 2\pi
\right) ^{2}}\frac{(1;\ K_{i}^{\mu _{1}};\ K_{iii}^{\mu _{12}};\
K_{iii}^{\mu _{123}})}{D_{i}}  \label{J1(ki)}
\end{equation}%
\begin{equation}
\left[ \bar{J}_{2}^{\left( 2\right) };\bar{J}_{2}^{\left( 2\right) \mu _{1}};%
\bar{J}_{2}^{\left( 2\right) \mu _{12}};\bar{J}_{2}^{\left( 2\right) \mu
_{123}},\bar{J}_{2}^{\left( 2\right) \mu _{1234}}\right] =\int \frac{\mathrm{%
d}^{2}k}{\left( 2\pi \right) ^{2}}\frac{(1;\ K_{1}^{\mu _{1}};\ K_{111}^{\mu
_{12}};\ K_{111}^{\mu _{123}};\ K_{1111}^{\mu _{1234}})}{D_{12}}.  \label{J2}
\end{equation}%
And at four dimensions, we define the functions with two and three
propagators%
\begin{eqnarray}
\left[ \bar{J}_{2}^{\left( 4\right) };\bar{J}_{2}^{\left( 4\right) \mu _{1}}%
\right] &=&\int \frac{\mathrm{d}^{4}k}{\left( 2\pi \right) ^{4}}\frac{(1;\
K_{i}^{\mu _{1}})}{D_{ij}}, \\
\left[ \bar{J}_{3}^{\left( 4\right) };\bar{J}_{3}^{\left( 4\right) \mu _{1}};%
\bar{J}_{3}^{\left( 4\right) \mu _{12}}\right] &=&\int \frac{\mathrm{d}^{4}k%
}{\left( 2\pi \right) ^{4}}\frac{(1;\ K_{1}^{\mu _{1}};\ K_{1}^{\mu
_{1}}K_{1}^{\mu _{2}})}{D_{123}}.
\end{eqnarray}%
We use the conventions $D_{12\cdots i}=D_{1}D_{2}\cdots D_{i},$ and $%
K_{i}=k+k_{i}$ with $K_{a_{1}\cdots a_{n}}^{\nu _{a_{1}}\cdots \nu
_{a_{n}}}=K_{a_{1}}^{\nu _{a_{1}}}\cdots K_{a_{n}}^{\nu _{a_{n}}}$, where $%
a_{i}\in \left\{ 1,\cdots ,n\right\} $. For the case of integrals with fewer
propagators of each dimension, it is necessary to specify the momenta.

\subsection{Two Dimensions}

The power counting of $n$-point integrals associated with the chiral anomaly
from odd amplitudes in two dimensions are%
\begin{equation}
\left\{ 
\begin{array}{c}
\omega (J_{2}^{\left( 2\right) })=-2 \\ 
\omega (J_{2}^{\left( 2\right) \mu _{1}})=-1 \\ 
\omega (J_{2}^{\left( 2\right) \mu _{12}})=0%
\end{array}%
\right. ;\quad \left\{ 
\begin{array}{c}
\omega (J_{1}^{\left( 2\right) })=0 \\ 
\omega (J_{1}^{\left( 2\right) \mu _{1}})=1%
\end{array}%
\right. ;  \label{w(2D)}
\end{equation}%
The power counting for integrals associated with derivative coupling for $n$%
-point integrals%
\begin{equation}
\left\{ 
\begin{array}{c}
\omega (J_{2}^{\left( 2\right) \mu _{123}})=1 \\ 
\omega (J_{2}^{\left( 2\right) \mu _{1234}})=2%
\end{array}%
\right. ;\quad \left\{ 
\begin{array}{c}
\omega (J_{1}^{\left( 2\right) \mu _{12}})=2 \\ 
\omega (J_{1}^{\left( 2\right) \mu _{123}})=3%
\end{array}%
\right. ;
\end{equation}

Some integrals contain finite and divergent parts, so we adopt the overbar
to indicate such a feature. For instance, in $2n$ dimensions, the integral $%
\bar{J}_{n}^{\left( 2n\right) }$ contains a diverging object and finite
contributions labeled as $J_{n}^{\left( 2n\right) }$. The presence of the
overbar distinguishes the complete integral from its finite content. That
also means they coincide for strictly finite integrals, namely $\bar{J}%
_{n+1}^{\left( 2n\right) \mu _{1}}=J_{n+1}^{\left( 2n\right) \mu _{1}}$ and $%
\bar{J}_{n+1}^{\left( 2n\right) }=J_{n+1}^{\left( 2n\right) }$.

The one-point integrals in (\ref{w(2D)}), are obtained using the identity (%
\ref{id}) with $N=1$ 
\begin{equation}
\frac{1}{D_{i}}=\frac{1}{D_{\lambda }}-\frac{A_{i}}{D_{\lambda }^{2}}+\frac{%
A_{i}^{2}}{D_{\lambda }^{2}D_{i}}.  \label{N=1}
\end{equation}%
When Integrating the finite parts and identifying the divergent objects as (%
\ref{Ilog}) and (\ref{delta2})%
\begin{eqnarray}
\bar{J}_{1}^{\left( 2\right) }\left( k_{i}\right) &=&I_{\log }^{\left(
2\right) }\left( \lambda ^{2}\right) -\frac{i}{4\pi }\log \frac{m_{i}^{2}}{%
\lambda ^{2}}  \label{2dJ1} \\
\bar{J}_{1\mu _{1}}^{\left( 2\right) }\left( k_{i}\right) &=&-k_{i}^{\nu
_{1}}\Delta _{2\nu _{1}\mu _{1}}^{\left( 2\right) }\left( \lambda
^{2}\right) .
\end{eqnarray}%
The two integrals show logarithmic divergence. The last one corresponds to a
pure surface term. The argument of $I_{\log }\left( \lambda ^{2}\right) $
object may be transformed by 
\begin{equation}
\frac{1}{\left( k^{2}-\lambda ^{2}\right) }=\frac{1}{\left(
k^{2}-m_{i}^{2}\right) }-\frac{\left( m_{i}^{2}-\lambda ^{2}\right) }{\left(
k^{2}-\lambda ^{2}\right) \left( k^{2}-m_{i}^{2}\right) }.
\end{equation}%
This identification implies a scale relation between the divergent and
finite part%
\begin{equation}
I_{\log }^{\left( 2\right) }\left( \lambda ^{2}\right) =I_{\log }^{\left(
2\right) }\left( m_{i}^{2}\right) +\frac{i}{\left( 4\pi \right) }\log \frac{%
m_{i}^{2}}{\lambda ^{2}}.  \label{scRel}
\end{equation}%
The scalar one be written as $\bar{J}_{1}^{\left( 2\right) }\left(
k_{i}\right) =I_{\log }^{\left( 2\right) }\left( m_{i}^{2}\right) $. For
more details, see Appendix (\ref{AppInt2D}).

For the two-point integrals with the power counting given by (\ref{w(2D)}),
we have%
\begin{eqnarray}
J_{2}^{\left( 2\right) } &=&\frac{i}{4\pi }[Z_{0}^{\left( -1\right) }\left(
q,m_{2},m_{1}\right) ]  \label{2DJ2} \\
J_{2}^{\left( 2\right) \mu _{1}} &=&\frac{i}{4\pi }[-q^{\mu
_{1}}Z_{1}^{\left( -1\right) }]  \label{2DJ2mu1} \\
J_{2}^{\left( 2\right) \mu _{1}\mu _{2}} &=&\frac{i}{4\pi }\left[ -\frac{1}{2%
}g^{\mu _{1}\mu _{2}}Z_{0}^{\left( 0\right) }+q^{\mu _{1}}q^{\mu
_{2}}Z_{2}^{\left( -1\right) }\right] \\
\bar{J}_{2}^{\left( 2\right) \mu _{1}\mu _{2}} &=&J_{2}^{\left( 2\right) \mu
_{1}\mu _{2}}+\frac{1}{2}\left[ \Delta _{2}^{\left( 2\right) \mu _{1}\mu
_{2}}+g^{\mu _{1}\mu _{2}}I_{\log }^{\left( 2\right) }\right] .
\label{2DJ2C}
\end{eqnarray}%
Arguments were omitted since they are the same for all integrals. The
two-point divergent integral is obtained by applying the identity (\ref{id})
with $N=0$; its complete calculation is performed in the Appendix (\ref%
{AppCJ22}).

\textbf{Reductions }$\mathbf{2D}$\textbf{: }For Chapters (\ref{2Dim2Pt}) and
(\ref{2masses}), we will need the reductions listed above%
\begin{eqnarray}
2q_{\mu _{1}}J_{2}^{\left( 2\right) \mu _{12}} &=&-\left(
q^{2}+m_{1}^{2}-m_{2}^{2}\right) J_{2}^{\left( 2\right) \mu _{2}}-\frac{i}{%
4\pi }q^{\mu _{2}}\log \left( m_{2}^{2}/\lambda ^{2}\right)  \label{qJ22} \\
g_{\mu _{12}}J_{2}^{\left( 2\right) \mu _{12}} &=&\left( \frac{i}{4\pi }%
+m_{1}^{2}J_{2}^{\left( 2\right) }\right) -\frac{i}{4\pi }\log \left(
m_{2}^{2}/\lambda ^{2}\right)  \label{gJ2} \\
2q_{\mu _{1}}J_{2}^{\left( 2\right) \mu _{1}} &=&-\left(
q^{2}+m_{1}^{2}-m_{2}^{2}\right) J_{2}^{\left( 2\right) }+\frac{i}{4\pi }%
\log \left( m_{2}^{2}/m_{1}^{2}\right)  \label{qJ21} \\
q^{2}(2J_{2\mu _{2}}^{\left( 2\right) }+q_{\mu _{2}}J_{2}^{\left( 2\right)
}) &=&-q_{\mu _{2}}\left( m_{1}^{2}-m_{2}^{2}\right) J_{2}^{\left( 2\right)
}-\frac{i}{4\pi }q_{\mu _{2}}\log \left( m_{1}^{2}/m_{2}^{2}\right) .
\label{J1J0}
\end{eqnarray}

In Chapter (\ref{modlDef}), in addition to the functions introduced above,
it is necessary to the single mass of 3rd-rank integral, obtained by
applying the identity with $N=1$:%
\begin{equation}
\bar{J}_{2\mu _{123}}^{\left( 2\right) }=J_{2\mu _{123}}^{\left( 2\right) }-%
\frac{1}{4}P^{\nu _{1}}W_{3\mu _{123}\nu _{1}}+\frac{1}{4}\left( P-q\right)
_{(\mu _{1}}\Delta _{2\mu _{23})}-\frac{1}{4}q_{(\mu _{1}}g_{\mu
_{23})}I_{\log }  \label{J2mu123B}
\end{equation}%
\begin{equation}
J_{2\mu _{123}}^{\left( 2\right) }=-\frac{i}{4\pi }\left[ -\frac{1}{2}%
q_{(\mu _{1}}g_{\mu _{23})}Z_{1}^{\left( 0\right) }+q_{\mu
_{123}}Z_{3}^{\left( -1\right) }\right] .
\end{equation}%
And 4th-rank integral, using $N=2$ in (\ref{id}):%
\begin{eqnarray}
\bar{J}_{2\mu _{12}\alpha _{12}}^{\left( 2\right) } &=&J_{2\mu _{12}\alpha
_{12}}^{\left( 2\right) }+\frac{1}{4}W_{2\mu _{12}\alpha _{12}}+\frac{1}{4}%
g_{(\mu _{12}\alpha _{12})}I_{\mathrm{quad}}  \label{J2mu1234} \\
&&-\frac{1}{24}\left[ q^{2}g_{(\mu _{12}\alpha _{12})}-4q_{(\mu
_{12}}g_{\alpha _{12})}\right] I_{\log }  \notag \\
&&+\frac{1}{48}\left( 3P^{\nu _{12}}+q^{\nu _{12}}\right) W_{4\mu
_{12}\alpha _{12}\nu _{12}}  \notag \\
&&-\frac{1}{16}\left( P^{2}+q^{2}\right) W_{3\mu _{12}\alpha _{12}}-\frac{1}{%
8}P^{\nu _{1}}\left( P-q\right) _{(\mu _{1}}W_{3\mu _{2}\alpha _{12})\nu
_{1}}  \notag \\
&&+\frac{1}{8}\left[ \left( P-q\right) _{\mu _{1}}\left( P-q\right)
_{(\alpha _{1}}\Delta _{2\alpha _{2})\mu _{2}}+\left( P-q\right) _{\mu
_{2}}\left( P-q\right) _{(\alpha _{1}}\Delta _{2\alpha _{2})\mu _{1}}\right]
\notag \\
&&+\frac{1}{8}\left( P-q\right) _{\alpha _{1}}\left( P-q\right) _{\alpha
_{2}}\Delta _{2\mu _{12}}+\frac{1}{8}\left( P-q\right) _{\mu _{1}}\left(
P-q\right) _{\mu _{2}}\Delta _{2\alpha _{12}}  \notag
\end{eqnarray}%
\begin{equation}
J_{2\mu _{12}\alpha _{12}}^{\left( 2\right) }=\frac{i}{4\pi }\left\{ \frac{1%
}{4}g_{(\mu _{12}}g_{\alpha _{12})}\left[ Z_{0}^{\left( 1\right) }-m^{2}%
\right] -\frac{1}{2}g_{(\mu _{12}}q_{\alpha _{12})}Z_{2}^{\left( 0\right)
}+q_{\mu _{12}}q_{\alpha _{12}}Z_{4}^{\left( -1\right) }\right\} .
\end{equation}%
We use index condensation notation for momentum, $q_{\mu _{1}}...q_{\mu
_{n}}=q_{\mu _{1}...\mu _{n}},$ as well as for metric $g_{\mu _{1}\mu
_{2}}=g_{\mu _{12}}.$ Remembering that $q_{(\mu _{12}}g_{\alpha _{12})}$ is
the symmetric combination.

Using the reduction of the last section, we derive the identities%
\begin{eqnarray}
2q^{\mu _{1}}J_{2\mu _{123}}^{\left( 2\right) } &=&-q^{2}J_{2\mu _{2}\mu
_{3}}^{\left( 2\right) } \\
2q^{\mu _{1}}J_{2\mu _{1234}}^{\left( 2\right) } &=&-q^{2}J_{2\mu
_{234}}^{\left( 2\right) }.
\end{eqnarray}%
And the contraction with the metric tensor given by%
\begin{eqnarray}
2g_{\mu _{12}}J_{2}^{\left( 2\right) \mu _{123}} &=&2m^{2}J_{2}^{\left(
2\right) \mu _{3}}-\frac{i}{4\pi }q^{\mu _{3}}  \label{trJ23} \\
2g_{\mu _{12}}J_{2}^{\left( 2\right) \mu _{1234}} &=&2m^{2}J_{2}^{\left(
2\right) \mu _{34}}+\frac{i}{4\pi }\frac{1}{6}\left[ 3q^{\mu _{34}}-\theta
^{\mu _{34}}\left( q\right) \right] .  \label{trJ24}
\end{eqnarray}%
\qquad

\subsection{Four Dimensions\label{Sub4DIn}}

As to the four-dimensional integral, we have the following power counting%
\begin{equation}
\left\{ 
\begin{array}{c}
\omega (J_{3}^{\left( 4\right) })=-2 \\ 
\omega (J_{3}^{\left( 4\right) \mu _{1}})=-1 \\ 
\omega (J_{3}^{\left( 4\right) \mu _{12}})=0%
\end{array}%
\right. ;\quad \left\{ 
\begin{array}{c}
\omega (J_{2}^{\left( 4\right) })=0 \\ 
\omega (J_{2}^{\left( 4\right) \mu _{1}})=1%
\end{array}%
\right. ;
\end{equation}%
The scalar and vector three-point functions are finite: $\bar{J}_{3}^{\left(
4\right) }=J_{3}^{\left( 4\right) }$and $\bar{J}_{3}^{\left( 4\right) \mu
_{1}}=J_{3}^{\left( 4\right) \mu _{1}}$. We compute the case with the
highest power counting to illustrate some features of our treatment. The
four-dimensional vector two-point integral,%
\begin{equation}
\bar{J}_{2}^{\left( 4\right) \mu _{1}}=\int \frac{\mathrm{d}^{4}k}{\left(
2\pi \right) ^{4}}\frac{K_{i}^{\mu _{1}}}{D_{ij}}
\end{equation}%
has linear power counting, which requires using the identity (\ref{id}) with 
$N=1,$ as (\ref{N=1}). Its replacement allows rewriting the integrand%
\begin{eqnarray}
\frac{K_{i}^{\mu _{1}}}{D_{ij}} &=&\frac{K_{i}^{\mu _{1}}}{D_{\lambda }^{2}}-%
\frac{\left( A_{i}+A_{j}\right) K_{i}^{\mu _{1}}}{D_{\lambda }^{3}} \\
&&+\left[ \frac{A_{i}A_{j}}{D_{\lambda }^{4}}+\frac{A_{i}^{2}}{D_{\lambda
}^{3}D_{i}}+\frac{A_{j}^{2}}{D_{\lambda }^{3}D_{j}}-\frac{A_{i}A_{j}^{2}}{%
D_{\lambda }^{4}D_{j}}-\frac{A_{j}A_{i}^{2}}{D_{\lambda }^{4}D_{i}}+\frac{%
A_{i}^{2}A_{j}^{2}}{D_{\lambda }^{4}D_{ij}}\right] K_{i}^{\mu _{1}}.  \notag
\end{eqnarray}%
After applying the integration sign, we gather the purely divergent
integrals and integrate the remaining finite integrals.

This result exhibits all elements presented before. We organize the local
divergences through surface terms and irreducible scalars,%
\begin{equation}
\bar{J}_{2}^{\left( 4\right) \mu _{1}}=J_{2}^{\left( 4\right) \mu
_{1}}\left( p_{ji}\right) -\frac{1}{2}[P_{ji}^{\nu _{1}}\Delta _{3\nu
_{1}}^{\left( 4\right) \mu _{1}}+p_{ji}^{\mu _{1}}I_{\log }^{\left( 4\right)
}],  \label{J2bar4D}
\end{equation}%
while integrating the finite part without restrictions,%
\begin{equation}
\ J_{2}^{\left( 4\right) \mu _{1}}\left( p_{ji}\right) =\frac{i}{\left( 4\pi
\right) ^{2}}p_{ji}^{\mu _{1}}Z_{1}^{\left( 0\right) }(p_{ij}^{2},m^{2}),
\end{equation}%
where $p_{ij}=k_{i}-k_{j}$ and $P_{ij}=k_{i}+k_{j}$ (\ref{pij}-\ref{Pij}).
For completeness, the scalar integral,%
\begin{equation}
\bar{J}_{2}^{\left( 4\right) }=I_{\log }^{\left( 4\right) }+J_{2}^{\left(
4\right) }\left( p_{ji}\right) .
\end{equation}%
Following our organization, its finite part is given by%
\begin{equation*}
J_{2}^{\left( 4\right) }\left( p_{ij}\right) =-\frac{i}{\left( 4\pi \right)
^{2}}Z_{0}^{\left( 0\right) }(p_{ij}^{2},m^{2}).
\end{equation*}

\textbf{Three-Point: }We need scalar, vector, and tensor integrals.\textbf{\ 
}%
\begin{eqnarray}
J_{3}^{\left( 4\right) } &=&i\left( 4\pi \right) ^{-2}[Z_{00}^{\left(
-1\right) }\left( p,q\right) ] \\
J_{3\mu _{1}}^{\left( 4\right) } &=&i\left( 4\pi \right) ^{-2}[-p_{\mu
_{1}}Z_{10}^{\left( -1\right) }-q_{\mu _{1}}Z_{01}^{\left( -1\right) }] \\
J_{3\mu _{1}\mu _{2}}^{\left( 4\right) } &=&i\left( 4\pi \right) ^{-2}\left[
p_{\mu _{12}}Z_{20}^{\left( -1\right) }+q_{\mu _{12}}Z_{02}^{\left(
-1\right) }+p_{(\mu _{1}}q_{\mu _{2})}Z_{11}^{\left( -1\right) }-\frac{1}{2}%
g_{\mu _{12}}Z_{00}^{\left( 0\right) }\right]
\end{eqnarray}%
\begin{equation}
\bar{J}_{3\mu _{1}\mu _{2}}^{\left( 4\right) }=J_{3\mu _{1}\mu _{2}}^{\left(
4\right) }+\frac{1}{4}(\Delta _{3\mu _{12}}^{\left( 4\right) }+g_{\mu
_{12}}I_{\log }^{\left( 4\right) }).
\end{equation}%
Of these, only the tensor integral is divergent, where we used $N=0$, in (%
\ref{id}). It is worth mentioning that the arguments $p$ and $q$ are only
general variables that tag the entries of the functions; they must be
carefully substituted for the ones that appear in a particular part of the
investigation.\ In four dimensions, we will adopt $p=p_{21}$ and $q=p_{31}$.

\textbf{Reductions }$\mathbf{4D}$\textbf{:} The three points obey the
reductions of the previous section as the two-point functions. Therefore it
is possible to show that the tensors $J$ satisfy%
\begin{eqnarray}
2p^{\mu _{1}}J_{3\mu _{1}}^{\left( 4\right) } &=&-p^{2}J_{3}^{\left(
4\right) }+[J_{2}^{\left( 4\right) }\left( q\right) -J_{2}^{\left( 4\right)
}\left( q-p\right) ] \\
2q^{\mu _{1}}J_{3\mu _{1}}^{\left( 4\right) } &=&-q^{2}J_{3}^{\left(
4\right) }+[J_{2}^{\left( 4\right) }\left( p\right) -J_{2}^{\left( 4\right)
}\left( q-p\right) ].
\end{eqnarray}%
And for the tensor integrals%
\begin{eqnarray}
2p^{\mu _{1}}J_{3\mu _{1}\mu _{2}}^{\left( 4\right) } &=&-p^{2}J_{3\mu
_{2}}^{\left( 4\right) }+[J_{2\mu _{2}}^{\left( 4\right) }\left( q\right)
+J_{2\mu _{2}}^{\left( 4\right) }\left( q-p\right) +q_{\mu
_{2}}J_{2}^{\left( 4\right) }\left( q-p\right) ] \\
2q^{\mu _{1}}J_{3\mu _{1}\mu _{2}}^{\left( 4\right) } &=&-q^{2}J_{3\mu
_{2}}^{\left( 4\right) }+[J_{2\mu _{2}}^{\left( 4\right) }\left( p\right)
+J_{2\mu _{2}}^{\left( 4\right) }\left( q-p\right) +q_{\mu
_{2}}J_{2}^{\left( 4\right) }\left( q-p\right) ].
\end{eqnarray}%
In addition to the trace contraction%
\begin{equation}
g^{\mu _{1}\mu _{2}}J_{3\mu _{1}\mu _{2}}^{\left( 4\right)
}=m^{2}J_{3}^{\left( 4\right) }+\frac{i}{2\left( 4\pi \right) ^{2}}%
+J_{2}^{\left( 4\right) }\left( q-p\right) .
\end{equation}%
In sections where a specific dimension is handled, we drop the super-index
in $J^{\left( d\right) }$ integrals.

We will also need the reductions of the $Z$-functions for the case of
different masses

\begin{eqnarray}
Z_{0}^{\left( 1\right) } &=&\left( -m_{2}^{2}\right) \left[ \log \frac{%
m_{2}^{2}}{\lambda ^{2}}-1\right] +2p^{2}Z_{2}^{\left( 0\right) }-\left(
p^{2}+m_{1}^{2}-m_{2}^{2}\right) Z_{1}^{\left( 0\right) } \\
2q^{2}Z_{1}^{\left( 0\right) } &=&\left( q^{2}+m_{1}^{2}-m_{2}^{2}\right)
Z_{0}^{\left( 0\right) }+m_{2}^{2}\log \frac{m_{2}^{2}}{\lambda ^{2}}%
-m_{1}^{2}\log \frac{m_{1}^{2}}{\lambda ^{2}}+\left(
m_{1}^{2}-m_{2}^{2}\right) \\
q^{2}Z_{n+2}^{\left( 0\right) } &=&\frac{\left( n+2\right) }{\left(
n+3\right) }\left( q^{2}+m_{1}^{2}-m_{2}^{2}\right) Z_{n+1}^{\left( 0\right)
}-\frac{\left( n+1\right) }{\left( n+3\right) }m_{1}^{2}Z_{n}^{\left(
0\right) } \\
&&+\frac{1}{\left( n+3\right) }m_{2}^{2}\log \frac{m_{2}^{2}}{\lambda ^{2}}-%
\frac{1}{\left( n+3\right) \left( n+2\right) }\left[ \frac{\left( n+1\right) 
}{\left( n+3\right) }q^{2}+\left( m_{2}^{2}-m_{1}^{2}\right) \right] .
\end{eqnarray}

All the results of this Session also will be used to determine under what
conditions the Einstein and Weyl anomalies manifest themselves in the
gravitational amplitudes. However, in the following two Chapters, we will
verify the explicit form of the odd two-dimensional and four-dimensional
abelian chiral amplitudes. After doing this, we will extend the results to
the two-dimensional gravitational case.

\chapter{Two-Dimensional $AV$-$VA$ Functions}

\label{2Dim2Pt}In this section, we compute amplitudes of two Lorentz indices
to establish the connection between linearity, symmetries, and low-energy
implications, which materialize through Relations Among Green Functions
(RAGFs) and Ward Identities (WIs). It is also defined what we mean by
uniqueness, exploring examples that evoke this concept. Since the involved
amplitudes exhibit logarithmic power counting, they depend only on the
difference between routings and not on the arbitrary sums; then, we adopt $%
q=p_{21}=k_{2}-k_{1}$.

Our first step, therefore, is to clarify the mentioned connection. After
introducing the model \ref{ModlDef}, we showed how to establish identities
among the amplitudes integrands (\ref{RAGF1})-(\ref{RAGF2}). The integration
should produce RAGFs for the vector and axial vertexes 
\begin{eqnarray}
q^{\mu _{2}}T_{\mu _{12}}^{AV} &=&T_{\mu _{1}}^{A}\left( 1\right) -T_{\mu
_{1}}^{A}\left( 2\right)  \label{p1AV} \\
q^{\mu _{1}}T_{\mu _{12}}^{VA} &=&T_{\mu _{2}}^{A}\left( 1\right) -T_{\mu
_{2}}^{A}\left( 2\right) .  \label{p1VA}
\end{eqnarray}%
\begin{eqnarray}
q^{\mu _{1}}T_{\mu _{12}}^{AV} &=&T_{\mu _{2}}^{A}\left( 1\right) -T_{\mu
_{2}}^{A}\left( 2\right) -2mT_{\mu _{2}}^{PV}  \label{p2AV} \\
q^{\mu _{2}}T_{\mu _{12}}^{VA} &=&T_{\mu _{1}}^{A}\left( 1\right) -T_{\mu
_{1}}^{A}\left( 2\right) +2mT_{\mu _{1}}^{VP}.  \label{p2VA}
\end{eqnarray}%
These contractions are direct implications of the integral linearity, and
conditions to their validity are the subject of the first subsection.
Meanwhile, WIs require vanishing the axial one-point functions above. That
occurs because the formal current-conservation equations require it (\ref%
{VWI} and \ref{AWI}).

Moreover, if these symmetry constraints are valid, the general structure of
these amplitudes as odd tensors implies kinematic properties to the scalar
invariants $F_{i}$ as,%
\begin{equation}
T_{\mu _{12}}^{AV}=\varepsilon _{\mu _{1}\mu _{2}}F_{1}+\varepsilon _{\mu
_{1}\nu }q^{\nu }q_{\mu _{2}}F_{2}+\varepsilon _{\mu _{2}\nu }q^{\nu }q_{\mu
_{1}}F_{3}.  \label{AVForm}
\end{equation}
Contracting with the external momenta in the respective indexes yields%
\begin{eqnarray}
q^{\mu _{2}}T_{\mu _{12}}^{AV} &=&\varepsilon _{\mu _{1}\nu }q^{\nu
}(q^{2}F_{2}+F_{1}),  \label{q2ContAV} \\
q^{\mu _{1}}T_{\mu _{12}}^{AV} &=&\varepsilon _{\mu _{2}\nu }q^{\nu
}(q^{2}F_{3}-F_{1}).  \label{q1ContAV}
\end{eqnarray}

The vector conservation in the first equation implies $F_{1}=-q^{2}F_{2},$
whose replacement in the second equation produces 
\begin{equation}
q^{\mu _{1}}T_{\mu _{12}}^{AV}=\varepsilon _{\mu _{2}\nu }q^{\nu
}q^{2}(F_{3}+F_{2}).
\end{equation}%
Hence, if invariants do not have poles in $q^{2}=0$, we have a low-energy
implication for axial contraction. If axial WI is satisfied, this
implication falls on the $PV$ amplitude 
\begin{equation}
\left. q^{\mu _{1}}T_{\mu _{12}}^{AV}\right\vert _{q^{2}=0}=0=\left.
-2mT_{\mu _{2}}^{PV}\right\vert _{q^{2}=0}=:\varepsilon _{\mu _{2}\nu
}q^{\nu }\Omega ^{PV}(q^{2}=0),  \label{lowAV}
\end{equation}%
being $\Omega ^{PV}$ is the form factor associated with $PV$. The deduction
of this last behavior requires the validity of both WIs, so it has the same
status as a symmetry property. The reciprocal form of this statement appears
by exchanging the order of the arguments. If the axial WI is selected first,
it implies $F_{1}=q^{2}F_{3}-\Omega ^{PV}$ in (\ref{q1ContAV}). Its
replacement in the vector contraction (\ref{q2ContAV}) gives the low-energy
implication for the contraction with the index of the vector current 
\begin{equation}
\left. q^{\mu _{2}}T_{\mu _{12}}^{AV}\right\vert _{q^{2}=0}=-\varepsilon
_{\mu _{1}\nu }q^{\nu }\Omega ^{PV}(q^{2}=0).  \label{lowAV2}
\end{equation}

With this scenario in hand, our objective is their analysis in the light of
explicit integration (\ref{T}). From definition (\ref{t}), the general
integrand of two-point amplitudes is%
\begin{eqnarray}
t^{\Gamma _{1}\Gamma _{2}} &=&K_{12}^{\nu _{12}}\text{\textrm{tr}}[\Gamma
_{1}\gamma _{\nu _{1}}\Gamma _{2}\gamma _{\nu _{2}}]\frac{1}{D_{12}}+m^{2}%
\text{\textrm{tr}}[\Gamma _{1}\Gamma _{2}]\frac{1}{D_{12}},  \notag \\
&&+mK_{1}^{\nu _{1}}\text{\textrm{tr}}[\Gamma _{1}\gamma _{\nu _{1}}\Gamma
_{2}]\frac{1}{D_{12}}+mK_{2}^{\nu _{1}}\text{\textrm{tr}}[\Gamma _{1}\Gamma
_{2}\gamma _{\nu _{1}}]\frac{1}{D_{12}}.  \label{2ptexp}
\end{eqnarray}%
Specific versions emerge after choosing the vertices and keeping the
non-zero traces: 
\begin{eqnarray}
t_{\mu _{12}}^{AV} &=&K_{12}^{\nu _{12}}\mathrm{tr}(\gamma _{\ast }\gamma
_{\mu _{1}\nu _{1}\mu _{2}\nu _{2}})\frac{1}{D_{12}}+m^{2}\mathrm{tr}(\gamma
_{\ast }\gamma _{\mu _{1}\mu _{2}})\frac{1}{D_{12}}, \\
t_{\mu _{12}}^{VA} &=&K_{12}^{\nu _{12}}\mathrm{tr}(\gamma _{\ast }\gamma
_{\mu _{1}\nu _{1}\mu _{2}\nu _{2}})\frac{1}{D_{12}}-m^{2}\mathrm{tr}(\gamma
_{\ast }\gamma _{\mu _{1}\mu _{2}})\frac{1}{D_{12}}.
\end{eqnarray}

As the trace of four gamma matrices is a linear combination of the metric
and the Levi-Civita tensor, various expressions emerge through substitutions
involving the following versions of the identity (\ref{Chiral-Id}):%
\begin{eqnarray}
2\gamma _{\ast } &=&\varepsilon _{\nu _{12}}\gamma ^{\nu _{12}};  \label{id1}
\\
\gamma _{\ast }\gamma _{\mu _{1}} &=&-\varepsilon _{\mu _{1}\nu _{1}}\gamma
^{\nu _{1}};  \label{id2} \\
\gamma _{\ast }\gamma _{\left[ \mu _{1}\mu _{2}\right] } &=&-\varepsilon
_{\mu _{1}\mu _{2}}.
\end{eqnarray}%
They lead to expressions that are not automatically equal after integration.
To unfold this rationale, let us apply the chiral matrix definition in form $%
2\gamma _{\ast }=\varepsilon ^{ef}\gamma _{ef}$\ to write%
\begin{eqnarray}
\mathrm{tr}(\gamma _{\ast }\gamma _{abcd}) &=&\frac{1}{2}\varepsilon ^{ef}%
\mathrm{tr}(\gamma _{efabcd}) \\
&=&2[-g_{ab}\varepsilon _{cd}+g_{ac}\varepsilon _{bd}-g_{ad}\varepsilon
_{bc}-g_{bc}\varepsilon _{ad}+g_{bd}\varepsilon _{ac}-g_{cd}\varepsilon
_{ab}].
\end{eqnarray}%
We explore two equivalent sorting of indices $\left( a,b,c,d\right) =\left(
\mu _{1},\nu _{1},\mu _{2},\nu _{2}\right) $ and $\left( a,b,c,d\right)
=\left( \mu _{2},\nu _{2},\mu _{1},\nu _{1}\right) $, corresponding to the
substitution of the chiral matrix definition around the first and second
vertices. The traces differ by signs of terms but are equivalent. To study
them, we perform the contractions with $K_{12}^{\nu _{12}}=K_{1}^{\nu
_{1}}K_{2}^{\nu _{2}}$ and write the equations%
\begin{eqnarray}
K_{12}^{\nu _{12}}\mathrm{tr}(\gamma _{\ast }\gamma _{\mu _{1}}\gamma _{\nu
_{1}}\gamma _{\mu _{2}}\gamma _{\nu _{2}}) &=&-2\varepsilon _{\mu _{1}\nu
_{1}}\left( K_{1\mu _{2}}K_{2}^{\nu _{1}}+K_{2\mu _{2}}K_{1}^{\nu
_{1}}\right) -2\varepsilon _{\mu _{2}\nu _{1}}\left( K_{1\mu _{1}}K_{2}^{\nu
_{1}}-K_{2\mu _{1}}K_{1}^{\nu _{1}}\right)  \notag \\
&&+2\varepsilon _{\mu _{1}\mu _{2}}\left( K_{1}\cdot K_{2}\right) +2g_{\mu
_{1}\mu _{2}}\varepsilon _{\nu _{1}\nu _{2}}K_{12}^{\nu _{12}}, \\
K_{12}^{\nu _{12}}\mathrm{tr}(\gamma _{\ast }\gamma _{\mu _{2}}\gamma _{\nu
_{2}}\gamma _{\mu _{1}}\gamma _{\nu _{1}}) &=&+2\varepsilon _{\mu _{1}\nu
_{1}}\left( K_{1\mu _{2}}K_{2}^{\nu _{1}}-K_{2\mu _{2}}K_{1}^{\nu
_{1}}\right) -2\varepsilon _{\mu _{2}\nu _{1}}\left( K_{1\mu _{1}}K_{2}^{\nu
_{1}}+K_{2\mu _{1}}K_{1}^{\nu _{1}}\right)  \notag \\
&&-2\varepsilon _{\mu _{1}\mu _{2}}\left( K_{1}\cdot K_{2}\right) -2g_{\mu
_{1}\mu _{2}}\varepsilon _{\nu _{1}\nu _{2}}K_{12}^{\nu _{12}}.
\end{eqnarray}

The general form (\ref{2ptexp}) shows that combining the bilinears with mass
terms associated with $\mathrm{tr}(\gamma _{\ast }\gamma _{\mu
_{12}})=-2\varepsilon _{\mu _{12}}$ leads to scalar two-point amplitudes
identified as 
\begin{eqnarray}
t^{PP} &=&q^{2}\frac{1}{D_{12}}-\frac{1}{D_{1}}-\frac{1}{D_{2}},  \label{PP}
\\
t^{SS} &=&(4m^{2}-q^{2})\frac{1}{D_{12}}+\frac{1}{D_{1}}+\frac{1}{D_{2}}.
\label{SS}
\end{eqnarray}%
The following reduction was used for these integrands%
\begin{equation}
S_{ij}=K_{i}\cdot K_{j}-m^{2}=\frac{1}{2}(D_{i}+D_{j}-p_{ij}^{2}).
\label{Sij}
\end{equation}

It is possible to express all other contributions in terms of the same
object, a standard tensor present similarly in all explored dimensions 
\begin{equation}
t_{\mu _{2}}^{\left( s_{1}\right) \nu _{1}}=\left( K_{1\mu _{2}}K_{2}^{\nu
_{1}}+s_{1}K_{2\mu _{2}}K_{1}^{\nu _{1}}\right) \frac{1}{D_{12}}.
\label{t(s)}
\end{equation}%
where the $s_{1}=$ $\pm $. The tensors that arise from the expression above
are given by%
\begin{eqnarray}
t_{\mu _{12}}^{\left( +\right) } &=&2\frac{K_{1\mu _{1}}K_{1\mu _{2}}}{D_{12}%
}+q_{(\mu _{1}}K_{1\mu _{2})}\frac{1}{D_{12}}  \label{t(s1)} \\
t_{\mu _{12}}^{\left( -\right) } &=&q_{[\mu _{2}}K_{1\mu _{1}]}\frac{1}{%
D_{12}}.
\end{eqnarray}%
Nevertheless, anticipating a connection with higher dimensions, we opt to
write the last term as a pseudo-scalar function 
\begin{equation*}
t^{SP}=-t^{PS}=\varepsilon _{\nu _{1}\nu _{2}}t^{\left( -\right) \nu _{1}\nu
_{2}}=2\frac{\varepsilon _{\nu _{1}\nu _{2}}K_{1}^{\nu _{1}}K_{2}^{\nu _{2}}%
}{D_{12}}
\end{equation*}%
using $\varepsilon _{\nu _{1}\nu _{2}}K_{1}^{\nu _{1}}K_{2}^{\nu
_{2}}=\varepsilon _{\nu _{1}\nu _{2}}p_{21}^{\nu _{2}}K_{1}^{\nu _{1}}$ and
then the definition of the vector integral for equal masses, proportional to 
$p_{21}^{\nu _{1}}Z_{1},$ results in%
\begin{equation*}
T^{SP}=2\varepsilon _{\nu _{1}\nu _{2}}q^{\nu _{2}}J_{2}^{\nu _{1}}=0.
\end{equation*}

Therefore, given both versions for the four-matrix trace, we have the
corresponding versions for the $AV$ amplitude%
\begin{eqnarray}
(t_{\mu _{12}}^{AV})_{1} &=&-2\varepsilon _{\mu _{1}\nu _{1}}t_{\mu
_{2}}^{\left( +\right) \nu _{1}}-\varepsilon _{\mu _{1}\mu
_{2}}t^{PP}-2\varepsilon _{\mu _{2}\nu _{1}}t_{\mu _{1}}^{\left( -\right)
\nu _{1}}+2g_{\mu _{1}\mu _{2}}t^{SP},  \label{av1full} \\
(t_{\mu _{12}}^{AV})_{2} &=&-2\varepsilon _{\mu _{2}\nu _{1}}t_{\mu
_{1}}^{\left( +\right) \nu _{1}}-\varepsilon _{\mu _{1}\mu
_{2}}t^{SS}+2\varepsilon _{\mu _{1}\nu _{1}}t_{\mu _{2}}^{\left( -\right)
\nu _{1}}-2g_{\mu _{1}\mu _{2}}t^{SP}.  \label{av2full}
\end{eqnarray}%
The same happens to the $VA$ amplitude%
\begin{eqnarray}
(t_{\mu _{12}}^{VA})_{1} &=&-2\varepsilon _{\mu _{1}\nu _{1}}t_{\mu
_{2}}^{\left( +\right) \nu _{1}}+\varepsilon _{\mu _{1}\mu
_{2}}t^{SS}-2\varepsilon _{\mu _{2}\nu _{1}}t_{\mu _{1}}^{\left( -\right)
\nu _{1}}+g_{\mu _{1}\mu _{2}}t^{SP} \\
(t_{\mu _{12}}^{VA})_{2} &=&-2\varepsilon _{\mu _{2}\nu _{1}}t_{\mu
_{1}}^{\left( +\right) \nu _{1}}+\varepsilon _{\mu _{1}\mu
_{2}}t^{PP}+2\varepsilon _{\mu _{1}\nu _{1}}t_{\mu _{2}}^{\left( -\right)
\nu _{1}}-g_{\mu _{1}\mu _{2}}t^{SP}.
\end{eqnarray}

As mentioned at the beginning of the section, integrated amplitudes depend
exclusively on the external momentum $q$. That precludes the construction of
some 2nd-order tensors, which cancels out terms like $t^{\left( -\right) }$
and $SP$. Further examination of the general form (\ref{2ptexp}) allows the
identification of even amplitudes%
\begin{eqnarray}
t_{\mu _{1}\mu _{2}}^{VV} &=&(2t_{\mu _{1}\mu _{2}}^{\left( +\right)
}+g_{\mu _{1}\mu _{2}}t^{PP})  \label{vv} \\
t_{\mu _{1}\mu _{2}}^{AA} &=&(2t_{\mu _{1}\mu _{2}}^{\left( +\right)
}-g_{\mu _{1}\mu _{2}}t^{SS}).  \label{aa}
\end{eqnarray}%
Hence, the integration provides the relations among odd and even amplitudes%
\begin{eqnarray}
(T_{\mu _{12}}^{AV})_{1} &=&-\varepsilon _{\mu _{1}}^{\quad \nu _{1}}(T_{\nu
_{1}\mu _{2}}^{VV});\text{\qquad }(T_{\mu _{12}}^{AV})_{2}=-\varepsilon
_{\mu _{2}}^{\quad \nu _{1}}(T_{\mu _{1}\nu _{1}}^{AA});  \label{AV1 and AV2}
\\
(T_{\mu _{12}}^{VA})_{1} &=&-\varepsilon _{\mu _{1}}^{\quad \nu _{1}}(T_{\nu
_{1}\mu _{2}}^{AA});\text{\qquad }(T_{\mu _{12}}^{VA})_{2}=-\varepsilon
_{\mu _{2}}^{\quad \nu _{1}}(T_{\mu _{1}\nu _{1}}^{VV}).
\end{eqnarray}%
Although we did not detail, following the same steps produced both $VA$
versions. These associations are directly achieved at the integrand level
using (\ref{id2}), the identity $\gamma _{\ast }\gamma _{\mu
_{i}}=-\varepsilon _{\mu _{i}}^{\quad \nu _{1}}\gamma _{\nu _{1}}$ in the
adequate position. We need a clear distinction among versions since their
comparison is not automatic for integrated amplitudes due to their diverging
character.

We also use the last identity $\gamma _{\ast }\gamma _{\left[ \mu _{1}\mu
_{2}\right] }=-\varepsilon _{\mu _{1}\mu _{2}}$ to introduce the third
version for the discussed amplitudes. Replacing the form $\gamma _{\ast
}\gamma _{\mu _{1}}\gamma _{\nu _{1}}=-\varepsilon _{\mu _{1}\nu
_{1}}+g_{\mu _{1}\nu _{1}}\gamma _{\ast }$ in the traces produces the results%
\begin{eqnarray}
(t_{\mu _{12}}^{AV})_{3} &=&-\frac{1}{2}[\varepsilon _{\mu _{1}}^{\hspace{7pt%
}\nu _{1}}(t_{\nu _{1}\mu _{2}}^{VV})+\varepsilon _{\mu _{2}}^{\hspace{7pt}%
\nu _{1}}(t_{\mu _{1}\nu _{1}}^{AA})]-\varepsilon _{\mu _{2}\nu _{1}}t_{\mu
_{1}}^{\left( -\right) \nu _{1}}+\varepsilon _{\mu _{1}\nu _{1}}t_{\mu
_{2}}^{\left( -\right) \nu _{1}}, \\
(t_{\mu _{12}}^{VA})_{3} &=&-\frac{1}{2}[\varepsilon _{\mu _{1}}^{\hspace{7pt%
}\nu _{1}}(t_{\nu _{1}\mu _{2}}^{AA})+\varepsilon _{\mu _{2}}^{\hspace{7pt}%
\nu _{1}}(t_{\mu _{1}\nu _{1}}^{VV})]-\varepsilon _{\mu _{2}\nu _{1}}t_{\mu
_{1}}^{\left( -\right) \nu _{1}}+\varepsilon _{\mu _{1}\nu _{1}}t_{\mu
_{2}}^{\left( -\right) \nu _{1}}.
\end{eqnarray}%
Since $t_{\mu }^{\left( -\right) \nu }$ tensors vanish after integration,
different versions with each other as follows%
\begin{equation}
(T_{\mu _{12}}^{AV})_{3}=\frac{1}{2}[(T_{\mu _{12}}^{AV})_{1}+(T_{\mu
_{12}}^{AV})_{2}];\quad (T_{\mu _{12}}^{VA})_{3}=\frac{1}{2}[(T_{\mu
_{12}}^{VA})_{1}+(T_{\mu _{12}}^{VA})_{2}].  \label{AV3}
\end{equation}%
This particular aspect receives further attention in the section (\ref%
{4Dim3Pt}). The investigation developed by the article \cite{Battistel2004}
uses this version in equation (85). It illustrates how any possible
expression follows from versions one and two.

Before proceeding, we need integrated expressions. Their obtainment occurs
by replacing the results of appendix (\ref{AppInt2D}) in the integrated
versions of structures (\ref{PP}),(\ref{SS}), and (\ref{t(s)}). The scalar
two-point functions assume the forms 
\begin{eqnarray}
T^{PP} &=&q^{2}J_{2}-2I_{\log }, \\
T^{SS} &=&\left( 4m^{2}-p^{2}\right) J_{2}+2I_{\log }.
\end{eqnarray}%
And the symmetric sign tensor is 
\begin{eqnarray}
T_{\mu _{12}}^{\left( +\right) } &=&2(\bar{J}_{2\mu _{12}}+q_{\mu
_{1}}J_{2\mu _{2}}) \\
&=&2\theta _{\mu _{12}}\left( q\right) \left( m^{2}J_{2}+\frac{i}{4\pi }%
\right) -\frac{1}{2}g_{\mu _{12}}q^{2}J_{2}+(\Delta _{2\mu _{12}}+g_{\mu
_{12}}I_{\log }),
\end{eqnarray}%
where $\theta _{\alpha \lambda }\left( q\right) =\left( g_{\alpha \lambda
}q^{2}-q_{\alpha }q_{\lambda }\right) /q^{2}$ is the transversal projector.
We put these pieces together to compound 2nd-order even tensors 
\begin{eqnarray}
T_{\mu _{1}\mu _{2}}^{VV} &=&2\Delta _{2\mu _{1}\mu _{2}}+4\theta _{\mu
_{1}\mu _{2}}\left( m^{2}J_{2}+\frac{i}{4\pi }\right) ,  \label{VV} \\
T_{\mu _{1}\mu _{2}}^{AA} &=&2\Delta _{2\mu _{1}\mu _{2}}+4\theta _{\mu
_{1}\mu _{2}}\left( m^{2}J_{2}+\frac{i}{4\pi }\right) -g_{\mu _{1}\mu
_{2}}\left( 4m^{2}J_{2}\right) ,
\end{eqnarray}%
which lead to the versions for the $AV$ amplitude%
\begin{eqnarray}
(T_{\mu _{12}}^{AV})_{1} &=&-2\varepsilon _{\mu _{1}}^{\quad \nu }\Delta
_{2\mu _{2}\nu }-4\varepsilon _{\mu _{1}\nu }\theta _{\mu _{2}}^{\nu }\left(
m^{2}J_{2}+\frac{i}{4\pi }\right) ,  \label{AV1} \\
(T_{\mu _{12}}^{AV})_{2} &=&-2\varepsilon _{\mu _{2}}^{\quad \nu }\Delta
_{2\mu _{1}\nu }-4\varepsilon _{\mu _{2}\nu }\theta _{\mu _{1}}^{\nu }\left(
m^{2}J_{2}+\frac{i}{4\pi }\right) -\varepsilon _{\mu _{1}\mu _{2}}\left(
4m^{2}J_{2}\right) .  \label{AV2}
\end{eqnarray}

Two-point functions within axial RAGFs are finite and related through the
expressions%
\begin{eqnarray}
T_{\mu }^{PV} &=&-T_{\mu }^{VP}=\varepsilon _{\mu \nu }q^{\nu }\left[
-2mJ_{2}\left( q\right) \right] ,  \label{PV} \\
T_{\mu }^{PA} &=&-T_{\mu }^{AP}=-\varepsilon _{\mu \nu }(T^{PV})^{\nu }.
\end{eqnarray}%
Whereas one-point functions are pure surface terms proportional to the
routing $k_{i}$ ,%
\begin{equation}
T_{\mu }^{A}\left( i\right) =-\varepsilon _{\mu }^{\quad \nu _{1}}T_{\nu
_{1}}^{V}\left( i\right) =2\varepsilon _{\mu }^{\quad \nu _{1}}k_{i}^{\nu
_{2}}\Delta _{2\nu _{1}\nu _{2}}.  \label{A}
\end{equation}

Even though the integrands are equivalent, the same does not apply to
integrated functions. In the case of even amplitudes ($VV$ and $AA$),
expressions depend on the prescription adopted for evaluating divergences.
That also occurs for odd amplitudes ($AV$ and $VA$), but they rely on the
version for the trace. Using the\ chiral matrix definition around the first
or the second vertexes brings implications for the index arrangement in
finite and divergent parts. This perspective produced identities originally,
but now the connection is not automatic. That becomes clear when we subtract
the $AV$ expressions%
\begin{eqnarray}
(T_{\mu _{12}}^{AV})_{1}-(T_{\mu _{12}}^{AV})_{2} &=&-2(\varepsilon _{\mu
_{1}\nu }\Delta _{2\mu _{2}}^{\nu }-\varepsilon _{\mu _{2}\nu }\Delta _{2\mu
_{1}}^{\nu }) \\
&&-4(\varepsilon _{\mu _{1}\nu _{1}}\theta _{\mu _{2}}^{\nu }-\varepsilon
_{\mu _{2}\nu _{1}}\theta _{\mu _{1}}^{\nu })\left( m^{2}J_{2}+\frac{i}{4\pi 
}\right) +4\varepsilon _{\mu _{1}\mu _{2}}m^{2}J_{2}.  \notag
\end{eqnarray}%
We use Schouten identities in 2D to rearrange indexes in the finite part and
in surface terms. Through the antisymmetry of the Levi-Civita tensor, we
have explicitly%
\begin{eqnarray}
\varepsilon _{\mu _{1}\nu }\Delta _{2\mu _{2}}^{\nu }+\varepsilon _{\mu
_{2}\mu _{1}}\Delta _{2\nu }^{\nu }+\varepsilon _{\nu \mu _{2}}\Delta _{2\mu
_{1}}^{\nu } &=&0=\varepsilon _{\lbrack \mu _{1}\nu }\Delta _{2\mu
_{2}]}^{\nu },  \label{SchDiv} \\
\varepsilon _{\mu _{1}\nu }\theta _{\mu _{2}}^{\nu }+\varepsilon _{\mu
_{2}\mu _{1}}\theta _{\nu }^{\nu }+\varepsilon _{\nu \mu _{2}}\theta _{\mu
_{1}}^{\nu } &=&0=\varepsilon _{\lbrack \mu _{1}\nu }\theta _{\mu
_{2}]}^{\nu }.  \label{SchTeta}
\end{eqnarray}%
So, the difference reduces to%
\begin{equation}
(T_{\mu _{12}}^{AV})_{1}-(T_{\mu _{12}}^{AV})_{2}=-\varepsilon _{\mu _{1}\mu
_{2}}\left( 2\Delta _{2\alpha }^{\alpha }+\frac{i}{\pi }\right) .
\label{Uni-2D}
\end{equation}

The integration linearity requires this difference to vanish identically,
constraining the value of $\Delta _{2\alpha }^{\alpha }$. That represents a
link between linearity and the uniqueness of perturbative solutions. Now, we
analyze the role the surface terms play regarding the RAGFs.

\section{Verification and Consequences of the RAGFs}

We perform contractions with momentum for the integrated amplitudes to
analyze the RAGFs, starting with even functions because they relate to the
odd ones. These operations produce the difference between vector one-point
functions (\ref{RAGF1}), and that occurs identically. After contracting the
integrated $VV$, finite parts cancel out due to $q^{\mu _{2}}\theta _{\mu
_{2}}^{\nu }=0$, and only a surface term remains. The comparison with the $V$
function (\ref{A}) leads directly to the expected relation 
\begin{eqnarray}
q^{\mu _{1}}T_{\mu _{12}}^{VV} &=&2q^{\nu _{1}}\Delta _{2\mu _{2}\nu
_{1}}=[T_{\mu _{2}}^{V}\left( 1\right) -T_{\mu _{2}}^{V}\left( 2\right) ]
\label{pVV} \\
q^{\mu _{1}}T_{\mu _{12}}^{AA}+2mT_{\mu _{2}}^{PA} &=&2q^{\nu _{1}}\Delta
_{2\mu _{2}\nu _{1}}=[T_{\mu _{2}}^{V}\left( 1\right) -T_{\mu
_{2}}^{V}\left( 2\right) ].
\end{eqnarray}%
The same occurs with the $AA$. In this case, finite function $PA$ and
surface term appear.

Now, we turn our attention to relations for odd amplitudes (\ref{p1AV})-(\ref%
{p2VA}). Taking first version of $AV$ (\ref{AV1}), the contraction with
vector vertex yields%
\begin{equation}
q^{\mu _{2}}(T_{\mu _{12}}^{AV})_{1}=-2\varepsilon _{\mu _{1}\nu _{1}}q^{\nu
_{2}}\Delta _{2\nu _{2}}^{\nu _{1}}=[T_{\mu _{1}}^{A}\left( 1\right) -T_{\mu
_{1}}^{A}\left( 2\right) ].  \label{qAV1}
\end{equation}%
Again, identifying the axial amplitude (\ref{A}) is straightforward and does
not require conditions. That differs from the axial contraction, which needs
the rearranging of indexes, 
\begin{equation}
q^{\mu _{1}}(T_{\mu _{12}}^{AV})_{1}=-2q^{\mu _{1}}\varepsilon _{\mu _{1}\nu
}\Delta _{2\mu _{2}}^{\nu }-4q^{\mu _{1}}\varepsilon _{\mu _{1}\nu }\theta
_{\mu _{2}}^{\nu }\left( m^{2}J_{2}+\frac{i}{4\pi }\right) .
\end{equation}%
After employing (\ref{SchDiv})-(\ref{SchTeta}), reminding that $\theta _{\nu
}^{\nu }=1$, we have%
\begin{equation}
q^{\mu _{1}}(T_{\mu _{12}}^{AV})_{1}=[T_{\mu _{2}}^{A}\left( k_{1}\right)
-T_{\mu _{2}}^{A}\left( k_{2}\right) ]-2mT_{\mu _{2}}^{PV}+\varepsilon _{\mu
_{2}\nu _{1}}q^{\nu _{1}}\left( 2\Delta _{2\alpha }^{\alpha }+\frac{i}{\pi }%
\right) ,  \label{pAV1}
\end{equation}%
where $PV$ has the form (\ref{PV}). The last term prevents automatic
satisfaction of this relation, conditioning the value assumed by the surface
term. This situation also occurs for the second version (\ref{AV2});
however, the additional term is on the vector contraction%
\begin{eqnarray}
q^{\mu _{2}}(T_{\mu _{12}}^{AV})_{2} &=&[T_{\mu _{1}}^{A}\left( k_{1}\right)
-T_{\mu _{1}}^{A}\left( k_{2}\right) ]+\varepsilon _{\mu _{1}\nu }q^{\nu
}\left( 2\Delta _{2\alpha }^{\alpha }+\frac{i}{\pi }\right)  \label{pAV2} \\
q^{\mu _{1}}(T_{\mu _{12}}^{AV})_{2} &=&[T_{\mu _{2}}^{A}\left( k_{1}\right)
-T_{\mu _{2}}^{A}\left( k_{2}\right) ]-2mT_{\mu _{2}}^{PV}.
\end{eqnarray}%
This pattern repeats for the $VA$ amplitude: additional terms arise in the
same contractions 
\begin{eqnarray}
q^{\mu _{1}}(T_{\mu _{12}}^{VA})_{1} &=&\varepsilon _{\mu _{2}\nu
_{1}}q^{\nu _{1}}\left( 2\Delta _{2\alpha }^{\alpha }+\frac{i}{\pi }\right)
+T_{\mu _{2}}^{A}\left( 1\right) -T_{\mu _{2}}^{A}\left( 2\right) \\
q^{\mu _{2}}(T_{\mu _{12}}^{VA})_{2} &=&\varepsilon _{\mu _{1}\nu
_{1}}q^{\nu _{1}}\left( 2\Delta _{2\alpha }^{\alpha }+\frac{i}{\pi }\right)
+T_{\mu _{1}}^{A}\left( 1\right) -T_{\mu _{1}}^{A}\left( 2\right) +2mT_{\mu
_{1}}^{VP}.
\end{eqnarray}

RAGFs, deduced as identities for integrands, represent integration linearity
within this context. Even amplitudes automatically satisfy the relations
since they do not depend on the surface term value. On the other hand, odd
amplitudes require the condition\footnote{%
Since the third version is a combination, see (\ref{AV3}), all vertices have
potentially violated terms.}%
\begin{equation}
\Delta _{2\alpha }^{\alpha }=-i\left( 2\pi \right) ^{-1}.  \label{finite1}
\end{equation}%
This term emerges for the contraction with the vertex that defines the
amplitude version (the position of use of the chiral matrix definition).
Besides, choosing this finite value for surface terms ensures that the $AV$%
's are equal (\ref{Uni-2D}), clarifying the relation between linearity and
uniqueness. Any formula to the Dirac traces leads to one unique answer that
respects the linearity of integration. Nevertheless, this condition sets
non-zero values for one-point functions (\ref{A}), affecting symmetry
implications through WIs. That occurs for all relations in this subsection
since amplitudes depend on the surface term. This subject receives attention
in the sequence.

\section{Ward Identities\label{WI}}

In the model, we discussed the divergence of axial and vector currents (\ref%
{AWI})-(\ref{VWI}), indicating implications through WIs for perturbative
amplitudes. The adopted strategy translates these implications as
restrictions over RAGFs, which link linearity and symmetries. This
subsection analyses such connection with particular attention to the
anomalous amplitudes, known for the impossibility of satisfying all WIs
simultaneously.

Adopting a prescription that eliminates surface terms reduces all RAGFs for
even amplitudes to the corresponding WIs. For odd amplitudes, this condition
satisfies those WIs corresponding to automatic RAGFs while violating the
others. Observe the first version of $AV$ to clarify this statement.
Identifying the relations was automatic to the vector RAGF; however, the
axial RAGF gets an additional term. Hence, the zero value for the surface
term satisfies the vector WI while violating the axial WI. We see the
opposite for the second version, which breaks vector WI. Both identities are
disregarded for the third version since it is a composition of the first
two. See all the results in the Table \ref{tab2d}. The same arguments are
applied to the $VA$. Under this perspective, selecting an amplitude version
would choose the vertex for symmetry violation. Furthermore, this value for
surface terms breaks the integration linearity (in anomalous case). 
\begin{table}[h]
\caption{Violations for vanishing surface term in each version.}
\label{tab2d}\centering\renewcommand{\baselinestretch}{1.4}{\normalsize {\ }%
\ }$%
\begin{tabular}{|l|l|}
\hline
$q^{\mu _{1}}(T_{\mu _{12}}^{AV})_{1}=-2mT_{\mu _{2}}^{PV}+\left( i/\pi
\right) \varepsilon _{\mu _{2}\nu _{1}}q^{\nu _{1}}$ & $q^{\mu _{2}}(T_{\mu
_{12}}^{AV})_{1}=0$ \\ \hline
$q^{\mu _{1}}(T_{\mu _{12}}^{AV})_{2}=-2mT_{\mu _{2}}^{PV}$ & $q^{\mu
_{2}}(T_{\mu _{12}}^{AV})_{2}=\left( i/\pi \right) \varepsilon _{\mu _{1}\nu
_{1}}q^{\nu _{1}}$ \\ \hline
$q^{\mu _{1}}(T_{\mu _{12}}^{AV})_{3}=-2mT_{\mu _{2}}^{PV}+\left( i/2\pi
\right) \varepsilon _{\mu _{2}\nu _{1}}q^{\nu _{1}}$ & $q^{\mu _{2}}(T_{\mu
_{12}}^{AV})_{3}=\left( i/2\pi \right) \varepsilon _{\mu _{1}\nu _{1}}q^{\nu
_{1}}$ \\ \hline
$q^{\mu _{1}}T_{\mu _{12}}^{VV}=0$ & $q^{\mu _{2}}T_{\mu _{12}}^{VV}=0$ \\ 
\hline
$q^{\mu _{1}}T_{\mu _{12}}^{AA}=-2mT_{\mu _{2}}^{PA}$ & $q^{\mu _{2}}T_{\mu
_{12}}^{AA}=2mT_{\mu _{2}}^{AP}$ \\ \hline
\end{tabular}%
$%
\end{table}

In contrast, by choosing the value that preserves linearity (\ref{finite1}),
different amplitude versions collapse into one unique form\footnote{%
The version $\left( AV\right) _{3}$ happens to be independent of value of
the surface term. Parametrizing $\Delta _{2\mu \nu }=ag_{\mu \nu }$ in its
equation, we get an expression independent of coefficient $a$ and equal to
the unique form.} (\ref{Uni-2D}). However, that violates all WIs for odd and
even amplitudes since they depend on the value of the surface term; see
Table \ref{tabuniq}. 
\begin{table}[h]
\caption{Violations for unique amplitudes}
\label{tabuniq}\centering\renewcommand{\baselinestretch}{1.0}{\normalsize {\ 
}\ }$%
\begin{tabular}{|l|l|}
\hline
$q^{\mu _{1}}T_{\mu _{12}}^{AV}=-2mT_{\mu _{2}}^{PV}+\left( i/2\pi \right)
\varepsilon _{\mu _{2}\nu }q^{\nu }$ & $q^{\mu _{2}}T_{\mu
_{12}}^{AV}=\left( i/2\pi \right) \varepsilon _{\mu _{1}\nu }q^{\nu }$ \\ 
\hline
$q^{\mu _{1}}T_{\mu _{12}}^{VV}=-\left( i/2\pi \right) q_{\mu _{2}}$ & $%
q^{\mu _{2}}T_{\mu _{12}}^{VV}=-\left( i/2\pi \right) q_{\mu _{2}}$ \\ \hline
$q^{\mu _{1}}T_{\mu _{12}}^{AA}=-2mT_{\mu _{2}}^{PA}-\left( i/2\pi \right)
q_{\mu _{2}}$ & $q^{\mu _{2}}T_{\mu _{12}}^{AA}=2mT_{\mu _{2}}^{AP}-\left(
i/2\pi \right) q_{\mu _{2}}$ \\ \hline
\end{tabular}%
$%
\end{table}

Low-energy properties of finite functions are fundamental to this analysis.
Under the hypothesis that both WIs for the $AV$ amplitude apply, we
established the kinematical behavior in zero of $\Omega ^{PV}$ as being zero
(\ref{lowAV}). Nevertheless, employing the $PV$ expression (\ref{PV}) and
the limit (\ref{LetZ2D}), we have 
\begin{equation}
\Omega ^{PV}\left( 0\right) =\left. 4m^{2}J_{2}\right\vert _{0}=\frac{i}{\pi 
}m^{2}Z_{0}^{\left( -1\right) }\left( 0\right) =-\frac{i}{\pi }.
\label{PV(0)}
\end{equation}%
That means the hypothesis is false. Hence, when satisfying the vector WI,
the axial WI violation is the value corresponding to the negative of $\Omega
^{PV}(0)$. The other expectation (\ref{lowAV2}) leads to the reciprocal:
satisfying the axial WI implies violating the vector WI.

The scenario can be understood by noting a general 2nd-order odd tensor 
\begin{equation}
F_{\mu _{1}\mu _{2}}=\varepsilon _{\mu _{1}\mu _{2}}F_{1}+\varepsilon _{\mu
_{1}\nu }q^{\nu }q_{\mu _{2}}F_{2}+\varepsilon _{\mu _{2}\nu }q^{\nu }q_{\mu
_{1}}F_{3},
\end{equation}%
exhibits a feature when contracted with the momentum: we get two equations
that are strict consequences of its tensor properties%
\begin{eqnarray}
q^{\mu _{1}}F_{\mu _{1}\mu _{2}} &=&\varepsilon _{\mu _{2}\nu }q^{\nu
}V_{1}\left( q^{2}\right) =\varepsilon _{\mu _{2}\nu }q^{\nu }\left(
q^{2}F_{3}-F_{1}\right) \\
q^{\mu _{2}}F_{\mu _{1}\mu _{2}} &=&\varepsilon _{\mu _{1}\nu }q^{\nu
}V_{2}\left( q^{2}\right) =\varepsilon _{\mu _{1}\nu }q^{\nu }\left(
q^{2}F_{2}+F_{1}\right) .
\end{eqnarray}%
If form factors are free of kinematic singularities observed in the explicit
forms of the amplitudes, we have the implication at zero 
\begin{equation}
V_{1}\left( 0\right) +V_{2}\left( 0\right) =0.
\end{equation}%
If one of the terms vanishes, the other must do so. Otherwise, if one of the 
$V_{i}\left( q^{2}\right) $ relates to a finite function ($PV$ or $VP$), an
additional constant must appear as compensation within the last equation.
Nevertheless, these statements are inconsistent with the satisfaction of
both WIs, which only occurs if linearity of integration holds with null
surface terms. Thus, the low-energy behavior of these finite functions is
the source of anomalous terms in amplitudes ($AV$-$VA$) and not their
perturbative ambiguity.

But ambiguities relate to the low-energy implications. Under the condition
of linearity and considering surface terms in the general tensor, this limit
implies the constraint $2\Delta _{2\alpha }^{\alpha }=\Omega ^{PV}(0)$. Such
an aspect will be fully explored in the section considering odd triangles in
the physical dimension. Conclusions similar to those drawn here anticipate
the presence of anomalies and linearity breaking in this new circumstances.
However, now we will explore the same two-dimensional scenario but consider
a model where different species of massive fermions interact and what
generalities we can obtain from this context.

\chapter{The $AV$ of Two Distinct Masses\label{2masses}}

To show that the behavior of amplitudes is independent of masses, let us
explore the universe where different species of massive fermions interact.
At the end of this Chapter, we answer the question: Can amplitudes be
obtained as consistent with their expected symmetry properties? The
generalization of this work is published in the paper \cite{Ebani2018}.

The $n$-point fermionic functions with different masses follow (\ref{t}),
where the mass indexes follow the momentum; In this scenario, the argument
of the propagator $i$ accounts for the routing and the mass running in the
internal lines, viz., $S\left( i\right) \equiv S\left( K_{i},m_{i}\right)
=\left( \slashed{K}_{i}-m_{i}\right) ^{-1}$. The expansion in terms of traces is
given by 
\begin{eqnarray}
t^{\Gamma _{1}\Gamma _{2}} &=&K_{12}^{\nu _{12}}\text{\textrm{tr}}[\Gamma
_{1}\gamma _{\nu _{1}}\Gamma _{2}\gamma _{\nu _{2}}]\frac{1}{D_{12}}%
+m_{1}m_{2}\text{\textrm{tr}}[\Gamma _{1}\Gamma _{2}]\frac{1}{D_{12}} \\
&&+m_{2}K_{1}^{\nu _{1}}\text{\textrm{tr}}[\Gamma _{1}\gamma _{\nu
_{1}}\Gamma _{2}]\frac{1}{D_{12}}+m_{1}K_{2}^{\nu _{1}}\text{\textrm{tr}}%
[\Gamma _{1}\Gamma _{2}\gamma _{\nu _{1}}]\frac{1}{D_{12}}.  \notag
\end{eqnarray}

The first relevant point concerns versions one and two as independent
equations for odd amplitudes, just as for equal masses. The expressions
established in (\ref{AV1 and AV2}) also apply, 
\begin{equation}
(T_{\mu _{1}\mu _{2}}^{AV})_{1}=-\varepsilon _{\mu _{1}}^{~~\nu _{1}}T_{\nu
_{1}\mu _{2}}^{VV}\qquad (T_{\mu _{1}\mu _{2}}^{AV})_{2}=-\varepsilon _{\mu
_{2}}^{~~\nu _{1}}T_{\mu _{1}\nu _{1}}^{AA}.
\end{equation}%
That happens to two masses since the $T^{SP}$ function and tensor $T_{\mu
_{2}\mu _{2}}^{\left( -\right) }$ are identically zero. They are
proportional to the vector integral $J_{2}^{\nu _{1}}=-i\left( 4\pi \right)
^{-1}q^{\nu _{1}}Z_{1}\left( q,m_{1},m_{2}\right) $. Explicitly,%
\begin{eqnarray}
T_{\mu _{12}}^{\left( -\right) } &=&q_{[\mu _{2}}J_{2\mu _{1}]}\left(
q,m_{1},m_{2}\right) =0 \\
T^{SP} &=&2\varepsilon _{\nu _{1}\nu _{2}}q^{\nu _{2}}J_{2}^{\nu _{1}}\left(
q,m_{1},m_{2}\right) =0.
\end{eqnarray}%
Effectively amounts to the validity for different masses regarding the
general expression obtainable through $\gamma _{\ast }$ definition, as (\ref%
{av1full}) and (\ref{av2full}).

Expressions to 2nd-order tensors are written through scalar sub-amplitudes $%
\Gamma _{1}\Gamma _{2}=SS$ and $\Gamma _{1}\Gamma _{2}=PP$. To obtain these
structures, we use the identity for the distinct fermions,%
\begin{equation}
2K_{2}\cdot K_{1}=D_{1}+D_{2}+\left( m_{1}^{2}+m_{2}^{2}-q^{2}\right) .
\label{red2masses}
\end{equation}%
Employing (\ref{2dJ1}) to one-point integrals\footnote{%
See $D_{1}$ and $D_{2}$ in the expression (\ref{red2masses}); when we
substitute this identity, these terms always cancel one of the propagators,
reducing the function from two to one-point.}, we have%
\begin{eqnarray}
T^{PP} &=&[+q^{2}-\left( m_{1}-m_{2}\right) ^{2}]J_{2}-\left[ 2I_{\log
}(\lambda ^{2})\right] +\frac{i}{4\pi }\left[ \log \left( m_{1}^{2}/\lambda
^{2}\right) +\log \left( m_{2}^{2}/\lambda ^{2}\right) \right]  \label{PP2m}
\\
T^{SS} &=&[-q^{2}+\left( m_{1}+m_{2}\right) ^{2}]J_{2}+\left[ 2I_{\log
}(\lambda ^{2})\right] -\frac{i}{4\pi }\left[ \log \left( m_{1}^{2}/\lambda
^{2}\right) +\log \left( m_{2}^{2}/\lambda ^{2}\right) \right] .
\label{SS2m}
\end{eqnarray}%
From the equations above, a relation that connects the sub-amplitudes is 
\begin{equation*}
T^{PP}+T^{SS}=4m_{1}m_{2}J_{2}.
\end{equation*}
While the tensorial part is compiled in the sign tensor (\ref{t(s1)}),%
\begin{equation}
T_{\mu _{12}}^{\left( +\right) }=2\bar{J}_{2\mu _{1}\mu _{2}}+q_{(\mu
_{1}}J_{2\mu _{2})}=2\bar{J}_{2\mu _{1}\mu _{2}}+2q_{\mu _{1}}J_{2\mu _{2}},
\end{equation}%
Evoking (\ref{2DJ2C}), we get the functional structure to equal masses, 
\begin{equation}
2T_{\mu _{12}}^{\left( +\right) }=4(J_{2\mu _{1}\mu _{2}}+q_{\mu
_{1}}J_{2\mu _{2}})+2\Delta _{2\mu _{12}}(\lambda ^{2})+2g_{\mu
_{12}}I_{\log }(\lambda ^{2}).
\end{equation}%
However, differences emerge in reducing the basic functions of two masses.

With these tools in hand, it is straightforward to express 2nd-order tensor
amplitudes: The first one is the Double-Vector ($VV$), given by%
\begin{eqnarray}
T_{\mu _{1}\mu _{2}}^{VV} &=&2T_{\mu _{1}\mu _{2}}^{\left( +\right) }+g_{\mu
_{12}}T^{PP}  \label{VV2m} \\
&=&2\left[ \Delta _{2\mu _{12}}\left( \lambda ^{2}\right) \right] +4(J_{2\mu
_{1}\mu _{2}}+q_{\mu _{2}}J_{2\mu _{1}})+g_{\mu _{12}}[q^{2}-\left(
m_{1}-m_{2}\right) ^{2}]J_{2}  \notag \\
&&+\frac{i}{4\pi }g_{\mu _{12}}\left[ \log \left( m_{1}^{2}/\lambda
^{2}\right) +\log \left( m_{2}^{2}/\lambda ^{2}\right) \right] .  \notag
\end{eqnarray}%
To show the elegance of the method, we also can write the amplitude in terms
of $Z_{n}^{\left( -1\right) }$,%
\begin{eqnarray}
T_{\mu _{1}\mu _{2}}^{VV} &=&2[\Delta _{2\mu _{12}}\left( \lambda
^{2}\right) ]+\frac{i}{\pi }\theta _{\mu _{12}}[1+m_{1}^{2}Z_{0}^{\left(
-1\right) }-\left( m_{1}^{2}-m_{2}^{2}\right) Z_{1}^{\left( -1\right) }] 
\notag \\
&&+\frac{i}{2\pi }g_{\mu _{12}}\left( m_{1}-m_{2}\right) [\left(
m_{1}+m_{2}\right) Z_{1}^{\left( -1\right) }-m_{1}Z_{0}^{\left( -1\right) }].
\notag
\end{eqnarray}%
It used reductions for $Z_{k}^{\left( n\right) }$ that are complementary to
using $J$-integrals. They occur when we perform contractions to investigate
symmetry relations. The expression for the Double-Axial Green Function ($AA$%
) is%
\begin{eqnarray}
T_{\mu _{1}\mu _{2}}^{AA} &=&T_{\mu _{1}\mu _{2}}^{VV}-g_{\mu _{1}\mu
_{2}}\left( T^{SS}+T^{PP}\right)  \label{AA2M} \\
&=&+2\Delta _{2\mu _{12}}+4(J_{2\mu _{1}\mu _{2}}+q_{\mu _{2}}J_{2\mu
_{1}})+g_{\mu _{12}}[q^{2}-\left( m_{1}+m_{2}\right) ^{2}]J_{2}  \notag \\
&&+\frac{i}{4\pi }g_{\mu _{12}}\left[ \log \left( m_{1}^{2}/\lambda
^{2}\right) +\log \left( m_{2}^{2}/\lambda ^{2}\right) \right] .  \notag
\end{eqnarray}

From even amplitudes can be to express the odd ones: the first version and
the second version for distinct masses are%
\begin{eqnarray}
(T_{\mu _{1}\mu _{2}}^{AV})_{1} &=&-2\varepsilon _{\mu _{1}\nu _{1}}\Delta
_{2\mu _{2}}^{\nu _{1}}-4\varepsilon _{\mu _{1}\nu _{1}}(J_{2\mu _{2}}^{\nu
_{1}}+q_{\mu _{2}}J_{2}^{\nu _{1}})-\varepsilon _{\mu _{1}\mu
_{2}}[q^{2}-\left( m_{1}-m_{2}\right) ^{2}]J_{2}  \notag \\
&&-\frac{i}{4\pi }\varepsilon _{\mu _{1}\mu _{2}}\left[ \log \left(
m_{1}^{2}/\lambda ^{2}\right) +\log \left( m_{2}^{2}/\lambda ^{2}\right) %
\right]  \label{AV1(2M)}
\end{eqnarray}%
\begin{eqnarray}
(T_{\mu _{1}\mu _{2}}^{AV})_{2} &=&-2\varepsilon _{\mu _{2}\nu _{1}}\Delta
_{2\mu _{1}}^{\nu _{1}}-4\varepsilon _{\mu _{2}\nu _{1}}(J_{2\mu _{1}}^{\nu
_{1}}+q_{\mu _{1}}J_{2}^{\nu _{1}})+\varepsilon _{\mu _{1}\mu
_{2}}[q^{2}-\left( m_{1}+m_{2}\right) ^{2}]J_{2}  \notag \\
&&+\frac{i}{4\pi }\varepsilon _{\mu _{1}\mu _{2}}\left[ \log \left(
m_{1}^{2}/\lambda ^{2}\right) +\log \left( m_{2}^{2}/\lambda ^{2}\right) %
\right] .  \label{AV2(2M)}
\end{eqnarray}

One-index two-point amplitudes coming from RAGFs for odd amplitudes:
Performing the traces and writing $K_{2}=K_{1}+q$\ to get the integrand for $%
\Gamma _{1}\Gamma _{2}=AS$ and $\Gamma _{1}\Gamma _{2}=PV.$ Thus, by our
defintions, we get the finite amplitudes%
\begin{eqnarray}
T_{\mu _{2}}^{PV} &=&2\varepsilon _{\mu _{2}\nu }\left[ \left(
m_{2}-m_{1}\right) J_{2}^{\nu }-m_{1}q^{\nu }J_{2}\right] =-T_{\mu _{2}}^{VP}
\label{PVm2} \\
T_{\mu _{1}}^{AS} &=&-2\varepsilon _{\mu _{1}\nu }\left[ \left(
m_{1}+m_{2}\right) J_{2}^{\nu }+m_{1}q^{\nu }J_{2}\right] =T_{\mu _{1}}^{SA}.
\label{ASm2}
\end{eqnarray}%
The same procedure applies to the two amplitudes coming from RAGFs for even
ones%
\begin{eqnarray}
T_{\mu _{2}}^{SV} &=&2\left[ \left( m_{1}+m_{2}\right) J_{2\mu
_{2}}+m_{1}q_{\mu _{2}}J_{2}\right] =T_{\mu _{2}}^{VS} \\
T_{\mu _{2}}^{PA} &=&-2\left[ \left( m_{2}-m_{1}\right) J_{2\mu
_{2}}-m_{1}q_{\mu _{2}}J_{2}\right] =-T_{\mu _{2}}^{AP}.  \label{AP(m1m2)}
\end{eqnarray}

A last point is the ubiquitous presence of the one-point differences; to
them, we adopt one more notation to simplify the expressions. They are the
same as the equal mass case because they are proportional to $\bar{J}_{1\mu
}\left( k_{1}\right) $ that remain a pure surface-term 
\begin{eqnarray}
T_{\left( -\right) \mu _{i}}^{A} &=&T_{\mu _{i}}^{A}\left( k_{1}\right)
-T_{\mu _{i}}^{A}\left( k_{2}\right) =-2\varepsilon _{\mu _{i}\nu
_{1}}q^{\nu _{2}}\Delta _{2\nu _{2}}^{\nu _{1}}  \label{TA(-)mi} \\
T_{\left( -\right) \mu _{i}}^{V} &=&T_{\mu _{i}}^{V}\left( k_{1}\right)
-T_{\mu _{i}}^{V}\left( k_{2}\right) =2q^{\nu _{1}}\Delta _{2\nu \mu _{i}}.
\end{eqnarray}%
Where we first time define the difference between axial one-point functions
as $T_{\left( -\right) \mu _{i}}^{A}=T_{\mu _{i}}^{A}\left( k_{1}\right)
-T_{\mu _{i}}^{A}\left( k_{2}\right) $. The other one-point function that
appears is the scalar one 
\begin{equation}
T^{S}\left( k_{i}\right) =2m_{i}\bar{J}_{1}\left( k_{i}\right)
=2m_{i}I_{\log }\left( m_{i}^{2}\right) =2m_{i}\left[ I_{\log }\left(
\lambda ^{2}\right) -\left( i/4\pi \right) \log \left( m_{i}^{2}/\lambda
^{2}\right) \right] .  \label{Smi}
\end{equation}

Following this, we will study RAGFs to odd and even amplitudes and the
effects over these relations due to two species of massive fermions in the
currents; since the divergent of the vector current is connected to the
scalar density, it is not strictly conserved now. Later, an expansion of the
discussion of the low-energy theorem to the $AV$ amplitude and its relation
to WI and integration linearity is exposed.

\section{Relations Among Green Functions}

RAFGs will be used as fundamental mathematical tools to provide essential
insights into the behavior of the amplitudes in question and how their
properties relate.

\textbf{Odd amplitudes}: To explore the mechanism, take the definition 
\begin{equation}
t_{\mu _{12}}^{AV}=\mathrm{tr}[\gamma _{\ast }\gamma _{\mu _{1}}S\left(
1\right) \gamma _{\mu _{2}}S\left( 2\right) ]
\end{equation}%
and contract with $q^{\mu _{2}}.$ Next, is it possible to apply the identity 
\begin{equation}
\slashed{q}=\left( \slashed{K}_{2}-m_{2}\right) -\left( \slashed{K}_{1}-m_{1}\right)
+\left( m_{2}-m_{1}\right) .
\end{equation}%
We yield a relation between one- and two-point amplitudes%
\begin{eqnarray}
q^{\mu _{2}}t_{\mu _{12}}^{AV} &=&\mathrm{tr}[\gamma _{\ast }\gamma _{\mu
_{1}}S\left( 1\right) ]-\mathrm{tr}[\gamma _{\ast }\gamma _{\mu _{1}}S\left(
2\right) ]+\left( m_{2}-m_{1}\right) \mathrm{tr}[\gamma _{\ast }\gamma _{\mu
_{1}}S\left( 1\right) S\left( 2\right) ] \\
&=&t_{\left( -\right) \mu _{1}}^{A}+\left( m_{2}-m_{1}\right) t_{\mu
_{1}}^{AS}.
\end{eqnarray}%
The procedure to obtain the vector contraction is similar, namely%
\begin{equation}
q^{\mu _{1}}t_{\mu _{12}}^{AV}=t_{\left( -\right) \mu _{2}}^{A}-\left(
m_{1}+m_{2}\right) t_{\mu _{2}}^{PV}.
\end{equation}%
With further exploration, let us introduce the second contractions for
amplitudes,%
\begin{eqnarray}
q^{\mu _{2}}q^{\mu _{1}}t_{\mu _{12}}^{AV} &=&q^{\mu _{2}}t_{\left( -\right)
\mu _{2}}^{A}-\left( m_{1}+m_{2}\right) t^{PP} \\
q^{\mu _{1}}q^{\mu _{2}}t_{\mu _{12}}^{AV} &=&q^{\mu _{1}}t_{\left( -\right)
\mu _{1}}^{A}+\left( m_{2}-m_{1}\right) t^{SS}.
\end{eqnarray}

In parallel to the equal mass scenario, we have RAGFs for even tensors.
Regarding these RAGFs, we have two-point functions that are not present for
equal masses since they are proportional to the mass difference,%
\begin{eqnarray}
q^{\mu _{1}}t_{\mu _{1}\mu _{2}}^{VV} &=&t_{\left( -\right) \mu
_{2}}^{V}+\left( m_{2}-m_{1}\right) t_{\mu _{2}}^{SV} \\
q^{\mu _{2}}t_{\mu _{1}\mu _{2}}^{VV} &=&t_{\left( -\right) \mu
_{1}}^{V}+\left( m_{2}-m_{1}\right) t_{\mu _{1}}^{VS}.
\end{eqnarray}
We have an additional term proportional to the contraction with $SV$ for two
contractions%
\begin{equation}
q^{\mu _{2}}q^{\mu _{1}}t_{\mu _{1}\mu _{2}}^{VV}=q^{\mu _{2}}t_{\left(
-\right) \mu _{2}}^{V}+\left( m_{2}-m_{1}\right) q^{\mu _{2}}t_{\mu
_{2}}^{SV}.  \label{qqVV}
\end{equation}
For the double-axial one, the simple and double contraction with the
momentum obeys%
\begin{eqnarray}
q^{\mu _{1}}t_{\mu _{1}\mu _{2}}^{AA} &=&t_{\left( -\right) \mu
_{2}}^{V}-\left( m_{1}+m_{2}\right) t_{\mu _{2}}^{PA} \\
q^{\mu _{2}}t_{\mu _{1}\mu _{2}}^{AA} &=&t_{\left( -\right) \mu
_{1}}^{V}+\left( m_{2}+m_{1}\right) t_{\mu _{1}}^{AP}.
\end{eqnarray}
\begin{equation}
q^{\mu _{2}}q^{\mu _{1}}t_{\mu _{1}\mu _{2}}^{AA}=q^{\mu _{2}}t_{\left(
-\right) \mu _{2}}^{V}-\left( m_{1}+m_{2}\right) q^{\mu _{2}}t_{\mu
_{2}}^{PA}.  \label{qqAA}
\end{equation}

\textbf{RAGF Verification}: The axial amplitudes exhibit a nontrivial
behavior, as is expected, since equal masses are a particular case. Here,
the vector and the axial currents are not conserved and are proportional to
a difference and the sum of the masses, 
\begin{eqnarray}
\partial _{\mu }J^{\mu } &=&i\left( m_{a}-m_{b}\right) \bar{\psi}_{a}\psi
_{b} \\
\partial _{\mu }J_{\ast }^{\mu } &=&-i\left( m_{a}+m_{b}\right) \bar{\psi}%
_{a}\psi _{b}.
\end{eqnarray}%
So, in these amplitudes, we will focus our attention now.

\textbf{Version one: }Contracting the expression (\ref{AV1(2M)}), terms
proportional to the vector integral vanishes by the symmetry of indices $%
\varepsilon _{\nu _{1}\nu _{2}}q^{\nu _{2}}J_{2}^{\nu _{1}}=0,$ so we have 
\begin{eqnarray}
q^{\mu _{1}}(T_{\mu _{1}\mu _{2}}^{AV})_{1} &=&2q^{\nu _{2}}\varepsilon
_{\nu _{1}\nu _{2}}\Delta _{2\mu _{2}}^{\nu _{1}}+4\varepsilon _{\nu _{1}\nu
_{2}}q^{\nu _{2}}J_{2\mu _{2}}^{\nu _{1}}+\varepsilon _{\mu _{2}\nu }q^{\nu
}[q^{2}-\left( m_{1}-m_{2}\right) ^{2}]J_{2}  \label{q1TAV(2m)} \\
&&+\left( i/4\pi \right) \varepsilon _{\mu _{2}\nu }q^{\nu }[\log \left(
m_{1}^{2}/\lambda ^{2}\right) +\log \left( m_{2}^{2}/\lambda ^{2}\right) ]. 
\notag
\end{eqnarray}%
We need to exchange the indices in $J_{2\mu _{2}}^{\nu _{1}}$ (in the first
line) employing Schouten identity as 
\begin{equation}
\varepsilon _{\nu _{1}\nu _{2}}q^{\nu _{2}}J_{2\mu _{2}}^{\nu
_{1}}+\varepsilon _{\mu _{2}\nu _{1}}q^{\nu _{2}}J_{2\nu _{2}}^{\nu
_{1}}+\varepsilon _{\nu _{2}\mu _{2}}q^{\nu _{2}}J_{2\nu _{1}}^{\nu _{1}}=0.
\end{equation}%
Two types of contractions arise from equations (\ref{qJ22})-(\ref{gJ2})
introduced in Section (\ref{BasisFI}),%
\begin{eqnarray}
2q^{\nu _{2}}J_{2\nu _{2}}^{\nu _{1}} &=&-\left(
q^{2}+m_{1}^{2}-m_{2}^{2}\right) J_{2}^{\nu _{1}}-\left( i/4\pi \right)
q^{\nu _{1}}\log \left( m_{2}^{2}/\lambda ^{2}\right) \\
g_{\nu _{12}}J_{2}^{\nu _{12}} &=&i/4\pi +m_{1}^{2}J_{2}-\left( i/4\pi
\right) \log \left( m_{2}^{2}/\lambda ^{2}\right) .  \notag
\end{eqnarray}%
Using the results above, we lead to the expression:%
\begin{eqnarray}
q^{\mu _{1}}(T_{\mu _{1}\mu _{2}}^{AV})_{1} &=&2q^{\nu _{2}}\varepsilon
_{\nu _{1}\nu _{2}}\Delta _{2\mu _{2}}^{\nu _{1}}+\frac{i}{\pi }\varepsilon
_{\mu _{2}\nu _{1}}q^{\nu _{1}} \\
&&+\varepsilon _{\mu _{2}\nu _{1}}q^{2}(2J_{2}^{\nu _{1}}+q^{\nu
_{1}}J_{2})+\left( i/4\pi \right) \varepsilon _{\mu _{2}\nu }q^{\nu }\log
\left( m_{1}^{2}/m_{2}^{2}\right)  \notag \\
&&+\varepsilon _{\mu _{2}\nu _{1}}\{2(m_{1}^{2}-m_{2}^{2})J_{2}^{\nu
_{1}}+q^{\nu _{1}}[4m_{1}^{2}-\left( m_{1}-m_{2}\right) ^{2}]J_{2}\}.  \notag
\end{eqnarray}

The identity $\varepsilon _{\lbrack \nu _{1}\nu _{2}}\Delta _{2\mu
_{2}]}^{\nu _{1}}=0$ allows adjusting indices and recognizing one-point
functions together with relation for finite vectors and scalar two-point
integrals of two masses%
\begin{equation}
q^{2}\left( 2J_{2}^{\nu }+q^{\nu }J_{2}\right) =-q^{\nu }\left(
m_{1}^{2}-m_{2}^{2}\right) J_{2}-\left( i/4\pi \right) q^{\nu }\log \left(
m_{1}^{2}/m_{2}^{2}\right) .
\end{equation}%
Doing it some more algebraic operations, we produce the result for this
contraction, 
\begin{eqnarray}
q^{\mu _{1}}(T_{\mu _{1}\mu _{2}}^{AV})_{1} &=&-2\varepsilon _{\mu _{1}\nu
_{1}}q^{\nu _{2}}\Delta _{2\nu _{2}}^{\nu _{1}}+\varepsilon _{\mu _{1}\nu
_{2}}q^{\nu _{2}}(2\Delta _{2\nu _{1}}^{\nu _{1}}+i/\pi ) \\
&&+2\left( m_{1}+m_{2}\right) \varepsilon _{\mu _{2}\nu
_{1}}[(m_{1}-m_{2})J_{2}^{\nu _{1}}+q^{\nu _{1}}m_{1}J_{2}].  \notag
\end{eqnarray}%
Recalling the $PV$ functions of two masses and one-point differences means%
\begin{equation}
q^{\mu _{1}}(T_{\mu _{1}\mu _{2}}^{AV})_{1}=T_{\left( -\right) \mu
_{2}}^{A}-(m_{1}+m_{2})T_{\mu _{2}}^{PV}+\varepsilon _{\mu _{2}\nu
_{2}}q^{\nu _{2}}(2\Delta _{2\nu _{1}}^{\nu _{1}}+i/\pi ).  \label{q1AV(2m)}
\end{equation}

The contraction with the second vertex in the same version starts with%
\begin{eqnarray}
q^{\mu _{2}}(T_{\mu _{1}\mu _{2}}^{AV})_{1} &=&-2\varepsilon _{\mu _{1}\nu
_{1}}q^{\nu _{2}}\Delta _{2\nu _{2}}^{\nu _{1}}-2\varepsilon _{\mu _{1}\nu
_{1}}(2q^{\nu _{2}}J_{2\nu _{2}}^{\nu _{1}}+2q^{2}J_{2}^{\nu _{1}}) \\
&&-\varepsilon _{\mu _{1}\nu _{1}}q^{\nu }[q^{2}-\left( m_{1}-m_{2}\right)
^{2}]J_{2}  \notag \\
&&-\left( i/4\pi \right) \varepsilon _{\mu _{1}\mu _{2}}q^{\mu _{2}}\left[
\log \left( m_{1}^{2}/\lambda ^{2}\right) +\log \left( m_{2}^{2}/\lambda
^{2}\right) \right] ;  \notag
\end{eqnarray}%
here, the reductions occur directly, see $q^{\nu _{2}}J_{2\nu _{2}}^{\nu
_{1}}.$ Using (\ref{qJ22}), we get 
\begin{equation}
q^{\mu _{2}}(T_{\mu _{1}\mu _{2}}^{AV})_{1}=-2\varepsilon _{\mu _{1}\nu
_{1}}q^{\nu _{2}}\Delta _{2\nu _{2}}^{\nu _{1}}-2\varepsilon _{\mu _{1}\nu
_{1}}\left( m_{2}-m_{1}\right) \left[ \left( m_{1}+m_{2}\right) J_{2}^{\nu
_{1}}+m_{1}q^{\nu _{1}}J_{2}\right] ,
\end{equation}%
where all the elements of the RAGF can be identified in the final result, 
\begin{equation}
q^{\mu _{2}}(T_{\mu _{1}\mu _{2}}^{AV})_{1}=T_{\left( -\right) \mu
_{1}}^{A}+\left( m_{2}-m_{1}\right) T_{\mu _{1}}^{AS}.
\end{equation}%
Note that RAGF is automatically satisfied and does not have an additional
term as (\ref{q1AV(2m)}).

\textbf{Version two: }To the second one apply the same considerations:
Starting with $q^{\mu _{1}},$ 
\begin{eqnarray}
q^{\mu _{1}}(T_{\mu _{1}\mu _{2}}^{AV})_{2} &=&-2\varepsilon _{\mu _{2}\nu
_{1}}q^{\mu _{1}}\Delta _{2\mu _{1}}^{\nu _{1}}-\left( i/4\pi \right)
\varepsilon _{\mu _{2}\nu }q^{\nu }\left[ \log \left( m_{1}^{2}/\lambda
^{2}\right) +\log \left( m_{2}^{2}/\lambda ^{2}\right) \right]  \notag \\
&&-4\varepsilon _{\mu _{2}\nu _{1}}(q^{\mu _{1}}J_{2\mu _{1}}^{\nu
_{1}}+q^{2}J_{2}^{\nu _{1}})-\varepsilon _{\mu _{2}\nu }q^{\nu
}[q^{2}-\left( m_{1}+m_{2}\right) ^{2}]J_{2}.
\end{eqnarray}%
Reducing the integrals in a direct way as $q^{\mu _{1}}J_{2\mu _{1}}^{\nu
_{1}}$ and recognizing the terms follows%
\begin{equation}
q^{\mu _{1}}(T_{\mu _{1}\mu _{2}}^{AV})_{1}=T_{\left( -\right) \mu
_{2}}^{A}-\left( m_{1}+m_{2}\right) T_{\mu _{2}}^{PV}.
\end{equation}%
The relation in the second vertex (vectorial) appears to have the same
behavior as the equation (\ref{q1TAV(2m)}). The terms can not be identified
directly; see the equation below 
\begin{eqnarray}
q^{\mu _{2}}(T_{\mu _{1}\mu _{2}}^{AV})_{2} &=&-2q^{\nu _{2}}\varepsilon
_{\nu _{1}\nu _{2}}\Delta _{2\mu _{1}}^{\nu _{1}}+4\varepsilon _{\nu _{1}\nu
_{2}}(J_{2\mu _{1}}^{\nu _{1}}+q_{\mu _{1}}J_{2}^{\nu _{1}})+\varepsilon
_{\mu _{1}\nu }q^{\nu }[q^{2}-\left( m_{1}+m_{2}\right) ^{2}]J_{2}  \notag \\
&&+\left( i/4\pi \right) \varepsilon _{\mu _{1}\nu }q^{\nu }\left[ \log
\left( m_{1}^{2}/\lambda ^{2}\right) +\log \left( m_{2}^{2}/\lambda
^{2}\right) \right]
\end{eqnarray}%
Again, we have to switch the indices of place what will amount to the
apperance of a conditioning factor in its RAGFs, namely, 
\begin{equation}
q^{\mu _{2}}(T_{\mu _{1}\mu _{2}}^{AV})_{2}=T_{\left( -\right) \mu
_{1}}^{A}+\left( m_{2}-m_{1}\right) T_{\mu _{1}}^{AS}+\varepsilon _{\mu
_{1}\nu _{1}}q^{\nu _{1}}(2\Delta _{2\nu _{2}}^{\nu _{2}}+i/\pi ).
\end{equation}

\textbf{Equivalence}: To be complete, we must evaluate the difference
between the versions (\ref{AV1(2M)}) and (\ref{AV2(2M)}). Taking their full
expression and subtracting one from another%
\begin{eqnarray}
(T_{\mu _{1}\mu _{2}}^{AV})_{1}-(T_{\mu _{1}\mu _{2}}^{AV})_{2}
&=&2[\varepsilon _{\mu _{2}\nu _{1}}\Delta _{2\mu _{1}}^{\nu
_{1}}-\varepsilon _{\mu _{1}\nu _{1}}\Delta _{2\mu _{2}}^{\nu
_{1}}]-2\varepsilon _{\mu _{1}\mu _{2}}[q^{2}-\left(
m_{1}^{2}+m_{2}^{2}\right) ]J_{2}  \notag \\
&&+4[\varepsilon _{\mu _{2}\nu _{1}}(J_{2\mu _{1}}^{\nu _{1}}+q_{\mu
_{1}}J_{2}^{\nu _{1}})-\varepsilon _{\mu _{1}\nu _{1}}(J_{2\mu _{2}}^{\nu
_{1}}+q_{\mu _{2}}J_{2}^{\nu _{1}})]  \notag \\
&&-(i/2\pi )\varepsilon _{\mu _{1}\mu _{2}}[\log (m_{1}^{2}/\lambda
^{2})+\log (m_{2}^{2}/\lambda ^{2})];
\end{eqnarray}%
thereby employing the Schouten identity in the second line above, we have 
\begin{equation*}
4\varepsilon _{\mu _{2}\nu _{1}}(J_{2\mu _{1}}^{\nu _{1}}+q_{\mu
_{1}}J_{2}^{\nu _{1}})-4\varepsilon _{\mu _{1}\nu _{1}}(J_{2\mu _{2}}^{\nu
_{1}}+q_{\mu _{2}}J_{2}^{\nu _{1}})=4\varepsilon _{\mu _{2}\mu _{1}}(J_{2\nu
_{1}}^{\nu _{1}}+q_{\nu _{1}}J_{2}^{\nu _{1}}).
\end{equation*}%
With the help of reductions, it is relatively easy to show exactly%
\begin{equation}
(T_{\mu _{1}\mu _{2}}^{AV})_{1}-(T_{\mu _{1}\mu
_{2}}^{AV})_{2}=2[\varepsilon _{\mu _{2}\nu _{1}}\Delta _{2\mu _{1}}^{\nu
_{1}}-\varepsilon _{\mu _{1}\nu _{1}}\Delta _{2\mu _{2}}^{\nu _{1}}]+\left(
i/\pi \right) \varepsilon _{\mu _{2}\mu _{1}}.
\end{equation}%
Apllying $\varepsilon _{\lbrack \mu _{2}\nu _{1}}\Delta _{2\mu _{1}]}^{\nu
_{1}}=0$, this result naturally also may be expressed as%
\begin{equation}
(T_{\mu _{1}\mu _{2}}^{AV})_{1}-(T_{\mu _{1}\mu _{2}}^{AV})_{2}=\varepsilon
_{\mu _{2}\mu _{1}}(2\Delta _{2\nu _{1}}^{\nu _{1}}+i/\pi ).
\end{equation}

Another way to systematize the RAGFs that will be used in Chapter (\ref%
{modlDef}) is to notice that every time the index is contracted with the one
remaining in the even amplitude, the relation is satisfied. Therefore we can
use the above relation to exchange the versions when contracting with the
index in the vertex used to define the version%
\begin{eqnarray}
q^{\mu _{1}}(T_{\mu _{1}\mu _{2}}^{AV})_{1} &=&q^{\mu _{1}}(T_{\mu _{1}\mu
_{2}}^{AV})_{2}+\varepsilon _{\mu _{2}\mu _{1}}q^{\mu _{1}}\left( 2\Delta
_{2\nu }^{\nu }+i/\pi \right)  \notag \\
&=&T_{\left( -\right) \mu _{2}}^{A}-2\left( m_{1}+m_{2}\right) T_{\mu
_{2}}^{PV}+\varepsilon _{\mu _{2}\nu }q^{\nu }\left( 2\Delta _{2\alpha
}^{\alpha }+i/\pi \right)
\end{eqnarray}%
\begin{eqnarray}
q^{\mu _{2}}(T_{\mu _{1}\mu _{2}}^{AV})_{2} &=&q^{\mu _{2}}(T_{\mu _{1}\mu
_{2}}^{AV})_{1}-\varepsilon _{\mu _{2}\mu _{1}}q^{\mu _{2}}(2\Delta
_{2\alpha }^{\alpha }+i/\pi )  \notag \\
&=&T_{\left( -\right) \mu _{2}}^{A}-2(m_{1}-m_{2})T_{\mu
_{2}}^{AS}+\varepsilon _{\mu _{1}\nu }q^{\nu }(2\Delta _{2\alpha }^{\alpha
}+i/\pi ).
\end{eqnarray}%
These features are notable in two dimensions. In four dimensions, we also
establish relations among versions (three of them). However, in that
scenario, the odd amplitudes do not collapse in a direct connection to even
ones. We have to check the RAGFs explicitly.

\textbf{Even Amplitudes}: The relations to the even amplitudes are easy to
check, 
\begin{eqnarray}
q^{\mu _{1}}T_{\mu _{1}\mu _{2}}^{VV} &=&2q^{\mu _{1}}\Delta _{2\mu
_{12}}+\left( i/4\pi \right) q_{\mu _{2}}\left[ \log \left(
m_{1}^{2}/\lambda ^{2}\right) +\log \left( m_{2}^{2}/\lambda ^{2}\right) %
\right] \\
&&+4\left( q^{\nu _{1}}J_{2\mu _{1}\nu _{1}}+q_{\mu _{2}}q^{\nu _{1}}J_{2\nu
_{1}}\right) +q_{\mu _{2}}[q^{2}-\left( m_{1}-m_{2}\right) ^{2}]J_{2}. 
\notag
\end{eqnarray}%
Using the same operations in $J_{2}$-integrals as applied to the odd
amplitudes follows%
\begin{eqnarray}
q^{\mu _{1}}T_{\mu _{1}\mu _{2}}^{VV} &=&T_{\left( -\right) \mu
_{2}}^{V}-2\left( m_{1}-m_{2}\right) \left[ \left( m_{1}+m_{2}\right)
J_{2\mu _{2}}+m_{1}q_{\mu _{2}}J_{2}\right] \\
&=&T_{\left( -\right) \mu _{2}}^{V}+\left( m_{2}-m_{1}\right) T_{\mu
_{2}}^{SV}  \notag \\
q^{\mu _{2}}T_{\mu _{1}\mu _{2}}^{VV} &=&T_{\left( -\right) \mu
_{1}}^{V}+\left( m_{2}-m_{1}\right) T_{\mu _{1}}^{VS}.
\end{eqnarray}%
For the $AA$-amplitude (\ref{AA2M}), the two relations follows by%
\begin{eqnarray}
q^{\mu _{1}}T_{\mu _{1}\mu _{2}}^{AA} &=&T_{\left( -\right) \mu
_{2}}^{V}-\left( m_{1}+m_{2}\right) T_{\mu _{2}}^{PA}  \notag \\
q^{\mu _{2}}T_{\mu _{1}\mu _{2}}^{AA} &=&T_{\left( -\right) \mu
_{2}}^{V}+\left( m_{2}+m_{1}\right) T_{\mu _{1}}^{AP}.
\end{eqnarray}%
See $PA$ in (\ref{AP(m1m2)}); we could have expressed only in term of one
since they differ by a sign.

The double-contraction for the even amplitudes (\ref{qqVV}) and (\ref{qqAA})
is associated with finite one-rank amplitudes. By themselves their relations
are%
\begin{eqnarray}
q^{\mu _{1}}t_{\mu _{1}}^{VS} &=&+\left( m_{2}-m_{1}\right)
t^{SS}+[t^{S}\left( 1\right) -t^{S}\left( 2\right) ]  \label{qVS2m} \\
q^{\mu _{1}}t_{\mu _{1}}^{AP} &=&-\left( m_{2}+m_{1}\right)
t^{PP}-[t^{S}\left( 1\right) +t^{S}\left( 2\right) ].  \label{qAP2m}
\end{eqnarray}%
The LHS is finite, but the RHS shows a log-divergent object $I_{\log }$.
Nonetheless, in our strategy, it is an exact and straightforward algebraic
step to verify them. Using as an example the following equation%
\begin{equation}
q^{\mu _{1}}T_{\mu _{1}}^{VS}=2\left( m_{1}+m_{2}\right) q^{\mu _{1}}J_{2\mu
_{1}}+2m_{1}q^{2}J_{2}.
\end{equation}%
Applying Eq. (\ref{qJ21}) in order to reduce the two-masses vector integral,
we have%
\begin{equation}
q^{\mu _{1}}T_{\mu _{1}}^{VS}=-\left( m_{2}-m_{1}\right) [q^{2}-\left(
m_{1}+m_{2}\right) ^{2}]J_{2}+\left( i/4\pi \right) \left(
m_{1}+m_{2}\right) \log \left( m_{2}^{2}/m_{1}^{2}\right) .
\end{equation}%
The last term can be manipulated by the scale relation (\ref{scRel}), viz., 
\begin{equation}
\left( i/4\pi \right) \log \left( m_{2}^{2}/m_{1}^{2}\right) =I_{\log
}\left( m_{1}^{2}\right) -I_{\log }\left( m_{2}^{2}\right) ,
\end{equation}%
which through an organization of the terms produces the following expression%
\begin{eqnarray}
q^{\mu _{1}}T_{\mu _{1}}^{VS} &=&\left( m_{2}-m_{1}\right) \{[-q^{2}+\left(
m_{1}+m_{2}\right) ^{2}]J_{2}+\left[ I_{\log }(m_{1}^{2})+I_{\log
}(m_{2}^{2})\right] \} \\
&&+2m_{1}I_{\log }(m_{1}^{2})-2m_{2}I_{\log }\left( m_{2}^{2}\right) . 
\notag
\end{eqnarray}%
We can rewrite the first term as the $SS$ amplitude (\ref{SS2m}) and
organize the result%
\begin{eqnarray}
q^{\mu _{1}}T_{\mu _{1}}^{VS} &=&\left( m_{2}-m_{1}\right) T^{SS} \\
&&+2m_{1}[I_{\log }(\lambda ^{2})-(i/4\pi )\log \left( m_{1}^{2}/\lambda
^{2}\right) ]  \notag \\
&&-2m_{2}[I_{\log }(\lambda ^{2})-(i/4\pi )\log \left( m_{2}^{2}/\lambda
^{2}\right) ].  \notag
\end{eqnarray}
The scalar one-point function is given in (\ref{Smi}). Hence we verify that
the two last lines correspond to the difference between them, representing
the satisfaction of its RAGF,%
\begin{equation}
q^{\mu _{1}}T_{\mu _{1}}^{VS}=\left( m_{2}-m_{1}\right) T^{SS}+T^{S}\left(
1\right) -T^{S}\left( 2\right) .  \label{qVS}
\end{equation}%
Note that in these case, the difference between scalar one-point functions
does not cancel and depends on the individual masses.

The $q^{\mu }T_{\mu }^{AP}$ works under the same manipulations used in $VS$,
starting with 
\begin{equation}
q^{\mu _{1}}T_{\mu _{1}}^{AP}=2\left( m_{2}-m_{1}\right) q^{\mu _{1}}J_{2\mu
_{1}}-2m_{1}q^{2}J_{2}.
\end{equation}%
Through of the relation estabilish in (\ref{qJ21}), the equation above
results in%
\begin{equation}
q^{\mu _{1}}T_{\mu _{1}}^{AP}=\left( m_{1}+m_{2}\right) [\left(
m_{1}-m_{2}\right) ^{2}-q^{2}]J_{2}+\left( i/4\pi \right) \left(
m_{2}-m_{1}\right) \log \left( m_{2}^{2}/m_{1}^{2}\right) .
\end{equation}%
Rewriten the first term by (\ref{PP2m}) and organize the result%
\begin{eqnarray}
q^{\mu _{1}}T_{\mu _{1}}^{AP} &=&-\left( m_{1}+m_{2}\right) T^{PP} \\
&&-2m_{1}[I_{\log }(\lambda ^{2})-\left( i/4\pi \right) \log \left(
m_{1}^{2}/\lambda ^{2}\right) ] \\
&&-2m_{2}[I_{\log }(\lambda ^{2})-\left( i/4\pi \right) \log \left(
m_{2}^{2}/\lambda ^{2}\right) ].  \notag
\end{eqnarray}%
The two last lines now appear as the sum of scalar one-point functions,
namely%
\begin{equation}
q^{\mu _{1}}T_{\mu _{1}}^{AP}=-\left( m_{1}+m_{2}\right) T^{PP}-[T^{S}\left(
1\right) +T^{S}\left( 2\right) ].  \label{qAP}
\end{equation}%
For equal masses, the term to one-point functions is proportional to the
masses' sum.

As explored in the chapter for equal masses, it is possible to obtain
properties for the amplitudes by combining their general tensor structures
with their symmetry relations or Ward's identities. These results are not
restricted to perturbative solutions and should remain valid even for exact
solutions. The $VS$ function is constructed from a vector with an external
vector, $T_{\mu }^{VS}=q_{\mu }F_{1}(q^{2}),$ where $F_{1}(q^{2})$ is an
invariant function. This form allows us to state a low-energy limit for this
amplitude contracting the equation, viz, $q^{\mu }T_{\mu
}^{VS}=q^{2}F_{1}(q^{2}).$ Then, $\left. q^{\mu }T_{\mu }^{VS}\right\vert
_{q^{2}=0}=0$, since $F_{1}(q^{2})$ does not poles at $q^{2}=0.$

In this way, to obtain an interpretation relation from RHS of relations (\ref%
{qVS}) and (\ref{qAP}), let us analyze the $SV$ and $AP$-amplitudes in the
limit in kinematical point. We have that $\lim_{q^{2}\rightarrow 0}(q^{\mu
_{1}}T_{\mu _{1}}^{VS})=0$ is satisfied, since the \thinspace $J_{2}=i/4\pi
Z_{0}^{\left( -1\right) }$ where the function $Z_{0}^{\left( -1\right) }$ in
this point is given by 
\begin{equation}
Z_{0}^{\left( -1\right) }\left( 0,m_{2},m_{1}\right) =\frac{1}{\left(
m_{1}^{2}-m_{2}^{2}\right) }\log \frac{m_{2}^{2}}{m_{1}^{2}}.
\end{equation}%
\ From Eq (\ref{qVS}) and the explicit result (\ref{SS2m}), follows%
\begin{eqnarray}
\left. (m_{2}-m_{1})T^{SS}\right\vert _{q^{2}=0} &=&2(m_{2}-m_{1})I_{\log
}(\lambda ^{2})+\left( i/2\pi \right) [m_{1}\log (m_{1}^{2}/\lambda
^{2})-m_{2}\log (m_{2}^{2}/\lambda ^{2})] \\
&=&-\left[ T^{S}\left( 1\right) -T^{S}\left( 2\right) \right] ,
\end{eqnarray}%
therefore $\left. q^{\mu _{1}}T_{\mu _{1}}^{VS}\right\vert _{q^{2}=0}=0.$
The low-energy theorem for $T_{\mu _{1}}^{AP}$ is also fulfilled because the
same operations leads us to%
\begin{equation}
\left. \left( m_{1}+m_{2}\right) T^{PP}\right\vert _{q^{2}=0}=-\left[
T^{S}\left( 1\right) +T^{S}\left( 2\right) \right] .
\end{equation}

We saw that the one-point functions were indispensable for satisfying the
deduced kinematical implication based on the tensor structure for amplitude
with one Lorentz index. That is the opposite of the situation for amplitudes
with two indices. The reason for the need for scalar one-point functions can
be understood by analyzing the canonical structure of WIs for multiple
masses. There, the meaning of these terms finds a justification.

\section{Ward Identities: Two Masses}

Here we will argue why the scalar one-point functions are part of WIs from
one-index two-point functions. We take free fields that generate our
amplitudes, of particular interest to our purposes, obeying the equal-time
anticommutation relation%
\begin{equation}
\{\psi _{i}^{\alpha }\left( y\right) ,\psi _{j}^{\dagger \kappa }\left(
x\right) \}=\delta _{ij}\delta ^{\alpha \kappa }\delta \left( \mathbf{x}-%
\mathbf{y}\right) ,
\end{equation}%
where $i$ and $j$ refer to different species of fermions ($\psi _{1}$ and $%
\psi _{2}$), all other anticommutators are null. Fermionic densities,
defined as a set of bilinear in the fermions, are 
\begin{equation}
J^{\Gamma _{i}}=\bar{\psi}_{2}\Gamma _{i}\psi _{1},\quad \text{and }\quad
J^{\Gamma _{i}\dagger }=\bar{\psi}_{1}(\gamma _{0}\Gamma _{i}^{\dagger
}\gamma _{0})\psi _{2},
\end{equation}%
where $\Gamma _{i}$ belong to set of the vertices given by (\ref%
{SetofVertexes}). Explicitly we have 
\begin{equation*}
V^{\mu }=\left( \bar{\psi}_{2}\gamma ^{\mu }\psi _{1}\right) ,\quad A_{\ast
}^{\mu }=\left( \bar{\psi}_{2}\gamma _{\ast }\gamma ^{\mu }\psi _{1}\right)
,\quad S=\left( \bar{\psi}_{2}\psi _{1}\right) ,\quad P=\left( \bar{\psi}%
_{2}\gamma _{\ast }\psi _{1}\right) .
\end{equation*}%
The adjoints yield the same matrices $\gamma _{0}\Gamma _{i}^{\dagger
}\gamma _{0}=\Gamma _{i}$ with the exception of pseudo-scalar one $\gamma
_{0}\gamma _{\ast }\gamma _{0}=-\gamma _{\ast }$. We adopted a different
notation here to avoid confusion with $J$-integrals. Two-point functions can
be seen in position space as%
\begin{equation}
T^{\Gamma _{1}\Gamma _{2}}\left( x-y\right) =\mathrm{tr}\left[ \Gamma
_{1}S_{F}\left( x-y,m_{1}\right) \Gamma _{2}S_{F}\left( y-x,m_{2}\right) %
\right] =-\left\langle J^{\Gamma _{1}}\left( x\right) J^{\Gamma _{2}\dagger
}\left( y\right) \right\rangle .  \label{ps2PT}
\end{equation}%
The minus sign occurs because Wick contraction yields $i$ times our
propagator definition, and $\left\langle \cdot \right\rangle =\left\langle
0\left\vert \mathcal{T}\left\{ \cdot \right\} \right\vert 0\right\rangle $
is an abbreviation for a time-ordered product. We recovered the letter for
the Feynman propagator to not mistake it for scalar density.

To clarify the WIs for two-point functions with one-index, we use Dirac
equations, 
\begin{equation}
\gamma ^{\mu }\partial _{\mu }\psi _{1}=-im_{1}\psi _{1};\quad \left(
\partial _{\mu }\bar{\psi}_{2}\right) \gamma ^{\mu }=im_{2}\bar{\psi}_{2}.
\end{equation}%
Through them, we obtain that the vector and axial currents satisfy%
\begin{eqnarray}
\partial _{\mu }V^{\mu } &=&+i\left( m_{2}-m_{1}\right) S\left( x\right)
=i\left( m_{2}-m_{1}\right) \bar{\psi}_{2}\psi _{1} \\
\partial _{\mu }A^{\mu } &=&-i\left( m_{2}+m_{1}\right) P\left( x\right)
=-i\left( m_{2}+m_{1}\right) \bar{\psi}_{2}\gamma _{\ast }\psi .
\end{eqnarray}%
The next step is to notice that when we perform space-time derivatives in
the time ordering for densities carrying Lorentz indices, equal-time
commutators will appear; to them, we will use the identity 
\begin{equation}
\left[ AB,CD\right] =-AC\left\{ B,D\right\} +A\left\{ B,C\right\} D-C\left\{
D,A\right\} B+\left\{ C,A\right\} DB.
\end{equation}%
Necessary formal commutators arise to time components, but in general, we
will have%
\begin{eqnarray}
\left[ J^{\Gamma _{1}}\left( x\right) ,J^{\Gamma _{2}\dagger }\left(
y\right) \right] _{x^{0}=y^{0}} &=&\left[ \bar{\psi}_{2}\left( x\right)
\Gamma ^{1}\psi _{1}\left( x\right) ,\bar{\psi}_{1}\left( y\right) \Gamma
^{2}\psi _{2}\left( y\right) \right] \\
&=&\left[ \bar{\psi}_{2}\left( x\right) \Gamma ^{1}\gamma ^{0}\Gamma
^{2}\psi _{2}\left( y\right) -\bar{\psi}_{1}\left( y\right) \Gamma
^{2}\gamma ^{0}\Gamma ^{1}\psi _{1}\left( x\right) \right] \delta ^{2}\left(
x-y\right) .  \notag
\end{eqnarray}

The commutators necessary to point out the differences between symmetry
relations of two and one indices two-point functions (satisfied for $VS$ and 
$AP$ amplitudes) are%
\begin{eqnarray}
\left[ V_{0}\left( x\right) ,V^{\nu \dagger }\left( y\right) \right] &=&%
\left[ \bar{\psi}_{2}\left( x\right) \gamma ^{\nu }\psi _{2}\left( y\right) -%
\bar{\psi}_{1}\left( y\right) \gamma ^{\nu }\psi _{1}\left( x\right) \right]
\delta ^{2}\left( x-y\right)  \label{VVC} \\
\left[ A_{0}\left( x\right) ,V^{\nu \dagger }\left( y\right) \right] &=&%
\left[ \bar{\psi}_{2}\left( x\right) \gamma _{\ast }\gamma ^{\nu }\psi
_{2}\left( y\right) -\bar{\psi}_{1}\left( y\right) \gamma _{\ast }\gamma
^{\nu }\psi _{1}\left( x\right) \right] \delta ^{2}\left( x-y\right)
\label{AVC} \\
\left[ V_{0}\left( x\right) ,S^{\dagger }\left( y\right) \right] &=&\left[ 
\bar{\psi}_{2}\left( x\right) \psi _{2}\left( y\right) -\bar{\psi}_{1}\left(
y\right) \psi _{1}\left( x\right) \right] \delta ^{2}\left( x-y\right)
\label{VSC} \\
\left[ A_{0}\left( x\right) ,P^{\dagger }\left( y\right) \right] &=&\left[ -%
\bar{\psi}_{2}\left( x\right) \psi _{2}\left( y\right) -\bar{\psi}_{1}\left(
y\right) \psi _{1}\left( x\right) \right] \delta ^{2}\left( x-y\right) ,
\end{eqnarray}%
all evaluated in $x_{0}=y_{0}$. Observe that densities in LHS carry two
distinct masses, and the RHS bilinears appear with only one mass, though the
two terms carry a distinct mass.

Taking the derivative of $VV$, using the motion's equation to the currents,
and observing the commutator at equal times (\ref{VVC}), we get the formal
result 
\begin{eqnarray}
\partial _{x}^{\mu }\left\langle V_{\mu }\left( x\right) V_{\nu }^{\dagger
}\left( y\right) \right\rangle &=&i\left( m_{2}-m_{1}\right) \left\langle
S\left( x\right) V_{\nu }^{\dagger }\left( y\right) \right\rangle \\
&&+\left[ \left\langle \bar{\psi}_{2}\left( x\right) \gamma _{\nu }\psi
_{2}\left( y\right) \right\rangle -\left\langle \bar{\psi}_{1}\left(
y\right) \gamma _{\nu }\psi _{1}\left( x\right) \right\rangle \right] \delta
^{2}\left( x-y\right) ,  \notag
\end{eqnarray}%
where $\partial _{x}^{\mu }=\partial /\partial x_{\mu }$. The Ward identity
for equal masses came from cancellation in the last line since the terms
become equal, and we are ignoring Schwinger's terms. As for two masses, it
arises from Lorentz symmetry that implies the vanishing of one-point vector
function individually, e.g., $\left\langle 0\left\vert \bar{\psi}_{1}\left(
x\right) \gamma ^{\nu }\psi _{1}\left( y\right) \right\vert 0\right\rangle
=0 $. It is understood by using the generator of translations in a vector
operator $O^{\mu }\left( x\right) $, 
\begin{equation}
\left\langle 0\left\vert O^{\mu }\left( x\right) \right\vert 0\right\rangle
=\left\langle 0\left\vert e^{-iP\cdot x}O^{\mu }\left( 0\right) e^{iP\cdot
x}\right\vert 0\right\rangle =\left\langle 0\left\vert O^{\mu }\left(
0\right) \right\vert 0\right\rangle =0.
\end{equation}
Furthermore, because o Lorentz symmetry, such a constant vector must vanish.
Note that this constraint may not be valid perturbatively. Putting aside
that, the proposed WI is%
\begin{equation}
\partial _{x}^{\mu }\left\langle V_{\mu }\left( x\right) V_{\nu }^{\dagger
}\left( y\right) \right\rangle =i\left( m_{2}-m_{1}\right) \left\langle
S\left( x\right) V_{\nu }^{\dagger }\left( y\right) \right\rangle .
\end{equation}%
In it, only the contribution of motion's equations plays a part;
additionally, if the correlator involves one axial and one vector current,
the argument for vanishing the one-point amplitudes in (\ref{AVC}) is the
same.

\textit{The situation is quite different for }$VS$\textit{\ and }$AP$\textit{%
\ functions}; symmetry constraints pass%
\begin{eqnarray}
\partial _{\mu }^{x}\left\langle V^{\mu }\left( x\right) S^{\dagger }\left(
y\right) \right\rangle &=&i\left( m_{2}-m_{1}\right) \left\langle S\left(
x\right) S^{\dagger }\left( y\right) \right\rangle \\
&&+\left\langle \bar{\psi}_{2}\left( x\right) \psi _{2}\left( y\right) -\bar{%
\psi}_{1}\left( y\right) \psi _{1}\left( x\right) \right\rangle \delta
^{2}\left( x-y\right) ,  \notag
\end{eqnarray}%
where the commutator $\left[ V_{0}\left( x\right) ,S^{\dagger }\left(
y\right) \right] =($\ref{VSC}$)$ generates one-point scalar functions that
formally cancel each other for equal masses, but in that case, the $VS$%
-amplitude is null. Nonetheless, in $AP$ (or $PA$), they appear in a
non-canceling way 
\begin{eqnarray}
\partial _{\mu }^{x}\left\langle A^{\mu }\left( x\right) P^{\dagger }\left(
y\right) \right\rangle &=&-i\left( m_{1}+m_{2}\right) \left\langle P\left(
x\right) P^{\dagger }\left( y\right) \right\rangle \\
&&-\left\langle \bar{\psi}_{2}\left( x\right) \psi _{2}\left( y\right) +\bar{%
\psi}_{1}\left( y\right) \psi _{1}\left( x\right) \right\rangle \delta
^{2}\left( x-y\right) .  \notag
\end{eqnarray}%
The commutator yields a sum, not a cancellation, for equal masses. So the
canonical commutator terms appear and may not be zero due to other symmetry
arguments.

As in the two masses scenario, the scalar one-point functions are not
removed from expression to Ward identities and are an integral part of them.
For one species of fermions, the commutator of vector (and axial) densities
being zero is a particular phenomenon; this term comes from canonical
algebra. Their eliminations are to be accounted for by additional arguments,
e.g., Lorentz invariance. Such statements are not present against scalar
densities that, in turn, guarantee a low-energy theorem to the $VS$ and $AP$
amplitudes.

To visualize consequences of this reasoning line and connect it with
calculated expression, let us remind that Wick contractions yield $i$ times
our definition of the propagator, 
\begin{equation}
\left\langle 0\left\vert T\psi ^{\alpha }\left( x\right) \bar{\psi}^{\kappa
}\left( y\right) \right\vert 0\right\rangle =iS_{F}^{\alpha \kappa }\left(
x-y,m_{i}\right) .
\end{equation}%
Therefore, Fourier transforming the two-point functions (\ref{ps2PT}), 
\begin{eqnarray}
T^{\Gamma _{1}\Gamma _{2}}\left( q\right) &=&\int \mathrm{d}^{2}ze^{-iq\cdot
z}\left[ T^{\Gamma _{1}\Gamma _{2}}\left( z\right) \right] =-\int \mathrm{d}%
^{2}ze^{-iq\cdot z}\left\langle J^{\Gamma _{1}}\left( x\right) J^{\Gamma
_{2}\dagger }\left( y\right) \right\rangle \\
&=&\int \frac{\mathrm{d}^{2}k}{\left( 2\pi \right) ^{2}}\mathrm{tr}\left[
\Gamma _{1}S_{F}\left( k+k_{1},m_{1}\right) \Gamma _{2}S_{F}\left(
k+k_{2},m_{2}\right) \right] ,
\end{eqnarray}%
where $z=x-y$ and $k_{2}-k_{1}=q$. In the case of double-vector $VV$ (\ref%
{VV2m}) and $VS$, we may write the motion's equation, and the commutation
relations furnish the formal equations%
\begin{eqnarray}
\partial _{z}^{\mu }T_{\mu \nu }^{VV}\left( z\right) &=&i\left(
m_{2}-m_{1}\right) T_{\nu }^{SV}\left( z\right) \\
&&+i\mathrm{tr}\left[ \gamma _{\nu }S_{F}\left( z,m_{1}\right) \right]
\delta ^{2}\left( z\right) -i\mathrm{tr}\left[ \gamma _{\nu }S_{F}\left(
-z,m_{2}\right) \right] \delta ^{2}\left( z\right) ,  \notag \\
\partial _{z}^{\mu }T_{\mu }^{VS}\left( z\right) &=&i\left(
m_{2}-m_{1}\right) T^{SS}\left( z\right) \\
&&+i\mathrm{tr}\left[ S_{F}\left( z,m_{1}\right) \right] \delta ^{2}\left(
z\right) -i\mathrm{tr}\left[ S_{F}\left( -z,m_{2}\right) \right] \delta
^{2}\left( z\right) ,  \notag
\end{eqnarray}%
whose Fourier transform returns an expression where we do not neglect any
term,%
\begin{eqnarray}
q^{\mu }T_{\mu \nu }^{VV}\left( k_{1},k_{2}\right) &=&\left(
m_{2}-m_{1}\right) T_{\nu }^{SV} \\
&&+\int \frac{\mathrm{d}^{2}k}{\left( 2\pi \right) ^{2}}\left\{ \mathrm{tr}%
\left[ \gamma ^{\nu }S_{F}\left( K_{1},m_{1}\right) \right] -\mathrm{tr}%
\left[ \gamma ^{\nu }S_{F}\left( K_{2},m_{2}\right) \right] \right\}  \notag
\\
q^{\mu }T_{\mu }^{VS}\left( k_{1},k_{2}\right) &=&\left( m_{2}-m_{1}\right)
T^{SS} \\
&&+\int \frac{\mathrm{d}^{2}k}{\left( 2\pi \right) ^{2}}\left\{ \mathrm{tr}%
\left[ S_{F}\left( K_{1},m_{1}\right) \right] -\mathrm{tr}\left[ S_{F}\left(
K_{2},m_{2}\right) \right] \right\} .  \notag
\end{eqnarray}

Recapitulating the facts, the parts from the time component of the
commutator of currents with vector and axial currents formally cancel for
one species of massive fermions. We got a WI whose contribution comes only
from motion equations. On the other hand, for two masses, formal Lorentz
invariance requires the vector and axial one-point functions to vanish as
well, and thus they are not part of the WI. Indeed using our strategy, we
saw in momentum space that they become pure surface-term that can be made
zero. Additionally, the anomalies of the odd amplitudes are related to the
impossibility of the formal/canonical WI being realized, which we establish
as a consequence of a Low energy implication from a finite function; see the
next section where that point is discussed and the relation with the
linearity of integration.

In contrast, the commutator of the time component of the currents with
scalar densities, or pseudo-scalar ones, giving rise to scalar one-point
functions, besides the term coming from the motion's equations, is not
necessarily zero. The point is that when the masses are equal, that
difference of amplitudes vanishes in pairs for $SV$ and sum for $AP$. They
do not cancel in any situation for distinct masses and can not be zero
because they are not a constant function of their mass parameters.

One way to see the difference between the two situations is to take into
account that for even dimension, there is a matrix such that $C\gamma _{\mu
}C^{-1}=-\gamma _{\mu }^{T}$, the charge conjugation matrix. This matrix
implies a behavior to the vertexes, viz.,%
\begin{equation}
C\left[ 1,\gamma _{\ast },\gamma _{\mu },\gamma _{\ast }\gamma _{\mu }\right]
C^{-1}=[1,-\gamma _{\ast }^{T},-\gamma _{\mu }^{T},-\left( \gamma _{\ast
}\gamma _{\mu }\right) ^{T}].
\end{equation}%
It is direct to see that the propagator obeys $%
CS_{F}(K_{i},m_{i})C^{-1}=S_{F}^{T}(-K_{i},m_{i})$. Applying it to the
definition of one-point function, we have 
\begin{equation}
t^{\Gamma _{1}}=\mathrm{tr}[\Gamma _{1}S_{F}\left( K_{i},m_{i}\right) ]=%
\mathrm{tr}[C\Gamma _{1}C^{-1}CS_{F}\left( K_{i},m_{i}\right) C^{-1}].
\end{equation}%
Using the trace properties and as well the relation for matrices 
\begin{equation}
\mathrm{tr}\left( B_{1}^{T}\cdots B_{n}^{T}\right) =\mathrm{tr}\left(
B_{n}\cdots B_{1}\right) ^{T}=\mathrm{tr}\left( B_{n}\cdots B_{1}\right) ,
\end{equation}%
we may write from general considerations established above 
\begin{equation*}
t^{\Gamma _{1}}=\mathrm{tr}[\left( C\Gamma _{1}C^{-1}\right) ^{T}S_{F}\left(
-K_{i},m_{i}\right) ].
\end{equation*}

At this point, note that there is a sign change to the pseudo-scalar,
vector, and axial vertices. Then integrating the result above, we have%
\begin{equation}
T^{\Gamma _{1}}=-\int \frac{\mathrm{d}^{2}k}{\left( 2\pi \right) ^{2}}%
\mathrm{tr}[\Gamma _{1}S_{F}\left( -k-k_{i},m_{i}\right) ].
\end{equation}%
Reflecting on the integration variable and shifting, as the hypothesis, we
get%
\begin{equation}
T^{\Gamma _{1}}\left( k_{i}\right) =-\int \frac{\mathrm{d}^{2}k}{\left( 2\pi
\right) ^{2}}\mathrm{tr}[\Gamma _{1}S_{F}\left( K_{i},m_{i}\right)
]=-T^{\Gamma _{1}}\left( k_{i}\right) .
\end{equation}%
That implies that axial and vector one-point functions must vanish
identically, as $T^{P}=0$ already in the trace level. As trivial as it may
appear, this is not a direct consequence of Feynman's rules; the possibility
of shifting is coded in the intrinsic surface term present in the
amplitudes, which is why the $T^{V},$ $T^{A}$ are only surface terms.
Nevertheless, it does not mean these parts in the amplitudes could not be
non-zero and violate WIs.

For instance, the scalar function may have surface terms in 4D, but it is
not obliged to be identically zero by translational invariance. In that
case, the above equation picks up a positive sign. Those $T^{S}\left(
k_{i},m_{i}\right) $ amplitudes show a masse dependence through a logarithm.
Since they are proportional to the basic divergent object, taking its
derivative,%
\begin{equation}
\frac{\partial I_{\log }\left( m_{i}^{2}\right) }{\partial m_{i}^{2}}=-\frac{%
i}{4\pi }\frac{1}{m_{i}^{2}}.
\end{equation}%
The integration picks up an arbitrary constant $I_{\log }\left(
m_{i}^{2}\right) =\left( i4\pi \right) ^{-1}\log \left( m_{i}^{2}/\lambda
_{0}^{2}\right) $ that could help with cancellations; however, in
combinations, this is not possible, see%
\begin{equation}
I_{\log }(m_{i}^{2})-I_{\log }(m_{j}^{2})=-\frac{i}{4\pi }\log \frac{%
m_{i}^{2}}{m_{j}^{2}}.
\end{equation}%
However, the scalar-one cancels each other for equal masses when they arise
from a commutator of vector currents. When the masses are unequal, there is
no reason for them to disappear in the perturbative expression. They are
integral parts of WI and necessary for their consistency. The low-energy
theorem derived for them requires that part to occur 
\begin{equation}
q^{\mu }T_{\mu }^{VS}=q^{2}F\left( q^{2}\right) =0.
\end{equation}

Next, in addition to the paper \cite{Ebani2018}, we will have to present the
construction of a low-energy theorem, ultimately responsible for violations
associated with the chiral anomaly in the odd amplitude where the vector
current as the axial are not classically conserved.

\section{Low-Energy Theorem and RAGFs}

As observed, WIs to $AV$-versions can not both simultaneously hold. Firstly,
vanishing the surface term eliminates the one-point functions; however, it
implies linearity breaking, and an additional constant can not get rid of by
any other choice. On the other hand, if the non-zero value corresponding to
the maintenance of RAGFs (linearity) is chosen, axial one-point functions
violate WIs in any case. In the scenario where the surface term could be
arbitrary through some device or interpretation, the violation does not give
up. To understand this state of affairs, we have resorted to an explanation
only utilizing properties that are immune to choices and do not privilege
one symmetry over another: \textit{the kinematical behavior of }$PV$\textit{%
\ function.}

We return to the last claims of the Chapter (\ref{2Dim2Pt}), assuming the
general tensor for odd amplitudes (\ref{AVForm}). In 2D, the amplitude has
Feynman integrals of power counting zero, one of which is a tensor integral.
These types of integrals, in any dimension, indeed own surface terms,
notwithstanding the coefficient of them only depending on the difference of
routings; they are intrinsic to Feynman diagrams, not only when the power
counting is linear. These features must be considered when stating general
theorems about kinematical properties and their relations to the symmetry
content of amplitudes coming from Feynman's rules. In 4D, we will have a
more complex scenario: the surface terms appear with ambiguous combinations
of routing sums, see Sections (\ref{LE4D}) and (\ref{LED4DSTS}).

Only external momenta imply that preserving divergent content intact follows
an expression to general tensor structure that accounts for the presence of
surface terms because, in the last instance, they contribute a coefficient
proportional to the metric, 
\begin{equation}
F_{\mu _{1}\mu _{2}}=\varepsilon _{\mu _{1}\mu _{2}}F_{1}+\varepsilon _{\mu
_{1}\nu }q^{\nu }q_{\mu _{2}}F_{2}+\varepsilon _{\mu _{2}\nu }q^{\nu }q_{\mu
_{1}}F_{3}.
\end{equation}%
The path often trailed to study symmetry violations is to perform
contractions and use some symmetry constraints to derive implications over
others. Nonetheless, we shall derive a device that prescinds from the choice
of some, \textit{a priori}, selected symmetry. Performing contractions and
identifying two invariant functions constructed with form factors $F_{i}$,
viz.,%
\begin{eqnarray}
q^{\mu _{1}}F_{\mu _{1}\mu _{2}} &=&:\varepsilon _{\mu _{2}\nu }q^{\nu
}V_{1}\left( q^{2}\right) \\
q^{\mu _{2}}F_{\mu _{1}\mu _{2}} &=&:\varepsilon _{\mu _{1}\nu }q^{\nu
}V_{2}\left( q^{2}\right) .
\end{eqnarray}%
We got two equations that are strict and intrinsic consequences of tensor
properties. If we sum them, $F_{1}$ drops, and an independent equation
emerges 
\begin{equation}
V_{1}\left( q^{2}\right) +V_{2}\left( q^{2}\right) =q^{2}\left(
F_{3}+F_{2}\right) .
\end{equation}%
For $F_{2}$ and $F_{3}\,$sufficiently regular in the point $q^{2}=0$ this
equation becomes%
\begin{equation}
V_{1}\left( 0\right) +V_{2}\left( 0\right) =0.
\end{equation}%
From it, being aware of its generality, we establish some computational-free
conclusions. First, suppose the general tensor is chosen to correspond with
the axial-vector amplitude and function of two masses, i.e., $F_{\mu \nu
}=T_{\mu \nu }^{AV}$. In that case, we may inquire about expected amplitudes
related to the hypothesis of WIs.

The systematization of 2pt, 1st-rank amplitude arising from contraction $%
q^{\mu _{i}}$ starts with 
\begin{equation}
(q^{\mu _{i}}T_{\mu _{12}}^{\Gamma _{12}})^{\mathrm{2pt}}=\varepsilon _{\mu
_{k}\nu }q^{\nu }\Omega _{i}^{\left( \mathrm{2pt}\right) },\ \quad \left\{
i,k\right\} =\left\{ 1,3\right\} ,\text{ }k\not=i.
\end{equation}%
That is a form to compare standard identifications with consequences of
tensor structure in the LHS. It denotes the 2pt functions (finite) coming
from the $i$-th contraction. They can be zero to some contractions, e.g.,
vector contraction for equal masses. Particularly, 
\begin{eqnarray}
\varepsilon _{\mu _{i}\nu }q^{\nu }\Omega ^{PV} &=&-\left(
m_{1}+m_{2}\right) T_{\mu _{i}}^{PV} \\
\varepsilon _{\mu _{i}\nu }q^{\nu }\Omega ^{AS} &=&+\left(
m_{2}-m_{1}\right) T_{\mu _{i}}^{AS},
\end{eqnarray}%
given by (\ref{PVm2}), (\ref{ASm2}). The vector and scalar integrals (\ref%
{2DJ2})-(\ref{2DJ2mu1}) enable to write%
\begin{eqnarray}
\Omega ^{AS} &=&\frac{i}{2\pi }[\left( m_{2}^{2}-m_{1}^{2}\right)
Z_{1}^{\left( -1\right) }-\left( m_{2}-m_{1}\right) m_{1}Z_{0}^{\left(
-1\right) }] \\
\Omega ^{PV} &=&\frac{i}{2\pi }[\left( m_{2}^{2}-m_{1}^{2}\right)
Z_{1}^{\left( -1\right) }+\left( m_{1}+m_{2}\right) m_{1}Z_{0}^{\left(
-1\right) }].
\end{eqnarray}%
Summing them, we have from combination (\ref{LetZ2M}), a result independent
of masses, 
\begin{equation}
\left( \Omega ^{AS}+\Omega ^{PV}\right) \left( 0\right) =-\frac{i}{\pi }%
\left. [\left( m_{1}^{2}-m_{2}^{2}\right) Z_{1}^{\left( -1\right)
}-m_{1}^{2}Z_{0}^{\left( -1\right) }]\right\vert _{q^{2}=0}=-\frac{i}{\pi }.
\end{equation}

A moment of reflection shows that anomalous amplitudes share this
combination. As it is incompatible with the low-energy theorem, we derived a
general parity-odd second-rank tensor of mass dimension zero. That is an
inviolable property if it is free of kinematical singularities. We have
anomalies in the vertices, which themselves can be arbitrary,%
\begin{equation}
V_{1}\left( 0\right) +V_{2}\left( 0\right) =0\not=-\frac{i}{\pi }=\left(
\Omega ^{AS}+\Omega ^{PV}\right) \left( 0\right) .
\end{equation}%
Hence, we at least can write $V_{i}\left( q^{2}\right) =\Omega _{i}\left(
q^{2}\right) +\mathcal{A}_{i}$, where the additional parameter will be
constrained by the equation above%
\begin{equation}
\mathcal{A}_{1}+\mathcal{A}_{2}=\frac{i}{\pi }.
\end{equation}%
That represents the restriction of arbitrary anomalies in the axial and
vector vertices. This kinematical implication has an important consequence
over the RAGFs as well.

\subsection{RAGFs: Linearity and Low-Energy Implications}

The surface terms appear in explicit computations and are the only type of
non-finite structures for the 2nd-rank amplitudes. Also, we have observed
that they conditioned the RAGFs. Nonetheless, we needed to establish in the
absolute how they do it. Besides the exciting fact that versions one and two
are the only independent possibilities, the answer to how this appears to be
so must be constructed. Therefore, we explicit this intrinsic part of
perturbative amplitudes; first, we split the general representation in 
\begin{equation}
F_{\mu _{1}\mu _{2}}=F_{\mu _{1}\mu _{2}}^{\Delta }+\hat{F}_{\mu _{1}\mu
_{2}},
\end{equation}%
where $\hat{F}_{\mu _{1}\mu _{2}}$ encodes the finite parts. The term $%
F_{\mu _{1}\mu _{2}}^{\Delta }$ stands for the most general combination of
surface terms, given by the equation 
\begin{equation*}
F_{\mu _{1}\mu _{2}}^{\Delta }=a\varepsilon _{\mu _{1}\nu }\Delta _{2\mu
_{2}}^{\nu }+b\varepsilon _{\mu _{2}\nu }\Delta _{2\mu _{1}}^{\nu
}+c\varepsilon _{\mu _{1}\mu _{2}}\Delta _{2\nu }^{\nu }.
\end{equation*}%
Since there is a linear relation in such tensor due to the vanishing of
3rd-rank complete antisymmetric tensor in $2D$, $\varepsilon _{\lbrack \mu
_{1}\mu _{2}}\Delta _{2\nu ]}^{\nu }=0$, we have a redefinition $%
a_{1}=\left( a+c\right) $ and $a_{2}=\left( b-c\right) $ of the
coefficients. Henceforth, the general structure assumes the form%
\begin{eqnarray}
F_{\mu _{1}\mu _{2}} &=&a_{1}\varepsilon _{\mu _{1}\nu }\Delta _{2\mu
_{2}}^{\nu }+a_{2}\varepsilon _{\mu _{2}\nu }\Delta _{2\mu _{1}}^{\nu } \\
&&+\varepsilon _{\mu _{1}\mu _{2}}\hat{F}_{1}+\varepsilon _{\mu _{1}\nu
}q^{\nu }q_{\mu _{2}}\hat{F}_{2}+\varepsilon _{\mu _{2}\nu }q^{\nu }q_{\mu
_{1}}\hat{F}_{3}.  \notag
\end{eqnarray}

The equation that represents the satisfaction of RAGFs can be systematized
through 
\begin{equation}
q^{\mu _{i}}T_{\mu _{12}}^{\Gamma _{12}}=T_{\left( -\right) \mu
_{k}}^{A}+\varepsilon _{\mu _{k}\nu }\Omega _{i}.
\end{equation}%
Remember the notation for the one-point differences (\ref{TA(-)mi}). The
condition of linearity of integration is embodied in the following equations
when performing the contractions,%
\begin{eqnarray}
q^{\mu _{1}}F_{\mu _{1}\mu _{2}} &=&a_{1}q^{\mu _{1}}\varepsilon _{\mu
_{1}\nu }\Delta _{2\mu _{2}}^{\nu }+a_{2}\varepsilon _{\mu _{2}\nu }q^{\mu
_{1}}\Delta _{2\mu _{1}}^{\nu }+\varepsilon _{\mu _{2}\nu }q^{\nu }(q^{2}%
\hat{F}_{3}-\hat{F}_{1}) \\
q^{\mu _{2}}F_{\mu _{1}\mu _{2}} &=&a_{1}\varepsilon _{\mu _{1}\nu }q^{\mu
_{2}}\Delta _{2\mu _{2}}^{\nu }+a_{2}q^{\mu _{2}}\varepsilon _{\mu _{2}\nu
}\Delta _{2\mu _{1}}^{\nu }+\varepsilon _{\mu _{1}\nu }q^{\nu }(q^{2}\hat{F}%
_{2}+\hat{F}_{1}).
\end{eqnarray}%
\qquad We rearrange their indices and recognize the one-point functions%
\begin{eqnarray}
q^{\mu _{1}}F_{\mu _{1}\mu _{2}} &=&-\frac{1}{2}\left( a_{1}+a_{2}\right)
T_{\left( -\right) \mu _{2}}^{A}+\varepsilon _{\mu _{2}\nu }q^{\nu }(q^{2}%
\hat{F}_{3}-\hat{F}_{1}-a_{1}\Delta _{2\alpha }^{\alpha }) \\
q^{\mu _{2}}F_{\mu _{1}\mu _{2}} &=&-\frac{1}{2}\left( a_{1}+a_{2}\right)
T_{\left( -\right) \mu _{1}}^{A}+\varepsilon _{\mu _{1}\nu }q^{\nu }(q^{2}%
\hat{F}_{2}+\hat{F}_{1}-a_{2}\Delta _{2\alpha }^{\alpha }).
\end{eqnarray}

The RAGFs require for the first terms $a_{1}+a_{2}=-2,$ and the other part
must comply with the 2pt functions, $\Omega ^{PV}$ and $\Omega ^{AS}$, which
means%
\begin{eqnarray}
\Omega ^{PV} &=&q^{2}\hat{F}_{3}-\hat{F}_{1}-a_{1}\Delta _{2\alpha }^{\alpha
} \\
\Omega ^{AS} &=&q^{2}\hat{F}_{2}+\hat{F}_{1}-a_{2}\Delta _{2\alpha }^{\alpha
}.
\end{eqnarray}%
Eliminating $\hat{F}_{1}$ and considering the first condition $%
a_{1}+a_{2}=-2 $, we obtain%
\begin{equation}
2\Delta _{2\alpha }^{\alpha }=\Omega ^{PV}+\Omega ^{AS}-q^{2}\hat{F}%
_{2}-q^{2}\hat{F}_{3}.
\end{equation}%
In the point $q^{2}=0$ follows the low-energy implication of the finite
amplitudes over the integration linearity (RAGFs)%
\begin{equation}
2\Delta _{2\alpha }^{\alpha }=\Omega ^{PV}\left( 0\right) +\Omega
^{AS}\left( 0\right) =-\frac{i}{\pi }.
\end{equation}

\textbf{Consequences:} The coefficients $a_{1}$ and $a_{2}$ may be
arbitrary, but once one is selected to satisfy one RAGF in automatic form,
the other must be zero. This unique solution signifies that most RAGFs found
without conditions are achieved by the basic versions we have defined. This
fact is independent of explicit computations through the traces of four
Dirac matrices and continues to happen in four dimensions. Another
consequence is that the satisfaction of all RAGFs is conditioned through
kinematical features of finite functions that require a non-zero and
specific amount value to the surface terms, implying that shifts in the
integration variable and linearity of integration are incompatible. The $%
T_{\mu }^{A}$ functions depend on the routings, and their subtraction is
zero if shifts are possible; only their difference is a function of the
external momentum. This aspect is peculiar to this dimension; nonetheless,
the restrictions from low-energy implications are precisely mirrored in four
dimensions. Simultaneously satisfaction of RAGFs and translational
invariance in momentum space is prohibited by the low-energy behavior of
finite functions.

\chapter{Four-Dimensional Three-Point Functions}

\label{4Dim3Pt}The analysis developed in the physical dimension focuses on
odd amplitudes that are rank-3 tensors, namely $AVV$, $VAV$, $VVA$, and $AAA$%
. Their mathematical structures follow the same features seen in two
dimensions. They depend on the trace involving six Dirac matrices plus the
chiral one, whose computation yields products between the Levi-Civita symbol
and metric tensor. After the integration, that generates expressions that
differ in their dependence on surface terms and finite parts. We want to
verify these prospects by evaluating the triangles' basic versions\footnote{%
To this aim, we compute twenty-four triangles of rank-one. Twelve
parity-even triangles: $VPP$, $ASP$, $VSS$, and their permutations. Twelve
parity-odd tensors: $ASS$, $APP$, $VPS$, and their permutations. Besides, we
identify three standard tensors in a similar fashion for two dimensions.}.
Once these resources are clear, we study how symmetries, linearity of
integration, and uniqueness manifest.

From Eqs. (\ref{t}) and (\ref{T}), integrated three-point amplitudes are
denoted through capital letters $T^{\Gamma _{1}\Gamma _{2}\Gamma _{3}}$ and
exhibit the integrand%
\begin{equation}
t^{\Gamma _{1}\Gamma _{2}\Gamma _{3}}=\text{\textrm{tr}}\left[ \Gamma
_{1}S\left( 1\right) \Gamma _{2}S\left( 2\right) \Gamma _{3}S\left( 3\right) %
\right] .  \label{t3}
\end{equation}%
Thus, after replacing vertex operators and disregarding vanishing traces,
3rd-order amplitudes assume the forms%
\begin{eqnarray}
t_{\mu _{123}}^{AVV} &=&[K_{123}^{\nu _{123}}\mathrm{tr}(\gamma _{\ast \mu
_{1}\nu _{1}\mu _{2}\nu _{2}\mu _{3}\nu _{3}})+m^{2}\mathrm{tr}(\gamma
_{\ast \mu _{1}\mu _{2}\mu _{3}\nu _{1}})(K_{1}^{\nu _{1}}-K_{2}^{\nu
_{1}}+K_{3}^{\nu _{1}})]\frac{1}{D_{123}}  \label{AVVexp} \\
t_{\mu _{123}}^{VAV} &=&[K_{123}^{\nu _{123}}\mathrm{tr}(\gamma _{\ast \mu
_{1}\nu _{1}\mu _{2}\nu _{2}\mu _{3}\nu _{3}})+m^{2}\mathrm{tr}(\gamma
_{\ast \mu _{1}\mu _{2}\mu _{3}\nu _{1}})(K_{1}^{\nu _{1}}+K_{2}^{\nu
_{1}}-K_{3}^{\nu _{1}})]\frac{1}{D_{123}}  \label{VAVexp} \\
t_{\mu _{123}}^{VVA} &=&[K_{123}^{\nu _{123}}\mathrm{tr}(\gamma _{\ast \mu
_{1}\nu _{1}\mu _{2}\nu _{2}\mu _{3}\nu _{3}})-m^{2}\mathrm{tr}(\gamma
_{\ast \mu _{1}\mu _{2}\mu _{3}\nu _{1}})(K_{1}^{\nu _{1}}-K_{2}^{\nu
_{1}}-K_{3}^{\nu _{1}})]\frac{1}{D_{123}}  \label{VVAexp} \\
t_{\mu _{123}}^{AAA} &=&[K_{123}^{\nu _{123}}\mathrm{tr}(\gamma _{\ast \mu
_{1}\nu _{1}\mu _{2}\nu _{2}\mu _{3}\nu _{3}})-m^{2}\mathrm{tr}(\gamma
_{\ast \mu _{1}\mu _{2}\mu _{3}\nu _{1}})(K_{1}^{\nu _{1}}+K_{2}^{\nu
_{1}}+K_{3}^{\nu _{1}})]\frac{1}{D_{123}},  \label{AAAexp}
\end{eqnarray}%
where we recall the conventions $K_{123}^{\nu _{123}}=K_{1}^{\nu
_{1}}K_{2}^{\nu _{2}}K_{3}^{\nu _{3}}$ and $D_{123}=D_{1}D_{2}D_{3}$.

Although the trace involving four Dirac matrices plus the chiral one is
univocal, different expressions are attributed to the leading trace when
considering identities (\ref{Chiral-Id}). Since Appendix (\ref{Tr6G4D})
shows that forms achieved through definition $\gamma _{\ast }=i\varepsilon
_{\nu _{1234}}\gamma ^{\nu _{1234}}/4!$ are enough to compound any other,
our starting point is on their structure 
\begin{eqnarray}
\left( 4i\right) ^{-1}\mathrm{tr}(\gamma _{\ast abcdef})
&=&+g_{ab}\varepsilon _{cdef}+g_{ad}\varepsilon _{bcef}+g_{af}\varepsilon
_{bcde}  \label{Mold6} \\
&&+g_{bc}\varepsilon _{adef}+g_{cd}\varepsilon _{abef}+g_{cf}\varepsilon
_{abde}  \notag \\
&&+g_{be}\varepsilon _{acdf}+g_{de}\varepsilon _{abcf}+g_{ef}\varepsilon
_{abcd}  \notag \\
&&-g_{bd}\varepsilon _{acef}-g_{df}\varepsilon _{abce}-g_{bf}\varepsilon
_{acde}  \notag \\
&&-g_{ac}\varepsilon _{bdef}-g_{ce}\varepsilon _{abdf}-g_{ae}\varepsilon
_{bcdf}.  \notag
\end{eqnarray}%
There are three basic versions, each corresponding to replacing the chiral
matrix near a specific vertex operator. We introduce a numeric label to
distinguish them: 
\begin{equation}
\lbrack \mathrm{tr}(\gamma _{\ast \mu _{1}\nu _{1}\mu _{2}\nu _{2}\mu
_{3}\nu _{3}})]_{1}=[\mathrm{tr}(\gamma _{\ast \mu _{2}\nu _{2}\mu _{3}\nu
_{3}\mu _{1}\nu _{1}})]_{2}=[\mathrm{tr}(\gamma _{\ast \mu _{3}\nu _{3}\mu
_{1}\nu _{1}\mu _{2}\nu _{2}})]_{3}.  \label{identraces}
\end{equation}%
They arise when setting the index configuration in the trace above (\ref%
{Mold6}), differing in the signs of terms. We cast their contraction with $%
K_{123}^{\nu _{123}}$ in the sequence. Their integration leads to three not
(automatically) equivalent expressions for each triangle. 
\begin{eqnarray}
\lbrack K_{123}^{\nu _{123}}\mathrm{tr}(\gamma _{\ast \mu _{1}\nu _{1}\mu
_{2}\nu _{2}\mu _{3}\nu _{3}})]_{1} &=&-4i\varepsilon _{\mu _{23}\nu
_{12}}[K_{1\mu _{1}}K_{23}^{\nu _{12}}-K_{2\mu _{1}}K_{13}^{\nu
_{12}}+K_{3\mu _{1}}K_{12}^{\nu _{12}}]  \label{tr1} \\
&&-4i\varepsilon _{\mu _{13}\nu _{12}}[K_{1\mu _{2}}K_{23}^{\nu
_{12}}+K_{2\mu _{2}}K_{13}^{\nu _{12}}-K_{3\mu _{2}}K_{12}^{\nu _{12}}] 
\notag \\
&&+4i\varepsilon _{\mu _{12}\nu _{12}}[K_{1\mu _{3}}K_{23}^{\nu
_{12}}-K_{2\mu _{3}}K_{13}^{\nu _{12}}-K_{3\mu _{3}}K_{12}^{\nu _{12}}] 
\notag \\
&&-4i\varepsilon _{\mu _{123}\nu _{1}}[K_{1}^{\nu _{1}}(K_{2}\cdot
K_{3})-K_{2}^{\nu _{1}}(K_{1}\cdot K_{3})+K_{3}^{\nu _{1}}(K_{1}\cdot K_{2})]
\notag \\
&&+4i[-g_{\mu _{12}}\varepsilon _{\mu _{3}\nu _{123}}-g_{\mu
_{23}}\varepsilon _{\mu _{1}\nu _{123}}+g_{\mu _{13}}\varepsilon _{\mu
_{2}\nu _{123}}]K_{123}^{\nu _{123}}  \notag
\end{eqnarray}%
\begin{eqnarray}
\lbrack K_{123}^{\nu _{123}}\mathrm{tr}(\gamma _{\ast \mu _{2}\nu _{2}\mu
_{3}\nu _{3}\mu _{1}\nu _{1}})]_{2} &=&+4i\varepsilon _{\mu _{13}\nu
_{12}}[K_{1\mu _{2}}K_{23}^{\nu _{12}}-K_{2\mu _{2}}K_{13}^{\nu
_{12}}+K_{3\mu _{2}}K_{12}^{\nu _{12}}]  \label{tr2} \\
&&-4i\varepsilon _{\mu _{12}\nu _{12}}[K_{1\mu _{3}}K_{23}^{\nu
_{23}}+K_{2\mu _{3}}K_{13}^{\nu _{13}}+K_{3\mu _{3}}K_{12}^{\nu _{12}}] 
\notag \\
&&-4i\varepsilon _{\mu _{23}\nu _{12}}[K_{1\mu _{1}}K_{23}^{\nu
_{23}}+K_{2\mu _{1}}K_{13}^{\nu _{13}}-K_{3\mu _{1}}K_{12}^{\nu _{12}}] 
\notag \\
&&-4i\varepsilon _{\mu _{123}\nu _{1}}[K_{1}^{\nu _{1}}(K_{2}\cdot
K_{3})+K_{2}^{\nu _{1}}(K_{1}\cdot K_{3})-K_{3}^{\nu _{1}}(K_{1}\cdot K_{2})]
\notag \\
&&+4i[g_{\mu _{12}}\varepsilon _{\mu _{3}\nu _{123}}-g_{\mu
_{13}}\varepsilon _{\mu _{2}\nu _{123}}-g_{\mu _{23}}\varepsilon _{\mu
_{1}\nu _{123}}]K_{123}^{\nu _{123}}  \notag
\end{eqnarray}%
\begin{eqnarray}
\lbrack K_{123}^{\nu _{123}}\mathrm{tr}(\gamma _{\ast \mu _{3}\nu _{3}\mu
_{1}\nu _{1}\mu _{2}\nu _{2}})]_{3} &=&-4i\varepsilon _{\mu _{12}\nu
_{12}}[K_{1\mu _{3}}K_{23}^{\nu _{12}}-K_{2\mu _{3}}K_{13}^{\nu
_{12}}+K_{3\mu _{3}}K_{12}^{\nu _{12}}]  \label{tr3} \\
&&-4i\varepsilon _{\mu _{23}\nu _{12}}[K_{1\mu _{1}}K_{23}^{\nu
_{12}}-K_{2\mu _{1}}K_{13}^{\nu _{12}}-K_{3\mu _{1}}K_{12}^{\nu _{12}}] 
\notag \\
&&-4i\varepsilon _{\mu _{13}\nu _{12}}[K_{1\mu _{2}}K_{23}^{\nu
_{12}}+K_{2\mu _{2}}K_{13}^{\nu _{12}}+K_{3\mu _{2}}K_{12}^{\nu _{12}}] 
\notag \\
&&+4i\varepsilon _{\mu _{123}\nu _{1}}[K_{1}^{\nu _{1}}(K_{2}\cdot
K_{3})-K_{2}^{\nu _{1}}(K_{1}\cdot K_{3})-K_{3}^{\nu _{1}}(K_{1}\cdot K_{2})]
\notag \\
&&+4i[-g_{\mu _{12}}\varepsilon _{\mu _{3}\nu _{123}}-g_{\mu
_{13}}\varepsilon _{\mu _{2}\nu _{123}}+g_{\mu _{23}}\varepsilon _{\mu
_{1}\nu _{123}}]K_{123}^{\nu _{123}}  \notag
\end{eqnarray}

Analogously to two-dimensional calculations, our next task consists of
organizing and integrating the complete expressions. As the three first rows
of the above equations are similar to the object (\ref{t(s)}), we define the
tensors%
\begin{equation}
\varepsilon _{\mu _{ab}\nu _{12}}t_{\mu _{c}}^{\nu _{12}\left(
s_{1}s_{2}\right) }=\varepsilon _{\mu _{ab}\nu _{12}}\left( K_{1\mu
_{c}}K_{23}^{\nu _{12}}+s_{1}K_{2\mu _{c}}K_{13}^{\nu _{12}}+s_{2}K_{3\mu
_{c}}K_{12}^{\nu _{12}}\right) \frac{1}{D_{123}}
\end{equation}%
where $s_{i}=\pm 1$. We rewrite this equation using $K_{i}=K_{j}+p_{ij}$ and 
$\varepsilon _{\mu _{ab}\nu _{12}}K_{ij}^{\nu _{12}}=\varepsilon _{\mu
_{ab}\nu _{12}}p_{ji}^{\nu _{2}}K_{i}^{\nu _{1}}$ to achieve the structures
introduced in Section (\ref{BasisFI}): 
\begin{eqnarray}
\varepsilon _{\mu _{ab}\nu _{12}}t_{\mu _{c}}^{\nu _{12}\left(
s_{1}s_{2}\right) } &=&\varepsilon _{\mu _{ab}\nu
_{12}}[(1+s_{1})p_{31}^{\nu _{2}}-(1-s_{2})p_{21}^{\nu _{2}}]K_{1}^{\nu
_{1}}K_{1\mu _{c}}\frac{1}{D_{123}} \\
&&+\varepsilon _{\mu _{ab}\nu _{12}}[p_{21}^{\nu _{1}}p_{32}^{\nu
_{2}}K_{1\mu _{c}}+(s_{1}p_{21\mu _{c}}p_{31}^{\nu _{2}}+s_{2}p_{31\mu
_{c}}p_{21}^{\nu _{2}})K_{1}^{\nu _{1}}]\frac{1}{D_{123}}.  \notag
\end{eqnarray}%
Hence, final expressions arise directly by replacing vector and tensor
Feynman integrals from Subsection (\ref{Sub4DIn}). Although four sign
configurations are available, the expression taking $s_{1}=-1$ and $s_{2}=1$
cancels out. That is straightforward for the first row, but a closer look at
the composition of the following integral is necessary to analyze the
second: 
\begin{equation}
\overline{J}_{3}^{\mu }=J_{3}^{\mu }=i\left( 4\pi \right) ^{-2}[-p_{21}^{\mu
}Z_{10}^{\left( -1\right) }\left( p_{21},p_{31}\right) -p_{31}^{\mu
}Z_{01}^{\left( -1\right) }\left( p_{21},p_{31}\right) ].  \label{4DJ3mu}
\end{equation}%
Since it is proportional to external momenta, it leads to symmetric tensors
that vanish when contracted with Levi-Civita symbol. We cast all sign
configurations in the sequence: 
\begin{eqnarray}
2\varepsilon _{\mu _{ab}\nu _{12}}T_{\mu _{c}}^{\nu _{12}\left( -+\right) }
&=&2\varepsilon _{\mu _{ab}\nu _{12}}[p_{21}^{\nu _{1}}p_{32}^{\nu
_{2}}J_{3\mu _{c}}+(-p_{21\mu _{c}}p_{31}^{\nu _{2}}+p_{31\mu
_{c}}p_{21}^{\nu _{2}})J_{3}^{\nu _{1}}]\equiv 0,  \label{T-+} \\
2\varepsilon _{\mu _{ab}\nu _{12}}T_{\mu _{c}}^{\nu _{12}(+-)}
&=&4\varepsilon _{\mu _{ab}\nu _{12}}[p_{31}^{\nu _{2}}(J_{3\mu _{c}}^{\nu
_{1}}+p_{21\mu _{c}}J_{3}^{\nu _{1}})-p_{21}^{\nu _{2}}(J_{3\mu _{c}}^{\nu
_{1}}+p_{31\mu _{c}}J_{3}^{\nu _{1}})]  \label{T+-} \\
&&+(\varepsilon _{\mu _{ab}\nu _{12}}p_{32}^{\nu _{2}}\Delta _{3\mu
_{c}}^{\nu _{1}}+\varepsilon _{\mu _{abc}\nu _{1}}p_{32}^{\nu _{1}}I_{\log
}),  \notag \\
2\varepsilon _{\mu _{ab}\nu _{12}}T_{\mu _{c}}^{\nu _{12}\left( --\right) }
&=&-4\varepsilon _{\mu _{ab}\nu _{12}}p_{21}^{\nu _{2}}(J_{3\mu _{c}}^{\nu
_{1}}+p_{31\mu _{c}}J_{3}^{\nu _{1}})  \label{T--} \\
&&-(\varepsilon _{\mu _{ab}\nu _{12}}p_{21}^{\nu _{2}}\Delta _{3\mu
_{c}}^{\nu _{1}}+\varepsilon _{\mu _{abc}\nu _{1}}p_{21}^{\nu _{1}}I_{\log
}),  \notag \\
2\varepsilon _{\mu _{ab}\nu _{12}}T_{\mu _{c}}^{\nu _{12}\left( ++\right) }
&=&+4\varepsilon _{\mu _{ab}\nu _{12}}p_{31}^{\nu _{2}}(J_{3\mu _{c}}^{\nu
_{1}}+p_{21\mu _{c}}J_{3}^{\nu _{1}})  \label{T++} \\
&&+(\varepsilon _{\mu _{ab}\nu _{12}}p_{31}^{\nu _{2}}\Delta _{3\mu
_{c}}^{\nu _{1}}+\varepsilon _{\mu _{abc}\nu _{1}}p_{31}^{\nu _{1}}I_{\log
}).  \notag
\end{eqnarray}

Different tensor contributions appear for each trace version from (\ref{tr1}%
)-(\ref{tr3}). Thus, after disregarding the vanishing contribution, we
identify the corresponding combinations%
\begin{eqnarray}
C_{1\mu _{123}} &=&-\varepsilon _{\mu _{13}\nu _{12}}T_{\mu _{2}}^{\nu
_{12}\left( +-\right) }+\varepsilon _{\mu _{12}\nu _{12}}T_{\mu _{3}}^{\nu
_{12}\left( --\right) }  \label{C1} \\
C_{2\mu _{123}} &=&-\varepsilon _{\mu _{12}\nu _{12}}T_{\mu _{3}}^{\nu
_{12}\left( ++\right) }-\varepsilon _{\mu _{23}\nu _{12}}T_{\mu _{1}}^{\nu
_{12}\left( +-\right) } \\
C_{3\mu _{123}} &=&-\varepsilon _{\mu _{23}\nu _{12}}T_{\mu _{1}}^{\nu
_{12}\left( --\right) }-\varepsilon _{\mu _{13}\nu _{12}}T_{\mu _{2}}^{\nu
_{12}\left( ++\right) }.  \label{C3}
\end{eqnarray}%
The sampling of indexes reflects the absence of the index $\mu _{i}$ of the
vertex $\Gamma _{i}$ in the sign tensors of the $C_{i\mu _{123}}$, enabling
the anticipation of violations of either WIs or RAGFs. That occurs because
this specific index appears in the tensor $\varepsilon _{\mu _{ab}\nu
_{12}}T_{\mu _{i}}^{\nu _{12}\left( -,+\right) }$, which is finite and
identically zero, present in each of the above expressions before
integration.

Let us return to the last row of Eqs. (\ref{tr1})-(\ref{tr3}), which
corresponds to 1st-order odd triangles. The precise identifications among
the possibilities occur when replacing the vertex configurations in the
general integrand (\ref{t3}); however, all of them are proportional to $ASS$
amplitude:%
\begin{equation}
t_{\mu _{i}}^{ASS}=4i\varepsilon _{\mu _{i}\nu _{123}}K_{123}^{\nu _{123}}%
\frac{1}{D_{123}}=4i\varepsilon _{\mu _{i}\nu _{123}}p_{21}^{\nu
_{2}}p_{31}^{\nu _{3}}K_{1}^{\nu _{1}}\frac{1}{D_{123}}.
\end{equation}%
We already performed some simplifications through the same resources from
the tensor discussion (beginning of the previous paragraph). After
integration, this function depends on the Feynman integral $J_{3}^{\nu _{1}}$%
. Since this object is a finite tensor proportional to external momenta $%
p_{ij}$, the contraction with the Levi-Civita symbol necessarily vanishes%
\begin{equation}
T_{\mu _{i}}^{ASS}=4i\varepsilon _{\mu _{i}\nu _{123}}p_{21}^{\nu
_{2}}p_{31}^{\nu _{3}}J_{3}^{\nu _{1}}=0.  \label{ASS}
\end{equation}%
For this reason, we omit this class of amplitudes from the final triangles.

We left the fourth line of (\ref{tr1})-(\ref{tr3}) for last since bilinears
get summed with mass terms from the remaining trace. Each investigated case
leads to a subamplitude identified after comparing vertex arrangements in (%
\ref{t3}). This result is general: besides $C_{i\mu _{123}}$ tensors,
different rank-1 even subamplitudes appear inside each version of rank-3 odd
amplitudes. Table \ref{tabversions} accounts for all of these possibilities,
while Appendix (\ref{AppSub}) presents explicit expressions for
subamplitudes. Let us consider the first version of $AVV$ to illustrate.
After combining mass terms from Eq. (\ref{AVVexp}) with bilinears from Eq. (%
\ref{tr1}), we find the $VPP$ subamplitude 
\begin{equation}
\mathrm{sub}(t_{\mu _{123}}^{AVV})_{1}=i\varepsilon _{\mu _{123}\nu
_{1}}(t^{VPP})^{\nu _{1}}.
\end{equation}%
The integrand of this correlator has the structure%
\begin{equation}
(t^{VPP})^{\nu _{1}}=\text{\textrm{tr}}[\gamma ^{\nu _{1}}S\left( 1\right)
\gamma _{\ast }S\left( 2\right) \gamma _{\ast }S\left( 3\right)
]=4(-K_{1}^{\nu _{1}}S_{23}+K_{2}^{\nu _{1}}S_{13}-K_{3}^{\nu _{1}}S_{12})%
\frac{1}{D_{123}},
\end{equation}%
where the combination $S_{ij}=K_{i}\cdot K_{j}-m^{2}$ comes from definition (%
\ref{Sij}). After reducing the denominator, we perform the integration%
\begin{eqnarray}
(T^{VPP})^{\nu _{1}} &=&2[P_{31}^{\nu _{2}}\Delta _{3\nu _{2}}^{\nu
_{1}}+(p_{21}^{\nu _{1}}-p_{32}^{\nu _{1}})I_{\log }]-4\left( p_{21}\cdot
p_{32}\right) J_{3}^{\nu _{1}}  \label{TVPP} \\
&&+2[(p_{31}^{\nu _{1}}p_{21}^{2}-p_{21}^{\nu
_{1}}p_{31}^{2})J_{3}+p_{21}^{\nu _{1}}J_{2}\left( p_{21}\right)
-p_{32}^{\nu _{1}}J_{2}\left( p_{32}\right) ].  \notag
\end{eqnarray}%
\begin{table}[tbph]
\caption{Even sub-amplitudes related to each version of 3rd-order odd
amplitudes.}
\label{tabversions}\centering\renewcommand{\baselinestretch}{1.4}%
{\normalsize {\ }}$%
\begin{tabular}{|c|c|c|c|c|}
\hline
$\text{Version/Type}$ & $AVV$ & $VAV$ & $VVA$ & $AAA$ \\ \hline
1 & $+VPP$ & $+ASP$ & $-APS$ & $-VSS$ \\ \hline
2 & $-SAP$ & $+PVP$ & $+PAS$ & $-SVS$ \\ \hline
3 & $+SPA$ & $-PSA$ & $+PPV$ & $-SSV$ \\ \hline
\end{tabular}%
${\normalsize \ }
\end{table}

Since all pieces are known, compounding triangle amplitudes is possible. For
instance, the $i$-th version of the $AVV$ arises as a combination involving
the $C_{i}$-tensor and the corresponding vector subamplitude. Thus,
consulting Table \ref{tabversions} leads to the following associations 
\begin{eqnarray}
(T_{\mu _{123}}^{AVV})_{1} &=&4iC_{1\mu _{123}}+i\varepsilon _{\mu _{123}\nu
_{1}}(T^{VPP})^{\nu _{1}},  \label{AVV1} \\
(T_{\mu _{123}}^{AVV})_{2} &=&4iC_{2\mu _{123}}-i\varepsilon _{\mu _{123}\nu
_{1}}(T^{SAP})^{\nu _{1}}, \\
(T_{\mu _{123}}^{AVV})_{3} &=&4iC_{3\mu _{123}}+i\varepsilon _{\mu _{123}\nu
_{1}}(T^{SPA})^{\nu _{1}}.
\end{eqnarray}%
The generalization for $VAV$, $VVA$, and $AAA$ is straightforward:%
\begin{equation}
(T_{\mu _{123}}^{\Gamma _{1}\Gamma _{2}\Gamma _{3}})_{i}=4iC_{i,\mu
_{123}}\pm i\varepsilon _{\mu _{123}\nu _{1}}\left( \text{Corresponding
sub-amplitude}\right) ^{\nu _{1}}.
\end{equation}

We still want to detail some important points about these amplitudes. To
illustrate this subject, we use tools developed in this section to build up
the first version of $AVV,$ 
\begin{eqnarray}
(T_{\mu _{123}}^{AVV})_{1} &=&S_{1\mu _{123}}-8i\varepsilon _{\mu _{12}\nu
_{12}}p_{21}^{\nu _{2}}(J_{3\mu _{3}}^{\nu _{1}}+p_{31\mu _{3}}J_{3}^{\nu
_{1}})  \label{AVV1complete} \\
&&-8i\varepsilon _{\mu _{13}\nu _{12}}[p_{31}^{\nu _{2}}(J_{3\mu _{2}}^{\nu
_{1}}+p_{21\mu _{2}}J_{3}^{\nu _{1}})-p_{21}^{\nu _{2}}(J_{3\mu _{2}}^{\nu
_{1}}+p_{31\mu _{2}}J_{3}^{\nu _{1}})]  \notag \\
&&-4i\varepsilon _{\mu _{123}\nu _{1}}(p_{21}\cdot p_{32})J_{3}^{\nu
_{1}}+2i\varepsilon _{\mu _{123}\nu _{1}}[(p_{31}^{\nu
_{1}}p_{21}^{2}-p_{21}^{\nu _{1}}p_{31}^{2})]J_{3}  \notag \\
&&+2i\varepsilon _{\mu _{123}\nu _{1}}[p_{21}^{\nu _{1}}J_{2}\left(
p_{21}\right) -p_{32}^{\nu _{1}}J_{2}\left( p_{32}\right) ].  \notag
\end{eqnarray}%
The divergent part of the tensor (\ref{C1}) comes from Eqs. (\ref{T+-}) and (%
\ref{T--}) as 
\begin{equation*}
4iC_{1\mu _{123}}=-2i[\varepsilon _{\mu _{13}\nu _{12}}p_{32}^{\nu
_{2}}\Delta _{3\mu _{2}}^{\nu _{1}}+\varepsilon _{\mu _{12}\nu
_{12}}p_{21}^{\nu _{2}}\Delta _{3\mu _{3}}^{\nu _{1}}+\varepsilon _{\mu
_{123}\nu _{1}}(p_{21}^{\nu _{1}}-p_{32}^{\nu _{1}})I_{\log }].
\end{equation*}%
When combined with the $VPP$ subamplitude, we acknowledge the exact
cancellation of the object $I_{\log }$ as it occurs for all investigated
versions. Thus, surface terms compound the whole structure of divergences%
\begin{equation}
S_{1\mu _{123}}=-2i(\varepsilon _{\mu _{13}\nu _{12}}p_{32}^{\nu _{2}}\Delta
_{3\mu _{2}}^{\nu _{1}}+\varepsilon _{\mu _{12}\nu _{12}}p_{21}^{\nu
_{2}}\Delta _{3\mu _{3}}^{\nu _{1}})+2i\varepsilon _{\mu _{123}\nu
_{1}}P_{31}^{\nu _{2}}\Delta _{3\nu _{2}}^{\nu _{1}}.  \label{ST1}
\end{equation}%
Moreover, contributions from vector subamplitudes exhibit arbitrary momenta $%
P_{ij}=k_{i}+k_{j}$ as coefficients. We stress that the divergent content is
shared; the first version of amplitudes $AVV$, $VAV$, $VVA$, and $AAA$
contains the same structure (\ref{ST1}). That is a feature of the specific
version and not on the vertex content of the diagram. For later use, we
define the other sets of surface terms%
\begin{eqnarray}
S_{2\mu _{123}} &=&-2i(\varepsilon _{\mu _{12}\nu _{12}}p_{31}^{\nu
_{2}}\Delta _{3\mu _{3}}^{\nu _{1}}+\varepsilon _{\mu _{23}\nu
_{12}}p_{32}^{\nu _{2}}\Delta _{3\mu _{1}}^{\nu _{1}})+2i\varepsilon _{\mu
_{123}\nu _{1}}P_{21}^{\nu _{2}}\Delta _{3\nu _{2}}^{\nu _{1}},  \label{ST2}
\\
S_{3\mu _{123}} &=&-2i(\varepsilon _{\mu _{13}\nu _{12}}p_{31}^{\nu
_{2}}\Delta _{3\mu _{2}}^{\nu _{1}}-\varepsilon _{\mu _{23}\nu
_{12}}p_{21}^{\nu _{2}}\Delta _{3\mu _{1}}^{\nu _{1}})+2i\varepsilon _{\mu
_{123}\nu _{1}}P_{32}^{\nu _{2}}\Delta _{3\nu _{2}}^{\nu _{1}}.  \label{ST3}
\end{eqnarray}%
That concludes the preliminary discussion on rank-3 triangles, so
investigating RAGFs is possible. That is the subject of the following
sections.

\section{Relations Among Green Functions and Uniqueness\label{unique}}

The next step is to perform momenta contractions that lead to RAGFs
following the recipes in (\ref{RAGF1}) and (\ref{RAGF2}). Although they are
algebraic identities at the integrand level, their satisfaction is not
automatic after integration. In parallel to what we saw in the
two-dimensional case, possibilities for Dirac traces and values of surface
terms have important implications for this analysis.%
\begin{eqnarray}
p_{31}^{\mu _{1}}t_{\mu _{123}}^{AVV} &=&t_{\mu _{32}}^{AV}\left( 1,2\right)
-t_{\mu _{23}}^{AV}\left( 2,3\right) -2mt_{\mu _{23}}^{PVV}  \label{AVVragfs}
\\
p_{21}^{\mu _{2}}t_{\mu _{123}}^{AVV} &=&t_{\mu _{13}}^{AV}\left( 1,3\right)
-t_{\mu _{13}}^{AV}\left( 2,3\right)  \notag \\
p_{32}^{\mu _{3}}t_{\mu _{123}}^{AVV} &=&t_{\mu _{12}}^{AV}\left( 1,2\right)
-t_{\mu _{12}}^{AV}\left( 1,3\right)  \notag
\end{eqnarray}%
\begin{eqnarray}
p_{31}^{\mu _{1}}t_{\mu _{123}}^{VAV} &=&t_{\mu _{23}}^{AV}\left( 2,1\right)
-t_{\mu _{23}}^{AV}\left( 2,3\right)  \label{VAVragfs} \\
p_{21}^{\mu _{2}}t_{\mu _{123}}^{VAV} &=&t_{\mu _{31}}^{AV}\left( 3,1\right)
-t_{\mu _{13}}^{AV}\left( 2,3\right) +2mt_{\mu _{13}}^{VPV}  \notag \\
p_{32}^{\mu _{3}}t_{\mu _{123}}^{VAV} &=&t_{\mu _{21}}^{AV}\left( 2,1\right)
-t_{\mu _{21}}^{AV}\left( 3,1\right)  \notag
\end{eqnarray}%
\begin{eqnarray}
p_{31}^{\mu _{1}}t_{\mu _{123}}^{VVA} &=&t_{\mu _{32}}^{AV}\left( 1,2\right)
-t_{\mu _{32}}^{AV}\left( 3,2\right)  \label{VVAragfs} \\
p_{21}^{\mu _{2}}t_{\mu _{123}}^{VVA} &=&t_{\mu _{31}}^{AV}\left( 3,1\right)
-t_{\mu _{31}}^{AV}\left( 3,2\right)  \notag \\
p_{32}^{\mu _{3}}t_{\mu _{123}}^{VVA} &=&t_{\mu _{12}}^{AV}\left( 1,2\right)
-t_{\mu _{21}}^{AV}\left( 3,1\right) +2mt_{\mu _{12}}^{VVP}  \notag
\end{eqnarray}%
\begin{eqnarray}
p_{31}^{\mu _{1}}t_{\mu _{123}}^{AAA} &=&t_{\mu _{23}}^{AV}\left( 2,1\right)
-t_{\mu _{32}}^{AV}\left( 3,2\right) -2mt_{\mu _{23}}^{PAA}  \label{AAAragfs}
\\
p_{21}^{\mu _{2}}t_{\mu _{123}}^{AAA} &=&t_{\mu _{13}}^{AV}\left( 1,3\right)
-t_{\mu _{31}}^{AV}\left( 3,2\right) +2mt_{\mu _{13}}^{APA}  \notag \\
p_{32}^{\mu _{3}}t_{\mu _{123}}^{AAA} &=&t_{\mu _{21}}^{AV}\left( 2,1\right)
-t_{\mu _{12}}^{AV}\left( 1,3\right) +2mt_{\mu _{12}}^{AAP}  \notag
\end{eqnarray}

Let us introduce the structures that emerged within the relations above.
First, the RHS's three-point functions are finite tensors external momenta
dependent. That is transparent due to their connection with finite Feynman
integrals introduced in Subsection (\ref{FinFcts4d}), so we only remove the
overbar notation from corresponding tensors $\bar{J}_{3}^{\nu
_{1}}=J_{3}^{\nu _{1}}$ and $\bar{J}_{3}=J_{3}$. We have for single axial
triangles%
\begin{eqnarray}
-2mT_{\mu _{23}}^{PVV} &=&\varepsilon _{\mu _{23}\nu _{12}}p_{21}^{\nu
_{1}}p_{32}^{\nu _{2}}(8im^{2}J_{3}),  \label{PVV} \\
2mT_{\mu _{13}}^{VPV} &=&\varepsilon _{\mu _{13}\nu _{12}}p_{21}^{\nu
_{1}}p_{32}^{\nu _{2}}(8im^{2}J_{3}), \\
2mT_{\mu _{12}}^{VVP} &=&\varepsilon _{\mu _{12}\nu _{12}}p_{21}^{\nu
_{1}}p_{32}^{\nu _{2}}(-8im^{2}J_{3}),
\end{eqnarray}%
while momenta contractions for the triple axial triangle lead to%
\begin{eqnarray}
-2mT_{\mu _{23}}^{PAA} &=&\varepsilon _{\mu _{23}\nu _{12}}p_{31}^{\nu
_{2}}[8im^{2}(2J_{3}^{\nu _{1}}+p_{21}^{\nu _{1}}J_{3})], \\
2mT_{\mu _{13}}^{APA} &=&\varepsilon _{\mu _{13}\nu _{12}}p_{21}^{\nu
_{2}}[-8im^{2}(2J_{3}^{\nu _{1}}+p_{31}^{\nu _{1}}J_{3})], \\
2mT_{\mu _{12}}^{AAP} &=&\varepsilon _{\mu _{12}\nu _{12}}p_{32}^{\nu
_{2}}[8im^{2}(2J_{3}^{\nu _{1}}+p_{21}^{\nu _{1}}J_{3})].  \label{AAP}
\end{eqnarray}%
These amplitudes have a low-energy behavior that we aim to explore in
connection with RAGFs in Sections (\ref{LE4D}) and (\ref{LED4DSTS}). Since
they depend on functions $Z_{n_{1}n_{2}}^{\left( -1\right) }$ (\ref{Znm(-1)}%
) through the scalar three-point integral $J_{3}=i(4\pi )^{-2}Z_{00}^{\left(
-1\right) }$ and the vector one (\ref{4DJ3mu}). We use (\ref{LetZ4D}) to
determine the behavior of these tensors when all bilinears in their momenta
are zero: 
\begin{eqnarray}
\left. -2mT_{\mu _{23}}^{PVV}\right\vert _{0} &=&\frac{1}{(2\pi )^{2}};\quad
\left. 2mT_{\mu _{13}}^{VPV}\right\vert _{0}=\frac{1}{(2\pi )^{2}};\quad
\left. 2mT_{\mu _{12}}^{VVP}\right\vert _{0}=-\frac{1}{(2\pi )^{2}};
\label{LEPVV} \\
\left. -2mT_{\mu _{23}}^{PAA}\right\vert _{0} &=&\frac{1}{3(2\pi )^{2}}%
;\quad \left. 2mT_{\mu _{13}}^{APA}\right\vert _{0}=\frac{1}{3(2\pi )^{2}}%
;\quad \left. 2mT_{\mu _{12}}^{AAP}\right\vert _{0}=-\frac{1}{3(2\pi )^{2}}.
\label{LEPAA}
\end{eqnarray}%
Each term above is multiplied by the corresponding tensor $\varepsilon _{\mu
_{kl}\nu _{12}}p_{21}^{\nu _{1}}p_{32}^{\nu _{2}}$ with $k<l$.

Second, the other structures that appeared in the RAGFs are $AV$ functions,
which are proportional to two-point vector integrals. Using the result (\ref%
{J2bar4D}), we achieve%
\begin{equation}
T_{\mu _{ij}}^{AV}\left( a,b\right) =-4i\varepsilon _{\mu _{i}\mu _{j}\nu
_{1}\nu _{2}}p_{ba}^{\nu _{2}}\bar{J}_{2}^{\nu _{1}}\left( a,b\right)
=2i\varepsilon _{\mu _{i}\mu _{j}\nu _{1}\nu _{2}}p_{ba}^{\nu
_{2}}P_{ab}^{\nu _{3}}\Delta _{3\nu _{3}}^{\nu _{1}}.  \label{AV4D}
\end{equation}%
As contributions (exclusively) on the external momentum cancel out in the
contraction, they are pure surface terms proportional to arbitrary label
combinations. After replacing the adequate labels ($k_{a}$ and $k_{b}$),
combinations seen in the RAGFs above arise: 
\begin{eqnarray}
T_{\mu _{32}}^{AV}\left( 1,2\right) -T_{\mu _{23}}^{AV}\left( 2,3\right)
&=&-2i\varepsilon _{\mu _{23}\nu _{12}}\left( p_{21}^{\nu _{2}}P_{12}^{\nu
_{3}}+p_{32}^{\nu _{2}}P_{32}^{\nu _{3}}\right) \Delta _{3\nu _{3}}^{\nu
_{1}}  \label{AV(-)1} \\
T_{\mu _{13}}^{AV}\left( 1,3\right) -T_{\mu _{13}}^{AV}\left( 2,3\right)
&=&-2i\varepsilon _{\mu _{13}\nu _{12}}\left( p_{32}^{\nu _{2}}P_{32}^{\nu
_{3}}-p_{31}^{\nu _{2}}P_{31}^{\nu _{3}}\right) \Delta _{3\nu _{3}}^{\nu
_{1}}  \label{AV(-)2} \\
T_{\mu _{12}}^{AV}\left( 1,2\right) -T_{\mu _{12}}^{AV}\left( 1,3\right)
&=&-2i\varepsilon _{\mu _{12}\nu _{12}}\left( p_{31}^{\nu _{2}}P_{31}^{\nu
_{3}}-p_{21}^{\nu _{2}}P_{21}^{\nu _{3}}\right) \Delta _{3\nu _{3}}^{\nu
_{1}}.  \label{AV(-)3}
\end{eqnarray}%
We stress that these forms depend only on the vertex contraction and not
specific amplitude ($AVV$, $VAV$, $VVA$, and $AAA$). That occurs because
there is a sign change in the $AV$ when permuting the position of free
indexes (see $\varepsilon _{\mu _{i}\mu _{j}\nu _{1}\nu _{2}}$) or changing
the role of routings (see $p_{ba}^{\nu _{2}}P_{ab}^{\nu _{3}}$). 
\begin{figure}[tbph]
\centering\includegraphics[scale=0.8]{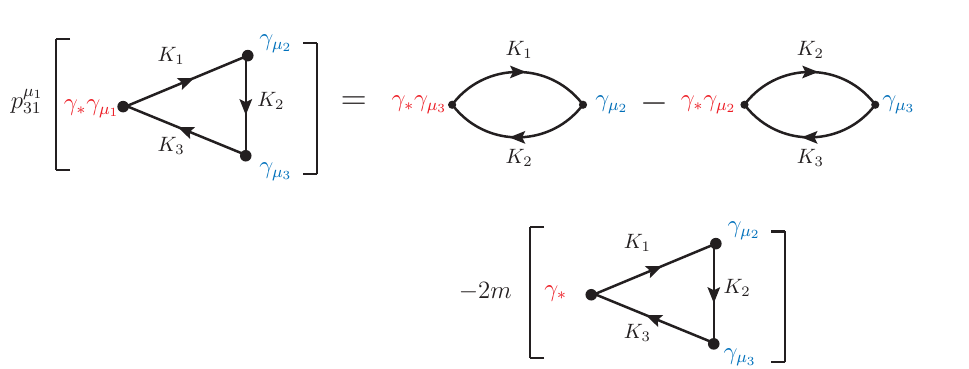}
\caption{The RAGF established for the contraction with momenta $q_{31}^{%
\protect\mu _{1}}T_{\protect\mu _{123}}^{AVV}.$}
\end{figure}

To verify RAGFs, we must contract external momenta with the explicit forms
of amplitudes. Observe the finite contributions displayed in the example (%
\ref{AVV1complete}) to clarify operations involving finite contributions.
These results use well-defined relations involving finite quantities. After
contracting with momenta, some terms vanish due to the Levi-Civita symbol.
Then, we manipulate the remaining terms using tools developed in Subsection (%
\ref{Sub4DIn}). The procedure involves reducing $J$-tensors to identify
finite 2nd-order amplitudes or achieve some cancellations. The referred
reductions are for tensor integrals 
\begin{eqnarray}
2p_{21}^{\nu _{2}}J_{3\nu _{2}}^{\nu _{1}} &=&-p_{21}^{2}J_{3}^{\nu
_{1}}+J_{2}^{\nu _{1}}\left( p_{31}\right) +J_{2}^{\nu _{1}}\left(
p_{32}\right) +p_{31}^{\nu _{1}}J_{2}\left( p_{32}\right) , \\
2p_{31}^{\nu _{2}}J_{3\nu _{2}}^{\nu _{1}} &=&-p_{31}^{2}J_{3}^{\nu
_{1}}+J_{2}^{\nu _{1}}\left( p_{21}\right) +J_{2}^{\nu _{1}}\left(
p_{32}\right) +p_{31}^{\nu _{1}}J_{2}\left( p_{32}\right) , \\
2J_{3\nu _{1}}^{\nu _{1}} &=&2m^{2}J_{3}+2J_{2}\left( p_{32}\right) +i\left(
4\pi \right) ^{-2},  \label{J3uu}
\end{eqnarray}%
and vector integrals%
\begin{eqnarray}
2p_{21\nu _{1}}J_{3}^{\nu _{1}} &=&-p_{21}^{2}J_{3}+J_{2}\left(
p_{31}\right) -J_{2}\left( p_{32}\right) , \\
2p_{31\nu _{1}}J_{3}^{\nu _{1}} &=&-p_{31}^{2}J_{3}+J_{2}\left(
p_{21}\right) -J_{2}\left( p_{32}\right) .
\end{eqnarray}

Although some reductions arise directly, other occurrences require further
algebraic manipulations. This circumstance manifests in cases where a $J$%
-tensor couples to the Levi-Civita symbol so that rearranging indexes is
necessary to find momenta contractions. For vector integrals, we consider
the identity $\varepsilon _{\lbrack \mu _{a}\mu _{b}\nu _{1}\nu _{2}}p_{\nu
_{3}]}J_{3}^{\nu _{1}}=0$ to achieve the formula\footnote{%
Two terms like $p_{a}\varepsilon _{b}\nu _{123}p_{21}^{\nu _{2}}p_{31}^{\nu
_{3}}J_{3}^{\nu _{1}}$ cancel due to triple contraction.}%
\begin{equation}
2\varepsilon _{\mu _{ab}\nu _{12}}\left[ p_{21}^{\nu _{2}}\left( p_{ij}\cdot
p_{31}\right) -p_{31}^{\nu _{2}}\left( p_{ij}\cdot p_{21}\right) \right]
J_{3}^{\nu _{1}}=-\varepsilon _{\mu _{ab}\nu _{23}}p_{21}^{\nu
_{2}}p_{31}^{\nu _{3}}\left[ 2p_{ij\nu _{1}}J_{3}^{\nu _{1}}\right] .
\label{SCH-2}
\end{equation}%
Similarly, we use $\varepsilon _{\lbrack \mu _{a}\nu _{1}\nu _{2}\nu
_{3}}J_{3\mu _{c}]}^{\nu _{1}}=0$ to reorganize terms involving the tensor
integral 
\begin{eqnarray}
&&2\varepsilon _{\mu _{b}\nu _{123}}p_{21}^{\nu _{2}}p_{31}^{\nu
_{3}}J_{3\mu _{a}}^{\nu _{1}}-2\varepsilon _{\mu _{a}\nu _{123}}p_{21}^{\nu
_{2}}p_{31}^{\nu _{3}}J_{3\mu _{b}}^{\nu _{1}}  \notag \\
&=&\varepsilon _{\mu _{ab}\nu _{13}}p_{31}^{\nu _{3}}\left[ 2p_{21}^{\nu
_{2}}J_{3\nu _{2}}^{\nu _{1}}\right] -\varepsilon _{\mu _{ab}\nu
_{12}}p_{21}^{\nu _{2}}\left[ 2p_{31}^{\nu _{3}}J_{3\nu _{3}}^{\nu _{1}}%
\right] -\varepsilon _{\mu _{ab}\nu _{23}}p_{21}^{\nu _{2}}p_{31}^{\nu _{3}}%
\left[ 2J_{3\nu _{1}}^{\nu _{1}}\right] .  \label{SCH-3}
\end{eqnarray}

In the amplitudes, we have two structures: standard tensors $C_{i\mu _{123}}$
(\ref{C1})-(\ref{C3}) and subamplitudes. The tensors are common to the
amplitudes versions and are comprised of the sign tensors (\ref{T-+})-(\ref%
{T++}). To illustrate the operations necessary for the RAGFs, let us take
the case%
\begin{eqnarray}
C_{1\mu _{123}}^{\mathrm{finite}} &=&-2\varepsilon _{\mu _{13}\nu
_{12}}[p_{31}^{\nu _{2}}(J_{3\mu _{2}}^{\nu _{1}}+p_{21\mu _{2}}J_{3}^{\nu
_{1}})-p_{21}^{\nu _{2}}(J_{3\mu _{2}}^{\nu _{1}}+p_{31\mu _{2}}J_{3}^{\nu
_{1}})] \\
&&-2\varepsilon _{\mu _{12}\nu _{12}}p_{21}^{\nu _{2}}(J_{3\mu _{3}}^{\nu
_{1}}+p_{31\mu _{3}}J_{3}^{\nu _{1}}).
\end{eqnarray}%
The first term in parenthesis cancels when contracting with $p_{31}^{\mu
_{1}}$, the remaining terms are%
\begin{equation}
p_{31}^{\mu _{1}}C_{1\mu _{123}}^{\mathrm{finite}}=-2[\varepsilon _{\mu
_{3}\nu _{123}}p_{21}^{\nu _{2}}p_{31}^{\nu _{3}}J_{3\mu _{2}}^{\nu
_{1}}-\varepsilon _{\mu _{2}\nu _{123}}p_{21}^{\nu _{2}}p_{31}^{\nu
_{3}}J_{3\mu _{3}}^{\nu _{1}}].
\end{equation}%
Then, we employ the identity (\ref{SCH-3}) to permute indexes and perform
reductions. That accomplishes our objective; furthermore, this rearrangement
implies the presence of Eq. (\ref{J3uu}), and that brings two additional
contributions: one proportional to squared mass and a numeric factor. That
differs from contractions $p_{21}^{\mu _{2}}$ and $p_{32}^{\mu _{3}}$, where
reductions of tensor integrals are immediate, and it is only necessary to
use (\ref{SCH-2}). The behavior of different contractions is not associated
with vertex content but with amplitude version.%
\begin{eqnarray}
p_{31}^{\mu _{1}}C_{1\mu _{123}}^{\mathrm{finite}} &=&\varepsilon _{\mu
_{23}\nu _{1}\nu _{2}}\{(p_{31}^{\nu _{2}}p_{21}^{2}-p_{21}^{\nu
_{2}}p_{31}^{2})J_{3}^{\nu _{1}}  \label{cont1} \\
&&+p_{21}^{\nu _{1}}p_{31}^{\nu _{2}}[2m^{2}J_{3}+i(4\pi
)^{-2}+J_{2}(p_{32})]\}  \notag \\
p_{21}^{\mu _{2}}C_{1\mu _{123}}^{\mathrm{finite}} &=&\frac{1}{2}\varepsilon
_{\mu _{13}\nu _{12}}p_{32}^{\nu _{2}}\left[ 2p_{21}^{2}\left( J_{3}^{\nu
_{1}}+p_{21}^{\nu _{1}}J_{3}\right) -p_{21}^{\nu _{1}}J_{2}\left(
p_{31}\right) \right]  \label{cont2} \\
p_{32}^{\mu _{3}}C_{1\mu _{123}}^{\mathrm{finite}} &=&\frac{1}{2}\varepsilon
_{\mu _{12}\nu _{1}\nu _{2}}p_{21}^{\nu _{2}}[-2p_{32}^{2}J_{3}^{\nu
_{1}}-p_{31}^{\nu _{1}}J_{2}\left( p_{31}\right) ]  \label{cont3}
\end{eqnarray}%
\begin{eqnarray}
p_{31}^{\mu _{1}}C_{2\mu _{123}}^{\mathrm{finite}} &=&\frac{1}{2}\varepsilon
_{\mu _{23}\nu _{12}}p_{32}^{\nu _{2}}[2p_{31}^{2}\left( J_{3}^{\nu
_{1}}+p_{21}^{\nu _{1}}J_{3}\right) -p_{21}^{\nu _{1}}J_{2}\left(
p_{21}\right) ] \\
p_{21}^{\mu _{2}}C_{2\mu _{123}}^{\mathrm{finite}} &=&\varepsilon _{\mu
_{13}\nu _{12}}\{(p_{31}^{\nu _{2}}p_{21}^{2}-p_{21}^{\nu
_{2}}p_{31}^{2})J_{3}^{\nu _{1}} \\
&&+p_{21}^{\nu _{1}}p_{31}^{\nu _{2}}[2m^{2}J_{3}+i(4\pi
)^{-2}+J_{2}(p_{32})]\}  \notag \\
p_{32}^{\mu _{3}}C_{2\mu _{123}}^{\mathrm{finite}} &=&\frac{1}{2}\varepsilon
_{\mu _{12}\nu _{12}}[2p_{31}^{\nu _{2}}p_{32}^{2}J_{3}^{\nu
_{1}}+p_{21}^{\nu _{1}}p_{31}^{\nu _{2}}J_{2}\left( p_{21}\right) ]
\end{eqnarray}%
\begin{eqnarray}
p_{31}^{\mu _{1}}C_{3\mu _{123}}^{\mathrm{finite}} &=&\frac{1}{2}\varepsilon
_{\mu _{23}\nu _{12}}p_{21}^{\nu _{2}}[2p_{31}^{2}J_{3}^{\nu
_{1}}+p_{31}^{\nu _{1}}J_{2}(p_{32})] \\
p_{21}^{\mu _{2}}C_{3\mu _{123}}^{\mathrm{finite}} &=&\frac{1}{2}\varepsilon
_{\mu _{13}\nu _{12}}p_{31}^{\nu _{2}}[-2p_{21}^{2}J_{3}^{\nu
_{1}}-p_{21}^{\nu _{1}}J_{2}(p_{32})] \\
p_{32}^{\mu _{3}}C_{3\mu _{123}}^{\mathrm{finite}} &=&\varepsilon _{\mu
_{12}\nu _{12}}\{\left( p_{21}^{\nu _{2}}p_{31}^{2}-p_{31}^{\nu
_{2}}p_{21}^{2}\right) J_{3}^{\nu _{1}}  \label{cont9} \\
&&-p_{21}^{\nu _{1}}p_{31}^{\nu _{2}}[2m^{2}J_{3}+i(4\pi
)^{-2}+J_{2}(p_{32})]\}.  \notag
\end{eqnarray}

We have to sum contributions from the subamplitudes to complete finite-parts
results. That requires the same resources discussed above, but only vector
integrals remain, and again we use Eq. (\ref{SCH-2}) to reduce these
integrals to scalar ones. Terms proportional to the squared mass arise from
a part of the common tensors and subamplitudes. They cancel in all
vector-vertex contractions and combine into the expected finite functions
for all axial-vertex contractions (\ref{PVV})-(\ref{AAP}). Lastly,
regardless of the specific amplitude, the additional term $i\left( 4\pi
\right) ^{-2}$ arises when the contracted index $\mu _{i}$ matches the $i$%
-th version.

To complete the RAGFs analysis, we recall Eqs. (\ref{ST1})-(\ref{ST3}). In
the set of surface terms $S_{i\mu _{123}}$, the index $\mu _{i}$ appears
only in the Levi-Civita tensor and not in $\Delta _{3\mu \nu }$.\textbf{\ }%
Hence, contracting other indexes leads to the expected differences (\ref%
{AV(-)1})-(\ref{AV(-)3}). Regardless of the particular triangle amplitude,
identifications are automatic whenever contractions with $S_{i\mu _{123}}$
consider the index $\mu _{j}$ with $i\neq j$. On the other hand, when the
contracted index corresponds to the vertex that defines the version ($i=j$),
the contraction between $p_{31}^{\mu _{1}}$ and $S_{1\mu _{123}}$ does not
produce the required index configuration since we do not find momenta
contractions with surface terms required to identify $AV$ functions. Thus,
in parallel to the procedure for 2nd-order $J$-tensors, indexes are
reorganized through the identity 
\begin{equation}
\varepsilon _{\mu _{1}\mu _{3}\nu _{1}\nu _{2}}\Delta _{3\mu _{2}}^{\nu
_{1}}-\varepsilon _{\mu _{1}\mu _{2}\nu _{1}\nu _{2}}\Delta _{3\mu
_{3}}^{\nu _{1}}=\varepsilon _{\mu _{2}\mu _{3}\nu _{1}\nu _{2}}\Delta
_{3\mu _{1}}^{\nu _{1}}+\varepsilon _{\mu _{1}\mu _{2}\mu _{3}\nu
_{1}}\Delta _{3\nu _{2}}^{\nu _{1}}-\varepsilon _{\mu _{1}\mu _{2}\mu
_{3}\nu _{2}}\Delta _{3\nu _{1}}^{\nu _{1}}.
\end{equation}%
After organizing the momenta by $p_{ij}=P_{ir}-P_{jr}$, these operations
yield (\ref{contS1}). Besides the expected contributions, note the presence
of an additional term on the trace $\Delta _{3\nu }^{\nu }$ resembling what
occurred for the finite part. 
\begin{eqnarray}
p_{31}^{\mu _{1}}S_{1\mu _{123}} &=&-2i\varepsilon _{\mu _{23}\nu
_{12}}\left( p_{21}^{\nu _{2}}P_{12}^{\nu _{3}}+p_{32}^{\nu _{2}}P_{32}^{\nu
_{3}}\right) \Delta _{3\nu _{3}}^{\nu _{1}}+2i\varepsilon _{\mu _{2}\mu
_{3}\nu _{2}\nu _{3}}p_{21}^{\nu _{2}}p_{31}^{\nu _{3}}\Delta _{3\nu
_{1}}^{\nu _{1}}  \label{contS1} \\
p_{21}^{\mu _{2}}S_{1\mu _{123}} &=&-2i\varepsilon _{\mu _{13}\nu
_{12}}(p_{32}^{\nu _{2}}P_{32}^{\nu _{3}}-p_{31}^{\nu _{2}}P_{31}^{\nu
_{3}})\Delta _{3\nu _{3}}^{\nu _{1}} \\
p_{32}^{\mu _{3}}S_{1\mu _{123}} &=&-2i\varepsilon _{\mu _{12}\nu
_{12}}\left( p_{31}^{\nu _{2}}P_{31}^{\nu _{3}}-p_{21}^{\nu _{2}}P_{21}^{\nu
_{3}}\right) \Delta _{3\nu _{3}}^{\nu _{1}}
\end{eqnarray}%
\begin{eqnarray}
p_{31}^{\mu _{1}}S_{2\mu _{123}} &=&-2i\varepsilon _{\mu _{23}\nu
_{12}}\left( p_{21}^{\nu _{2}}P_{12}^{\nu _{3}}+p_{32}^{\nu _{2}}P_{32}^{\nu
_{3}}\right) \Delta _{3\nu _{3}}^{\nu _{1}} \\
p_{21}^{\mu _{2}}S_{2\mu _{123}} &=&-2i\varepsilon _{\mu _{13}\nu
_{12}}\left( p_{32}^{\nu _{2}}P_{32}^{\nu _{3}}-p_{31}^{\nu _{2}}P_{31}^{\nu
_{3}}\right) \Delta _{3\nu _{3}}^{\nu _{1}}+2i\varepsilon _{\mu _{1}\mu
_{3}\nu _{2}\nu _{3}}p_{21}^{\nu _{2}}p_{31}^{\nu _{3}}\Delta _{3\nu
_{1}}^{\nu _{1}}  \label{contS2} \\
p_{32}^{\mu _{3}}S_{2\mu _{123}} &=&-2i\varepsilon _{\mu _{12}\nu
_{12}}\left( p_{31}^{\nu _{2}}P_{31}^{\nu _{3}}-p_{21}^{\nu _{2}}P_{21}^{\nu
_{3}}\right) \Delta _{3\nu _{3}}^{\nu _{1}}
\end{eqnarray}%
\begin{eqnarray}
p_{31}^{\mu _{1}}S_{3\mu _{123}} &=&-2i\varepsilon _{\mu _{23}\nu
_{12}}\left( p_{21}^{\nu _{2}}P_{12}^{\nu _{3}}+p_{32}^{\nu _{2}}P_{32}^{\nu
_{3}}\right) \Delta _{3\nu _{3}}^{\nu _{1}} \\
p_{21}^{\mu _{2}}S_{3\mu _{123}} &=&-2i\varepsilon _{\mu _{13}\nu
_{12}}\left( p_{32}^{\nu _{2}}P_{32}^{\nu _{3}}-p_{31}^{\nu _{2}}P_{31}^{\nu
_{3}}\right) \Delta _{3\nu _{3}}^{\nu _{1}} \\
p_{32}^{\mu _{3}}S_{3\mu _{123}} &=&-2i\varepsilon _{\mu _{12}\nu
_{12}}\left( p_{31}^{\nu _{2}}P_{31}^{\nu _{3}}-p_{21}^{\nu _{2}}P_{21}^{\nu
_{3}}\right) \Delta _{3\nu _{3}}^{\nu _{1}}-2i\varepsilon _{\mu _{1}\mu
_{2}\nu _{2}\nu _{3}}p_{21}^{\nu _{2}}p_{31}^{\nu _{3}}\Delta _{3\nu
_{1}}^{\nu _{1}}  \label{contS9}
\end{eqnarray}

With these properties in hands, we establish RAGFs for the explicit $(T_{\mu
_{123}}^{AVV})_{1},$ see (\ref{AVV1}) to illustrate how to proceed in any
case. The axial contraction comes from reducing the common tensor in Eq. (%
\ref{cont1}) plus the nonzero terms from subamplitude (\ref{TVPP})%
\begin{equation*}
i\varepsilon _{\mu _{123}\nu _{1}}p_{31}^{\mu _{1}}(T^{VPP})^{\nu
_{1}}=2i\varepsilon _{\mu _{23}\nu _{12}}p_{31}^{\nu _{2}}\left\{
2(p_{21}\cdot p_{32})J_{3}^{\nu _{1}}+p_{21}^{\nu
_{1}}[p_{31}^{2}J_{3}-J_{2}(p_{21})-J_{2}(p_{32})]\right\} .
\end{equation*}%
At this stage, we have when summing both contributions 
\begin{eqnarray}
p_{31}^{\mu _{1}}(T_{\mu _{123}}^{AVV})_{1} &=&p_{31}^{\mu _{1}}S_{1\mu
_{123}}+4i\varepsilon _{\mu _{23}\nu _{12}}p_{21}^{\nu _{1}}p_{31}^{\nu
_{2}}[2m^{2}J_{3}+i\left( 4\pi \right) ^{-2}] \\
&&+4i\varepsilon _{\mu _{23}\nu _{12}}\left[ p_{31}^{\nu _{2}}\left(
p_{21}\cdot p_{31}\right) -p_{21}^{\nu _{2}}p_{31}^{2}\right] J_{3}^{\nu
_{1}}  \notag \\
&&+2i\varepsilon _{\mu _{23}\nu _{12}}p_{21}^{\nu _{1}}p_{31}^{\nu
_{2}}[p_{31}^{2}J_{3}+J_{2}\left( p_{32}\right) -J_{2}\left( p_{21}\right) ].
\notag
\end{eqnarray}%
To find reductions in terms like the second row, we use (\ref{SCH-2}) to
identify the needed contraction and obtain a cancellation 
\begin{equation}
p_{31}^{\mu _{1}}(T_{\mu _{123}}^{AVV})_{1}=p_{31}^{\mu _{1}}S_{1\mu
_{123}}+4i\varepsilon _{\mu _{23}\nu _{12}}p_{21}^{\nu _{1}}p_{31}^{\nu
_{2}}[2m^{2}J_{3}+i\left( 4\pi \right) ^{-2}].
\end{equation}%
After contracting surface terms using (\ref{contS1}) and identifying the $%
PVV $ (\ref{PVV}), we write%
\begin{equation}
p_{31}^{\mu _{1}}(T_{\mu _{123}}^{AVV})_{1}=T_{\mu _{32}}^{AV}\left(
1,2\right) -T_{\mu _{23}}^{AV}\left( 2,3\right) -2mT_{\mu _{23}}^{PVV}+%
\underline{2i\varepsilon _{\mu _{23}\nu _{12}}p_{21}^{\nu _{1}}p_{31}^{\nu
_{2}}[\Delta _{3\alpha }^{\alpha }+2i\left( 4\pi \right) ^{-2}]}.
\end{equation}%
Similarly, RAGFs coming from vector vertices use (\ref{cont2})-(\ref{cont3})
for the common tensor and identity (\ref{SCH-2}). They imply the vanishing
of finite parts, while the remaining parts correspond to $AV$ differences:%
\begin{eqnarray}
p_{21}^{\mu _{2}}(T_{\mu _{123}}^{AVV})_{1} &=&p_{21}^{\mu _{2}}S_{1\mu
_{123}}=T_{\mu _{13}}^{AV}\left( 1,3\right) -T_{\mu _{13}}^{AV}\left(
2,3\right) \\
p_{32}^{\mu _{3}}(T_{\mu _{123}}^{AVV})_{1} &=&p_{32}^{\mu _{3}}S_{1\mu
_{123}}=T_{\mu _{12}}^{AV}\left( 1,2\right) -T_{\mu _{12}}^{AV}\left(
1,3\right) .
\end{eqnarray}

This pattern repeats for the first version of the other amplitudes ($VAV$, $%
VVA$, and $AAA$). Whereas the contraction with first vertex exhibits the
additional term, the other RAGFs are satisfied without conditions. The
pattern changes to the second and third versions, for they show the
violating term in the second and third vertex independent of its nature:
axial or vector vertex.

\begin{figure}[tbph]
\centering\includegraphics[scale=0.8]{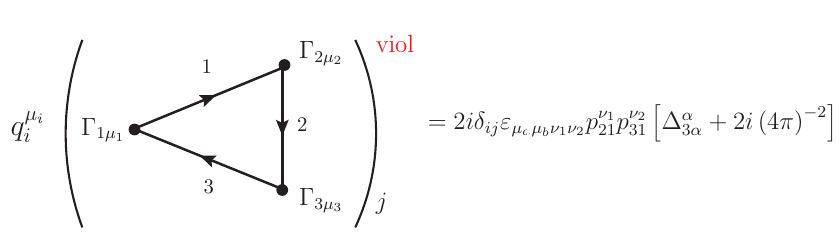}
\caption{The violation factor of the RAGF established for the contraction
with momenta $q_{i}^{\protect\mu _{1}}.$}
\label{diagviol}
\end{figure}

Following the developed steps, equations below subsume all potentially
offending terms, which emerge in momentum contractions where the version is
defined. We adopt the notation to the routing differences $q_{1}=p_{31}$, $%
q_{2}=p_{21}$, and $q_{3}=p_{32}$ to mark a convention for first, second,
and third vertices. The notation has already appeared in Figure \ref{diag1}
for the general diagram. In addition, the symbol $\Gamma _{123}\equiv \Gamma
_{1}\Gamma _{2}\Gamma _{3}$ is an abbreviation for all combinations of
vertices $\Gamma _{i}\in \{A,V\}$ we are investigating. 
\begin{eqnarray}
q_{1}^{\mu _{1}}(T_{\mu _{123}}^{\Gamma _{123}})_{1}^{\mathrm{viol}}
&=&+2i\varepsilon _{\mu _{23}\nu _{12}}q_{2}^{\nu _{1}}q_{3}^{\nu
_{2}}[\Delta _{3\alpha }^{\alpha }+2i\left( 4\pi \right) ^{-2}]
\label{ragfViol4D} \\
q_{2}^{\mu _{2}}(T_{\mu _{123}}^{\Gamma _{123}})_{2}^{\mathrm{viol}}
&=&+2i\varepsilon _{\mu _{13}\nu _{12}}q_{2}^{\nu _{1}}q_{3}^{\nu
_{2}}[\Delta _{3\alpha }^{\alpha }+2i\left( 4\pi \right) ^{-2}]  \notag \\
q_{3}^{\mu _{3}}(T_{\mu _{123}}^{\Gamma _{123}})_{3}^{\mathrm{viol}}
&=&-2i\varepsilon _{\mu _{12}\nu _{12}}q_{2}^{\nu _{1}}q_{3}^{\nu
_{2}}[\Delta _{3\alpha }^{\alpha }+2i\left( 4\pi \right) ^{-2}].  \notag
\end{eqnarray}%
\ The other vertices (to each version) have their RAGFs identically
satisfied. To visualize this violation pattern, we offer the schematic graph
in Figure \ref{diagviol}. 

RAGFs are not automatic as they require further explorations regarding
values accessible to surface terms, meaning they only apply under the
constraint%
\begin{equation}
\Delta _{3\alpha }^{\alpha }=-\frac{2i}{\left( 4\pi \right) ^{2}}.
\label{Daa}
\end{equation}%
From another perspective, if these relations apply identically, we could
satisfy all Ward identities by nullifying surface terms (this works channel
by channel). That is not the case because it requires conflicting
interpretations of surface terms: zero for the momentum-space translational
invariance and nonzero for the linearity of integration. Thence, these
properties do not hold simultaneously. General tensor properties and the
low-energy behavior of $PVV$-$PAA$ and permutations show these conclusions
are inescapable in Section (\ref{LED4DSTS}). That is independent of any
possible trace.

Once the RAGFs are clear, we would like to deepen the discussion about
different versions of amplitudes. The investigated integrands are
well-defined tensors and obey $(t_{\mu _{123}}^{\Gamma _{123}})_{i}=(t_{\mu
_{123}}^{\Gamma _{123}})_{j}$. Even if we separate expressions in finite and
divergent sectors without commitment to the divergences, after integration,
the sampling of indexes makes the results of finite parts and tensor surface
terms different. We highlight differences among the three main versions to
elucidate this point:%
\begin{eqnarray}
(T_{\mu _{123}}^{\Gamma _{123}})_{1}-(T_{\mu _{123}}^{\Gamma _{123}})_{2}
&=&+2i\varepsilon _{\mu _{123}\nu _{1}}p_{32}^{\nu _{1}}[\Delta _{3\alpha
}^{\alpha }+2i\left( 4\pi \right) ^{-2}], \\
(T_{\mu _{123}}^{\Gamma _{123}})_{1}-(T_{\mu _{123}}^{\Gamma _{123}})_{3}
&=&-2i\varepsilon _{\mu _{123}\nu _{1}}p_{21}^{\nu _{1}}[\Delta _{3\alpha
}^{\alpha }+2i\left( 4\pi \right) ^{-2}], \\
(T_{\mu _{123}}^{\Gamma _{123}})_{2}-(T_{\mu _{123}}^{\Gamma _{123}})_{3}
&=&-2i\varepsilon _{\mu _{123}\nu _{1}}p_{31}^{\nu _{1}}[\Delta _{3\alpha
}^{\alpha }+2i\left( 4\pi \right) ^{-2}].
\end{eqnarray}%
After subtracting two versions, we reorganized indexes to identify
reductions of finite functions and recognize the same potentially violating
term acknowledged in (\ref{ragfViol4D}). At this point, we define the
meaning of uniqueness adopted within this investigation: any possible form
to compute the same expression returns the same result. Canceling the RHS of
these equations would be required to achieve this property. That only
happens when adopting the same prescription seen above $\Delta _{3\alpha
}^{\alpha }=-2i\left( 4\pi \right) ^{-2}$. This notion of uniqueness implies
that an amplitude does not depend on Dirac traces. Nevertheless, unlike in
the two-dimensional context, the nonzero surface terms required by this
notion allow dependence on ambiguous combinations of arbitrary internal
momenta. In this sense, there is no unique expression in the external
momenta.

The trace of six matrices is the unique place\ where the amplitude versions
differ. Achieving traces different from those starting this argumentation is
possible through other identities involving the chiral matrix, Eq. (\ref%
{Chiral-Id}). Nonetheless, as detailed in Appendix (\ref{Tr6G4D}), versions
that are linear combinations of them arise. Observe the form%
\begin{equation}
\lbrack T_{\mu _{123}}^{\Gamma _{123}}]_{i;j}=[(T_{\mu _{123}}^{\Gamma
_{123}})_{i}+(T_{\mu _{123}}^{\Gamma _{123}})_{j}]/2,  \label{Tij}
\end{equation}%
which manifests potentially violating terms in RAGFs for both vertices $%
\Gamma _{i}$ and $\Gamma _{j}$. The three independent combinations (setting $%
i$ and $j$) are enough to reproduce any expressions achieved through the
referred identities. That justifies taking $(T_{\mu _{123}}^{\Gamma
_{123}})_{i}$ as the basic versions; moreover, \textit{they have the maximum
number of RAGFs identically satisfied}, see Section (\ref{LED4DSTS}). For
instance, the expression associated with the substitution 
\begin{equation}
\gamma _{\ast }\gamma _{\mu _{i}\nu _{i}\mu _{i+1}}=i\varepsilon _{\mu
_{i}\nu _{i}\mu _{i+1}\nu }\gamma ^{\nu }+\gamma _{\ast }(g_{\nu _{i}\mu
_{i+1}}\gamma _{\mu _{i}}-g_{\mu _{i}\mu _{i+1}}\gamma _{\nu _{i}}+g_{\mu
_{i}\nu _{i}}\gamma _{\mu _{i+1}})
\end{equation}%
has an integrand differing from $[T_{\mu _{123}}^{\Gamma _{123}}]_{i;i+1}$%
\footnote{%
Note that when $i=3$ the notation means $[T_{\mu _{123}}^{\Gamma
_{123}}]_{3,1}$, or $\gamma _{\ast }\gamma _{\mu _{2}\nu _{2}\mu _{1}}$ in
the identity used.} in terms that have finite and identically vanishing
integrals (\ref{T-+}) and (\ref{ASS}). Using this identity or combining
traces of basic versions before integration makes expressions exhibit the
same terms when integrated, divergent and finite parts. As another example,
employing the identity $\gamma _{\ast }\gamma _{\mu _{i}}=\varepsilon _{\mu
_{i}\nu _{123}}\gamma ^{\nu _{123}}/3!$ expresses the trace through ten
monomials. Even without some index configurations, the integrated expression
coincides with the $i$-th version. That means the chiral matrix definition
has no special role compared to other identities.

With these facts in mind, we define linear combinations that reproduce any
possible expression with the building-block versions 
\begin{equation}
\lbrack T_{\mu _{123}}^{\Gamma _{123}}]_{\left\{ r_{1}r_{2}r_{3}\right\} }=%
\frac{1}{r_{1}+r_{2}+r_{3}}\sum_{i=1}^{3}r_{i}(T_{\mu _{123}}^{\Gamma
_{123}})_{i},  \label{r123}
\end{equation}%
where $r_{1}+r_{2}+r_{3}\not=0$. They have equivalent integrands as it
occurs for combinations (\ref{Tij}). This general form compiles all involved
arbitrariness, accounting for any choices regarding routings or Dirac
traces. From this formula, assuming zero surface terms after the
integration, we identify an infinity set of amplitudes that violate RAGFs by
arbitrary amounts. That is useful for obtaining different violation values
in the literature, e.g., \cite{Wu2006}.

We have shown how traces and surface terms interfere with the investigated
tensors' linearity of integration and uniqueness. In the subsequent
subsections, we demonstrate that these properties are unavoidable since
conditions for RAGFs arise without explicit computations of the primary
amplitudes.

\section{A Low-Energy Theorem and its Relation with Ward Identities \label%
{LE4D}}

This section proposes a structure depending only on external momenta to
formulate a low-energy implication for a tensor representing three-point
amplitudes. That does not mean we ignore the possible presence of ambiguous
routing combinations because these terms can be transformed into linear
covariant combinations of physical momenta. The structure is a general
3rd-order tensor having odd parity:%
\begin{eqnarray}
F_{\mu _{123}} &=&\varepsilon _{\mu _{123}\nu }(q_{2}^{\nu }F_{1}+q_{3}^{\nu
}F_{2})+\varepsilon _{\mu _{12}\nu _{12}}q_{2}^{\nu _{1}}q_{3}^{\nu
_{2}}(q_{2\mu _{3}}G_{1}+q_{3\mu _{3}}G_{2})  \label{GForm} \\
&&+\varepsilon _{\mu _{13}\nu _{12}}q_{2}^{\nu _{1}}q_{3}^{\nu _{2}}\left(
q_{2\mu _{2}}G_{3}+q_{3\mu _{2}}G_{4}\right) +\varepsilon _{\mu _{23}\nu
_{12}}q_{2}^{\nu _{1}}q_{3}^{\nu _{2}}(q_{2\mu _{1}}G_{5}+q_{3\mu
_{1}}G_{6}).  \notag
\end{eqnarray}%
That is a function of two variables: the incoming external momenta $q_{2}$
and $q_{3}$ associated with vertices $\Gamma _{2}$ and $\Gamma _{3}$.
Conservation sets the relation $q_{1}=q_{2}+q_{3}$ with the outcoming
momentum of the vertex $\Gamma _{1}$.

After performing the momenta contractions, one identifies the arrangements $%
q_{i}^{\mu _{i}}F_{\mu _{123}}=\varepsilon _{\mu _{kl}\nu _{12}}q_{2}^{\nu
_{1}}q_{3}^{\nu _{2}}V_{i}$ with $k<l\not=i$. These operations lead to three
functions written regarding form factors of the general tensor%
\begin{eqnarray}
V_{1} &=&-F_{1}+F_{2}+\left( q_{1}\cdot q_{2}\right) G_{5}+\left( q_{1}\cdot
q_{3}\right) G_{6}, \\
V_{2} &=&-F_{2}+q_{2}^{2}G_{3}+\left( q_{2}\cdot q_{3}\right) G_{4}, \\
V_{3} &=&-F_{1}+q_{3}^{2}G_{2}+\left( q_{2}\cdot q_{3}\right) G_{1}.
\end{eqnarray}%
At the kinematical point where all bilinears are zero $q_{i}\cdot q_{j}=0$,
if $G_{i}$ are regular or at most discontinuous, we have the relations 
\begin{equation*}
V_{1}\left( 0\right) =F_{2}-F_{1},\quad V_{2}\left( 0\right) =-F_{2},\quad
V_{3}\left( 0\right) =-F_{1}.
\end{equation*}%
From the steps above, we derive the following equation among invariants%
\begin{equation}
V_{1}\left( 0\right) +V_{2}\left( 0\right) -V_{3}\left( 0\right) =0.
\label{VertsZero}
\end{equation}%
This relation contains information about symmetries or their violations at
the zero limit, even if no particular symmetry is needed for its deduction.
That occurs because it represents a constraint over three-point structures
arising in the RHS of proposed WIs.

To illustrate this resource, suppose that the axial contraction with the $%
AVV $ connects to\textbf{\ }the amplitude coming from the pseudo-scalar
density 
\begin{equation}
\varepsilon _{\mu _{23}\nu _{12}}q_{2}^{\nu _{1}}q_{3}^{\nu _{2}}V_{1}\left(
0\right) =-2mT_{\mu _{23}}^{PVV}\left( 0\right) =:\varepsilon _{\mu _{23}\nu
_{12}}q_{2}^{\nu _{1}}q_{3}^{\nu _{2}}\Omega _{1}^{PVV}\left( 0\right) ,
\end{equation}%
with the behavior (\ref{LEPVV}) leading to the value for the first invariant 
$V_{1}\left( 0\right) =1/(2\pi )^{2}$. Since the constraint above prevents
the simultaneous vanishing of both other invariants $V_{2}\left( 0\right)
=V_{3}\left( 0\right) =0$, at least one vector WI is violated. On the other
hand, supposing that both vector WIs apply implies violating the axial one.
That occurs because parameters defining the considered tensor and regularity
require the existence of an additional term $V_{1}\left( 0\right) =1/(2\pi
)^{2}+\mathcal{A}$, the anomaly. Thus, $\mathcal{A}=-\Omega _{1}^{PVV}\left(
0\right) $, relating a property of the finite amplitude and the symmetry
content of a rank-3 amplitude. Satisfying the symmetry at this point does
not guarantee invariance for all points; however, its violation at zero
implies symmetry violation.

That is the starting point of the violation pattern in anomalous amplitudes.
Numerical values presented above for invariants $V_{i}$ at zero represent
the preservation of corresponding WIs. Nevertheless, their co-occurrence
implies a violation of the linear-algebra type solution (\ref{VertsZero}).
No tensor, independent of its origin, can connect to the PVV and
simultaneously have vanishing contractions with momenta $q_{2}^{\mu _{2}}$
and $q_{3}^{\mu _{3}}$. Whenever an axial-vertex contraction is connected to
an amplitude coming from the pseudo-scalar density (anomalously or not),
there will be an anomaly in at least one of the vertices; the same
conclusion stands for other diagrams. These facts are known; however, the
form we raise is general.\textbf{\ }The low-energy theorem invoking vector
WIs is only one of the solutions, as in Section (4.2) of \cite{Treiman1985}.
The built equation is an exclusive and inviolable consequence of properties
assumed to the 3rd-order tensor\textbf{,} and symmetry violations occur when
the RHS terms of WIs do not behave accordingly.

The explicit computation of\textbf{\ }perturbative expressions corroborates
these assertions. Moreover, the RAGFs furnish an exact connection among
ultraviolet and infrared features of amplitudes, namely $\Omega
_{1}^{PVV}\left( 0\right) =2i\Delta _{3\alpha }^{\alpha }$\label{IRUV}. That
is the requirement for linearity seen after evaluating the RAGFs, and it
will be derived in the next subsection.\textbf{\ }There\textbf{, }we assume
the form $V_{i}=\Omega _{i}+\mathcal{A}_{i}$ and demonstrate\textbf{\ }the
implication 
\begin{equation}
\Omega _{1}\left( 0\right) +\Omega _{2}\left( 0\right) -\Omega _{3}\left(
0\right) =(2\pi )^{-2},  \label{GamatZero}
\end{equation}%
where we suppress superindexes in $\Omega _{i}$ coming from finite functions
(e.g., $PVV$-$PAA$), see (\ref{GR}). The equation above holds even to
classically non-conserved vector currents or amplitudes with three arbitrary
masses running in the loop. Albeit rank-2 amplitudes of multiple masses are
complicated functions of these masses, the relation at the point zero is
ever the finite constant above.

Independently of divergent aspects, the last equation is incompatible with (%
\ref{VertsZero}); therefore, characterizing violations for rank-3 triangles
under the form (\ref{GForm}). Hence, anomalous terms coming from different
vertices $\mathcal{A}_{i}$ obey the general constraint%
\begin{equation}
\mathcal{A}_{1}+\mathcal{A}_{2}-\mathcal{A}_{3}=-(2\pi )^{-2},  \label{Ans}
\end{equation}%
This equation shows that the value of axial anomaly is unique by preserving
two vector WIs. Likewise, any explicit tensor\footnote{%
This tensor can be obtained via regularization or not. See the approach of
G. Scharf (\cite{Scharf2010}) in Section 5.1, using causal perturbation
theory. The analogous to $PVV$ is not computed until the very end. Instead,
the authors study analogous differences between the contraction of $AVV$ and
the $PVV$ without Feynman diagrams.} having WIs violated by any quantity
obeys this equation if $\mathcal{A}_{i}$ relates to finite amplitudes from
Feynman's rules. The crossed channel of finite amplitudes brings a
multiplicative factor 2 in the last couple of equations.

It is possible to anticipate restrictions over surface terms based on the
general dependence that 3rd-order tensors have on such terms and preserving
the independence and arbitrariness of internal momenta sums. That is
achieved through the connection with $AV$\ functions via integration
linearity\textbf{.} In the next section, this reasoning leads to the
proposition $\Omega _{1}^{PVV}\left( 0\right) =2i\Delta _{3\alpha }^{\alpha
} $ and Eq. (\ref{GamatZero}).

\section{RAGFs and Kinematical Behavior of Amplitudes \label{LED4DSTS}}

In Section (\ref{unique}), we performed explicit calculations related to
different amplitude versions. When satisfying all RAGFs, a condition
connecting the surface term with a finite contribution emerged in at least
one of the relations (\ref{Daa}). This condition appeared without explicitly
calculating surface terms, inferring it from potentially violating terms.
Furthermore, these additional terms arise in RAGFs associated with the
vertex that defines the version (\ref{ragfViol4D}). Here, we will show
generality how the constraints based on linearity are obtained by carefully
analyzing the most general tensor structure of 3pt-amplitudes without using
any specific traces. The meaning of the basic version emerges as the one
that automatically satisfies the most possible RAGFs but not all. Also, we
will consider that when the contractions are done, a set of results is
generated that can only be restricted by linearity for arbitrary and
independent internal momenta. Such a condition shows how the finite
amplitudes in the RHS of the RAGFs determine the surface terms.

From the explicit calculation, we can write the general equation for
linearity as%
\begin{equation}
q_{i}^{\mu _{i}}T_{\mu _{123}}^{\Gamma _{123}}=T_{i\left( -\right) \mu
_{kl}}^{AV}+\varepsilon _{\mu _{kl}\nu _{12}}q_{2}^{\nu _{1}}q_{3}^{\nu
_{2}}\Omega _{i},  \label{GR}
\end{equation}%
the ordering of indexes is always by $k<l\not=i$. The first term of the RHS
is the differences (\ref{AV(-)1})-(\ref{AV(-)3}). The second one has the
invariants corresponding to the rank-2 amplitudes in RAGFs. Note that some
are zero to vertices of specific diagrams. Expressing the three independent
differences of $AV$ functions in terms of $P_{ij}$, we have%
\begin{eqnarray}
T_{1\left( -\right) \mu _{23}}^{AV} &=&-2i\varepsilon _{\mu _{23}\nu _{12}} 
\left[ -P_{21}^{\nu _{2}}P_{32}^{\nu _{3}}+P_{31}^{\nu _{2}}\left(
P_{32}^{\nu _{3}}-P_{21}^{\nu _{3}}\right) +P_{32}^{\nu _{2}}P_{21}^{\nu
_{3}}\right] \Delta _{3\nu _{3}}^{\nu _{1}} \\
T_{2\left( -\right) \mu _{13}}^{AV} &=&-2i\varepsilon _{\mu _{13}\nu _{12}} 
\left[ +P_{21}^{\nu _{2}}\left( P_{31}^{\nu _{3}}-P_{32}^{\nu _{3}}\right)
+P_{31}^{\nu _{2}}P_{32}^{\nu _{3}}-P_{32}^{\nu _{2}}P_{31}^{\nu _{3}}\right]
\Delta _{3\nu _{3}}^{\nu _{1}} \\
T_{3\left( -\right) \mu _{12}}^{AV} &=&-2i\varepsilon _{\mu _{12}\nu _{12}} 
\left[ -P_{21}^{\nu _{2}}P_{31}^{\nu _{3}}+P_{31}^{\nu _{2}}P_{21}^{\nu
_{3}}+P_{32}^{\nu _{2}}\left( P_{31}^{\nu _{3}}-P_{21}^{\nu _{3}}\right) %
\right] \Delta _{3\nu _{3}}^{\nu _{1}}.
\end{eqnarray}%
The notation $T_{i\left( -\right) }^{AV}$ is used to remember it came from
the RAGF where we contracted with $q^{\mu _{i}}$\textit{\ }in the integrand.
These equations preserve the arbitrary label for the internal lines and the
value of the surface term and do not depend on the traces used because there
is no ambiguity in expressing the trace of four Dirac matrices and a chiral
one.

Due to the tensor integral of power counting zero e vector with power
counting one, it must be expected from the expression to depend on surface
term with physical as well ambiguous momenta. On the other hand, the
routings present are not obliged to be written as external momenta, as we
assumed in the previous section. The general tensor must consider that the
perturbative amplitudes are a function of the six variables: the sums and
differences of routings; the last ones are restricted by momentum
conservation, notwithstanding the sums are arbitrary, reducing for five
variables. In turn, with the sums, we generate the differences; thereby, the
number of variables is three. Nevertheless, the summation of routings
appears multiplied necessarily and only by surface terms.

Since central amplitudes are linear-diverging tensors, they have mass one
and\ depend on the arbitrary momenta and surface terms, as $q_{i}$ vectors
are differences of the routings $k_{i}$ (but not the opposite), we replace
the former with the latter. Then using the combinations $P_{ij}=k_{i}+k_{j}$%
, the most general tensor of these variables under the stated conditions is%
\begin{eqnarray}
F_{\mu _{123}}^{\Delta } &=&+\varepsilon _{\mu _{23}\nu _{12}}\left(
a_{11}P_{21}+a_{12}P_{31}+a_{13}P_{32}\right) ^{\nu _{2}}\Delta _{3\mu
_{1}}^{\nu _{1}}  \label{Fdiv} \\
&&+\varepsilon _{\mu _{13}\nu _{12}}\left(
a_{21}P_{21}+a_{22}P_{31}+a_{23}P_{32}\right) ^{\nu _{2}}\Delta _{3\mu
_{2}}^{\nu _{1}}  \notag \\
&&+\varepsilon _{\mu _{12}\nu _{12}}\left(
a_{31}P_{21}+a_{32}P_{31}+a_{33}P_{32}\right) ^{\nu _{2}}\Delta _{3\mu
_{3}}^{\nu _{1}}  \notag \\
&&+\varepsilon _{\mu _{123}\nu _{1}}\left(
b_{1}P_{21}+b_{2}P_{31}+b_{3}P_{32}\right) ^{\nu _{2}}\Delta _{3\nu
_{2}}^{\nu _{1}}.  \notag
\end{eqnarray}%
Finite parts are handled separately. The $a_{ij}$ and $b_{j}$ are twelve
arbitrary constants that summarize all the freedom of such tensor: Function
of three variables of the diagram routings, rank, parity, and power
counting. The $j$ captures the $P$ momenta in the order $\left(
P_{21},P_{31},P_{32}\right) $, and the index $i$ links to the index $\mu
_{i} $ associated with the vertex in the amplitudes $T_{\mu _{123}}^{\Gamma
_{123}}$. Contracting (\ref{Fdiv}) with the routing differences, for this
tensor to be related to the $AV$ tensors, we used the identity\footnote{%
These structures have indices of surface terms contracted with the
coefficient and the epsilon tensor and no trace of the surface term, by
example, $q_{2}^{\mu _{2}}T_{\mu _{123}}^{AVV}=T_{\mu _{13}}^{AV}\left(
1,3\right) -T_{\mu _{13}}^{AV}\left( 2,3\right) $.} $\varepsilon _{\lbrack
\mu _{1}\mu _{2}\mu _{3}\nu _{1}}\Delta _{3\nu _{2}]}^{\nu _{2}}=0$ to cast
the tensor. That reduces, without losing information, the number of
arbitrary parameters.

Now the question is: Performing the three contractions with the vertices
momenta, is it possible to identify all of them with the two-point functions
without additional conditions? That means they must be simultaneously valid
for any value of the surface term. The answer is no, as we show that
requiring two RAGF satisfied without conditions over surface term determines
all coefficients $a_{ij}$ and $b_{i}$. The other relation belongs to an
incompatible solution for these coefficients. We will see as the finite
amplitudes condition the satisfaction of all RAGFs.

Beginning by contracting $F_{\mu _{123}}^{\Delta }$ with $q_{1}^{\mu
_{1}}=p_{31}^{\mu _{1}},$ we have the expression 
\begin{eqnarray}
p_{31}^{\mu _{1}}F_{\mu _{123}}^{\Delta } &=&+\varepsilon _{\mu _{3}\nu
_{123}}[-(a_{21}+a_{23})P_{21}^{\nu _{2}}P_{32}^{\nu _{3}}+a_{22}P_{31}^{\nu
_{2}}(P_{21}^{\nu _{3}}-P_{32}^{\nu _{3}})]\Delta _{3\mu _{2}}^{\nu _{1}} \\
&&+\varepsilon _{\mu _{2}\nu _{123}}[-(a_{31}+a_{33})P_{21}^{\nu
_{2}}P_{32}^{\nu _{3}}+a_{32}P_{31}^{\nu _{2}}(P_{21}^{\nu _{3}}-P_{32}^{\nu
_{3}})]\Delta _{3\mu _{3}}^{\nu _{1}}  \notag \\
&&+\varepsilon _{\mu _{23}\nu _{12}}[-(a_{11}-b_{1})P_{21}^{\nu
_{2}}P_{21}^{\nu _{3}}+(a_{13}-b_{3})P_{32}^{\nu _{2}}P_{32}^{\nu
_{3}}]\Delta _{3\nu _{3}}^{\nu _{1}}  \notag \\
&&+\varepsilon _{\mu _{23}\nu _{12}}[+(a_{11}+b_{3})P_{21}^{\nu
_{2}}P_{32}^{\nu _{3}}-a_{12}P_{31}^{\nu _{2}}(P_{21}^{\nu _{3}}-P_{32}^{\nu
_{3}})]\Delta _{3\nu _{3}}^{\nu _{1}}  \notag \\
&&+\varepsilon _{\mu _{23}\nu _{12}}[-(a_{13}+b_{1})P_{32}^{\nu
_{2}}P_{21}^{\nu _{3}}+b_{2}(P_{21}^{\nu _{2}}-P_{32}^{\nu _{2}})P_{31}^{\nu
_{3}}]\Delta _{3\nu _{3}}^{\nu _{1}}.  \notag
\end{eqnarray}%
From the first two rows, $\boldsymbol{a}_{2}=\left( -a_{23},0,a_{23}\right) $
and $\boldsymbol{a}_{3}=\left( -a_{33},0,a_{33}\right) $, the remaining
compared with $T_{1\left( -\right) \mu _{23}}^{AV},$ we have $%
a_{11}+b_{3}=2i;\ a_{12}=-2i;\ a_{13}+b_{1}=2i;\ b_{2}=0;$ and $%
b_{3}=2i-b_{1}$. In vector notation, the full solution is%
\begin{equation}
\left( 
\begin{array}{c}
\boldsymbol{b} \\ 
\boldsymbol{a}_{1} \\ 
\boldsymbol{a}_{2} \\ 
\boldsymbol{a}_{3}%
\end{array}%
\right) _{1}=%
\begin{pmatrix}
b_{1} & 0 & 2i-b_{1} \\ 
b_{1} & -2i & 2i-b_{1} \\ 
-a_{23} & 0 & a_{23} \\ 
-a_{33} & 0 & a_{33}%
\end{pmatrix}%
.  \label{Sol1}
\end{equation}%
Note the reduction from twelve parameters to just three $\left\{
a_{23},a_{33},b_{1}\right\} $ by requiring just one of the relations to be
satisfied. Repeating the analysis to $q^{\mu _{2}}F_{\mu _{123}}^{\Delta }$
with $q_{2}^{\mu _{2}}=p_{21}^{\mu _{2}}$ and forming the system of linear
equation by comparing with (\ref{AV(-)2}), follows the solution%
\begin{equation}
\left( 
\begin{array}{c}
\boldsymbol{b} \\ 
\boldsymbol{a}_{1} \\ 
\boldsymbol{a}_{2} \\ 
\boldsymbol{a}_{3}%
\end{array}%
\right) _{2}=%
\begin{pmatrix}
0 & b_{2} & 2i-b_{2} \\ 
0 & -a_{13} & a_{13} \\ 
2i & -b_{2} & b_{2}-2i \\ 
0 & -a_{33} & a_{33}%
\end{pmatrix}%
,  \label{Sol2}
\end{equation}%
for the RAGF in the second vertex. The conditions for $q_{3}^{\mu
_{3}}F_{\mu _{123}}^{\Delta }=T_{3\left( -\right) \mu _{12}}^{AV}$ with $%
q_{3}^{\mu _{3}}=p_{32}^{\mu _{3}}$, follows that the solution to the
automatic satisfaction of the RAGF is 
\begin{equation}
\left( 
\begin{array}{c}
\boldsymbol{b} \\ 
\boldsymbol{a}_{1} \\ 
\boldsymbol{a}_{2} \\ 
\boldsymbol{a}_{3}%
\end{array}%
\right) _{3}=%
\begin{pmatrix}
b_{1} & 2i-b_{1} & 0 \\ 
a_{11} & -a_{11} & 0 \\ 
a_{21} & -a_{21} & 0 \\ 
b_{1} & 2i-b_{1} & -2i%
\end{pmatrix}%
.  \label{Sol3}
\end{equation}

The intersection of (\ref{Sol1}) and (\ref{Sol2}), the ones that
automatically satisfy the RAGFs coming from the contraction with $q_{1}^{\mu
_{1}}$ and $q_{2}^{\mu _{2}}$, leads to a unique solution with $b_{1}=0$, $%
b_{2}=0$, $b_{3}=2i$, and all the other coefficients are also determined.
Replacing in the tensor,%
\begin{equation}
(F_{\mu _{123}}^{\Delta })_{12}=-2i[\varepsilon _{\mu _{23}\nu
_{12}}(P_{32}^{\nu _{2}}-P_{31}^{\nu _{2}})\Delta _{3\mu _{1}}^{\nu
_{1}}+\varepsilon _{\mu _{13}\nu _{12}}(P_{21}^{\nu _{2}}-P_{32}^{\nu
_{2}})\Delta _{3\mu _{2}}^{\nu _{1}}+\varepsilon _{\mu _{123}\nu
_{1}}P_{32}^{\nu _{2}}\Delta _{3\nu _{2}}^{\nu _{1}}],
\end{equation}%
where $p_{ij}=P_{il}-P_{jl}$. Sub-index $ij$ in $(F_{\mu _{123}}^{\Delta
})_{ij}$ stands for the vertices where the RAGFs are satisfied without
further assumptions. As the relations above depend on three parameters and
are compatible in pairs, the coefficients solution is unique once one pair
of two RAGFs is determined. Complementary contraction is always an
incompatible solution; coefficients are different for each solution $(F_{\mu
_{123}}^{\Delta })_{ij}$. The pair solutions for at most two RAGFs
identically satisfied correspond to the amplitudes versions computed
explicitly. See (\ref{ST1})-(\ref{ST3}), namely%
\begin{equation*}
(F_{\mu _{123}}^{\Delta })_{23}=S_{1\mu _{123}};\quad (F_{\mu
_{123}}^{\Delta })_{13}=S_{2\mu _{123}};\quad (F_{\mu _{123}}^{\Delta
})_{12}=S_{3\mu _{123}},
\end{equation*}%
the trace of the surface term separates from the difference of $AV$ in one
of the contractions.

\textbf{Consequences}: With this derivation in hand, we draw a similar
conclusion to the one stated in the Subsection (\ref{LE4D}). The value at
zero of $PVV$ had consequences over symmetries. Here this amplitude will
establish a connection between linearity in the RAGFs and the low-energy
behavior of the same $PVV$.

For this, we have to read this result in light of form factors in (\ref%
{GForm}), taken as the \textit{finite parts}. Choosing the solution
satisfying the RAGFs in vertices two and three%
\begin{equation}
T_{\mu _{123}}^{\Gamma _{123}}=F_{\mu _{123}}+(F_{\mu _{123}}^{\Delta
})_{23}=F_{\mu _{123}}+S_{1\mu _{123}},
\end{equation}%
to any vertices combination. Let $\Omega _{i}$ represent the finite scalar
invariants of 2nd-order tensors from RAGFs; writing the equations of the
hypothesis of satisfaction, (\ref{GR}),%
\begin{eqnarray}
q_{1}^{\mu _{1}}T_{\mu _{123}}^{\Gamma _{123}}-T_{1\left( -\right) \mu
_{23}}^{AV} &=&\varepsilon _{\mu _{23}\nu _{12}}q_{2}^{\nu _{1}}q_{3}^{\nu
_{2}}\Omega _{1}=\varepsilon _{\mu _{23}\nu _{12}}q_{2}^{\nu _{1}}q_{3}^{\nu
_{2}}(V_{1}+2i\Delta _{3\alpha }^{\alpha }) \\
q_{2}^{\mu _{2}}T_{\mu _{123}}^{\Gamma _{123}}-T_{2\left( -\right) \mu
_{13}}^{AV} &=&\varepsilon _{\mu _{13}\nu _{12}}q_{2}^{\nu _{1}}q_{3}^{\nu
_{2}}\Omega _{2}=\varepsilon _{\mu _{13}\nu _{12}}q_{2}^{\nu _{1}}q_{3}^{\nu
_{2}}V_{2} \\
q_{3}^{\mu _{3}}T_{\mu _{123}}^{\Gamma _{123}}-T_{3\left( -\right) \mu
_{12}}^{AV} &=&\varepsilon _{\mu _{12}\nu _{12}}q_{2}^{\nu _{1}}q_{3}^{\nu
_{2}}\Omega _{3}=\varepsilon _{\mu _{12}\nu _{12}}q_{2}^{\nu _{1}}q_{3}^{\nu
_{2}}V_{3}.
\end{eqnarray}

Using the previous results, we see that the trace of the surface term must
be put together with the finite part of the first contraction due to the Eq.
(\ref{contS1}),%
\begin{equation}
q_{1}^{\mu _{1}}S_{1\mu _{123}}=T_{1\left( -\right) \mu
_{23}}^{AV}+\varepsilon _{\mu _{23}\nu _{23}}q_{2}^{\nu _{2}}q_{3}^{\nu
_{3}}(2i\Delta _{3\nu _{1}}^{\nu _{1}}).
\end{equation}%
We wrote the $AV$ structures on LHS to focus on the non-trivial part of the
relations. We get the final condition: $V_{1}+2i\Delta _{3\alpha }^{\alpha
}=\Omega _{1};$ $V_{2}=\Omega _{2};$ and $V_{3}=\Omega _{3}.$ Observing the
formulas%
\begin{eqnarray}
V_{1} &=&-F_{1}+F_{2}+q_{2}^{2}G_{5}+q_{3}^{2}G_{6}+\left( q_{2}\cdot
q_{3}\right) \left( G_{5}+G_{6}\right) \\
V_{2} &=&-F_{2}+q_{2}^{2}G_{3}+\left( q_{2}\cdot q_{3}\right) G_{4} \\
V_{3} &=&-F_{1}+q_{3}^{2}G_{2}+\left( q_{2}\cdot q_{3}\right) G_{1}.
\end{eqnarray}%
It is possible to eliminate the $F_{i}$ form factors to reach at%
\begin{eqnarray}
2i\Delta _{3\alpha }^{\alpha }+\Omega _{3}-\Omega _{2}-\Omega _{1}
&=&-q_{2}^{2}\left( G_{3}+G_{5}\right) +q_{3}^{2}\left( G_{2}-G_{6}\right) \\
&&+\left( q_{2}\cdot q_{3}\right) \left( G_{1}-G_{4}-G_{5}-G_{6}\right) . 
\notag
\end{eqnarray}%
Under the condition that $G_{i}$ functions are regular at zero\footnote{%
The functions $Z_{nm}^{\left( 0\right) }$, $Z_{nm}^{\left( -1\right) }$, $%
Z_{n}^{\left( 0\right) }$ that comprise the finite part of any of these
amplitudes do not have kinematical singularities at the point $q_{i}\cdot
q_{j}=0$.}, follows%
\begin{equation}
2i\Delta _{3\alpha }^{\alpha }=\Omega _{1}\left( 0\right) +\Omega _{2}\left(
0\right) -\Omega _{3}\left( 0\right) .  \label{ME}
\end{equation}

The equation is true irrespective of the choice of which relation is
satisfied without restriction. Suppose one starts with a version with $%
S_{2\mu _{123}}$ that satisfies the RAGFs in the first and third vertex. To
this tensor, the term $\Delta _{3\alpha }^{\alpha }$ will appear in $%
q_{2}^{\mu _{2}}S_{2\mu _{123}}$, see Eq. (\ref{contS2}). From $V_{1}=\Omega
_{1}$ and $V_{3}=\Omega _{3},$ and trading the $F_{1}$ and $F_{2}$ by $G_{i}$
plus finite functions, again in zero, we retrieve the previous result. That
is a proper relation between a low-energy property and surface terms stated
in the former section in (\ref{IRUV}). The hypotheses were a tensor with two
RAGFs satisfied without restriction, connected to $AV$differences and $PVV$/$%
PAA$-like amplitudes. From that, the zero value of rank-2 amplitudes bound
the third RAGF. It is always possible to achieve these hypotheses in
explicit computations.

When assessing $\Omega _{i}(0)$, see (\ref{LEPVV}), $\Omega ^{PVV}=\Omega
^{VPV}=-\Omega ^{VVP}=(2\pi )^{-2}$, we find out%
\begin{equation}
\Omega _{1}\left( 0\right) +\Omega _{2}\left( 0\right) -\Omega _{3}\left(
0\right) =(2\pi )^{-2},
\end{equation}%
Notice that for the $AVV$, $VAV$, and $VVA,$ two of the $\Omega _{i}$ are
zero to each amplitude, which means the result above represents three
situations. The same happens to the $AAA$ triangle. In this case, the three
contractions of the same amplitude relate to $PAA$, $APA$, and $AAP$.
Combining the constants cast in Eq. (\ref{LEPAA}), we have%
\begin{equation}
\Omega _{1}^{PAA}\left( 0\right) +\Omega _{2}^{APA}\left( 0\right) -\Omega
_{3}^{AAP}\left( 0\right) =(2\pi )^{-2}.
\end{equation}%
Since the $AV$ differences depend only on the contractions with the momenta,
but the correlators with the $P$ density are distinct, it could be that
distinct diagrams would require different numerical values to the surface
term, despite that one always find%
\begin{equation}
\text{RAGF}\Leftrightarrow 2\Delta _{3\alpha }^{\alpha }=-i(2\pi )^{-2}.
\label{LiUni4D}
\end{equation}%
Constraint remains for amplitudes where three distinct masses run in the
internal lines.

Let us consider an example of this scenario for the $AVV$. The propagator's
indexes now account for the masses too, $S\left( a\right) =(\slashed{K}%
_{a}-m_{a})^{-1}$. Using the standard identity [$\slashed{p}%
_{ij}=S^{-1}\left( i\right) -S^{-1}\left( j\right) +(m_{i}-m_{j})$] to
derive the RAGFs expressed in Eqs. (\ref{AVVragfs}), the terms associated
with the three-point functions are \ now%
\begin{equation*}
-(m_{1}+m_{3})T_{\mu _{23}}^{PVV};\quad (m_{2}-m_{1})T_{\mu _{13}}^{ASV};%
\text{ and}\quad (m_{3}-m_{2})T_{\mu _{12}}^{AVS},
\end{equation*}%
coming from verteces $\Gamma _{1},$ $\Gamma _{2},$ and $\Gamma _{3}$
respectively. In this scenario, vector currents are not classically
conserved. However, $ASV$, $AVS$, and $PVV$ will not comply the Eq. (\ref%
{VertsZero}), and their relations are identical to the ones (\ref{GamatZero}%
). For the three-point rank-2 amplitudes,%
\begin{eqnarray*}
T_{\mu _{23}}^{PVV} &=&\varepsilon _{\mu _{23}\nu _{12}}p_{21}^{\nu
_{1}}p_{32}^{\nu _{2}}[\left( m_{1}-m_{2}\right) Z_{10}^{\left( -1\right)
}+\left( m_{1}-m_{3}\right) Z_{01}^{\left( -1\right) }-m_{1}Z_{00}^{\left(
-1\right) }] \\
T_{\mu _{13}}^{ASV} &=&\varepsilon _{\mu _{13}\nu _{12}}p_{21}^{\nu
_{1}}p_{32}^{\nu _{2}}[\left( m_{1}+m_{2}\right) Z_{10}^{\left( -1\right)
}+\left( m_{1}+m_{3}\right) Z_{01}^{\left( -1\right) }-m_{1}Z_{00}^{\left(
-1\right) }] \\
T_{\mu _{12}}^{AVS} &=&\varepsilon _{\mu _{12}\nu _{12}}p_{21}^{\nu
_{1}}p_{32}^{\nu _{2}}[\left( m_{2}-m_{1}\right) Z_{10}^{\left( -1\right)
}-\left( m_{3}+m_{1}\right) Z_{01}^{\left( -1\right) }+m_{1}Z_{00}^{\left(
-1\right) }],
\end{eqnarray*}%
it is possible to identify the form factor through the relation%
\begin{eqnarray*}
\varepsilon _{\mu _{23}\nu _{12}}p_{21}^{\nu _{1}}p_{32}^{\nu _{2}}\Omega
_{1}^{PVV} &=&-(m_{1}+m_{3})T_{\mu _{23}}^{PVV} \\
\varepsilon _{\mu _{13}\nu _{12}}p_{21}^{\nu _{1}}p_{32}^{\nu _{2}}\Omega
_{2}^{ASV} &=&+(m_{2}-m_{1})T_{\mu _{13}}^{ASV} \\
\varepsilon _{\mu _{12}\nu _{12}}p_{21}^{\nu _{1}}p_{32}^{\nu _{2}}\Omega
_{3}^{AVS} &=&+(m_{3}-m_{2})T_{\mu _{12}}^{AVS}.
\end{eqnarray*}%
By combining them as done in the other cases, we have%
\begin{equation*}
\Omega _{1}^{PVV}+\Omega _{2}^{ASV}-\Omega _{3}^{AVS}=2(2\pi
)^{-2}[(m_{1}^{2}-m_{2}^{2})Z_{10}^{\left( -1\right)
}+(m_{1}^{2}-m_{3}^{2})Z_{01}^{\left( -1\right) }-m_{1}^{2}Z_{00}^{\left(
-1\right) }].
\end{equation*}%
Since in the definition, the $Q$ polynomial for distinct masses\footnote{%
To arbitrary masses, the Feynman polynomial for the function involved in
this derivation reads 
\begin{equation*}
Q=q_{1}^{2}x_{1}\left( 1-x_{1}\right) +q_{2}^{2}x_{2}\left( 1-x_{2}\right)
-2q_{1}\cdot q_{2}x_{1}x_{2}+\left( m_{1}^{2}-m_{2}^{2}\right) x_{1}+\left(
m_{1}^{2}-m_{3}^{2}\right) x_{2}-m_{1}^{2}.
\end{equation*}%
And the function is given by 
\begin{equation*}
Z_{rs}^{\left( -1\right) }=\int_{0}^{1}\mathrm{d}x_{1}\int_{0}^{1-x_{1}}%
\mathrm{d}x_{2}\frac{x_{1}^{r}x_{2}^{s}}{Q\left(
q_{i}^{2},m_{1}^{2},m_{2}^{2},m_{3}^{2}\right) }.
\end{equation*}%
In the kinematical point the the polynomial assumes the form $Q\left(
0\right) =\left( m_{1}^{2}-m_{2}^{2}\right) x_{1}+\left(
m_{1}^{2}-m_{3}^{2}\right) x_{2}-m_{1}^{2}.$}, hence the relation is 
\begin{equation}
\left[ (m_{1}^{2}-m_{2}^{2})Z_{10}^{\left( -1\right)
}+(m_{1}^{2}-m_{3}^{2})Z_{01}^{\left( -1\right) }-m_{1}^{2}Z_{00}^{\left(
-1\right) }\right] _{q_{i}\cdot q_{j}=0}=1/2.
\end{equation}%
Finally, in the limit studied follows $(\Omega _{1}^{PVV}+\Omega
_{2}^{ASV}-\Omega _{3}^{AVS})|_{0}=(2\pi )^{-2}$. The integrals with various
masses are laborious, but integrating all these functions explicitly in the
limit under consideration follows the result.

The kinematical limits of all rank-2 amplitudes are incompatible with the
satisfaction of all Ward identities since they ask for additional constants
to be compatible with the tensor structure of rank-3 amplitudes, as already
established in the 2D. Although these claims are implicit in the discussion
of these tensors, often, the focus is the regularization properties. In this
way, when we write the internal momenta as covariant combinations
(non-covariant combinations amount to Lorentz violations), we must have%
\begin{eqnarray*}
&&\left[ V_{1}^{AVV}\left( m_{1},m_{2},m_{3}\right) -V_{2}^{AVV}\left(
m_{1},m_{2},m_{3}\right) -V_{3}^{AVV}\left( m_{1},m_{2},m_{3}\right) \right]
\left( 0\right) \\
&=&(\Omega _{1}^{PVV}+\Omega _{2}^{ASV}-\Omega _{3}^{AVS})|_{0}+\left( 
\mathcal{A}_{1}^{AVV}-\mathcal{A}_{2}^{AVV}-\mathcal{A}_{3}^{AVV}\right) =0.
\end{eqnarray*}%
That means we can not simultaneously make all $\mathcal{A}_{i}=0$ by reasons
unrelated to divergences. Utilizing this equation to study the symmetries,
we have the scenario. If eventually is not found symmetry violation in that
point, it does not mean they could not be in other points. However, finding
a problem in zero implies a violation.

\section{General Parameters to the Violations\protect\footnote{%
Throughout this section, we factored out three-point rank-two finite
amplitudes from the discussion.}}

Summarizing the last sections: (i) Integration linearity holds if and only
if the surface terms are nonzero (\ref{LiUni4D}). Simultaneously the results
are independent of Dirac traces for the same value, which saves linearity.
(ii) Since some surface-terms coefficients are ambiguous combinations of the
routings, we must make choices for them. iii) From (ii), if a procedure
nullifies that terms, the linearity is violated by $\sim \pm (2\pi )^{-2}$;
see these results in (\ref{ragfViol4D}). There is an equilibrium between
routing and trace ambiguities organized by the surface term's value. Let us
see the parameter space for this competition.

Combining versions that save the most RAGFs with no condition on the surface
term\footnote{%
This claim is independent of explicit computations performed in the previous
section.},%
\begin{equation}
\lbrack t_{\mu _{123}}^{\Gamma _{123}}]_{\left\{ r_{1}r_{2}r_{3}\right\} }=%
\frac{1}{R}[r_{1}(t_{\mu _{123}}^{\Gamma _{123}})_{1}+r_{2}(t_{\mu
_{123}}^{\Gamma _{123}})_{2}+r_{3}(t_{\mu _{123}}^{\Gamma _{123}})_{3}],
\label{R123}
\end{equation}%
where $R=r_{1}+r_{2}+r_{3}\not=0$. As discussed at the end of Section (\ref%
{unique}), they are identical before integration. However, when $\Delta
_{3\mu \nu }=0$, they become an infinity set of different tensors. In
particular, they reproduce any tensor through our strategy using any
identity for the chiral matrix. For zero surface terms, their symmetry
violations are in the $i$-th vertex and get a factor of $r_{i}/R$,
satisfying the equation determined to its anomalies (\ref{Ans}) due to
kinematic properties of finite amplitudes.

If we have considered the surface term as an arbitrary parameter given by a
constant $c_{1}$, equal to one for the satisfaction of RAGFs or zero for the
momentum-space translational invariance. Parametrizing internal lines by
choosing any of the sums $P_{ij}=k_{i}+k_{j}$, we have $%
P_{31}=c_{2}q_{2}+c_{3}q_{3}\rightarrow P_{21}=c_{2}q_{2}+\left(
c_{3}-1\right) q_{3}$, and $P_{32}=\left( c_{2}+1\right) q_{2}+c_{3}q_{3},$
with%
\begin{equation}
2\Delta _{3\mu _{12}}=-ic_{1}(4\pi )^{-2}g_{\mu _{12}},
\end{equation}%
the $AV$ functions, see Section (\ref{unique}), Eqs. (\ref{AV(-)1})-(\ref%
{AV(-)3}), are written as function of $c_{1}$, $c_{2}$,and $c_{3}$, and also
violations of RAGFs, Eqs. (\ref{ragfViol4D}).Those parameters express any
possible values to the contractions of basic versions. With the caveat that
only in the contraction of $i$-th version with $q_{i}^{\mu _{i}}$, both the
two-point functions and the linearity-breaking term contributes. For this
version, the contraction with $q_{j}$, $j\not=i$, only $AV$'s contribute.

Modulus finite amplitudes, the combination defined in Eq. (\ref{R123}) has
the properties%
\begin{eqnarray}
q_{1}^{\mu _{1}}[T_{\mu _{123}}^{\Gamma _{123}}]_{\left\{
r_{1}r_{2}r_{3}\right\} } &=&\varepsilon _{\mu _{23}\nu _{12}}q_{2}^{\nu
_{1}}q_{3}^{\nu _{2}}\{[4R\left( 2\pi \right) ^{2}]^{-1}[4r_{1}\left(
c_{1}-1\right) +Rc_{1}\left( c_{3}-c_{2}-2\right) ]\}  \notag \\
&=&\varepsilon _{\mu _{23}\nu _{12}}q_{2}^{\nu _{1}}q_{3}^{\nu _{2}}\mathcal{%
A}_{1}  \label{genViol} \\
q_{2}^{\mu _{2}}[T_{\mu _{123}}^{\Gamma _{123}}]_{\left\{
r_{1}r_{2}r_{3}\right\} } &=&\varepsilon _{\mu _{13}\nu _{12}}q_{2}^{\nu
_{1}}q_{3}^{\nu _{2}}\{[4R\left( 2\pi \right) ^{2}]^{-1}[4r_{2}\left(
c_{1}-1\right) -Rc_{1}\left( c_{3}+1\right) ]\}  \notag \\
&=&\varepsilon _{\mu _{13}\nu _{12}}q_{2}^{\nu _{1}}q_{3}^{\nu _{2}}\mathcal{%
A}_{2} \\
q_{3}^{\mu _{3}}[T_{\mu _{123}}^{\Gamma _{123}}]_{\left\{
r_{1}r_{2}r_{3}\right\} } &=&\varepsilon _{\mu _{12}\nu _{12}}q_{2}^{\nu
_{1}}q_{3}^{\nu _{2}}\{[4R\left( 2\pi \right) ^{2}]^{-1}[4r_{3}\left(
1-c_{1}\right) -Rc_{1}\left( c_{2}-1\right) ]\}  \notag \\
&=&\varepsilon _{\mu _{12}\nu _{12}}q_{2}^{\nu _{1}}q_{3}^{\nu _{2}}\mathcal{%
A}_{3}.
\end{eqnarray}%
Parameters combination implies $\mathcal{A}_{1}=\mathcal{A}_{3}-\mathcal{A}%
_{2}-(2\pi )^{-2},$ decreasing the number of independent variables for two.
So when we have numerical amounts of two violations, no matter the path
leading them, the third arises without ambiguity. Derived in the previous
sections based only on finite functions and when the internal momenta as
covariant functions of external ones.

If $c_{1}=1$, there is no dependence in $r_{i}$, we have the unique solution
that satisfies linearity but is not a function of the external momenta. If $%
c_{1}=0$, there will be no dependence in $c_{2}$ and $c_{3}$, and the
tensors are functions of the external momenta but not unique. These
parameters are the full range of possibilities. The crossed diagrams add
more parameters to the discussion but have the same behavior: linearity
break, ambiguities, and symmetries violation. The crucial factor is the
kinematic behavior of finite functions that code amplitudes for
pseudo-scalar density. In the massless limit, this aspect falls in the
values to the residue of poles of form factors, which are regular in the
massive case. Breaking linearity has a function in divergent amplitudes that
corroborates with the low-energy value of finite amplitude $PV^{n}$ in
dimension $d=2n$. If it does not occur, shifts in the integration variable
are allowed by removing surface terms. Hence the $AV$ functions through (\ref%
{VertsZero}) relate the $V_{i}$, and the finite amplitudes would have to be
zero at the point where the bilinears vanish.

The situation happens when integrating an identically zero tensor; it is
obtained a nonzero result. Take the identity for the integrand of the
Feynman integral $\bar{J}_{3\mu \nu }$,%
\begin{equation*}
\lbrack K_{1}^{\mu _{5}}(\varepsilon _{\mu _{5123}}K_{1\mu _{4}}+\varepsilon
_{\mu _{4512}}K_{1\mu _{3}}+\varepsilon _{\mu _{3451}}K_{1\mu
_{2}}+\varepsilon _{\mu _{2345}}K_{1\mu _{1}})+\varepsilon _{\mu
_{1234}}m^{2}]\frac{1}{D_{123}}=-\varepsilon _{\mu _{1234}}\frac{1}{D_{23}}
\end{equation*}%
the equation comes from $\varepsilon _{\lbrack \mu _{1234}}K_{1\mu _{5}]}=0$%
, multiplying by $K_{1}^{\mu _{5}}/D_{123}$, and using $%
K_{1}^{2}=D_{1}+m^{2} $. When integrated, the identity is only valid for
just one surface-term value. The critical step arises when we separate the
finite and divergent parts, explicitly 
\begin{eqnarray*}
\bar{J}_{2}\left( 2,3\right) &=&J_{2}\left( p_{32}\right) +I_{\log } \\
\bar{J}_{3}\left( 1,2,3\right) &=&J_{3}\left( p_{21},p_{31}\right) \\
\bar{J}_{3\mu \nu }\left( 1,2,3\right) &=&J_{3\mu \nu }\left(
p_{21},p_{31}\right) +\left( \Delta _{3\mu \nu }+g_{\mu \nu }I_{\log
}\right) /4, \\
J_{3\alpha }^{\alpha }\left( p_{21},p_{31}\right) &=&m^{2}J_{3}\left(
p_{21},p_{31}\right) +J_{2}\left( p_{32}\right) +i[2\left( 4\pi \right)
^{2}]^{-1}.
\end{eqnarray*}%
This step is performed using $\varepsilon _{\lbrack \mu _{1235}}\Delta
_{3\mu _{5}]}^{\mu _{5}}=0$ and $\varepsilon _{\lbrack \mu _{1235}}J_{3\mu
_{5}]}^{\mu _{5}}=0$. Then, the initial identity gets transformed in a
condition to the linearity breaking $\varepsilon _{\mu _{1234}}[\Delta
_{3\mu _{5}}^{\mu _{5}}+2i/\left( 4\pi \right) ^{2}]=0.$ Now, the identity
for the surface term is consistent to any value, constrained only by $\Delta
_{3\mu \nu }=[g^{\alpha \beta }\Delta _{3\alpha \beta }]/4$, however the
same is not true to the bare integral $\bar{J}_{3\mu \nu }$. The identity is
respected if and only if $\Delta _{3\alpha }^{\alpha }=-2i/(4\pi )^{2}$,
derived without explicitly manipulating divergent integrals. As a part of
the Feynman integrals, the satisfaction of the Schouten identity to any
surface-term value is not enough to make it valid for the entire integrals.
We used the results of Section (\ref{Sub4DIn}).

We must mention that the violation by an evanescent term that occurs in
dimensional methods\footnote{%
See \cite{EliasMckeon1983}\cite{Chowdhury1986} for this type of view.} does
not affect linearity breaking. The finite value we demonstrate to be
necessary is not a function of the dimension, and it corresponds to the
low-energy limit of the integral $J_{3}$. No limiting process can change
that value and, if not adopted, violates the linearity and uniqueness of
these perturbative amplitudes.

\chapter{Gravitational Perturbative Amplitudes}

\label{modlDef}

The quantization of fermionic fields is according to the canonical rules of
Quantum Field Theory. To introduce these fields in a curved space, we
associate to space-time a Lorentz manifold, in which each point has a plane
space tangent to it. The connection between the two spaces is through
vielbein fields defined by 
\begin{eqnarray}
g_{\mu \nu }\left( x\right) &=&\eta _{ab}e_{\mu }^{a}\left( x\right) e_{\nu
}^{b}\left( x\right) \\
\eta _{ab} &=&\mathrm{diag}\left( 1,-1-1-1\right) \\
e_{a}^{\mu }e_{\mu }^{b} &=&\delta _{a}^{b};\qquad e_{a}^{\mu }e_{\nu
}^{a}=\delta _{\nu }^{\mu }.
\end{eqnarray}%
These fields work in such a way as to transform the coordinate basis into an
orthonormal basis. Through that basis, it is possible to introduce locally
the Clifford algebra whose representations the spinor field can be defined.
The algebra acquires a local character, 
\begin{eqnarray}
\gamma ^{\mu }\left( x\right) &:&=e_{a}^{\mu }\left( x\right) \gamma ^{a} \\
\left\{ \gamma _{\mu }\left( x\right) ,\gamma _{\nu }\left( x\right)
\right\} &=&2g_{\mu \nu }\left( x\right) \\
\left\{ \gamma ^{a},\gamma ^{b}\right\} &=&2\eta ^{ab} \\
\gamma _{\left[ ab\right] } &=&\frac{1}{2}\left[ \gamma _{a},\gamma _{b}%
\right] ;
\end{eqnarray}%
the last term $\gamma _{\left[ ab\right] }/2$ corresponds to the spinor
generator to the Lorentz group.

In this way, we will introduce a covariant generalization of the equations
formulated in flat spacetime to introduce fermions coupled to a spacetime
with arbitrary metrics. The action $S$ must be invariant by Lorentz
transformations and general transformations of coordinates. We start by
considering the flat-space real Lagrangian 
\begin{equation}
\mathcal{L}=\frac{1}{2}[i\bar{\psi}\gamma ^{\mu }\partial _{\mu }\psi
-i\left( \partial _{\mu }\bar{\psi}\right) \gamma ^{\mu }\psi ]=\frac{1}{2}[i%
\bar{\psi}\gamma ^{\mu }\partial _{\mu }\psi +\left( i\bar{\psi}\gamma ^{\mu
}\partial _{\mu }\psi \right) ^{\dagger }],
\end{equation}%
and replace the covariant for the flat-spacetime metric in coordinate and
orthonormal frame $\partial _{\mu }$ (cartesian one) by the spinor covariant
derivative in an arbitrary coordinate frame (but still flat geometry), we
have 
\begin{equation}
\nabla _{\mu }\psi :=\partial _{\mu }\psi +\frac{1}{4}\omega _{\mu
}^{~~ab}\gamma _{\left[ ab\right] }\psi ;\qquad \text{and\qquad }\nabla
_{\mu }\bar{\psi}:=\partial _{\mu }\bar{\psi}-\frac{1}{4}\bar{\psi}\omega
_{\mu }^{~~ab}\gamma _{\left[ ab\right] }.
\end{equation}%
We used $\gamma _{0}\gamma _{\left[ ab\right] }^{\dagger }\gamma
_{0}=-\gamma _{\left[ ab\right] }$ in defining last equation; $\omega _{\mu
}^{~~ab}\ $are components of metric-compatible spin connection 
\begin{eqnarray}
\omega _{\quad \mu }^{ab} &=&\omega _{\quad c\mu }^{a}\eta ^{cb} \\
\omega _{\quad c\mu }^{a} &=&e_{\quad \nu }^{a}\partial _{\mu }e_{c}^{\quad
\nu }+e_{c}^{\quad \nu }e_{\quad \lambda }^{a}\Gamma _{\mu \nu }^{\lambda },
\end{eqnarray}%
being $\Gamma _{\mu \nu }^{\lambda }$ the components of the connection in
the coordinate basis. Then, we allow the metric to correspond to a curved
background geometry, and thereby, the fermion propagation will be
classically given by 
\begin{equation}
S=\int_{\mathcal{M}}\mathrm{d}^{2}xe\left( x\right) \frac{i}{2}\left[ \bar{%
\psi}\gamma ^{\mu }\nabla _{\mu }\psi -\left( \nabla _{\mu }\bar{\psi}%
\right) \gamma ^{\mu }\psi \right] ,
\end{equation}%
where we introduced the scalar density $e\left( x\right) =\sqrt{\left\vert
g\left( x\right) \right\vert }$ in the volume 2-form \textrm{d}$V=\sqrt{%
\left\vert g\left( x\right) \right\vert }\mathrm{d}x_{1}\mathrm{d}x_{2}$, $%
g\left( x\right) =\det g_{\mu \nu }$, and modulus is due to the Lorentz
signature.

The extremization of action yields the motion's equations: $\nabla _{\mu
}\psi =0$ and $\nabla _{\mu }\bar{\psi}=0.$ Additionally, in 2D, the term
coupling to the spin-connection drops out from the action 
\begin{equation}
\frac{1}{4}\omega _{\mu }^{~~ab}e_{c}^{\mu }\bar{\psi}\{\gamma ^{c},\gamma _{%
\left[ ab\right] }\}\psi =0,
\end{equation}
due to the in this dimension $\gamma _{\left[ ab\right] }=-i\gamma _{\ast
}\varepsilon _{ab}$ and $\left\{ \gamma ^{c},\gamma _{\ast }\right\} =0$.
Therefore, we adopt Weyl fermions henceforth, and the action simplifies to 
\begin{equation}
S=\frac{i}{2}\int_{\mathcal{M}}\mathrm{d}^{2}xe\left( x\right) e_{a}^{\mu }[%
\bar{\psi}\gamma ^{a}\overleftrightarrow{\partial }_{\mu }P_{\pm }\psi ],
\end{equation}%
where the chiral projectors are given by $P_{\pm }=\left( 1\pm \gamma _{\ast
}\right) /2,$ being that the chiral matrix (\ref{gammastar}) is $\gamma
_{\ast }=\varepsilon _{ab}\gamma ^{a}\gamma ^{b}$ and the 'flat' Levi-Cevita
symbol is normalized by $\varepsilon ^{01}=1$ (it is a tensor density with
world indices).

The gravitational field appears only as a background field, without being
necessarily quantized and without associated dynamics. Then, we consider the
approximation expanding in powers of $h_{\mu \nu }$ around the Minkowski
metric%
\begin{eqnarray}
g_{\mu \nu } &=&\eta _{\mu \nu }+\kappa h_{\mu \nu } \\
g^{\mu \nu } &=&\eta ^{\mu \nu }-\kappa h_{\mu \nu }+\mathcal{O}\left(
\kappa ^{2}\right) . \\
e_{\mu }^{a} &=&\delta _{\mu }^{a}+\frac{1}{2}\kappa h_{\mu }^{a};\text{%
\qquad }e_{a}^{\mu }=\delta _{a}^{\mu }-\frac{1}{2}\kappa h_{a}^{\mu };\text{%
\qquad }e=1+\frac{1}{2}\kappa h_{\mu }^{\mu }.
\end{eqnarray}%
We may expand in $e\left( x\right) $ and inverse vielbein $e_{a}^{\mu }$
independently; in this way, we would get%
\begin{equation}
\frac{i}{2}e\left( x\right) e_{a}^{\mu }[\bar{\psi}\gamma ^{a}%
\overleftrightarrow{\partial }_{\mu }\psi ]=\frac{i}{2}\bar{\psi}\gamma
^{\mu }\overleftrightarrow{\partial }_{\mu }\psi -\frac{1}{2}h^{\mu \nu }%
\left[ \frac{i}{4}\bar{\psi}\gamma _{(\nu }\overleftrightarrow{\partial }%
_{\mu )}\psi -\frac{i}{2}\eta _{\mu \nu }\bar{\psi}\overleftrightarrow{\slashed%
{\partial}}\psi \right] +\mathcal{O}\left( h^{2}\right) ,
\end{equation}%
where $\overleftrightarrow{\slashed{\partial}}=\gamma ^{\rho }%
\overleftrightarrow{\partial }_{\rho }$. The energy-momentum tensor, in this
linearized approximation, reads%
\begin{equation}
T_{\mu \nu }^{\prime }=\frac{i}{4}(\bar{\psi}\gamma _{\nu }%
\overleftrightarrow{\partial }_{\mu }\psi +\bar{\psi}\gamma _{\mu }%
\overleftrightarrow{\partial }_{\nu }\psi )-\frac{i}{2}\eta _{\mu \nu }\bar{%
\psi}\overleftrightarrow{\slashed{\partial}}\psi .
\end{equation}%
Alternatively, we can absorb the $e=\sqrt{\left\vert g\right\vert }$ into a
redefinition of $e^{1/2}\psi =\Psi $ see Bonara et al. (\cite{Bonora2016})
in Appendix B of that reference. Therefore, we have%
\begin{equation}
S=\frac{i}{2}\int \mathrm{d}^{2}xe_{a}^{\mu }(\bar{\Psi}\gamma ^{a}%
\overleftrightarrow{\partial }_{\mu }\Psi )=\frac{i}{2}\int \mathrm{d}^{2}x[%
\bar{\Psi}\gamma ^{\mu }\overleftrightarrow{\partial }_{\mu }\Psi +\mathcal{L%
}_{\mathrm{int}}\left( h,\Psi \right) +\mathcal{O}\left( h^{2}\right) ].
\end{equation}%
In this way, the interaction Lagrangian $\mathcal{L}_{\mathrm{int}}$ is
still defined as%
\begin{equation}
\mathcal{L}_{\mathrm{int}}\left( h,\Psi \right) =-\frac{1}{2}h^{\mu \nu }%
\left[ \frac{i}{4}(\bar{\Psi}e_{a\mu }\gamma ^{a}\overleftrightarrow{%
\partial }_{\nu }P_{\pm }\Psi +\bar{\Psi}e_{av}\gamma ^{a}%
\overleftrightarrow{\partial }_{\mu }P_{\pm }\Psi )\right] =-\frac{1}{2}%
h^{\mu \nu }T_{\mu \nu }.  \label{tensorem-em}
\end{equation}%
Then the linearized approximation of the energy-momentum tensor definition
follows as 
\begin{equation}
T_{\mu \nu }=\frac{i}{4}\bar{\Psi}\gamma _{(\mu }\overleftrightarrow{%
\partial }_{\nu )}P_{\pm }\Psi .
\end{equation}

\begin{figure}[tbph]
\centering\includegraphics[scale=1.0]{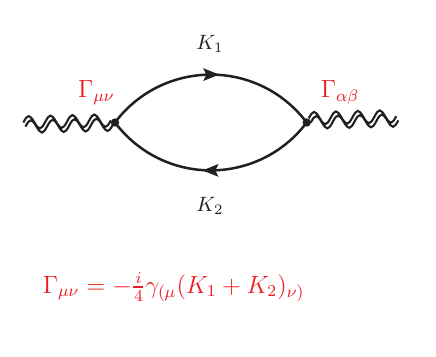}
\caption{The diagram for two-point function of the the linearized energy
momentum tensor.}
\label{GravB}
\end{figure}

From interaction Lagrangian follows the Feynman rules that will be used in
this work. The two-point gravitational amplitude is%
\begin{equation}
T_{\mu \nu \alpha \beta }^{G}\left( q\right) =i\int \mathrm{d}%
^{2}xe^{iq\cdot x}\left\langle 0\right\vert T\left[ T_{\mu \nu }\left(
x\right) ,T_{\alpha \beta }\left( 0\right) \right] \left\vert 0\right\rangle
.  \label{TGreen}
\end{equation}%
Moreover, the vertices of the perturbative amplitudes relative to the
interaction between the graviton and a fermion-antifermion pair are%
\begin{equation}
\Gamma _{\mu \nu }^{G}=-\frac{i}{4}[\gamma _{\mu }\left( K_{1}+K_{2}\right)
_{\nu }+\gamma _{\nu }\left( K_{1}+K_{2}\right) _{\mu }]P_{\pm }.
\end{equation}%
At the trace level, the gravitational amplitude of our interest is, see the
figure \ref{GravB}, 
\begin{equation}
t_{\mu \nu \alpha \beta }^{G}=\mathrm{tr}[\Gamma _{\mu \nu }^{G}S\left(
1\right) \Gamma _{\alpha \beta }^{G}S\left( 2\right) ].
\end{equation}%
After integration, we will call $T_{\mu \nu \alpha \beta }^{G}$. The total
amplitude with massive propagators is%
\begin{eqnarray}
\left( i64\right) T_{\mu \nu \alpha \beta }^{G}\left( q\right) &=&\int \frac{%
\mathrm{d}^{2}k}{\left( 2\pi \right) ^{2}}\mathrm{tr}[(1\pm \gamma _{\ast
})\gamma _{(\mu }(K_{1}+K_{2})_{\nu )}S\left( 1\right) \\
&&\times (1\pm \gamma _{\ast })\gamma _{(\alpha }(K_{1}+K_{2})_{\beta
)}S\left( 2\right) ].  \notag
\end{eqnarray}%
We recall that the fermionic propagator is given by (\ref{Prop}).

We will offer some layers of notations to devise an organizational scheme to
deal with this amplitude, as our approach presents multiple characteristics
and complexities. For the first one, let us break it down into four basic
permutations, given by 
\begin{equation}
t_{\mu \nu \alpha \beta }^{G}=-\frac{i}{64}\left( \hat{t}_{\mu \nu \alpha
\beta }^{G}+\hat{t}_{\mu \nu \beta \alpha }^{G}+\hat{t}_{\nu \mu \alpha
\beta }^{G}+\hat{t}_{\nu \mu \beta \alpha }^{G}\right) .  \label{tG}
\end{equation}%
The structures presented above can be identified as%
\begin{equation}
\hat{t}_{\mu \nu \alpha \beta }^{G}=(K_{1\nu }+K_{2\nu })(K_{1\beta
}+K_{2\beta })\mathrm{tr}[(1\pm \gamma _{\ast })\gamma _{\mu }S\left(
1\right) (1\pm \gamma _{\ast })\gamma _{\alpha }S\left( 2\right) ].
\label{tGtil}
\end{equation}%
The other three tensors come from the permutation $\mu \leftrightarrow
\alpha ,$ followed by $\nu \leftrightarrow \beta $.

Any computational element developed to this permutation can be mirrored in
the others. Second step: expanding the products like $(1\pm \gamma _{\ast
})\gamma _{\mu }$, we identify the integrand of typical fermionic amplitudes
as the one explored in the previous chapters. Explicitly%
\begin{equation}
\hat{t}_{\mu \nu \alpha \beta }^{G}=(K_{1\nu }+K_{2\nu })(K_{1\beta
}+K_{2\beta })\left[ t_{\mu \alpha }^{VV}+t_{\mu \alpha }^{AA}\pm t_{\mu
\alpha }^{AV}\pm t_{\mu \alpha }^{VA}\right] .
\end{equation}%
When integrated, we recognize another element in this decomposition layer,
allowing us to write the basic permutation for the structure below%
\begin{equation}
\mathcal{T}_{\mu \alpha \nu \beta }^{\Gamma _{1}\Gamma _{2}}=\int \frac{%
\mathrm{d}^{2}k}{\left( 2\pi \right) ^{2}}\left( K_{1}+K_{2}\right) _{\nu
}\left( K_{1}+K_{2}\right) _{\beta }[t_{\mu \alpha }^{\Gamma _{1}\Gamma
_{2}}\left( k_{1},k_{2}\right) ],  \label{Tgamma2}
\end{equation}%
where the vertices are $\Gamma _{i}\in \{1,\gamma _{\ast },\gamma _{\mu
},\gamma _{\ast }\gamma _{\mu }\}$, see (\ref{SetofVertexes}). This last
equation will be constructed explicitly in the next chapter since it
comprises even more fundamental components. To cast these components, we
observe that $K_{2}-K_{1}=q\rightarrow K_{1}+K_{2}=2K_{1}+q.$

Furthermore, expanding the Eq. (\ref{Tgamma2}) we write this combination%
\begin{equation}
\mathcal{T}_{\mu \alpha \nu \beta }^{\Gamma _{12}}=4T_{\mu \alpha ;\nu \beta
}^{\Gamma _{12}}+2q_{\nu }T_{\mu \alpha ;\beta }^{\Gamma _{12}}+2q_{\beta
}T_{\mu \alpha ;\nu }^{\Gamma _{12}}+q_{\nu }q_{\beta }T_{\mu \alpha
}^{\Gamma _{12}}.  \label{T12}
\end{equation}%
We must define what we mean by $T_{\mu \alpha ;\nu \beta }^{\Gamma _{12}}$, $%
T_{\mu \alpha ;\beta }^{\Gamma _{12}}$, and $T_{\mu \alpha ;\nu }^{\Gamma
_{12}}$, which we call derivative amplitudes for the sake of simplicity. As
an example, we have%
\begin{equation}
T_{\mu \alpha ;\nu \beta }^{VV}=\int \frac{\mathrm{d}^{2}k}{\left( 2\pi
\right) ^{2}}t_{\mu \alpha ;\nu \beta }^{VV}=\int \frac{\mathrm{d}^{2}k}{%
\left( 2\pi \right) ^{2}}K_{1\nu }K_{1\beta }\left( t_{\mu \alpha
}^{VV}\right) .
\end{equation}%
Derivative two-point amplitudes are defined even to $\Gamma _{i}$ that do
not carry Lorentz indexes,%
\begin{eqnarray}
T_{\_;\alpha _{1}}^{\Gamma _{1}\Gamma _{2}} &=&\int \frac{\mathrm{d}^{2}k}{%
\left( 2\pi \right) ^{2}}t_{\_;\alpha _{1}}^{\Gamma _{1}\Gamma _{2}}=\int 
\frac{\mathrm{d}^{2}k}{\left( 2\pi \right) ^{2}}K_{1\alpha _{1}}\mathrm{tr}%
[\Gamma _{1}S\left( 1\right) \Gamma _{2}S\left( 2\right) ]  \label{tGG1} \\
T_{\_;\alpha _{1}\alpha _{2}}^{\Gamma _{1}\Gamma _{2}} &=&\int \frac{\mathrm{%
d}^{2}k}{\left( 2\pi \right) ^{2}}t_{\_;\alpha _{1}\alpha _{2}}^{\Gamma
_{1}\Gamma _{2}}=\int \frac{\mathrm{d}^{2}k}{\left( 2\pi \right) ^{2}}%
K_{1\alpha _{1}}K_{1\alpha _{2}}\mathrm{tr}[\Gamma _{1}S\left( 1\right)
\Gamma _{2}S\left( 2\right) ].  \label{tGG3}
\end{eqnarray}%
When vertices to the matrix $\Gamma _{i}$ have Lorentz indices, the notation
will carry such indices in the position we left a blank space. Indexes $%
\alpha _{i}$ attached to factor $K_{1}$ are derivative indexes. The $%
\mathcal{T}_{\mu \nu \alpha \beta }^{\Gamma _{12}}$ contain only a subset of
general amplitudes we have defined in our last layer. Typical amplitudes
associated with $T^{\Gamma _{1}\Gamma _{2}}$ are the ones investigated in
Chapter (\ref{2Dim2Pt}). On the other hand, amplitudes as (\ref{tGG1})-(\ref%
{tGG3}) carrying derivative indices are the new ingredients to comprise
two-point functions of the energy-momentum tensor.

To illustrate the notation, let us take a derivative amplitude that is not
part of the permutations $\mathcal{T}_{\mu \nu \alpha \beta }^{\Gamma _{12}}$%
, by example selecting $\Gamma _{1}=S$ and $\Gamma _{2}=V$, we have%
\begin{equation}
t_{\mu ;\nu }^{SV}=K_{1\nu }t_{\mu }^{SV}=K_{1\nu }\mathrm{tr}[S\left(
1\right) \gamma _{\mu }\left( 1\right) ].
\end{equation}%
Note that the index $\nu $ appearing after the semicolon is a derivative
index. It may happen that integration, through our technique, returns an
expression symmetric in the indices, being this amplitude an example $T_{\mu
;\nu }^{SV}=$ $T_{\nu ;\mu }^{SV}$ as we will see. Besides these comments,
introducing these general definitions is crucial because they are all
related through RAGFs.

Relations relevant to this chapter arise from two types of momentum
contraction and traces, e.g., $g^{\mu \alpha }t_{\mu \nu ;\alpha \beta
}^{VV}=mt_{\nu ;\beta }^{SV}+t_{\nu ;\beta }^{V}\left( k_{2}\right) $. The
one-point functions are part of the set: 
\begin{eqnarray}
t^{\Gamma _{1}} &=&\mathrm{tr}\left[ \Gamma _{1}S\left( k_{i}\right) \right]
;  \label{T1def} \\
t_{\_;\alpha _{1}}^{\Gamma _{1}} &=&K_{1\alpha _{1}}\mathrm{tr}\left[ \Gamma
_{1}S\left( k_{i}\right) \right] ;  \label{T2def} \\
t_{\_;\alpha _{1}\alpha _{2}}^{\Gamma _{1}} &=&K_{1\alpha _{1}}K_{1\alpha
_{2}}\mathrm{tr}\left[ \Gamma _{1}S\left( k_{i}\right) \right] .
\label{T3def}
\end{eqnarray}%
Amplitudes $t^{\Gamma _{1}}$ and their integrals are the ones used for RAGF
investigations, fully developed in Chapter (\ref{2Dim2Pt}). When integrated,
they get a capital letter also.

To systematically analyze $\hat{t}_{\mu \nu \alpha \beta }^{G}$, we split it
in even and odd tensors: amplitudes with two vector vertices, called $VV$,
and two axial vertices, called $AA$, are even, and amplitudes with composite
vertices, $AV$ and $VA$, are odd. For this permutation of indices, we get 
\begin{equation}
\hat{T}_{\mu \nu \alpha \beta }^{G}=\hat{T}_{\mu \nu \alpha \beta }^{V}+\hat{%
T}_{\mu \nu \alpha \beta }^{A},  \label{tGtilV-A}
\end{equation}%
where each of the sectors above has the following combination of amplitudes,%
\begin{eqnarray}
\hat{T}_{\mu \nu \alpha \beta }^{V} &=&\mathcal{T}_{\mu \alpha \nu \beta
}^{VV}+\mathcal{T}_{\mu \alpha \nu \beta }^{AA}  \label{ttillv} \\
\hat{T}_{\mu \nu \alpha \beta }^{A} &=&\pm \left( \mathcal{T}_{\mu \alpha
\nu \beta }^{AV}+\mathcal{T}_{\mu \alpha \nu \beta }^{VA}\right) .
\label{ttila}
\end{eqnarray}%
The disposition of indices can be a trick to avoid confusion. Observe the
indexes in $\hat{T}_{\mu \nu \alpha \beta }^{V}$, we chose the sequence $\mu
\nu \alpha \beta $ since they come from $T^{G}$, however in $\mathcal{T}%
_{\mu \alpha ;\nu \beta }^{VV}$ the disposition emphasizes that the last two
indices correspond to derivative type, what is quite helpful in the
calculations. The basic permutations above ($\mathcal{T}_{\mu \alpha \nu
\beta }^{\Gamma _{1}\Gamma _{2}}$) are shown here to make clear the
expansion in terms of derivatives structures%
\begin{eqnarray}
\mathcal{T}_{\mu \alpha \nu \beta }^{VV} &=&2\left( 2T_{\mu \alpha ;\nu
\beta }^{VV}+q_{\nu }T_{\mu \alpha ;\beta }^{VV}\right) +q_{\beta }\left(
2T_{\mu \alpha ;\nu }^{VV}+q_{\nu }T_{\mu \alpha }^{VV}\right)
\label{VV-basic} \\
\mathcal{T}_{\mu \alpha \nu \beta }^{AA} &=&2\left( 2T_{\mu \alpha ;\nu
\beta }^{AA}+q_{\nu }T_{\mu \alpha ;\beta }^{AA}\right) +q_{\beta }\left(
2T_{\mu \alpha ;\nu }^{AA}+q_{\nu }T_{\mu \alpha }^{AA}\right) \\
\mathcal{T}_{\mu \alpha \nu \beta }^{AV} &=&2\left( 2T_{\mu \alpha ;\nu
\beta }^{AV}+q_{\nu }T_{\mu \alpha ;\beta }^{AV}\right) +q_{\beta }\left(
2T_{\mu \alpha ;\nu }^{AV}+q_{\nu }T_{\mu \alpha }^{AV}\right) \\
\mathcal{T}_{\mu \alpha \nu \beta }^{VA} &=&2\left( 2T_{\mu \alpha ;\nu
\beta }^{VA}+q_{\nu }T_{\mu \alpha ;\beta }^{VA}\right) +q_{\beta }\left(
2T_{\mu \alpha ;\nu }^{VA}+q_{\nu }T_{\mu \alpha }^{VA}\right) .
\label{VA-basic}
\end{eqnarray}%
Summing the four permutations, we get%
\begin{eqnarray}
\mathcal{T}_{\mu \nu \alpha \beta }^{V} &=&\hat{T}_{\mu \nu \alpha \beta
}^{V}+\hat{T}_{\alpha \nu \mu \beta }^{V}+\hat{T}_{\mu \beta \alpha \nu
}^{V}+\hat{T}_{\alpha \beta \mu \nu }^{V} \\
\mathcal{T}_{\mu \nu \alpha \beta }^{A} &=&\hat{T}_{\mu \nu \alpha \beta
}^{A}+\hat{T}_{\alpha \nu \mu \beta }^{A}+\hat{T}_{\mu \beta \alpha \nu
}^{A}+\hat{T}_{\alpha \beta \mu \nu }^{A}.
\end{eqnarray}%
Finally inserting in the definition it was given above (\ref{tG}), we have 
\begin{equation}
T_{\mu _{1}\mu _{2}\sigma _{1}\sigma _{2}}^{G}=-\frac{i}{64}\{[\mathcal{T}%
_{\mu _{12}\sigma _{12}}^{\mathrm{V}}]+[\mathcal{T}_{\mu _{12}\sigma _{12}}^{%
\mathrm{A}}]\}.  \label{Tgravfull}
\end{equation}

From these elaborations, we can identify that we have already exposed the
amplitudes with two Lorentz indices $T_{\mu \nu }^{\Gamma _{1}\Gamma _{2}}$
in the Chapter (\ref{2Dim2Pt}). So our task boils down to calculating only
typical fermionic amplitudes with three and four indices as the following
sequence.

\textbf{Ward Identities:} The symmetries role is crucial for understanding a
QFT because we have an anomaly in quantum theory when there is a symmetry
violation of the action or the classical conservation law. However, in some
cases, we can avoid these anomalies by imposing severe restrictions on the
physical content of the approach. In this section, we will establish
symmetries and general restrictions that will guide the consistency of the
method and the interpretation of the presence of anomalies.

Classically, the energy-momentum tensor defined in (\ref{tensorem-em}) has
symmetry properties, $T_{\mu \nu }=T_{\nu \mu },$ current conservation, \ $%
\nabla ^{\mu }T_{\mu \nu }=0$, and null trace, $T_{\mu }^{\mu }=0,$ see \cite%
{Bertlmann2001a}. These would lead us to the identities for the green
function defined in (\ref{TGreen}) 
\begin{eqnarray}
T_{\mu \nu \alpha \beta }^{G}\left( q\right) &=&T_{\nu \mu \alpha \beta
}^{G}\left( q\right) ; \\
q^{\mu }T_{\mu \nu \alpha \beta }^{G}\left( q\right) &=&0;  \label{pT=0} \\
g^{\mu \alpha }T_{\mu \nu \alpha \beta }^{G}\left( q\right) &=&0.
\label{gT=0}
\end{eqnarray}%
However, the literature shows gravitation as a gauge theory. Therefore these
canonical identities are not necessarily satisfied. We will have an Einstein
anomaly in the violation of general coordinate transformations
(diffeomorphisms) and Lorentz anomalies that imply an antisymmetric part in
the first equation above. In the case of conformal transformations (Weyl
transformations) violations, we will have a Weyl anomaly.

In the context of Einstein and Weyl invariances, we obtain consistency tests
before the symmetry analysis. They arise when we perform $q^{\mu }T_{\mu \nu
\alpha \beta }^{G}$ and $g^{\mu \nu }T_{\mu \nu \alpha \beta }^{G}$ to their
integrands and obtain relations (based on integration linearity) among the
set of structures defined above, i.e., through RAGFs. Since decomposition (%
\ref{tGtilV-A}) can be done, writing a basic permutation of gravitational
amplitude, in terms of amplitudes with vertices analogous to those of vector
and axial currents, these can be studied individually, as they will present
well-defined relations among them. Their complete introduction and detailed
verification occur in the Section to even amplitudes (\ref{evenragfsec}) and
(\ref{oddragfsec}).

\section{$VV$-$AA$: Even Amplitudes\label{AmpInt}}

We aim to determine all components that integrate gravitational amplitudes
while assuming no choice in intermediate steps. In this way, it is possible
to systematize all odd amplitudes in terms of even amplitudes $VV$'s: their
divergent properties are functions of divergent parts from $VV$-amplitudes,
and their finite parts gain an additional term proportional to the mass
squared. So, we will focus on this amplitude, finding a set of definitions
that makes their discussion viable. Otherwise, it would be too long due to
the number of surface terms within the \textit{IReg strategy}. From here on,
all the time, metric symbol $g_{\mu \nu }$ means flat metric $g_{\mu \nu
}=\eta _{\mu \nu }$.

As we saw in (\ref{VV}), the expression for the\ amplitude $VV$ is given by%
\begin{equation*}
T_{\mu _{1}\mu _{2}}^{VV}=2\Delta _{2\mu _{1}\mu _{2}}+\theta _{\mu _{1}\mu
_{2}}\left( 4m^{2}J_{2}+i/\pi \right) .
\end{equation*}%
That also can be written in closed form by Feynman integrals basis, see
Section (\ref{BasisFI}),%
\begin{equation}
T_{\mu _{1}\mu _{2}}^{VV}=\mathcal{D}_{\mu _{1}\mu _{2}}^{VV}+4J_{2\mu
_{1}\mu _{2}}+2q_{(\mu _{1}}J_{2\mu _{2})}+g_{\mu _{1}\mu _{2}}q^{2}J_{2}.
\end{equation}%
Finite parts come from definitions $J_{2},$ $J_{2\mu _{i},}$ and $J_{2\mu
_{1}\mu _{2}}$ as combinations of $Z_{k}^{\left( n\right) }$. As for the
divergent part, we collect all divergent terms and combine them in the
definition%
\begin{equation}
\mathcal{D}_{\mu _{1}\mu _{2}}^{VV}=2\Delta _{2\mu _{1}\mu _{2}}.
\label{DivVV2}
\end{equation}

Amplitudes with additional factors $K_{1\alpha _{i}}$ follow the operations
of those without derivative indices. The effect of this factor is to produce
an algebraic structure similar to $J$-integrals but with higher tensor
degrees. From previous definitions, 
\begin{eqnarray}
t_{\mu _{1}\mu _{2};\alpha _{1}}^{VV} &=&K_{1\alpha _{1}}t_{\mu _{1}\mu
_{2}}^{VV} \\
t_{\mu _{1}\mu _{2};\alpha _{1}\alpha _{2}}^{VV} &=&K_{1\alpha
_{1}}K_{1\alpha _{2}}t_{\mu _{1}\mu _{2}}^{VV}.
\end{eqnarray}%
Therefore, amplitudes will have a greater degree of divergence, implying
that finite and divergent parts are more complex and lengthier. These are
expressed in Section (\ref{BasisFI}).

Expressions appear as a standard tensor plus a PP amplitude; see (\ref{vv}),
thus 
\begin{eqnarray}
t_{\mu _{12};\alpha _{1}}^{VV} &=&2t_{\mu _{1}\mu _{2};\alpha _{1}}^{\left(
+\right) }+g_{\mu _{1}\mu _{2}}t_{\alpha _{1}}^{PP}  \label{t(s2)} \\
t_{\mu _{12};\alpha _{1}\alpha _{2}}^{VV} &=&2t_{\mu _{1}\mu _{2};\alpha
_{1}\alpha _{2}}^{\left( +\right) }+g_{\mu _{1}\mu _{2}}t_{\alpha _{1}\alpha
_{2}}^{PP}.  \label{t(s3)}
\end{eqnarray}
Tensors $t_{\mu _{1}\mu _{2};\alpha _{1}}^{\left( +\right) }$ and $t_{\mu
_{1}\mu _{2};\alpha _{1}\alpha _{2}}^{\left( +\right) }$ appearing above are
particular cases of general tensors:%
\begin{eqnarray}
t_{\mu _{12};\alpha _{1}}^{\left( s_{1}\right) } &=&K_{1\alpha _{1}}\left(
K_{1\mu _{1}}K_{2\mu _{2}}+s_{1}K_{1\mu _{2}}K_{2\mu _{1}}\right) \frac{1}{%
D_{12}}  \label{T3} \\
t_{\mu _{12};\alpha _{12}}^{\left( s_{1}\right) } &=&K_{1\alpha _{1}}t_{\mu
_{12};\alpha _{1}}^{\left( s_{1}\right) };  \label{T4}
\end{eqnarray}%
where $s_{1}=$ $\pm $, see (\ref{t(s)}). For example, in the tensor of
3rd-order, two cases assume are%
\begin{eqnarray}
t_{\mu _{12};\alpha _{1}}^{\left( +\right) } &=&2\frac{K_{1\alpha
_{1}}K_{1\mu _{1}}K_{1\mu _{2}}}{D_{12}}+\frac{q_{(\mu _{1}}K_{1\mu
_{2})}K_{1\alpha _{1}}}{D_{12}}  \label{t+3} \\
t_{\mu _{12};\alpha _{1}}^{\left( -\right) } &=&\frac{q_{[\mu _{2}}K_{1\mu
_{1}]}K_{1\alpha _{1}}}{D_{12}}.
\end{eqnarray}%
Moreover, 4th-rank ones naturally get one more $K_{1\alpha _{2}}$ factor. To 
$PP$ amplitude (\ref{PP}), we add a $K_{1\alpha _{1}}$ according to our
definitions 
\begin{equation}
t_{\alpha _{1}}^{PP}=K_{1\alpha _{1}}\left( q^{2}\frac{1}{D_{12}}-\frac{1}{%
D_{1}}-\frac{1}{D_{2}}\right) =K_{1\alpha _{1}}t^{PP},  \label{pp1}
\end{equation}%
and with two indices $t_{\alpha _{1}\alpha _{2}}^{PP}=K_{1\alpha
_{2}}t_{\alpha _{1}}^{PP}$.

Integrating (\ref{t(s2)}) using (\ref{t+3}) and (\ref{pp1}), derivative $VV$
with three indices become%
\begin{eqnarray}
T_{\mu _{12};\alpha _{1}}^{VV} &=&4\bar{J}_{2\mu _{12}\alpha _{1}}+2q_{(\mu
_{1}}\bar{J}_{2\mu _{2})\alpha _{1}}+g_{\mu _{1}\mu _{2}}q^{2}J_{2\alpha
_{1}} \\
&&-g_{\mu _{12}}[\bar{J}_{1\alpha _{1}}\left( k_{2}\right) +\bar{J}_{1\alpha
_{1}}\left( k_{1}\right) ]+g_{\mu _{12}}q_{\alpha _{1}}\bar{J}_{1}\left(
k_{2}\right) .  \notag
\end{eqnarray}%
For $VV$ with four indices (\ref{t(s3)}), we have when integrating the
tensor (\ref{T4}) and $K_{1\alpha _{2}}t_{\alpha _{1}}^{PP}$,%
\begin{eqnarray}
T_{\mu _{12};\alpha _{12}}^{VV} &=&4\bar{J}_{2\mu _{12}\alpha
_{12}}+2q_{(\mu _{1}}\bar{J}_{2\mu _{2})\alpha _{12}}+g_{\mu _{12}}q^{2}\bar{%
J}_{2\alpha _{12}} \\
&&+g_{\mu _{12}}\left[ \bar{J}_{1\alpha _{12}}\left( k_{2}\right) -\bar{J}%
_{1\alpha _{12}}\left( k_{1}\right) \right] -g_{\mu _{12}}\left[ q_{(\alpha
_{1}}\bar{J}_{1\alpha _{2})}\left( k_{2}\right) -q_{\alpha _{12}}\bar{J}%
_{1}\left( k_{2}\right) \right] .  \notag
\end{eqnarray}%
Additional terms in the $k_{2}$ that appear in the $J$'s (with only one
propagator) come from the translations $K_{1}=K_{2}-q$ used to define
functions in Section (\ref{DivTerms}). Remember that barred $J$'s have
finite and divergent parts.

To simplify the exposition of finite and divergent parts from equations (\ref%
{2DJ2mu1}), (\ref{2DJ2C}), (\ref{J2mu123B}), and (\ref{J2mu1234}), it is
possible to write the results as%
\begin{eqnarray}
T_{\mu _{12};\alpha _{1}}^{VV} &=&4J_{2\mu _{12}\alpha _{1}}+2q_{(\mu
_{1}}J_{2\mu _{2})\alpha _{1}}+g_{\mu _{12}}q^{2}J_{2\alpha _{1}}+\mathcal{D}%
_{\mu _{12};\alpha _{1}}^{VV}  \label{TVV3} \\
T_{\mu _{12};\alpha _{12}}^{VV} &=&4J_{2\mu _{12}\alpha _{12}}+2q_{(\mu
_{1}}J_{2\mu _{2})\alpha _{12}}+g_{\mu _{12}}q^{2}J_{2\alpha _{12}}+\mathcal{%
D}_{\mu _{12};\alpha _{12}}^{VV}.  \label{TVV4}
\end{eqnarray}%
In these cases, all divergent terms of integrals and define the 3rd-order
tensor 
\begin{equation}
\mathcal{D}_{\mu _{12};\alpha _{1}}^{VV}=-P^{\nu _{1}}W_{3\mu _{12}\alpha
_{1}\nu _{1}}+P_{(\mu _{1}}\Delta _{2\mu _{2}\alpha _{1})}+g_{\mu
_{12}}P^{\nu _{1}}\Delta _{2\alpha _{1}\nu _{1}}-q_{\alpha _{1}}\Delta
_{2\mu _{12}},  \label{DivVV3}
\end{equation}%
and the 4th-order tensor%
\begin{eqnarray}
\mathcal{D}_{\mu _{12};\alpha _{12}}^{VV} &=&+(W_{2\mu _{12}\alpha
_{12}}-g_{\mu _{12}}\Delta _{1\alpha _{12}})+g_{\mu _{1}(\alpha
_{1}}g_{\alpha _{2})\mu _{2}}I_{\text{\textrm{quad}}}+\frac{\Omega _{\mu
_{12}\alpha _{12}}}{6q^{2}}I_{\text{\textrm{log}}}  \label{DivVV4} \\
&&+\frac{1}{12}(3P^{\nu _{12}}+q^{\nu _{12}})W_{4\mu _{12}\alpha _{12}\nu
_{12}}-\frac{1}{4}(P^{\nu _{12}}+q^{\nu _{12}})g_{\mu _{12}}W_{3\alpha
_{12}\nu _{12}}  \notag \\
&&-\frac{1}{2}P^{\nu _{1}}\left( P_{\mu _{1}}W_{3\mu _{2}\alpha _{12}\nu
_{1}}+P_{\mu _{2}}W_{3\mu _{1}\alpha _{12}\nu _{1}}\right) +\frac{1}{2}%
P^{\nu _{1}}g_{\mu _{12}}\left( P-q\right) _{(\alpha _{1}}\Delta _{2\alpha
_{2})\nu _{1}}  \notag \\
&&-\frac{1}{4}(P^{2}+q^{2})W_{3\mu _{12}\alpha _{12}}-\frac{1}{2}P^{\nu
_{1}}\left( P-q\right) _{(\alpha _{1}}W_{3\alpha _{2})\mu _{12}\nu _{1}} 
\notag \\
&&+\frac{1}{4}[2(\theta _{\mu _{12}}+P_{\mu _{12}})+g_{\mu
_{12}}(P^{2}+q^{2})]\Delta _{2\alpha _{12}}+\frac{1}{2}(P-q)_{\alpha
_{1}}(P-q)_{\alpha _{2}}\Delta _{2\mu _{12}}  \notag \\
&&+\frac{1}{2}P_{\mu _{2}}\left( P-q\right) _{(\alpha _{1}}\Delta _{2\alpha
_{2})\mu _{1}}+\frac{1}{2}P_{\mu _{1}}(P-q)_{(\alpha _{1}}\Delta _{2\alpha
_{2})\mu _{2}}.  \notag
\end{eqnarray}%
We use definition of projectors $\theta _{\mu _{12}}$ and $\Omega _{\mu
_{12}\alpha _{12}}$ as 
\begin{eqnarray}
\theta _{\mu _{12}} &=&g_{\mu _{1}\mu _{2}}q^{2}-q_{\mu _{1}}q_{\mu _{2}} \\
\Omega _{\mu _{12}\alpha _{12}}\left( q\right) &=&2\theta _{\mu _{12}}\theta
_{\alpha _{12}}-\left( \theta _{\mu _{1}\alpha _{1}}\theta _{\mu _{2}\alpha
_{2}}+\theta _{\mu _{1}\alpha _{2}}\theta _{\mu _{2}\alpha _{1}}\right) ;
\end{eqnarray}%
both are transverse; additionally, $\Omega $ is traceless in all its
indices. Note that here the projector $\theta _{\mu \nu }$ \textit{is not
dimensionless} as in Chapter (\ref{2Dim2Pt}); it has mass dimension two.

The finite part also can be expressed from explicit functions plus $\mathcal{%
D}$-tensor%
\begin{equation}
T_{\mu _{12};\alpha _{1}}^{VV}=\frac{i}{2\pi }q_{\alpha _{1}}\theta _{\mu
_{12}}(Z_{2}^{\left( -1\right) }-Z_{1}^{\left( -1\right) })+\mathcal{D}_{\mu
_{12};\alpha _{1}}^{VV}.
\end{equation}%
The four-index amplitude is more complicated but can be written in the
projectors%
\begin{eqnarray}
T_{\mu _{12};\alpha _{12}}^{VV} &=&\frac{i}{4\pi }\frac{1}{q^{2}}\left[
-\Omega _{\mu _{12}\alpha _{12}}(2Z_{2}^{\left( 0\right) }-Z_{1}^{\left(
0\right) })+2\theta _{\mu _{12}}\theta _{\alpha _{12}}(3Z_{2}^{\left(
0\right) }-2Z_{1}^{\left( 0\right) })\right] \\
&&-\frac{i}{4\pi }q_{\alpha _{12}}\theta _{\mu _{12}}(Z_{2}^{\left(
-1\right) }-Z_{1}^{\left( -1\right) })+\mathcal{D}_{\mu _{12};\alpha
_{12}}^{VV}.  \notag
\end{eqnarray}%
It is possible to maintain closed form in $J$'s, as we will see in RAGF,
through reductions as in Section (\ref{FinFcts}). In this way, we find
leading amplitudes as a substructure of $T_{\mu _{12}}^{VV},$%
\begin{equation}
T_{\mu _{12};\alpha _{1}}^{VV}=-\frac{1}{2}q_{\alpha _{1}}T_{\mu _{12}}^{VV}+%
\mathcal{D}_{\mu _{12};\alpha _{1}}^{VV}+\frac{1}{2}q_{\alpha _{1}}\mathcal{D%
}_{\mu _{12}}^{VV}.
\end{equation}%
Moreover, the same is true for the 4th-order amplitude%
\begin{eqnarray}
T_{\mu _{12};\alpha _{12}}^{VV} &=&\frac{1}{4}q_{\alpha _{12}}T_{\mu
_{12}}^{VV}+\mathcal{D}_{\mu _{12};\alpha _{12}}^{VV}-\frac{1}{4}q_{\alpha
_{12}}\mathcal{D}_{\mu _{12}}^{VV} \\
&&+\frac{i}{4\pi }\frac{1}{q^{2}}\left[ -\Omega _{\mu _{12}\alpha
_{12}}(2Z_{2}^{\left( 0\right) }-Z_{1}^{\left( 0\right) })+2\theta _{\mu
_{12}}\theta _{\alpha _{12}}(3Z_{2}^{\left( 0\right) }-2Z_{1}^{\left(
0\right) })\right] .  \notag
\end{eqnarray}

The next amplitude to be calculated is the $AA$. Like $VV$, this amplitude
will contribute to the even sector of the gravitational amplitude in (\ref%
{ttillv}). From the chapter on equal masses, after traces, we have expressed
it exactly as (\ref{aa}). However, writing this result in terms of amplitude 
$t_{\mu _{12}}^{VV}$ plus a scalar function proportional to the metric, is
feasible%
\begin{equation}
t_{\mu _{12}}^{AA}=t_{\mu _{12}}^{VV}-4m^{2}g_{\mu _{12}}\frac{1}{D_{12}}.
\end{equation}%
This form allows us to write equations directly from definitions for
derivative amplitudes%
\begin{eqnarray}
t_{\mu _{12};\alpha _{1}}^{AA} &=&t_{\mu _{12};\alpha
_{1}}^{VV}-4m^{2}g_{\mu _{12}}\frac{K_{1\alpha _{1}}}{D_{12}} \\
t_{\mu _{12};\alpha _{12}}^{AA} &=&t_{\mu _{12};\alpha
_{12}}^{VV}-4m^{2}g_{\mu _{12}}\frac{K_{1\alpha _{1}}K_{1\alpha _{2}}}{D_{12}%
},
\end{eqnarray}%
and their integrals 
\begin{eqnarray}
T_{\mu _{12}}^{AA} &=&T_{\mu _{12}}^{VV}-4m^{2}g_{\mu _{12}}J_{2}
\label{AA-VV} \\
T_{\mu _{12};\alpha _{1}}^{AA} &=&T_{\mu _{12};\alpha
_{1}}^{VV}-4m^{2}g_{\mu _{12}}J_{2\alpha _{1}}  \label{AA-VV1} \\
T_{\mu _{12};\alpha _{12}}^{AA} &=&T_{\mu _{12};\alpha
_{12}}^{VV}-4m^{2}g_{\mu _{12}}\bar{J}_{2\alpha _{12}}.  \label{AA-VV2}
\end{eqnarray}%
Additional contributions of massive terms present in this amplitude are
worth noting. For the divergent part, only the amplitude with four indices
has a non-zero term in $\bar{J}_{2\alpha _{12}}$, see (\ref{2DJ2C}).
Integrals appearing in the amplitudes of fewer indices contribute only to
the finite part. However, the 4th-rank amplitude has an additional
contribution as a surface term and $I_{\log }$. The final result is
identical to that obtained from the first form presented.

\section{$AV$-$VA$: Odd amplitudes}

We will calculate all odd parts of gravitational amplitude. As seen in (\ref%
{AV1 and AV2}), we wrote two-index functions in terms of even ones using
general identity for $2D$, $\gamma _{\ast }\gamma _{\mu _{1}}=-\varepsilon
_{\mu _{1}\nu _{1}}\gamma ^{\nu _{1}}$, present in (\ref{id2}). For
higher-rank amplitudes, traces operate in the same way but add indices to
the integrals:%
\begin{equation}
(T_{\mu _{12};\alpha _{1}}^{AV})_{1}=-\varepsilon _{\mu _{1}}^{\quad \nu
_{1}}T_{\nu _{1}\mu _{2};\alpha _{1}}^{VV};\qquad (T_{\mu _{12};\alpha
_{1}}^{AV})_{2}=-\varepsilon _{\mu _{2}}^{\quad \nu _{1}}T_{\mu _{1}\nu
_{1};\alpha _{1}}^{AA}  \label{AV32}
\end{equation}%
\begin{equation}
(T_{\mu _{12};\alpha _{12}}^{AV})_{1}=-\varepsilon _{\mu _{1}}^{\quad \nu
_{1}}T_{\nu _{1}\mu _{2};\alpha _{12}}^{VV};\qquad (T_{\mu _{12};\alpha
_{12}}^{AV})_{2}=-\varepsilon _{\mu _{2}}^{\quad \nu _{1}}T_{\mu _{1}\nu
_{1};\alpha _{12}}^{AA}.  \label{AV42}
\end{equation}%
To complete odd amplitudes, we cast the analogous VA equations:%
\begin{eqnarray}
(T_{\mu _{12};\alpha _{1}}^{VA})_{1} &=&-\varepsilon _{\mu _{1}}^{\text{%
\quad }\nu _{1}}T_{\nu _{1}\mu _{2};\alpha _{1}}^{AA}\quad (T_{\mu
_{12};\alpha _{1}}^{VA})_{2}=-\varepsilon _{\mu _{2}}^{\text{\quad }\nu
_{1}}T_{\mu _{1}\nu _{1};\alpha _{1}}^{VV}; \\
(T_{\mu _{12};\alpha _{12}}^{VA})_{1} &=&-\varepsilon _{\mu _{1}}^{\text{%
\quad }\nu _{1}}T_{\nu _{1}\mu _{2};\alpha _{12}}^{AA}\quad (T_{\mu
_{12};\alpha _{12}}^{VA})_{2}=-\varepsilon _{\mu _{2}}^{\text{\quad }\nu
_{1}}T_{\mu _{1}\nu _{1};\alpha _{12}}^{VV}.
\end{eqnarray}%
The same considerations can be made when using the chiral matrix definition (%
\ref{id1}) directly in the Dirac traces. By considering expressions for
amplitudes with additional terms, as in (\ref{av1full}) and (\ref{av2full}),
for amplitudes with derivative vertices, we have 
\begin{eqnarray}
(t_{\mu _{12};\alpha _{1}}^{AV})_{1} &=&-\varepsilon _{\mu _{1}}^{\text{%
\quad }\nu _{1}}t_{\nu _{1}\mu _{2};\alpha _{1}}^{VV}+2\varepsilon _{\mu
_{2}}^{\text{\quad }\nu _{1}}t_{\mu _{1}\nu _{1};\alpha _{1}}^{\left(
-\right) }+g_{\mu _{12}}t_{\alpha _{1}}^{SP} \\
(t_{\mu _{12};\alpha _{1}}^{AV})_{2} &=&-\varepsilon _{\mu _{2}}^{\text{%
\quad }\nu _{1}}t_{\mu _{1}\nu _{1};\alpha _{1}}^{AA}+2\varepsilon _{\mu
_{1}}^{\text{\quad }\nu _{1}}t_{\nu _{1}\mu _{2};\alpha _{1}}^{\left(
-\right) }-g_{\mu _{12}}t_{\alpha _{1}}^{SP}
\end{eqnarray}%
\begin{eqnarray}
(t_{\mu _{12};\alpha _{12}}^{AV})_{1} &=&-\varepsilon _{\mu _{1}}^{\text{%
\quad }\nu _{1}}t_{\nu _{1}\mu _{2};\alpha _{12}}^{VV}+2\varepsilon _{\mu
_{2}}^{\text{\quad }\nu _{1}}t_{\mu _{1}\nu _{1};\alpha _{12}}^{\left(
-\right) }+g_{\mu _{12}}t_{\alpha _{12}}^{SP} \\
(t_{\mu _{12};\alpha _{12}}^{AV})_{2} &=&-\varepsilon _{\mu _{2}}^{\text{%
\quad }\nu _{1}}t_{\mu _{1}\nu _{1};\alpha _{12}}^{AA}+2\varepsilon _{\mu
_{1}}^{\text{\quad }\nu _{1}}t_{\nu _{1}\mu _{2};\alpha _{12}}^{\left(
-\right) }-g_{\mu _{12}}t_{\alpha _{12}}^{SP}.
\end{eqnarray}

Additional terms combine and cancel out when integrated, so the equations
above reduce to those given in (\ref{AV32})-(\ref{AV42}). Let us demonstrate
this fact, using the definition (\ref{T3}) to $t^{\left( -\right) }$ at the
beginning of the last section. Thus we have 
\begin{equation}
2\varepsilon _{\mu _{2}}^{\text{\quad }\nu _{1}}T_{\mu _{1}\nu _{1};\alpha
_{1}}^{\left( -\right) }+g_{\mu _{12}}T_{\alpha _{1}}^{SP}=2\varepsilon
_{\mu _{2}\nu _{1}}q_{\mu _{1}}\bar{J}_{2\alpha _{1}}^{\nu
_{1}}-2\varepsilon _{\mu _{2}\nu _{1}}q^{\nu _{1}}\bar{J}_{2\mu _{1}\alpha
_{1}}+2g_{\mu _{1}\mu _{2}}\varepsilon _{\nu _{1}\nu _{2}}q^{\nu _{2}}\bar{J}%
_{2\alpha _{1}}^{\nu _{1}}.
\end{equation}%
We applied our definitions of $J_{2}$ integrals, and employed the identity
below in the las term%
\begin{equation}
\varepsilon _{\nu _{1}\nu _{2}}g_{\mu _{2}\mu _{1}}+\varepsilon _{\mu
_{2}\nu _{1}}g_{\nu _{2}\mu _{1}}+\varepsilon _{\nu _{2}\mu _{2}}g_{\nu
_{1}\mu _{1}}=0.
\end{equation}%
It is direct to observe the exact cancellation of the first two terms%
\begin{equation}
2\varepsilon _{\mu _{2}}^{\text{\quad }\nu _{1}}T_{\mu _{1}\nu _{1};\alpha
_{1}}^{\left( -\right) }+g_{\mu _{12}}T_{\alpha _{1}}^{SP}=0.
\end{equation}%
That occurs independently of divergent content of $\bar{J}_{2\mu \nu }$. It
is easy to see that the same happens to the analogous terms in the 4th-rank
amplitude's version, 
\begin{equation}
2\varepsilon _{\mu _{2}}^{\text{\quad }\nu _{1}}T_{\mu _{1}\nu _{1};\alpha
_{12}}^{\left( -\right) }+g_{\mu _{12}}T_{\alpha _{1}\alpha _{2}}^{SP}=0.
\end{equation}

Definitions for the $VA$ computed with the definition of the chiral matrix
were not present because the logic and result are the same. As for the
relation between $VV$ and $AA$ amplitudes, we write from the integrand level%
\begin{eqnarray*}
(T_{\mu _{12};\alpha _{1}}^{VA})_{1} &=&-\varepsilon _{\mu _{1}}^{\text{%
\quad }\nu _{1}}T_{\nu _{1}\mu _{2};\alpha _{1}}^{AA}=-\varepsilon _{\mu
_{1}}^{\text{\quad }\nu _{1}}(T_{\nu _{1}\mu _{2};\alpha
_{1}}^{VV}-4m^{2}g_{\nu _{1}\mu _{2}}J_{2\alpha _{1}}) \\
&=&(T_{\mu _{1}\mu _{2};\alpha _{1}}^{AV})_{1}+4m^{2}\varepsilon _{\mu
_{1}\mu _{2}}J_{2\alpha _{1}}.
\end{eqnarray*}%
This relation is satisfied without any conditions. In general, we have 
\begin{eqnarray}
(T_{\mu _{12}}^{VA})_{i} &=&(T_{\mu _{12}}^{AV})_{i}+4m^{2}\varepsilon _{\mu
_{1}\mu _{2}}J_{2}  \label{VA-AVt0} \\
(T_{\mu _{12};\alpha _{1}}^{VA})_{i} &=&(T_{\mu _{12};\alpha
_{1}}^{AV})_{i}+4m^{2}\varepsilon _{\mu _{1}\mu _{2}}J_{2\alpha _{1}} \\
(T_{\mu _{12};\alpha _{12}}^{VA})_{i} &=&(T_{\mu _{12};\alpha
_{12}}^{AV})_{i}+4m^{2}\varepsilon _{\mu _{1}\mu _{2}}\bar{J}_{2\alpha
_{12}},  \label{VA-AVt2}
\end{eqnarray}%
where the index $i=1,2$ is associated with versions, and the Eqs (\ref%
{VA-AVt0})-(\ref{VA-AVt2}) will often be used to reduce manipulations
required for the gravitational anomaly.

On the other hand, basic and independent versions one and two are only
strictly equivalent with conditions. This fact was worked in Chapters (\ref%
{2Dim2Pt}) and (\ref{2masses}), where a single mass and two masses in odd
amplitudes were handled. Let us retrieve the explicitly computed result to
establish general results to be used in the sequel%
\begin{eqnarray}
(T_{\mu _{12}}^{AV})_{1}-(T_{\mu _{12}}^{AV})_{2} &=&-2(\varepsilon _{\mu
_{1}\nu }\Delta _{2\mu _{2}}^{\nu }-\varepsilon _{\mu _{2}\nu }\Delta _{2\mu
_{1}}^{\nu }) \\
&&-(\varepsilon _{\mu _{1}\nu _{1}}\theta _{\mu _{2}}^{\nu }-\varepsilon
_{\mu _{2}\nu _{1}}\theta _{\mu _{1}}^{\nu })\frac{1}{q^{2}}\left(
4m^{2}J_{2}+i/\pi \right) +4\varepsilon _{\mu _{1}\mu _{2}}m^{2}J_{2}. 
\notag
\end{eqnarray}%
We rearrange the finite part using $\varepsilon _{\lbrack \mu _{1}\nu
}\theta _{\mu _{2}]}^{\nu }=0$ and surface terms $\varepsilon _{\lbrack \mu
_{1}\nu }\Delta _{2\mu _{2}]}^{\nu }=0$,%
\begin{eqnarray}
\varepsilon _{\mu _{1}\nu }\Delta _{2\mu _{2}}^{\nu }+\varepsilon _{\mu
_{2}\mu _{1}}\Delta _{2\nu }^{\nu }+\varepsilon _{\nu \mu _{2}}\Delta _{2\mu
_{1}}^{\nu } &=&0=\varepsilon _{\lbrack \mu _{1}\nu }\Delta _{2\mu
_{2}]}^{\nu }, \\
\varepsilon _{\mu _{1}\nu }\theta _{\mu _{2}}^{\nu }+\varepsilon _{\mu
_{2}\mu _{1}}\theta _{\nu }^{\nu }+\varepsilon _{\nu \mu _{2}}\theta _{\mu
_{1}}^{\nu } &=&0=\varepsilon _{\lbrack \mu _{1}\nu }\theta _{\mu
_{2}]}^{\nu };
\end{eqnarray}%
hence, the difference between the two versions reduces to%
\begin{equation}
(T_{\mu _{12}}^{AV})_{1}-(T_{\mu _{12}}^{AV})_{2}=-\varepsilon _{\mu _{1}\mu
_{2}}\left( 2\Delta _{2\alpha }^{\alpha }+i/\pi \right) .
\end{equation}

Here we clarify how this result can be written systematically. It boils down
to using the definitions and caveat that each term present represents
complete amplitudes, 
\begin{eqnarray}
T_{\mu _{1}\mu _{2}}^{VV} &=&2\Delta _{2\mu _{1}\mu _{2}}+\frac{\theta _{\mu
_{1}\mu _{2}}}{q^{2}}\left( 4m^{2}J_{2}+i/\pi \right) , \\
T_{\mu _{1}\mu _{2}}^{AA} &=&2\Delta _{2\mu _{1}\mu _{2}}+\frac{\theta _{\mu
_{1}\mu _{2}}}{q^{2}}\left( 4m^{2}J_{2}+i/\pi \right) -g_{\mu _{1}\mu
_{2}}\left( 4m^{2}J_{2}\right) .
\end{eqnarray}%
Using (\ref{AV1 and AV2}), the versions for $AV$-amplitudes arise%
\begin{eqnarray}
(T_{\mu _{12}}^{AV})_{1} &=&-2\varepsilon _{\mu _{1}}^{\quad \nu }\Delta
_{2\mu _{2}\nu }-\frac{\varepsilon _{\mu _{1}\nu }\theta _{\mu _{2}}^{\nu }}{%
q^{2}}\left( 4m^{2}J_{2}+i/\pi \right) , \\
(T_{\mu _{12}}^{AV})_{2} &=&-2\varepsilon _{\mu _{2}}^{\quad \nu }\Delta
_{2\mu _{1}\nu }-\frac{\varepsilon _{\mu _{2}\nu }\theta _{\mu _{1}}^{\nu }}{%
q^{2}}\left( 4m^{2}J_{2}+i/\pi \right) -\varepsilon _{\mu _{1}\mu
_{2}}\left( 4m^{2}J_{2}\right) .
\end{eqnarray}

After writing the difference between them%
\begin{equation}
(T_{\mu _{12}}^{AV})_{1}-(T_{\mu _{12}}^{AV})_{2}=-\varepsilon _{\mu
_{1}}^{\quad \nu _{1}}T_{\nu _{1}\mu _{2}}^{VV}+\varepsilon _{\mu
_{2}}^{\quad \nu _{1}}T_{\mu _{1}\nu _{1}}^{AA},
\end{equation}%
we take into account identity among $AA$ and $VV$ (\ref{AA-VV}):%
\begin{equation}
(T_{\mu _{12}}^{AV})_{1}-(T_{\mu _{12}}^{AV})_{2}=-\varepsilon _{\mu
_{1}}^{\quad \nu _{1}}T_{\nu _{1}\mu _{2}}^{VV}+\varepsilon _{\mu
_{2}}^{\quad \nu _{1}}T_{\mu _{1}\nu _{1}}^{VV}-4m^{2}\varepsilon _{\mu
_{2}\mu _{1}}J_{2}.
\end{equation}%
Lastly, employ%
\begin{equation*}
\varepsilon _{\lbrack \mu _{1}\nu _{1}}\left( T^{VV}\right) _{\quad \mu
_{2}]}^{\nu _{1}}=0\Leftrightarrow -\varepsilon _{\mu _{1}}^{\quad \nu
_{1}}T_{\nu _{1}\mu _{2}}^{VV}+\varepsilon _{\mu _{2}}^{\quad \nu
_{1}}T_{\mu _{1}\nu _{1}}^{VV}=\varepsilon _{\mu _{2}\mu _{1}}\left( g^{\nu
_{12}}T_{\nu _{1}\nu _{2}}^{VV}\right) ,
\end{equation*}%
to reach an expression equivalent to work term by term on the amplitude, 
\begin{equation}
(T_{\mu _{12}}^{AV})_{1}-(T_{\mu _{12}}^{AV})_{2}=\varepsilon _{\mu _{2}\mu
_{1}}\left( g^{\nu _{12}}T_{\nu _{1}\nu _{2}}^{VV}-4m^{2}J_{2}\right)
=:\Upsilon .
\end{equation}%
With the help of explicit expression, follows%
\begin{equation}
\Upsilon =2\Delta _{2\rho }^{\rho }+\left( 4m^{2}J_{2}+i/\pi \right)
-4m^{2}J_{2}=2\Delta _{2\rho }^{\rho }+i/\pi .  \label{Ups0}
\end{equation}%
As it must be, this condition is equal to that deduced to the equivalence of
basic (\ref{Uni-2D}). For these amplitudes, the equality among independent
expressions is obtained through any possible way to employ the trace of four
gamma matrices and a chiral one.

It is a direct task to identify this condition to higher-rank amplitudes,
namely,%
\begin{eqnarray}
(T_{\mu _{12}}^{AV})_{1}-(T_{\mu _{12}}^{AV})_{2} &=&\varepsilon _{\mu
_{2}\mu _{1}}\Upsilon  \label{Uni} \\
(T_{\mu _{12};\alpha _{1}}^{AV})_{1}-(T_{\mu _{12};\alpha _{1}}^{AV})_{2}
&=&\varepsilon _{\mu _{2}\mu _{1}}\Upsilon _{\alpha _{1}}  \label{Un1} \\
(T_{\mu _{12};\alpha _{12}}^{AV})_{1}-(T_{\mu _{12};\alpha _{12}}^{AV})_{2}
&=&\varepsilon _{\mu _{2}\mu _{1}}\Upsilon _{\alpha _{12}}.  \label{Un2}
\end{eqnarray}%
Due to the relevance of these terms, we present the following definition%
\begin{eqnarray}
\Upsilon &=&\left( g^{\nu _{12}}T_{\nu _{1}\nu _{2}}^{VV}-4m^{2}J_{2}\right)
\label{U1} \\
\Upsilon _{\alpha _{1}} &=&\left( g^{\nu _{12}}T_{\nu _{1}\nu _{2};\alpha
_{1}}^{VV}-4m^{2}J_{2\alpha _{1}}\right)  \label{U2} \\
\Upsilon _{\alpha _{12}} &=&\left( g^{\nu _{12}}T_{\nu _{1}\nu _{2};\alpha
_{12}}^{VV}-4m^{2}J_{2\alpha _{12}}\right) .  \label{U3}
\end{eqnarray}%
At the end of calculations, identities of this type must be used in surface
terms and finite parts of amplitudes. This approach simplifies the
conclusions that can be given by exposing hundreds of terms that build up
some of these amplitudes, making that path prohibitively long to be exposed.
The identities only express the vanishing of a complete antisymmetric tensor
of degree three in two dimensions.

The last section exposed detailed results for finite and divergent parts of
core component $VV$ amplitudes that appear in RHS of (\ref{U2}) and (\ref{U3}%
). Thus, we take expressions (\ref{TVV3}) and (\ref{TVV4}) into account to
write%
\begin{eqnarray}
g^{\nu _{12}}T_{\nu _{12};\alpha _{1}}^{VV} &=&4g^{\nu _{12}}J_{2\nu
_{12}\alpha _{1}}+2(2q^{\nu _{1}}J_{2\nu _{1}\alpha _{1}}+q^{2}J_{2\alpha
_{1}})+g^{\nu _{12}}\mathcal{D}_{\nu _{12};\alpha _{1}}^{VV}  \label{gT} \\
g^{\nu _{12}}T_{\nu _{12};\alpha _{12}}^{VV} &=&4g^{\nu _{12}}J_{2\nu
_{12}\alpha _{12}}+2(2q^{\nu _{1}}J_{2\nu _{1}\alpha _{12}}+q^{2}J_{2\alpha
_{12}})+g^{\nu _{12}}\mathcal{D}_{\nu _{12};\alpha _{12}}^{VV}.  \label{gT1}
\end{eqnarray}%
Observe that $J$-functions comprise the entire finite part while $\mathcal{D}%
^{VV}$-tensor accounts for divergent terms. Therefore, these calculations
require the traces 
\begin{eqnarray}
4g^{\nu _{12}}J_{2\nu _{12}\alpha _{1}} &=&4m^{2}J_{2\alpha _{1}}-\frac{i}{%
2\pi }q_{\alpha _{1}} \\
4g^{\nu _{12}}J_{2\nu _{12}\alpha _{12}} &=&4m^{2}J_{2\alpha _{12}}-\frac{i}{%
12\pi }[\theta _{\alpha _{1}\alpha _{2}}\left( q\right) -3q_{\alpha
_{1}}q_{\alpha _{2}}],
\end{eqnarray}%
and relations coming from momentum contractions%
\begin{eqnarray}
2q^{\nu _{1}}J_{2\nu _{1}\alpha _{1}}+q^{2}J_{2\alpha _{1}} &=&0 \\
2q^{\nu _{1}}J_{2\nu _{1}\alpha _{1}\alpha _{2}}+q^{2}J_{2\alpha _{1}\alpha
_{2}} &=&0;
\end{eqnarray}%
results derived in Sections (\ref{FinFcts}) and (\ref{BasisFI}).
Substituting in (\ref{gT}) and (\ref{gT1}) yields%
\begin{eqnarray}
g^{\nu _{12}}T_{\nu _{12};\alpha _{1}}^{VV} &=&4m^{2}J_{2\alpha _{1}}-\frac{i%
}{2\pi }q_{\alpha _{1}}+g^{\nu _{12}}\mathcal{D}_{\nu _{12};\alpha _{1}}^{VV}
\\
g^{\nu _{12}}T_{\nu _{12};\alpha _{12}}^{VV} &=&4m^{2}J_{2\alpha _{12}}-%
\frac{i}{12\pi }(\theta _{\alpha _{1}\alpha _{2}}-3q_{\alpha _{1}}q_{\alpha
_{2}})+g^{\nu _{12}}\mathcal{D}_{\nu _{12};\alpha _{12}}^{VV}.
\end{eqnarray}

The trace of $\mathcal{D}^{VV}$-tensor, their explicit forms from (\ref%
{DivVV3}) and (\ref{DivVV4}). For the one derivative index, the divergent
terms have only logarithmic divergent surface terms%
\begin{equation}
g^{\nu _{12}}\mathcal{D}_{\nu _{12};\alpha _{1}}^{VV}=-\frac{1}{2}P^{\nu
_{1}}(2W_{3\rho \alpha _{1}\nu _{1}}^{\rho }-8\Delta _{2\alpha _{1}\nu
_{1}})+P_{\alpha _{1}}\Delta _{2\rho }^{\rho }-q_{\alpha _{1}}\Delta _{2\rho
}^{\rho }.
\end{equation}%
As for the trace of the two-derivative indices tensor, its divergent part is
more complex. It presents a relation involving the trace of quadratically
divergent objects, as seen in the first line of the following equation 
\begin{eqnarray}
g^{\nu _{12}}\mathcal{D}_{\nu _{12};\alpha _{12}}^{VV} &=&(W_{2\rho \alpha
_{12}}^{\rho }-2\Delta _{1\alpha _{12}})+2g_{\alpha _{1}\alpha _{2}}I_{\text{%
\textrm{quad}}}+\frac{1}{6p^{2}}g^{\mu _{12}}\Omega _{\mu _{12}\alpha
_{12}}I_{\text{\textrm{log}}} \\
&&+\frac{1}{36}(3P^{\nu _{12}}+q^{\nu _{12}})(3W_{4\rho \alpha _{12}\nu
_{12}}^{\rho }-18W_{3\alpha _{12}\nu _{12}})  \notag \\
&&-\frac{1}{4}P^{\nu _{1}}(P_{\alpha _{2}}-q_{\alpha _{2}})(2W_{3\rho \alpha
_{1}\nu _{1}}^{\rho }-8\Delta _{2\alpha _{1}\nu _{1}})  \notag \\
&&-\frac{1}{4}P^{\nu _{1}}(P_{\alpha _{1}}-q_{\alpha _{1}})(2W_{3\rho \alpha
_{2}\nu _{1}}^{\rho }-8\Delta _{2\alpha _{2}\nu _{1}})  \notag \\
&&-\frac{1}{8}(P^{2}+q^{2})(2W_{3\rho \alpha _{12}}^{\rho }-8\Delta
_{2\alpha _{12}})  \notag \\
&&+\frac{1}{2}(P_{\alpha _{1}}-q_{\alpha _{1}})(P_{\alpha _{2}}-q_{\alpha
_{2}})\Delta _{2\rho }^{\rho }.  \notag
\end{eqnarray}%
Identities involving $W_{4\rho \alpha _{12}\nu _{12}}^{\rho }$, $W_{3\rho
\alpha _{2}\nu _{1}}^{\rho }$ and $\Delta _{2\rho }^{\rho }$ are a valuable
way to write the results. They arise from taking the trace of $W$'s and
applying combinatorial analysis in their definition as linear expansions of
surface terms, which was performed in Section (\ref{DivTerms}), Eqs (\ref%
{trW3})-(\ref{trW4}). They are 
\begin{equation}
2W_{3\rho \mu _{12}}^{\rho }-8\Delta _{2\mu _{12}}=[2(\square _{3\rho \mu
_{12}}^{\rho }-\Delta _{2\mu _{12}})-g_{\mu _{12}}\Delta _{2\rho }^{\rho
}]+2g_{\mu _{12}}\Delta _{2\rho }^{\rho },
\end{equation}%
\begin{eqnarray}
3W_{4\rho \mu _{1234}}^{\rho }-18W_{3\mu _{1234}} &=&[3\Sigma _{4\rho \mu
_{1234}}^{\rho }-8\square _{3\mu _{1234}}-g_{(\mu _{12}}g_{\mu _{34})}\Delta
_{2\rho }^{\rho }] \\
&&+g_{(\mu _{12}}[\square _{3\rho \mu _{34})}^{\rho }-\Delta _{2\mu _{34})}-%
\frac{1}{2}g_{\mu _{34})}\Delta _{2\rho }^{\rho }]+3g_{(\mu _{12}}g_{\mu
_{34})}\Delta _{2\rho }^{\rho }.  \notag
\end{eqnarray}%
The use of these relations will become apparent in the course of the
investigation.

To get an explicit expression for terms that make versions of amplitudes
distinct, see (\ref{U2}) and (\ref{U3}), we join the results $g^{\nu _{12}}%
\mathcal{D}_{\nu _{12};\alpha _{1}}^{VV}$ and $g^{\nu _{12}}\mathcal{D}_{\nu
_{12};\alpha _{12}}^{VV}$ with finite part previously calculated, which
allow us to write: 
\begin{eqnarray}
\Upsilon _{\alpha _{1}} &=&\left( g^{\nu _{12}}T_{\nu _{12};\alpha
_{1}}^{VV}-4m^{2}J_{2\alpha _{1}}\right)  \label{Ups1} \\
&=&-\frac{1}{2}P^{\nu _{1}}[2(\square _{3\rho \alpha _{1}\nu _{1}}^{\rho
}-\Delta _{2\alpha _{1}\nu _{1}})-g_{\alpha _{1}\nu _{1}}\Delta _{2\rho
}^{\rho }]-q_{\alpha _{1}}(\Delta _{2\rho }^{\rho }+i/2\pi )  \notag
\end{eqnarray}%
\begin{eqnarray}
\Upsilon _{\alpha _{1}\alpha _{2}} &=&\left( g^{\nu _{12}}T_{\nu
_{12};\alpha _{12}}^{VV}-4m^{2}\bar{J}_{2\alpha _{12}}\right)  \label{Ups2}
\\
&=&-\frac{1}{6}\left( \theta _{\alpha _{1}\alpha _{2}}-3q_{\alpha
_{1}}q_{\alpha _{2}}\right) (\Delta _{2\rho }^{\rho }+i/2\pi )  \notag \\
&&+\frac{1}{36}(3P^{\nu _{12}}+q^{\nu _{12}})[3\Sigma _{4\rho \alpha
_{12}\nu _{12}}^{\rho }-8\square _{3\alpha _{12}\nu _{12}}-g_{(\alpha
_{12}}g_{\nu _{12})}\Delta _{2\rho }^{\rho }]  \notag \\
&&+\frac{1}{72}(3P^{\nu _{12}}+q^{\nu _{12}})g_{(\alpha _{12}}[2\square
_{3\nu _{12})\rho }^{\rho }-2\Delta _{2\nu _{12})}-g_{\nu _{12})}\Delta
_{2\rho }^{\rho }]  \notag \\
&&-\frac{1}{4}P^{\nu _{1}}\left( P_{\alpha _{2}}-q_{\alpha _{2}}\right)
[2(\square _{3\rho \alpha _{1}\nu _{1}}^{\rho }-\Delta _{2\alpha _{1}\nu
_{1}})-g_{\alpha _{1}\nu _{1}}\Delta _{2\rho }^{\rho }]  \notag \\
&&-\frac{1}{4}P^{\nu _{1}}\left( P_{\alpha _{1}}-q_{\alpha _{1}}\right)
[2(\square _{3\rho \alpha _{2}\nu _{1}}^{\rho }-\Delta _{2\alpha _{2}\nu
_{1}})-g_{\alpha _{2}\nu _{1}}\Delta _{2\rho }^{\rho }]  \notag \\
&&-\frac{1}{8}(P^{2}+q^{2})[2(\square _{3\rho \alpha _{12}}^{\rho }-\Delta
_{2\alpha _{12}})-g_{\alpha _{12}}\Delta _{2\rho }^{\rho }]  \notag \\
&&+(W_{2\rho \alpha _{12}}^{\rho }-2\Delta _{1\alpha _{12}})+2g_{\alpha
_{1}\alpha _{2}}I_{\text{\textrm{quad}}}-2m^{2}\left( \Delta _{2\alpha
_{12}}+g_{\alpha _{12}}I_{\log }\right) .  \notag
\end{eqnarray}%
For the last relation, we defined the complete two-point tensor integral%
\begin{equation*}
\bar{J}_{2\alpha _{12}}=\frac{1}{2}\left( \Delta _{2\alpha _{12}}+g_{\alpha
_{12}}I_{\log }\right) +J_{2}.
\end{equation*}

We already have all expressions that make up gravitational amplitude.
However, we also need to know how they manifest in RAGFs, a subject we will
address next. In a second step, we will analyze its consequences for
symmetries of keeping these relations preserved and whether it is possible
to determine them independently of amplitudes context.

\section{Even Amplitudes: RAGFs\label{evenragfsec}}

Now, we will explore RAGFs for even amplitudes. In Chapters (\ref{2Dim2Pt})
and (\ref{2masses}), relations served as a bridge to establish how they
operate in odd amplitudes since contractions related to vertex indices
(called \textit{internal indices}) are trivially satisfied. Beyond the
relation%
\begin{equation}
q^{\mu _{1}}t_{\mu _{12}}^{VV}=t_{\mu _{2}}^{V}\left( k_{1}\right) -t_{\mu
_{2}}^{V}\left( k_{2}\right) =t_{\left( -\right) \mu _{2}}^{V},
\label{Relmu0}
\end{equation}%
already verified in (\ref{pVV}), we need relations for amplitudes
derivative: 
\begin{eqnarray}
q^{\mu _{1}}t_{\mu _{12};\alpha _{1}}^{VV} &=&t_{\mu _{2};\alpha
_{1}}^{V}\left( k_{1}\right) -t_{\mu _{2};\alpha _{1}}^{V}\left(
k_{2}\right) =t_{\left( -\right) \mu _{2};\alpha _{1}}^{V}  \label{Relmu1} \\
q^{\mu _{1}}t_{\mu _{12};\alpha _{12}}^{VV} &=&t_{\mu _{2};\alpha
_{12}}^{V}\left( k_{1}\right) -t_{\mu _{2};\alpha _{12}}^{V}\left(
k_{2}\right) =t_{\left( -\right) \mu _{2};\alpha _{12}}^{V},  \label{Relmu2}
\end{eqnarray}%
where $t_{\left( -\right) \mu _{2};\alpha _{1}}^{V}$ and $t_{\mu _{2};\alpha
_{12}}^{V}\left( k_{1}\right) $ denotes the difference of vectorial
one-point functions.

In addition to relations for internal indices, the contractions with
derivative indices momentum (called \textit{external indices}) also produce
relations for the gravitational amplitudes. They are obtained using the
following identity inside the Dirac trace 
\begin{equation}
2q^{\alpha _{1}}K_{1\alpha _{1}}=[S^{-1}\left( K_{2}\right) \slashed{q}+\slashed%
{q}S^{-1}\left( K_{1}\right) +2m\slashed{q}-q^{2}],  \label{qK}
\end{equation}%
in two distinct positions: around the first or second vertex. For example,
we apply it in front of the first vertex and split terms in the sum as%
\begin{eqnarray}
q^{\alpha _{1}}t_{\mu _{12};\alpha _{1}}^{VV} &=&-\frac{1}{2}q^{2}t_{\mu
_{12}}^{VV}+m\mathrm{tr}\left[ \slashed{q}\gamma _{\mu _{1}}S\left( K_{1}\right)
\gamma _{\mu _{2}}S\left( K_{2}\right) \right] \\
&&+\frac{1}{2}\mathrm{tr}\left[ \slashed{q}S^{-1}\left( K_{1}\right) \gamma
_{\mu _{1}}S\left( K_{1}\right) \gamma _{\mu _{2}}S\left( K_{2}\right) %
\right]  \notag \\
&&+\frac{1}{2}\mathrm{tr}\left[ S^{-1}\left( K_{2}\right) \slashed{q}\gamma
_{\mu _{1}}S\left( K_{1}\right) \gamma _{\mu _{2}}S\left( K_{2}\right) %
\right] .  \notag
\end{eqnarray}%
Substituting in the second line of the relation above 
\begin{equation*}
S^{-1}\left( K_{1}\right) \gamma _{\mu _{1}}S\left( K_{1}\right) =2K_{1\mu
_{1}}S\left( K_{1}\right) -\gamma _{\mu _{1}}-2m\gamma _{\mu _{1}}S\left(
K_{1}\right) ,
\end{equation*}%
the mass term was canceled. Applying $\slashed{q}=[S^{-1}\left( K_{2}\right)
-S^{-1}\left( K_{1}\right) ]$ leads to the difference of the one-point
functions with derivative indices, $t_{\left( -\right) \mu _{2};\mu
_{1}}^{V} $. We use the (anti)-commutations among $\gamma _{\mu _{i}}$ and $%
\slashed{q}$ matrices, 
\begin{eqnarray}
2\gamma _{\mu _{1}}\gamma _{\mu _{2}} &=&\{\gamma _{\mu _{1}},\gamma _{\mu
_{2}}\}+[\gamma _{\mu _{1}},\gamma _{\mu _{2}}] \\
\lbrack \gamma _{\mu _{1}},\gamma _{\mu _{2}}] &=&-2\varepsilon _{\mu
_{1}\mu _{2}}\gamma _{\ast } \\
2g_{\mu _{12}}\slashed{q} &=&\slashed{q}\gamma _{\mu _{2}}\gamma _{\mu _{1}}+\gamma
_{\mu _{2}}\gamma _{\mu _{1}}\slashed{q}.
\end{eqnarray}

The systematic procedure gives back the identity given by%
\begin{equation}
\mathrm{tr}\left[ \gamma _{\mu _{2}}\slashed{q}\gamma _{\mu _{1}}S\left(
K_{1}\right) \right] -\mathrm{tr}\left[ \slashed{q}\gamma _{\mu _{1}}\gamma
_{\mu _{2}}S\left( K_{2}\right) \right] =q_{\mu _{1}}t_{\left( -\right) \mu
_{2}}^{V}+q_{\mu _{2}}t_{\left( +\right) \mu _{1}}^{V}-g_{\mu _{12}}q^{\nu
}t_{\left( +\right) \nu }^{V}
\end{equation}%
where the notation $t_{\left( +\right) \mu _{2};\alpha _{1}}^{V},$ is
associated with the sum of vectorial one-point functions%
\begin{equation}
t_{\left( +\right) \mu _{2};\alpha _{1}}^{V}=t_{\mu _{2};\alpha
_{1}}^{V}\left( k_{1}\right) +t_{\mu _{2};\alpha _{1}}^{V}\left(
k_{2}\right) ,
\end{equation}%
similarly to $t_{\left( +\right) \mu _{1}}^{V}$. The operations described
above leads the relation for $q^{\alpha }$ contraction, 
\begin{equation}
q^{\alpha _{1}}t_{\mu _{12};\alpha _{1}}^{VV}=-\frac{1}{2}q^{2}t_{\mu
_{12}}^{VV}+t_{\left( -\right) \mu _{2};\mu _{1}}^{V}+\frac{1}{2}q_{\mu
_{1}}t_{\left( -\right) \mu _{2}}^{V}+\frac{1}{2}q_{\mu _{2}}t_{\left(
+\right) \mu _{1}}^{V}-\frac{1}{2}g_{\mu _{12}}q^{\alpha _{1}}t_{\left(
+\right) \alpha _{1}}^{V}.  \label{qaVV3}
\end{equation}%
Starting with the initial identity (\ref{qK}) close to the right of vertex $%
\gamma _{\mu _{2}},$ the relation obtained is equal to the previous one
after interchanging $\mu _{1}\leftrightarrow \mu _{2}$ on the RHS. The
relation is the same for four indices amplitude, just adding a derived index
on the amplitudes.

Amplitudes with derivative indices also account for trace identities, which
will later be necessary to characterize Weyl anomalies. The result emerges
directly by using $\slashed{K}_{1}=S^{-1}\left( k_{1}\right) +m$, so relations
for the two amplitudes are given by%
\begin{eqnarray}
g^{\mu _{1}\alpha _{1}}t_{\mu _{12};\alpha _{1}}^{VV} &=&t_{\mu
_{2}}^{V}\left( k_{2}\right) +mt_{\mu _{2}}^{SV}  \label{gTVV1} \\
g^{\mu _{1}\alpha _{1}}t_{\mu _{12};\alpha _{12}}^{VV} &=&t_{\mu _{2};\alpha
_{2}}^{V}\left( k_{2}\right) +mt_{\mu _{2};\alpha _{2}}^{SV}.  \label{gTVV2}
\end{eqnarray}%
These relations are symmetric for $\mu _{1}\leftrightarrow \mu _{2}$
exchanges.

To finalize the exposure of RAGFs for even amplitudes, we extend the
procedure adopted for the $VV$s to $AA$. For the momentum contraction $%
q^{\mu _{i}}$ (internal contractions):%
\begin{eqnarray}
q^{\mu _{1}}t_{\mu _{12};\alpha _{1}}^{AA} &=&t_{\mu _{2};\alpha
_{1}}^{V}\left( k_{1}\right) -t_{\mu _{2};\alpha _{1}}^{V}\left(
k_{2}\right) -2mt_{\mu _{2};\alpha _{1}}^{PA}=t_{\left( -\right) \mu
_{2};\alpha _{1}}^{V}-2mt_{\mu _{2};\alpha _{1}}^{PA}  \label{qAA3} \\
q^{\mu _{1}}t_{\mu _{12};\alpha _{12}}^{AA} &=&t_{\mu _{2};\alpha
_{12}}^{V}\left( k_{1}\right) -t_{\mu _{2};\alpha _{12}}^{V}\left(
k_{2}\right) -2mt_{\mu _{2};\alpha _{12}}^{PA}=t_{\left( -\right) \mu
_{2};\alpha _{12}}^{V}-2mt_{\mu _{2};\alpha _{12}}^{PA}.  \label{qAA4}
\end{eqnarray}%
For the momentum contraction $q^{\alpha _{i}}$ (external contractions):%
\begin{eqnarray}
q^{\alpha _{1}}t_{\mu _{12};\alpha _{1}}^{AA} &=&-\frac{1}{2}q^{2}t_{\mu
_{12}}^{AA}+t_{\left( -\right) \mu _{2};\mu _{1}}^{V}+\frac{1}{2}q_{\mu
_{1}}t_{\left( -\right) \mu _{2}}^{V}+\frac{1}{2}q_{\mu _{2}}t_{\left(
+\right) \mu _{1}}^{V}-\frac{1}{2}g_{\mu _{12}}q^{\alpha _{1}}t_{\left(
+\right) \alpha _{1}}^{V}  \label{qaa} \\
&&+mg_{\mu _{12}}[t^{S}\left( k_{2}\right) -t^{S}\left( k_{1}\right)
]+m\varepsilon _{\mu _{12}}[t^{P}\left( k_{2}\right) +t^{P}\left(
k_{1}\right) ].  \notag
\end{eqnarray}%
The last line has two additional terms compared to (\ref{qaVV3}). These
terms do not contribute for three-index amplitudes; however, those with four
indices have $[t_{\alpha _{2}}^{S}\left( k_{2}\right) -t_{\alpha
_{2}}^{S}\left( k_{1}\right) ]\neq 0$ when integrated. And for the trace
Contractions:%
\begin{eqnarray}
g^{\mu _{1}\alpha _{1}}t_{\mu _{12};\alpha _{1}}^{AA} &=&t_{\mu
_{2}}^{V}\left( k_{2}\right) +mt_{\mu _{2}}^{PA} \\
g^{\mu _{1}\alpha _{1}}t_{\mu _{12};\alpha _{12}}^{AA} &=&t_{\mu _{2};\alpha
_{2}}^{V}\left( k_{2}\right) +mt_{\mu _{2};\alpha _{2}}^{PA}.
\end{eqnarray}

The following subsections pursue links between the finite and divergent
parts that will guide us in studying even and odd parts of Einstein and Weyl
anomalies. Some passages are detailed to explain that all mathematical
operations carried out follow rigorously.

\subsection{Internal contractions: $q^{\protect\mu }T_{\protect\mu \protect%
\nu ;\protect\sigma }^{VV}$ and $q^{\protect\mu }T_{\protect\mu \protect\nu ;%
\protect\sigma \protect\lambda }^{VV}$}

From detailed results for amplitudes, we can proceed to the verification of
RAGFs, starting with those involving vertex-index contractions (\ref{Relmu1}%
) and (\ref{Relmu2}). We expect them to remain valid to ensure the linearity
of integration operation:%
\begin{eqnarray}
q^{\mu _{1}}T_{\mu _{12};\alpha _{1}}^{VV} &=&T_{\mu _{2};\alpha
_{1}}^{V}\left( k_{1}\right) -T_{\mu _{2};\alpha _{1}}^{V}\left(
k_{2}\right) =T_{\left( -\right) \mu _{2};\alpha _{1}}^{V}  \label{pTVV3} \\
q^{\mu _{1}}T_{\mu _{12};\alpha _{12}}^{VV} &=&T_{\mu _{2};\alpha
_{12}}^{V}\left( k_{1}\right) -T_{\mu _{2};\alpha _{12}}^{V}\left(
k_{2}\right) =T_{\left( -\right) \mu _{2};\alpha _{12}}^{V}.  \label{pTVV4}
\end{eqnarray}%
As they involve differences of vector one-point functions from (\ref{T2def})
and (\ref{T3def}), we need to calculate these values $\{T_{\mu _{1};\alpha
_{1}}^{V}\left( k_{i}\right) ;\quad T_{\mu _{1};\alpha _{12}}^{V}\left(
k_{i}\right) \}.$ These amplitudes are expressed in terms of one-point
integrals $\bar{J}_{1\mu _{1}}\left( k_{i}\right) $, $\bar{J}_{1\mu
_{12}}\left( k_{i}\right) $ and $\bar{J}_{1\mu _{123}}\left( k_{i}\right) $
in Eq's (\ref{J1mu1B})-(\ref{J1(123)}) in Appendix (\ref{AppInt2D}). For the
amplitudes with the label $k_{1}$ as the reference momentum, expressions
follow directly from definitions used in the $J$-integrals, namely, 
\begin{eqnarray}
T_{\mu _{1;}\alpha _{1}}^{V}\left( k_{1}\right) &=&2\bar{J}_{1\mu _{1}\alpha
_{1}}\left( k_{1}\right) \\
T_{\mu _{2};\alpha _{12}}^{V}\left( k_{1}\right) &=&2\bar{J}_{1\mu
_{1}\alpha _{12}}\left( k_{1}\right) .
\end{eqnarray}%
The amplitudes with the label $k_{2}$ momentum require the translation $%
K_{1}\rightarrow k+k_{2}-q$, just for convenience because $J$'s functions
were defined using this convention, so%
\begin{eqnarray}
T_{\mu _{1};\alpha _{1}}^{V}\left( k_{2}\right) &=&2\bar{J}_{1\mu _{1}\alpha
_{1}}\left( k_{2}\right) -2q_{\alpha _{1}}\bar{J}_{1\mu _{1}}\left(
k_{2}\right) \\
T_{\mu _{2};\alpha _{12}}^{V}\left( k_{2}\right) &=&2\bar{J}_{1\mu
_{2}\alpha _{12}}\left( k_{2}\right) -2q_{(\alpha _{1}}\bar{J}_{1\alpha
_{2})\mu _{2}}\left( k_{2}\right) +2q_{\alpha _{12}}\bar{J}_{1\mu
_{2}}\left( k_{2}\right) .
\end{eqnarray}

Differences of one-point vectorial functions with one and two derivative
indices are%
\begin{eqnarray}
T_{\left( -\right) \mu _{2};\alpha _{1}}^{V} &=&-q^{\nu _{2}}P^{\nu
_{1}}W_{3\alpha _{1}\mu _{2}\nu _{12}}+(P_{\mu _{2}}q^{\nu _{1}}+P^{\nu
_{1}}q_{\mu _{2}})\Delta _{3\alpha _{1}\nu _{1}}  \label{TV2-TV2} \\
&&+\left( P\cdot q\right) \Delta _{3\mu _{2}\alpha _{1}}+\left( P_{\alpha
_{1}}-q_{\alpha _{1}}\right) q^{\nu _{1}}\Delta _{2\mu _{2}\nu _{1}},  \notag
\end{eqnarray}%
\begin{eqnarray}
T_{\left( -\right) \mu _{2};\alpha _{12}}^{V} &=&q^{\nu _{1}}W_{2\mu
_{2}\alpha _{12}\nu _{1}}-q_{\mu _{2}}\Delta _{1\alpha _{12}}+q_{(\alpha
_{1}}g_{\alpha _{2})\mu _{2}}I_{\text{\textrm{quad}}}  \label{TV3-TV3} \\
&&+\frac{1}{12}[P^{(\nu _{12}}q^{\nu _{3})}+q^{\nu _{123}}]W_{4\mu
_{2}\alpha _{12}\nu _{123}}  \notag \\
&&-\frac{1}{4}[2q^{\nu _{1}}P^{\nu _{2}}P_{\mu _{2}}+(P^{\nu _{12}}+q^{\nu
_{12}})q_{\mu _{2}}]W_{3\alpha _{12}\nu _{12}}  \notag \\
&&-\frac{1}{4}\left[ 2\left( P\cdot q\right) P^{\nu _{1}}+q^{\nu
_{1}}(P^{2}+q^{2})\right] W_{3\mu _{2}\alpha _{12}\nu _{1}}  \notag \\
&&-\frac{1}{2}q^{\nu _{2}}P^{\nu _{1}}(P-q)_{(\alpha _{1}}W_{3\alpha
_{2})\mu _{2}\nu _{12}}]  \notag \\
&&+\frac{1}{4}\left[ 2(P\cdot q)P_{\mu _{2}}+(P^{2}+q^{2})q_{\mu _{2}}\right]
\Delta _{2\alpha _{12}}+\frac{1}{2}(P\cdot q)(P-q)_{(\alpha _{1}}\Delta _{2%
\text{ }\alpha _{2})\mu _{2}}  \notag \\
&&+\frac{1}{2}(P_{\mu }q^{\nu _{1}}+q_{\mu }P^{\nu _{1}})(P-q)_{(\alpha
_{1}}\Delta _{2\text{ }\alpha _{2})\nu _{1}}+\frac{1}{2}(P-q)_{\alpha
_{1}}(P-q)_{\alpha _{2}}q^{\nu _{1}}\Delta _{2\mu _{2}\nu _{1}}.  \notag
\end{eqnarray}

We will begin verifying relations obtained for even amplitudes. From the
relation for 2nd-order VV amplitude in (\ref{pVV}), the verification for
derivative amplitudes follows the same procedure. We have for the 3rd and
4th-order $VV$ amplitude%
\begin{equation}
q^{\mu _{1}}T_{\mu _{12};\alpha _{1}}^{VV}=2(2q^{\mu _{1}}J_{2\mu
_{12}\alpha _{1}}+q^{2}J_{2\mu _{2}\alpha _{1}})+q_{\mu _{2}}(2q^{\mu
_{1}}J_{2\mu _{1}\alpha _{1}}+q^{2}J_{2\alpha _{1}})+q^{\mu _{1}}\mathcal{D}%
_{\mu _{12};\alpha _{1}}^{VV}
\end{equation}%
\begin{equation}
q^{\mu _{1}}T_{\mu _{12};\alpha _{12}}^{VV}=2(2q^{\mu _{1}}J_{2\mu
_{12}\alpha _{12}}+q^{2}J_{2\mu _{2}\alpha _{12}})+q_{\mu _{2}}(2q^{\mu
_{1}}J_{2\mu _{1}\alpha _{12}}+q^{2}J_{2\alpha _{12}})+q^{\mu _{1}}\mathcal{D%
}_{\mu _{12};\alpha _{12}}^{VV}.
\end{equation}%
from Section (\ref{BasisFI}), which are of the same type used in
establishing constraints over odd amplitudes (by example $2q^{\mu
_{1}}J_{2\mu _{12}\alpha _{1}}=-q^{2}J_{2\mu _{2}\alpha _{1}})$, we have
that finite part vanishes and divergent factors $q^{\mu _{1}}\mathcal{D}%
_{\mu _{12};\alpha _{1}}^{VV}$ and $q^{\mu _{1}}\mathcal{D}_{\mu
_{12};\alpha _{12}}^{VV}$ satisfy identically 
\begin{eqnarray}
q^{\mu _{1}}\mathcal{D}_{\mu _{12};\alpha _{1}}^{VV} &=&T_{\left( -\right)
\mu _{2};\alpha _{1}}^{V} \\
q^{\mu _{1}}\mathcal{D}_{\mu _{12};\alpha _{12}}^{VV} &=&T_{\left( -\right)
\mu _{2};\alpha _{12}}^{V}.
\end{eqnarray}

Due to the definitions of tensors $W_{4\mu _{123456}}$ and $W_{3\mu _{1234}}$%
, see Section (\ref{DivTerms}) there are hundreds of surface terms in the
last relation. Although it seems complicated to verify such equality, its
satisfaction follows from the observation that each of the lines that we
arrange for tensor $\mathcal{D}_{\mu _{12};\alpha _{12}}^{VV}$ in (\ref%
{DivVV4}) will correspond to one of the lines expressed by difference $%
T_{\left( -\right) \mu _{2};\alpha _{12}}^{V}$ in (\ref{TV3-TV3}), when
contracting with momentum. We facilitate these identifications by
classifying surface terms, following criteria regarding the divergence
degree, tensor rank, and contraction type. For example, it is necessary to
note that index $\mu _{1}$ becomes a contracted index, $3q^{\mu _{1}}P^{\nu
_{12}}W_{4\mu _{12}\alpha _{12}\nu _{12}}=P^{(\nu _{12}}q^{\nu _{3})}W_{4\mu
_{2}\alpha _{12}\nu _{123}}.$ As the tensor $W_{4\mu _{123456}}$ is fully
symmetric, terms are identical, and so on for all others. Expanding $W$
combinations in primary surface terms is not necessary. In this way,
relations for amplitudes at the trace level incorporate integration
linearity established in (\ref{pTVV4}) and are satisfied without restriction
on the divergent parts of expressions.

\subsection{External Contractions: $q^{\protect\sigma }T_{\protect\mu 
\protect\nu ;\protect\sigma }^{VV}$ and $q^{\protect\sigma }T_{\protect\mu 
\protect\nu ;\protect\sigma \protect\lambda }^{VV}$}

We have one more momentum contraction to check regarding amplitudes $T_{\mu
_{12};\alpha _{1}}^{VV}$ and $T_{\mu _{12};\alpha _{12}}^{VV}$: they are $%
q^{\alpha _{1}}T_{\mu _{12};\alpha _{1}}^{VV}$ and $q^{\alpha _{1}}T_{\mu
_{12};\alpha _{12}}^{VV}$ from (\ref{qaVV3}). It can be made by contracting
the amplitude and identifying the function of the RHS. Nevertheless, we
proceed through an alternative route, using manipulations to reorganize
integrands of amplitudes. Effectively these indices exchange from Dirac
matrices $\mu _{i}$ with indices from derivative factors $\alpha _{i}$. In
this way, if previously verified relations (\ref{pTVV4}) are satisfied, they
will also be satisfied since they come from Dirac traces. We will detail
calculations for relations involving amplitude with a derivative index. At
the end of the operations, we expect to obtain (\ref{qaVV3}) integrated. We
will extend this result to amplitude with two derivatives, drawing attention
to their differences. These two amplitudes with exchange indexes will be the
basis for calculating the relations for the other even and odd amplitudes.

From definition for the amplitude $t_{\mu _{12}}^{VV}$ and $t_{\mu
_{12};\alpha _{1}}^{VV},$ see Eqs (\ref{t(s)}) and (\ref{t(s2)}), we obtain 
\begin{eqnarray}
2t_{\mu _{12}}^{\left( +\right) } &=&t_{\mu _{12}}^{VV}-g_{\mu _{12}}t^{PP}
\label{t+TVV-gTPP0} \\
2t_{\mu _{12};\alpha _{1}}^{\left( +\right) } &=&t_{\mu _{12};\alpha
_{1}}^{VV}-g_{\mu _{12}}t_{\alpha _{1}}^{PP}.  \label{t+TVV-gTPP1}
\end{eqnarray}%
It can be noted that the role of indexes position in the second tensor is%
\begin{eqnarray}
2t_{\mu _{12};\alpha _{1}}^{\left( +\right) } &=&2K_{1\alpha _{1}}(K_{1\mu
_{1}}K_{2\mu _{2}}+K_{2\mu _{1}}K_{1\mu _{2}})\frac{1}{D_{12}} \\
2t_{\alpha _{1}\mu _{2};\mu _{1}}^{\left( +\right) } &=&2K_{1\mu _{1}}\left(
K_{1\alpha _{1}}K_{2\mu _{2}}+K_{2\alpha _{1}}K_{1\mu _{2}}\right) \frac{1}{%
D_{12}},
\end{eqnarray}%
where the outside term in parentheses comes from derivative contribution.
Manipulating the expression for $t_{\mu _{12};\alpha _{1}}^{\left( +\right)
} $ using $K_{2}=K_{1}+q$ relate both tensors, changing the role of indices $%
\mu _{1}\leftrightarrow \alpha _{1}$. We use the notation to represent the
antisymmetry of indices $[$ $\ ]$:%
\begin{equation}
t_{\mu _{12};\alpha _{1}}^{\left( +\right) }=t_{\alpha _{1}\mu _{2};\mu
_{1}}^{\left( +\right) }-q_{[\alpha _{1}}K_{1\mu _{1}]}K_{1\mu _{2}}\frac{1}{%
D_{12}}.  \label{T+}
\end{equation}%
The tensors $t_{\mu _{12};\alpha _{1}}^{\left( +\right) }$ and $t_{\alpha
_{1}\mu _{2};\mu _{1}}^{\left( +\right) }$ differ by an additional tensor
from translation of $K_{1}$ momentum. Expressing $t^{\left( +\right) }$
parts in terms of $VV$ and $PP$ amplitudes leads to%
\begin{equation}
t_{\mu _{12};\alpha _{1}}^{VV}=t_{\alpha _{1}\mu _{2};\mu _{1}}^{VV}+g_{\mu
_{12}}t_{\alpha _{1}}^{PP}-g_{\alpha _{1}\mu _{2}}t_{\mu
_{1}}^{PP}-2q_{[\alpha _{1}}K_{1\mu _{1}]}K_{1\mu _{2}}\frac{1}{D_{12}}.
\end{equation}%
Furthermore, the last term with this equation also has a form in terms of $%
VV $-amplitude%
\begin{equation}
2q_{[\alpha _{1}}K_{1\mu _{1}]}K_{1\mu _{2}}\frac{1}{D_{12}}=\frac{1}{2}%
(q_{\alpha _{1}}t_{\mu _{12}}^{VV}-q_{\mu _{1}}t_{\alpha _{1}\mu _{2}}^{VV})-%
\frac{1}{2}q_{[\alpha _{1}}g_{\mu _{1}]\mu _{2}}t^{PP}-q_{\mu
_{2}}q_{[\alpha _{1}}K_{1\mu _{1}]}\frac{1}{D_{12}}.
\end{equation}%
Starting from the definition of amplitude $T_{\alpha _{1}}^{PP}$ (\ref{pp1})
and using $K_{1}=K_{2}-q$, the sum of one-point vector functions appears
straightforwardly%
\begin{equation}
t_{\alpha _{1}}^{PP}=q^{2}\frac{K_{1\alpha _{1}}}{D_{12}}+q_{\alpha _{1}}%
\frac{1}{D_{2}}-\frac{1}{2}[t_{\alpha _{1}}^{V}\left( k_{1}\right)
+t_{\alpha _{1}}^{V}\left( k_{2}\right) ].  \label{tppa1}
\end{equation}

These observations, we obtain an identity representing the exchanging of
indices that facilitate the study of this relation coming from contractions
involving derivatives indices%
\begin{equation}
t_{\mu _{12};\alpha _{1}}^{VV}=-\frac{1}{2}q_{\alpha _{1}}t_{\mu
_{12}}^{VV}+t_{\alpha _{1}\mu _{2};\mu _{1}}^{VV}+\frac{1}{2}q_{\mu
_{1}}t_{\alpha _{1}\mu _{2}}^{VV}-\frac{1}{2}g_{\mu _{12}}t_{\left( +\right)
\alpha _{1}}^{V}+\frac{1}{2}g_{\alpha _{1}\mu _{2}}t_{\left( +\right) \mu
_{1}}^{V}+r_{\mu _{12};\alpha _{1}}.  \label{IDtroca1}
\end{equation}%
Here, $r_{\mu _{12};\alpha _{1}}$ is a residual term anissymetric in $\mu
_{1}$ and $\alpha _{1}$ 
\begin{equation}
r_{\mu _{12};\alpha _{1}}=\frac{1}{2}q_{[\alpha _{1}}g_{\mu _{1}]\mu
_{2}}\left( q^{2}\frac{1}{D_{12}}+\frac{1}{D_{2}}-\frac{1}{D_{1}}\right)
-\theta _{\mu _{2}[\alpha _{1}}K_{1\mu _{1}]}\frac{1}{D_{12}},  \label{R1}
\end{equation}%
whose integration yields 
\begin{eqnarray}
R_{\mu _{12};\alpha _{1}} &=&\frac{1}{2}q_{[\alpha _{1}}g_{\mu _{1}]\mu
_{2}}q^{2}J_{2}-q^{2}g_{\mu _{2}[\alpha _{1}}J_{2\mu _{1}]}+q_{\mu
_{2}}q_{[\alpha _{1}}J_{2\mu _{1}]} \\
&&+\frac{1}{2}q_{[\alpha _{1}}g_{\mu _{1}]\mu _{2}}q^{2}[\bar{J}_{1}(k_{2})-%
\bar{J}_{1}(k_{1})].  \notag
\end{eqnarray}%
After substitutions of $J_{2\mu },$ $J_{2}$ and $\bar{J}_{1}\left(
k_{i}\right) ,$ this term is null, $R_{\mu _{12};\alpha _{1}}=0.$ It is
essential to mention that we carry out passive operations. Rearranging
amplitude terms does not represent any operations performed on the original
amplitude. The full expression is 
\begin{equation}
T_{\mu _{12};\alpha _{1}}^{VV}=-\frac{1}{2}q_{\alpha _{1}}T_{\mu
_{12}}^{VV}+T_{\alpha _{1}\mu _{2};\mu _{1}}^{VV}+\frac{1}{2}q_{\mu
_{1}}T_{\alpha _{1}\mu _{2}}^{VV}-\frac{1}{2}g_{\mu _{12}}T_{\left( +\right)
\alpha _{1}}^{V}+\frac{1}{2}g_{\alpha _{1}\mu _{2}}T_{\left( +\right) \mu
_{1}}^{V}.
\end{equation}%
Let us analyze $q^{\alpha _{1}}$ contractions. We have already verified that
RAGF is satisfied with matrix indices. Thus, we have automatic satisfaction
of contractions with $\alpha _{1}$ index 
\begin{equation}
q^{\alpha _{1}}T_{\mu _{12};\alpha _{1}}^{VV}=-\frac{1}{2}q^{2}T_{\mu
_{12}}^{VV}+T_{\left( -\right) \mu _{2};\mu _{1}}^{V}+\frac{1}{2}q_{\mu
_{1}}T_{\left( -\right) \mu _{2}}^{V}+\frac{1}{2}q_{\mu _{2}}T_{\left(
+\right) \mu _{1}}^{V}-\frac{1}{2}g_{\mu _{12}}q^{\alpha _{1}}T_{\left(
+\right) \alpha _{1}}^{V}.  \label{IDtroca11}
\end{equation}

Adding one more factor $K_{1\alpha _{2}}$ in (\ref{IDtroca1}), the structure
is the same as the previous one, 
\begin{eqnarray}
t_{\mu _{12};\alpha _{12}}^{VV} &=&-\frac{1}{2}q_{\alpha _{1}}t_{\mu
_{12};\alpha _{2}}^{VV}+t_{\alpha _{1}\mu _{2};\mu _{1}\alpha _{2}}^{VV}+%
\frac{1}{2}q_{\mu _{1}}t_{\alpha _{1}\mu _{2};\alpha _{2}}^{VV}
\label{IDtroca2} \\
&&-\frac{1}{2}g_{\mu _{12}}t_{\left( +\right) \alpha _{1};\alpha _{2}}^{V}+%
\frac{1}{2}g_{\alpha _{1}\mu _{2}}t_{\left( +\right) \mu _{1};\alpha
_{2}}^{V}+r_{\mu _{12};\alpha _{12}}.  \notag
\end{eqnarray}
However, we need to analyze the effect on the term $r_{\mu _{12};\alpha
_{12}}=K_{1\alpha _{2}}r_{\mu _{12};\alpha _{1}}$ from (\ref{R1}), 
\begin{equation}
r_{\mu _{12};\alpha _{12}}=\frac{1}{2}q_{[\alpha _{1}}g_{\mu _{1}]\mu
_{2}}K_{1\alpha _{2}}\left( q^{2}\frac{1}{D_{12}}+\frac{1}{D_{2}}-\frac{1}{%
D_{1}}\right) -\theta _{\mu _{2}[\alpha _{1}}K_{1\mu _{1}]}\frac{K_{1\alpha
_{2}}}{D_{12}}.
\end{equation}%
After being integrated, the residual terms can be organized as%
\begin{eqnarray}
R_{\mu _{12};\alpha _{12}} &=&\frac{1}{2}q_{[\alpha _{1}}g_{\mu _{1}]\mu
_{2}}\left[ q^{2}J_{2\alpha _{2}}-J_{1\alpha _{2}}\left( k_{1}\right)
+J_{1\alpha _{2}}\left( k_{2}\right) -q_{\alpha _{2}}J_{1}\left(
k_{2}\right) \right] -\theta _{\mu _{2}[\alpha _{1}}\bar{J}_{2\mu
_{1}]\alpha _{2}}  \notag \\
&=&\frac{1}{2}q_{[\alpha _{1}}g_{\mu _{1}]\mu _{2}}(q^{2}J_{2\alpha
_{2}}-q^{\nu _{1}}\Delta _{2\alpha _{2}\nu _{1}}-q_{\alpha _{2}}I_{\log
})-\theta _{\mu _{2}[\alpha _{1}}\bar{J}_{2\mu _{1}]\alpha _{2}}.  \label{R4}
\end{eqnarray}%
This term does not cancel itself when integrated; nonetheless, its
contractions do not contribute to the relations%
\begin{equation}
2q^{\alpha _{1}}R_{\mu _{12};\alpha _{12}}=\theta _{\mu _{12}}\left(
2q^{\alpha _{1}}\bar{J}_{2\alpha _{12}}+q^{2}J_{2\alpha _{2}}-q^{\nu
_{1}}\Delta _{2\alpha _{2}\nu _{1}}-q_{\alpha _{2}}I_{\log }\right) =0.
\label{qR}
\end{equation}%
Furthermore, we find the same outcome for the trace%
\begin{equation}
g^{\mu _{1}\alpha _{1}}R_{\mu _{12};\alpha _{12}}=\theta _{\mu _{2}}^{\alpha
_{1}}\bar{J}_{2\alpha _{1}\alpha _{2}}-\theta _{\mu _{2}}^{\alpha _{1}}\bar{J%
}_{2\alpha _{1}\alpha _{2}}=0.
\end{equation}%
Thus, it will not contribute to any of the contractions that remain to be
verified.%
\begin{eqnarray}
q^{\mu _{1}}R_{\mu _{12};\alpha _{12}} &=&0\quad q^{\mu _{2}}R_{\mu
_{12};\alpha _{12}}=0\text{\quad }q^{\alpha _{1}}R_{\mu _{12};\alpha _{12}}=0
\\
g^{\mu _{2}\alpha _{2}}R_{\mu _{12};\alpha _{12}} &=&0\quad g^{\mu
_{1}\alpha _{1}}R_{\mu _{12};\alpha _{12}}=0\text{\quad }g^{\mu _{1}\mu
_{2}}R_{\mu _{12};\alpha _{12}}=0.
\end{eqnarray}

The complete expression is given by%
\begin{eqnarray}
T_{\mu _{12};\alpha _{12}}^{VV} &=&-\frac{1}{2}q_{\alpha _{1}}T_{\mu
_{12};\alpha _{2}}^{VV}+T_{\alpha _{1}\mu _{2};\mu _{1}\alpha _{2}}^{VV}+%
\frac{1}{2}q_{\mu _{1}}T_{\alpha _{1}\mu _{2};\alpha _{2}}^{VV}
\label{IDtroca22} \\
&&-\frac{1}{2}g_{\mu _{12}}T_{\left( +\right) \alpha _{1};\alpha _{2}}^{V}+%
\frac{1}{2}g_{\alpha _{1}\mu _{2}}T_{\left( +\right) \mu _{1};\alpha
_{2}}^{V}+R_{\mu _{12};\alpha _{12}}.  \notag
\end{eqnarray}%
Divergent parts are not restricted to any values. Contracting the equation
and using (\ref{qR}), we have the relation (\ref{qaVV3}) satisfied for this
amplitude.

\subsection{Metric Contractions: $g^{\protect\mu \protect\sigma }T_{\protect%
\mu \protect\nu ;\protect\sigma }^{VV}$ and $g^{\protect\mu \protect\sigma %
}T_{\protect\mu \protect\nu ;\protect\sigma \protect\lambda }^{VV}$\label%
{metriceven}}

Relations from metric contraction (\ref{gTVV1}) and (\ref{gTVV2}) can be
rewritten as%
\begin{eqnarray}
g^{\mu _{1}\alpha _{1}}t_{\mu _{1}\mu _{2};\alpha _{1}}^{VV}-t_{\mu
_{2}}^{V}\left( k_{2}\right) &=&mt_{\mu _{2}}^{SV}  \label{RelMetVV1} \\
g^{\mu _{1}\alpha _{1}}t_{\mu _{1}\mu _{2};\alpha _{12}}^{VV}-t_{\mu
_{2};\alpha _{2}}^{V}\left( k_{2}\right) &=&mt_{\mu _{2};\alpha _{2}}^{SV}.
\label{RelMetVV2}
\end{eqnarray}%
They can be reformulated based on what was discussed for contractions
involving derivative indices---in this case, exchanging $\mu
_{1}\leftrightarrow \alpha _{1}$ to get the relation. That is also valid for
the permutation $\mu _{2}\leftrightarrow \alpha _{1}$ since two matrix
indices $\mu $'s are symmetric, and for the second expression, the same is
valid for indexes $\alpha $'s. We have to the integrated (\ref{IDtroca11})%
\begin{equation}
2g^{\mu _{1}\alpha _{1}}T_{\mu _{12};\alpha _{1}}^{VV}=g^{\nu _{12}}[2T_{\nu
_{12};\mu _{2}}^{VV}+q_{\mu _{2}}T_{\nu _{12}}^{VV}]+[T_{\left( +\right) \mu
_{2}}^{V}-q^{\mu _{1}}T_{\mu _{12}}^{VV}].
\end{equation}%
The argument follows the previous case: if the relation (\ref{pVV}) is
valid, then%
\begin{equation}
2[g^{\mu _{1}\alpha _{1}}T_{\mu _{12};\alpha _{1}}^{VV}-T_{\mu
_{2}}^{V}\left( k_{2}\right) ]=g^{\nu _{12}}[2T_{\nu _{12};\mu
_{2}}^{VV}+q_{\mu _{2}}T_{\nu _{12}}^{VV}].
\end{equation}%
Moreover, the contraction of (\ref{IDtroca22}) is condionated by
satisfaction of (\ref{pTVV3}), therefore 
\begin{equation}
2[g^{\mu _{1}\alpha _{1}}T_{\mu _{12};\alpha _{12}}^{VV}-T_{\mu _{2};\alpha
_{2}}^{V}\left( k_{2}\right) ]=g^{\nu _{12}}[2T_{\nu _{12};\mu _{2}\alpha
_{2}}^{VV}+q_{\mu _{2}}T_{\nu _{12};\alpha _{2}}^{VV}].
\end{equation}

If we compare these expressions with the integrated ones (\ref{RelMetVV1})
and (\ref{RelMetVV2}), showing their equivalence is doable. In this way, the
RHS can be written as 
\begin{eqnarray}
g^{\nu _{12}}[2T_{\nu _{12};\mu _{2}}^{VV}+q_{\mu _{2}}T_{\nu _{12}}^{VV}]
&=&2mT_{\mu _{2}}^{SV}  \label{trace=SV1} \\
g^{\nu _{12}}[2T_{\nu _{12};\mu _{2}\alpha _{2}}^{VV}+q_{\mu _{2}}T_{\nu
_{12};\alpha _{2}}^{VV}] &=&2mT_{\mu _{2};\alpha _{2}}^{SV}.
\label{trace=SV2}
\end{eqnarray}%
We need traces of the $VV$ to verify if these relations are satisfied since
divergent terms will be contained in traces of $\mathcal{D}^{VV}$-parts.
Nonetheless, there is a path using exclusively $\Upsilon _{\mu _{2}}$ and $%
\Upsilon _{\alpha _{2}\mu _{2}}$ that emerged in constraint of equivalence
among odd amplitudes. Explicit forms of $SV$-amplitudes regard $J_{2}$'s
integrals; therefore, let us write the results%
\begin{equation}
T_{\mu _{2}}^{SV}=2m(2J_{2\mu _{2}}+q_{\mu _{2}}J_{2})=0.  \label{SV}
\end{equation}%
Even if it is identically zero due to relations among finite integrals of
equal masses, we will use its terms separately in the sequel. The other%
\begin{equation}
T_{\mu _{2};\alpha _{2}}^{SV}=2m(2\bar{J}_{2\mu _{2}\alpha _{2}}+q_{\mu
_{2}}J_{2\alpha _{2}})=2m(\Delta _{2\mu _{2}\alpha _{2}}+g_{\mu _{2}\alpha
_{2}}I_{\log })+2m(2J_{2\mu _{2}\alpha _{2}}+q_{\mu _{2}}J_{2\alpha _{2}}).
\end{equation}

Beginning with Eq. (\ref{trace=SV1}), we write%
\begin{equation}
2(g^{\nu _{12}}T_{\nu _{12};\mu _{2}}^{VV}-4m^{2}J_{2\mu _{2}})+q_{\mu
_{2}}(g^{\nu _{12}}T_{\nu _{12}}^{VV}-4m^{2}J_{2})=0.
\end{equation}%
It is a matter of recognizing $\Upsilon $-factors; consult their explicit
expressions in Eqs. (\ref{Ups0}) and (\ref{Ups1}) to write $2\Upsilon _{\mu
_{2}}+q_{\mu _{2}}\Upsilon =0.$ That is a condition for compliance with RAGF
derived through the metric contraction. Extending this construction to Eq. (%
\ref{trace=SV2}),%
\begin{equation}
2(g^{\nu _{12}}T_{\nu _{12};\mu _{2}\alpha _{2}}^{VV}-4m^{2}\bar{J}_{2\mu
_{2}\alpha _{2}})+q_{\mu _{2}}(g^{\nu _{12}}T_{\nu _{12};\alpha
_{2}}^{VV}-4m^{2}J_{2\alpha _{2}})=0.
\end{equation}%
That means $2\Upsilon _{\mu _{2}\alpha _{2}}+q_{\mu _{2}}\Upsilon _{\alpha
_{2}}=0$ due to the definition already given, see (\ref{Ups2}) for the
explicit expression of $\Upsilon _{\mu _{2}\alpha _{2}}$. Hence, metric
RAGFs are not automatically satisfied also for even amplitudes. Owing
derivations until this point, we can lay down the equations:%
\begin{eqnarray}
g^{\nu _{12}}[2T_{\nu _{12};\mu _{2}}^{VV}+q_{\mu _{2}}T_{\nu _{12}}^{VV}]
&=&2mT_{\mu _{2}}^{SV}+2\Upsilon _{\mu _{2}}+q_{\mu _{2}}\Upsilon \\
g^{\nu _{12}}[2T_{\nu _{12};\mu _{2}\alpha _{2}}^{VV}+q_{\mu _{2}}T_{\nu
_{12};\alpha _{2}}^{VV}] &=&2mT_{\mu _{2};\alpha _{2}}^{SV}+2\Upsilon _{\mu
_{2}\alpha _{2}}+q_{\mu _{2}}\Upsilon _{\alpha _{2}}.
\end{eqnarray}%
Alternatively, we can express them in the way it was derived%
\begin{eqnarray}
2g^{\mu _{1}\alpha _{1}}T_{\mu _{12};\alpha _{1}}^{VV} &=&2T_{\mu
_{2}}^{V}\left( k_{2}\right) +2mT_{\mu _{2}}^{SV}+\left( 2\Upsilon _{\mu
_{2}}+q_{\mu _{2}}\Upsilon \right) .  \label{traceVV1} \\
2g^{\mu _{1}\alpha _{1}}T_{\mu _{12};\alpha _{12}}^{VV} &=&2T_{\mu
_{2};\alpha _{2}}^{V}\left( k_{2}\right) +2mT_{\mu _{2};\alpha
_{2}}^{SV}+(2\Upsilon _{\mu _{2}\alpha _{2}}+q_{\mu _{2}}\Upsilon _{\alpha
_{2}}).  \label{traceVV2}
\end{eqnarray}

The vanishing of individual violating terms $\Upsilon $ is enough to satisfy
these relations. This constraint preserves all RAGFs in all amplitudes;
however, in (\ref{traceVV1}), combinations of violating terms can be made
zero without canceling each term. That is the only place this happens; they
always arise individually in other relations. Two-index combination requires
that terms cancel independently. Calling for the full results (\ref{Ups0})
and (\ref{Ups1}), it is clear that violating terms in three-indices relation%
\begin{equation}
2\Upsilon _{\mu _{2}}+q_{\mu _{2}}\Upsilon =-P^{\nu _{1}}[2(\square _{3\rho
\mu _{2}\nu _{1}}^{\rho }-\Delta _{2\mu _{2}\nu _{1}})-g_{\mu _{2}\nu
_{1}}\Delta _{2\rho }^{\rho }].
\end{equation}%
It can be restricted to zero without each component being zero
independently. As a last comment, violating factors come from suitably
complex functions of momenta, physical $q$, or ambiguous $P$. Nonetheless,
they are local polynomials in these variables, which can be asserted from
their expressions. The remaining appears in (\ref{Ups2}).

Discussing if violating terms are null and the consequences of this property
is a crucial point of this investigation and what perspective we can
establish from conditions for RAGF satisfaction in odd amplitudes context.

\subsection{Internal Contractions: $q^{\protect\mu }T_{\protect\mu \protect%
\nu ;\protect\sigma }^{AA}$ and $q^{\protect\mu }T_{\protect\mu \protect\nu ;%
\protect\sigma \protect\lambda }^{AA}$}

We must analyze RAGF for two-point amplitudes with two axial vertexes to
complete relations for even amplitudes; see (\ref{qAA3}) and (\ref{qAA4}).
These relations differ from those associated with vector amplitudes by an
additional term given by $PA$-amplitudes,%
\begin{eqnarray}
T_{\mu _{2}}^{PA} &=&2mq_{\mu _{2}}J_{2}  \label{PA1} \\
T_{\mu _{2};\alpha _{1}}^{PA} &=&2mq_{\mu _{2}}J_{2\alpha _{1}}  \label{PA2}
\\
T_{\mu _{2};\alpha _{12}}^{PA} &=&2mq_{\mu _{2}}\bar{J}_{2\alpha _{12}}.
\label{PA3}
\end{eqnarray}%
As they exactly match the additional terms through connection with the $VV$%
-amplitudes, we have when contracting the expressions (\ref{AA-VV})%
\begin{eqnarray}
q^{\mu _{1}}T_{\mu _{12}}^{AA} &=&q^{\mu _{1}}T_{\mu _{12}}^{VV}-2mT_{\mu
_{2}}^{PA}=T_{\left( -\right) \mu _{2}}^{V}-2mT_{\mu _{2}}^{PA}
\label{pMUTAA0} \\
q^{\mu _{1}}T_{\mu _{12};\alpha _{1}}^{AA} &=&q^{\mu _{1}}T_{\mu
_{12};\alpha _{1}}^{VV}-2mT_{\mu _{2};\alpha _{1}}^{PA}=T_{\left( -\right)
\mu _{2};\alpha _{1}}^{V}-2mT_{\mu _{2};\alpha _{1}}^{PA}  \label{pMUTAA1} \\
q^{\mu _{1}}T_{\mu _{12};\alpha _{12}}^{AA} &=&q^{\mu _{1}}T_{\mu
_{12};\alpha _{12}}^{VV}-2mT_{\mu _{2};\alpha _{1}\alpha
_{2}}^{PA}=T_{\left( -\right) \mu _{2};\alpha _{12}}^{V}-2mT_{\mu
_{2};\alpha _{1}\alpha _{2}}^{PA}.  \label{pMUTAA2}
\end{eqnarray}%
We have unconditional RAGF, the satisfaction established for $VV$%
-amplitudes, followed by the satisfaction of these for $AA$-amplitudes.

\subsection{External Contractions: $q^{\protect\sigma }T_{\protect\mu 
\protect\nu ;\protect\sigma }^{AA}$ and $q^{\protect\sigma }T_{\protect\mu 
\protect\nu ;\protect\sigma \protect\lambda }^{AA}$}

To extend the results obtained in (\ref{IDtroca1}) and (\ref{IDtroca2}) for $%
AA$-amplitudes, use the relation connecting even amplitudes (\ref{AA-VV}), (%
\ref{AA-VV1}), and (\ref{AA-VV2}),%
\begin{eqnarray}
T_{\mu _{12};\alpha _{1}}^{AA} &=&-\frac{1}{2}q_{\alpha _{1}}T_{\mu
_{12}}^{AA}+T_{\alpha _{1}\mu _{1};\mu _{2}}^{AA}+\frac{1}{2}q_{\mu
_{2}}T_{\alpha _{1}\mu _{1}}^{AA}+\frac{1}{2}g_{\alpha _{1}\mu
_{1}}T_{\left( +\right) \mu _{2}}^{V}-\frac{1}{2}g_{\mu _{12}}T_{\left(
+\right) \alpha _{1}}^{V}  \label{AA1exp} \\
&&+2m^{2}\left[ g_{\mu _{1}\alpha _{1}}\left( 2J_{2\mu _{2}}+q_{\mu
_{2}}J_{2}\right) -g_{\mu _{12}}\left( 2J_{2\alpha _{1}}+q_{\alpha
_{1}}J_{2}\right) \right] .  \notag
\end{eqnarray}%
The combination $2J_{2\mu _{2}}+q_{\mu _{2}}J_{2}=0$ cancels out the last
two terms. Using (\ref{pMUTAA0}) and (\ref{pMUTAA1}) allows us to show that
the relation with indices $\alpha _{i}$ are also automatically satisfied:%
\begin{eqnarray}
q^{\alpha _{1}}T_{\mu _{12};\alpha _{1}}^{AA} &=&-\frac{1}{2}q^{2}T_{\mu
_{12}}^{AA}+T_{\left( -\right) \mu _{2};\mu _{1}}^{V}+\frac{1}{2}q_{\mu
_{1}}T_{\left( -\right) \mu _{2}}^{V}+\frac{1}{2}q_{\mu _{2}}T_{\left(
+\right) \mu _{1}}^{V}  \label{pa1AA3} \\
&&-\frac{1}{2}g_{\mu _{12}}q^{\nu _{1}}T_{\left( +\right) \nu
_{1}}^{V}+mg_{\mu _{12}}[T^{S}\left( k_{2}\right) -T^{S}\left( k_{1}\right)
].  \notag
\end{eqnarray}%
We have two additional terms corresponding to $PA$-amplitudes from RAGF with 
$q^{\mu _{1}}$ contraction. Using (\ref{PA1}) and (\ref{PA2}) is easy to see
which combination 
\begin{equation*}
2T_{\mu _{1};\mu _{2}}^{PA}+q_{\mu _{2}}T_{\mu _{1}}^{PA}=2mq_{\mu
_{2}}\left( 2J_{2\mu _{1}}+q_{\mu _{1}}J_{2}\right) =0.
\end{equation*}%
We have the cancelation $T^{P}\left( k_{i}\right) =0$, and the difference
between one-point functions also vanishes $T^{S}\left( k_{2}\right)
-T^{S}\left( k_{1}\right) =0,$ satisfying relation (\ref{qaa}).

For the expression with four indices, we have%
\begin{eqnarray}
T_{\mu _{12};\alpha _{12}}^{AA} &=&-\frac{1}{2}q_{\alpha _{1}}T_{\mu
_{12};\alpha _{2}}^{AA}+T_{\alpha _{1}\mu _{2};\mu _{1}\alpha _{2}}^{AA}+%
\frac{1}{2}q_{\mu _{1}}T_{\alpha _{1}\mu _{2};\alpha _{2}}^{AA}
\label{AAexc} \\
&&+\frac{1}{2}g_{\alpha _{1}\mu _{2}}T_{\left( +\right) \mu _{1};\alpha
_{2}}^{V}-\frac{1}{2}g_{\mu _{12}}T_{\left( +\right) \alpha _{1};\alpha
_{2}}^{V}+R_{\mu _{21};\alpha _{21}}  \notag \\
&&+2m^{2}\left[ g_{\mu _{2}\alpha _{1}}\left( 2\bar{J}_{2\mu _{1}\alpha
_{2}}+q_{\mu _{1}}J_{2\alpha _{2}}\right) -g_{\mu _{12}}\left( 2\bar{J}%
_{2\alpha _{12}}+q_{\alpha _{1}}J_{2\alpha _{2}}\right) \right] ,  \notag
\end{eqnarray}%
where $R_{\mu _{21};\alpha _{21}}$ is defined in (\ref{R4}). Eq. (\ref{qR})
shows that contracting the form above with q produces a null result.
Considering the relations (\ref{pMUTAA1})-(\ref{pMUTAA2}),%
\begin{eqnarray}
q^{\alpha _{1}}T_{\mu _{12};\alpha _{12}}^{AA} &=&-\frac{1}{2}q^{2}T_{\mu
_{12};\alpha _{2}}^{AA}+T_{\left( -\right) \mu _{2};\mu _{1}\alpha _{2}}^{V}+%
\frac{1}{2}q_{\mu _{1}}T_{\left( -\right) \mu _{2};\alpha _{1}}^{V}+\frac{1}{%
2}q_{\mu _{2}}T_{\left( +\right) \mu _{1};\alpha _{2}}^{V}-\frac{1}{2}g_{\mu
_{12}}q^{\nu _{1}}T_{\left( +\right) \nu _{1};\alpha _{2}}^{V}  \notag \\
&&-m\left[ (2T_{\mu _{2};\mu _{1}\alpha _{2}}^{PA}+q_{\mu _{1}}T_{\mu
_{2};\alpha _{1}}^{PA})-2mq_{\mu _{2}}\left( 2\bar{J}_{2\mu _{1}\alpha
_{2}}+q_{\mu _{1}}J_{2\alpha _{2}}\right) \right]  \notag \\
&&-2m^{2}g_{\mu _{12}}\left( \Delta _{2\alpha _{2}\nu _{1}}+g_{\alpha
_{2}\nu _{1}}I_{\log }\right) -2m^{2}g_{\mu _{12}}\left( 2q^{\alpha
_{1}}J_{2\alpha _{12}}+q^{2}J_{2\alpha _{2}}\right) .
\end{eqnarray}%
Using $2q^{\alpha _{1}}J_{2\alpha _{12}}=-q^{2}J_{2\alpha _{2}}$, last term
is null. Still, identifying other null combinations 
\begin{equation}
(2T_{\mu _{2};\mu _{1}\alpha _{2}}^{PA}+q_{\mu _{1}}T_{\mu _{2};\alpha
_{2}}^{PA})=2mq_{\mu _{2}}\left( 2\bar{J}_{2\mu _{1}\alpha _{2}}+q_{\mu
_{1}}J_{2\alpha _{2}}\right) .
\end{equation}%
The difference between one-point scalar functions, using (\ref{J1B}), (\ref%
{J1mu1B}), and (\ref{J1mu12}), 
\begin{equation}
T_{\alpha }^{S}\left( k_{2}\right) -T_{\alpha }^{S}\left( k_{1}\right) =2m[%
\bar{J}_{1\alpha }\left( k_{2}\right) -\bar{J}_{1\alpha }\left( k_{1}\right)
-q_{\alpha }\bar{J}_{1}\left( k_{2}\right) ]=-2mq^{\nu }(\Delta _{2\alpha
\nu }+g_{\alpha \nu }I_{\log }).
\end{equation}%
The relation is satisfied directly, such that 
\begin{eqnarray}
q^{\alpha _{1}}T_{\mu _{12};\alpha _{12}}^{AA} &=&-\frac{1}{2}q^{2}T_{\mu
_{12};\alpha _{2}}^{AA}+T_{\left( -\right) \mu _{2};\mu _{1}\alpha _{2}}^{V}+%
\frac{1}{2}q_{\mu _{1}}T_{\left( -\right) \mu _{2};\alpha _{1}}^{V}+\frac{1}{%
2}q_{\mu _{2}}T_{\left( +\right) \mu _{1};\alpha _{2}}^{V} \\
&&-\frac{1}{2}g_{\mu _{12}}q^{\nu _{1}}T_{\left( +\right) \nu _{1};\alpha
_{2}}^{V}+mg_{\mu _{12}}[T_{\alpha _{2}}^{S}\left( k_{2}\right) -T_{\alpha
_{2}}^{S}\left( k_{1}\right) ].  \notag
\end{eqnarray}

\subsection{Metric Contractions: $g^{\protect\mu \protect\sigma }T_{\protect%
\mu \protect\nu ;\protect\sigma }^{AA}$ and $g^{\protect\mu \protect\sigma %
}T_{\protect\mu \protect\nu ;\protect\sigma \protect\lambda }^{AA}$}

The same conditions as $VV$-amplitudes will constrain relations involving
traces, 
\begin{eqnarray}
g^{\mu _{1}\alpha _{1}}T_{\mu _{12};\alpha _{1}}^{AA} &=&T_{\mu
_{2}}^{V}\left( k_{2}\right) +mT_{\mu _{2}}^{PA}+\frac{1}{2}\left( 2\Upsilon
_{\mu _{2}}+q_{\mu _{2}}\Upsilon \right)  \label{gAA1} \\
g^{\mu _{1}\alpha _{1}}T_{\mu _{12};\alpha _{12}}^{AA} &=&T_{\mu _{2};\alpha
_{2}}^{V}\left( k_{2}\right) +mT_{\mu _{2};\alpha _{2}}^{PA}+\frac{1}{2}%
\left( 2\Upsilon _{\mu _{2}\alpha _{2}}+q_{\mu _{2}}\Upsilon _{\alpha
_{2}}\right) .
\end{eqnarray}%
Requiring that tensors calculated on (\ref{Ups0}), (\ref{Ups1}), and (\ref%
{Ups2}) being zero leads to its satisfaction. All relations deduced for even
amplitudes are symmetric by exchanges $\mu _{1}\leftrightarrow \mu _{2}$ and 
$\alpha _{1}\leftrightarrow \alpha _{2}$. To make this part complete must be
noticed that if we contract with the second index $\mu _{2}$ and one
derivative index, we get a superficially different expression; however,
two-point amplitudes in the RHS obey $T^{AP}=-T^{PA}$.

For instance, to obtain (\ref{gAA1}) one may use (\ref{AA-VV1}),%
\begin{equation}
2g^{\mu _{1}\alpha _{1}}T_{\mu _{12};\alpha _{1}}^{AA}=2T_{\mu
_{2}}^{V}\left( k_{2}\right) +2m(T_{\mu _{2}}^{SV}-4mJ_{2\mu _{2}})+\left(
2\Upsilon _{\mu _{2}}+q_{\mu _{2}}\Upsilon \right) .
\end{equation}%
Furthermore, notice that the identity $(T_{\mu _{2}}^{SV}-4mJ_{2\mu
_{2}})=2mq_{\mu _{2}}J_{2}=T_{\mu _{2}}^{PA}$ returns the first equation we
showed. The deduction steps for two derivative indices are unchanged. One
could also invoke Eq. (\ref{AA1exp}) for trading between one derivative and
one matrix index; thus, taking the trace, there will appear a RAGF to inner
contractions (with matrix indices), which in turn are identically satisfied
as demonstrated previously. Therefore, we employ that derivation in the
equation below 
\begin{equation}
2g^{\mu _{1}\alpha _{1}}T_{\mu _{12};\alpha _{1}}^{AA}=g^{\mu _{1}\alpha
_{1}}(2T_{\alpha _{1}\mu _{1};\mu _{2}}^{AA}+q_{\mu _{2}}T_{\alpha _{1}\mu
_{1}}^{AA})-q^{\mu _{1}}T_{\mu _{12}}^{AA}+T_{\left( +\right) \mu
_{2}}^{V}+4m^{2}(2J_{2\mu _{2}}+q_{\mu _{2}}J_{2}).
\end{equation}%
Reminding that $T_{\mu _{2}}^{SV}=2m\left( 2J_{2\mu _{2}}+q_{\mu
_{2}}J_{2}\right) $, final expression assumes the form%
\begin{equation}
2g^{\mu _{1}\alpha _{1}}T_{\mu _{12};\alpha _{1}}^{AA}=g^{\nu _{12}}(2T_{\nu
_{12};\mu _{2}}^{AA}+q_{\mu _{2}}T_{\nu _{12}}^{AA})+2T_{\mu _{2}}^{V}\left(
k_{2}\right) +2mT_{\mu _{2}}^{PA}+2mT_{\mu _{2}}^{SV}.
\end{equation}%
After that, we transform $AA$ into $VV$ on the LHS following (\ref{gAA1}).

We finished calculating all the amplitudes and RAGF of even amplitudes. The
relations involving momentum with matrix indices and derivatives are all
automatically satisfied. However, in the case of traces, we saw that two
groups of amplitudes presented violations by the same terms.

\section{Odd Amplitudes: RAGFs\label{oddragfsec}}

For odd amplitudes $AV$-$VA$, internal contractions are different by the
vertex character; specifying the contraction with the axial vertex is
necessary. As we saw, these relations are not satisfied without restriction,
and the presence of an anomalous term is due to the existence of a chiral
anomaly at this vertex,%
\begin{eqnarray}
q^{\mu _{1}}t_{\mu _{12};\alpha _{1}}^{AV} &=&[t_{\mu _{2};\alpha
_{1}}^{A}\left( k_{1}\right) -t_{\mu _{2};\alpha _{1}}^{A}\left(
k_{2}\right) ]-2mt_{\mu _{2};\alpha _{1}}^{PV}=t_{\left( -\right) \mu
_{2};\alpha _{1}}^{A}-2mt_{\mu _{2};\alpha _{1}}^{PV} \\
q^{\mu _{1}}t_{\mu _{12};\alpha _{12}}^{AV} &=&[t_{\mu _{2};\alpha
_{12}}^{A}\left( k_{1}\right) -t_{\mu _{2};\alpha _{12}}^{A}\left(
k_{2}\right) ]-2mt_{\mu _{2};\alpha _{12}}^{PV}=t_{\left( -\right) \mu
_{2};\alpha _{12}}^{A}-2mt_{\mu _{2};\alpha _{12}}^{PV},
\end{eqnarray}%
where $t_{\left( -\right) \mu _{2};\alpha _{1}}^{A}$ and $t_{\left( -\right)
\mu _{2};\alpha _{12}}^{A}$ are associated with difference of axial
one-point function,%
\begin{eqnarray}
t_{\left( -\right) \mu _{2};\alpha _{1}}^{A} &=&t_{\mu _{2};\alpha
_{1}}^{A}\left( k_{1}\right) -t_{\mu _{2};\alpha _{1}}^{A}\left( k_{2}\right)
\\
t_{\left( -\right) \mu _{2};\alpha _{12}}^{A} &=&t_{\mu _{2};\alpha
_{12}}^{A}\left( k_{1}\right) -t_{\mu _{2};\alpha _{12}}^{A}\left(
k_{2}\right) .
\end{eqnarray}%
Relations for vectorial vertexes are given by%
\begin{eqnarray}
q^{\mu _{2}}t_{\mu _{12};\alpha _{1}}^{AV} &=&t_{\mu _{1};\alpha
_{1}}^{A}\left( k_{1}\right) -t_{\mu _{1};\alpha _{1}}^{A}\left(
k_{2}\right) =t_{\left( -\right) \mu _{1};\alpha _{1}}^{A} \\
q^{\mu _{2}}t_{\mu _{12};\alpha _{12}}^{AV} &=&t_{\mu _{1};\alpha
_{12}}^{A}\left( k_{1}\right) -t_{\mu _{1};\alpha _{12}}^{A}\left(
k_{2}\right) =t_{\left( -\right) \mu _{1};\alpha _{12}}^{A}.
\end{eqnarray}

Two identities can be constructed in external contractions, as explored in
the even ones. If we insert the factor (\ref{qK}) next to the first vertex
we will obtain%
\begin{eqnarray}
q^{\alpha _{1}}t_{\mu _{12};\alpha _{1}}^{AV} &=&-\frac{1}{2}q^{2}t_{\mu
_{12}}^{AV}+t_{\left( -\right) \mu _{2};\mu _{1}}^{A}+\frac{1}{2}q_{\mu
_{1}}t_{\left( -\right) \mu _{2}}^{A}+\frac{1}{2}q_{\mu _{2}}t_{\left(
+\right) \mu _{1}}^{A}-\frac{1}{2}g_{\mu _{12}}q^{\alpha _{1}}t_{\left(
+\right) \alpha _{1}}^{A}  \label{pa1AV2} \\
&&+m\varepsilon _{\mu _{12}}[t^{S}\left( k_{2}\right) -t^{S}\left(
k_{1}\right) ]+mg_{\mu _{12}}[t^{P}\left( k_{2}\right) +t^{P}\left(
k_{1}\right) ].  \notag
\end{eqnarray}%
The notation $t_{\left( +\right) \mu _{1}}^{A}$ is associated with the sum
of the axial one-point function, namely%
\begin{equation}
t_{\left( +\right) \mu _{1}}^{A}=t_{\mu _{1}}^{A}\left( k_{1}\right) +t_{\mu
_{1}}^{A}\left( k_{2}\right) .
\end{equation}%
But if we use the same identity around the second vertex, the relations are%
\begin{equation}
q^{\alpha _{1}}t_{\mu _{12};\alpha _{1}}^{AV}=-\frac{1}{2}q^{2}t_{\mu
_{12}}^{AV}+t_{\left( -\right) \mu _{1};\mu _{2}}^{A}+\frac{1}{2}q_{\mu
_{2}}t_{\left( -\right) \mu _{1}}^{A}+\frac{1}{2}q_{\mu _{1}}t_{\left(
+\right) \mu _{2}}^{A}-\frac{1}{2}g_{\mu _{12}}q^{\rho }t_{\left( +\right)
\rho }^{A}.  \label{pa1AV3}
\end{equation}%
The same to the four-indexes amplitudes, adding one index more. In addition
to the relations (\ref{pa1AV2}) having additional terms when compared to (%
\ref{pa1AV3}). The roles of indices $\mu _{1}$ and $\mu _{2}$ are different.
We will see its consequences in the course of this investigation.

In contractions with the metric, the indices $\mu _{1}$ and $\mu _{2}$ give
us different relations:%
\begin{eqnarray}
g^{\mu _{1}\alpha _{1}}t_{\mu _{12};\alpha _{1}}^{AV} &=&t_{\mu
_{2}}^{A}\left( k_{2}\right) +mt_{\mu _{2}}^{PV} \\
g^{\mu _{1}\alpha _{1}}t_{\mu _{12};\alpha _{12}}^{AV} &=&t_{\mu _{2};\alpha
_{2}}^{A}\left( k_{2}\right) +mt_{\mu _{2};\alpha _{2}}^{PV}.
\end{eqnarray}%
\begin{eqnarray}
g^{\mu _{2}\alpha _{1}}t_{\mu _{12};\alpha _{1}}^{AV} &=&t_{\mu
_{1}}^{A}\left( k_{2}\right) +mt_{\mu _{1}}^{AS} \\
g^{\mu _{2}\alpha _{2}}t_{\mu _{12};\alpha _{12}}^{AV} &=&t_{\mu _{1};\alpha
_{1}}^{A}\left( k_{2}\right) +mt_{\mu _{1};\alpha _{1}}^{AS}.
\end{eqnarray}%
The relations for $VA$ amplitudes are analogous and complementary.

These relations, it is possible to establish all relations that come from
the contractions for the complete expression of Gravitational Amplitude, see
(\ref{Tgravfull}). Their violations or satisfactions are closely related to
the symmetries to be determined. From the view of our strategy, these
relations establish a minimum consistency test of amplitudes after
integration. In other words, if they are satisfied, the linearity of the
integration operation is maintained. Since we expect that when we explicitly
calculate an amplitude, whatever calculation procedure is used, the
contraction of the final result with the external momentum for each
amplitude vertex should reproduce the expected RAGF. Otherwise, we can
establish some relations of amplitude violations.

As we have seen in sections for even amplitudes, relations with momenta
contractions are unconditionally satisfied. It was not necessary to impose
any condition regarding divergent content. However, the case is somewhat
different for odd amplitudes. This relation type is not trivially satisfied.
Furthermore, we will show that presence of terms (\ref{Ups0}), (\ref{Ups1}),
and (\ref{Ups2}) violate different contractions depending on $AV$-versions.

\subsection{Internal Contractions: $q^{\protect\mu }T_{\protect\mu \protect%
\nu ;\protect\sigma }^{AV}$ and $q^{\protect\mu }T_{\protect\mu \protect\nu ;%
\protect\sigma \protect\lambda }^{AV}$ and $V\leftrightarrow A$}

Derived in Chapter (\ref{2Dim2Pt}), we have that contraction with the axial
vertex for the first version of $AV$-amplitudes in (\ref{pAV1}) is violated.
The mechanism develops similarly for $q^{\mu _{2}}$ contraction; the index
meets the index inside $VV$-amplitude and, through its identities, implies
automatic preservation of RAGF,%
\begin{eqnarray}
q^{\mu _{1}}(T_{\mu _{12}}^{AV})_{1} &=&T_{\left( -\right) \mu
_{2}}^{A}-2mT_{\mu _{2}}^{PV}+\varepsilon _{\mu _{2}\mu _{1}}q^{\mu
_{1}}\Upsilon  \label{q1AV1} \\
q^{\mu _{2}}(T_{\mu _{12}}^{AV})_{1} &=&T_{\left( -\right) \mu _{1}}^{A}.
\end{eqnarray}%
The second version works oppositely and satisfies relations established for $%
q^{\mu _{1}}.$ Just because the $AA$ automatically satisfies its RAGF, the
relation for index $\mu _{2}$ follows with an additional term, as expected.
To see this, we use the link connecting versions and obtain%
\begin{eqnarray}
q^{\mu _{1}}(T_{\mu _{12}}^{AV})_{2} &=&T_{\left( -\right) \mu
_{2}}^{A}-2mT_{\mu _{2}}^{PV} \\
q^{\mu _{2}}(T_{\mu _{12}}^{AV})_{2} &=&T_{\left( -\right) \mu
_{1}}^{A}+\varepsilon _{\mu _{1}\nu }q^{\nu }\Upsilon .
\end{eqnarray}%
Hence, to this relation type and for amplitudes with derivative indices
also, the RAGF coming from $q^{\mu _{i}}$ contraction is directly verified
if a version is $j=i$ and needs manipulation in its indices given by
relations among versions (\ref{Uni}) if $i=j$. In the second case arises
factors $\{\Upsilon ,\Upsilon _{\alpha },\Upsilon _{\alpha _{1}\alpha
_{2}}\} $ that we developed as specific tensors connecting two basic
versions.

Elements that we have elaborated on are enough to establish relations for
both contractions $q^{\mu _{i}}$ and both versions $\left\{ \left( AV\right)
_{i},\left( VA\right) _{i}\right\} $ and any number of derivative indices.
To do this, first, we call attention to specific results $T_{\mu
}^{PA}=-T_{\mu }^{AP}$ and $T_{\mu }^{PV}=-T_{\mu }^{VP}$. This result is
valid irrespective of their finite character since they do not depend on the
traces employed in their calculation. Therefore, they are also helpful for
structures with more indices. The required results are listed below 
\begin{eqnarray}
-\varepsilon _{\mu _{2}}^{\text{\quad }\nu _{1}}T_{\nu _{1}}^{PA} &=&T_{\mu
_{2}}^{PV}=-\varepsilon _{\mu _{2}}^{\text{\quad }\nu _{1}}(2mq_{\nu
_{1}}J_{2}) \\
-\varepsilon _{\mu _{2}}^{\text{\quad }\nu _{1}}T_{\nu _{1};\alpha
_{1}}^{PA} &=&T_{\mu _{2};\alpha _{1}}^{PV}=-\varepsilon _{\mu _{2}}^{\text{%
\quad }\nu _{1}}(2mq_{\nu _{1}}J_{2\alpha _{1}}) \\
-\varepsilon _{\mu _{2}}^{\text{\quad }\nu _{1}}T_{\nu _{1};\alpha
_{12}}^{PA} &=&T_{\mu _{2};\alpha _{12}}^{PV}=-\varepsilon _{\mu _{2}}^{%
\text{\quad }\nu _{1}}(2mq_{\nu _{1}}\bar{J}_{2\alpha _{12}}).
\end{eqnarray}

General structures of RAGFs are obtained by explicitly calculating all
amplitudes%
\begin{eqnarray}
q^{\mu _{i}}(T_{\mu _{12}}^{AV})_{j} &=&T_{\left( -\right) \mu
_{k}}^{A}-\delta _{1,i}(2mT_{\mu _{2}}^{PV})+\delta _{i,j}\left( \varepsilon
_{\mu _{k}\nu }q^{\nu }\Upsilon \right)  \label{qAV1(gen)} \\
q^{\mu _{i}}(T_{\mu _{12};\alpha _{1}}^{AV})_{j} &=&T_{\left( -\right) \mu
_{k};\alpha _{1}}^{A}-\delta _{1,i}(2mT_{\mu _{2};\alpha _{1}}^{PV})+\delta
_{i,j}\left( \varepsilon _{\mu _{k}\nu }q^{\nu }\Upsilon _{\alpha
_{1}}\right)\label{qmu2AV11} \\
q^{\mu _{i}}(T_{\mu _{12};\alpha _{12}}^{AV})_{j} &=&T_{\left( -\right) \mu
_{k};\alpha _{12}}^{A}-\delta _{1,i}(2mT_{\mu _{2};\alpha
_{12}}^{PV})+\delta _{i,j}\left( \varepsilon _{\mu _{k}\nu }q^{\nu }\Upsilon
_{\alpha _{12}}\right) ,  \label{qAV3(gen)}
\end{eqnarray}
$i,j,k=\left\{ 1,2\right\} $ with $k\not=i$, and $\delta _{ij}$ is Kronecker
delta equal to one if $i=j$ and zero otherwise. The formulae encode when one
contracts with $q^{\mu _{i}}$ the version $j=i$, i.e., with vertex index
where the version was defined, there is a $\Upsilon $-factor, not if there
is no match $i\not=j$, $\delta _{ij}$ encodes these behaviors; it also
captures if contraction has a $PV$ function (see $\delta _{1,i}$). Note that
when $i\not=j,$ there is no constraint over surface terms; in complementary
cases, constraints are to be studied. They happen over the same $\Upsilon $%
-factors as even amplitude traces; however, not in combination as in
Subsection (\ref{metriceven}).

To complete, we ought to remind condition-less relations among $VA$ and $AV$%
-tensors:%
\begin{eqnarray}
T_{\mu _{12}}^{VA} &=&T_{\mu _{12}}^{AV}+4m^{2}\varepsilon _{\mu _{1}\mu
_{2}}J_{2} \\
T_{\mu _{12};\alpha _{1}}^{VA} &=&T_{\mu _{12};\alpha
_{1}}^{AV}+4m^{2}\varepsilon _{\mu _{1}\mu _{2}}J_{2\alpha _{1}}  \notag \\
T_{\mu _{12};\alpha _{12}}^{VA} &=&T_{\mu _{12};\alpha
_{12}}^{AV}+4m^{2}\varepsilon _{\mu _{1}\mu _{2}}\bar{J}_{2\alpha _{12}}. 
\notag
\end{eqnarray}%
As they are valid for any version, we did not use indices. Despite this, we
could also study the unicity relations $(T_{\mu _{12}}^{VA})_{2}-(T_{\mu
_{12}}^{VA})_{1}=-\varepsilon _{\mu _{21}}\Upsilon ,$ and so on for higher
rank. In parallel to previous deductions, we can cast the pattern of
contractions related to the RAGFs explicitly and in a systematic form as $AV$
versions:%
\begin{eqnarray}
q^{\mu _{i}}(T_{\mu _{12}}^{VA})_{j} &=&T_{\left( -\right) \mu
_{k}}^{A}+\delta _{2,i}(2mT_{\mu _{2}}^{VP})+\delta _{i,j}\left( \varepsilon
_{\mu _{k}\nu }q^{\nu }\Upsilon \right) \\
q^{\mu _{i}}(T_{\mu _{12};\alpha _{1}}^{VA})_{j} &=&T_{\left( -\right) \mu
_{k};\alpha _{1}}^{A}+\delta _{2,i}(2mT_{\mu _{2};\alpha _{1}}^{VP})+\delta
_{i,j}\left( \varepsilon _{\mu _{k}\nu }q^{\nu }\Upsilon _{\alpha
_{1}}\right) \\
q^{\mu _{i}}(T_{\mu _{12};\alpha _{12}}^{VA})_{j} &=&T_{\left( -\right) \mu
_{k};\alpha _{12}}^{A}+\delta _{2,i}(2mT_{\mu _{2};\alpha
_{12}}^{VP})+\delta _{i,j}\left( \varepsilon _{\mu _{k}\nu }q^{\nu }\Upsilon
_{\alpha _{12}}\right) .
\end{eqnarray}%
It is worth noticing that $\delta _{2,i}$ makes precise VP functions appear
in $q^{\mu _{2}}$-relations. Once more, this is a summary of the results; an
important point is the appearance of conditioning factors in relations
corresponding to the vertices around those we used the chiral matrix
definition. As demonstrated in sections, that is equivalent to substituting (%
\ref{id2}).

\subsection{External Contractions: $q^{\protect\sigma }T_{\protect\mu 
\protect\nu ;\protect\sigma }^{AV}$ and $q^{\protect\sigma }T_{\protect\mu 
\protect\nu ;\protect\sigma \protect\lambda }^{AV}$ and $V\leftrightarrow A$}

Treating relations involving derivative indices as we did for the even case
is possible. The $VV$ amplitudes can be manipulated and written through the
identities (\ref{IDtroca1}) and (\ref{IDtroca2}); when we exchange any
derivative index for a matrix index, 
\begin{equation*}
t_{\mu _{12};\alpha _{1}}^{VV}=-\frac{1}{2}q_{\alpha _{1}}t_{\mu
_{12}}^{VV}+t_{\alpha _{1}\mu _{2};\mu _{1}}^{VV}+\frac{1}{2}q_{\mu
_{1}}t_{\alpha _{1}\mu _{2}}^{VV}+\frac{1}{2}g_{\alpha _{1}\mu
_{2}}t_{\left( +\right) \mu _{1}}^{V}-\frac{1}{2}g_{\mu _{12}}t_{\left(
+\right) \alpha _{1}}^{V}+r_{\mu _{12};\alpha _{1}}.
\end{equation*}%
The exchange effect is equally valid for $\alpha _{1}\leftrightarrow \mu
_{2} $, resulting $\mu _{1}\leftrightarrow \mu _{2}$ in the equation above.

We can get relations for odd amplitudes obtained of $VV$-amplitudes.
Appropriately exchanging indices and multiplying by tensor $-\varepsilon
_{\mu _{i}}^{\text{ \ }\nu _{1}}$ leads us to unconditional identities%
\begin{eqnarray}
(T_{\mu _{12};\alpha _{1}}^{AV})_{1} &=&-\frac{1}{2}q_{\alpha _{1}}(T_{\mu
_{12}}^{AV})_{1}+(T_{\mu _{1}\alpha _{1};\mu _{2}}^{AV})_{1}+\frac{1}{2}%
q_{\mu _{2}}(T_{\mu _{1}\alpha _{1}}^{AV})_{1}  \label{TAV1.3} \\
&&+\frac{1}{2}\varepsilon _{\mu _{1}\mu _{2}}T_{\left( +\right) \alpha
_{1}}^{V}-\frac{1}{2}\varepsilon _{\mu _{1}\alpha _{1}}T_{\left( +\right)
\mu _{2}}^{V}  \notag
\end{eqnarray}%
\begin{eqnarray}
(T_{\mu _{12};\alpha _{1}}^{VA})_{2} &=&-\frac{1}{2}q_{\alpha _{1}}(T_{\mu
_{12}}^{VA})_{2}+(T_{\alpha _{1}\mu _{2};\mu _{1}}^{VA})_{2}+\frac{1}{2}%
q_{\mu _{1}}(T_{\alpha _{1}\mu _{2}}^{VA})_{2} \\
&&-\frac{1}{2}\varepsilon _{\mu _{1}\mu _{2}}T_{\left( +\right) \alpha
_{1}}^{V}-\frac{1}{2}\varepsilon _{\mu _{2}\alpha _{1}}T_{\left( +\right)
\mu _{1}}^{V}.  \notag
\end{eqnarray}%
It is necessary to remember the versions of amplitudes in terms of $AA$ (\ref%
{AA1exp}) and (\ref{AAexc}). Follow the other identities satisfied by odd
amplitudes,%
\begin{eqnarray}
(T_{\mu _{12};\alpha _{1}}^{AV})_{2} &=&-\frac{1}{2}q_{\alpha _{1}}(T_{\mu
_{12}}^{AV})_{2}+(T_{\alpha _{1}\mu _{2};\mu _{1}}^{AV})_{2}+\frac{1}{2}%
q_{\mu _{1}}(T_{\alpha _{1}\mu _{2}}^{AV})_{2}  \label{TAV2ex} \\
&&-\frac{1}{2}\varepsilon _{\mu _{1}\mu _{2}}T_{\left( +\right) \alpha
_{1}}^{V}-\frac{1}{2}\varepsilon _{\mu _{2}\alpha _{1}}T_{\left( +\right)
\mu _{1}}^{V}  \notag
\end{eqnarray}%
\begin{eqnarray}
(T_{\mu _{12};\alpha _{1}}^{VA})_{1} &=&-\frac{1}{2}q_{\alpha _{1}}(T_{\mu
_{12}}^{VA})_{1}+(T_{\mu _{1}\alpha _{1};\mu _{2}}^{VA})_{1}+\frac{1}{2}%
q_{\mu _{2}}(T_{\mu _{1}\alpha _{1}}^{VA})_{1} \\
&&+\frac{1}{2}\varepsilon _{\mu _{1}\mu _{2}}T_{\left( +\right) \alpha
_{1}}^{V}-\frac{1}{2}\varepsilon _{\mu _{1}\alpha _{1}}T_{\left( +\right)
\mu _{2}}^{V}.  \notag
\end{eqnarray}

By construction, we will see that these identities will always be satisfied.
Starting to analyze this trajectory by the first version. From expression (%
\ref{TAV1.3}), we have%
\begin{eqnarray}
q^{\alpha _{1}}(T_{\mu _{12};\alpha _{1}}^{AV})_{1} &=&-\frac{1}{2}%
q^{2}(T_{\mu _{12}}^{AV})_{1}+q^{\alpha _{1}}(T_{\mu _{1}\alpha _{1};\mu
_{2}}^{AV})_{1}+\frac{1}{2}q_{\mu _{2}}[q^{\alpha _{1}}(T_{\mu _{1}\alpha
_{1}}^{AV})_{1}] \\
&&+\frac{1}{2}\varepsilon _{\mu _{1}\mu _{2}}q^{\alpha _{1}}T_{\left(
+\right) \alpha _{1}}^{V}-\frac{1}{2}\varepsilon _{\mu _{1}\alpha
_{1}}q^{\alpha _{1}}T_{\left( +\right) \mu _{2}}^{V}.  \notag
\end{eqnarray}%
Identifying relations with internal indices that are satisfied for version
one yields%
\begin{eqnarray}
q^{\alpha _{1}}(T_{\mu _{12};\alpha _{1}}^{AV})_{1} &=&-\frac{1}{2}%
q^{2}(T_{\mu _{12}}^{AV})_{1}+T_{\left( -\right) \mu _{1};\mu _{2}}^{A}+%
\frac{1}{2}q_{\mu _{2}}T_{\left( -\right) \mu _{1}}^{A} \\
&&+\frac{1}{2}\varepsilon _{\mu _{1}\mu _{2}}q^{\alpha _{1}}T_{\left(
+\right) \alpha _{1}}^{V}-\frac{1}{2}\varepsilon _{\mu _{1}\alpha
_{1}}q^{\alpha _{1}}T_{\left( +\right) \mu _{2}}^{V}.  \notag
\end{eqnarray}

As in the last line, there is no direct identification of one-point
vectorial functions with axial ones. We need to use the Schouten identity
just like 
\begin{equation}
\lbrack \varepsilon _{\mu _{12}}q^{\nu _{2}}-q^{\nu _{1}}\varepsilon _{\mu
_{1}\nu _{1}}\delta _{\mu _{2}}^{\nu _{2}}]T_{\left( +\right) \nu
_{2}}^{V}=q_{\mu _{1}}T_{\left( +\right) \mu _{2}}^{A}-g_{\mu _{12}}q^{\nu
_{1}}T_{\left( +\right) \nu _{1}}^{A}.
\end{equation}%
Thus, replacing in equation above, we obtain 
\begin{eqnarray}
q^{\alpha _{1}}(T_{\mu _{12};\alpha _{1}}^{AV})_{1} &=&-\frac{1}{2}%
q^{2}(T_{\mu _{12}}^{AV})_{1}+T_{\left( -\right) \mu _{1};\mu _{2}}^{A}+%
\frac{1}{2}q_{\mu _{2}}T_{\left( -\right) \mu _{1}}^{A} \\
&&-\frac{1}{2}g_{\mu _{12}}q^{\nu _{1}}T_{\left( +\right) \nu _{1}}^{A}+%
\frac{1}{2}q_{\mu _{1}}T_{\left( +\right) \mu _{2}}^{A}.  \notag
\end{eqnarray}

That is the relation obtained around the second vertex (\ref{pa1AV3}). The
reason for satisfaction is that index replaced by $\alpha _{i}$ always
appears as the one amplitude version, and $q^{\alpha _{i}}$ is always
complimentary. In the case of $q^{\alpha _{1}}(T_{\mu _{1}\alpha _{1};\mu
_{2}}^{AV})_{1}$ and $q^{\alpha _{1}}(T_{\mu _{1}\alpha _{1}}^{AV})_{1}$,
the RAGF for vectorial indices are automatically satisfied. The same happens
contraction for $(T_{\mu _{12};\alpha _{1}}^{VA})_{1}$: contractions with
axial indices are satisfied, and the additional finite part cancels out 
\begin{equation}
-m(2T_{\mu _{2};\mu _{1}}^{PV}+q_{\mu _{1}}T_{\mu
_{2}}^{PV})=2m^{2}\varepsilon _{\mu _{2}\nu _{1}}q^{\nu _{1}}(2J_{2\mu
_{1}}+q_{\mu _{1}}J_{2})=0.  \label{relPVs}
\end{equation}%
So, we have the RAGF satisfied around the second vertex.

Violations occur precisely in relations established around the vertex
associated with version: first vertex, thus first version, second vertex,
second version. For example, the same manipulations lead to%
\begin{eqnarray}
q^{\alpha _{1}}(T_{\mu _{12};\alpha _{1}}^{AV})_{2} &=&-\frac{1}{2}%
q^{2}(T_{\mu _{12}}^{AV})_{2}+T_{\left( -\right) \mu _{2};\mu _{1}}^{A}+%
\frac{1}{2}q_{\mu _{1}}T_{\left( -\right) \mu _{2}}^{A}  \label{qAV23} \\
&&+\frac{1}{2}[\varepsilon _{\mu _{21}}q^{\nu _{2}}-\varepsilon _{\mu
_{2}\nu _{1}}\delta _{\mu _{1}}^{\nu _{2}}q^{\nu _{1}}]T_{\left( +\right)
\nu _{2}}^{V}+m(2T_{\mu _{2};\mu _{1}}^{PV}+q_{\mu _{1}}T_{\mu _{2}}^{PV}). 
\notag
\end{eqnarray}%
Applying Schouten identity in the last line and canceling out additional
finite parts,%
\begin{eqnarray}
q^{\alpha _{1}}(T_{\mu _{12};\alpha _{1}}^{AV})_{2} &=&-\frac{1}{2}%
q^{2}(T_{\mu _{12}}^{AV})_{2}+T_{\left( -\right) \mu _{2};\mu _{1}}^{A}+%
\frac{1}{2}q_{\mu _{1}}T_{\left( -\right) \mu _{2}}^{A}  \label{paAV2} \\
&&+\frac{1}{2}q_{\mu _{2}}T_{\left( +\right) \mu _{1}}^{A}-\frac{1}{2}g_{\mu
_{12}}q^{\nu _{1}}T_{\left( +\right) \nu _{1}}^{A}.  \notag
\end{eqnarray}%
It satisfies the relation deduced around the first vertex (\ref{pa1AV2}) but
does not satisfy the relation deduced around the second (\ref{pa1AV3}).
Remembering that massive terms do not contribute because they are null for
these amplitudes.

Taking advantage of equations (\ref{Uni}) and (\ref{Un1}) incorporate
uniqueness conditions and invariably connect them, we will have the possible
violating term: 
\begin{equation*}
\left. q^{\alpha _{1}}(T_{\mu _{12};\alpha _{1}}^{AV})_{1}\right\vert ^{%
\mathrm{viol}}=-\frac{1}{2}\varepsilon _{\mu _{12}}q^{\alpha _{1}}\left(
2\Upsilon _{\alpha _{1}}+q_{\alpha _{1}}\Upsilon \right) .
\end{equation*}%
However, let us consider that the expression obtained around the second
vertex is valid (\ref{pa1AV3}). The same type of violation will be present
in the second version, and the first will be automatically satisfied.

For amplitude with two derivative factors, the calculation follows equation (%
\ref{AAexc}),%
\begin{eqnarray}
(T_{\mu _{12};\alpha _{12}}^{AV})_{2} &=&-\frac{1}{2}q_{\alpha _{1}}(T_{\mu
_{12};\alpha _{2}}^{AV})_{2}+(T_{\alpha _{1}\mu _{2};\mu _{1}\alpha
_{2}}^{AV})_{2}+\frac{1}{2}q_{\mu _{1}}(T_{\alpha _{1}\mu _{2};\alpha
_{2}}^{AV})_{2}  \notag \\
&&-\frac{1}{2}\varepsilon _{\mu _{2}\alpha _{1}}T_{\left( +\right) \mu
_{1};\alpha _{2}}^{V}-\frac{1}{2}\varepsilon _{\mu _{1}\mu _{2}}T_{\left(
+\right) \alpha _{1};\alpha _{2}}^{V}-\varepsilon _{\mu _{2}}^{\text{ \ \ }%
\nu _{1}}R_{\mu _{1}\nu _{1};\alpha _{12}}  \notag \\
&&-2m^{2}\varepsilon _{\mu _{2}\alpha _{1}}(2\bar{J}_{2\mu _{1}\alpha
_{2}}+q_{\mu _{1}}J_{2\alpha _{2}})-2m^{2}\varepsilon _{\mu _{12}}(2\bar{J}%
_{2\alpha _{1}\alpha _{2}}+q_{\alpha _{1}}J_{2\alpha _{2}}),
\end{eqnarray}%
where $R_{\mu _{1}\nu _{1};\alpha _{12}}$ is defined in (\ref{R4}) and null
by contraction. It is simple to show that version one, using (\ref{Un2}),
the possible violating term is given by 
\begin{equation}
\left. q^{\alpha _{1}}(T_{\mu _{12};\alpha _{12}}^{AV})_{1}\right\vert ^{%
\mathrm{viol}}=-\frac{1}{2}\varepsilon _{\mu _{12}}q^{\nu }\left( 2\Upsilon
_{\alpha _{2}\nu }+q_{\nu }\Upsilon _{\alpha _{2}}\right) .
\end{equation}%
The same analysis leads to similar conclusions for the second version of
amplitudes if the relation around the second vertex is the reference.

\subsection{Metric Contractions: $g^{\protect\mu \protect\sigma }T_{\protect%
\mu \protect\nu ;\protect\sigma }^{AV}$ and $g^{\protect\mu \protect\sigma %
}T_{\protect\mu \protect\nu ;\protect\sigma \protect\lambda }^{AV}$\ }

Finally, the last relation we need to calculate. Once again, we will make
use of relations through a reorganization of terms that can be seen from%
\begin{eqnarray}
(T_{\mu _{12};\alpha _{1}}^{AV})_{1} &=&-\frac{1}{2}q_{\alpha _{1}}(T_{\mu
_{12}}^{AV})_{1}+(T_{\mu _{1}\alpha _{1};\mu _{2}}^{AV})_{1}+\frac{1}{2}%
q_{\mu _{2}}(T_{\mu _{1}\alpha _{1}}^{AV})_{1}  \label{TAV111} \\
&&+\frac{1}{2}\varepsilon _{\mu _{1}\mu _{2}}T_{\left( +\right) \alpha
_{1}}^{V}-\frac{1}{2}\varepsilon _{\mu _{1}\alpha _{1}}T_{\left( +\right)
\mu _{2}}^{V}.  \notag
\end{eqnarray}%
Starting by contracting the expression above with $g^{\mu _{1}\alpha _{1}}$,%
\begin{equation}
g^{\mu _{1}\alpha _{1}}(T_{\mu _{12};\alpha _{1}}^{AV})_{1}=\frac{1}{2}%
[-q^{\mu _{1}}(T_{\mu _{12}}^{AV})_{1}+T_{\left( +\right) \mu _{2}}^{A}]+%
\frac{1}{2}g^{\mu _{1}\alpha _{1}}[2(T_{\mu _{1}\alpha _{1};\mu
_{2}}^{AV})_{1}+q_{\mu _{2}}(T_{\mu _{1}\alpha _{1}}^{AV})_{1}].
\end{equation}%
At this point, it is straightforward to note that the $AV$ amplitude can be
written as%
\begin{equation}
g^{\mu _{1}\alpha _{1}}[2(T_{\mu _{1}\alpha _{1};\mu _{2}}^{AV})_{1}+q_{\mu
_{2}}(T_{\mu _{1}\alpha _{1}}^{AV})_{1}]=-\varepsilon ^{\alpha _{1}\nu
_{1}}[2T_{\nu _{1}\alpha _{1};\mu _{2}}^{VV}+q_{\mu _{2}}T_{\nu _{1}\alpha
_{1}}^{VV}]=0.  \label{gAV=eVV=0}
\end{equation}%
It is canceled because complete $VV$-amplitudes are symmetric in its first
indices, finite and non-finite parts. Using (\ref{q1AV1}) for $q^{\mu
_{1}}(T_{\mu _{12}}^{AV})_{1}$, where appear $\Upsilon =(2\Delta _{2\rho
}^{\rho }+i/\pi )$, follows%
\begin{equation}
g^{\mu _{1}\alpha _{1}}(T_{\mu _{12};\alpha _{1}}^{AV})_{1}=mT_{\mu
_{2}}^{PV}+T_{\mu _{2}}^{A}\left( k_{2}\right) -\frac{1}{2}\varepsilon _{\mu
_{2}\nu }q^{\nu }\Upsilon .  \label{gTAV11}
\end{equation}

For $g^{\mu _{1}\alpha _{1}}(T_{\mu _{12};\alpha _{1}}^{AV})_{2}$ we also
find this relation conditioned. Using the equation that connects two
versions (\ref{Uni}) and (\ref{Un1}), we obtain the desired relation%
\begin{equation}
g^{\mu _{1}\alpha _{1}}(T_{\mu _{12};\alpha _{1}}^{AV})_{2}=mT_{\mu
_{2}}^{PV}+T_{\mu _{2}}^{A}\left( k_{2}\right) -\frac{1}{2}\varepsilon _{\mu
_{2}\nu }\left( 2\Upsilon ^{\nu }+q^{\nu }\Upsilon \right) .
\end{equation}%
An alternative way to extract this information, valid whenever the index of
inner vertices is not the one used to define the version, is to invoke the
equation derived from $VV$ or $AA$ functions and multiply them by an
adequate tensor. Explicitly, we multiplied the equation below by $%
-\varepsilon _{\mu _{2}}^{\quad \nu },$%
\begin{equation}
2g^{\mu _{1}\alpha _{1}}T_{\mu _{1}\nu ;\alpha _{1}}^{AA}-2T_{\nu
}^{V}\left( k_{2}\right) =2mT_{\nu }^{PA}+\left( 2\Upsilon _{\nu }+q_{\nu
}\Upsilon \right) .
\end{equation}%
By definition, it follows%
\begin{equation}
2g^{\mu _{1}\alpha _{1}}(T_{\mu _{1}\mu _{2};\alpha _{1}}^{AV})_{2}=2mT_{\mu
_{2}}^{PV}+2T_{\mu _{2}}^{A}\left( k_{2}\right) -\varepsilon _{\mu _{2}\nu
}\left( 2\Upsilon ^{\nu }+q^{\nu }\Upsilon \right) .
\end{equation}

Contracting with $g^{\mu _{2}\alpha _{1}}$, the application of equation (\ref%
{TAV2ex}) leads to%
\begin{equation}
2g^{\mu _{2}\alpha _{1}}(T_{\mu _{12};\alpha _{1}}^{AV})_{2}=-q^{\mu
_{2}}(T_{\mu _{12}}^{AV})_{2}+T_{\left( +\right) \mu _{1}}^{A}+g^{\nu
_{2}\nu _{1}}[2(T_{\nu _{12};\mu _{1}}^{AV})_{2}+q_{\mu _{1}}(T_{\nu
_{12}}^{AV})_{2}].  \notag
\end{equation}%
The last line drops out by index symmetry in the $AA$-amplitudes. Then using
(\ref{qmu2AV11}) and finite piece $T_{\mu _{1}}^{AS}=0$, follows%
\begin{equation}
g^{\mu _{2}\alpha _{1}}(T_{\mu _{12};\alpha _{1}}^{AV})_{2}=mT_{\mu
_{1}}^{AS}+T_{\mu _{1}}^{A}\left( k_{2}\right) -\frac{1}{2}\varepsilon _{\mu
_{1}\nu }q^{\nu }\Upsilon
\end{equation}%
For version one, we also find this relation violated%
\begin{equation}
g^{\mu _{2}\alpha _{1}}(T_{\mu _{12};\alpha _{1}}^{AV})_{1}=mT_{\mu
_{1}}^{AS}+T_{\mu _{1}}^{A}\left( k_{2}\right) -\frac{1}{2}\varepsilon _{\mu
_{1}\nu }\left( 2\Upsilon ^{\nu }+q^{\nu }\Upsilon \right) .
\end{equation}

The 4th-rank amplitudes with two external indices are easily obtained
following the same steps. Thus, we have the list of equations below, 
\begin{eqnarray}
g^{\mu _{i}\alpha _{1}}(T_{\mu _{12};\alpha _{1}}^{AV})_{j} &=&T_{\mu
_{k}}^{A}\left( k_{2}\right) +m(\delta _{i,1}T_{\mu _{k}}^{PV}+\delta
_{i,2}T_{\mu _{k}}^{AS}) \\
&&-\frac{1}{2}\varepsilon _{\mu _{k}\nu }\left[ q^{\nu }\Upsilon +2\left(
1-\delta _{i,j}\right) \Upsilon ^{\nu }\right]  \notag \\
g^{\mu _{i}\alpha _{1}}(T_{\mu _{12};\alpha _{12}}^{AV})_{j} &=&T_{\mu
_{k};\alpha _{2}}^{A}\left( k_{2}\right) +m(\delta _{i,1}T_{\mu _{k};\alpha
_{2}}^{PV}+\delta _{i,2}T_{\mu _{k};\alpha _{2}}^{AS}) \\
&&-\frac{1}{2}\varepsilon _{\mu _{k}\nu }\left[ q^{\nu }\Upsilon _{\alpha
_{2}}+2\left( 1-\delta _{i,j}\right) \Upsilon _{\alpha _{2}}^{\nu }\right] ,
\notag
\end{eqnarray}
where $\left\{ i,j,k\right\} =\left\{ 1,2\right\} $, $k\not=i.$ The
Kronecker delta guarantees that only correct terms appear in each equation;
note that they reproduce all the previous equations. Additionally, for the $%
VA$ amplitude, we have 
\begin{eqnarray}
g^{\mu _{i}\alpha _{1}}(T_{\mu _{12};\alpha _{1}}^{VA})_{j} &=&T_{\mu
_{k}}^{A}\left( k_{2}\right) +m(\delta _{i,1}T_{\mu _{2}}^{AS}-\delta
_{i,2}T_{\mu _{1}}^{VP}) \\
&&-\frac{1}{2}\varepsilon _{\mu _{k}\nu }\left[ q^{\nu }\Upsilon +2\left(
1-\delta _{i,j}\right) \Upsilon ^{\nu }\right]  \notag \\
g^{\mu _{i}\alpha _{1}}(T_{\mu _{12};\alpha _{12}}^{VA})_{j} &=&T_{\mu
_{k};\alpha _{2}}^{A}\left( k_{2}\right) +m(\delta _{i,1}T_{\mu _{2};\alpha
_{2}}^{AS}-\delta _{i,2}T_{\mu _{1};\alpha _{1}}^{VP}) \\
&&-\frac{1}{2}\varepsilon _{\mu _{k}\nu }\left[ q^{\nu }\Upsilon _{\alpha
_{2}}+2\left( 1-\delta _{i,j}\right) \Upsilon _{\alpha _{2}}^{\nu }\right] ,
\notag
\end{eqnarray}%
where $T_{\mu _{1};\alpha _{1}}^{AS}=-2m\varepsilon _{\mu _{1}}^{\text{\quad 
}\nu _{1}}(2\bar{J}_{2\alpha _{1}\nu _{1}}+q_{\nu _{1}}J_{2\alpha
_{1}})=T_{\mu _{1};\alpha _{1}}^{SA}.$

We have seen in this chapter that terms that may violate the RAGFs are local
polynomials in $P$ and $q$ momenta. These violating terms have values
determined from the set (\ref{Ups0}), (\ref{Ups1}), and (\ref{Ups2}). We
will see that choosing to save the linearity of integration operation,
manifested in the satisfaction of RAGF, will force us to establish finite
values for surface terms present in amplitudes. From now on, we will analyze
the results' consequences and their implications for Einstein and Weyl
anomalies.

\chapter{Gravitational Anomalies\label{GAno}}

This chapter will list the formulas and general results developed in the
previous chapter as a form of organization. They are used in the sequence to
track the possible violating terms of the RAGFs that appear when we combine
the core elements in the permutations contributing to the full two-point
functions of the energy-momentum tensor. As we will adopt the following set
of indices $\left\langle T_{\mu _{1}\mu _{2}}\left( x\right) T_{\alpha
_{1}\alpha _{2}}\left( 0\right) \right\rangle $ to the energy-momentum
tensors in the correlator, see Eq. (\ref{Tgravfull}), we will have 
\begin{equation}
T_{\mu _{1}\mu _{2}\alpha _{1}\alpha _{2}}^{G}=-\frac{i}{64}\{[\mathcal{T}%
_{\mu _{12}\alpha _{12}}^{V}]+[\mathcal{T}_{\mu _{12}\alpha _{12}}^{A}]\}.
\end{equation}%
Hence, the formulas from the previous deductions have indices for even and
odd amplitudes arranged according to the sequence below%
\begin{eqnarray*}
\hat{T}_{\mu _{1}\mu _{2}\alpha _{1}\alpha _{2}}^{V} &=&\mathcal{T}_{\mu
_{1}\alpha _{1}\mu _{2}\alpha _{2}}^{VV}+\mathcal{T}_{\mu _{1}\alpha _{1}\mu
_{2}\alpha _{2}}^{AA} \\
\hat{T}_{\mu _{1}\mu _{2}\alpha _{1}\alpha _{2}}^{A} &=&\mathcal{T}_{\mu
_{1}\alpha _{1}\mu _{2}\alpha _{2}}^{AV}+\mathcal{T}_{\mu _{1}\alpha _{1}\mu
_{2}\alpha _{2}}^{VA}.
\end{eqnarray*}%
The sum of permutations $\mu _{1}\leftrightarrow \mu _{2}$ and from the
result $\alpha _{1}\leftrightarrow \alpha _{2}$ deliver the vector and axial
part of the gravitational amplitude. 
\begin{eqnarray}
\lbrack \mathcal{T}_{\mu _{12}\alpha _{12}}^{V}] &=&[\hat{T}_{\mu _{1}\mu
_{2}\alpha _{1}\alpha _{2}}^{V}]+[\hat{T}_{\mu _{2}\mu _{1}\alpha _{1}\alpha
_{2}}^{V}]+[\hat{T}_{\mu _{1}\mu _{2}\alpha _{2}\alpha _{1}}^{V}]+[\hat{T}%
_{\mu _{2}\mu _{1}\alpha _{2}\alpha _{1}}^{V}] \\
\lbrack \mathcal{T}_{\mu _{12}\alpha _{12}}^{A}] &=&[\hat{T}_{\mu _{1}\mu
_{2}\alpha _{1}\alpha _{2}}^{A}]+[\hat{T}_{\mu _{2}\mu _{1}\alpha _{1}\alpha
_{2}}^{A}]+[\hat{T}_{\mu _{1}\mu _{2}\alpha _{2}\alpha _{1}}^{A}]+[\hat{T}%
_{\mu _{2}\mu _{1}\alpha _{2}\alpha _{1}}^{A}].
\end{eqnarray}

\textbf{Basic Permutations }$\mathcal{T}_{\mu \alpha \sigma \rho }^{\Gamma
_{12}}$: As elaborated at the beginning of the previous chapter, the next
task after computing all the equations satisfied to the amplitudes is to
explore the basic permutations. Through their definition, we expanded our
definitions for derivative amplitudes accordingly. We have%
\begin{equation}
\mathcal{T}_{\mu _{1}\alpha _{1}\mu _{2}\alpha _{2}}^{\Gamma _{1}\Gamma
_{2}}=4T_{\mu _{1}\alpha _{1};\mu _{2}\alpha _{2}}^{\Gamma _{1}\Gamma
_{2}}+2q_{\mu _{2}}T_{\mu _{1}\alpha _{1};\alpha _{2}}^{\Gamma _{1}\Gamma
_{2}}+q_{\alpha _{2}}(2T_{\mu _{1}\alpha _{1};\mu _{2}}^{\Gamma _{1}\Gamma
_{2}}+q_{\mu _{2}}T_{\mu _{1}\alpha _{1}}^{\Gamma _{1}\Gamma _{2}}).
\end{equation}

We must call attention to two features of the notation: The placement of
indices in $\hat{T}_{\mu _{1}\mu _{2}\alpha _{1}\alpha _{2}}^{V}$ is chosen
to mirror the ones from $T_{\mu _{1}\mu _{2}\alpha _{1}\alpha _{2}}^{G}$,
however in $\mathcal{T}_{\mu _{1}\alpha _{1}\mu _{2}\alpha _{2}}^{VV}$ the
disposition emphasizes that the last two indices correspond to derivative
type. This attitude is helpful in the calculations to distinguish their
origin, either as the matrix or derivative indices. Another point in the
calligraphic letter $\mathcal{T}_{\mu \alpha \sigma \rho }^{\Gamma _{12}}$
is to contrast the 4th-rank derivative amplitude that comes with a
semi-colon and the basic permutation involves four terms.

The basic permutations regarding derivatives structures were listed in (\ref%
{VV-basic})-(\ref{VA-basic}). We resume them with the indexes: 
\begin{eqnarray}
\mathcal{T}_{\mu _{1}\alpha _{1}\mu _{2}\alpha _{2}}^{VV} &=&2(2T_{\mu
_{1}\alpha _{1};\mu _{2}\alpha _{2}}^{VV}+q_{\mu _{2}}T_{\mu _{1}\alpha
_{1};\alpha _{2}}^{VV})+q_{\alpha _{2}}(2T_{\mu _{1}\alpha _{1};\mu
_{2}}^{VV}+q_{\mu _{2}}T_{\mu _{1}\alpha _{1}}^{VV}) \\
\mathcal{T}_{\mu _{1}\alpha _{1}\mu _{2}\alpha _{2}}^{AA} &=&2(2T_{\mu
_{1}\alpha _{1};\mu _{2}\alpha _{2}}^{AA}+q_{\mu _{2}}T_{\mu _{1}\alpha
_{1};\alpha _{2}}^{AA})+q_{\alpha _{2}}(2T_{\mu _{1}\alpha _{1};\mu
_{2}}^{AA}+q_{\mu _{2}}T_{\mu _{1}\alpha _{1}}^{AA}) \\
\mathcal{T}_{\mu _{1}\alpha _{1}\mu _{2}\alpha _{2}}^{AV} &=&2(2T_{\mu
_{1}\alpha _{1};\mu _{2}\alpha _{2}}^{AV}+q_{\mu _{2}}T_{\mu _{1}\alpha
_{1};\alpha _{2}}^{AV})+q_{\alpha _{2}}(2T_{\mu _{1}\alpha _{1};\mu
_{2}}^{AV}+q_{\mu _{2}}T_{\mu _{1}\alpha _{1}}^{AV}) \\
\mathcal{T}_{\mu _{1}\alpha _{1}\mu _{2}\alpha _{2}}^{VA} &=&2(2T_{\mu
_{1}\alpha _{1};\mu _{2}\alpha _{2}}^{VA}+q_{\mu _{2}}T_{\mu _{1}\alpha
_{1};\alpha _{2}}^{VA})+q_{\alpha _{2}}(2T_{\mu _{1}\alpha _{1};\mu
_{2}}^{VA}+q_{\mu _{2}}T_{\mu _{1}\alpha _{1}}^{VA}).
\end{eqnarray}

In the RAGFs, combinations of the basic derivative amplitudes from momenta
and metric contractions often arise:%
\begin{eqnarray}
\mathcal{B}_{\alpha _{1};\alpha _{2}} &=&\int \frac{\mathrm{d}^{2}k}{\left(
2\pi \right) ^{2}}\left( K_{1}+K_{2}\right) _{\alpha _{2}}\left[ t_{\alpha
_{1}}^{V}\left( k_{1}\right) +t_{\alpha _{1}}^{V}\left( k_{2}\right) \right]
\\
\mathcal{S}_{\left( -\right) \alpha _{1};\mu _{2}\alpha _{2}}^{\Gamma _{1}}
&=&\int \frac{\mathrm{d}^{2}k}{\left( 2\pi \right) ^{2}}\left(
K_{1}+K_{2}\right) _{\mu _{2}}\left( K_{1}+K_{2}\right) _{\alpha
_{2}}[t_{\alpha _{1}}^{\Gamma _{1}}\left( k_{1}\right) -t_{\alpha
_{1}}^{\Gamma _{1}}\left( k_{2}\right) ],
\end{eqnarray}%
where $\Gamma _{1}=\left\{ V,A\right\} $. By projecting $K_{2}=K_{1}+q,$ we
may decompose them in%
\begin{eqnarray}
\mathcal{B}_{\alpha _{1};\alpha _{2}} &=&2T_{\left( +\right) \alpha
_{1};\alpha _{2}}^{V}+q_{\alpha _{2}}T_{\left( +\right) \alpha _{1}}^{V}
\label{tensor-B} \\
\mathcal{S}_{\left( -\right) \alpha ;\rho \sigma }^{\Gamma _{1}}
&=&4T_{\left( -\right) \alpha ;\rho \sigma }^{\Gamma _{1}}+2q_{\sigma
}T_{\left( -\right) \alpha ;\rho }^{\Gamma _{1}}+2q_{\rho }T_{\left(
-\right) \alpha ;\sigma }^{\Gamma _{1}}+q_{\rho }q_{\sigma }T_{\left(
-\right) \alpha }^{\Gamma _{1}},  \label{tensor-S}
\end{eqnarray}%
being careful to remind that $T_{\left( \pm \right) }^{\Gamma _{1}}$ stands
for the difference or sum of one-point functions%
\begin{equation*}
T_{\left( \pm \right) }^{\Gamma _{1}}=T^{\Gamma _{1}}\left( k_{1}\right) \pm
T^{\Gamma _{1}}\left( k_{2}\right) .
\end{equation*}%
For contractions with metric, only $\mathcal{B}_{\alpha _{1};\alpha _{2}}$
arises; for momentum contraction in matrix indices, only $\mathcal{S}%
_{\left( -\right) \alpha _{1};\mu _{2}\alpha _{2}}^{\Gamma _{1}}$ is
present. In contrast, for derivatives indexes, there arises both.

In the course of the previous chapter, we dealt with a set of finite
functions that are identically zero due to relations among the scalar and
vector $J_{2}$-integrals of for equal masses (\ref{J1J0}) coming from the
reduction for $Z_{1}^{\left( -1\right) }$ (\ref{Z1Z0}). Here we list them to
make it easier to follow the next stages of derivations.%
\begin{eqnarray}
T_{\mu }^{SV} &=&+2m\left( 2J_{2\mu }+q_{\mu }J_{2}\right) \equiv 0
\label{SV=0} \\
T_{\mu }^{SA} &=&-2m\varepsilon _{\mu \nu }\left( 2J_{2}^{\nu }+q^{\nu
}J_{2}\right) \equiv 0 \\
2T_{\mu ;\alpha }^{PV}+q_{\alpha }T_{\mu }^{PV} &=&-2m\varepsilon _{\mu \nu
}q^{\nu }\left( 2J_{2\alpha }+q_{\alpha }J_{2}\right) \equiv 0
\label{D1PV=0} \\
2T_{\mu ;\alpha }^{PA}+q_{\alpha }T_{\mu }^{PA} &=&+2mq_{\mu }\left(
2J_{2\alpha }+q_{\alpha }J_{2}\right) \equiv 0  \label{D1PA=0}
\end{eqnarray}

Amplitudes with non-negative power counting that we meet by studying the RHS
of RAGFs are combinations of the set $\left\{ SV,SA,PV,PA\right\} $ and
contain one or two derivative indices. Among those amplitudes is a set of
relevant identities fully used to systematize the final results.%
\begin{eqnarray}
T_{\mu _{2};\alpha _{2}}^{SV} &=&T_{\mu _{2};\alpha _{2}}^{VS}=2m(2\bar{J}%
_{2\mu _{2}\alpha _{2}}+q_{\mu _{2}}J_{2\alpha _{2}})  \label{D1SV} \\
T_{\mu _{1};\mu _{2}}^{SV} &=&2m(\Delta _{2\mu _{1}\mu _{2}}+g_{\mu _{1}\mu
_{2}}I_{\log })-\frac{im}{2\pi }\theta _{\mu _{1}\mu _{2}}[2Z_{2}^{\left(
-1\right) }-Z_{1}^{\left( -1\right) }] \\
T_{\mu _{1};\mu _{2}}^{AS} &=&T_{\mu _{1};\mu _{2}}^{SA}=-2m\varepsilon
_{\mu _{1}\nu }(2\bar{J}_{2\mu _{2}}^{\nu }+q^{\nu }J_{2\mu _{2}}) \\
T_{\mu _{2};\alpha _{2}}^{AS} &=&-\varepsilon _{\mu _{2}}^{\quad \nu }T_{\nu
;\alpha _{2}}^{VS} \\
2T_{\alpha _{1};\mu _{2}\alpha _{2}}^{PA}+q_{\mu _{2}}T_{\alpha _{1};\alpha
_{2}}^{PA} &=&q_{\alpha _{1}}T_{\mu _{2};\alpha _{2}}^{SV}  \label{D2PA=SV}
\\
2T_{\alpha _{1};\mu _{2}\alpha _{2}}^{PV}+q_{\mu _{2}}T_{\alpha _{1};\alpha
_{2}}^{PV} &=&-\varepsilon _{\alpha _{1}\nu }q^{\nu }T_{\mu _{2};\alpha
_{2}}^{SV}  \label{D2PV=-eSV}
\end{eqnarray}

All the 4th-order tensors corresponding to a $VV$-$AA$ and $AV$-$VA$ can be
expressed as%
\begin{eqnarray}
\mathcal{T}_{\mu _{1}\alpha _{1}\mu _{2}\alpha _{2}}^{VV} &=&+\frac{4\Omega
_{\mu _{1}\alpha _{1}\mu _{2}\alpha _{2}}}{q^{2}}\left\{ -\frac{i}{\left(
4\pi \right) }[2Z_{2}^{\left( 0\right) }-Z_{1}^{\left( 0\right) }]+\frac{1}{6%
}I_{\text{\textrm{log}}}\right\} \\
&&+\frac{i}{\left( 4\pi \right) }\frac{8\theta _{\mu _{1}\alpha _{1}}\theta
_{\mu _{2}\alpha _{2}}}{q^{2}}[3Z_{2}^{\left( 0\right) }-2Z_{1}^{\left(
0\right) }]+\mathcal{D}_{\mu _{1}\alpha _{1}\mu _{2}\alpha _{2}}^{VV}, 
\notag
\end{eqnarray}%
with attention to their finite parts.

To express the relations due to contractions with derivative indices we list
the identities needed for the exchange indices and reduce the verification
to the contractions with the matrix indices (coming from $\Gamma _{i}$):%
\begin{eqnarray*}
2T_{\mu _{1}\alpha _{1};\mu _{2}}^{VV}+q_{\mu _{2}}T_{\mu _{1}\alpha
_{1}}^{VV} &=&2T_{\mu _{2}\alpha _{1};\mu _{1}}^{VV}+q_{\mu _{1}}T_{\mu
_{2}\alpha _{1}}^{VV}+g_{\mu _{2}\alpha _{1}}T_{\left( +\right) \mu
_{1}}^{V}-g_{\mu _{1}\alpha _{1}}T_{\left( +\right) \mu _{2}}^{V} \\
2T_{\mu _{1}\alpha _{1};\mu _{2}\alpha _{2}}^{VV}+q_{\mu _{2}}T_{\mu
_{1}\alpha _{1};\alpha _{2}}^{VV} &=&2T_{\mu _{2}\alpha _{1};\mu _{1}\alpha
_{2}}^{VV}+q_{\mu _{1}}T_{\mu _{2}\alpha _{1};\alpha _{2}}^{VV}+2R_{\mu
_{1}\alpha _{1};\mu _{2}\alpha _{2}} \\
&&+g_{\mu _{2}\alpha _{1}}T_{\left( +\right) \mu _{1};\alpha
_{2}}^{V}-g_{\mu _{1}\alpha _{1}}T_{\left( +\right) \mu _{2};\alpha
_{2}}^{V}.
\end{eqnarray*}%
Multiplying by two the second identity and summing both, we have an
expression of basic permutation given by 
\begin{equation}
\mathcal{T}_{\mu _{1}\alpha _{1}\mu _{2}\alpha _{2}}^{VV}=\mathcal{T}_{\mu
_{2}\alpha _{1};\mu _{1}\alpha _{2}}^{VV}+g_{\mu _{2}\alpha _{1}}\mathcal{B}%
_{\mu _{1};\alpha _{2}}-g_{\mu _{1}\alpha _{1}}\mathcal{B}_{\mu _{2};\alpha
_{2}}+4R_{\mu _{1}\alpha _{1};\mu _{2}\alpha _{2}}.  \notag
\end{equation}%
Double axial amplitudes follows (\ref{AA-VV})-(\ref{AA-VV2}).

\textbf{Odd amplitudes}: The $AV$-amplitudes:%
\begin{eqnarray*}
2(T_{\mu _{1}\alpha _{1};\mu _{2}}^{AV})_{1}+q_{\mu _{2}}(T_{\mu _{1}\alpha
_{1}}^{AV})_{1} &=&2(T_{\mu _{1}\mu _{2};\alpha _{1}}^{AV})_{1}+q_{\alpha
_{1}}(T_{\mu _{1}\mu _{2}}^{AV})_{1}+\varepsilon _{\mu _{1}\alpha
_{1}}T_{\left( +\right) \mu _{2}}^{V}-\varepsilon _{\mu _{1}\mu
_{2}}T_{\left( +\right) \alpha _{1}}^{V} \\
2(T_{\mu _{1}\alpha _{1};\mu _{2}}^{AV})_{2}+q_{\mu _{2}}(T_{\mu _{1}\alpha
_{1}}^{AV})_{2} &=&2(T_{\mu _{2}\alpha _{1};\mu _{1}}^{AV})_{2}+q_{\mu
_{1}}(T_{\mu _{2}\alpha _{1}}^{AV})_{2}-\varepsilon _{\alpha _{1}\mu
_{2}}T_{\left( +\right) \mu _{1}}^{V}+\varepsilon _{\alpha _{1}\mu
_{1}}T_{\left( +\right) \mu _{2}}^{V}
\end{eqnarray*}%
\begin{eqnarray*}
2(T_{\mu _{1}\alpha _{1};\mu _{2}\alpha _{2}}^{AV})_{1}+q_{\mu _{2}}(T_{\mu
_{1}\alpha _{1};\alpha _{2}}^{AV})_{1} &=&2(T_{\mu _{1}\mu _{2};\alpha
_{1}\alpha _{2}}^{AV})_{1}+q_{\alpha _{1}}(T_{\mu _{1}\mu _{2};\alpha
_{2}}^{AV})_{1}-2\varepsilon _{\mu _{1}}^{\quad \nu }R_{\alpha _{1}\nu ;\mu
_{2}\alpha _{2}} \\
&&+\varepsilon _{\mu _{1}\alpha _{1}}T_{\left( +\right) \mu _{2};\alpha
_{2}}^{V}-\varepsilon _{\mu _{1}\mu _{2}}T_{\left( +\right) \alpha
_{1};\alpha _{2}}^{V} \\
2(T_{\mu _{1}\alpha _{1};\alpha _{2}\mu _{2}}^{AV})_{2}+q_{\mu _{2}}(T_{\mu
_{1}\alpha _{1};\alpha _{2}}^{AV})_{2} &=&2(T_{\mu _{2}\alpha _{1};\mu
_{1}\alpha _{2}}^{AV})_{2}+q_{\mu _{1}}(T_{\mu _{2}\alpha _{1};\alpha
_{2}}^{AV})_{2}-2\varepsilon _{\alpha _{1}}^{\quad \nu _{1}}R_{\mu _{1}\nu
_{1};\mu _{2}\alpha _{2}} \\
&&-\varepsilon _{\alpha _{1}\mu _{2}}T_{\left( +\right) \mu _{1};\alpha
_{2}}^{V}+\varepsilon _{\alpha _{1}\mu _{1}}T_{\left( +\right) \mu
_{2};\alpha _{2}}^{V} \\
&&-2m[\varepsilon _{\alpha _{1}\mu _{2}}T_{\mu _{1};\alpha
_{2}}^{SV}-\varepsilon _{\alpha _{1}\mu _{1}}T_{\alpha _{2};\mu _{2}}^{SV}]
\end{eqnarray*}%
We have omitted the $VA$ formulas because, as was seen in the previous
chapter, they are perfectly retrievable from $AV$ ones.

\section{Table of RAGFs}

\textbf{Even Amplitudes:}%
\begin{eqnarray}
q^{\mu _{1}}T_{\mu _{1}\alpha _{1}}^{VV} &=&T_{\left( +\right) \alpha
_{1}}^{V} \\
2q^{\mu _{1}}T_{\mu _{1}\alpha _{1};\alpha _{2}}^{VV} &=&2T_{\left( +\right)
\alpha _{1};\alpha _{2}}^{V} \\
4q^{\mu _{1}}T_{\mu _{1}\alpha _{1};\mu _{2}\alpha _{2}}^{VV} &=&4T_{\left(
+\right) \alpha _{1};\alpha _{2}\mu _{2}}^{V}
\end{eqnarray}%
\begin{eqnarray}
2g^{\mu _{12}}T_{\mu _{1}\alpha _{1};\mu _{2}}^{VV} &=&2\left[ mT_{\alpha
_{1}}^{SV}+T_{\alpha _{1}}^{V}\left( k_{2}\right) \right] +\left( 2\Upsilon
_{\alpha _{1}}+q_{\alpha _{1}}\Upsilon \right) \\
2g^{\mu _{12}}T_{\mu _{1}\alpha _{1};\mu _{2}\alpha _{2}}^{VV} &=&2\left[
mT_{\alpha _{1};\alpha _{2}}^{SV}+T_{\alpha _{1};\alpha _{2}}^{V}\left(
k_{2}\right) \right] +\left( 2\Upsilon _{\alpha _{1}\alpha _{2}}+q_{\alpha
_{1}}\Upsilon _{\alpha _{2}}\right)
\end{eqnarray}%
The contractions with $g^{\alpha _{12}}$ have the same results.

\textbf{Odd amplitudes:} \ 
\begin{eqnarray}
q^{\mu _{1}}(T_{\mu _{1}\alpha _{1}}^{AV})_{1} &=&T_{\left( -\right) \alpha
_{1}}^{A}-2mT_{\alpha _{1}}^{PV}+\varepsilon _{\alpha _{1}\nu _{1}}q^{\nu
_{1}}\Upsilon \\
2q^{\mu _{1}}(T_{\mu _{1}\alpha _{1};\mu _{2}}^{AV})_{1} &=&2T_{\left(
-\right) \alpha _{1};\mu _{2}}^{A}-4mT_{\alpha _{1};\mu
_{2}}^{PV}+2\varepsilon _{\alpha _{1}\nu _{1}}q^{\nu _{1}}\Upsilon _{\mu
_{2}} \\
4q^{\mu _{1}}(T_{\mu _{1}\alpha _{1};\mu _{2}\alpha _{2}}^{AV})_{1}
&=&4T_{\left( -\right) \alpha _{1};\mu _{2}\alpha _{2}}^{A}-8mT_{\alpha
_{1};\mu _{2}\alpha _{2}}^{PV}+4\varepsilon _{\alpha _{1}\nu _{1}}q^{\nu
_{1}}\Upsilon _{\mu _{2}\alpha _{2}},
\end{eqnarray}%
remmember that $T_{\left( -\right) \alpha _{1}}^{A}=[T_{\alpha
_{1}}^{A}\left( k_{1}\right) -T_{\alpha _{1}}^{A}\left( k_{2}\right) ]$. The
other relations for $q^{\alpha _{1}}$-contraction, 
\begin{eqnarray}
q^{\alpha _{1}}(T_{\mu _{1}\alpha _{1}}^{AV})_{1} &=&T_{\left( -\right) \mu
_{1}}^{A} \\
2q^{\alpha _{1}}(T_{\mu _{1}\alpha _{1};\mu _{2}}^{AV})_{1} &=&2T_{\left(
-\right) \mu _{1};\mu _{2}}^{A} \\
4q^{\alpha _{1}}(T_{\mu _{1}\alpha _{1};\mu _{2}\alpha _{2}}^{AV})_{1}
&=&4T_{\left( -\right) \mu _{1};\mu _{2}\alpha _{2}}^{A}.
\end{eqnarray}%
Organizing the trace relations in the form they appear in this part: 
\begin{eqnarray}
2g^{\mu _{1}\mu _{2}}(T_{\mu _{1}\alpha _{1};\mu _{2}}^{AV})_{1}
&=&2mT_{\alpha _{1}}^{PV}+2T_{\alpha _{1}}^{A}\left( k_{2}\right)
-\varepsilon _{\alpha _{1}\nu }q^{\nu }\Upsilon \\
2g^{\mu _{1}\mu _{2}}(T_{\mu _{1}\alpha _{1};\mu _{2}\alpha _{2}}^{AV})_{1}
&=&2mT_{\alpha _{1};\alpha _{2}}^{PV}+2T_{\alpha _{1};\alpha _{2}}^{A}\left(
k_{2}\right) -\varepsilon _{\alpha _{1}\nu }q^{\nu }\Upsilon _{\alpha _{2}}
\\
2g^{\mu _{1}\mu _{2}}(T_{\mu _{1}\alpha _{1};\mu _{2}}^{AV})_{2}
&=&2mT_{\alpha _{1}}^{PV}+2T_{\alpha _{1}}^{A}\left( k_{2}\right)
-\varepsilon _{\alpha _{1}\nu }\left( 2\Upsilon ^{\nu }+q^{\nu }\Upsilon
\right) \\
2g^{\mu _{1}\mu _{2}}(T_{\mu _{1}\alpha _{1};\mu _{2}\alpha _{2}}^{AV})_{2}
&=&2mT_{\alpha _{1};\alpha _{2}}^{PV}+2T_{\alpha _{1};\alpha _{2}}^{A}\left(
k_{2}\right) -\varepsilon _{\alpha _{1}\nu }\left( 2\Upsilon _{\alpha
_{2}}^{\nu }+q^{\nu }\Upsilon _{\alpha _{2}}\right)
\end{eqnarray}%
\begin{eqnarray}
2g^{\alpha _{1}\alpha _{2}}(T_{\mu _{1}\alpha _{1};\alpha _{2}}^{AV})_{1}
&=&-\varepsilon _{\mu _{1}\nu }\left( 2\Upsilon ^{\nu }+q^{\nu }\Upsilon
\right) +2mT_{\mu _{1}}^{AS}+2T_{\mu _{1}}^{A}\left( k_{2}\right) \\
2g^{\alpha _{1}\alpha _{2}}(T_{\mu _{1}\alpha _{1};\mu _{2}\alpha
_{2}}^{AV})_{1} &=&-\varepsilon _{\mu _{1}\nu }(2\Upsilon _{\mu _{2}}^{\nu
}+q^{\nu }\Upsilon _{\mu _{2}})+2mT_{\mu _{1};\mu _{2}}^{AS}+2T_{\mu
_{1};\mu _{2}}^{A}\left( k_{2}\right) \\
2g^{\alpha _{1}\alpha _{2}}(T_{\mu _{1}\alpha _{1};\alpha _{2}}^{AV})_{2}
&=&-\varepsilon _{\mu _{1}\nu }q^{\nu }\Upsilon +2mT_{\mu _{1}}^{AS}+2T_{\mu
_{1}}^{A}\left( k_{2}\right) \\
2g^{\alpha _{1}\alpha _{2}}(T_{\mu _{1}\alpha _{1};\mu _{2}\alpha
_{2}}^{AV})_{2} &=&-\varepsilon _{\mu _{1}\nu }q^{\nu }\Upsilon _{\mu
_{2}}+2mT_{\mu _{1};\mu _{2}}^{AS}+2T_{\mu _{1};\mu _{2}}^{A}\left(
k_{2}\right)
\end{eqnarray}

\subsection{Even amplitudes: $(\mathcal{T}_{\protect\mu \protect\alpha 
\protect\sigma \protect\rho }^{VV})$ and $(\mathcal{T}_{\protect\mu \protect%
\alpha \protect\sigma \protect\rho }^{AA})$}

From now on, we will systematically explore all the results from the
amplitude combinations that effectively appear in the relations for the
gravitational amplitude. Starting by (\ref{AA-VV})-(\ref{AA-VV2}) follows 
\begin{equation}
\mathcal{T}_{\mu _{1}\alpha _{1}\mu _{2}\alpha _{2}}^{AA}=\mathcal{T}_{\mu
_{1}\alpha _{1}\mu _{2}\alpha _{2}}^{VV}-2mg_{\mu _{1}\alpha _{1}}(4\bar{J}%
_{2\mu _{2}\alpha _{2}}+2q_{\alpha _{2}}J_{2\mu _{2}}),
\end{equation}%
The terms corresponding to the $J$-vector and $J$-scalar functions do not
appear because their combination is null. The relation is given by 
\begin{equation}
\mathcal{T}_{\mu _{1}\alpha _{1}\mu _{2}\alpha _{2}}^{AA}=\mathcal{T}_{\mu
_{1}\alpha _{1}\mu _{2}\alpha _{2}}^{VV}-4mg_{\mu _{1}\alpha _{1}}T_{\mu
_{2};\alpha _{2}}^{SV}.  \label{BAA-VV}
\end{equation}%
That amounts to replacing double axial structures for the double vector
diminishing the number of operations necessary to express the relevant
results.

\subsubsection{Internal Contractions}

The contractions with internal indices for these amplitudes follow from the
definition 
\begin{equation}
q^{\mu _{1}}\mathcal{T}_{\mu _{1}\alpha _{1}\mu _{2}\alpha
_{2}}^{VV}=2q^{\mu _{1}}(2T_{\mu _{1}\alpha _{1};\mu _{2}\alpha
_{2}}^{VV}+q_{\mu _{2}}T_{\mu _{1}\alpha _{1};\alpha _{2}}^{VV})+q^{\mu
_{1}}(2q_{\alpha _{2}}T_{\mu _{1}\alpha _{1};\mu _{2}}^{VV}+q_{\alpha
_{2}}q_{\mu _{2}}T_{\mu _{1}\alpha _{1}}^{VV}).
\end{equation}%
The index of $q^{\mu _{1}}$ hits only the matrix vertex of the amplitude,
and the consequence is that only the difference of one-point functions
appears, see (\ref{Relmu1}), (\ref{pTVV3}) and (\ref{pTVV4}). Hence,
employing our definition 
\begin{equation}
\mathcal{S}_{\left( -\right) \alpha _{1};\alpha _{2}\mu
_{2}}^{V}=2[2T_{\left( -\right) \alpha _{1};\mu _{2}\alpha _{2}}^{V}+q_{\mu
_{2}}T_{\left( -\right) \alpha _{1};\alpha _{2}}^{V}]+q_{\alpha
_{2}}[2T_{\left( -\right) \alpha _{1};\mu _{2}}^{V}+q_{\alpha _{2}}T_{\left(
-\right) \alpha _{1}}^{V}],
\end{equation}%
the equation obtained reads 
\begin{equation}
q^{\mu _{1}}\mathcal{T}_{\mu _{1}\alpha _{1}\mu _{2}\alpha _{2}}^{VV}=%
\mathcal{S}_{\left( -\right) \alpha _{1};\alpha _{2}\mu _{2}}^{V}=\mathcal{S}%
_{\alpha _{1};\mu _{2}\alpha _{2}}^{V}\left( k_{1}\right) -\mathcal{S}%
_{\alpha _{1};\mu _{2}\alpha _{2}}^{V}\left( k_{2}\right) .
\end{equation}%
Note the symmetry in the indices corresponding to derivatives, $\mathcal{S}%
_{\alpha _{1};\mu _{2}\alpha _{2}}^{\Gamma _{1}}=\mathcal{S}_{\alpha
_{1};\alpha _{2}\mu _{2}}^{\Gamma _{1}}$.

For the $\mathcal{T}^{AA}$, we could either use for its contraction the $PA$%
's as in (\ref{pMUTAA0})-(\ref{pMUTAA2}),%
\begin{equation}
q^{\mu _{1}}\mathcal{T}_{\mu _{1}\alpha _{1}\mu _{2}\alpha
_{2}}^{AA}=-2m[4T_{\alpha _{1};\mu _{2}\alpha _{2}}^{PA}+2q_{\mu
_{2}}T_{\alpha _{1};\alpha _{2}}^{PA}+q_{\alpha _{2}}(2T_{\alpha _{1};\mu
_{2}}^{PA}+q_{\mu _{2}}T_{\alpha _{1}}^{PA})]+\mathcal{S}_{\left( -\right)
\alpha _{1};\mu _{2}\alpha _{2}}^{V}
\end{equation}%
which is their composition of RAGFs. Using the connection with $\mathcal{T}%
^{VV}$ (\ref{BAA-VV}), we have 
\begin{equation}
q^{\mu _{1}}\mathcal{T}_{\mu _{1}\alpha _{1}\mu _{2}\alpha _{2}}^{AA}=%
\mathcal{S}_{\left( -\right) \alpha _{1};\mu _{2}\alpha
_{2}}^{V}-4mq_{\alpha _{1}}T_{\mu _{2};\alpha _{2}}^{SV}.
\end{equation}%
The $PA$ amplitudes did not appear since they are related to derivative $SV$
\ through (\ref{D1PA=0}) and (\ref{D2PA=SV}). In this amplitude, if the
operation is done in $\alpha _{1},$ the RHS shows $AP$-structures, however,
with the opposite sign. As $T_{\mu }^{AP}=-T_{\mu }^{PA}$ and so on for more
indices, hence the results written in terms of $T_{\mu _{2};\alpha
_{2}}^{SV} $ amplitude have the same functional form.

\subsubsection{External Contractions}

Terms from relations involving the derivative indices organize in the tensor 
$\mathcal{B}_{\mu ;\alpha }$ besides $S_{\alpha _{1};\alpha _{2}\mu
_{1}}^{V},$ see Eqs. (\ref{tensor-B}) and (\ref{tensor-S}). To see this, we
combine the identities used to trade a derivative for a matrix index, as in (%
\ref{IDtroca1}) and (\ref{IDtroca2}). We have,%
\begin{equation}
\mathcal{T}_{\mu _{1}\alpha _{1};\mu _{2}\alpha _{2}}^{VV}=\mathcal{T}_{\mu
_{2}\alpha _{1};\mu _{1}\alpha _{2}}^{VV}+g_{\mu _{2}\alpha _{1}}\mathcal{B}%
_{\mu _{1};\alpha _{2}}-g_{\mu _{1}\alpha _{1}}\mathcal{B}_{\mu _{2};\alpha
_{2}}+2R_{\mu _{1}\alpha _{1};\mu _{2}\alpha _{2}}.  \notag
\end{equation}%
Note the presence of $B_{\sigma ;\rho }$ and the residual term, which always
vanishes under contraction. Contracting with $q^{\mu _{2}}$, the first term
in the RHS, we have $\mu _{2}$ index in the position of a matrix index,
whose result we developed previously. Follows the compact result 
\begin{equation}
q^{\mu _{2}}\mathcal{T}_{\mu _{1}\alpha _{1};\mu _{2}\alpha _{2}}^{VV}=%
\mathcal{S}_{\left( -\right) \alpha _{1};\mu _{1}\alpha _{2}}^{V}+q_{\alpha
_{1}}\mathcal{B}_{\mu _{1};\alpha _{2}}-g_{\mu _{1}\alpha _{1}}q^{\nu }%
\mathcal{B}_{\nu ;\alpha _{2}}.
\end{equation}

For the $\mathcal{T}^{AA}$, we substitute the equation (\ref{BAA-VV}) into
the last one, what implies in%
\begin{equation}
q^{\mu _{2}}\mathcal{T}_{\mu _{1}\alpha _{1}\mu _{2}\alpha _{2}}^{AA}=%
\mathcal{S}_{\left( -\right) \alpha _{1};\mu _{1}\alpha _{2}}^{V}+q_{\alpha
_{1}}\mathcal{B}_{\mu _{1};\alpha _{2}}-g_{\mu _{1}\alpha _{1}}q^{\nu }%
\mathcal{B}_{\nu ;\alpha _{2}}-4mg_{\mu _{1}\alpha _{1}}T_{\left( -\right)
\alpha _{2}}^{S}.
\end{equation}%
remember that $q^{\nu }T_{\nu ;\alpha _{2}}^{SV}=T_{\alpha _{2}}^{S}\left(
k_{1}\right) -T_{\alpha _{2}}^{S}\left( k_{2}\right) =T_{\left( -\right)
\alpha _{2}}^{S}.$ There does not exist any condition for the momentum
RAGFs. A different scenario occurs to the metric RAGFs.

\subsubsection{Metric Contractions}

These relations combine the metric relations of the basic derivative
amplitudes and the momentum relations for the matrix indices. Make explicit
this property by%
\begin{equation}
g^{\mu _{12}}\mathcal{T}_{\mu _{1}\alpha _{1}\mu _{2}\alpha
_{2}}^{VV}=2g^{\mu _{12}}(2T_{\mu _{1}\alpha _{1};\mu _{2}\alpha
_{2}}^{VV}+q_{\alpha _{2}}T_{\mu _{1}\alpha _{1};\mu _{2}}^{VV})+q^{\mu
_{1}}(2T_{\mu _{1}\alpha _{1};\alpha _{2}}^{VV}+q_{\alpha _{2}}T_{\mu
_{1}\alpha _{1}}^{VV}).
\end{equation}%
The next stage is observing that momentum RAGFs in even amplitudes are
automatically satisfied. Replacing them and e summing with the equations for
metric contractions (\ref{traceVV1}) and (\ref{traceVV2}), we arrive at%
\begin{eqnarray}
g^{\mu _{12}}\mathcal{T}_{\mu _{1}\alpha _{1}\mu _{2}\alpha _{2}}^{VV}
&=&+4mT_{\alpha _{1};\alpha _{2}}^{SV}+[2T_{\left( +\right) \alpha
_{1};\alpha _{2}}^{V}+q_{\alpha _{2}}T_{\left( +\right) \alpha _{1}}^{V}] \\
&&+4\Upsilon _{\alpha _{1}\alpha _{2}}+2q_{\alpha _{1}}\Upsilon _{\alpha
_{2}}+2q_{\alpha _{2}}\Upsilon _{\alpha _{1}}+q_{\alpha _{2}}q_{\alpha
_{1}}\Upsilon ,  \notag
\end{eqnarray}%
where we used the pattern that appears in one-point functions, $T_{\left(
-\right) \alpha _{1}}^{V}+2T_{\alpha _{1}}^{V}\left( k_{2}\right) =T_{\left(
+\right) \alpha _{1}}^{V}$. We dropped the $T_{\alpha _{1}}^{SV}=0$ term.

The conditioning factors $\left\{ \Upsilon ,\Upsilon _{\alpha _{1}},\Upsilon
_{\alpha _{1}\alpha _{2}}\right\} $ were combined in a fundamental tensor
called uniqueness factor; it will encompass the conditions for satisfaction
of all RAGFs as well the equivalence of the odd-amplitude versions. Because
of its importance, we define it as 
\begin{equation}
U_{\alpha _{1}\alpha _{2}}=4\Upsilon _{\alpha _{1}\alpha _{2}}+2q_{\alpha
_{1}}\Upsilon _{\alpha _{2}}+2q_{\alpha _{2}}\Upsilon _{\alpha
_{1}}+q_{\alpha _{2}}q_{\alpha _{1}}\Upsilon .
\end{equation}%
The investigation of values assumed to this tensor and its connection to the
finite part and surface terms will be developed soon. Thus, we have the
compact expression 
\begin{equation}
g^{\mu _{12}}\mathcal{T}_{\mu _{1}\alpha _{1}\mu _{2}\alpha
_{2}}^{VV}=4mT_{\alpha _{1};\alpha _{2}}^{SV}+\mathcal{B}_{\alpha
_{1};\alpha _{2}}+U_{\alpha _{1}\alpha _{2}}.
\end{equation}%
The relations for $g^{\alpha _{12}}$-contraction are identical, changing the
indices $\mu _{12}\leftrightarrow \alpha _{12}.$

Calculating directly or using the relation (\ref{BAA-VV}) between $\mathcal{T%
}^{AA}$ and $\mathcal{T}^{VV}$, follows%
\begin{eqnarray}
g^{\mu _{12}}\mathcal{T}_{\mu _{1}\alpha _{1}\mu _{2}\alpha _{2}}^{AA} &=&%
\mathcal{B}_{\alpha _{1};\alpha _{2}}+U_{\alpha _{1}\alpha _{2}}, \\
g^{\alpha _{12}}\mathcal{T}_{\mu _{1}\alpha _{1}\mu _{2}\alpha _{2}}^{AA} &=&%
\mathcal{B}_{\mu _{1};\mu _{2}}+U_{\mu _{1}\mu _{2}}.
\end{eqnarray}

\textbf{Uniqueness factor: }The definitions follow in (\ref{Ups0}), (\ref%
{Ups1}) and (\ref{Ups2}), thus%
\begin{eqnarray}
U_{\alpha _{1}\alpha _{2}} &=&-\frac{1}{3}\theta _{\alpha _{1}\alpha
_{2}}\Upsilon  \label{U} \\
&&+\frac{1}{9}(3P^{\nu _{12}}+q^{\nu _{12}})\left[ 3\Sigma _{4\rho \alpha
_{12}\nu _{12}}^{\rho }-8\square _{3\alpha _{12}\nu _{12}}-g_{(\alpha
_{12}}g_{\nu _{12})}\Delta _{2\rho }^{\rho }\right]  \notag \\
&&+\frac{1}{18}(3P^{\nu _{12}}+q^{\nu _{12}})g_{(\alpha _{12}}[2\square
_{3\nu _{12})\rho }^{\rho }-2\Delta _{2\nu _{12})}-g_{\nu _{12})}\Delta
_{2\rho }^{\rho }]  \notag \\
&&-P_{\alpha _{2}}P^{\nu }[2(\square _{3\rho \alpha _{1}\nu }^{\rho }-\Delta
_{2\alpha _{1}\nu })-g_{\alpha _{1}\nu }\Delta _{2\rho }^{\rho }]  \notag \\
&&-P_{\alpha _{1}}P^{\nu }[2(\square _{3\rho \alpha _{2}\nu }^{\rho }-\Delta
_{2\alpha _{2}\nu })-g_{\alpha _{2}\nu }\Delta _{2\rho }^{\rho }]  \notag \\
&&-\frac{1}{2}(P^{2}+q^{2})[2(\square _{3\rho \alpha _{12}}^{\rho }-\Delta
_{2\alpha _{12}})-g_{\alpha _{12}}\Delta _{2\rho }^{\rho }]  \notag \\
&&+4[(W_{2\rho \alpha _{12}}^{\rho }-2\Delta _{1\alpha _{12}})+2g_{\alpha
_{1}\alpha _{2}}I_{\text{\textrm{quad}}}-2m^{2}\left( \Delta _{2\alpha
_{12}}+g_{\alpha _{12}}I_{\log }\right) ].  \notag
\end{eqnarray}

\subsection{Odd Amplitudes: $(\mathcal{T}_{\protect\mu \protect\alpha 
\protect\sigma \protect\rho }^{AV})$ and $(\mathcal{T}_{\protect\mu \protect%
\alpha \protect\sigma \protect\rho }^{VA})$}

In this part, a series of considerations are in order. The decomposition in
derivatives was taken to the most basic level; a set of possibilities from
Dirac traces is fully exploited. We came out with two independent forms,
version one and two, as we called them. Now, for any term of the basic
permutation, an arbitrary version choice must be made because the choice of
traces employed is arbitrary. Nonetheless, even if the analysis can be
performed in the most general scenario, we will adopt the position of
considering the uniform version, where $\mathcal{T}_{\mu \alpha \sigma \rho
}^{\Gamma _{12}}$ is an odd tensor. Then we will have the notation 
\begin{eqnarray}
(\mathcal{T}_{\mu _{1}\alpha _{1}\mu _{2}\alpha _{2}}^{AV})_{i}
&=&2[2(T_{\mu _{1}\alpha _{1};\mu _{2}\alpha _{2}}^{AV})_{i}+2q_{\mu
_{2}}(T_{\mu _{1}\alpha _{1};\alpha _{2}}^{AV})_{i}]+  \label{defBAVi} \\
&&+q_{\alpha _{2}}[2(T_{\mu _{1}\alpha _{1};\mu _{2}}^{AV})_{i}+q_{\mu
_{2}}(T_{\mu _{1}\alpha _{1}}^{AV})_{i}]  \notag \\
(\mathcal{T}_{\mu _{1}\alpha _{1}\mu _{2}\alpha _{2}}^{VA})_{i}
&=&2[2(T_{\mu _{1}\alpha _{1};\mu _{2}\alpha _{2}}^{VA})_{i}+q_{\mu
_{2}}(T_{\mu _{1}\alpha _{1};\alpha _{2}}^{VA})_{i}]+  \label{defBVAi} \\
&&+q_{\alpha _{2}}[2(T_{\mu _{1}\alpha _{1};\mu _{2}}^{VA})_{i}+q_{\mu
_{2}}(T_{\mu _{1}\alpha _{1}}^{VA})_{i}],  \notag
\end{eqnarray}%
with $i=1,2$. In this moment we may use the transition equations (\ref{Uni}%
)-(\ref{Un2}) to derive the relations among what we call basic permutations 
\begin{equation}
(\mathcal{T}_{\mu _{1}\alpha _{1}\mu _{2}\alpha _{2}}^{AV})_{2}=(\mathcal{T}%
_{\mu _{1}\alpha _{1}\mu _{2}\alpha _{2}}^{AV})_{1}-\varepsilon _{\alpha
_{1}\mu _{1}}\left( 4\Upsilon _{\mu _{2}\alpha _{2}}+2q_{\mu _{2}}\Upsilon
_{\alpha _{2}}+2q_{\alpha _{2}}\Upsilon _{\mu _{2}}+q_{\alpha _{2}}q_{\mu
_{2}}\Upsilon \right) .
\end{equation}%
In the RHS appear, the $U$-factor, making it simpler to express the
uniqueness relation as%
\begin{equation}
(\mathcal{T}_{\mu _{1}\alpha _{1}\mu _{2}\alpha _{2}}^{AV})_{2}=(\mathcal{T}%
_{\mu _{1}\alpha _{1}\mu _{2}\alpha _{2}}^{AV})_{1}-\varepsilon _{\alpha
_{1}\mu _{1}}U_{\mu _{2}\alpha _{2}}.  \label{BAV2-1}
\end{equation}

Analogously the transition between $AA$-$VV$, the amplitude $\mathcal{T}%
_{\mu _{1}\alpha _{1}\mu _{2}\alpha _{2}}^{VA}$ can be written in term of $%
\mathcal{T}_{\mu _{1}\alpha _{1}\mu _{2}\alpha _{2}}^{AV}$, in a way
independent of traces employed. See Eqs. (\ref{VA-AVt0})-(\ref{VA-AVt2}) to
derive the relation%
\begin{equation}
(\mathcal{T}_{\mu _{1}\alpha _{1}\mu _{2}\alpha _{2}}^{VA})_{i}=(\mathcal{T}%
_{\mu _{1}\alpha _{1}\mu _{2}\alpha _{2}}^{AV})_{i}+4m^{2}\varepsilon _{\mu
_{1}\alpha _{1}}\left[ 4\bar{J}_{2\mu _{2}\alpha _{2}}+2q_{\alpha
_{2}}J_{2\mu _{2}}+q_{\mu _{2}}\left( 2J_{2\alpha _{2}}+q_{\alpha
_{2}}J_{2}\right) \right] ,
\end{equation}%
using (\ref{SV=0})-(\ref{D1SV}) to identify the integrals as amplitudes,\ we
obtain $VA$-$AV$ connection%
\begin{equation}
(\mathcal{T}_{\mu _{1}\alpha _{1}\mu _{2}\alpha _{2}}^{VA})_{i}=(\mathcal{T}%
_{\mu _{1}\alpha _{1}\mu _{2}\alpha _{2}}^{AV})_{i}+4m\varepsilon _{\mu
_{1}\alpha _{1}}T_{\mu _{2};\alpha _{2}}^{SV}.  \label{BVA-AVt}
\end{equation}%
This enables us to study only the versions $(\mathcal{T}^{AV})_{1}$.

\subsection{Permutation's versions: $(\mathcal{T}_{\protect\mu \protect%
\alpha \protect\sigma \protect\rho }^{AV})_{1}$ and $(\mathcal{T}_{\protect%
\mu \protect\alpha \protect\sigma \protect\rho }^{AV})_{2}$}

\subsubsection{Momentum: Internal Contractions}

To make apparent the notation's use, let us explore the internal contraction
with $q^{\mu }(\mathcal{T}_{\mu \alpha \sigma \rho }^{AV})_{i}$. We begin
with the definition (\ref{defBAVi}) and the formulas generalized in (\ref%
{qAV1(gen)})-(\ref{qAV3(gen)}). Notice that those relations turn up with $%
\Upsilon $ factors; summing the contributions,%
\begin{eqnarray}
q^{\mu _{1}}(\mathcal{T}_{\mu _{1}\alpha _{1}\mu _{2}\alpha _{2}}^{AV})_{1}
&=&-2m[4T_{\alpha _{1};\mu _{2}\alpha _{2}}^{PV}+2q_{\mu _{2}}T_{\alpha
_{1};\alpha _{2}}^{PV}+2q_{\alpha _{2}}T_{\alpha _{1};\mu
_{2}}^{PV}+q_{\alpha _{2}}q_{\mu _{2}}T_{\alpha _{1}}^{PV}] \\
&&+\mathcal{S}_{\alpha _{1};\alpha _{2}\mu _{2}}^{A}\left( k_{1}\right) -%
\mathcal{S}_{\alpha _{1};\alpha _{2}\mu _{2}}^{A}\left( k_{2}\right)
+\varepsilon _{\alpha _{1}\nu _{1}}q^{\nu _{1}}U_{\mu _{2}\alpha _{2}}. 
\notag
\end{eqnarray}%
We gathered the one-point functions in our definition of $\mathcal{S}%
_{\alpha _{1};\alpha _{2}\mu _{2}}^{A}$. The identities (\ref{D2PV=-eSV})
and (\ref{D1PV=0}) involving the $PV$ enables one to write the result%
\begin{equation}
q^{\mu _{1}}(\mathcal{T}_{\mu _{1}\alpha _{1}\mu _{2}\alpha
_{2}}^{AV})_{1}=4m\varepsilon _{\alpha _{1}\nu }q^{\nu }T_{\mu _{2};\alpha
_{2}}^{SV}+\mathcal{S}_{\left( -\right) \alpha _{1};\alpha _{2}\mu
_{2}}^{A}+\varepsilon _{\alpha _{1}\nu _{1}}q^{\nu _{1}}U_{\mu _{2}\alpha
_{2}}.
\end{equation}%
For the contraction with $q^{\alpha _{1}},$ the relations to the component
amplitudes are identically satisfied. Hence there are no $\Upsilon $
factors, namely%
\begin{equation}
q^{\alpha _{1}}(\mathcal{T}_{\mu _{1}\alpha _{1}\mu _{2}\alpha
_{2}}^{AV})_{1}=\mathcal{S}_{\left( -\right) \mu _{1};\mu _{2}\alpha
_{2}}^{A}=\mathcal{S}_{\mu _{1};\mu _{2}\alpha _{2}}^{A}\left( k_{1}\right) -%
\mathcal{S}_{\mu _{1};\mu _{2}\alpha _{2}}^{A}\left( k_{2}\right) .
\end{equation}%
The other form of the basic permutation will readily comply with the
equations%
\begin{eqnarray}
q^{\mu _{1}}(\mathcal{T}_{\mu _{1}\alpha _{1}\mu _{2}\alpha _{2}}^{AV})_{2}
&=&\mathcal{S}_{\left( -\right) \alpha _{1};\alpha _{2}\mu
_{2}}^{A}+4m\varepsilon _{\alpha _{1}\nu }q^{\nu }T_{\mu _{2};\alpha
_{2}}^{SV} \\
q^{\alpha _{1}}(\mathcal{T}_{\mu _{1}\alpha _{1}\mu _{2}\alpha
_{2}}^{AV})_{2} &=&\mathcal{S}_{\left( -\right) \mu _{1};\mu _{2}\alpha
_{2}}^{A}+\varepsilon _{\mu _{1}\nu }q^{\nu }U_{\mu _{2}\alpha _{2}}.
\end{eqnarray}

\subsubsection{Momentum: External contractions}

We have one identity automatically satisfied and one with $U$-factor.
Beginning by 
\begin{eqnarray}
q^{\mu _{2}}(\mathcal{T}_{\mu _{1}\alpha _{1}\mu _{2}\alpha _{2}}^{AV})_{1}
&=&4q^{\mu _{2}}(T_{\mu _{1}\alpha _{1};\mu _{2}\alpha
_{2}}^{AV})_{1}+2q^{2}(T_{\mu _{1}\alpha _{1};\alpha _{2}}^{AV})_{1} \\
&&+2q_{\alpha _{2}}q^{\mu _{2}}(T_{\mu _{1}\alpha _{1};\mu
_{2}}^{AV})_{1}+q_{\alpha _{2}}q^{2}(T_{\mu _{1}\alpha _{1}}^{AV})_{1}. 
\notag
\end{eqnarray}%
The equation below can be written in compact form through the use of
formulae developed before that do not require any new ingredient but careful
application, 
\begin{equation}
q^{\mu _{2}}(\mathcal{T}_{\mu _{1}\alpha _{1}\mu _{2}\alpha _{2}}^{AV})_{1}=%
\mathcal{S}_{\left( -\right) \mu _{1};\alpha _{1}\alpha
_{2}}^{A}+\varepsilon _{\mu _{1}\alpha _{1}}q^{\nu }\mathcal{B}_{\nu ;\alpha
_{2}}-\varepsilon _{\mu _{1}\nu }q^{\nu }\mathcal{B}_{\alpha _{1};\alpha
_{2}}.
\end{equation}%
Making one more manipulation by using $\varepsilon _{\lbrack \mu _{1}\alpha
_{1}}\mathcal{B}_{\nu ];\alpha _{2}}=0$, follows the final form%
\begin{equation}
q^{\mu _{2}}(\mathcal{T}_{\mu _{1}\alpha _{1}\mu _{2}\alpha _{2}}^{AV})_{1}=%
\mathcal{S}_{\left( -\right) \mu _{1};\alpha _{1}\alpha
_{2}}^{A}-\varepsilon _{\alpha _{1}\nu }q^{\nu }\mathcal{B}_{\mu _{1};\alpha
_{2}}.
\end{equation}

The version $(\mathcal{T}_{\mu _{1}\alpha _{1}\mu _{2}\alpha _{2}}^{AV})_{2}$
also have a relation which is satisfied by construction, namely,%
\begin{equation}
q^{\mu _{2}}(\mathcal{T}_{\mu _{1}\alpha _{1}\mu _{2}\alpha _{2}}^{AV})_{2}=%
\mathcal{S}_{\left( -\right) \alpha _{1};\alpha _{2}\mu
_{1}}^{A}-\varepsilon _{\mu _{1}\nu }q^{\nu }\mathcal{B}_{\alpha _{1};\alpha
_{2}}-4m\varepsilon _{\mu _{1}\alpha _{1}}\left[ T_{\alpha _{2}}^{S}\left(
k_{1}\right) -T_{\alpha _{2}}^{S}\left( k_{2}\right) \right] .
\end{equation}%
For this, we have observed the combination of two-point functions (\ref%
{D1PV=0}) and (\ref{D2PV=-eSV}).

Now, the relations where arises the $\Upsilon $ factors came from the use of
the equation that exists between the versions (\ref{BAV2-1}). They furnish 
\begin{eqnarray}
q^{\mu _{2}}(\mathcal{T}_{\mu _{1}\alpha _{1}\mu _{2}\alpha _{2}}^{AV})_{1}
&=&\mathcal{S}_{\left( -\right) \alpha _{1};\alpha _{2}\mu
_{1}}^{A}-\varepsilon _{\mu _{1}\nu }q^{\nu }\mathcal{B}_{\alpha _{1};\alpha
_{2}}+\varepsilon _{\alpha _{1}\mu _{1}}q^{\nu }U_{\nu \alpha
_{2}}-4m\varepsilon _{\mu _{1}\alpha _{1}}T_{\left( -\right) \alpha _{2}}^{S}
\\
q^{\mu _{2}}(\mathcal{T}_{\mu _{1}\alpha _{1}\mu _{2}\alpha _{2}}^{AV})_{2}
&=&\mathcal{S}_{\left( -\right) \mu _{1};\alpha _{1}\alpha
_{2}}^{A}-\varepsilon _{\alpha _{1}\nu }q^{\nu }\mathcal{B}_{\mu _{1};\alpha
_{2}}-\varepsilon _{\alpha _{1}\mu _{1}}q^{\nu }U_{\nu \alpha _{2}}.
\end{eqnarray}%
Two forms obtained for these relations are equivalent. As we saw, we always
kept intact all terms where the results could deviate. Therefore is
straightforward to see that they ought to be equal. Moreover, the ones with
violating terms are obtained by employing those free of $U$-term, using an
identity again. Even so, if one desires to check such a statement
explicitly, the path is reasonably long but feasible. Here we give the
directions; start by using $\mathcal{S}_{\sigma ;\alpha \rho
}^{A}=-\varepsilon _{\sigma }^{\nu }\mathcal{S}_{\nu ;\alpha \rho }^{V}$,
then subtract the identities without $U$ and with $U$,%
\begin{eqnarray}
q^{\mu _{2}}[(\mathcal{T}_{\mu _{1}\alpha _{1}\mu _{2}\alpha
_{2}}^{AV})_{1}-(\mathcal{T}_{\mu _{1}\alpha _{1}\mu _{2}\alpha
_{2}}^{AV})_{1}] &=&\varepsilon _{\alpha _{1}}^{\quad \nu }\mathcal{S}%
_{\left( -\right) \nu ;\alpha _{2}\mu _{1}}^{V}-\varepsilon _{\mu
_{1}}^{\quad \nu }\mathcal{S}_{\left( -\right) \nu ;\alpha
_{12}}^{V}-\varepsilon _{\alpha _{1}\mu _{1}}q^{\nu }U_{\nu \alpha _{2}} \\
&&-\varepsilon _{\alpha _{1}\nu }q^{\nu }\mathcal{B}_{\mu _{1};\alpha
_{2}}+\varepsilon _{\mu _{1}\nu }q^{\nu }\mathcal{B}_{\alpha _{1};\alpha
_{2}}-4m\varepsilon _{\alpha _{1}\mu _{1}}T_{\left( -\right) \alpha
_{2}}^{S},  \notag
\end{eqnarray}%
employing the identities $g^{\nu \rho }\mathcal{S}_{\nu ;\rho \lbrack \alpha
}^{V}\varepsilon _{\sigma \rho ]}=0$ and $\varepsilon _{\lbrack \alpha
_{1}\nu }\mathcal{B}_{\mu _{1}];\alpha _{2}}=0,$ we obtain an expression
where everything is known and whose summation cancels without any conditions,%
\begin{equation}
\varepsilon _{\alpha _{1}\mu _{1}}\{g^{\nu \rho }\mathcal{S}_{\left(
-\right) \nu ;\rho \alpha _{2}}^{V}-q^{\nu }\mathcal{B}_{\nu ;\alpha
_{2}}-4mT_{\left( -\right) \alpha _{2}}^{S}-q^{\nu }U_{\nu \alpha
_{2}}\}\equiv 0.
\end{equation}

\subsubsection{Metric Contractions}

We use the form $(\mathcal{T}^{AV})_{i}$ and perform the analysis for $%
g^{\mu _{12}}$ and $g^{\alpha _{12}}\,$. First, we have%
\begin{eqnarray}
g^{\mu _{12}}(\mathcal{T}_{\mu _{1}\alpha _{1}\mu _{2}\alpha _{2}}^{AV})_{1}
&=&4g^{\mu _{12}}(T_{\mu _{1}\alpha _{1};\mu _{2}\alpha
_{2}}^{AV})_{1}+2q^{\mu _{1}}(T_{\mu _{1}\alpha _{1};\alpha _{2}}^{AV})_{1}
\\
&&+q_{\alpha _{2}}[2g^{\mu _{12}}(T_{\mu _{1}\alpha _{1};\mu
_{2}}^{AV})_{1}+q^{\mu _{1}}(T_{\mu _{1}\alpha _{1}}^{AV})_{1}],  \notag
\end{eqnarray}%
then, recollecting the formulas for traces and gathering the contributions
for momentum contractions, the $PV$ functions from both sectors cancel each
other and the conditioning $\Upsilon $ factors. The remaining $T^{A}$
amplitudes arrange themselves as%
\begin{equation}
g^{\mu _{12}}(\mathcal{T}_{\mu _{1}\alpha _{1}\mu _{2}\alpha
_{2}}^{AV})_{1}=2T_{\left( +\right) \alpha _{1};\alpha _{2}}^{A}+q_{\alpha
_{2}}T_{\left( +\right) \alpha _{1}}^{A}=-\varepsilon _{\alpha _{1}}^{\quad
\nu }\mathcal{B}_{\nu ;\alpha _{2}}.  \notag
\end{equation}%
These amplitudes are precisely related to $T^{V}$ ones.

The equation satisfied by $g^{\alpha _{12}}(\mathcal{T}_{\mu _{1}\alpha
_{1}\mu _{2}\alpha _{2}}^{AV})_{1}$ starts with%
\begin{eqnarray}
g^{\alpha _{12}}(\mathcal{T}_{\mu _{1}\alpha _{1}\mu _{2}\alpha
_{2}}^{AV})_{1} &=&4g^{\alpha _{12}}(T_{\mu _{1}\alpha _{1};\mu _{2}\alpha
_{2}}^{AV})_{1}+2q_{\mu _{2}}g^{\alpha _{12}}(T_{\mu _{1}\alpha _{1};\alpha
_{2}}^{AV})_{1} \\
&&+2q^{\alpha _{1}}(T_{\mu _{1}\alpha _{1};\mu _{2}}^{AV})_{1}+q_{\mu
_{2}}q^{\alpha _{1}}(T_{\mu _{1}\alpha _{1}}^{AV})_{1}.  \notag
\end{eqnarray}%
The first line is the only one with conditioning factors; the momentum
contraction is identically satisfied because the relation appears for the
second vertex (specifically a vector one) and in the first version. Lumping
together all these considerations, we get%
\begin{equation}
g^{\alpha _{12}}(\mathcal{T}_{\mu _{1}\alpha _{1}\mu _{2}\alpha
_{2}}^{AV})_{1}=4mT_{\mu _{1};\mu _{2}}^{AS}-\varepsilon _{\mu _{1}}^{\quad
\nu }\mathcal{B}_{\nu ;\mu _{2}}-\varepsilon _{\mu _{1}\nu }U_{\mu
_{2}}^{\nu },
\end{equation}
$U_{\mu _{2}}^{\nu }$ is a term common to all relations with a constraint.
For version two, 
\begin{eqnarray}
g^{\mu _{12}}(\mathcal{T}_{\mu _{1}\alpha _{1}\mu _{2}\alpha _{2}}^{AV})_{2}
&=&-\varepsilon _{\alpha _{1}}^{\quad \nu }\mathcal{B}_{\nu ;\alpha
_{2}}-\varepsilon _{\alpha _{1}}^{\quad \nu _{1}}U_{\nu _{1}\alpha _{2}} \\
g^{\alpha _{12}}(\mathcal{T}_{\mu _{1}\alpha _{1}\mu _{2}\alpha
_{2}}^{AV})_{2} &=&4mT_{\mu _{1};\mu _{2}}^{AS}-\varepsilon _{\mu
_{1}}^{\quad \nu }\mathcal{B}_{\nu ;\mu _{2}}.
\end{eqnarray}

Concerning $VA$ as it can be expressed in $AV$\ terms without conditions
from (\ref{BVA-AVt}), 
\begin{eqnarray}
g^{\mu _{12}}(\mathcal{T}_{\mu _{1}\alpha _{1}\mu _{2}\alpha _{2}}^{VA})_{1}
&=&4mT_{\alpha _{1};\alpha _{2}}^{AS}-\varepsilon _{\alpha _{1}}^{\quad \nu
}B_{\nu ;\alpha _{2}} \\
g^{\mu _{12}}(\mathcal{T}_{\mu _{1}\alpha _{1}\mu _{2}\alpha _{2}}^{VA})_{2}
&=&4mT_{\alpha _{1};\alpha _{2}}^{AS}-\varepsilon _{\alpha _{1}}^{\quad \nu }%
\mathcal{B}_{\nu ;\alpha _{2}}-\varepsilon _{\alpha _{1}}^{\quad \nu
_{1}}U_{\nu _{1}\alpha _{2}} \\
g^{\alpha _{12}}(\mathcal{T}_{\mu _{1}\alpha _{1}\mu _{2}\alpha
_{2}}^{VA})_{1} &=&-\varepsilon _{\mu _{1}}^{\quad \nu }\mathcal{B}_{\nu
;\mu _{2}}-\varepsilon _{\mu _{1}\nu }U_{\mu _{2}}^{\nu } \\
g^{\alpha _{12}}(\mathcal{T}_{\mu _{1}\alpha _{1}\mu _{2}\alpha
_{2}}^{VA})_{2} &=&-\varepsilon _{\mu _{1}}^{\quad \nu }\mathcal{B}_{\nu
;\mu _{2}}.
\end{eqnarray}%
Different from momentum relations, when an index is the one that defines the
version, then $U$-factor appears in the complementary contraction, $%
g^{\alpha _{12}}(\mathcal{T}_{\mu _{1}\alpha _{1}\mu _{2}\alpha
_{2}}^{AV})_{1}=-\varepsilon _{\mu _{1}}^{\nu }g^{\alpha _{12}}(\mathcal{T}%
_{\nu \alpha _{1}\mu _{2}\alpha _{2}}^{VV})$ shows a possible violation, as
opposed to $q^{\alpha _{1}}(\mathcal{T}_{\mu _{1}\alpha _{1}\mu _{2}\alpha
_{2}}^{AV})_{1}$ which is identically satisfied.

\section{Summing all permutations: $[\hat{T}^{V}]$ and $[\hat{T}^{A}]_{ij}$}

In preparation for summing all contributions, that will constitute the
two-point function of the stress tensor, it is necessary to establish a
point of view about the odd part. In the preceding expressions, we adopted a
uniform version to $\{(\mathcal{T}_{\mu _{1}\alpha _{1}\mu _{2}\alpha
_{2}}^{AV})_{i};(\mathcal{T}_{\mu _{1}\alpha _{1}\mu _{2}\alpha
_{2}}^{VA})_{i}\}$, signifying the same version of derivatives amplitudes
were chosen. For the permutation $\mu _{1}\leftrightarrow \mu _{2}$ and
subsequently $\alpha _{1}\leftrightarrow \alpha _{2}$, it is entirely free
which combinations to use in this step. In this work, we will explore a
subset of possibilities,%
\begin{equation}
\lbrack \hat{T}_{\mu _{1}\mu _{2}\alpha _{1}\alpha _{2}}^{A}]_{ij}=(\mathcal{%
T}_{\mu _{1}\alpha _{1}\mu _{2}\alpha _{2}}^{AV})_{i}+(\mathcal{T}_{\mu
_{1}\alpha _{1}\mu _{2}\alpha _{2}}^{VA})_{j},
\end{equation}%
with $i,j=\left\{ 1,2\right\} $, amounting to four combinations in
principle. Permutations do not change this choice as it could be done.

The even sector works as $[\hat{T}_{\mu _{1}\mu _{2}\alpha _{1}\alpha
_{2}}^{V}]=\mathcal{T}_{\mu _{1}\alpha _{1}\mu _{2}\alpha _{2}}^{VV}+%
\mathcal{T}_{\mu _{1}\alpha _{1}\mu _{2}\alpha _{2}}^{AA}.$ To get the total
contribution, it is necessary to sum the permutation $\mu
_{1}\leftrightarrow \mu _{2}$ and then $\alpha _{1}\leftrightarrow \alpha
_{2}$ of that result. In the even sector, we use (\ref{BAA-VV}) and to have
the systematic formula%
\begin{eqnarray}
\lbrack \mathcal{T}_{\mu _{12}\alpha _{12}}^{V}] &=&2[\mathcal{T}_{\mu
_{1}\mu _{2}\alpha _{1}\alpha _{2}}^{VV}+\mathcal{T}_{\mu _{2}\mu _{1}\alpha
_{1}\alpha _{2}}^{VV}+\mathcal{T}_{\mu _{1}\mu _{2}\alpha _{2}\alpha
_{1}}^{VV}+\mathcal{T}_{\mu _{2}\mu _{1}\alpha _{2}\alpha _{1}}^{VV}]
\label{FVec} \\
&&-4m[g_{\mu _{1}\alpha _{1}}T_{\mu _{2};\alpha _{2}}^{SV}+g_{\mu _{2}\alpha
_{1}}T_{\mu _{1};\alpha _{2}}^{SV}+g_{\mu _{1}\alpha _{2}}T_{\mu _{2};\alpha
_{1}}^{SV}+g_{\mu _{2}\alpha _{2}}T_{\mu _{1};\alpha _{1}}^{SV}].  \notag
\end{eqnarray}

For the odd sector, we go in search of a simplification in the operations;
for that, 
\begin{eqnarray}
(\mathcal{T}_{\mu _{1}\alpha _{1}\mu _{2}\alpha _{2}}^{VA})_{j} &=&(\mathcal{%
T}_{\mu _{1}\alpha _{1}\mu _{2}\alpha _{2}}^{AV})_{j}+4m\varepsilon _{\mu
_{1}\alpha _{1}}T_{\mu _{2};\alpha _{2}}^{SV} \\
(\mathcal{T}_{\mu _{1}\alpha _{1}\mu _{2}\alpha _{2}}^{AV})_{i} &=&(\mathcal{%
T}_{\mu _{1}\alpha _{1}\mu _{2}\alpha _{2}}^{AV})_{1}-\delta
_{i,2}\varepsilon _{\alpha _{1}\mu _{1}}U_{\mu _{2}\alpha _{2}},
\end{eqnarray}%
where $\delta _{i,2}$. Its function is to capture only version two, given
that the second term is zero if it already has version one. The above
equations allow us to write the result%
\begin{equation}
\lbrack \hat{T}_{\mu _{1}\mu _{2}\alpha _{1}\alpha _{2}}^{A}]_{ij}=2(%
\mathcal{T}_{\mu _{1}\alpha _{1}\mu _{2}\alpha _{2}}^{AV})_{1}+4m\varepsilon
_{\mu _{1}\alpha _{1}}T_{\mu _{2};\alpha _{2}}^{SV}-\left( \delta
_{i,2}+\delta _{j,2}\right) \varepsilon _{\alpha _{1}\mu _{1}}U_{\mu
_{2}\alpha _{2}}.
\end{equation}%
These arguments have the consequence that it is also possible to write%
\begin{eqnarray}
\lbrack \mathcal{T}_{\mu _{12}\alpha _{12}}^{A}]_{ij} &=&2(\mathcal{T}_{\mu
_{1}\alpha _{1}\mu _{2}\alpha _{2}}^{AV})_{1}+2(\mathcal{T}_{\mu _{2}\alpha
_{1}\mu _{1}\alpha _{2}}^{AV})_{1}+2(\mathcal{T}_{\mu _{1}\alpha _{2}\mu
_{2}\alpha _{1}}^{AV})_{1}+2(\mathcal{T}_{\mu _{2}\alpha _{2}\mu _{1}\alpha
_{1}}^{AV})_{1}  \label{FAxialij} \\
&&+4m[\varepsilon _{\mu _{1}\alpha _{1}}T_{\mu _{2};\alpha
_{2}}^{SV}+\varepsilon _{\mu _{2}\alpha _{1}}T_{\mu _{1};\alpha
_{2}}^{SV}+\varepsilon _{\mu _{1}\alpha _{2}}T_{\mu _{2};\alpha
_{1}}^{SV}+\varepsilon _{\mu _{2}\alpha _{2}}T_{\mu _{1};\alpha _{1}}^{SV}] 
\notag \\
&&-(\delta _{i,2}+\delta _{j,2})[\varepsilon _{\alpha _{1}\mu _{1}}U_{\mu
_{2}\alpha _{2}}+\varepsilon _{\alpha _{1}\mu _{2}}U_{\mu _{1}\alpha
_{2}}+\varepsilon _{\alpha _{2}\mu _{1}}U_{\mu _{2}\alpha _{1}}+\varepsilon
_{\alpha _{2}\mu _{2}}U_{\mu _{1}\alpha _{1}}].  \notag
\end{eqnarray}

In this way, we can sum the Eqs. (\ref{FVec}) and (\ref{FAxialij})
corresponding to the odd and the even part to obtain the two-point
correlator of the stress tensor reads 
\begin{equation}
T_{\mu _{1}\mu _{2}\alpha _{1}\alpha _{2}}^{G}=-\frac{i}{64}\{[\mathcal{T}%
_{\mu _{12}\alpha _{12}}^{V}]+[\mathcal{T}_{\mu _{12}\alpha
_{12}}^{A}]_{ij}\}.
\end{equation}%
Now it is easy to organize all the contractions obtained by sector from this
tensor.

\subsection{Even Part}

We must observe from the permutations sum $\mu _{1}\leftrightarrow \mu _{2}$%
; that the index $\mu _{1}$ occupies the positions in such a way that
contraction with $q^{\mu _{1}}$ corresponds to the two types of momentum
relations (in the matrix and derivative index positions). Hence we get%
\begin{eqnarray}
q^{\mu _{1}}[\hat{T}_{\mu _{1}\mu _{2}\alpha _{1}\alpha _{2}}^{V}] &=&q^{\mu
_{1}}\mathcal{T}_{\mu _{1}\alpha _{1}\mu _{2}\alpha _{2}}^{VV}+q^{\mu _{1}}%
\mathcal{T}_{\mu _{1}\alpha _{1}\mu _{2}\alpha _{2}}^{AA}=2\mathcal{S}%
_{\left( -\right) \alpha _{1};\alpha _{2}\mu _{2}}^{V}-4mq_{\alpha
_{1}}T_{\mu _{2};\alpha _{2}}^{SV}  \notag \\
q^{\mu _{1}}[\hat{T}_{\mu _{2}\mu _{1}\alpha _{1}\alpha _{2}}^{V}] &=&2%
\mathcal{S}_{\left( -\right) \alpha _{1};\alpha _{2}\mu _{2}}^{V}+2q_{\alpha
_{1}}\mathcal{B}_{\mu _{2};\alpha _{2}}-2g_{\mu _{2}\alpha _{1}}q^{\nu }%
\mathcal{B}_{\nu ;\alpha _{2}}-4mg_{\mu _{2}\alpha _{1}}T_{\left( -\right)
\alpha _{2}}^{S}.
\end{eqnarray}%
Summing the permutation $\alpha _{1}\leftrightarrow \alpha _{2}$ of these
contributions symmetrize\footnote{%
Our definition of symmetrization and unit coeffient: $\mathcal{S}_{\left(
-\right) \left( \alpha _{1};\alpha _{2}\right) \mu _{2}}^{V}=\mathcal{S}%
_{\left( -\right) \alpha _{1};\alpha _{2}\mu _{2}}^{V}+\mathcal{S}_{\left(
-\right) \alpha _{2};\alpha _{1}\mu _{2}}^{V}$} the final expression in
these last indices. The complete result of the vector part of gravitational
amplitude is%
\begin{eqnarray}
q^{\mu _{1}}[\mathcal{T}_{\mu _{12}\alpha _{12}}^{V}] &=&-4m[q_{\alpha
_{1}}T_{\mu _{2};\alpha _{2}}^{SV}+q_{\alpha _{2}}T_{\mu _{2};\alpha
_{1}}^{SV}]-4m[g_{\mu _{2}\alpha _{1}}T_{\left( -\right) \alpha
_{2}}^{S}+g_{\mu _{2}\alpha _{2}}T_{\left( -\right) \alpha _{1}}^{S}]
\label{CqV} \\
&&+4\mathcal{S}_{\left( -\right) \left( \alpha _{1};\alpha _{2}\right) \mu
_{2}}^{V}+2[q_{\alpha _{1}}\mathcal{B}_{\mu _{2};\alpha _{2}}+q_{\alpha _{2}}%
\mathcal{B}_{\mu _{2};\alpha _{1}}-g_{\mu _{2}\alpha _{1}}q^{\nu }\mathcal{B}%
_{\nu ;\alpha _{2}}-g_{\mu _{2}\alpha _{2}}q^{\nu }\mathcal{B}_{\nu ;\alpha
_{1}}].  \notag
\end{eqnarray}%
Notably, the distinction of derivative or matrix indices gets dissolved in
the complete expression. Due to this equation's symmetries and unique form,
we do not show the other contractions, as they may be extracted simply by
substituting the convenient indices.

The compilation of the identities involving the traces is given by%
\begin{equation}
g^{\mu _{12}}[\hat{T}_{\mu _{1}\mu _{2}\alpha _{1}\alpha _{2}}^{V}]=g^{\mu
_{12}}\mathcal{T}_{\mu _{1}\alpha _{1}\mu _{2}\alpha _{2}}^{VV}+g^{\mu _{12}}%
\mathcal{T}_{\mu _{1}\alpha _{1}\mu _{2}\alpha _{2}}^{AA}=4mT_{\alpha
_{1};\alpha _{2}}^{SV}+2\mathcal{B}_{\alpha _{1};\alpha _{2}}+2U_{\alpha
_{1}\alpha _{2}}.
\end{equation}%
Noticing that the trace $g^{\mu _{12}}\hat{T}_{\mu _{2}\mu _{1}\alpha
_{1}\alpha _{2}}^{V}$ is equal. The symmetrization brought about by $\alpha
_{1}\leftrightarrow \alpha _{2}$ furnishes the complete result%
\begin{eqnarray}
g^{\mu _{12}}[\mathcal{T}_{\mu _{12}\alpha _{12}}^{V}] &=&8mT_{\left( \alpha
_{1};\alpha _{2}\right) }^{SV}+4\mathcal{B}_{\left( \alpha _{1};\alpha
_{2}\right) }+8U_{\alpha _{1}\alpha _{2}}  \label{CgV} \\
g^{\alpha _{12}}[\mathcal{T}_{\mu _{12}\alpha _{12}}^{V}] &=&8mT_{\left( \mu
_{1};\mu _{2}\right) }^{SV}+4\mathcal{B}_{\left( \mu _{1};\mu _{2}\right)
}+8U_{\mu _{1}\mu _{2}},
\end{eqnarray}%
where identical arguments implies to the second equation.

\subsection{Odd Part}

To discuss the more intricate odd part in combinations seen in equation (\ref%
{FAxialij}), we only need results for the basic permutation of version one.
Nonetheless, different from the even sector, the odd part allows for an
extensive set of possibilities whose contractions with $q^{\mu _{1}}$, $%
q^{\mu _{2}}$, $q^{\alpha _{1}},$ and $q^{\alpha _{2}}$ may be, in
principle, all unrelated. However, to our adopted representatives, only
independent contractions with momentum are with $q^{\mu _{1}}$ and $%
q^{\alpha _{1}}$.

To express the first relation, we recall that version one has a $U$-term
when index $\mu _{1}$ is in the first position $(\mathcal{T}_{\mu _{1}\alpha
_{1}\mu _{2}\alpha _{2}}^{AV})_{1}$, but in permutation $(\mathcal{T}_{\mu
_{2}\alpha _{1}\mu _{1}\alpha _{2}}^{AV})_{1}$ it corresponds to an external
contraction that has two forms. Selecting a convenient expression follows 
\begin{eqnarray}
q^{\mu _{1}}[(\mathcal{T}_{\mu _{1}\alpha _{1}\mu _{2}\alpha
_{2}}^{AV})_{1}+(\mathcal{T}_{\mu _{2}\alpha _{1}\mu _{1}\alpha
_{2}}^{AV})_{1}] &=&4m[\varepsilon _{\alpha _{1}\nu }q^{\nu }T_{\mu
_{2};\alpha _{2}}^{SV}-\varepsilon _{\mu _{2}\alpha _{1}}T_{\left( -\right)
\alpha _{2}}^{S}] \\
&&+2\mathcal{S}_{\left( -\right) \alpha _{1};\alpha _{2}\mu
_{2}}^{A}-\varepsilon _{\mu _{2}\nu }q^{\nu }\mathcal{B}_{\alpha _{1};\alpha
_{2}}  \notag \\
&&+\varepsilon _{\alpha _{1}\mu _{2}}q^{\nu }U_{\nu \alpha _{2}}+\varepsilon
_{\alpha _{1}\nu }q^{\nu }U_{\mu _{2}\alpha _{2}}.  \notag
\end{eqnarray}%
Finally, summing with the above equation the permutations in $\alpha _{i}$,
we arrive at 
\begin{eqnarray}
2q^{\mu _{1}}[(\mathcal{T}_{\mu _{1}\alpha _{1}\mu _{2}\alpha
_{2}}^{AV})_{1}+\text{3-perm}] &=&8m[\varepsilon _{\alpha _{1}\nu }q^{\nu
}T_{\mu _{2};\alpha _{2}}^{SV}+\varepsilon _{\alpha _{2}\nu }q^{\nu }T_{\mu
_{2};\alpha _{1}}^{SV}] \\
&&-8m[\varepsilon _{\mu _{2}\alpha _{1}}T_{\left( -\right) \alpha
_{2}}^{S}+\varepsilon _{\mu _{2}\alpha _{2}}T_{\left( -\right) \alpha
_{1}}^{S}]  \notag \\
&&+4\mathcal{S}_{\left( -\right) \left( \alpha _{1};\alpha _{2}\right) \mu
_{2}}^{A}-2\varepsilon _{\mu _{2}\nu }q^{\nu }\mathcal{B}_{\left( \alpha
_{1};\alpha _{2}\right) }  \notag \\
&&+2q^{\nu }\left( \varepsilon _{\alpha _{1}\mu _{2}}U_{\nu \alpha
_{2}}+\varepsilon _{\alpha _{2}\mu _{2}}U_{\nu \alpha _{1}}\right)  \notag \\
&&+2q^{\nu }\left( \varepsilon _{\alpha _{1}\nu }U_{\mu _{2}\alpha
_{2}}+\varepsilon _{\alpha _{2}\nu }U_{\mu _{2}\alpha _{1}}\right) .  \notag
\end{eqnarray}%
Remaining contributions are easy to be dealt with 
\begin{equation*}
q^{\mu _{1}}[4m\varepsilon _{\mu _{1}\alpha _{1}}T_{\mu _{2};\alpha
_{2}}^{SV}-\left( \delta _{i,2}+\delta _{j,2}\right) \varepsilon _{\alpha
_{1}\mu _{1}}U_{\mu _{2}\alpha _{2}}+\text{3-perm}].
\end{equation*}
When added to the previous equation, it follows one of the important results
of this section%
\begin{eqnarray}
q^{\mu _{1}}[\mathcal{T}_{\mu _{12}\alpha _{12}}^{A}]_{ij} &=&4m[\varepsilon
_{\alpha _{1}\nu }q^{\nu }T_{\mu _{2};\alpha _{2}}^{SV}+\varepsilon _{\alpha
_{2}\nu }q^{\nu }T_{\mu _{2};\alpha _{1}}^{SV}]  \label{CAij1} \\
&&-4m[\varepsilon _{\mu _{2}\alpha _{1}}T_{\left( -\right) \alpha
_{2}}^{S}+\varepsilon _{\mu _{2}\alpha _{2}}T_{\left( -\right) \alpha
_{1}}^{S}]  \notag \\
&&+4\mathcal{S}_{\left( -\right) \left( \alpha _{1};\alpha _{2}\right) \mu
_{2}}^{A}-2\varepsilon _{\mu _{2}\nu }q^{\nu }\mathcal{B}_{\left( \alpha
_{1};\alpha _{2}\right) }  \notag \\
&&-\left( 2-\delta _{i,2}-\delta _{j,2}\right) q^{\nu }(\varepsilon _{\mu
_{2}\alpha _{1}}U_{\nu \alpha _{2}}+\varepsilon _{\mu _{2}\alpha _{2}}U_{\nu
\alpha _{1}})  \notag \\
&&+\left( 2-\delta _{i,2}-\delta _{j,2}\right) q^{\nu }(\varepsilon _{\alpha
_{1}\nu }U_{\mu _{2}\alpha _{2}}+\varepsilon _{\alpha _{2}\nu }U_{\mu
_{2}\alpha _{1}}).  \notag
\end{eqnarray}%
The results to $q^{\mu _{2}}$ come from permuting $\mu _{2}$ by $\mu _{1}$
because, among other things, they hit the contracted indices that become
dummy ones in an equivalent position.

As concerning $q^{\alpha _{1}}[\mathcal{T}_{\mu _{12}\mu _{12}}^{A}]_{ij}$
contraction, we exploit the permutation 
\begin{equation*}
q^{\alpha _{1}}[(\mathcal{T}_{\mu _{1}\alpha _{1}\mu _{2}\alpha
_{2}}^{AV})_{1}+(\mathcal{T}_{\mu _{1}\alpha _{2}\mu _{2}\alpha
_{1}}^{AV})_{1}]=2\mathcal{S}_{\left( -\right) \mu _{1};\mu _{2}\alpha
_{2}}^{A}-\varepsilon _{\alpha _{2}\nu }q^{\nu }\mathcal{B}_{\mu _{1};\mu
_{2}}.
\end{equation*}%
We are choosing formulas for the external contraction without $U$-term. The
contraction in the second vertex of version one has an automatically
satisfied RAGF using an appropriate form of relation with the derivative
index is suitable. Adding the permutation in $\mu _{i}$, we have a
symmetrization of these indices. The last part of this derivation needs%
\begin{equation*}
q^{\alpha _{1}}[4m\varepsilon _{\mu _{1}\alpha _{1}}T_{\alpha _{2};\mu
_{2}}^{SV}-\left( \delta _{i,2}+\delta _{j,2}\right) \varepsilon _{\alpha
_{1}\mu _{1}}U_{\mu _{2}\alpha _{2}}+\text{3-perm}].
\end{equation*}%
They organize the final expression as 
\begin{eqnarray}
q^{\alpha _{1}}[\mathcal{T}_{\mu _{12}\alpha _{12}}^{A}]_{ij}
&=&+4m[\varepsilon _{\mu _{1}\nu }q^{\nu }T_{\alpha _{2};\mu
_{2}}^{SV}+\varepsilon _{\mu _{2}\nu }q^{\nu }T_{\alpha _{2};\mu _{1}}^{SV}]
\label{CAij2} \\
&&-4m[\varepsilon _{\alpha _{2}\mu _{1}}T_{\left( -\right) \mu
_{2}}^{S}+\varepsilon _{\alpha _{2}\mu _{2}}T_{\left( -\right) \mu _{1}}^{S}]
\notag \\
&&+4\mathcal{S}_{\left( -\right) \left( \mu _{1};\mu _{2}\right) \alpha
_{2}}^{A}-2\varepsilon _{\alpha _{2}\nu }q^{\nu }\mathcal{B}_{\left( \mu
_{1};\mu _{2}\right) }  \notag \\
&&-\left( \delta _{i,2}+\delta _{j,2}\right) q^{\nu }[\varepsilon _{\alpha
_{2}\mu _{1}}U_{\nu \mu _{2}}+\varepsilon _{\alpha _{2}\mu _{2}}U_{\nu \mu
_{1}}]  \notag \\
&&+\left( \delta _{i,2}+\delta _{j,2}\right) q^{\nu }[\varepsilon _{\mu
_{2}\nu }U_{\mu _{1}\alpha _{2}}+\varepsilon _{\mu _{1}\nu }U_{\mu
_{2}\alpha _{2}}].  \notag
\end{eqnarray}

The trace equation has interesting properties compared with momentum
contraction: through analysis of basic permutation, conditioning factors
appear in a complementary set of indexes. First, we have for the trace of
the combination 
\begin{equation*}
2g^{\mu _{12}}[(\mathcal{T}_{\mu _{1}\alpha _{1}\mu _{2}\alpha
_{2}}^{AV})_{1}+(\mathcal{T}_{\mu _{2}\alpha _{1}\mu _{1}\alpha
_{2}}^{AV})_{1}]=4g^{\mu _{12}}(\mathcal{T}_{\mu _{1}\alpha _{1}\mu
_{2}\alpha _{2}}^{AV})_{1}=-4\varepsilon _{\alpha _{1}}^{\quad \nu }\mathcal{%
B}_{\nu ;\alpha _{2}}.
\end{equation*}%
Summing up all terms with the exchange of indices $\alpha
_{1}\leftrightarrow \alpha _{2}$ with the remaining components leaves us
with a final expression given by 
\begin{eqnarray}
g^{\mu _{12}}[\mathcal{T}_{\mu _{12}\alpha _{12}}^{A}]_{ij} &=&8mT_{\left(
\alpha _{1};\alpha _{2}\right) }^{SA}-4\varepsilon _{\alpha _{1}}^{\quad \nu
}\mathcal{B}_{\nu ;\alpha _{2}}-4\varepsilon _{\alpha _{2}}^{\quad \nu }%
\mathcal{B}_{\nu ;\alpha _{1}}  \label{CgA} \\
&&-2\left( \delta _{i,2}+\delta _{j,2}\right) (\varepsilon _{\alpha _{1}\nu
}U_{\alpha _{2}}^{\nu }+\varepsilon _{\alpha _{2}\nu }U_{\alpha _{1}}^{\nu
}).  \notag
\end{eqnarray}%
We utilized the relation $-\varepsilon _{\alpha _{1}}^{\quad \nu }T_{\nu
;\alpha _{2}}^{SV}=T_{\alpha _{1};\alpha _{2}}^{SA}$. So remember, version
one is automatically satisfied. However, $U$-contribution came from the
equation between versions one and two.

Another trace independent is with $g^{\alpha _{12}}$; the conditioning
factors coming from 
\begin{eqnarray}
2g^{\alpha _{12}}[(\mathcal{T}_{\mu _{1}\alpha _{1}\mu _{2}\alpha
_{2}}^{AV})_{1}+(\mathcal{T}_{\mu _{1}\alpha _{2}\mu _{2}\alpha
_{1}}^{AV})_{1}] &=&4g^{\alpha _{12}}(\mathcal{T}_{\mu _{1}\alpha _{1}\mu
_{2}\alpha _{2}}^{AV})_{1} \\
&=&16mT_{\mu _{1};\mu _{2}}^{AS}-4\varepsilon _{\mu _{1}}^{\quad \nu }%
\mathcal{B}_{\nu ;\mu _{2}}-4\varepsilon _{\mu _{1}\nu }U_{\mu _{2}}^{\nu }.
\end{eqnarray}%
Thus, symmetrizing in $\mu _{i}$ and adding the remaining contributions, we
arrive at%
\begin{eqnarray}
g^{\alpha _{12}}[\mathcal{T}_{\mu _{12}\alpha _{12}}^{A}]_{ij}
&=&8mT_{\left( \mu _{1};\mu _{2}\right) }^{AS}-4\varepsilon _{\mu
_{1}}^{\quad \nu }\mathcal{B}_{\nu ;\mu _{2}}-4\varepsilon _{\mu
_{2}}^{\quad \nu }\mathcal{B}_{\nu ;\mu _{1}} \\
&&-2[2-\left( \delta _{i,2}+\delta _{j,2}\right) ](\varepsilon _{\mu _{1}\nu
}U_{\mu _{2}}^{\nu }+\varepsilon _{\mu _{2}\nu }U_{\mu _{1}}^{\nu }).  \notag
\end{eqnarray}%
The only difference is for the coefficients of violating terms. One
immediate consequence is the existence of operations with the Dirac traces
and surface terms where such terms do not arise. That is thoroughly argued
in the next part, where the surface terms in these expressions are
investigated. After that, the Weyl and Einstein anomalies are discussed.

To illustrate how they look like when everything is put together, see a
trace relation associated do the Weyl anomaly: 
\begin{eqnarray}
\left( 64i\right) g^{\alpha _{12}}[T_{\mu _{12}\alpha _{12}}^{G}]_{ij}
&=&8mT_{\left( \mu _{1};\mu _{2}\right) }^{SV}+8mT_{\left( \mu _{1};\mu
_{2}\right) }^{AS}+4\mathcal{B}_{\left( \mu _{1};\mu _{2}\right)
}-4\varepsilon _{\mu _{1}}^{\quad \nu }\mathcal{B}_{\nu ;\mu
_{2}}-4\varepsilon _{\mu _{2}}^{\quad \nu }\mathcal{B}_{\nu ;\mu _{1}} 
\notag \\
&&-2[2-\left( \delta _{i,2}+\delta _{j,2}\right) ](\varepsilon _{\mu _{1}\nu
}U_{\mu _{2}}^{\nu }+\varepsilon _{\mu _{2}\nu }U_{\mu _{1}}^{\nu })+8U_{\mu
_{1}\mu _{2}} \\
\left( 64i\right) g^{\mu _{12}}[T_{\mu _{12}\alpha _{12}}^{G}]_{ij}
&=&8mT_{\left( \alpha _{1};\alpha _{2}\right) }^{SV}+8mT_{\left( \alpha
_{1};\alpha _{2}\right) }^{SA}+4\mathcal{B}_{\left( \alpha _{1};\alpha
_{2}\right) }-4\varepsilon _{\alpha _{1}}^{\quad \nu }\mathcal{B}_{\nu
;\alpha _{2}}-4\varepsilon _{\alpha _{2}}^{\quad \nu }\mathcal{B}_{\nu
;\alpha _{1}}  \notag \\
&&-2\left( \delta _{i,2}+\delta _{j,2}\right) (\varepsilon _{\alpha _{1}\nu
}U_{\alpha _{2}}^{\nu }+\varepsilon _{\alpha _{2}\nu }U_{\alpha _{1}}^{\nu
})+8U_{\alpha _{1}\alpha _{2}}.  \notag
\end{eqnarray}%
And a momentum equation related to the Einstein anomaly:%
\begin{eqnarray}
\left( 64i\right) q^{\mu _{1}}[T_{\mu _{12}\alpha _{12}}^{G}]_{ij}
&=&4m[\varepsilon _{\alpha _{1}\nu }q^{\nu }T_{\mu _{2};\alpha
_{2}}^{SV}+\varepsilon _{\alpha _{2}\nu }q^{\nu }T_{\mu _{2};\alpha
_{1}}^{SV}-q_{\alpha _{1}}T_{\mu _{2};\alpha _{2}}^{SV}-q_{\alpha
_{2}}T_{\mu _{2};\alpha _{1}}^{SV}]  \notag \\
&&-4m[\varepsilon _{\mu _{2}\alpha _{1}}T_{\left( -\right) \alpha
_{2}}^{S}+\varepsilon _{\mu _{2}\alpha _{2}}T_{\left( -\right) \alpha
_{1}}^{S}+g_{\mu _{2}\alpha _{1}}T_{\left( -\right) \alpha _{2}}^{S}+g_{\mu
_{2}\alpha _{2}}T_{\left( -\right) \alpha _{1}}^{S}]  \notag \\
&&+4\mathcal{S}_{\left( -\right) \left( \alpha _{1};\alpha _{2}\right) \mu
_{2}}^{A}+4\mathcal{S}_{\left( -\right) \left( \alpha _{1};\alpha
_{2}\right) \mu _{2}}^{V}-2\varepsilon _{\mu _{2}\nu }q^{\nu }\mathcal{B}%
_{\left( \alpha _{1};\alpha _{2}\right) } \\
&&+2[q_{\alpha _{1}}\mathcal{B}_{\mu _{2};\alpha _{2}}+q_{\alpha _{2}}%
\mathcal{B}_{\mu _{2};\alpha _{1}}-g_{\mu _{2}\alpha _{1}}q^{\nu }\mathcal{B}%
_{\nu ;\alpha _{2}}-g_{\mu _{2}\alpha _{2}}q^{\nu }\mathcal{B}_{\nu ;\alpha
_{1}}]  \notag \\
&&-\left( 2-\delta _{i,2}-\delta _{j,2}\right) q^{\nu }(\varepsilon _{\mu
_{2}\alpha _{1}}U_{\nu \alpha _{2}}+\varepsilon _{\mu _{2}\alpha _{2}}U_{\nu
\alpha _{1}})  \notag \\
&&+\left( 2-\delta _{i,2}-\delta _{j,2}\right) q^{\nu }(\varepsilon _{\alpha
_{1}\nu }U_{\mu _{2}\alpha _{2}}+\varepsilon _{\alpha _{2}\nu }U_{\mu
_{2}\alpha _{1}}).  \notag
\end{eqnarray}

\section{Constraints: The Matter of RAGFs Satisfaction}

RAGFs for derivative amplitudes as a whole require that $\left\{ \Upsilon
,\Upsilon _{\alpha _{1}},\Upsilon _{\alpha _{1}\alpha _{2}}\right\} =0$
holds independently. We already composed them into $U_{\alpha _{1}\alpha
_{2}}$, which arises in the final form of gravitational amplitude. We will
recover their explicit expression by simplifying the investigation but with
some notation to relevant structures. Combinations of surface terms, which
we carefully introduced and managed since the first chapter, are given by%
\begin{eqnarray}
\Xi _{\mu _{1}\mu _{2}}^{\left( a\right) } &=&[2(\square _{3\rho \mu _{1}\mu
_{2}}^{\rho }-\Delta _{2\mu _{1}\mu _{2}})-g_{\mu _{1}\mu _{2}}\Delta
_{2\rho }^{\rho }]=ag_{\mu _{1}\mu _{2}}  \label{CsiA} \\
\Xi _{\mu _{1}\mu _{2}\mu _{3}\mu _{4}}^{\left( b\right) } &=&\left[ 3\Sigma
_{4\rho \mu _{1234}}^{\rho }-8\square _{3\mu _{1234}}-g_{(\mu _{1}\mu
_{2}}g_{\mu _{3}\mu _{4})}\Delta _{2\rho }^{\rho }\right] =bg_{(\mu _{1}\mu
_{2}}g_{\mu _{3}\mu _{4})}  \label{CsiB} \\
\Xi _{\rho a_{1}\alpha _{2}}^{\rho \text{\textrm{quad}}} &=&(W_{2\rho \alpha
_{12}}^{\rho }-2\Delta _{1\alpha _{12}})+2g_{\alpha _{1}\alpha _{2}}I_{\text{%
\textrm{quad}}}-2m^{2}\left( \Delta _{2\alpha _{12}}+g_{\alpha _{12}}I_{\log
}\right)
\end{eqnarray}%
The importance of this attitude is two-fold: one, it reduces the size of
expressions, and two, if bilinears are reduced in the integrand, these
tensors become convergent surface terms that identically vanish; see
Appendix (\ref{Uni-Red}). Moreover, their integrands are typical of 4D
integrals. On the other hand, all the following analyses do not use such an
operation.

Evoking Eqs. (\ref{Ups0}), (\ref{Ups1}) and (\ref{Ups2}), we have the set 
\begin{eqnarray}
\Upsilon &=&2\Delta _{2\rho }^{\rho }+i/\pi \\
\Upsilon _{\alpha _{1}} &=&-\frac{1}{2}P^{\nu _{1}}\Xi _{\alpha _{1}\nu
}^{\left( a\right) }-\frac{1}{2}q_{\alpha _{1}}\Upsilon  \notag \\
\Upsilon _{\alpha _{1}\alpha _{2}} &=&-\frac{1}{12}\left( \theta _{\alpha
_{12}}-3q_{\alpha _{12}}\right) \Upsilon -\frac{1}{4}P^{\nu }[\left(
P_{\alpha _{2}}-q_{\alpha _{2}}\right) \Xi _{\alpha _{1}\nu }^{\left(
a\right) }+\left( P_{\alpha _{1}}-q_{\alpha _{1}}\right) \Xi _{\alpha
_{2}\nu }^{\left( a\right) }]+\Xi _{a_{12}}^{\text{\textrm{quad}}}  \notag \\
&&+\frac{1}{72}(3P^{\nu _{12}}+q^{\nu _{12}})g_{(\alpha _{12}}\Xi _{\nu
_{12})}^{\left( a\right) }-\frac{1}{8}(P^{2}+q^{2})\Xi _{\alpha
_{12}}^{\left( a\right) }+\frac{1}{36}(3P^{\nu _{12}}+q^{\nu _{12}})\Xi
_{\alpha _{12}\nu _{12}}^{\left( b\right) }.  \notag
\end{eqnarray}%
Now, as the variables $\left\{ P;q\right\} $ or the routings $\left\{
k_{1};k_{2}\right\} $ are linearly independent, only solution for their
vanishing is $\Upsilon =0,$ $\Xi _{\mu _{1}\mu _{2}}^{\left( a\right) }=0,$
and $\Xi _{\mu _{1}\mu _{2}\mu _{3}\mu _{4}}^{\left( b\right) }=0$. For
quadratic terms, we have%
\begin{equation*}
\Xi _{a_{12}}^{\text{\textrm{quad}}}=(W_{2\rho \alpha _{12}}^{\rho }-2\Delta
_{1\alpha _{12}})+2g_{\alpha _{1}\alpha _{2}}I_{\text{\textrm{quad}}%
}-2m^{2}\left( \Delta _{2\alpha _{12}}+g_{\alpha _{12}}I_{\log }\right) =0.
\end{equation*}%
This happens because if $\Upsilon =0$ and $\Upsilon _{\alpha
_{1}}=0\Rightarrow \Xi _{\alpha _{1}\nu }^{\left( a\right) }=0$, that
substituted in $\Upsilon _{\alpha _{1}\alpha _{2}}$ oblige other terms to
vanish. If one takes $\Upsilon _{\alpha _{1}\alpha _{2}}$ alone, it has
crossed terms $q_{\alpha _{i}}P^{\nu }$ that requires its coefficient $\Xi
_{\alpha _{1}\nu }^{\left( a\right) }$ to be zero and the term $P^{\nu
_{12}}\Xi _{\alpha _{12}\nu _{12}}^{\left( b\right) }$ in the only remnant
of arbitrary $P$-variable, hence this tensor will have to be zero and
subsequently $\Upsilon =0$ as well. In any case, we have conditions stated.
Additionally, the condition $\Upsilon _{\mu _{2}}=0$ alone would be the same
since for arbitrary $P$ and $q$, both terms, $\Xi _{\alpha _{1}\nu }^{\left(
a\right) }$ and $\Upsilon $, must vanish.

In the last statement, we have the exception of the places whose violating
terms sum into $2\Upsilon _{\alpha _{1}}+q_{\alpha _{1}}\Upsilon $, that
occur exactly for combinations $[2T_{\mu _{12};\alpha _{1}}^{\Gamma
_{12}}+q_{\alpha _{1}}T_{\mu _{12}}^{\Gamma _{12}}]$. However, if finite,
this combination ought to vanish. Why? Because in 2D for vértices $\Gamma
_{i}=\{\gamma _{\mu },\gamma _{\ast }\gamma _{\mu }\}$ the charge
conjugation $C$ matrix implies $C\Gamma _{i}C^{-1}=-\Gamma _{i}^{T}$ and for
the propagator 
\begin{equation}
CS\left( K_{i}\right) C^{-1}=(C\slashed{K}_{i}C^{-1}+m)/D_{i}=S^{T}\left(
-K_{i}\right) .
\end{equation}

Expliciting the structure $[2T_{\mu _{12};\alpha _{1}}^{\Gamma
_{12}}+q_{\alpha _{1}}T_{\mu _{12}}^{\Gamma _{12}}]$ can be written as 
\begin{eqnarray}
\lbrack 2T_{\mu _{12};\alpha _{1}}^{\Gamma _{12}}+q_{\alpha _{1}}T_{\mu
_{12}}^{\Gamma _{12}}] &=&\int \frac{\mathrm{d}^{2}k}{\left( 2\pi \right)
^{2}}\left( K_{1}+K_{2}\right) _{\alpha _{1}}\mathrm{tr}[\Gamma _{1}S\left(
K_{1}\right) \Gamma _{2}S\left( K_{2}\right) ] \\
&=&\int \frac{\mathrm{d}^{2}k}{\left( 2\pi \right) ^{2}}\left(
K_{1}+K_{2}\right) _{\alpha _{1}}t^{\Gamma _{1}\Gamma _{2}},
\end{eqnarray}%
where integrand $t^{\Gamma _{1}\Gamma _{2}}$ s the function without
derivative index. It readily obeys%
\begin{eqnarray}
t^{\Gamma _{1}\Gamma _{2}} &=&\mathrm{tr}\{[C\Gamma _{1}C^{-1}][CS\left(
K_{1}\right) C^{-1}][C\Gamma _{2}C^{-1}][CS\left( K_{2}\right) C^{-1}]\} \\
&=&\left( -1\right) ^{2}\mathrm{tr}[S\left( -K_{2}\right) \Gamma _{2}S\left(
-K_{1}\right) \Gamma _{1}]^{T} \\
&=&\mathrm{tr}[\Gamma _{1}S\left( -k-k_{2}\right) \Gamma _{2}S\left(
-k-k_{1}\right) ].
\end{eqnarray}%
Under integration, reflecting the integration variable $k\rightarrow -k$
after shifting it by $k\rightarrow k+k_{1}+k_{2}$, the arguments of $%
t^{\Gamma _{1}\Gamma _{2}}$ return to their starting configuration. However,
the factor $(K_{1}+K_{2})$ picks up a minus sign $-(K_{1}+K_{2})$, and the
derivative vertex behaves like it had negative parity. These steps are valid
as hypotheses; observe that at the beginning that we mentioned, if finite,
we can do the operations listed. Therefore, we would get 
\begin{equation}
2T_{\mu _{12};\alpha _{1}}^{\Gamma _{12}}+q_{\alpha _{1}}T_{\mu
_{12}}^{\Gamma _{12}}=\left( -1\right) [T_{\mu _{12};\alpha _{1}}^{\Gamma
_{12}}+q_{\alpha _{1}}T_{\mu _{12}}^{\Gamma _{12}}].
\end{equation}%
If shifts can be done, the result must vanish. As the surface terms violate
this hypothesis, the non-polynomial sector of the finite part disappears,
which depends on external momentum $q=k_{2}-k_{1}$. The leftover part, in
general, is a local polynomial in $q$ and $P$ momenta and surface terms,
with a degree up to power counting of amplitude.

That fact naturally can be checked in their explicit forms, where no shift
of the loop momentum was performed. For instance, see the combination above
between $VV$'s, 
\begin{eqnarray}
2T_{\mu _{12};\alpha _{1}}^{VV}+q_{\alpha _{1}}T_{\mu _{12}}^{VV} &=&%
\mathcal{D}_{\mu _{12};\alpha _{1}}^{VV}+q_{\alpha _{1}}\mathcal{D}_{\mu
_{12}}^{VV}  \notag \\
&=&-2P^{\nu _{1}}W_{3\mu _{12}\alpha _{1}\nu _{1}}+2P_{(\mu _{1}}\Delta
_{2\mu _{2}\alpha _{1})}+2g_{\mu _{12}}P^{\nu _{1}}\Delta _{2\alpha _{1}\nu
_{1}}.
\end{eqnarray}%
That happens to odd amplitudes and also in its two basic modalities. Without
derivatives, the finite functions $T_{\mu }^{SV}=0$ and $T_{\mu }^{AS}=0$
have a vertex that picks a minus sign ($V$ e $A$, respectively). We always
expressed one part in the basic permutation the way we did because the most
complex part, finite ones, drops from calculations. For this subset of
amplitudes, the violating terms either are not present, as in 
\begin{equation}
g^{\mu _{1}\alpha _{1}}[2(T_{\mu _{12};\alpha _{1}}^{AV})_{1}+q_{\alpha
_{1}}(T_{\mu _{12}}^{AV})_{1}]=T_{\alpha _{1}}^{A}\left( k_{1}\right)
+T_{\alpha _{1}}^{A}\left( k_{2}\right) .
\end{equation}%
Alternatively, they are present and appear in the form%
\begin{equation}
g^{\mu _{1}\alpha _{1}}[2T_{\mu _{12};\alpha _{1}}^{VV}+q_{\alpha
_{1}}T_{\mu _{12}}^{VV}]=T_{\mu _{2}}^{V}\left( k_{1}\right) +T_{\mu
_{2}}^{V}\left( k_{2}\right) +2mT_{\mu _{2}}^{SV}+(2\Upsilon _{\mu
_{2}}+q_{\mu _{2}}\Upsilon ),
\end{equation}%
where $2\Upsilon _{\mu _{2}}+q_{\mu _{2}}\Upsilon =-P^{\nu _{1}}\Xi _{\mu
_{2}\nu }^{\left( a\right) }$ happens to vanish either for surface terms
corresponding to RAGFs satisfied or with zero value.

Therefore, back to the analysis, the constraints $(\Upsilon ,\Upsilon
_{\alpha _{1}},\Upsilon _{\alpha _{1}\alpha _{2}})=0,$ in addition to
satisfying all RAGFs imply in defined values for the tensors (\ref{Ups0}), (%
\ref{CsiA}) and (\ref{CsiB}) 
\begin{eqnarray}
\Upsilon &=&2\Delta _{2\rho }^{\rho }+i/\pi  \label{UP} \\
\Xi _{\alpha _{1}\nu }^{\left( a\right) } &=&2\square _{3\rho \alpha _{1}\nu
}^{\rho }-2g_{\alpha _{1}\nu }\Delta _{2\rho }^{\rho }=0 \\
\Xi _{\alpha _{12}\nu _{12}}^{\left( a\right) } &=&3\Sigma _{4\rho \alpha
_{12}\nu _{12}}^{\rho }-3g_{(\alpha _{1}\alpha _{2}}g_{\nu _{12})}\Delta
_{2\rho }^{\rho }=0.
\end{eqnarray}%
That choice, in turn, allows us to organize a ladder of restrictions on
surface terms:%
\begin{eqnarray}
\square _{3\rho \alpha _{1}\nu _{1}}^{\rho } &=&g_{\alpha _{1}\nu
_{1}}\Delta _{2\rho }^{\rho } \\
\square _{3\rho \alpha _{1}\nu _{1}}^{\rho } &=&cg^{\nu _{23}}g_{(\alpha
_{1}\nu _{1}}g_{\nu _{23})}=4cg_{\alpha _{1}\nu _{1}} \\
\square _{3\alpha _{12}\nu _{12}} &=&\frac{1}{4}g_{(\alpha _{12}}g_{\nu
_{12})}\Delta _{2\rho }^{\rho }.
\end{eqnarray}%
Notice that we adopted an utterly symmetric definition of surface terms. As
they are dimensionless, we got to determine their coefficients. The fourth
order will be given by%
\begin{eqnarray}
\Sigma _{4\rho \alpha _{12}\nu _{12}}^{\rho } &=&g_{(\alpha _{1}\alpha
_{2}}g_{\nu _{12})}\Delta _{2\rho }^{\rho } \\
\Sigma _{4\rho \alpha _{12}\nu _{12}}^{\rho } &=&dg^{\nu _{23}}g_{(\alpha
_{1}\alpha _{2}}g_{\nu _{12}}g_{\nu _{23})}=6dg_{(\alpha _{1}\alpha
_{2}}g_{\nu _{12})} \\
\Sigma _{4\alpha _{12}\nu _{12}\nu _{34}} &=&\frac{1}{6}g_{(\alpha
_{1}\alpha _{2}}g_{\nu _{12}}g_{\nu _{34})}\Delta _{2\rho }^{\rho },
\end{eqnarray}%
As the trace is $2\Delta _{2\rho }^{\rho }=-i/\pi $,$\,\ $see (\ref{UP}),\
there arise the values to the surface terms. Only the concepts of the RAGFs
and unicity are enough to determine the other values,%
\begin{eqnarray}
\Delta _{2\mu \nu } &=&-\frac{ig_{\mu \nu }}{4\pi }  \label{deltafin} \\
\square _{3\alpha _{12}\nu _{12}} &=&-\frac{ig_{(\alpha _{12}}g_{\nu _{12})}%
}{8\pi }  \label{boxfin} \\
\Sigma _{4\alpha _{12}\nu _{12}\nu _{34}} &=&-\frac{i}{12\pi }g_{(\alpha
_{1}\alpha _{2}}g_{\nu _{12}}g_{\nu _{34})}.  \label{sigmafin}
\end{eqnarray}

However, if the attitude towards the undetermined parts were to preserve
translational invariance in momentum space. The interpretation given to this
tensor should be 
\begin{equation*}
\square _{3\alpha _{12}\nu _{12}}=\Delta _{2\mu \nu }=\Sigma _{4\alpha
_{12}\nu _{12}\nu _{34}}=0,
\end{equation*}%
In this way, we have the complementary consequence in the tensors, 
\begin{equation}
\Upsilon =\frac{i}{\pi };\quad \Upsilon _{\alpha _{1}}=-\frac{1}{2}q_{\alpha
_{1}}\Upsilon ;\quad \Upsilon _{\alpha _{1}}=-\frac{1}{12}\left( \theta
_{\alpha _{1}\alpha _{2}}-3q_{\alpha _{1}}q_{\alpha _{2}}\right) \Upsilon .
\end{equation}%
And, about the U-tensor, if the vanishing surface terms, we break
integration linearity by 
\begin{equation}
U_{\alpha _{1}\alpha _{2}}=-\frac{1}{3}\left( \frac{i}{\pi }\right) \theta
_{\alpha _{1}\alpha _{2}}.  \label{Uvalue}
\end{equation}%
In parallel, if RAGFs hold or the odd amplitudes are unique or independent
of intermediary steps of the calculation, e.g., Dirac traces used. Using the
results to $\Upsilon $ in this scenario, we have $U_{\alpha _{1}\alpha
_{2}}=0.$ To clarify that conditions are exactly equal for the U-factor
since the crossed term $qP$ drops out, it may be possible that other linear
combinations of $\Xi $'s could cancel the RAGF's violator.

Once more, the explicit expression for $U$, in terms of (\ref{CsiA}) and (%
\ref{CsiB}), is 
\begin{eqnarray}
U_{\alpha _{1}\alpha _{2}} &=&-\frac{1}{3}\theta _{\alpha _{1}\alpha
_{2}}\Upsilon +\frac{1}{18}q^{\nu _{12}}[2\Xi _{2\alpha _{12}\nu
_{12}}+g_{(\alpha _{12}}\Xi _{1\nu _{1}\nu _{2})}]-\frac{1}{2}q^{2}\Xi
_{1\alpha _{1}\alpha _{2}}+4\Xi _{a_{1}\alpha _{2}}^{\text{\textrm{quad}}} \\
&&+\frac{1}{6}P^{\nu _{12}}[2\Xi _{2\alpha _{12}\nu _{12}}^{\left( b\right)
}+g_{(\alpha _{12}}\Xi _{1\nu _{1}\nu _{2})}^{\left( a\right) }]-\frac{1}{2}%
P^{2}\Xi _{1\alpha _{1}\alpha _{2}}-P_{\alpha _{2}}P^{\nu _{1}}\Xi _{1\alpha
_{1}\nu _{1}}-P_{\alpha _{1}}P^{\nu _{1}}\Xi _{1\alpha _{2}\nu _{1}}.  \notag
\end{eqnarray}%
Expanding in its coefficients and using the arbitrary internal momenta, we
get 
\begin{eqnarray}
U_{\alpha _{1}\alpha _{2}} &=&+\frac{1}{9}\left( 4b-5a-3\Upsilon \right)
g_{\alpha _{1}\alpha _{2}}\left( k_{1}^{2}+k_{2}^{2}\right) \\
&&+\frac{1}{9}\left( 4b+4a+6\Upsilon \right) g_{\alpha _{1}\alpha
_{2}}\left( k_{1}\cdot k_{2}\right)  \notag \\
&&+\frac{1}{9}\left( 8b-10a+3\Upsilon \right) \left( k_{1\alpha
_{1}}k_{1\alpha _{2}}+k_{2\alpha _{1}}k_{2\alpha _{2}}\right)  \notag \\
&&+\frac{1}{9}\left( 4b-14a-3\Upsilon \right) \left( k_{1\alpha
_{1}}k_{2\alpha _{2}}+k_{2\alpha _{1}}k_{1\alpha _{2}}\right) =0.
\end{eqnarray}%
As each row corresponds to linearly independent tensors, the only solution
to the system is $a=b=\Upsilon =0$. That is the unique solution we have
discussed so far.

To deep down into the reasons, as demonstrated in the Appendix (\ref{Uni-Red}%
), if one accepts a natural reduction in the integrand, it leads to, by
example,%
\begin{eqnarray}
\Xi _{\mu _{1}\mu _{2}\mu _{3}\mu _{4}}^{\left( b\right) } &=&\left[ 3\Sigma
_{4\rho \mu _{1}\mu _{2}\mu _{3}\mu _{4}}^{\rho }-8\square _{3\mu _{1}\mu
_{2}\mu _{3}\mu _{4}}-g_{(\mu _{1}\mu _{2}}g_{\mu _{3}\mu _{4})}\Delta
_{2\rho }^{\rho }\right] \\
&=&m^{2}\int \frac{\mathrm{d}^{2}k}{\left( 2\pi \right) ^{2}}\left\{
\sum_{i=1}^{4}\frac{\partial }{\partial k^{\mu _{i}}}\frac{-6k_{\mu
_{1}\cdots \hat{\mu}_{i}\cdots \mu _{4}}}{D_{\lambda }^{3}}-g_{(\mu _{3}\mu
_{4}}\frac{\partial }{\partial k^{\mu _{5}}}\frac{k_{\mu _{6})}}{D_{\lambda
}^{2}}\right\} =0.
\end{eqnarray}%
Hence, this corresponds to a convergent integral that vanishes.
Nevertheless, we established this result based on the RAGFs without this
manipulation.

It is worthwhile to call attention to that $\Upsilon =i/\pi $-factor emerged
in the description of the chiral anomaly (from $T_{\mu _{12}}^{AV}$). It
uses methods that allow variable integration shifts, 
\begin{equation}
q^{\mu _{1}}(T_{\mu _{1}\mu _{2}}^{AV})_{1}=-2mT_{\mu _{2}}^{PV}+\varepsilon
_{\mu _{2}\nu }q^{\nu }\Upsilon \text{ and }q^{\mu _{2}}(T_{\mu _{1}\mu
_{2}}^{AV})_{2}=\varepsilon _{\mu _{1}\nu }q^{\nu }\Upsilon ;
\end{equation}%
while the other Ward Identities are fulfilled in and equal to zero.

The combination of the quadratic surface terms $\Xi _{\alpha _{1}\alpha
_{2}}^{\text{\textrm{quad}}}$ may be organized in the form 
\begin{eqnarray}
\Xi _{\alpha _{1}\alpha _{2}}^{\text{\textrm{quad}}} &=&(W_{2\rho \alpha
_{12}}^{\rho }-2\Delta _{1\alpha _{12}})+2g_{\alpha _{1}\alpha _{2}}I_{\text{%
\textrm{quad}}}-2m^{2}\left( \Delta _{2\alpha _{12}}+g_{\alpha _{12}}I_{\log
}\right) \\
&=&\int \frac{\mathrm{d}^{2}k}{\left( 2\pi \right) ^{2}}\left( \frac{4\left(
k^{2}-m^{2}\right) k_{\alpha _{12}}}{D_{\lambda }^{2}}-\frac{4k_{\alpha
_{12}}}{D_{\lambda }}\right) =0.
\end{eqnarray}%
We chose the mass parameter such that $D_{\lambda }=k^{2}-m^{2}$. There are
three arguments, reducing bilinear in the integrand of the last line yields
an exact cancellation, or in the massless limit since it is proportional to
the mass that goes to zero. Thirdly, some prescriptions make this term zero
in various analytic regularization methods.

\section{Einstein and Weyl Anomalies}

We now turn to anomalies; we must take the massless limit. First, looking
into the results of contractions, for instance, $q^{\mu _{1}}$-contraction
of the vector part (\ref{CqV}), axial part (\ref{CAij1}), or with the metric
(\ref{CgV}) or (\ref{CgA}). There are terms proportional to the mass: the
two and one-point functions with mass as coefficient go to zero in this
limit: 
\begin{eqnarray}
4mT_{\mu _{1};\mu _{2}}^{SV} &=&8m^{2}\left( \Delta _{2\mu _{1}\mu
_{2}}+g_{\mu _{1}\mu _{2}}I_{\log }\right) -\frac{i}{\pi }2m^{2}\theta _{\mu
_{1}\mu _{2}}[2Z_{2}^{\left( -1\right) }-Z_{1}^{\left( -1\right) }] \\
4mT_{\left( -\right) \mu _{2}}^{S} &=&8m^{2}q^{\nu }\left( \Delta _{2\nu \mu
_{2}}+g_{\nu \mu _{2}}I_{\log }\right) .
\end{eqnarray}%
The last line can also be seen through $q^{\nu }T_{\nu ;\mu
_{2}}^{SV}=T_{\left( -\right) \mu _{2}}^{S}$. Thereby $\lim_{m^{2}%
\rightarrow 0}4mT_{\mu _{1};\mu _{2}}^{SV}=0$ and $\lim_{m^{2}\rightarrow
0}4mT_{\left( +\right) \alpha _{2}}^{S}=0.$ Furthermore, in this way, we
have only the vector and axial one-point functions and the RAGFs violating
factor $U_{\alpha \beta }$.

For these terms that remain, we consider two scenarios: One that derives
from the preservation of WI for $T_{\alpha \beta }^{VV}$ and $T_{\alpha
\beta }^{AA},$\ which requires vanishing of surface terms and preserves
momentum-space translational invariance. The other scenario exploited is
when surface terms are finite and determined by the constraint of RAGFs.%
\begin{eqnarray}
\left( 64i\right) q^{\alpha _{1}}[T_{\mu _{12}\alpha _{12}}^{G}]_{ij} &=&4%
\mathcal{S}_{\left( -\right) \left( \mu _{1};\mu _{2}\right) \alpha
_{2}}^{A}+4\mathcal{S}_{\left( -\right) \left( \mu _{1};\mu _{2}\right)
\alpha _{2}}^{V}-2\varepsilon _{\alpha _{2}\nu }q^{\nu }\mathcal{B}_{\left(
\mu _{1};\mu _{2}\right) } \\
&&+2[q_{\mu _{1}}\mathcal{B}_{\alpha _{2};\mu _{2}}+q_{\mu _{2}}\mathcal{B}%
_{\alpha _{2};\mu _{1}}-g_{\alpha _{2}\mu _{1}}q^{\nu }\mathcal{B}_{\nu ;\mu
_{2}}-g_{\alpha _{2}\mu _{2}}q^{\nu }\mathcal{B}_{\nu ;\mu _{1}}]  \notag \\
&&-\left( \delta _{i,2}+\delta _{j,2}\right) q^{\nu }[\varepsilon _{\alpha
_{2}\mu _{1}}U_{\nu \mu _{2}}+\varepsilon _{\alpha _{2}\mu _{2}}U_{\nu \mu
_{1}}]  \notag \\
&&+\left( \delta _{i,2}+\delta _{j,2}\right) q^{\nu }[\varepsilon _{\mu
_{2}\nu }U_{\mu _{1}\alpha _{2}}+\varepsilon _{\mu _{1}\nu }U_{\mu
_{2}\alpha _{2}}]  \notag
\end{eqnarray}%
\begin{eqnarray}
\left( 64i\right) q^{\mu _{1}}[T_{\mu _{12}\alpha _{12}}^{G}]_{ij} &=&4%
\mathcal{S}_{\left( -\right) \left( \alpha _{1};\alpha _{2}\right) \mu
_{2}}^{A}+4\mathcal{S}_{\left( -\right) \left( \alpha _{1};\alpha
_{2}\right) \mu _{2}}^{V}-2\varepsilon _{\mu _{2}\nu }q^{\nu }\mathcal{B}%
_{\left( \alpha _{1};\alpha _{2}\right) } \\
&&+2[q_{\alpha _{1}}\mathcal{B}_{\mu _{2};\alpha _{2}}+q_{\alpha _{2}}%
\mathcal{B}_{\mu _{2};\alpha _{1}}-g_{\mu _{2}\alpha _{1}}q^{\nu }\mathcal{B}%
_{\nu ;\alpha _{2}}-g_{\mu _{2}\alpha _{2}}q^{\nu }\mathcal{B}_{\nu ;\alpha
_{1}}]  \notag \\
&&-\left( 2-\delta _{i,2}-\delta _{j,2}\right) q^{\nu }(\varepsilon _{\mu
_{2}\alpha _{1}}U_{\nu \alpha _{2}}+\varepsilon _{\mu _{2}\alpha _{2}}U_{\nu
\alpha _{1}})  \notag \\
&&+\left( 2-\delta _{i,2}-\delta _{j,2}\right) q^{\nu }(\varepsilon _{\alpha
_{1}\nu }U_{\mu _{2}\alpha _{2}}+\varepsilon _{\alpha _{2}\nu }U_{\mu
_{2}\alpha _{1}}).  \notag
\end{eqnarray}%
\begin{eqnarray}
\left( 64i\right) g^{\alpha _{12}}[T_{\mu _{12}\alpha _{12}}^{G}]_{ij} &=&4%
\mathcal{B}_{\left( \mu _{1};\mu _{2}\right) }-4\varepsilon _{\mu
_{1}}^{\quad \nu }\mathcal{B}_{\nu ;\mu _{2}}-4\varepsilon _{\mu
_{2}}^{\quad \nu }\mathcal{B}_{\nu ;\mu _{1}} \\
&&-2[2-\left( \delta _{i,2}+\delta _{j,2}\right) ](\varepsilon _{\mu _{1}\nu
}U_{\mu _{2}}^{\nu }+\varepsilon _{\mu _{2}\nu }U_{\mu _{1}}^{\nu })+8U_{\mu
_{1}\mu _{2}}  \notag \\
\left( 64i\right) g^{\mu _{12}}[T_{\mu _{12}\alpha _{12}}^{G}]_{ij} &=&4%
\mathcal{B}_{\left( \alpha _{1};\alpha _{2}\right) }-4\varepsilon _{\alpha
_{1}}^{\quad \nu }\mathcal{B}_{\nu ;\alpha _{2}}-4\varepsilon _{\alpha
_{2}}^{\quad \nu }\mathcal{B}_{\nu ;\alpha _{1}} \\
&&-2\left( \delta _{i,2}+\delta _{j,2}\right) (\varepsilon _{\alpha _{1}\nu
}U_{\alpha _{2}}^{\nu }+\varepsilon _{\alpha _{2}\nu }U_{\alpha _{1}}^{\nu
})+8U_{\alpha _{1}\alpha _{2}}  \notag
\end{eqnarray}

\subsection{Vanishing Surface Terms: Violating RAGFs}

In the first scenario investigated, we adopt the interpretation of the
surfaces as 
\begin{equation*}
\Delta _{2\mu \nu }=0;\square _{3\alpha _{12}\nu _{12}}=0;\Sigma _{4\alpha
_{12}\nu _{12}\nu _{34}}=0.
\end{equation*}%
In the massless limit, dropping out the quadratic structures as they are
proportional to the mass is possible. The condition implies $W_{4}=W_{3}=0$
as well because these tensors are defined as a linear combination of the
previous ones (\ref{W3}-\ref{W4}). In tandem, this restriction sets the
result to the sum and differences of one-point functions $\mathcal{S}%
_{\left( -\right) }^{V}=\mathcal{S}_{\left( -\right) }^{A}=\mathcal{B}=0.$
The present interpretation for surface terms violates RAGFs, the amount
which the $U$-factor gives shown in the previous section, see (\ref{Uvalue}%
). We recover its value%
\begin{equation}
U_{\alpha \mu }=-\frac{1}{3}\theta _{\alpha \mu }\Upsilon =-\frac{1}{3}%
\left( \frac{i}{\pi }\right) \theta _{\alpha \mu }.
\end{equation}

\textbf{Einstein Anomaly:} They could appear in the vector and axial
sectors; however, in the current setting, the vector part vanishes. For this
symmetry, we only need to evaluate the results for one index, namely,%
\begin{equation*}
q^{\mu _{1}}[\mathcal{T}_{\mu _{12}\alpha _{12}}^{V}]=4\mathcal{S}_{\left(
-\right) \left( \alpha _{1};\alpha _{2}\right) \mu _{2}}^{V}+2[q_{\alpha
_{1}}\mathcal{B}_{\mu _{2};\alpha _{2}}+q_{\alpha _{2}}\mathcal{B}_{\mu
_{2};\alpha _{1}}-g_{\mu _{2}\alpha _{1}}q^{\nu }\mathcal{B}_{\nu ;\alpha
_{2}}-g_{\mu _{2}\alpha _{2}}q^{\nu }\mathcal{B}_{\nu ;\alpha _{1}}]=0.
\end{equation*}%
That is an interesting consequence of this perspective; however, it breaks
integration linearity if even and odd amplitudes should have a uniform
mathematical treatment. The other equations to be discussed get
contributions from the axial part and are%
\begin{eqnarray}
q^{\alpha _{1}}[T_{\mu _{12}\alpha _{12}}^{G}]_{ij} &=&-\frac{1}{96}\left( 
\frac{1}{\pi }\right) \left( \delta _{i,2}+\delta _{j,2}\right) \varepsilon
_{\mu _{1}\nu }q^{\nu }\theta _{\mu _{2}\alpha _{2}} \\
q^{\mu _{1}}[T_{\mu _{12}\alpha _{12}}^{G}]_{ij} &=&-\frac{1}{96}\left( 
\frac{1}{\pi }\right) \left( 2-\delta _{i,2}-\delta _{j,2}\right)
\varepsilon _{\alpha _{1}\nu }q^{\nu }\theta _{\mu _{2}\alpha _{2}},
\end{eqnarray}%
where was used the identity $\varepsilon _{\alpha _{2}\nu }q^{\nu }\theta
_{\mu _{2}\alpha _{1}}=\varepsilon _{\alpha _{1}\nu }q^{\nu }\theta _{\mu
_{2}\alpha _{2}}$.

It exhibits a richer structure because, for null surface terms, the axial
sector reveals a dependence on the version of trace with the chiral and four
Dirac matrices that are employed. After integration, the identities valid
for the integrand are transformed by the present interpretation in different
tensors. It implies that intermediary operations lead to many possibilities,
some of which are present above. The breaking of linearity makes the
versions unequal as the simpler $T_{\mu \nu }^{AV}$. The version $ij=\left\{
11,22\right\} $ only has anomalies in one set of indexes, $\mu _{i}$ or $%
\alpha _{i}$. A table of results can clarify these statements: 
\begin{equation*}
\left\{ 
\begin{array}{ll}
q^{\alpha _{1}}[T_{\mu _{12}\alpha _{12}}^{G}]_{11}=0 & q^{\alpha
_{1}}[T_{\mu _{12}\alpha _{12}}^{G}]_{22}=-\frac{1}{48}\left( \frac{1}{\pi }%
\right) \varepsilon _{\mu _{1}\nu }q^{\nu }\theta _{\mu _{2}\alpha _{2}} \\ 
&  \\ 
q^{\mu _{1}}[T_{\mu _{12}\alpha _{12}}^{G}]_{11}=-\frac{1}{48}\left( \frac{1%
}{\pi }\right) \varepsilon _{\alpha _{1}\nu }q^{\nu }\theta _{\mu _{2}\alpha
_{2}} & q^{\mu _{1}}[T_{\mu _{12}\alpha _{12}}^{G}]_{22}=0%
\end{array}%
\right\}
\end{equation*}%
In the case of $ij=\left\{ 12,21\right\} ,$ the mixed versions of the
anomaly appear equally distributed and are half of the other versions:%
\begin{equation*}
\left\{ 
\begin{array}{ll}
q^{\alpha _{1}}[T_{\mu _{12}\alpha _{12}}^{G}]_{12}=-\frac{1}{96}\left( 
\frac{1}{\pi }\right) \varepsilon _{\mu _{1}\nu }q^{\nu }\theta _{\mu
_{2}\alpha _{2}} & q^{\alpha _{1}}[T_{\mu _{12}\alpha _{12}}^{G}]_{21}=-%
\frac{1}{96}\left( \frac{1}{\pi }\right) \varepsilon _{\mu _{1}\nu }q^{\nu
}\theta _{\mu _{2}\alpha _{2}} \\ 
&  \\ 
q^{\mu _{1}}[T_{\mu _{12}\alpha _{12}}^{G}]_{12}=-\frac{1}{96}\left( \frac{1%
}{\pi }\right) \varepsilon _{\alpha _{1}\nu }q^{\nu }\theta _{\mu _{2}\alpha
_{2}} & q^{\mu _{1}}[T_{\mu _{12}\alpha _{12}}^{G}]_{21}=-\frac{1}{96}\left( 
\frac{1}{\pi }\right) \varepsilon _{\alpha _{1}\nu }q^{\nu }\theta _{\mu
_{2}\alpha _{2}}%
\end{array}%
\right\}
\end{equation*}%
The results above are the common finding in the literature. In other words,
we have options for expressing the $AV$/$VA$ functions in terms of the even $%
VV$/$AA$ amplitudes.

\textbf{Weyl Anomaly: }In the scenario of RAGFs violations, we get 
\begin{eqnarray}
g^{\alpha _{12}}[T_{\mu _{12}\alpha _{12}}^{G}]_{ij} &=&-\frac{1}{96\pi }%
\left[ 4\theta _{\mu _{1}\mu _{2}}-[2-\left( \delta _{i,2}+\delta
_{j,2}\right) ](\varepsilon _{\mu _{1}\nu }\theta _{\mu _{2}}^{\nu
}+\varepsilon _{\mu _{2}\nu }\theta _{\mu _{1}}^{\nu })\right] \\
g^{\mu _{12}}[T_{\mu _{12}\alpha _{12}}^{G}]_{ij} &=&-\frac{1}{96\pi }\left[
4\theta _{\alpha _{1}\alpha _{2}}-\left( \delta _{i,2}+\delta _{j,2}\right)
(\varepsilon _{\alpha _{1}\nu }\theta _{\alpha _{2}}^{\nu }+\varepsilon
_{\alpha _{2}\nu }\theta _{\alpha _{1}}^{\nu })\right] .
\end{eqnarray}%
As the equations are not unique, the odd part of Weyl anomaly is absent in
some versions,%
\begin{eqnarray}
g^{\alpha _{12}}[T_{\mu _{12}\alpha _{12}}^{G}]_{11} &=&-\frac{1}{48\pi }%
[2\theta _{\mu _{1}\mu _{2}}-(\varepsilon _{\mu _{1}\nu }\theta _{\mu
_{2}}^{\nu }+\varepsilon _{\mu _{2}\nu }\theta _{\mu _{1}}^{\nu })] \\
g^{\mu _{12}}[T_{\mu _{12}\alpha _{12}}^{G}]_{11} &=&-\frac{1}{24\pi }\theta
_{\alpha _{1}\alpha _{2}} \\
g^{\alpha _{12}}[T_{\mu _{12}\alpha _{12}}^{G}]_{22} &=&-\frac{1}{24\pi }%
\theta _{\mu _{1}\mu _{2}} \\
g^{\mu _{12}}[T_{\mu _{12}\alpha _{12}}^{G}]_{22} &=&-\frac{1}{48\pi }%
[2\theta _{\alpha _{1}\alpha _{2}}-(\varepsilon _{\alpha _{1}\nu }\theta
_{\alpha _{2}}^{\nu }+\varepsilon _{\alpha _{2}\nu }\theta _{\alpha
_{1}}^{\nu })].
\end{eqnarray}%
Note that the above equation expresses the possibility of not having
anomalies in one energy-momentum tensor occurring when that version has an
Einstein anomaly. The mixed versions show the same amount of violation in
all contractions%
\begin{eqnarray}
g^{\alpha _{12}}[T_{\mu _{12}\alpha _{12}}^{G}]_{12} &=&-\frac{1}{96\pi }%
\left[ 4\theta _{\mu _{1}\mu _{2}}-(\varepsilon _{\mu _{1}\nu }\theta _{\mu
_{2}}^{\nu }+\varepsilon _{\mu _{2}\nu }\theta _{\mu _{1}}^{\nu })\right] \\
g^{\mu _{12}}[T_{\mu _{12}\alpha _{12}}^{G}]_{12} &=&-\frac{1}{96\pi }\left[
4\theta _{\alpha _{1}\alpha _{2}}-(\varepsilon _{\alpha _{1}\nu }\theta
_{\alpha _{2}}^{\nu }+\varepsilon _{\alpha _{2}\nu }\theta _{\alpha
_{1}}^{\nu })\right] \\
g^{\mu _{12}}[T_{\mu _{12}\alpha _{12}}^{G}]_{21} &=&-\frac{1}{96\pi }\left[
4\theta _{\alpha _{1}\alpha _{2}}-(\varepsilon _{\alpha _{1}\nu }\theta
_{\alpha _{2}}^{\nu }+\varepsilon _{\alpha _{2}\nu }\theta _{\alpha
_{1}}^{\nu })\right] \\
g^{\alpha _{12}}[T_{\mu _{12}\alpha _{12}}^{G}]_{21} &=&-\frac{1}{96\pi }%
\left[ 4\theta _{\mu _{1}\mu _{2}}-(\varepsilon _{\mu _{1}\nu }\theta _{\mu
_{2}}^{\nu }+\varepsilon _{\mu _{2}\nu }\theta _{\mu _{1}}^{\nu })\right] .
\end{eqnarray}%
They show Einstein anomalies in all contractions as well.

For the sake of commentary, we rederived the finite part of the $U$-factor.
The finite part of the basic permutation may be written as 
\begin{equation}
\mathcal{T}_{\mu _{1}\alpha _{1}\mu _{2}\alpha _{2}}^{VV}=\left( \frac{i}{%
\pi }\right) \frac{1}{q^{2}}\left\{ 2\theta _{\mu _{1}\alpha _{1}}\theta
_{\mu _{2}\alpha _{2}}[3Z_{2}^{\left( 0\right) }-2Z_{1}^{\left( 0\right)
}]-\Omega _{\mu _{1}\alpha _{1}\mu _{2}\alpha _{2}}[2Z_{2}^{\left( 0\right)
}-Z_{1}^{\left( 0\right) }]\right\} .
\end{equation}%
The finite part of the $U$-factor comes from the equation below%
\begin{eqnarray}
U_{\alpha _{2}\mu _{2}} &=&(g^{\mu _{1}\alpha _{1}}\mathcal{T}_{\mu
_{1}\alpha _{1}\mu _{2}\alpha _{2}}^{VV}-4mT_{\mu _{2};\alpha _{2}}^{SV}) \\
&=&\frac{2i}{\pi }\theta _{\mu _{2}\alpha _{2}}\{[3Z_{2}^{\left( 0\right)
}-2Z_{1}^{\left( 0\right) }]+m^{2}[2Z_{2}^{\left( -1\right) }-Z_{1}^{\left(
-1\right) }]\}=-(i/3\pi )\theta _{\mu _{2}\alpha _{2}}.
\end{eqnarray}%
For the last equation, we have used the reductions above 
\begin{equation*}
3Z_{2}^{\left( 0\right) }-2Z_{1}^{\left( 0\right) }=-\frac{m^{2}}{q^{2}}%
Z_{0}^{\left( 0\right) }-\frac{1}{6};\qquad Z_{0}^{\left( 0\right)
}=q^{2}[2Z_{2}^{\left( -1\right) }-Z_{1}^{\left( -1\right) }].
\end{equation*}

\subsection{Finite Surface Terms: RAGFs satisfied}

Summarizing: In this scenario to be investigated, we adopt the
interpretation of surfaces as finite and their values determined by RAGFs, (%
\ref{deltafin})-(\ref{sigmafin}). Thus, all relations are satisfied, and odd
amplitudes become unique and independent of the trace prescription. However,
now the one-point functions take finite values while $U=0$.

The finite violating terms in the momentum contractions: to derive this term
in general, we remind that $q^{a}\mathcal{T}_{abcd}^{VV}=\mathcal{S}_{\left(
-\right) b;cd}^{V}$, where $\mathcal{S}_{\left( -\right) b;cd}^{V}$ is the
difference of combining the vectorial one-point functions defined in (\ref%
{tensor-S}). In the massless limit, the explicit contribution of the surface
term can be arranged as%
\begin{eqnarray}
\mathcal{S}_{\left( -\right) b;cd}^{V} &=&+P^{\nu _{12}}q^{\nu
_{3}}W_{4bcd\nu _{123}} \\
&&-2P^{\nu _{1}}q^{\nu _{2}}(P_{b}W_{3cd\nu _{12}}+P_{d}W_{3bc\nu
_{12}}+P_{c}W_{3bd\nu _{12}})-q_{b}P^{\nu _{12}}W_{3cd\nu _{12}}  \notag \\
&&+2P^{\nu _{1}}\left[ -\left( P\cdot q\right) W_{3bcd\nu _{1}}+q_{b}\left(
P_{d}\Delta _{2c\nu _{1}}+P_{c}\Delta _{2d\nu _{1}}\right) \right]  \notag \\
&&+q^{\nu _{1}}\left[ -P^{2}W_{3bcd\nu _{1}}+2\left( P_{b}P_{c}\Delta
_{2d\nu _{1}}+P_{b}P_{d}\Delta _{2c\nu _{1}}+P_{c}P_{d}\Delta _{2b\nu
_{1}}\right) \right]  \notag \\
&&+2\left( P\cdot q\right) \left( P_{b}\Delta _{2cd}+P_{c}\Delta
_{2bd}+P_{d}\Delta _{2bc}\right) +q_{b}P^{2}\Delta _{2cd}  \notag \\
&&+\frac{1}{3}q^{\nu _{12}}q^{\nu _{3}}W_{4bcd\nu _{123}}-q_{b}q^{\nu
_{12}}W_{3cd\nu _{12}}-q^{2}q^{\nu _{1}}W_{3bcd\nu _{1}}+q_{b}q^{2}\Delta
_{2cd}.  \notag
\end{eqnarray}%
Here we are using Latin letters in order to make index replacement
operational. The combination of surface terms defined in (\ref{delta2}), (%
\ref{W4}) and \ref{W3} assuming the values%
\begin{equation*}
W_{4abcd\nu _{12}}=-\frac{i}{4\pi }\frac{11}{6}g_{(ab}g_{cd}g_{\nu
_{12})};\quad W_{3abcd}=-\frac{i}{4\pi }\frac{3}{2}g_{(ab}g_{cd)};\quad
W_{2ab}=\Delta _{2ab}=-\frac{i}{4\pi }g_{ab}.
\end{equation*}%
And for the basic permutation as well, it is reasonable to get%
\begin{eqnarray}
-i\left( 4\pi \right) \mathcal{T}_{abcd}^{VV} &=&-\frac{1}{3}%
g_{(ab}g_{cd)}P^{2}+\frac{1}{2}g_{ab}g_{cd}P^{2}+\frac{1}{3}%
P_{(a}P_{b}g_{cd)}-P_{d}P_{c}g_{ab} \\
&&+\frac{8}{9}g_{(ab}g_{cd)}q^{2}-\frac{11}{9}%
q_{(a}q_{b}g_{cd)}+3g_{ab}q_{c}q_{d}-\frac{3}{2}%
g_{ab}g_{cd}q^{2}+2g_{cd}q_{a}q_{b}.  \notag
\end{eqnarray}%
where the symmetrization of the notation follows (the same for $%
q_{(a}q_{b}g_{cd)}$), 
\begin{equation}
P_{(a}P_{b}g_{cd)}=P_{a}P_{b}g_{cd}+P_{a}P_{c}g_{bd}+P_{a}P_{d}g_{bc}+P_{b}P_{c}g_{ad}+P_{b}P_{d}g_{ac}+P_{c}P_{d}g_{ab}.
\end{equation}

Now, we admit a covariant parameterization of the ambiguous momentum
concerning the external one. As an example, we have 
\begin{equation}
P_{\mu }=\left( k_{1\mu }+k_{2\mu }\right) =\chi q_{\mu }.  \label{P=xq}
\end{equation}%
Therefore one of the terms in the RAGFs can be expressed as%
\begin{eqnarray}
\mathcal{S}_{\left( -\right) b;cd}^{V} &=&\frac{i}{\left( 4\pi \right) }%
\frac{\chi ^{2}}{2}q_{b}\left( \theta _{cd}+q_{c}q_{d}\right) +\frac{i}{%
2\left( 4\pi \right) }q_{b}\theta _{cd} \\
&&+\frac{i}{6\left( 4\pi \right) }\left[ -2\left[ q_{d}\theta
_{bc}+q_{c}\theta _{bd}+q_{b}\theta _{cd}\right] -7q_{b}q_{c}q_{d}\right] , 
\notag
\end{eqnarray}%
inside the full contractions we get symmetrizations $\mathcal{S}_{\left(
b;c\right) d}^{V}$.

The factor that appears in the trace relations, defined (\ref{tensor-B}), is
developed in the form%
\begin{eqnarray}
\mathcal{B}_{\alpha _{1};\alpha _{2}} &=&2T_{\left( +\right) \alpha
_{1};\alpha _{2}}^{V}+q_{\alpha _{2}}T_{\left( +\right) \alpha _{1}}^{V} \\
&=&4\left( \Delta _{1\alpha _{1}\alpha _{2}}+g_{\alpha _{1}\alpha _{2}}I_{%
\mathrm{quad}}\right) +2q_{\alpha _{2}}q^{\nu _{1}}\Delta _{2\alpha _{1}\nu
_{1}} \\
&&+P^{\nu _{12}}W_{3\alpha _{1}\alpha _{2}\nu _{12}}-P^{2}\Delta _{2\alpha
_{1}\alpha _{2}}-2P^{\nu _{1}}\left( P_{\alpha _{1}}\Delta _{2\alpha _{2}\nu
_{1}}+P_{\alpha _{2}}\Delta _{2\alpha _{1}\nu _{1}}\right)  \notag \\
&&+q^{\nu _{12}}W_{3\alpha _{1}\alpha _{2}\nu _{12}}-q^{2}\Delta _{2\alpha
_{1}\alpha _{2}}-2q^{\nu _{1}}\left( q_{\alpha _{1}}\Delta _{2\alpha _{2}\nu
_{1}}+q_{\alpha _{2}}\Delta _{2\alpha _{1}\nu _{1}}\right) .  \notag
\end{eqnarray}%
In the symmetric limit (massless limit) and using the parametrization (\ref%
{P=xq}), we have%
\begin{eqnarray}
-i\left( 4\pi \right) \mathcal{B}_{\left( \alpha _{1};\alpha _{2}\right) }
&=&-\chi ^{2}\left( \theta _{\alpha _{1}\alpha _{2}}-q_{\alpha
_{2}}q_{\alpha _{1}}\right) -\left( \theta _{\alpha _{1}\alpha
_{2}}+3q_{\alpha _{2}}q_{\alpha _{1}}\right) \\
-i\left( 4\pi \right) q^{\alpha _{1}}\mathcal{B}_{\alpha _{1};\alpha _{2}}
&=&\frac{\left( \chi ^{2}-3\right) }{2}q_{\alpha _{2}}q^{2}.
\end{eqnarray}%
Axial combinations $\mathcal{S}_{a;bc}^{A}=-\varepsilon _{a}^{\text{ \ }\nu }%
\mathcal{S}_{a;bc}^{V}$ , symmetrizing these terms as in the final result%
\begin{eqnarray}
\mathcal{S}_{\left( -\right) \left( a;b\right) c}^{A} &=&-\frac{\chi ^{2}}{2}%
\varepsilon _{a\nu }[2q^{\nu }\theta _{bc}-q_{c}\theta _{b}^{\nu
}+q_{b}q_{c}q^{\nu }] \\
&&+\frac{1}{6}\varepsilon _{a\nu }[-5q_{c}\theta _{b}^{\nu }+4q_{b}\theta
_{c}^{\nu }-2q^{\nu }\theta _{bc}+5q_{b}q_{c}q^{\nu }].  \notag
\end{eqnarray}

\textbf{Einstein Anomaly}: The total contribution for the odd sector where
we can isolate one term that corresponds to the version $[\mathcal{T}_{\mu
_{12}\alpha _{12}}^{A}]_{12}$, 
\begin{equation}
q^{\mu _{1}}[\mathcal{T}_{\mu _{12}\alpha _{12}}^{A}]=-\frac{i\varepsilon
_{\alpha _{1}\nu }}{12\pi }\{8q^{\nu }\theta _{\alpha _{2}\mu _{2}}+(6\chi
^{2}-10)[q^{\nu }\theta _{\alpha _{2}\mu _{2}}-q_{\mu _{2}}\theta _{\alpha
_{2}}^{\nu }-q_{\alpha _{2}}\theta _{\mu _{2}}^{\nu }+q_{\alpha _{2}}q_{\mu
_{2}}q^{\nu }]\},  \label{quTAGodd}
\end{equation}%
therefore the choice $\chi ^{2}=5/3$ can recover that value. Despite that,
there is a choice of routings that can reproduce the values for a specific
version when surface terms are made null; the even part does not show such a
possibility, as can be seen in 
\begin{equation}
q^{\mu _{1}}[\mathcal{T}_{\mu _{12}\alpha _{12}}^{V}]=\frac{i}{6\pi }%
\{(6\chi ^{2}-10)q_{\alpha _{1}}q_{\alpha _{2}}q_{\mu _{2}}+2(q_{\alpha
_{1}}\theta _{\alpha _{2}\mu _{2}}+q_{\alpha _{2}}\theta _{\alpha _{1}\mu
_{2}}-2q_{\mu _{2}}\theta _{\alpha _{1}\alpha _{2}}-2q_{\alpha
_{1}}q_{\alpha _{2}}q_{\mu _{2}})\}.
\end{equation}%
This presents us with two features: it is impossible to use any choice of
routings to eliminate the anomaly, and the choice that makes the axial part
with a standard value implies in the equation above,%
\begin{equation}
q^{\mu _{1}}[\mathcal{T}_{\mu _{12}\alpha _{12}}^{V}]=\frac{i}{3\pi }%
(q_{\alpha _{1}}\theta _{\alpha _{2}\mu _{2}}+q_{\alpha _{2}}\theta _{\alpha
_{1}\mu _{2}}-2q_{\mu _{2}}\theta _{\alpha _{1}\alpha _{2}}-2q_{\alpha
_{1}}q_{\alpha _{2}}q_{\mu _{2}}).  \label{quTAGeven}
\end{equation}%
Summing the Eqs. (\ref{quTAGodd}) and (\ref{quTAGeven}), the gravitational
amplitude independent of the Dirac trace becomes%
\begin{eqnarray}
q^{\mu _{1}}T_{\mu _{1}\mu _{2}\alpha _{1}\alpha _{2}}^{G} &=&-\frac{1}{%
96\pi }\varepsilon _{\alpha _{1}\nu }q^{\nu }\theta _{\alpha _{2}\mu _{2}}+%
\frac{1}{192\pi }(q_{\alpha _{1}}\theta _{\alpha _{2}\mu _{2}}+q_{\alpha
_{2}}\theta _{\alpha _{1}\mu _{2}}-2q_{\mu _{2}}\theta _{\alpha _{1}\alpha
_{2}}-2q_{\alpha _{1}}q_{\alpha _{2}}q_{\mu _{2}})  \notag \\
&&+\frac{\left( 3\chi ^{2}-5\right) }{384\pi }\{2q_{\alpha _{1}}q_{\alpha
_{2}}q_{\mu _{2}}+\varepsilon _{\alpha _{1}}^{~~\nu }\left( q_{\mu
_{2}}\theta _{\nu \alpha _{2}}+q_{\alpha _{2}}\theta _{\nu \mu _{2}}-q_{\nu
}\theta _{\alpha _{2}\mu _{2}}-q_{\alpha _{2}}q_{\mu _{2}}q_{\nu }\right) \}.
\notag
\end{eqnarray}%
The vector part is irremovable through choices that are intrinsic elements
of Feynman's diagrammatic computation of this correlator.

\textbf{Weyl Anomaly:} The odd part of this symmetry violation arises from
tensor $\mathcal{B}_{\sigma ;\rho }$, 
\begin{eqnarray}
g^{\mu _{12}}[\mathcal{T}_{\mu _{12}\alpha _{12}}^{A}] &=&-4\varepsilon
_{\alpha _{1}}^{\quad \nu }\mathcal{B}_{\nu ;\alpha _{2}}-4\varepsilon
_{\alpha _{2}}^{\quad \nu }\mathcal{B}_{\nu ;\alpha _{1}} \\
g^{\alpha _{12}}[\mathcal{T}_{\mu _{12}\alpha _{12}}^{A}] &=&-4\varepsilon
_{\mu _{1}}^{\quad \nu }\mathcal{B}_{\nu ;\mu _{2}}-4\varepsilon _{\mu
_{2}}^{\quad \nu }\mathcal{B}_{\nu ;\mu _{1}}.
\end{eqnarray}%
Simple manipulation of indices yields the expressions%
\begin{equation}
g^{\mu _{12}}[\mathcal{T}_{\mu _{12}\alpha _{12}}^{A}]=-\frac{i}{\pi }\left(
\chi ^{2}-1\right) q^{\nu }\left( \varepsilon _{\alpha _{1}\nu }q_{\alpha
_{2}}+\varepsilon _{\alpha _{2}\nu }q_{\alpha _{1}}\right) ,
\end{equation}%
and analogously for the other trace. The odd part of the Weyl anomaly can be
removed, but this does not happen to the even part. If the parameter $\chi $
is chosen to make the Einstein anomaly with the standard form, we obtain an
equivalent result as%
\begin{equation}
g^{\mu _{12}}[\mathcal{T}_{\mu _{12}\alpha _{12}}^{A}]=-\frac{2i}{3\pi }%
q^{\nu }(\varepsilon _{\alpha _{1}\nu }q_{\alpha _{2}}+\varepsilon _{\alpha
_{2}\nu }q_{\alpha _{1}})-\frac{i}{3\pi }(3\chi ^{2}-5)q^{\nu }(\varepsilon
_{\alpha _{1}\nu }q_{\alpha _{2}}+\varepsilon _{\alpha _{2}\nu }q_{\alpha
_{1}}).
\end{equation}%
Since that constraint is given by $\chi ^{2}=5/3$.

Through the same line of reasoning, we obtain the even part%
\begin{equation}
g^{\mu _{12}}[\mathcal{T}_{\mu _{12}\alpha _{12}}^{V}]=4\mathcal{B}_{\left(
\alpha _{1};\alpha _{2}\right) }=-\frac{i}{\pi }\chi ^{2}\left( \theta
_{\alpha _{1}\alpha _{2}}-q_{\alpha _{2}}q_{\alpha _{1}}\right) -\frac{i}{%
\pi }\left( \theta _{\alpha _{1}\alpha _{2}}+3q_{\alpha _{2}}q_{\alpha
_{1}}\right) ,
\end{equation}%
similar to the other set of indices. However, now the constraint which
reproduced the standard result to the odd part furnishes a different
expression to the Weyl anomaly of the even part, namely, 
\begin{equation}
g^{\mu _{12}}[\mathcal{T}_{\mu _{12}\alpha _{12}}^{V}]=-\frac{4i}{3\pi }%
\left( 2\theta _{\alpha _{1}\alpha _{2}}+q_{\alpha _{2}}q_{\alpha
_{1}}\right) -\frac{i}{6\pi }\left( 6\chi ^{2}-10\right) \left( \theta
_{\alpha _{1}\alpha _{2}}-q_{\alpha _{2}}q_{\alpha _{1}}\right) .
\end{equation}%
Therefore, the total routing-dependent trace anomaly is given by%
\begin{eqnarray}
g^{\mu _{12}}[T_{\mu _{12}\alpha _{12}}^{G}] &=&-\frac{1}{96\pi }q^{\nu
}\left( \varepsilon _{\alpha _{1}\nu }q_{\alpha _{2}}+\varepsilon _{\alpha
_{2}\nu }q_{\alpha _{1}}\right) -\frac{1}{48\pi }\left( 2\theta _{\alpha
_{1}\alpha _{2}}+q_{\alpha _{2}}q_{\alpha _{1}}\right) \\
&&-\frac{1}{192\pi }\left( 3\chi ^{2}-5\right) \left[ q^{\nu }\left(
\varepsilon _{\alpha _{1}\nu }q_{\alpha _{2}}+\varepsilon _{\alpha _{2}\nu
}q_{\alpha _{1}}\right) +\left( \theta _{\alpha _{1}\alpha _{2}}-q_{\alpha
_{2}}q_{\alpha _{1}}\right) \right] .  \notag
\end{eqnarray}

In this context, where the integration linearity is maintained, and
intermediary operations on the Dirac traces have no effect, we have the
finiteness of the relevant surface terms as the constraint. However, this
also implies violations of the energy-momentum tensor symmetries and the
break of translational invariance (in momentum space, at least). To keep
Ward identities, which crucially depend on translational invariance, the
attitude often adopted is, by some regularization, to remove the surface
terms. The algebraic consequence is to spoil the RAGFs to odd-tensor
amplitudes, deduced without making any shifts whose unique hypothesis is the
linearity of integration. Equivalently, the uniqueness of these amplitudes
is lost as they come from the Feynman rules, thus opening the room for
multiple expressions that violate the symmetries under study anyway. Only a
subset of these possibilities is visualized in the literature.

\chapter{Final Remarks and Perspectives}

\label{finalremarks}

We performed a detailed probe of a significant number of pseudo-tensor
diagrams that correspond to anomalous amplitudes in two and four dimensions,
following a strategy to cope with the divergences introduced in the thesis
of O.A. Battistel. We apply this procedure to the bubbles (the gravitational
case is discussed in the sequel) and triangles with power counting
logarithmic and linear, respectively. The finite ones get integrated after
splitting off and organizing the divergent parts without further action. In
this point, the scalar objects $I_{\log }^{\left( 2n\right) }$ exactly
cancel, letting the final result as a sum of finite tensors and surface
terms, $\Delta _{n+1;\mu _{12}}^{\left( 2n\right) }$. This recipe relies on
the principle of the linearity of integration.

The role of that aspect emerges in the odd amplitudes in even dimensions;
see the e-print (\cite{Preprint}). Contracting with the external momenta
follows RAGFs that, after integration, incorporate the linearity of
integration. For the relevant two and three-point functions in the
respective dimensions, we wrote the equations (because they are not
automatically valid) representing that property as 
\begin{eqnarray}
q^{\mu _{i}}T_{\mu _{12}}^{\left( 2D\right) \Gamma _{1}\Gamma _{2}}
&=&T_{i\left( -\right) \mu _{a}}^{\left( 2D\right) A}+\varepsilon _{\mu
_{a}\nu }\Omega _{i}^{\left( 2pt\right) },\quad i,a=\left\{ 1,2\right\}
,i\not=a  \notag \\
q_{i}^{\mu _{i}}T_{\mu _{123}}^{\left( 4D\right) \Gamma _{1}\Gamma
_{2}\Gamma _{3}} &=&T_{i\left( -\right) \mu _{ab}}^{\left( 4D\right)
AV}+\varepsilon _{\mu _{ab}\nu _{12}}q_{2}^{\nu _{1}}q_{3}^{\nu _{2}}\Omega
_{i}^{\left( 3pt\right) },\quad i,a,b=\left\{ 1,2,3\right\} ,i\not=a<b
\label{3ragf}
\end{eqnarray}%
where the vertices $\Gamma _{i}\in (V;A)=(\gamma _{\mu };\gamma _{\ast
}\gamma _{\mu })$ and the notation $T_{\left( -\right) }^{\left( 2D\right)
A} $, $T_{i\left( -\right) }^{\left( 4D\right) AV}$ means the actual
differences that appear in (\ref{TA(-)mi}) and (\ref{AV(-)1}-\ref{AV(-)3}).
The explicit surface terms read%
\begin{eqnarray}
T_{\mu }^{\left( 2D\right) A}\left( k_{i}\right) &=&2\varepsilon _{\mu
\alpha }k_{i}^{\nu }\Delta _{2\nu }^{\left( 2\right) \alpha } \\
T_{\mu \nu }^{\left( 4D\right) AV}\left( k_{i},k_{j}\right) &=&2i\varepsilon
_{\mu \nu \alpha \sigma }\left( k_{j}-k_{i}\right) ^{\sigma }\left(
k_{i}+k_{j}\right) ^{\gamma }\Delta _{3\gamma }^{\left( 4\right) \alpha }.
\end{eqnarray}

Let us start with four dimensions and then back to two. There, if the three
equations for the RAGFs (\ref{3ragf}) hold at\ the same time and the
vanishing of $T_{\mu \nu }^{\left( 4D\right) AV}$ functions, or their
difference, were possible, then that would allow the vector and partial
axial symmetry to hold simultaneously. That signifies we can make shifts and
thus have momentum-space translational invariance since the only hypothesis
necessary to prove $T_{\mu \nu }^{\left( 4\right) AV}=-T_{\mu \nu }^{\left(
4\right) AV}=0$ is this symmetry. However, such structures depend on the
unphysical and arbitrary sum of routings and are proportional to surface
terms that can violate translational symmetry. If we were only searching to
cancel that terms, it would be seen that choosing routings is not possible
since we should have $P_{31}=P_{21}=P_{32}=0\rightarrow q_{i}=0$. A partial
solution is to make the surface term zero, then recover that symmetry.

Nevertheless, low-energy theorems demonstrated in Section (\ref{LE4D})
showed that a tensor with the characteristics of $AVV$, for example, a
function of the external momenta related to $PVV$ tensor, must satisfy, in
this case, $p_{31}^{\mu _{1}}T_{\mu _{123}}^{AVV}|_{0}=0\not=-2mT_{\mu
_{23}}^{PVV}|_{0}$. That is impossible since the finite $PVV$ does not
behave like that. In general, we demonstrated that assuming the most general
tensor (when written in terms of the physical momenta), without resorting to
a specific symmetry, we got to have%
\begin{equation}
q_{i}^{\mu _{i}}T_{\mu _{123}}^{\left( 4D\right) \Gamma _{123}}=\varepsilon
_{\mu _{ab}\nu _{12}}q_{2}^{\nu _{1}}q_{3}^{\nu _{2}}V_{i}\rightarrow \left.
\left( V_{1}+V_{2}-V_{3}\right) \right\vert _{0}=0.
\end{equation}%
On the other hand, computing the three-point form factors $\Omega _{i}$ from
the amplitudes $PVV$, $PAA$, and for amplitudes $AVS$ and $ASV$ with three
different masses, we find%
\begin{equation}
\Omega _{1}\left( 0\right) +\Omega _{2}\left( 0\right) -\Omega _{3}\left(
0\right) =1/\left( 2\pi \right) ^{2}.  \label{let1}
\end{equation}%
Thus, if the linearity of integration and translational symmetry were
simultaneously valid, we should have $V_{i}=\Omega _{i}$. Therefore, the two
last and independent equations above would be in contradiction. We can say
that the low-energy behavior of finite functions precludes these two
properties from living together. Writing $V_{i}=\Omega _{i}+\mathcal{A}_{i}$%
, we have a constraint over the anomalies $\mathcal{A}_{i}$ by finite
functions, stating that once two of them are fixed, the third is
unambiguously determined. At this point, we have that integration linearity
can not hold for any value of the surface term, in particular, not for the
vanishing one.

All the tensors we investigated show independent combinations of routings,
surface terms, and the $\varepsilon $-tensor. We took these elements as
hypotheses and general as allowed, not writing the internal through external
momenta since the former can also be non-covariant. Thus, by knowing the RHS
of the relations, we lay down: it is impossible without additional
conditions to satisfy all the RAGF. In other words, they are not valid for
any value of the surface term, see Section (\ref{LED4DSTS}). The
satisfaction of all the RAGFs makes the low-energy limit above (\ref{let1})
the value and the reason why the surface term can not vanish; see the
derivation of the equation in (\ref{ME}), as integration linearity requires%
\begin{equation}
2i\Delta _{3\rho }^{\left( 4\right) \rho }=1/\left( 2\pi \right) ^{2}=\Omega
_{1}\left( 0\right) +\Omega _{2}\left( 0\right) -\Omega _{3}\left( 0\right) .
\label{let2}
\end{equation}%
For this reason, we demonstrated that translational symmetry and linearity
of integration are incompatible properties for these perturbative
amplitudes. Furthermore, the same derivations clarify the nomenclature and
choice of the versions; they are the expressions that automatically satisfy
as many RAGFs as possible.

Returning to two dimensions: In this scenario, the 2pt functions do not show
linearly divergent integrals that are the assumed source of the symmetry
violations. However, they show power-counting zero and tensor integrals with
intrinsic surface terms, though the coefficients are the physical momenta.
In reality, in context with the one-point axial amplitudes $T_{\mu
}^{A}\left( k_{i}\right) $, we have linear power counting integrals, and
their shift invariance takes place in the discussion when establishing WIs.
The constraints on the differences $T_{\left( -\right) \mu
}^{A}=-\varepsilon _{\mu }^{~~\alpha }T_{\left( -\right) \alpha
}^{V}=2\varepsilon _{\mu \alpha }\left( k_{1}^{\nu }-k_{2}^{\nu }\right)
\Delta _{2\nu }^{\left( 2\right) \alpha }$ are formally necessary for the
WIs for even and odd amplitudes ($VV$-$AA$ and $AV$-$VA$), but we cannot
choose the arbitrary momenta as $k_{1}=k_{2}=0$ since this implies the
physical one is $q=0=k_{2}-k_{1}$, we must have $\Delta _{2\mu \nu }^{\left(
2\right) }=0$. Nonetheless, this is a premature conclusion once we know that
we must have both RAGFs and vanishing of surface terms. The non-concomitant
presence of these properties is due to the kinematical implications below
that we also showed without resorting to a particular symmetry, and for two
masses, 
\begin{equation}
q^{\mu _{i}}T_{\mu _{12}}^{\left( 2D\right) AV}=\varepsilon _{\mu _{a}\nu
}q^{\nu }V_{i}\rightarrow \left. \left( V_{1}+V_{2}\right) \right\vert
_{0}=0.
\end{equation}%
The kinematical theorem is incompatible with the low-energy limit of finite
functions%
\begin{equation}
\Omega _{1}^{PV}\left( 0\right) +\Omega _{1}^{AS}\left( 0\right) =-i/\pi .
\label{let3}
\end{equation}
Hence, the $V_{1}$ and $V_{2}$ functions are inevitably of the form $%
V_{i}=\Omega _{i}+\mathcal{A}_{i}$, with $\mathcal{A}_{1}+\mathcal{A}%
_{2}=i/\pi $.

Moreover, considering the surface terms for the expression to the general
tensor, an analogous condition is derived through the constraint of
algebraic property encoded by the RAGFs, viz., 
\begin{equation}
2\Delta _{2\alpha }^{\left( 2\right) \alpha }=-i/\pi =\Omega ^{PV}\left(
0\right) +\Omega ^{AS}\left( 0\right) .  \label{let4}
\end{equation}%
This constraint also makes the amplitudes unique concerning the Dirac traces
used. To four dimensions, this turns the amplitudes quantities subject to
routing choices. In contrast, to two dimensions, satisfying RAGFs leads to
Dirac-trace independent expressions that only depend on the physical
momentum.

The feature of Dirac traces appearing in all the treated amplitudes and the
analogous ones for $2n$ dimensions arises for the trace of $2n+2$ Dirac
matrices and an odd number of the chiral matrices. An assortment of
expressions is available when one writes the tensor representing that trace,
differing by the number of monomials and their signs, plus what subset of
its Lorentz indexes appear. Those expressions are equivalent under the
condition that surface terms have a value corresponding to the low-energy
limit of finite-functions combination (\ref{let1}) in $4D$ or (\ref{let3})
in $2D$.

Adopting the \textit{zero value} follows a set of expressions to each
amplitude that may keep at most two RAGFs in $4D$ or one in $2D$. These
expressions can be obtained either applying the definition of $\gamma _{\ast
}$, in some position along the trace or using the identity below in the
adjacent position of matrix $\gamma _{\mu _{i}},$%
\begin{eqnarray}
\left( 2n\right) &:&\gamma _{\ast }\gamma _{\mu _{i}}=\frac{i^{n+1}}{\left(
2n-1\right) !}\varepsilon _{\mu _{i}\nu _{2}\cdots \nu _{2n}}\gamma ^{\nu
_{2}\cdots \nu _{2n}} \\
\left( 2D\right) &:&\gamma _{\ast }\gamma _{\mu _{i}}=-\varepsilon _{\mu
_{i}\nu _{1}}\gamma ^{\nu _{1}}\text{\quad and\quad }\left( 4D\right)
:\gamma _{\ast }\gamma _{\mu _{i}}=\varepsilon _{\mu _{i}\nu _{123}}\gamma
^{\nu _{123}}/6.
\end{eqnarray}%
Thus, the tensors calculated for the amplitudes will correspond to the
versions defined as the main ingredients of the investigation. They violate
the RAGF for the vertex corresponding to $\gamma _{\mu _{i}}$, and the WI
gets violated in the same vertex. Two aspects must be noticed: (i) To have
all the indices present, or to use the definition of the chiral matrix, is
not exceptional since identities (above ones) yield fewer terms and deliver
the same integrated expressions. (ii) The specialty of these identities is
that they furnish the maximum number of RAGF automatically satisfied; hence
the last RAGF can not be met because we would be violating a low-energy
implication (\ref{let2}) in $4D$ and (\ref{let4}) in $2D$.

To sum up, adopting null surface terms makes the amplitudes depend on the
traces used. The Schouten identity inside the integral that connects the
integrands ceases to make it in the final integrated results. Ultimately,
this breaks the linearity of integration and violates the RAGFs. Different
formulae for the traces do not deliver identical tensors. The main elements
involved in the versions were that they correspond to the same integrand;
for instance, in $2D$ $(t_{\mu _{12}}^{AV})_{1}=(t_{\mu _{12}}^{AV})_{2}$.
However, after being integrated separately, we find their subtraction as%
\begin{equation}
(T_{\mu _{12}}^{\left( 2D\right) AV})_{1}-(T_{\mu _{12}}^{\left( 2D\right)
AV})_{2}=2\varepsilon _{\mu _{2}\mu _{1}}(2\Delta _{2\rho }^{\left( 2\right)
\rho }+i/\pi ).
\end{equation}%
Following the same argument, we build up the combination 
\begin{equation}
(t_{\mu _{12}}^{\left( 2D\right) AV})_{1}=(t_{\mu _{12}}^{\left( 2D\right)
AV})_{2}=\frac{1}{r_{1}+r_{2}}[r_{1}(t_{\mu _{12}}^{\left( 2D\right)
AV})_{1}+r_{2}(t_{\mu _{12}}^{\left( 2D\right) AV})_{2}],
\end{equation}%
with $r_{1}+r_{2}\not=0$ and otherwise arbitrary numbers; thus, after
integration and adoption of $\Delta _{2\mu \nu }^{\left( 2\right) }=0$ we
may write any other expression, in particular, the version $(T_{\mu
_{12}}^{\left( 2D\right) AV})_{3}$ discussed in Chapter (\ref{2Dim2Pt})
which is the linear combination above with $r_{1}=r_{2}=1$. In that chapter,
it was used one of the identities satisfied by the antisymmetric products of
Dirac matrices, viz., $\gamma _{\ast }\gamma _{\left[ \mu _{1}\mu _{2}\right]
}=-\varepsilon _{\mu _{1}\mu _{2}}$. In general, not only 2D, all
expressions obtainable utilizing those identities are a linear combination
of the basic versions. Once more because they satisfy the most RAGFs as
possible. With this algorithm in mind, we can build, if desired, the content
one needs, by example,%
\begin{equation}
(T_{\mu _{123}}^{AAA})_{\{1,1,1\}}=\frac{1}{3}[(T_{\mu
_{123}}^{AAA})_{1}+(T_{\mu _{123}}^{AAA})_{2}+(T_{\mu _{123}}^{AAA})_{3}]
\end{equation}%
has one-third of the anomaly in $(T_{\mu _{123}}^{AVV})_{1}$, for each
vertex.

About uniqueness, some definition is necessary. A criterion that makes the
amplitudes unique in a universal sense is impossible since they are
divergent quantities. After renormalization, they become dependent on an
arbitrary mass scale. We employed the definition: One expression coming from
the Feynman rules is unique if, for all intrinsic arbitrariness in
intermediary algebraic manipulations, as Dirac traces and arbitrary
routings, the final result is the same. This concept definition is well
defined in the odd and non-derivative amplitudes studied in $2D$ because we
got an expression depending on the external momentum and independent from
Dirac traces. To the amplitudes investigated in $4D$, the 'unique' answer is
a function of the routings taken as independent variables. Meaning one does
not have a unique amplitude of the external momenta.

As for rules, it makes the surface terms zero as done in even amplitudes and
by an intelligent choice of Dirac trace to obtain the symmetry content.
Notwithstanding, if RAGFs are respected, turning amplitudes unique functions
of their routings, this enables one to recover the symmetry content by
choice of the remaining ambiguities for the momenta labels $k_{i}$, except
2D; this can be done in all even dimensions to the tensors like $T_{\mu
_{1}\cdots \mu _{n+1}}^{\left( 2n\right) A^{2r+1}V^{n-2r}};r\leq \left[ n/2%
\right] .$

\textbf{Gravitation}: The situation changes drastically when the power
counting is higher than linear. For quadratic divergent gravitational
amplitude, by preserving the RAGFs, we have the finiteness of the relevant
surface terms as the constraint; see (\ref{deltafin},\ref{boxfin} and \ref%
{sigmafin}). Thus, it follows a unique form independent of manipulations in
the Dirac algebra but ambiguous in what refers to the routing of the
diagram. The results, in this scenario, for the Weyl anomaly is 
\begin{eqnarray}
\mathcal{W}_{\alpha _{1}\alpha _{2}} &:&=g^{\mu _{12}}T_{\mu _{12}\alpha
_{12}}^{G}=-\frac{1}{96\pi }q^{\nu }\left( \varepsilon _{\alpha _{1}\nu
}q_{\alpha _{2}}+\varepsilon _{\alpha _{2}\nu }q_{\alpha _{1}}\right) -\frac{%
1}{24\pi }\theta _{\alpha _{1}\alpha _{2}}-\frac{1}{48\pi }q_{\alpha
_{2}}q_{\alpha _{1}} \\
&&-\frac{1}{192\pi }\left( 3\chi ^{2}-5\right) \left[ q^{\nu }\left(
\varepsilon _{\alpha _{1}\nu }q_{\alpha _{2}}+\varepsilon _{\alpha _{2}\nu
}q_{\alpha _{1}}\right) +\left( \theta _{\alpha _{1}\alpha _{2}}-q_{\alpha
_{2}}q_{\alpha _{1}}\right) \right] .  \notag
\end{eqnarray}%
Furthermore, for the Einstein anomaly, we have the expression above%
\begin{eqnarray}
\mathcal{E}_{\mu _{2}\alpha _{1}\alpha _{2}} &:&=q^{\mu _{1}}T_{\mu _{1}\mu
_{2}\alpha _{1}\alpha _{2}}^{G}=-\frac{1}{96\pi }\varepsilon _{\alpha
_{1}\nu }q^{\nu }\theta _{\alpha _{2}\mu _{2}}+ \\
&&+\frac{1}{192\pi }(q_{\alpha _{1}}\theta _{\alpha _{2}\mu _{2}}+q_{\alpha
_{2}}\theta _{\alpha _{1}\mu _{2}}-2q_{\mu _{2}}\theta _{\alpha _{1}\alpha
_{2}}-2q_{\alpha _{1}}q_{\alpha _{2}}q_{\mu _{2}})  \notag \\
&&+\frac{\left( 3\chi ^{2}-5\right) }{384\pi }\{2q_{\alpha _{1}}q_{\alpha
_{2}}q_{\mu _{2}}+\varepsilon _{\alpha _{1}}^{~~\nu }\left( q_{\mu
_{2}}\theta _{\nu \alpha _{2}}+q_{\alpha _{2}}\theta _{\nu \mu _{2}}-q_{\nu
}\theta _{\alpha _{2}\mu _{2}}-q_{\alpha _{2}}q_{\mu _{2}}q_{\nu }\right) \}.
\notag
\end{eqnarray}%
The first terms of each expression correspond to the ones in Bertlmann and
Kohlprath \cite{Bertlmann2001a,Bertlmann2001b}. The result shows that apart
from the question of the origin of the additional terms as trivial anomalies
and which actions generate them. They are the product of preserving
algebraic operations determined without resorting to a specific evaluation
of divergent integrals, even though the representation of surface terms
appears in this fashion.

Distinctly from the chiral anomalies, and in a certain sense similar to the
vacuum polarization tensor of 4D quantum electrodynamics, the symmetry
content (or violation thereof) can not be recovered by choice of the
arbitrary internal momenta $k_{1}+k_{2}=\chi q$, at least for the even part
(we restrict ourselves to covariant choices). The odd part allows this for
the parameter $\chi ^{2}=5/3$, namely 
\begin{eqnarray}
\left. \mathcal{W}_{\alpha _{12}}\right\vert _{\chi ^{2}=5/3} &=&-\frac{1}{%
96\pi }[q^{\nu }\left( \varepsilon _{\alpha _{1}\nu }q_{\alpha
_{2}}+\varepsilon _{\alpha _{2}\nu }q_{\alpha _{1}}\right) +4\theta _{\alpha
_{12}}+2q_{\alpha _{12}})] \\
\left. \mathcal{E}_{\mu _{2}\alpha _{12}}\right\vert _{\chi ^{2}=5/3} &=&-%
\frac{1}{192\pi }[2\varepsilon _{\alpha _{1}\nu }q^{\nu }\theta _{\mu
_{2}\alpha _{2}}-q_{(\alpha _{1}}\theta _{\alpha _{2})\mu _{2}}+2q_{\mu
_{2}}\theta _{\alpha _{12}}+2q_{\mu _{2}}q_{\alpha _{12}})].
\end{eqnarray}%
There is no choice of $\chi $ which eliminates the vector part of the
Einstein anomaly for finite surface terms, nor the vector part of the Weyl
one. The only possibility to eliminate the even part of Einstein's anomaly
is to spoil the linearity of integration and turn off the surface terms.
This attitude brings a complex set of possibilities in the axial sector to
be discussed in the sequel. The axial part of the Weyl anomaly can be
eliminated by adopting $\chi ^{2}=1$. However, we did not explore the
aspect, which is interesting since adding the Bardeen-Zumino polynomial in
the stress tensor to change the consistent anomaly in the covariant one, the
odd part disappears; see the book of Bertlmann \cite{Bertlmann1996}, pg. 541
or the paper cited previously.

Turning to the scenario where surface terms vanish and thus freeing the even
part of the Einstein anomaly, the odd part, constituted of multiple terms,
allows the exploration of the traces in each component. It is a choice
available once the algebraic properties of the amplitudes are broken. In
this thesis, we restricted to simplifications where the expressions to each
of the four permutations ($\mu _{1}\leftrightarrow \mu _{2}$)$%
\leftrightarrow $($\alpha _{1}\leftrightarrow \alpha _{2}$) in the expansion
that follows have the same version for each term. 
\begin{equation}
\mathcal{T}_{\mu _{1}\alpha _{1}\mu _{2}\alpha _{2}}^{AV}=4T_{\mu _{1}\alpha
_{1};\mu _{2}\alpha _{2}}^{AV}+2q_{\mu _{2}}T_{\mu _{1}\alpha _{1};\alpha
_{2}}^{AV}+2q_{\alpha _{2}}T_{\mu _{1}\alpha _{1};\mu _{2}}^{AV}+q_{\alpha
_{2}}q_{\mu _{2}}T_{\mu _{1}\alpha _{1}}^{AV}.  \label{BPcon}
\end{equation}

We allowed other trace choices only for the partner $\mathcal{T}^{VA}$,
uniformly in its terms. We do not impose a priori symmetries in the indices,
exploiting just the freedom of the versions. Those symmetries are preserved
once the RAGFs are so, e.g., $T_{\mu _{1}\mu _{2}\alpha _{1}\alpha
_{2}}^{G}=T_{\alpha _{1}\alpha _{2}\mu _{1}\mu _{2}}^{G}$. In making the
selections stated, we arrive at a phenomenon already observed in the chiral
counterparts: the anomalies can migrate from contraction to contraction. The
compact formula for the Einstein anomalies becomes%
\begin{eqnarray}
\mathcal{E}_{\mu _{1}\mu _{2}\alpha _{r}}^{ij} &=&-\frac{1}{96\pi }\left(
\delta _{i,2}+\delta _{j,2}\right) \varepsilon _{\mu _{1}\nu }q^{\nu }\theta
_{\mu _{2}\alpha _{r}} \\
\mathcal{E}_{\mu _{r}\alpha _{1}\alpha _{2}}^{ij} &=&-\frac{1}{96\pi }\left(
2-\delta _{i,2}-\delta _{j,2}\right) \varepsilon _{\alpha _{1}\nu }q^{\nu
}\theta _{\alpha _{2}\mu _{r}}.
\end{eqnarray}%
They come from the contraction with $q^{\alpha _{1};\alpha _{2}}$ and $%
q^{\mu _{1};\mu _{2}}$, being that upper-indices in $\mathcal{E}^{ij}$
assumes $1$ or $2$ values. The Weyl ones are%
\begin{eqnarray}
\mathcal{W}_{\mu _{1}\mu _{2}}^{ij} &=&-\frac{1}{24\pi }\theta _{\mu _{1}\mu
_{2}}-\frac{1}{96\pi }(2-\delta _{i,2}-\delta _{j,2})q^{\nu }(\varepsilon
_{\mu _{1}\nu }q_{\mu _{2}}+\varepsilon _{\mu _{2}\nu }q_{\mu _{1}}) \\
\mathcal{W}_{\alpha _{1}\alpha _{2}}^{ij} &=&-\frac{1}{24\pi }\theta
_{\alpha _{1}\alpha _{2}}-\frac{1}{96\pi }\left( \delta _{i,2}+\delta
_{j,2}\right) q^{\nu }(\varepsilon _{\alpha _{1}\nu }q_{\alpha
_{2}}+\varepsilon _{\alpha _{2}\nu }q_{\alpha _{1}}).
\end{eqnarray}

Notice that when the Einstein anomaly (odd part) drops out in one group of
indices, the Weyl anomaly does so in the complementary set, occurring when $%
i=j$. In the combinations $ij=12$ or $ij=21$, none are zero and equal to
half of the results for the non-vanishing parts of $ij=11$ or $ij=22$. The
mixed versions have coefficients equal to the ones in Bertlmann \cite%
{Bertlmann2001a}, which is one particular result of our analysis.

Ultimately, the expression (\ref{BPcon}) above admits independent choices
for each term. As a consequence, the factor $\Upsilon ,\Upsilon _{\alpha
},\Upsilon _{\alpha \mu }$ (\ref{Ups0}-\ref{Ups1},\ref{Ups2}) do not combine
into the $U_{\alpha \mu }$-factor, and the other projector aside $\theta
_{\mu \alpha }$ ($\omega _{\mu \alpha }=q_{\mu }q_{\alpha }$) would arise
with a proliferation of coefficients. This scenario is allowed for once the
surface terms are interpreted as quantities that vanish. This element leads
to expressions that exhibit Lorentz anomaly. We deviated from this anomaly
once the same version was used when summing the basic permutations. Another
interesting point is to study a low-energy theorem in the gravitational
setting, as done for the chiral anomalies. Research along these lines is
underway.

As a final comment, the possibility of final and compact expressions that
preserve all the features of the computation is mainly due to the use of a
definition of the surface terms of rank four $\square _{3\mu \nu \alpha \rho
}$, and six $\Sigma _{4\mu \nu \alpha \rho \sigma \lambda }$, which are
explicitly total symmetric in the Lorentz indices. In addition, their
compilation into terms that may break the algebraic RAGFs, the objects $%
\{\Upsilon ,\Upsilon _{\alpha },\Upsilon _{\alpha \mu }\}$. In particular,
we call attention to the scalar one, $\Upsilon =2\Delta _{2\rho }^{\rho
}+i/\pi $, which in the last instance, determines the satisfaction or not of
all RAGFs for the energy-momentum two-point function. It is precisely the
same one that appears in the 2D chiral anomaly. The extension of these
protocols to four dimensions facilitates the investigations underway
associated with trace anomalies closely related to the recent publications
in Bonora \cite{Bonora2015} and \cite{Bonora2021}. The RAGFs will become
exceedingly complicated; as an example, we have%
\begin{eqnarray*}
&&2p_{31}^{\alpha }T_{\mu _{123};\alpha }^{AVV}+p_{31}^{2}T_{\mu
_{123}}^{AVV} \\
&=&-2m[T_{\mu _{31}\mu _{2}}^{\tilde{T}V}\left( 1,2\right) +T_{\mu _{12}\mu
_{3}}^{\tilde{T}V}\left( 2,3\right) ] \\
&&+i[\varepsilon _{\mu _{12}}^{\hspace{10pt}\nu _{12}}p_{31\nu _{2}}T_{\nu
_{1}\mu _{3}}^{VV}\left( 2,3\right) ]-i[\varepsilon _{\mu _{13}}^{\hspace{%
10pt}\nu _{12}}p_{31\nu _{2}}T_{\nu _{1}\mu _{2}}^{VV}\left( 1,2\right) ] \\
&&+2[T_{\mu _{32};\mu _{1}}^{AV}\left( 1,2\right) -T_{\mu _{23};\mu
_{1}}^{AV}\left( 2,3\right) ] \\
&&-p_{31}^{\nu _{1}}[g_{\mu _{1}\mu _{2}}T_{\nu _{1}\mu _{3}}^{AV}\left(
2,3\right) +g_{\mu _{1}\mu _{3}}T_{\nu _{1}\mu _{2}}^{AV}\left( 1,2\right) ]
\\
&&+[p_{31\mu _{1}}T_{\mu _{32}}^{AV}\left( 1,2\right) +p_{31\mu _{3}}T_{\mu
_{12}}^{AV}\left( 1,2\right) -p_{31\mu _{1}}T_{\mu _{23}}^{AV}\left(
2,3\right) +p_{31\mu _{2}}T_{\mu _{13}}^{AV}\left( 2,3\right) ],
\end{eqnarray*}%
where even arises a pseudo-tensor vertex $\tilde{T}=\gamma _{\ast }\gamma _{%
\left[ \mu \nu \right] }$. Nonetheless, by the systematization developed in
this thesis such task becomes feasible as well.

\appendix%

\chapter{Dirac Matrices and Traces}

Lets us introduce the Clifford algebra representation in terms of matrices $%
\left\{ \gamma _{\mu _{1}},\gamma _{\mu _{2}}\right\} =2g_{\mu _{12}}\mathbf{%
1,}$ the dimension of irreducible representations are $\mathrm{\dim }\left(
\gamma \right) =2^{\left[ d/2\right] }\times 2^{\left[ d/2\right] }$, and
the basic traces are%
\begin{eqnarray}
\mathrm{tr}\left( \gamma ^{\mu }\right) &=&0 \\
\mathrm{tr}\left\{ \gamma _{\alpha },\gamma _{\beta }\right\} &=&2g_{\alpha
\beta }\mathrm{tr}\left( \mathbf{1}_{2^{n}\times 2^{n}}\right) .
\end{eqnarray}%
For the two dimensional representation, we have: 
\begin{eqnarray}
\gamma _{0} &=&\sigma _{2};\gamma _{1}=i\sigma _{1};\gamma _{3}=\sigma _{3}
\\
\gamma _{0} &=&\sigma _{1};\gamma _{1}=i\sigma _{2};\gamma _{3}=-\sigma _{3}.
\notag
\end{eqnarray}

For even dimensions, $d=2n$, there is a matrix given by%
\begin{equation}
\gamma _{\ast }:=i^{n-1}\gamma _{0}\gamma _{1}\cdots \gamma _{2n-1}=\frac{%
i^{n-1}}{\left( 2n\right) !}\varepsilon _{\nu _{1}\cdots \nu _{2n}}\gamma
^{\nu _{1}\cdots \nu _{2n}}
\end{equation}%
that obeys $\left\{ \gamma _{\ast },\gamma _{\mu }\right\} =0,$ with $%
\varepsilon _{012\cdots d-1}=-1$. For four matrices, we have the trace%
\begin{equation}
\mathrm{tr}\left( \gamma ^{\mu _{1}\cdots \mu _{4}}\right) =\mathrm{tr}%
\left( \mathbf{1}_{2^{n}\times 2^{n}}\right) \left( g^{\mu _{1}\mu
_{2}}g^{\mu _{3}\mu _{4}}-g^{\mu _{1}\mu _{3}}g^{\mu _{2}\mu _{4}}+g^{\mu
_{1}\mu _{4}}g^{\mu _{2}\mu _{3}}\right) ,
\end{equation}%
the general formula is%
\begin{equation}
\mathrm{tr}\left( \gamma _{\mu _{1}\cdots \mu _{2n}}\right)
=\sum_{i=2}^{2n}\left( -1\right) ^{i}g_{\mu _{1}\mu _{i}}\mathrm{tr}\left(
\gamma _{\mu _{1}\cdots \hat{\mu}_{i}\cdots \mu _{2n}}\right) .
\end{equation}

The first non-zero trace with the chiral matrix in any even dimension is
given by%
\begin{equation}
\mathrm{tr}\left( \gamma _{\ast }\gamma _{\mu _{1}}\gamma _{\mu _{2}}\cdots
\gamma _{\mu _{2n}}\right) =2^{n}i^{n-1}\left( -1\right) ^{n}\varepsilon
_{\mu _{12\cdots }\mu _{2n}},
\end{equation}%
for $d=2n$ to the string of $2n+2$ gamma matrices plus $\gamma _{\ast }$
using its definition follows the formula%
\begin{equation}
\mathrm{tr}(\gamma _{\ast }\gamma _{a_{1}a_{2}\cdots
a_{2n+1}a_{2n+2}})=2^{n}i^{3n-1}\sum_{k=1}^{2n+1}\sum_{j=k+1}^{2n+2}\left(
-1\right) ^{j+k+1}g_{a_{k}a_{j}}\varepsilon _{a_{1}\cdots \hat{a}_{k}\cdots 
\hat{a}_{j}\cdots (2n+2)},
\end{equation}%
where we have used the abbreviation $\gamma _{a_{1}a_{2}\cdots
a_{2n+1}a_{2n+2}}=\prod_{j=1}^{2b+2}\gamma _{a_{j}}.$ The Latin index ought
to be substituted to whatever configuration of Lorentz indices is
scrutinized.

\section{Traces of a String of Six Gamma and the Chiral Matrix}

\label{Tr6G4D}

One uses the following identities to insert the Levi-Civita tensor in traces
with the chiral matrix%
\begin{equation*}
\gamma _{\ast }\gamma _{\left[ \mu _{1}\cdots \mu _{r}\right] }=\frac{%
i^{n-1+r\left( r+1\right) }}{\left( 2n-r\right) !}\varepsilon _{\mu
_{1}\cdots \mu _{r}\nu _{r+1}\cdots \nu _{2n}}\gamma ^{\left[ \nu
_{r+1}\cdots \nu _{2n}\right] },
\end{equation*}%
where the notation $\gamma _{\left[ \mu _{1}\cdots \mu _{r}\right] }$
indicates antisymmetrized products of gammas and the investigated dimension
is $2n=4$. This appendix uses this resource to achieve different trace
expressions and explore their relations.

\textbf{Trace using the definition }$\gamma _{\ast }=i\varepsilon _{\nu
_{1}\nu _{2}\nu _{3}\nu _{4}}\gamma ^{\nu _{1}\nu _{2}\nu _{3}\nu _{4}}/4!$%
\textbf{\ - }The three leading positions to substitute the definition are
around vertices $\Gamma _{1}$, $\Gamma _{2}$, and $\Gamma _{3}$. Even if
that brings six options, the same integrated expressions arise regardless of
replacing at the left or right. Thus, we cast the possibilities in the
sequence 
\begin{eqnarray*}
t_{1} &=&\mathrm{tr}(\gamma _{\ast }\gamma _{\mu _{1}\nu _{1}\mu _{2}\nu
_{2}\mu _{3}\nu _{3}})=i\varepsilon ^{\alpha _{1}\alpha _{2}\alpha
_{3}\alpha _{4}}\mathrm{tr}(\gamma _{\alpha _{1}\alpha _{2}\alpha _{3}\alpha
_{4}}\gamma _{\mu _{1}\nu _{1}\mu _{2}\nu _{2}\mu _{3}\nu _{3}})/4! \\
&=&+g_{\mu _{1}\nu _{1}}\varepsilon _{\mu _{2}\nu _{2}\mu _{3}\nu
_{3}}-g_{\mu _{1}\mu _{2}}\varepsilon _{\nu _{1}\nu _{2}\mu _{3}\nu
_{3}}+g_{\mu _{1}\nu _{2}}\varepsilon _{\nu _{1}\mu _{2}\mu _{3}\nu
_{3}}-g_{\mu _{1}\mu _{3}}\varepsilon _{\nu _{1}\mu _{2}\nu _{2}\nu
_{3}}+g_{\mu _{1}\nu _{3}}\varepsilon _{\nu _{1}\mu _{2}\nu _{2}\mu _{3}} \\
&&+g_{\nu _{1}\mu _{2}}\varepsilon _{\mu _{1}\nu _{2}\mu _{3}\nu
_{3}}-g_{\nu _{1}\nu _{2}}\varepsilon _{\mu _{1}\mu _{2}\mu _{3}\nu
_{3}}+g_{\nu _{1}\mu _{3}}\varepsilon _{\mu _{1}\mu _{2}\nu _{2}\nu
_{3}}-g_{\nu _{1}\nu _{3}}\varepsilon _{\mu _{1}\mu _{2}\nu _{2}\mu
_{3}}+g_{\mu _{2}\nu _{2}}\varepsilon _{\mu _{1}\nu _{1}\mu _{3}\nu _{3}} \\
&&-g_{\mu _{2}\mu _{3}}\varepsilon _{\mu _{1}\nu _{1}\nu _{2}\nu
_{3}}+g_{\mu _{2}\nu _{3}}\varepsilon _{\mu _{1}\nu _{1}\nu _{2}\mu
_{3}}+g_{\nu _{2}\mu _{3}}\varepsilon _{\mu _{1}\nu _{1}\mu _{2}\nu
_{3}}-g_{\nu _{2}\nu _{3}}\varepsilon _{\mu _{1}\nu _{1}\mu _{2}\mu
_{3}}+g_{\mu _{3}\nu _{3}}\varepsilon _{\mu _{1}\nu _{1}\mu _{2}\nu _{2}},
\end{eqnarray*}%
\begin{eqnarray*}
t_{2} &=&\mathrm{tr}(\gamma _{\mu _{1}\nu _{1}}\gamma _{\ast }\gamma _{\mu
_{2}\nu _{2}\mu _{3}\nu _{3}})=i\varepsilon ^{\alpha _{1}\alpha _{2}\alpha
_{3}\alpha _{4}}\mathrm{tr}(\gamma _{\mu _{1}\nu _{1}}\gamma _{\alpha
_{1}\alpha _{2}\alpha _{3}\alpha _{4}}\gamma _{\mu _{2}\nu _{2}\mu _{3}\nu
_{3}})/4! \\
&=&+g_{\mu _{1}\nu _{1}}\varepsilon _{\mu _{2}\nu _{2}\mu _{3}\nu
_{3}}+g_{\mu _{1}\mu _{2}}\varepsilon _{\nu _{1}\nu _{2}\mu _{3}\nu
_{3}}-g_{\mu _{1}\nu _{2}}\varepsilon _{\mu _{2}\nu _{2}\mu _{3}\nu
_{3}}+g_{\mu _{1}\mu _{3}}\varepsilon _{\nu _{1}\mu _{2}\nu _{2}\nu
_{3}}-g_{\mu _{1}\nu _{3}}\varepsilon _{\nu _{1}\mu _{2}\nu _{2}\mu _{3}} \\
&&-g_{\nu _{1}\mu _{2}}\varepsilon _{\mu _{1}\nu _{2}\mu _{3}\nu
_{3}}+g_{\nu _{1}\nu _{2}}\varepsilon _{\mu _{1}\mu _{2}\mu _{3}\nu
_{3}}-g_{\nu _{1}\mu _{3}}\varepsilon _{\mu _{1}\mu _{2}\nu _{2}\nu
_{3}}+g_{\nu _{1}\nu _{3}}\varepsilon _{\mu _{1}\mu _{2}\nu _{2}\mu
_{3}}+g_{\mu _{2}\nu _{2}}\varepsilon _{\mu _{1}\nu _{1}\mu _{3}\nu _{3}} \\
&&-g_{\mu _{2}\mu _{3}}\varepsilon _{\mu _{1}\nu _{1}\nu _{2}\nu
_{3}}+g_{\mu _{2}\nu _{3}}\varepsilon _{\mu _{1}\nu _{1}\nu _{2}\mu
_{3}}+g_{\nu _{2}\mu _{3}}\varepsilon _{\mu _{1}\nu _{1}\mu _{2}\nu
_{3}}-g_{\nu _{2}\nu _{3}}\varepsilon _{\mu _{1}\nu _{1}\mu _{2}\mu
_{3}}+g_{\mu _{3}\nu _{3}}\varepsilon _{\mu _{1}\nu _{1}\mu _{2}\nu _{2}},
\end{eqnarray*}%
\begin{eqnarray*}
t_{3} &=&\mathrm{tr}(\gamma _{\mu _{1}\nu _{1}\mu _{2}\nu _{2}}\gamma _{\ast
}\gamma _{\mu _{3}\nu _{3}})=i\varepsilon ^{\alpha _{1}\alpha _{2}\alpha
_{3}\alpha _{4}}\mathrm{tr}(\gamma _{\mu _{1}\nu _{1}\mu _{2}\nu _{2}}\gamma
_{\alpha _{1}\alpha _{2}\alpha _{3}\alpha _{4}}\gamma _{\mu _{3}\nu _{3}})/4!
\\
&=&+g_{\mu _{1}\nu _{1}}\varepsilon _{\mu _{2}\nu _{2}\mu _{3}\nu
_{3}}-g_{\mu _{1}\mu _{2}}\varepsilon _{\nu _{1}\nu _{2}\mu _{3}\nu
_{3}}+g_{\mu _{1}\nu _{2}}\varepsilon _{\mu _{2}\nu _{2}\mu _{3}\nu
_{3}}+g_{\mu _{1}\mu _{3}}\varepsilon _{\nu _{1}\mu _{2}\nu _{2}\nu
_{3}}-g_{\mu _{1}\nu _{3}}\varepsilon _{\nu _{1}\mu _{2}\nu _{2}\mu _{3}} \\
&&+g_{\nu _{1}\mu _{2}}\varepsilon _{\mu _{1}\nu _{2}\mu _{3}\nu
_{3}}-g_{\nu _{1}\nu _{2}}\varepsilon _{\mu _{1}\mu _{2}\mu _{3}\nu
_{3}}-g_{\nu _{1}\mu _{3}}\varepsilon _{\mu _{1}\mu _{2}\nu _{2}\nu
_{3}}+g_{\nu _{1}\nu _{3}}\varepsilon _{\mu _{1}\mu _{2}\nu _{2}\mu
_{3}}+g_{\mu _{2}\nu _{2}}\varepsilon _{\mu _{1}\nu _{1}\mu _{3}\nu _{3}} \\
&&+g_{\mu _{2}\mu _{3}}\varepsilon _{\mu _{1}\nu _{1}\nu _{2}\nu
_{3}}-g_{\mu _{2}\nu _{3}}\varepsilon _{\mu _{1}\nu _{1}\nu _{2}\mu
_{3}}-g_{\nu _{2}\mu _{3}}\varepsilon _{\mu _{1}\nu _{1}\mu _{2}\nu
_{3}}+g_{\nu _{2}\nu _{3}}\varepsilon _{\mu _{1}\nu _{1}\mu _{2}\mu
_{3}}+g_{\mu _{3}\nu _{3}}\varepsilon _{\mu _{1}\nu _{1}\mu _{2}\nu _{2}},
\end{eqnarray*}%
where we omit the global factor $4i$. Since each expression contains fifteen
monomials featuring all index configurations, different signs are the unique
distinguishing factor among them. That is also the reason why references
often name them symmetric or democratic \cite{AguilaVictoria1998, Wu2006,
Viglioni2016}.

These (main) versions play fundamental roles in this investigation as they
are enough to obtain any other result. If we use any other identity
constructed with the equations involving the antisymmetric products the
trace expressions relate directly to them or their combinations $%
t_{ij}=\left( t_{i}+t_{j}\right) /2$ only using sums and no other operation.
Consequently, any expression attributed to the investigated triangles is a
linear combination of those detailed in the main body of this work. All of
them produce the mentioned relations, so we cast some at the end of this
appendix.%
\begin{eqnarray*}
t_{12} &=&-g_{\mu _{1}\nu _{1}}\varepsilon _{\mu _{2}\mu _{3}\nu _{2}\nu
_{3}}-g_{\mu _{2}\nu _{2}}\varepsilon _{\mu _{1}\mu _{3}\nu _{1}\nu
_{3}}+g_{\mu _{2}\nu _{3}}\varepsilon _{\mu _{1}\mu _{3}\nu _{1}\nu _{2}} \\
&&-g_{\nu _{2}\mu _{3}}\varepsilon _{\mu _{1}\mu _{2}\nu _{1}\nu
_{3}}-g_{\mu _{3}\nu _{3}}\varepsilon _{\mu _{1}\mu _{2}\nu _{1}\nu
_{2}}-g_{\mu _{2}\mu _{3}}\varepsilon _{\mu _{1}\nu _{1}\nu _{2}\nu
_{3}}-g_{\nu _{2}\nu _{3}}\varepsilon _{\mu _{1}\mu _{2}\mu _{3}\nu _{1}},
\end{eqnarray*}%
\begin{eqnarray*}
t_{13} &=&-g_{\mu _{3}\nu _{3}}\varepsilon _{\mu _{1}\mu _{2}\nu _{1}\nu
_{2}}-g_{\mu _{1}\nu _{1}}\varepsilon _{\mu _{2}\mu _{3}\nu _{2}\nu
_{3}}+g_{\mu _{1}\nu _{2}}\varepsilon _{\mu _{2}\mu _{3}\nu _{1}\nu _{3}} \\
&&-g_{\nu _{1}\mu _{2}}\varepsilon _{\mu _{1}\mu _{3}\nu _{2}\nu
_{3}}-g_{\mu _{2}\nu _{2}}\varepsilon _{\mu _{1}\mu _{3}\nu _{1}\nu
_{3}}-g_{\mu _{1}\mu _{2}}\varepsilon _{\mu _{3}\nu _{1}\nu _{2}\nu
_{3}}-g_{\nu _{1}\nu _{2}}\varepsilon _{\mu _{1}\mu _{2}\mu _{3}\nu _{3}},
\end{eqnarray*}%
\begin{eqnarray*}
t_{23} &=&-g_{\mu _{2}\nu _{2}}\varepsilon _{\mu _{1}\mu _{3}\nu _{1}\nu
_{3}}-g_{\mu _{1}\nu _{1}}\varepsilon _{\mu _{2}\mu _{3}\nu _{2}\nu
_{3}}+g_{\mu _{1}\nu _{3}}\varepsilon _{\mu _{2}\mu _{3}\nu _{1}\nu _{2}} \\
&&-g_{\nu _{1}\mu _{3}}\varepsilon _{\mu _{1}\mu _{2}\nu _{2}\nu
_{3}}-g_{\mu _{3}\nu _{3}}\varepsilon _{\mu _{1}\mu _{2}\nu _{1}\nu
_{2}}-g_{\mu _{1}\mu _{3}}\varepsilon _{\mu _{2}\nu _{1}\nu _{2}\nu
_{3}}-g_{\nu _{1}\nu _{3}}\varepsilon _{\mu _{1}\mu _{2}\mu _{3}\nu _{2}},
\end{eqnarray*}

\textbf{Trace using} $\gamma _{\ast }\gamma _{a}=-i\varepsilon _{a\nu
_{1}\nu _{2}\nu _{3}}\gamma ^{\nu _{1}\nu _{2}\nu _{3}}/3!$ - After using
this identity for the chiral matrix and the first gamma, we write this trace
through ten monomials.%
\begin{equation*}
\eta _{1}\left( a\right) =\mathrm{tr}\left( \gamma _{\ast }\gamma
_{abcdef}\right) =-i\varepsilon _{a}^{\quad \nu _{1}\nu _{2}\nu _{3}}\mathrm{%
tr}\left( \gamma _{\nu _{1}\nu _{2}\nu _{3}}\gamma _{bcdef}\right) /6
\end{equation*}%
\begin{eqnarray*}
\eta _{1}\left( a\right) &=&g_{\text{$b$}c}\varepsilon _{\text{$a$}def}-g_{%
\text{$b$}d}\varepsilon _{\text{$a$}cef}+g_{\text{$b$}e}\varepsilon _{\text{$%
a$}cdf}-g_{\text{$b$}f}\varepsilon _{\text{$a$}cde}+g_{\text{$c$}%
d}\varepsilon _{\text{$a$}bef} \\
&&-g_{\text{$c$}e}\varepsilon _{\text{$a$}bdf}+g_{\text{$c$}f}\varepsilon _{%
\text{$a$}bde}+g_{de}\varepsilon _{abcf}+g_{ef}\varepsilon
_{abcd}-g_{df}\varepsilon _{abce}
\end{eqnarray*}

\textbf{Trace using} $\gamma _{\ast }\gamma _{\left[ ab\right]
}=-i\varepsilon _{ab\nu _{1}\nu _{2}}\gamma ^{\nu _{1}\nu _{2}}/2!$ - This
case requires expressing the ordinary product in terms of the
antisymmetrized one. We find seven monomials after taking the traces. 
\begin{equation*}
\gamma _{\ast }\gamma _{ab}=-\frac{1}{2}i\varepsilon _{ab\nu _{1}\nu
_{2}}\gamma ^{\nu _{1}\nu _{1}}+g_{ab}\gamma _{\ast }
\end{equation*}%
\begin{eqnarray*}
\eta _{2}\left( ab\right) =\mathrm{tr}\left( \gamma _{\ast }\gamma
_{abcdef}\right) &=&g_{ab}\varepsilon _{cdef}+g_{cd}\varepsilon
_{abef}-g_{ce}\varepsilon _{abdf}+g_{cf}\varepsilon _{abde} \\
&&+g_{d\text{$e$}}\varepsilon _{abcf}-g_{df}\varepsilon
_{abce}+g_{ef}\varepsilon _{abcd}
\end{eqnarray*}

\textbf{Trace using} $\gamma _{\ast }\gamma _{\left[ abc\right]
}=i\varepsilon _{abc\nu }\gamma ^{\nu }$ - Following a similar procedure we
find six monomials.%
\begin{equation*}
\gamma _{\ast }\gamma _{abc}=i\varepsilon _{abc\nu }\gamma ^{\nu }+\gamma
_{\ast }\left( g_{bc}\gamma _{a}-g_{ac}\gamma _{b}+g_{ab}\gamma _{c}\right)
\end{equation*}%
\begin{equation*}
\eta _{3}\left( abc\right) =\mathrm{tr}\left( \gamma _{\ast }\gamma
_{abcdef}\right) =g_{ab}\varepsilon _{cdef}-g_{ac}\varepsilon
_{bdef}+g_{bc}\varepsilon _{adef}+g_{de}\varepsilon
_{abcf}-g_{df}\varepsilon _{abce}+g_{ef}\varepsilon _{abcd}
\end{equation*}

\textbf{Trace using} $\gamma _{\ast }\gamma _{\left[ abcd\right]
}=i\varepsilon _{abcd}$ - This case also generates seven monomials. 
\begin{eqnarray*}
\gamma _{\ast }\gamma _{abcd} &=&i\varepsilon _{abcd}\mathbf{1}+g_{ab}\gamma
_{\ast }\gamma _{\left[ cd\right] }-g_{ac}\gamma _{\ast }\gamma _{\left[ bd%
\right] }+g_{ad}\gamma _{\ast }\gamma _{\left[ bc\right] } \\
&&+g_{bc}\gamma _{\ast }\gamma _{\left[ ad\right] }-g_{bd}\gamma _{\ast
}\gamma _{\left[ ac\right] }+g_{cd}\gamma _{\ast }\gamma _{\left[ ab\right]
}+\left( g_{ab}g_{cd}-g_{ac}g_{bd}+g_{ad}g_{bc}\right) \gamma _{\ast }
\end{eqnarray*}%
\begin{eqnarray*}
\eta _{4}\left( abcd\right) =\mathrm{tr}\left( \gamma _{\ast }\gamma
_{abcdef}\right) &=&g_{ab}\varepsilon _{cdef}-g_{ac}\varepsilon
_{bdef}+g_{ad}\varepsilon _{bcef}+g_{bc}\varepsilon _{adef} \\
&&-g_{bd}\varepsilon _{acef}+g_{cd}\varepsilon _{abef}+g_{ef}\varepsilon
_{abcd}
\end{eqnarray*}

\textbf{Interconnection among formulas: }When computing the difference
between two integrated versions of the same amplitude, we acknowledge two
situations. First, it cancels out identically as their integrands are
precisely equal, for example:%
\begin{equation*}
\left[ t_{12}-\eta _{2}\left( \mu _{1}\nu _{1}\right) \right] =0,\qquad %
\left[ t_{23}-\eta _{4}\left( \mu _{3}\nu _{3}\mu _{1}\nu _{1}\right) \right]
=0.
\end{equation*}%
Second, it vanishes in the integration because the explicit computation
corresponds to finite null integrals embodied into the $t^{\left( -+\right)
} $ tensor (\ref{T-+}) and the $ASS$ amplitude (\ref{ASS}). Some examples
are:%
\begin{eqnarray*}
\left[ t_{1}-\eta _{1}\left( \mu _{1}\right) \right] \frac{K_{123}^{\nu
_{123}}}{D_{123}} &=&\varepsilon _{\mu _{2}\mu _{3}\nu _{1}\nu _{2}}t_{\mu
_{1}}^{\left( -+\right) \nu _{12}}-g_{\mu _{1}\mu _{3}}t_{\mu
_{2}}^{ASS}+g_{\mu _{1}\mu _{2}}t_{\mu _{3}}^{ASS}, \\
\left[ t_{12}+\eta _{2}\left( \nu _{1}\mu _{2}\right) \right] \frac{%
K_{123}^{\nu _{123}}}{D_{123}} &=&-\varepsilon _{\mu _{2}\mu _{3}\nu _{1}\nu
_{2}}t_{\mu _{1}}^{\left( -+\right) \nu _{12}}+\varepsilon _{\mu _{1}\mu
_{3}\nu _{1}\nu _{2}}t_{\mu _{2}}^{\left( -+\right) \nu _{12}}-g_{\mu
_{2}\mu _{3}}t_{\mu _{1}}^{ASS}+g_{\mu _{1}\mu _{3}}t_{\mu _{2}}^{ASS}, \\
\left[ t_{13}+\eta _{4}\left( \nu _{1}\mu _{2}\nu _{2}\mu _{3}\right) \right]
\frac{K_{123}^{\nu _{123}}}{D_{123}} &=&-\varepsilon _{\mu _{2}\mu _{3}\nu
_{1}\nu _{2}}t_{\mu _{1}}^{\left( -+\right) \nu _{12}}-\varepsilon _{\mu
_{1}\mu _{2}\nu _{1}\nu _{2}}t_{\mu _{3}}^{\left( -+\right) \nu
_{12}}+g_{\mu _{2}\mu _{3}}t_{\mu _{1}}^{ASS}-g_{\mu _{1}\mu _{2}}t_{\mu
_{3}}^{ASS}.
\end{eqnarray*}%
\begin{eqnarray*}
\left[ t_{12}-\eta _{3}\left( \mu _{1}\nu _{1}\mu _{2}\right) \right] \frac{%
K_{123}^{\nu _{123}}}{D_{123}} &=&-g_{\mu _{2}\mu _{3}}t_{\mu
_{1}}^{ASS}+\varepsilon _{\mu _{13}\nu _{12}}t_{\mu _{2}}^{\left( -+\right)
\nu _{12}}+g_{\mu _{1}\mu _{2}}t_{\mu _{3}}^{ASS}, \\
\left[ t_{23}-\eta _{3}\left( \mu _{2}\nu _{2}\mu _{3}\right) \right] \frac{%
K_{123}^{\nu _{123}}}{D_{123}} &=&-g_{\mu _{3}\mu _{1}}t_{\mu
_{2}}^{ASS}-\varepsilon _{\mu _{12}\nu _{12}}t_{\mu _{3}}^{\left( -+\right)
\nu _{12}}+g_{\mu _{2}\mu _{3}}t_{\mu _{1}}^{ASS} \\
\left[ t_{31}-\eta _{3}\left( \mu _{3}\nu _{3}\mu _{1}\right) \right] \frac{%
K_{123}^{\nu _{123}}}{D_{123}} &=&-g_{\mu _{1}\mu _{2}}t_{\mu
_{3}}^{ASS}-\varepsilon _{\mu _{23}\nu _{12}}t_{\mu _{1}}^{\left( -+\right)
\nu _{12}}+g_{\mu _{3}\mu _{1}}t_{\mu _{2}}^{ASS}
\end{eqnarray*}%
We showed the forms that identically correspond here, not that all
differences are finite and vanishing. For example, the form obtained from $%
t_{12}$ is not identical without conditions to any $t_{i}$.

\chapter{Feynman Integrals}

\section{Feynman's parametrization}

Any integral that is explicitly evaluated in this work is well defined. To
operate, we combine the denominators that appear using Feynman
parametrization. The functions that occur after they have been split through
the formula (\ref{id}) share the form 
\begin{equation}
\frac{1}{D_{\lambda }^{N}D_{1}...D_{n}}.
\end{equation}%
They can be combined as%
\begin{equation}
\frac{1}{D_{\lambda }^{N}D_{1}...D_{n}}=\left( N\right) _{n}\int_{0}^{1}%
\mathrm{d}x_{1}\cdots \int_{0}^{1-x_{1}-...-x_{n-1}}\mathrm{d}x_{n}\frac{%
\left( 1-x_{1}-\cdots x_{N}\right) ^{N-1}}{\left[ \sum_{i=1}^{n}\left(
D_{i}-D_{\lambda }\right) x_{i}+D_{\lambda }\right] ^{n+N}},
\end{equation}%
where $\left( N\right) _{n}$ is the Pochhammer symbol $\left( N\right)
_{n}=\Gamma \left( N+n\right) /\Gamma \left( N\right) .$ It is a direct task
by induction to show that%
\begin{eqnarray}
\sum_{i=1}^{n}\left( D_{i}-D_{\lambda }\right) x_{i}+D_{\lambda }
&=&k^{2}-\lambda ^{2}+\sum_{i=1}^{n}\left( 2k\cdot k_{i}+k_{i}^{2}\right)
x_{i}+\sum_{i=1}^{n}\left( \lambda ^{2}-m_{i}^{2}\right) x_{i} \\
&=&\left( k+\sum_{i=1}^{n}k_{i}x_{i}\right) ^{2}+Q\left( \left\{
k_{i},m_{i}^{2}\right\} ;\lambda ^{2}\right) ,  \notag
\end{eqnarray}%
where we define the $Q$ polynomial%
\begin{equation}
Q\left( \left\{ k_{i},m_{i}^{2}\right\} ;\lambda ^{2}\right)
=\sum_{i=1}^{n}k_{i}^{2}x_{i}\left( 1-x_{i}\right) -2\sum_{j>i}^{n}\left(
k_{i}\cdot k_{j}\right) x_{i}x_{j}+\sum_{i=1}^{n}\left( \lambda
^{2}-m_{i}^{2}\right) x_{i}-\lambda ^{2}.
\end{equation}%
After integrating into the momentum $k$, we have a function of $Q$ whose
integral over adequate parameter delivers the integrals used in work. As of
the finite functions, they appear as%
\begin{equation}
\frac{1}{D_{1}...D_{n}}=\Gamma \left( n\right) \int_{0}^{1}\mathrm{d}%
x_{1}\cdots \int_{0}^{1-x_{1}-...-x_{n-2}}\mathrm{d}x_{n-1}\frac{1}{\left[
\sum_{i=1}^{n-1}\left( D_{i}-D_{1}\right) x_{i}+D_{1}\right] ^{n}}.
\end{equation}

An example to illustrate this is the finite integral%
\begin{equation}
I_{2}=\int \frac{\mathrm{d}^{2}k}{\left( 2\pi \right) ^{2}}\frac{1}{D_{12}}
\end{equation}%
the explicit $D_{i}$ are%
\begin{eqnarray}
D_{1} &=&\left( k+k_{1}\right) ^{2}-m_{1}^{2} \\
D_{2} &=&\left( k+k_{2}\right) ^{2}-m_{2}^{2}
\end{eqnarray}%
thus we identify%
\begin{eqnarray}
\left( D_{2}-D_{1}\right) x+D_{1} &=&k^{2}+2k\cdot \left[ \left(
k_{2}-k_{1}\right) x+k_{1}\right] +\left( k_{2}^{2}-k_{1}^{2}\right)
x+k_{1}^{2}+\left( m_{1}^{2}-m_{2}^{2}\right) x-m_{1}^{2} \\
&=&\left[ k+\left( k_{2}-k_{1}\right) x+k_{1}\right] ^{2}+\left(
k_{2}-k_{1}\right) ^{2}x\left( 1-x\right) +\left( m_{1}^{2}-m_{2}^{2}\right)
x-m_{1}^{2}
\end{eqnarray}%
and with $q=k_{2}-k_{1}$ the $Q$ polynomial%
\begin{equation}
Q=q^{2}x\left( 1-x\right) +\left( m_{1}^{2}-m_{2}^{2}\right) x-m_{1}^{2}.
\end{equation}%
When integrating the translation in the k variable 
\begin{equation}
k\rightarrow k-\left[ \left( k_{2}-k_{1}\right) x+k_{1}\right]
\end{equation}%
allows us to write the integral as%
\begin{equation}
J_{2}=\int_{0}^{1}\mathrm{d}z\int \frac{\mathrm{d}^{2}k}{\left( 2\pi \right)
^{2}}\frac{1}{\left( k^{2}+Q\right) ^{2}}.
\end{equation}%
The next step is integration in the momentum, where the next section derives
the necessary formulae.

\section{The $J_{2\protect\mu \protect\nu }^{\left( 2\right) }$ Integral}

\label{AppCJ22}

For non-negative power counting integrals, we must split them using the
identity (\ref{id}). Let us illustrate the type of operations needed to
integrate such integrals using as an example the fundamental tensor integral
with arbitrary masses in two dimensions%
\begin{equation}
\bar{J}_{2}^{\left( 2\right) \mu \nu }=\int \frac{\mathrm{d}^{2}k}{\left(
2\pi \right) ^{2}}\frac{K_{1}^{\mu }K_{1}^{\nu }}{D_{12}}.
\end{equation}%
Its integrand is decomposed in the form%
\begin{equation}
\frac{K_{1}^{\mu }K_{1}^{\nu }}{D_{12}}=\frac{K_{1}^{\mu }K_{1}^{\nu }}{%
D_{\lambda }^{2}}-\frac{K_{1}^{\mu }K_{1}^{\nu }A_{2}}{D_{\lambda }^{2}D_{2}}%
-\frac{K_{1}^{\mu }K_{1}^{\nu }A_{1}}{D_{\lambda }D_{12}}.
\end{equation}%
Then, the following integrals are required to perform%
\begin{eqnarray}
\bar{J}_{2}^{\left( 2\right) \mu \nu } &=&\int \frac{\mathrm{d}^{2}k}{\left(
2\pi \right) ^{2}}\frac{K_{1}^{\mu }K_{1}^{\nu }}{D_{12}}=\int \frac{\mathrm{%
d}^{2}k}{\left( 2\pi \right) ^{2}}\left\{ \frac{K_{1}^{\mu }K_{1}^{\nu }}{%
D_{\lambda }^{2}}-\frac{K_{1}^{\mu }K_{1}^{\nu }A_{2}}{D_{\lambda }^{2}D_{2}}%
-\frac{K_{1}^{\mu }K_{1}^{\nu }A_{1}}{D_{\lambda }D_{12}}\right\} \\
&=&\int \frac{\mathrm{d}^{2}k}{\left( 2\pi \right) ^{2}}\frac{K_{1}^{\mu
}K_{1}^{\nu }}{D_{\lambda }^{2}}-F_{b}^{\mu \nu }-F_{a}^{\mu \nu }.
\end{eqnarray}%
The final answer will be expressed as functional in $Q=q^{2}x\left(
1-x\right) +\left( m_{1}^{2}-m_{2}^{2}\right) x-m_{1}^{2}.$

To start with, we combine the denominators with Feynman parametrization for $%
F_{a}^{\mu \nu }$%
\begin{equation}
\frac{K_{1}^{\mu }K_{1}^{\nu }A_{1}}{D_{\lambda }D_{12}}=2\int_{0}^{1}%
\mathrm{d}x_{1}\int_{0}^{1-x_{1}}\mathrm{d}x_{2}\frac{1}{\left[ \left(
D_{2}-D_{\lambda }\right) x_{1}+\left( D_{1}-D_{\lambda }\right)
x_{2}+D_{\lambda }\right] ^{3}}.
\end{equation}%
Integrating into the loop momentum and making the shift $k\rightarrow
k-\left( k_{2}x_{1}+k_{1}x_{2}\right) $, we reach to%
\begin{equation}
F_{a}^{\mu \nu }=\int \frac{\mathrm{d}^{2}k}{\left( 2\pi \right) ^{2}}\frac{%
K_{1}^{\mu }K_{1}^{\nu }A_{1}}{D_{\lambda }D_{12}}=2\int_{0}^{1}\mathrm{d}%
x_{1}\int_{0}^{1-x_{1}}\mathrm{d}x_{2}\int \frac{\mathrm{d}^{2}k}{\left(
2\pi \right) ^{2}}\frac{\left[ K_{1}^{\mu }K_{1}^{\nu }A_{1}\right]
_{k-\left( k_{2}x_{1}+k_{1}x_{2}\right) }}{\left( k^{2}+Q\right) ^{3}},
\end{equation}%
where the $Q$ polynomial is given by%
\begin{eqnarray}
Q\left( k_{2},k_{1},x_{1},x_{2}\right) &=&k_{2}^{2}x_{1}\left(
1-x_{1}\right) +k_{1}^{2}x_{2}\left( 1-x_{2}\right) -2\left( k_{2}\cdot
k_{1}\right) x_{1}x_{2} \\
&&+\left( \lambda ^{2}-m_{2}^{2}\right) x_{1}+\left( \lambda
^{2}-m_{1}^{2}\right) x_{2}-\lambda ^{2}.  \notag
\end{eqnarray}%
The integration limits satisfies%
\begin{eqnarray}
Q\left( x_{1},1-x_{2}\right) &=&q^{2}x_{1}\left( 1-x_{1}\right) +\left(
m_{1}^{2}-m_{2}^{2}\right) x_{1}-m_{1}^{2} \\
Q\left( x_{1},0\right) &=&k_{2}^{2}x_{1}\left( 1-x_{1}\right) +\left(
\lambda ^{2}-m_{2}^{2}\right) x_{1}-\lambda ^{2}.
\end{eqnarray}%
\qquad Recovering definition of $A_{i}=2k\cdot k_{i}+k_{i}^{2}+\lambda
^{2}-m_{i}^{2}.$ After shifting, it assumes the form%
\begin{equation}
\left( A_{1}\right) _{k-\left( k_{2}x_{1}+k_{1}x_{2}\right) }=\left( 2k\cdot
k_{1}\right) +\frac{\partial Q}{\partial x_{2}}.
\end{equation}%
This feature will always happen to some $A_{i}$, which means one factor
becomes a sum of a bilinear and a derivative about the last integration
parameter. The next stage is to make partial integrations until all
derivatives are consumed.

For the vector $K_{1}$ that we used as reference (although any other could
be chosen) in definitions of the integral, under shifting, it turns into $%
\left( K_{1}\right) _{k-\left( k_{2}x_{1}+k_{1}x_{2}\right) }=k-\left(
k_{2}x_{1}+k_{1}x_{2}-k_{1}\right) .$ Moreover, in order to simplify and
organize, we define%
\begin{eqnarray}
L &=&\left( k_{2}x_{1}+k_{1}x_{2}-k_{1}\right) \\
L\left( x_{1},1-x_{1}\right) &=&\left( k_{2}-k_{1}\right) x_{1}=qx \\
L\left( x_{1},0\right) &=&L_{0}=\left( k_{2}x_{1}-k_{1}\right)
\end{eqnarray}%
Gathering all the elements, we are left with this expression to integrate%
\begin{equation}
F_{a}^{\mu \nu }=2\int_{0}^{1}\mathrm{d}x_{1}\int_{0}^{1-x_{1}}\mathrm{d}%
x_{2}\int \frac{\mathrm{d}^{2}k}{\left( 2\pi \right) ^{2}}\left\{ \left[
2k\cdot k_{1}+\frac{\partial Q}{\partial x_{2}}\right] \frac{\left(
k-L\right) ^{\mu }\left( k-L\right) ^{\nu }}{\left( k^{2}+Q\right) ^{3}}%
\right\} .
\end{equation}

At this point, we use the results that are elaborated in the sequel, namely%
\begin{eqnarray}
\int \frac{\mathrm{d}^{2}k}{\left( 2\pi \right) ^{2}}\frac{1}{\left(
k^{2}+Q\right) ^{3}} &=&\frac{i}{4\pi }\frac{1}{2Q^{2}} \\
\int \frac{\mathrm{d}^{2}k}{\left( 2\pi \right) ^{2}}\frac{k^{\mu }k^{\nu }}{%
\left( k^{2}+Q\right) ^{3}} &=&\frac{i}{4\pi }\frac{1}{2}g^{\mu \nu }\frac{1%
}{2Q},
\end{eqnarray}%
odd integrals drop from the expression, and we get%
\begin{equation}
F_{a}^{\mu \nu }=\frac{i}{4\pi }\int_{0}^{1}\mathrm{d}x_{1}\int_{0}^{1-x_{1}}%
\mathrm{d}x_{2}\left[ -\left( k_{1}^{\mu }L^{\nu }+k_{1}^{\nu }L^{\mu
}\right) \frac{1}{Q}+\frac{1}{2}g^{\mu \nu }\frac{\partial Q}{\partial x_{2}}%
\frac{1}{Q}+L^{\mu }L^{\nu }\frac{\partial Q}{\partial x_{2}}\frac{1}{Q^{2}}%
\right] .
\end{equation}%
Integrating by parts, we find a total derivative%
\begin{equation}
F_{a}^{\mu \nu }=\frac{i}{4\pi }\int_{0}^{1}\mathrm{d}x_{1}\int_{0}^{1-x_{1}}%
\mathrm{d}x_{2}\frac{\partial }{\partial x_{2}}\left[ \frac{1}{2}g^{\mu \nu
}\log \frac{Q}{-\lambda ^{2}}-L^{\mu }L^{\nu }\frac{1}{Q}\right]
\end{equation}%
that gives us%
\begin{eqnarray}
F_{a}^{\mu \nu } &=&\frac{i}{4\pi }\int_{0}^{1}\mathrm{d}x_{1}\left[ \frac{1%
}{2}g^{\mu \nu }\log \frac{Q\left( x_{1},1-x_{1}\right) }{-\lambda ^{2}}%
-q^{\mu }q^{\nu }\frac{x^{2}}{Q\left( x_{1},1-x_{1}\right) }\right] \\
&&-\frac{i}{4\pi }\int_{0}^{1}\mathrm{d}x_{1}\left[ \frac{1}{2}g^{\mu \nu
}\log \frac{Q\left( x_{1},0\right) }{-\lambda ^{2}}-L_{0}^{\mu }L_{0}^{\nu }%
\frac{1}{Q\left( x_{1},0\right) }\right]  \notag
\end{eqnarray}%
recalling that%
\begin{eqnarray}
Q\left( x_{1},1-x_{2}\right) &=&q^{2}x_{1}\left( 1-x_{1}\right) +\left(
m_{1}^{2}-m_{2}^{2}\right) x_{1}-m_{1}^{2} \\
Q\left( x_{1},0\right) &=&k_{2}^{2}x_{1}\left( 1-x_{1}\right) +\left(
\lambda ^{2}-m_{2}^{2}\right) x_{1}-\lambda ^{2}.
\end{eqnarray}

The other integral is easily expressed in the form%
\begin{equation}
F_{b}^{\mu \nu }=\frac{i}{4\pi }\int_{0}^{1}\mathrm{d}x_{1}\left(
1-x_{1}\right) \left[ -\left( k_{2}^{\mu }L_{0}^{\nu }+k_{2}^{\nu
}L_{0}^{\mu }\right) \frac{1}{Q}+\frac{1}{2}g^{\mu \nu }\frac{1}{Q}\frac{%
\partial Q}{\partial x_{1}}+L_{0}^{\mu }L_{0}^{\nu }\frac{1}{Q^{2}}\frac{%
\partial Q}{\partial x_{1}}\right] .
\end{equation}%
Here the argument of polynomial is $Q\left( x_{1},0\right) $. Thus, partial
integration follows%
\begin{equation}
F_{b}^{\mu \nu }=\frac{i}{4\pi }\int_{0}^{1}\mathrm{d}x_{1}\left[ \frac{1}{2}%
g^{\mu \nu }\log \frac{Q\left( x_{1},0\right) }{-\lambda ^{2}}-\frac{%
L_{0}^{\mu }L_{0}^{\nu }}{Q\left( x_{1},0\right) }\right] +\frac{i}{4\pi }%
\frac{k_{1}^{\mu }k_{1}^{\nu }}{\left( -\lambda ^{2}\right) },
\end{equation}%
again taking into account that $L\left( x_{1},0\right) =L_{0}=\left(
k_{2}x_{1}-k_{1}\right) $.

Finally, summing both contributions $F_{a}^{\mu \nu }$ and $F_{b}^{\mu \nu }$%
, plus a external-momentum independent finite piece%
\begin{equation}
\int \frac{\mathrm{d}^{2}k}{\left( 2\pi \right) ^{2}}\frac{k_{1}^{\mu
}k_{1}^{\nu }}{D_{\lambda }^{2}}=\frac{i}{4\pi }\frac{k_{1}^{\mu }k_{1}^{\nu
}}{\left( -\lambda ^{2}\right) },
\end{equation}%
follows the complete integration of finite parts. The organization of tensor
integral for general masses give us the result%
\begin{equation}
\bar{J}_{2}^{\mu \nu }=\frac{1}{2}\left[ \Delta _{2}^{\mu \nu }+g^{\mu \nu
}I_{\log }\left( \lambda ^{2}\right) \right] +\frac{i}{4\pi }\left[ -\frac{1%
}{2}g^{\mu \nu }Z_{0}^{\left( 0\right) }+q^{\mu }q^{\nu }Z_{2}^{\left(
-1\right) }\right] .
\end{equation}%
Any other integral in this thesis can be obtained with the computational
elements illustrated here.

\section{Integration in the loop momentum}

After Feynman parametrization, all integrals assume the form of the rational
functions%
\begin{equation}
\int \frac{\mathrm{d}^{n}k}{\left( 2\pi \right) ^{n}}\frac{\left( 1,k_{\mu
},k_{\mu \nu },k_{\mu \nu \rho },...\right) }{\left( k^{2}+Q\right) ^{\alpha
}}.
\end{equation}%
To solve the integral, we start with the form%
\begin{equation}
I\left( k,M,n\right) =\int \frac{\mathrm{d}^{n}k}{\left( 2\pi \right) ^{n}}%
\frac{1}{\left( k^{2}-M^{2}\right) ^{\alpha }}=\int \frac{\mathrm{d}^{n}k}{%
\left( 2\pi \right) ^{n}}\frac{1}{\left( k^{2}-2k\cdot q+Q\right) ^{\alpha }}%
,  \label{In}
\end{equation}%
where $2\alpha >n$ and $M^{2}=q^{2}-Q.$ The auxiliary variable $q$ helps to
develop the tensor integrals. The integration measure $\mathrm{d}^{n}k=%
\mathrm{d}^{n-1}k\mathrm{d}k_{0}$. The square the momentum loop $%
k^{2}=k_{0}^{2}-\mathbf{k}^{2};$ and $\mathbf{k}^{2}=%
\sum_{i=1}^{n-1}k_{i}^{2}.$ The integral (\ref{In}) only 
\begin{eqnarray}
I\left( Q,n\right) &=&\int \frac{\mathrm{d}^{n}k}{\left( 2\pi \right) ^{n}}%
\frac{1}{\left( k^{2}-M^{2}\right) ^{\alpha }}=\int \frac{\mathrm{d}^{n-1}k}{%
\left( 2\pi \right) ^{n}}\left[ \int_{-\infty }^{+\infty }\mathrm{d}%
k_{0}f\left( k_{0}\right) \right] \\
f\left( k_{0}\right) &=&\left[ k_{0}^{2}-(\sqrt{\mathbf{k}^{2}+M^{2}}%
-i\varepsilon )^{2}\right] ^{-\alpha }, \\
f\left( k_{0}\right) &\sim &\frac{1}{k_{0}^{2\alpha }},\text{ as }%
k_{0}^{2}\rightarrow \infty .
\end{eqnarray}%
The poles and prescription coming from Feynman propagators 
\begin{eqnarray}
k_{0}^{2} &=&\sqrt{\mathbf{k}^{2}+M^{2}}-i\varepsilon \\
k_{0}^{2} &=&-\sqrt{\mathbf{k}^{2}+M^{2}}+i\varepsilon .
\end{eqnarray}

To compute the integral, we extend the integration for $k_{0}\in \mathbb{C}$
and consider the following contour $C=C_{1}+C_{2}+C_{3}+C_{4}$ in the figure
below 
\begin{figure}[tbph]
\centering\includegraphics[scale=0.6]{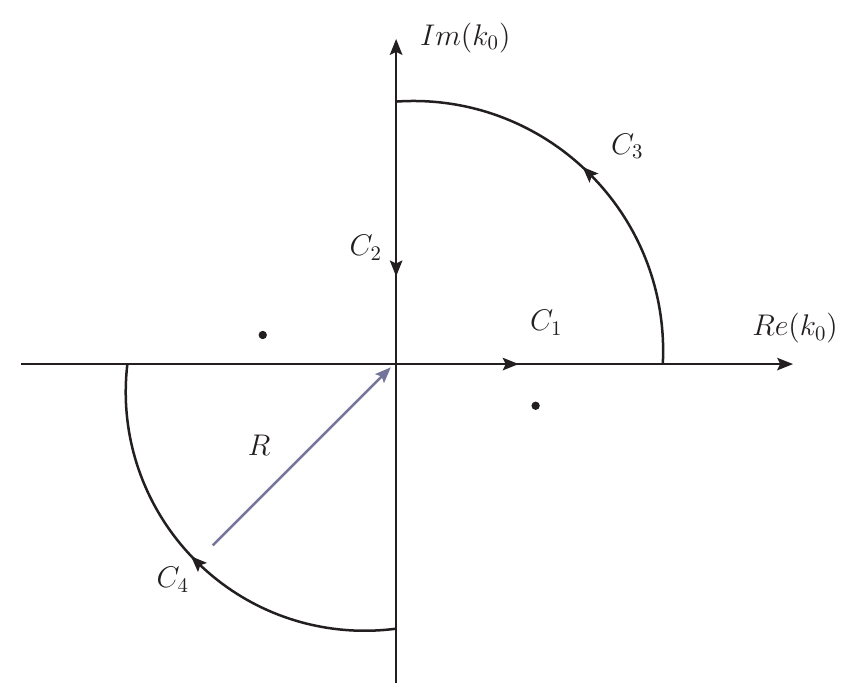}
\caption{Contour of integration}
\end{figure}
Then take the integral over that contour%
\begin{equation}
F_{C}(\mathbf{k}^{2},M^{2})=\int_{C}\mathrm{d}k_{0}f\left( k_{0}\right)
=\left( \int_{C_{1}}+\int_{C_{2}}+\int_{C_{3}}+\int_{C_{4}}\right) \mathrm{d}%
k_{0}f\left( k_{0}\right) =0
\end{equation}%
since there are no poles inside the closed path of integration. We write the
integral as $F_{C}=F_{1}+F_{2}+F_{3}+F_{4}$; the semi-circle contributions
vanish in the limit $\lim_{R\rightarrow \infty }\left(
F_{C_{3}}+F_{C_{4}}\right) =0.$ The reminder contribution gives the desired
relation%
\begin{equation}
\lim_{R\rightarrow \infty }F_{C_{1}}=-\lim_{R\rightarrow \infty
}F_{C_{2}}\rightarrow \int_{-\infty }^{\infty }\mathrm{d}k_{0}f\left(
k_{0}\right) =-\int_{+i\infty }^{-i\infty }\mathrm{d}k_{0}f\left(
k_{0}\right) .
\end{equation}

Changing the integration variable in the last integral over the imaginary
axis by adopting $k_{0}=ik_{0}^{\prime },$ we may write%
\begin{equation}
I\left( Q,n\right) =\int \frac{\mathrm{d}^{n-1}k}{\left( 2\pi \right) ^{n-1}}%
\int_{-\infty }^{\infty }\mathrm{d}k_{0}\frac{1}{\left( k_{0}^{2}-\mathbf{k}%
^{2}-M^{2}\right) ^{\alpha }}=\int \frac{\mathrm{d}^{n-1}k}{\left( 2\pi
\right) ^{n-1}}\int_{-\infty }^{\infty }\mathrm{d}k_{0}^{\prime }\frac{i}{%
\left( -k_{0}^{\prime 2}-\mathbf{k}^{2}-M^{2}\right) ^{\alpha }}
\end{equation}%
and effectively we have an euclidean signature $(k^{\prime 2}:=k_{0}^{\prime
2}+\sum_{i=1}^{n-1}k_{i}^{2})$ to perform the integral%
\begin{equation}
I\left( Q,n\right) =i\left( -1\right) ^{\alpha }\int \frac{\mathrm{d}%
^{n}k^{\prime }}{\left( 2\pi \right) ^{n}}\frac{1}{\left( k^{\prime
2}+M^{2}\right) ^{\alpha }}.
\end{equation}%
Now we introduce spherical coordinates to these variables and split the
radius and solid angle integrations%
\begin{equation}
I\left( Q,n\right) =\frac{i\left( -1\right) ^{\alpha }}{\left( 2\pi \right)
^{n}}\int_{S^{n-1}}\mathrm{d}\Omega \int_{0}^{\infty }\mathrm{d}rr^{n-1}%
\frac{1}{\left( r^{2}+M^{2}\right) ^{\alpha }}.
\end{equation}%
The solid angle furnish%
\begin{equation}
\frac{1}{\left( 2\pi \right) ^{n}}\int_{S^{n-1}}\mathrm{d}\Omega =\frac{2}{%
\left( 4\pi \right) ^{n/2}\Gamma \left( \frac{n}{2}\right) }
\end{equation}%
and simple manipulations bring the form%
\begin{equation}
I\left( Q,n\right) =\frac{2i\left( -1\right) ^{\alpha }}{\left( 4\pi \right)
^{n/2}\Gamma \left( \frac{n}{2}\right) }\frac{1}{2M^{2\left( \alpha
-n/2\right) }}\int_{0}^{\infty }\mathrm{d}\left( r^{\prime 2}\right) \left(
r^{\prime 2}\right) ^{\left( n-2\right) /2}\left( r^{\prime 2}+1\right)
^{-\alpha }.
\end{equation}

Another variables change $r^{\prime 2}=\left( 1-y\right) /y\rightarrow 
\mathrm{d}\left( r^{\prime 2}\right) =-\mathrm{d}y1/y^{2},$ the Beta
function is%
\begin{equation*}
B\left( \alpha -\frac{n}{2},\frac{n}{2}\right) =\int_{0}^{1}\mathrm{d}%
yy^{\left( \alpha -n/2\right) -1}\left( 1-y\right) ^{n/2-1}.
\end{equation*}%
We have%
\begin{eqnarray}
I\left( Q,n\right) &=&\frac{i\left( -1\right) ^{\alpha }}{\left( 4\pi
\right) ^{n/2}\Gamma \left( \frac{n}{2}\right) M^{2\left( \alpha -n/2\right)
}}\int_{0}^{1}\mathrm{d}yy^{\left( \alpha -n/2\right) -1}\left( 1-y\right)
^{n/2-1} \\
&=&\frac{i\left( -1\right) ^{\alpha }}{\left( 4\pi \right) ^{n/2}}\frac{%
\Gamma \left( \alpha -\frac{n}{2}\right) }{\Gamma \left( \alpha \right)
M^{2\left( \alpha -n/2\right) }},
\end{eqnarray}%
thus, from $M^{2}=q^{2}-Q$ follows%
\begin{equation}
I\left( Q,n\right) =\frac{i\left( -1\right) ^{\alpha }}{\left( 4\pi \right)
^{n/2}}\frac{\Gamma \left( \alpha -\frac{n}{2}\right) }{\Gamma \left( \alpha
\right) \left( q^{2}-Q\right) ^{\alpha -n/2}}.
\end{equation}

Now taking derivatives concerning the variable q on both sides and shifting
the parameters $\alpha \rightarrow \alpha -1$ in the form%
\begin{equation}
I\left( Q,n\right) =\frac{i}{\left( -4\pi \right) ^{n/2}}\frac{\Gamma \left(
\alpha -n/2\right) }{\Gamma \left( \alpha \right) \left( Q-q^{2}\right)
^{\left( \alpha -n/2\right) }},
\end{equation}%
the explicit derivative is%
\begin{equation}
\frac{\partial I}{\partial q^{\mu _{1}}}=\frac{i}{\left( -4\pi \right) ^{n/2}%
}\frac{2q_{\mu _{1}}\Gamma \left( \alpha -n/2+1\right) }{\Gamma \left(
\alpha \right) \left( Q-q^{2}\right) ^{\alpha +1-n/2}}=\int \frac{\mathrm{d}%
^{n}k}{\left( 2\pi \right) ^{n}}\frac{2\alpha k_{\mu _{1}}}{\left(
k^{2}-2k\cdot q+Q\right) ^{\alpha +1}},
\end{equation}%
follows the relation%
\begin{equation}
\int \frac{\mathrm{d}^{n}k}{\left( 2\pi \right) ^{n}}\frac{k_{\mu _{1}}}{%
\left( k^{2}-2k\cdot q+Q\right) ^{\alpha }}=\frac{i}{\left( -4\pi \right)
^{n/2}}q_{\mu _{1}}\frac{\Gamma \left( \alpha -n/2\right) }{\Gamma \left(
\alpha \right) \left( Q-q^{2}\right) ^{\alpha -n/2}}.
\end{equation}%
Recursively%
\begin{equation}
\int \frac{\mathrm{d}^{n}k}{\left( 2\pi \right) ^{n}}\frac{k_{\mu
_{2}}k_{\mu _{1}}}{\left( k^{2}+Q\right) ^{\alpha }}=\frac{i}{\left( -4\pi
\right) ^{n/2}}\frac{1}{2}g_{\mu _{12}}\frac{\Gamma \left( \alpha
-n/2-1\right) }{\Gamma \left( \alpha \right) Q^{\alpha -n/2-1}}.
\end{equation}

From the formulae presented, it is possible to obtain a general result,
adopting $n=2\omega $, which reads%
\begin{equation}
\int \frac{\mathrm{d}^{2\omega }k}{\left( 2\pi \right) ^{2\omega }}\frac{%
k_{\mu _{1}}\cdots k_{\mu _{2l+1}}}{\left( k^{2}+Q\right) ^{\alpha }}=0
\end{equation}%
\begin{equation}
\int \frac{\mathrm{d}^{2\omega }k}{\left( 2\pi \right) ^{2\omega }}\frac{%
k_{\mu _{1}}\cdots k_{\mu _{2l}}}{\left( k^{2}+Q\right) ^{\alpha }}=\frac{i}{%
\left( 4\pi \right) ^{\omega }}\frac{1}{2^{l}}g_{(\mu _{1}\mu _{2}}\cdots
g_{\mu _{2l-1},\mu _{2l})}\frac{\Gamma \left( \alpha -\omega -l\right) }{%
\Gamma \left( \alpha \right) Q^{\alpha -\omega -l}}
\end{equation}%
It is interesting to note that these results imply in the properties:%
\begin{eqnarray}
\int \frac{\mathrm{d}^{2\omega }k}{\left( 2\pi \right) ^{2\omega }}f\left(
k^{2}\right) k_{\mu } &=&0 \\
\int \frac{\mathrm{d}^{2\omega }k}{\left( 2\pi \right) ^{2\omega }}k_{\mu
}k_{\nu }f\left( k^{2}\right) &=&\frac{g_{\mu \nu }}{2\omega }\int \frac{%
\mathrm{d}^{2\omega }k}{\left( 2\pi \right) ^{2\omega }}k^{2}f\left(
k^{2}\right)  \label{RD6} \\
\int \frac{\mathrm{d}^{2\omega }k}{\left( 2\pi \right) ^{2\omega }}k_{\mu
\nu \alpha \beta }f\left( k^{2}\right) &=&\frac{\left( g_{\mu \nu \alpha
\beta }+g_{\mu \alpha \nu \beta }+g_{\mu \beta \nu \alpha }\right) }{4\left(
\omega +1\right) }\int \frac{\mathrm{d}^{2\omega }k}{\left( 2\pi \right)
^{2\omega }}k^{4}f\left( k^{2}\right) .
\end{eqnarray}

\chapter{The One point Integrals in Two Dimensions\label{AppInt2D}}

After performing the Dirac traces present in the definitions we established
for the perturbative amplitudes in (\ref{tGG1}) and (\ref{tGG3}), their
integrals naturally decompose in Feynman integrals that we define in the
equations (\ref{J1(ki)}) and (\ref{J2}). The calculations follow the IReg
method by applying the separation identity (\ref{id}) on the divergent
integrals. The finite part is integrated and projected in definitions (\ref%
{Zn(-1)}) and (\ref{Zn(0)}). The residual divergent part is projected onto
divergent objects of the set, expressed in (\ref{IntsDiv}) and their
relations in the session (\ref{DivTerms}).

We start with integrals that have only one propagator. These have only
divergent structures. The finite parts after separating the labels, cancel
out when they are integrated.

\textbf{Integral }$J_{1}:$ by power-counting this integral has a superficial
degree of divergence is logarithmic

\begin{equation}
\bar{J}_{1}\left( k_{i}\right) =I_{\log }  \label{J1B}
\end{equation}%
From the next integral, it is necessary to specify the $k_{1}$ and $k_{2}$
labels of the integral.

\textbf{Integral }$J_{1\mu _{1}}:$ superficial degree of divergence is linear

\begin{eqnarray}
2\bar{J}_{1\mu _{1}}\left( k_{1}\right) &=&-\left( P-q\right) ^{\nu
_{1}}\Delta _{2\mu _{1}\nu _{1}}  \label{J1mu1B} \\
2\bar{J}_{1\mu _{1}}\left( k_{2}\right) &=&-\left( P+q\right) ^{\nu
_{1}}\Delta _{2\mu _{1}\nu _{1}}  \label{J1mu12}
\end{eqnarray}

\textbf{Integral }$J_{1\mu _{12}}:$ superficial degree of divergence is
quadratic%
\begin{eqnarray}
\bar{J}_{1\mu _{12}}\left( k_{1}\right) &=&\frac{1}{2}\left( \Delta _{1\mu
_{12}}+g_{\mu _{12}}I_{\mathrm{quad}}\right) -\frac{1}{8}\left( P-q\right)
^{2}\Delta _{2\mu _{12}}  \label{J11} \\
&&+\frac{1}{8}\left( P-q\right) ^{\nu _{1}}\left[ \left( P-q\right) ^{\nu
_{2}}W_{3\mu _{12}\nu _{12}}-2\left( P-q\right) _{(\mu _{1}}\Delta _{2\mu
_{2})\nu _{1}}\right]  \notag
\end{eqnarray}%
\begin{eqnarray}
\bar{J}_{1\mu _{12}}\left( k_{2}\right) &=&\frac{1}{2}\left( \Delta _{1\mu
_{12}}+g_{\mu _{12}}I_{\mathrm{quad}}\right) -\frac{1}{8}\left( P+q\right)
^{2}\Delta _{2\mu _{12}}  \label{J122} \\
&&+\frac{1}{8}\left( P+q\right) ^{\nu _{1}}\left[ \left( P+q\right) ^{\nu
_{2}}W_{3\mu _{12}\nu _{12}}-2\left( P+q\right) _{(\mu _{1}}\Delta _{2\mu
_{2})\nu _{1}}\right]  \notag
\end{eqnarray}

\textbf{Integral }$J_{1\mu _{123}}:$ superficial degree of divergence is
cubic and are the integrals with the highest power-counting%
\begin{eqnarray}
J_{1\mu _{123}}\left( k_{1}\right) &=&-\frac{1}{4}\left( P-q\right) ^{\nu
_{1}}W_{2\mu _{123}\nu _{1}}+\frac{1}{4}\left( P-q\right) _{(\mu _{1}}\Delta
_{1\mu _{23})} \\
&&-\frac{1}{48}\left( P-q\right) ^{\nu _{1}}\left( P-q\right) ^{\nu
_{2}}\left( P-q\right) ^{\nu _{3}}W_{4\mu _{123}\nu _{123}}  \notag \\
&&+\frac{1}{16}\left( P-q\right) ^{\nu _{1}}\left( P-q\right) ^{\nu
_{2}}\left( P-q\right) _{(\mu _{1}}W_{3\mu _{23})\nu _{12}}  \notag \\
&&+\frac{1}{16}\left( P-q\right) ^{2}\left[ \left( P+p\right) ^{\nu
_{1}}W_{3\mu _{123}\nu _{1}}-\left( P-q\right) _{(\mu _{1}}\Delta _{2\mu
_{23})}\right]  \notag \\
&&-\frac{1}{8}\left( P-q\right) ^{\nu _{1}}\left( P-q\right) _{(\mu
_{1}}\left( P-q\right) _{\mu _{2}}\Delta _{2\mu _{3})\nu _{1}}  \notag
\end{eqnarray}%
\begin{eqnarray}
J_{1\mu _{123}}\left( k_{2}\right) &=&-\frac{1}{4}\left( P+q\right) ^{\nu
_{1}}W_{2\mu _{123}\nu _{1}}+\frac{1}{4}\left( P+q\right) _{(\mu _{1}}\Delta
_{1\mu _{23})}  \label{J1(123)} \\
&&-\frac{1}{48}\left( P+q\right) ^{\nu _{1}}\left( P+q\right) ^{\nu
_{2}}\left( P+q\right) ^{\nu _{3}}W_{4\mu _{123}\nu _{123}}  \notag \\
&&+\frac{1}{16}\left( P+q\right) ^{\nu _{1}}\left( P+q\right) ^{\nu
_{2}}\left( P+q\right) _{(\mu _{1}}W_{3\mu _{23})\nu _{12}}  \notag \\
&&+\frac{1}{16}\left( P+q\right) ^{2}\left[ \left( P+q\right) ^{\nu
_{1}}W_{3\mu _{123}\nu _{1}}-\left( P+q\right) _{(\mu _{1}}\Delta _{2\mu
_{23})}\right]  \notag \\
&&-\frac{1}{8}\left( P+q\right) ^{\nu _{1}}\left( P+q\right) _{(\mu
_{1}}\left( P+q\right) _{\mu _{2}}\Delta _{2\mu _{3})\nu _{1}}.  \notag
\end{eqnarray}

For instance, we calculated the $J_{1}^{\mu _{1}\mu _{2}}\left( k_{i}\right) 
$. The complete expression:

\begin{eqnarray}
J_{1}^{\mu _{1}\mu _{2}}\left( k_{i}\right) &=&\int \frac{\mathrm{d}^{2}k}{%
\left( 2\pi \right) ^{2}}\frac{k^{\mu _{1}}k^{\mu _{2}}}{D_{i}} \\
&&+\int \frac{\mathrm{d}^{2}k}{\left( 2\pi \right) ^{2}}\left( k_{i}^{\mu
_{1}}k^{\mu _{2}}+k_{i}^{\mu _{2}}k^{\mu _{1}}\right) \frac{1}{D_{i}}  \notag
\\
&&+k_{i}^{\mu _{1}}k_{i}^{\mu _{2}}\int \frac{\mathrm{d}^{2}k}{\left( 2\pi
\right) ^{2}}\frac{1}{D_{i}}.  \notag
\end{eqnarray}%
Using the expansions for two first integral above, we have%
\begin{eqnarray}
\left[ \frac{k^{\mu _{1}}k^{\mu _{2}}}{D_{i}}\right] _{\mathrm{even}} &=&%
\frac{k^{\mu _{1}}k^{\mu _{2}}}{D_{\lambda }}-k_{i}^{2}\frac{k^{\mu
_{1}}k^{\mu _{2}}}{D_{\lambda }^{2}}+4k_{i\nu _{12}}\frac{k^{\mu _{1}}k^{\mu
_{2}}k^{\nu _{1}}k^{\nu _{2}}}{D_{\lambda }^{3}} \\
\left[ \frac{k^{\mu _{1}}}{D_{i}}\right] _{\mathrm{even}} &=&-2k_{i\nu _{1}}%
\frac{k^{\mu _{1}}k^{\nu _{1}}}{D_{\lambda }^{2}}.
\end{eqnarray}%
So the expanded integral is given by%
\begin{eqnarray}
J_{1}^{\mu _{1}\mu _{2}}\left( k_{i}\right) &=&\int \frac{\mathrm{d}^{2}k}{%
\left( 2\pi \right) ^{2}}\left( \frac{k^{\mu _{1}}k^{\mu _{2}}}{D_{\lambda }}%
-k_{i}^{2}\frac{k^{\mu _{1}}k^{\mu _{2}}}{D_{\lambda }^{2}}+4k_{i\nu
_{1}}k_{i\nu _{2}}\frac{k^{\mu _{1}}k^{\mu _{2}}k^{\nu _{1}}k^{\nu _{2}}}{%
D_{\lambda }^{3}}\right)  \label{J1u1u2} \\
&&-2k_{i\nu _{1}}\int \frac{\mathrm{d}^{2}k}{\left( 2\pi \right) ^{2}}\left(
k_{i}^{\mu _{1}}\frac{k^{\mu _{2}}k^{\nu _{1}}}{D_{\lambda }^{2}}+k_{i}^{\mu
_{2}}\frac{k^{\mu _{1}}k^{\nu _{1}}}{D_{\lambda }^{2}}\right)  \notag \\
&&+k_{i}^{\mu _{1}}k_{i}^{\mu _{2}}\int \frac{\mathrm{d}^{2}k}{\left( 2\pi
\right) ^{2}}\frac{1}{D_{i}}.  \notag
\end{eqnarray}%
Identifying the divergent objects in Section (\ref{DivTerms})%
\begin{eqnarray*}
\int \frac{\mathrm{d}^{2}k}{\left( 2\pi \right) ^{2}}\frac{8k^{\mu _{1}\mu
_{2}\nu _{1}\nu _{2}}}{D_{\lambda }^{3}} &=&W_{3}^{\mu _{1}\mu _{2}\nu
_{1}\nu _{2}}+g^{\mu _{1}\mu _{2}\nu _{1}\nu _{2}}I_{\text{\textrm{log}}} \\
W_{3}^{\mu _{1}\mu _{2}\nu _{1}\nu _{2}} &=&\square _{3}^{\mu _{1}\mu
_{2}\nu _{1}\nu _{2}}+\frac{1}{2}g^{(\mu _{1}\nu _{1}}\Delta _{2}^{\mu
_{2}\nu _{2})} \\
\int \frac{\mathrm{d}^{2}k}{\left( 2\pi \right) ^{2}}\frac{2k^{\mu _{1}\mu
_{2}}}{D_{\lambda }^{2}} &=&\Delta _{2}^{\mu _{1}\mu _{2}}+g^{\mu _{1}\mu
_{2}}I_{\text{\textrm{log}}} \\
\int \frac{\mathrm{d}^{2}k}{\left( 2\pi \right) ^{2}}\frac{2k^{\mu _{1}\mu
_{2}}}{D_{\lambda }} &=&\Delta _{1}^{\mu _{1}\mu _{2}}+g^{\mu _{1}\mu
_{2}}I_{\text{\textrm{quad}}}.
\end{eqnarray*}%
Substituting in (\ref{J1u1u2}), we can see the scalars $I_{\log }$ cancel
and remains the final expression%
\begin{eqnarray}
J_{1}^{\mu _{1}\mu _{2}}\left( k_{i}\right) &=&\frac{1}{2}\left[ \Delta
_{1}^{\mu _{1}\mu _{2}}+g^{\mu _{1}\mu _{2}}I_{\mathrm{quad}}\right] +\frac{1%
}{2}k_{i\nu _{12}}W_{3}^{\mu _{1}\mu _{2}\nu _{1}\nu _{2}} \\
&&-k_{i}^{\mu _{1}}k_{i\nu _{1}}\Delta _{2}^{\mu _{2}\nu _{1}}-k_{i}^{\mu
_{2}}k_{i\nu _{1}}\Delta _{2}^{\mu _{1}\nu _{1}}-\frac{1}{2}k_{i}^{2}\Delta
_{2}^{\mu _{1}\mu _{2}}.  \notag
\end{eqnarray}%
The expression above can be written as (\ref{J11}) and \ref{J122} replacing
the routing $k_{i}$ by $2k_{1}=\left( P-q\right) $ or $2k_{2}=\left(
P+q\right) .$

\chapter{Function $Z_{k}^{\left( -1\right) }\left(
q^{2},m_{1}^{2},m_{2}^{2}\right) $}

As we saw throughout the text, it is sometimes interesting to consider
explicit forms of these functions due to their importance in discussing some
important aspects of amplitudes. So we consider the following function%
\begin{equation*}
Z_{k}^{\left( -1\right) }\left( q^{2},m_{1}^{2},m_{2}^{2}\right) \equiv
\int_{0}^{1}d\mathrm{z}\frac{z^{k}}{Q\left( q^{2},m_{1}^{2},m_{2}^{2}\right) 
},
\end{equation*}%
where $Q=q^{2}z\left( 1-z\right) +\left( m_{1}^{2}-m_{2}^{2}\right)
z-m_{1}^{2}$ is the polynomial form of denominator. Since all the functions $%
Z_{k}^{\left( -1\right) }$ can be put in terms of the functions $%
Z_{0}^{\left( -1\right) },$ we will consider in this appendix the
calculation explicitly only of the function, defined by 
\begin{equation}
Z_{0}^{\left( -1\right) }\left( q^{2},m_{1}^{2},m_{2}^{2}\right)
=\int_{0}^{1}\frac{1}{Q}.  \label{Z}
\end{equation}

One way to integrate is to write the polynomial present in the denominator
through its roots. We do%
\begin{equation}
Q=-q^{2}\left[ z^{2}-\frac{1}{q^{2}}\left( q^{2}+m_{1}^{2}-m_{2}^{2}\right)
z+\frac{m_{1}^{2}}{q^{2}}\right] =-q^{2}\left( z-\alpha \right) \left(
z-\beta \right) .  \label{Q}
\end{equation}%
Where the roots of the polynomial are $\alpha $ and $\beta $ given by%
\begin{eqnarray}
\alpha &=&\frac{\left( q^{2}+m_{1}^{2}-m_{2}^{2}\right) +\sqrt{\left(
q^{2}+m_{1}^{2}-m_{2}^{2}\right) ^{2}-4m_{1}^{2}q^{2}}}{2q^{2}}; \\
\beta &=&\frac{q^{2}+m_{1}^{2}-m_{2}^{2}-\sqrt{\left(
q^{2}+m_{1}^{2}-m_{2}^{2}\right) ^{2}-4m_{1}^{2}q^{2}}}{2q^{2}},
\end{eqnarray}%
where $\alpha $ and $\beta $ satisfy the following relations:%
\begin{eqnarray}
\alpha +\beta &=&\frac{\left( q^{2}+m_{1}^{2}-m_{2}^{2}\right) }{q^{2}}%
;\qquad \alpha \beta =\frac{m_{1}^{2}}{q^{2}} \\
\alpha -\beta &=&\frac{\sqrt{\left( q^{2}+m_{1}^{2}-m_{2}^{2}\right)
^{2}-4m_{1}^{2}q^{2}}}{q^{2}}.
\end{eqnarray}
Rewriting Eq. (\ref{Z}) as

\begin{equation}
Z_{0}^{\left( -1\right) }=-\frac{1}{q^{2}}\int_{0}^{1}\frac{1}{\left(
z-\alpha \right) \left( z-\beta \right) }=-\frac{1}{q^{2}}\frac{1}{\alpha
-\beta }\int_{0}^{1}d\mathrm{z}\left[ \frac{1}{\left( z-\alpha \right) }-%
\frac{1}{\left( z-\beta \right) }\right] .
\end{equation}%
Using the passage%
\begin{equation}
\int_{0}^{1}d\mathrm{z}\frac{1}{\left( z-\alpha \right) }=\ln \left(
1-\alpha \right) -\ln \left( -\alpha \right) =\ln \left( \frac{\alpha -1}{%
\alpha }\right) ,
\end{equation}%
we will have%
\begin{equation}
Z_{0}^{\left( -1\right) }=-\frac{1}{q^{2}}\frac{1}{\alpha -\beta }\left\{
\ln \left( \frac{\alpha -1}{\alpha }\right) -\ln \left( \frac{\beta -1}{%
\beta }\right) \right\} =\frac{1}{q^{2}}\frac{1}{\alpha -\beta }\left[ \ln
\left( \frac{\alpha -1}{\alpha }\right) \left( \frac{\beta }{\beta -1}%
\right) \right] .
\end{equation}

From that, we can write the explicit form for the function $Z_{0}^{\left(
-1\right) },$%
\begin{equation}
Z_{0}^{\left( -1\right) }=\frac{1}{\sqrt{\left(
q^{2}+m_{1}^{2}-m_{2}^{2}\right) ^{2}-4m_{1}^{2}q^{2}}}\ln \left[ \frac{%
\left( m_{1}^{2}+m_{2}^{2}-q^{2}\right) +\sqrt{\left(
q^{2}+m_{1}^{2}-m_{2}^{2}\right) ^{2}-4m_{1}^{2}q^{2}}}{\left(
m_{1}^{2}+m_{2}^{2}-q^{2}\right) -\sqrt{\left(
q^{2}+m_{1}^{2}-m_{2}^{2}\right) ^{2}-4m_{1}^{2}q^{2}}}\right]
\end{equation}%
In the kinematical limit, where $q^{2}\ll 1,$ we have the result%
\begin{equation}
Z_{0}^{\left( -1\right) }=\frac{1}{\left( m_{1}^{2}-m_{2}^{2}\right) }\ln %
\left[ \frac{\left( m_{1}^{2}-m_{2}^{2}\right) +\left(
m_{1}^{2}-m_{2}^{2}\right) }{\left( m_{1}^{2}-m_{2}^{2}\right) -\left(
m_{1}^{2}-m_{2}^{2}\right) }\right] .
\end{equation}

\chapter{Subamplitudes}

\label{AppSub}We cast vector subamplitudes in this appendix. They are
ordered following the amplitudes that originate them ($AVV$, $VAV$, $VVA$,
and $AAA$) and then grouped according to the version. That emphasizes
patterns attributed to each version and additional terms depending on the
squared mass.

\textbf{First version:}

\begin{eqnarray}
\left( t^{VPP}\right) ^{\nu _{1}} &=&\left[ -K_{1}^{\nu
_{1}}S_{23}+K_{2}^{\nu _{1}}S_{13}-K_{3}^{\nu _{1}}S_{12}\right] \frac{1}{%
D_{123}} \\
\left( t^{ASP}\right) ^{\nu _{1}} &=&\left[ -K^{\nu _{1}}S_{23}+K_{2}^{\nu
_{1}}\left( S_{13}+2m^{2}\right) -K_{3}^{\nu _{1}}\left(
S_{12}+2m^{2}\right) \right] \frac{1}{D_{123}} \\
\left( t^{APS}\right) ^{\nu _{1}} &=&\left[ K_{1}^{\nu _{1}}\left(
S_{23}+2m^{2}\right) -K_{2}^{\nu _{1}}\left( S_{13}+2m^{2}\right)
+K_{3}^{\nu _{1}}S_{12}\right] \frac{1}{D_{123}} \\
\left( t^{VSS}\right) ^{\nu _{1}} &=&\left[ K_{1}^{\nu _{1}}\left(
S_{23}+2m^{2}\right) -K_{2}^{\nu _{1}}S_{13}+K_{3}^{\nu _{1}}\left(
S_{12}+2m^{2}\right) \right] \frac{1}{D_{123}}
\end{eqnarray}%
\begin{eqnarray}
(T^{VPP})^{\nu _{1}} &=&2\left[ P_{31}^{\nu _{2}}\Delta _{3\nu _{2}}^{\nu
_{1}}+(p_{21}^{\nu _{1}}-p_{32}^{\nu _{1}})I_{\log }\right] -4\left(
p_{21}\cdot p_{32}\right) J_{3}^{\nu _{1}} \\
&&+2\left[ (p_{31}^{\nu _{1}}p_{21}^{2}-p_{21}^{\nu
_{1}}p_{31}^{2})J_{3}+p_{21}^{\nu _{1}}J_{2}\left( p_{21}\right)
-p_{32}^{\nu _{1}}J_{2}\left( p_{32}\right) \right]  \notag \\
\left( T^{ASP}\right) ^{\nu _{1}} &=&2\left[ P_{31}^{\nu _{2}}\Delta _{3\nu
_{2}}^{\nu _{1}}+\left( p_{21}^{\nu _{1}}-p_{32}^{\nu _{1}}\right) I_{\log }%
\right] -4\left( p_{21}\cdot p_{32}\right) J_{3}^{\nu _{1}} \\
&&+2\left[ \left( p_{31}^{\nu _{1}}p_{21}^{2}-p_{21}^{\nu
_{1}}p_{31}^{2}-4m^{2}p_{32}^{\nu _{1}}\right) J_{3}+p_{21}^{\nu
_{1}}J_{2}\left( p_{21}\right) -p_{32}^{\nu _{1}}J_{2}\left( p_{32}\right) %
\right]  \notag \\
-\left( T^{APS}\right) ^{\nu _{1}} &=&2\left[ P_{31}^{\nu _{2}}\Delta _{3\nu
_{2}}^{\nu _{1}}+(p_{21}^{\nu _{1}}-p_{32}^{\nu _{1}})I_{\log }\right]
-4\left( p_{21}\cdot p_{32}\right) J_{3}^{\nu _{1}} \\
&&+2\left[ \left( p_{31}^{\nu _{1}}p_{21}^{2}-p_{21}^{\nu
_{1}}p_{31}^{2}+4m^{2}p_{21}^{\nu _{1}}\right) J_{3}+p_{21}^{\nu
_{1}}J_{2}\left( p_{21}\right) -p_{32}^{\nu _{1}}J_{2}\left( p_{32}\right) %
\right]  \notag \\
-\left( T^{VSS}\right) ^{\nu _{1}} &=&2\left[ P_{31}^{\nu _{2}}\Delta _{3\nu
_{2}}^{\nu _{1}}+(p_{21}^{\nu _{1}}-p_{32}^{\nu _{1}})I_{\log }\right]
-4\left( p_{21}\cdot p_{32}+4m^{2}\right) J_{3}^{\nu _{1}} \\
&&+2\left[ \left( p_{31}^{\nu _{1}}p_{21}^{2}-p_{21}^{\nu
_{1}}p_{31}^{2}-4m^{2}p_{31}^{\nu _{1}}\right) J_{3}+p_{21}^{\nu
_{1}}J_{2}\left( p_{21}\right) -p_{32}^{\nu _{1}}J_{2}\left( p_{32}\right) %
\right]  \notag
\end{eqnarray}

\textbf{Second version:}%
\begin{eqnarray}
\left( t^{SAP}\right) ^{\nu _{1}} &=&\left[ K_{1}^{\nu
_{1}}S_{23}+K_{2}^{\nu _{1}}\left( S_{13}+2m^{2}\right) -K_{3}^{\nu
_{1}}\left( S_{12}+2m^{2}\right) \right] \frac{1}{D_{123}} \\
\left( t^{PVP}\right) ^{\nu _{1}} &=&\left[ -K_{1}^{\nu
_{1}}S_{23}-K_{2}^{\nu _{1}}S_{13}+K_{3}^{\nu _{1}}S_{12}\right] \frac{1}{%
D_{123}} \\
\left( t^{PAS}\right) ^{\nu _{1}} &=&-\left[ K_{1}^{\nu _{1}}\left(
S_{23}+2m^{2}\right) +K_{2}^{\nu _{1}}S_{13}-K_{3}^{\nu _{1}}\left(
S_{12}+2m^{2}\right) \right] \frac{1}{D_{123}} \\
\left( t^{SVS}\right) ^{\nu _{1}} &=&\left[ K_{1}^{\nu _{1}}\left(
S_{23}+2m^{2}\right) +K_{2}^{\nu _{1}}\left( S_{13}+2m^{2}\right)
-K_{3}^{\nu _{1}}S_{12}\right] \frac{1}{D_{123}}
\end{eqnarray}%
\begin{eqnarray}
-\left( T^{SAP}\right) ^{\nu _{1}} &=&2\left[ P_{21}^{\nu _{2}}\Delta _{3\nu
_{2}}^{\nu _{1}}+\left( p_{32}^{\nu _{1}}+p_{31}^{\nu _{1}}\right) I_{\log }%
\right] +4\left( p_{32}\cdot p_{31}\right) J_{3}^{\nu _{1}} \\
&&+2\left[ \left( p_{21}^{\nu _{1}}p_{31}^{2}-p_{31}^{\nu
_{1}}p_{21}^{2}+4m^{2}p_{32}^{\nu _{1}}\right) J_{3}+p_{32}^{\nu
_{1}}J_{2}\left( p_{32}\right) +p_{31}^{\nu _{1}}J_{2}\left( p_{31}\right) %
\right]  \notag \\
\left( T^{PVP}\right) ^{\nu _{1}} &=&2\left[ P_{21}^{\nu _{2}}\Delta _{3\nu
_{2}}^{\nu _{1}}+\left( p_{32}^{\nu _{1}}+p_{31}^{\nu _{1}}\right) I_{\log }%
\right] +4\left( p_{32}\cdot p_{31}\right) J_{3}^{\nu _{1}} \\
&&+2\left[ \left( p_{21}^{\nu _{1}}p_{31}^{2}-p_{31}^{\nu
_{1}}p_{21}^{2}\right) J_{3}+p_{32}^{\nu _{1}}J_{2}\left( p_{32}\right)
+p_{31}^{\nu _{1}}J_{2}\left( p_{31}\right) \right]  \notag \\
\left( T^{PAS}\right) ^{\nu _{1}} &=&2\left[ P_{21}^{\nu _{2}}\Delta _{3\nu
_{2}}^{\nu _{1}}+\left( p_{32}^{\nu _{1}}+p_{31}^{\nu _{1}}\right) I_{\log }%
\right] +4\left( p_{32}\cdot p_{31}\right) J_{3}^{\nu _{1}} \\
&&+2\left[ \left( p_{21}^{\nu _{1}}p_{31}^{2}-p_{31}^{\nu
_{1}}p_{21}^{2}+4m^{2}p_{31}^{\nu _{1}}\right) J_{3}+p_{32}^{\nu
_{1}}J_{2}\left( p_{32}\right) +p_{31}^{\nu _{1}}J_{2}\left( p_{31}\right) %
\right]  \notag \\
-\left( T^{SVS}\right) ^{\nu _{1}} &=&2\left[ P_{21}^{\nu _{2}}\Delta _{3\nu
_{2}}^{\nu _{1}}+\left( p_{32}^{\nu _{1}}+p_{31}^{\nu _{1}}\right) I_{\log }%
\right] +4\left( p_{32}\cdot p_{31}-4m^{2}\right) J_{3}^{\nu _{1}} \\
&&+2\left[ \left( p_{21}^{\nu _{1}}p_{31}^{2}-p_{31}^{\nu
_{1}}p_{21}^{2}-4m^{2}p_{21}^{\nu _{1}}\right) J_{3}+p_{32}^{\nu
_{1}}J_{2}\left( p_{32}\right) +p_{31}^{\nu _{1}}J_{2}\left( p_{31}\right) %
\right]  \notag
\end{eqnarray}

\textbf{Third version:}

\begin{eqnarray}
\left( t^{SPA}\right) ^{\nu _{1}} &=&\left[ K_{1}^{\nu _{1}}\left(
S_{23}+2m^{2}\right) -K_{2}^{\nu _{1}}\left( S_{13}+2m^{2}\right)
-K_{3}^{\nu _{1}}S_{12}\right] \frac{1}{D_{123}} \\
\left( t^{PSA}\right) ^{\nu _{1}} &=&\left[ -K_{1}^{\nu _{1}}\left(
S_{23}+2m^{2}\right) +K_{2}^{\nu _{1}}S_{13}+K_{3}^{\nu _{1}}\left(
S_{12}+2m^{2}\right) \right] \frac{1}{D_{123}} \\
\left( t^{PPV}\right) ^{\nu _{1}} &=&-\left[ -K_{1}^{\nu
_{1}}S_{23}+K_{2}^{\nu _{1}}S_{13}+K_{3}^{\nu _{1}}S_{12}\right] \frac{1}{%
D_{123}} \\
\left( t^{SSV}\right) ^{\nu _{1}} &=&\left[ -K_{1}^{\nu
_{1}}S_{23}+K_{2}^{\nu _{1}}\left( S_{13}+2m^{2}\right) +K_{3}^{\nu
_{1}}\left( S_{12}+2m^{2}\right) \right] \frac{1}{D_{123}}
\end{eqnarray}%
\begin{eqnarray}
\left( T^{SPA}\right) ^{\nu _{1}} &=&2\left[ P_{32}^{\nu _{2}}\Delta _{3\nu
_{2}}^{\nu _{1}}-\left( p_{21}^{\nu _{1}}+p_{31}^{\nu _{1}}\right) I_{\log }%
\right] +4\left( p_{21}\cdot p_{31}\right) J_{3}^{\nu _{1}} \\
&&+2\left[ \left( p_{31}^{\nu _{1}}p_{21}^{2}+p_{21}^{\nu
_{1}}p_{31}^{2}-4m^{2}p_{21}^{\nu _{1}}\right) J_{3}-p_{21}^{\nu
_{1}}J_{2}\left( p_{21}\right) -p_{31}^{\nu _{1}}J_{2}\left( p_{31}\right) %
\right]  \notag \\
-\left( T^{PSA}\right) ^{\nu _{1}} &=&2\left[ P_{32}^{\nu _{2}}\Delta _{3\nu
_{2}}^{\nu _{1}}-\left( p_{21}^{\nu _{1}}+p_{31}^{\nu _{1}}\right) I_{\log }%
\right] +4\left( p_{21}\cdot p_{31}\right) J_{3}^{\nu _{1}} \\
&&+2\left[ \left( p_{31}^{\nu _{1}}p_{21}^{2}+p_{21}^{\nu
_{1}}p_{31}^{2}-4m^{2}p_{31}^{\nu _{1}}\right) J_{3}-p_{21}^{\nu
_{1}}J_{2}\left( p_{21}\right) -p_{31}^{\nu _{1}}J_{2}\left( p_{31}\right) %
\right]  \notag \\
\left( T^{PPV}\right) ^{\nu _{1}} &=&2\left[ P_{32}^{\nu _{2}}\Delta _{3\nu
_{2}}^{\nu _{1}}-\left( p_{21}^{\nu _{1}}+p_{31}^{\nu _{1}}\right) I_{\log }%
\right] +4\left( p_{21}\cdot p_{31}\right) J_{3}^{\nu _{1}} \\
&&+2\left[ \left( p_{31}^{\nu _{1}}p_{21}^{2}+p_{21}^{\nu
_{1}}p_{31}^{2}\right) J_{3}-p_{21}^{\nu _{1}}J_{2}\left( p_{21}\right)
-p_{31}^{\nu _{1}}J_{2}\left( p_{31}\right) \right]  \notag \\
-\left( T^{SSV}\right) ^{\nu _{1}} &=&2\left[ P_{32}^{\nu _{2}}\Delta _{3\nu
_{2}}^{\nu _{1}}-\left( p_{21}^{\nu _{1}}+p_{31}^{\nu _{1}}\right) I_{\log }%
\right] +4\left( p_{21}\cdot p_{31}-4m^{2}\right) J_{3}^{\nu _{1}} \\
&&+2\left[ \left( p_{31}^{\nu _{1}}p_{21}^{2}+p_{21}^{\nu
_{1}}p_{31}^{2}-4m^{2}\left( p_{21}^{\nu _{1}}+p_{31}^{\nu _{1}}\right)
\right) J_{3}-p_{21}^{\nu _{1}}J_{2}\left( p_{21}\right) -p_{31}^{\nu
_{1}}J_{2}\left( p_{31}\right) \right]  \notag
\end{eqnarray}

\chapter{Surface Terms}

The surface terms used in this work appear in a totally symmetrical way in
the indices, for the first time treated from the point of view of the IReg
strategy. The meaning of the notation used is 
\begin{equation}
g_{(\mu _{12}}g_{\mu _{34})}=g_{\mu _{12}}g_{\mu _{34}}+g_{\mu _{13}}g_{\mu
_{24}}+g_{\mu _{14}}g_{\mu _{23}}.
\end{equation}
For instance, in the case of permutations involving six indices as the
product of the metrics by the logarithmically divergent object $\Delta
_{2\mu \nu },$ we have forty-five terms given by, 
\begin{eqnarray}
&&g_{(\mu _{12}}g_{\mu _{34}}  \notag \\
&=&\Delta _{2\mu _{12}}g_{(\mu _{34}}g_{\mu _{56})}+\Delta _{2\mu
_{13}}g_{(\mu _{24}}g_{\mu _{56})}+\Delta _{2\mu _{14}}g_{(\mu _{23}}g_{\mu
_{56})}+\Delta _{2\mu _{15}}g_{(\mu _{23}}g_{\mu _{46})}+\Delta _{2\mu
_{16}}g_{(\mu _{23}}g_{\mu _{45})}  \notag \\
&&+\Delta _{2\mu _{23}}g_{(\mu _{14}}g_{\mu _{56})}+\Delta _{2\mu
_{24}}g_{(\mu _{13}}g_{\mu _{56})}+\Delta _{2\mu _{25}}g_{(\mu _{13}}g_{\mu
_{46})}+\Delta _{2\mu _{26}}g_{(\mu _{13}}g_{\mu _{45})}  \notag \\
&&+\Delta _{2\mu _{34}}g_{(\mu _{12}}g_{\mu _{56})}+\Delta _{2\mu
_{35}}g_{(\mu _{12}}g_{\mu _{46})}+\Delta _{2\mu _{36}}g_{(\mu _{12}}g_{\mu
_{45})}  \notag \\
&&+\Delta _{2\mu _{45}}g_{(\mu _{12}}g_{\mu _{36})}+\Delta _{2\mu
_{46}}g_{(\mu _{12}}g_{\mu _{35})}  \notag \\
&&+\Delta _{2\mu _{56}}g_{(\mu _{12}}g_{\mu _{34})}.
\end{eqnarray}%
This can be written succinctly as%
\begin{equation}
g_{(\mu _{12}}g_{\mu _{34}}\Delta _{2\mu
_{56})}=\sum_{i_{2}>i_{1}=1}^{5}\Delta
_{2i_{1}i_{2}}g_{(i_{3}i_{4}}g_{i_{5}i_{6})}\text{ with }i_{n}\not=i_{m}
\end{equation}%
where $i_{n}$ denotes $\mu _{i_{n}}$. For the box terms we may also write%
\begin{equation}
g_{(\mu _{1}\mu _{2}}\square _{3\mu _{3456})}=\sum_{i_{2}>i_{1}=1}^{5}g_{\mu
_{i_{1}}\mu _{i_{2}}}\square _{3\mu _{i_{3}}\mu _{i_{4}}\mu _{i_{5}}\mu
_{i_{6}}}.
\end{equation}

\section{Uniqueness Factor: Combination of the violating terms}

\label{Uni-Red}

As we saw, surface terms violate several symmetry relations. However, if the
relations are satisfied, relations between surface terms emerge for their
traces and the finite part. Through the strategy (\ref{IREG}), we saw that
all the divergent objects were organized into standardized objects as to
their tensor degree and power counting. We have%
\begin{eqnarray}
\Xi _{\nu _{23}}^{\left( a\right) } &=&\left[ 2\square _{3\rho \nu
_{23}}^{\rho }-2\Delta _{2\nu _{23}}-g_{\nu _{23}}\Delta _{2\rho }^{\rho }%
\right] \\
\Xi _{\alpha _{12}\nu _{23}}^{\left( b\right) } &=&\left[ 3\Sigma _{4\rho
\alpha _{12}\nu _{23}}^{\rho }-8\square _{3\alpha _{12}\nu _{23}}-g_{\alpha
_{12}\nu _{23}}\Delta _{2\rho }^{\rho }\right] \\
\Xi _{\alpha _{1}\alpha _{2}}^{\mathrm{quad}} &=&\left[ W_{2\rho \alpha
_{1}\alpha _{2}}^{\rho }-2\Delta _{1\alpha _{12}}+2g_{\alpha _{12}}I_{\text{%
\textrm{quad}}}-2m^{2}\left( \Delta _{2\alpha _{12}}+g_{\alpha _{12}}I_{\log
}\right) \right] .
\end{eqnarray}%
In this way, this organization allows us to write the $U$-factor as%
\begin{eqnarray}
U_{\alpha _{1}\alpha _{2}} &=&-\frac{1}{3}\theta _{\alpha _{1}\alpha
_{2}}\left( 2\Delta _{2\rho }^{\rho }+i/\pi \right) +\frac{1}{9}\left(
3P^{\nu _{2}}P^{\nu _{3}}+q^{\nu _{23}}\right) \Xi _{\alpha _{1}\alpha
_{2}\nu _{23}}^{b} \\
&&+\frac{1}{18}\left( 3P^{\nu _{2}}P^{\nu _{3}}+q^{\nu _{23}}\right)
g_{(\alpha _{1}\alpha _{2}}\Xi _{\nu _{2}\nu _{3})}^{a}  \notag \\
&&-\frac{1}{2}\left( P^{2}+q^{2}\right) \Xi _{\alpha _{12}}^{a}-P^{\nu
_{1}}P_{(\alpha _{2}}\Xi _{\alpha _{1})\nu _{1}}^{a}+4\Xi ^{\mathrm{quad}} 
\notag
\end{eqnarray}%
The uniqueness factor that arises in the basic permutations%
\begin{equation}
U_{\alpha _{2}\nu _{1}}=\left( 4\Upsilon _{\nu _{1}\alpha _{2}}+2q_{\nu
_{1}}\Upsilon _{\alpha _{2}}+2q_{\alpha _{2}}\Upsilon _{\nu _{1}}+q_{\alpha
_{2}}q_{\nu _{1}}\Upsilon \right) ,
\end{equation}%
its explicit expression reads%
\begin{eqnarray}
U_{\alpha _{2}\nu _{1}} &=&-\frac{1}{3}\theta _{\nu _{1}\alpha _{2}}\left(
2\Delta _{2\rho }^{\rho }+i/\pi \right) \\
&&+\frac{1}{9}\left( 3P^{\nu _{2}}P^{\nu _{3}}+q^{\nu _{23}}\right) \left[
3\Sigma _{4\rho \nu _{1}\alpha _{2}\nu _{23}}^{\rho }-8\square _{3\nu
_{1}\alpha _{2}\nu _{23}}-g_{\nu _{1}\alpha _{2}\nu _{23}}\Delta _{2\rho
}^{\rho }\right]  \notag \\
&&+\frac{1}{18}\left( 3P^{\nu _{2}}P^{\nu _{3}}+q^{\nu _{23}}\right) \left[
g_{(\alpha _{1}\alpha _{2}}\left( 2\square _{3\rho \nu _{23})}^{\rho
}-2\Delta _{2\nu _{23})}-g_{\nu _{23})}\Delta _{2\rho }^{\rho }\right) %
\right]  \notag \\
&&-\frac{1}{2}\left( P^{2}+q^{2}\right) \left[ 2\left( \square _{3\rho \nu
_{1}\alpha _{2}}^{\rho }-\Delta _{2\nu _{1}\alpha _{2}}\right) -g_{\nu
_{1}\alpha _{2}}\Delta _{2\rho }^{\rho }\right]  \notag \\
&&-P_{\alpha _{2}}P^{\nu _{2}}\left[ 2\left( \square _{3\rho \nu _{1}\nu
_{2}}^{\rho }-\Delta _{2\nu _{1}\nu _{2}}\right) -g_{\nu _{1}\nu _{2}}\Delta
_{2\rho }^{\rho }\right]  \notag \\
&&-P_{\nu _{1}}P^{\nu _{2}}\left[ 2\left( \square _{3\rho \alpha _{2}\nu
_{2}}^{\rho }-\Delta _{2\alpha _{2}\nu _{2}}\right) -g_{\alpha _{2}\nu
_{2}}\Delta _{2\rho }^{\rho }\right]  \notag \\
&&+4\left[ W_{2\rho \nu _{1}\alpha _{2}}^{\rho }-2\Delta _{1\nu _{1}\alpha
_{2}}+2g_{\nu _{1}\alpha _{2}}I_{\text{\textrm{quad}}}-2m^{2}\left( \Delta
_{2\nu _{1}\alpha _{2}}+g_{\nu _{1}\alpha _{2}}I_{\log }\right) \right] . 
\notag
\end{eqnarray}

In the massless limit and independent of unique or vanishing surface terms%
\begin{equation}
U_{\alpha _{2}\nu _{1}}=-\frac{1}{3}\theta _{\nu _{1}\alpha _{2}}\left(
2\Delta _{2\rho }^{\rho }+\frac{i}{\pi }\right) =-\frac{1}{3}\theta _{\nu
_{1}\alpha _{2}}\Upsilon
\end{equation}

\begin{eqnarray}
U_{\alpha _{1}\alpha _{2}} &=&-\frac{1}{3}\theta _{\alpha _{1}\alpha
_{2}}\Upsilon +\frac{1}{9}\left( 3P^{\nu _{12}}+q^{\nu _{12}}\right) \Xi
_{\alpha _{1}\alpha _{2}\nu _{12}}^{\left( b\right) }-P^{\nu _{1}}P_{(\alpha
_{2}}\Xi _{\alpha _{1})\nu _{1}}^{\left( a\right) } \\
&&+\frac{1}{18}\left( 3P^{\nu _{12}}+q^{\nu _{12}}\right) g_{(\alpha
_{1}\alpha _{2}}\Xi _{1\nu _{12})}^{\left( a\right) }-\frac{1}{2}\left(
P^{2}+q^{2}\right) \Xi _{\alpha _{12}}^{\left( a\right) }+4\Xi ^{\mathrm{quad%
}},  \notag
\end{eqnarray}%
where the definitions%
\begin{eqnarray}
\Xi _{\alpha _{12}}^{\mathrm{quad}} &=&\square _{2\rho \alpha _{12}}^{\rho }+%
\frac{1}{2}g_{\alpha _{12}}\Delta _{1\rho }^{\rho }+\Delta _{1\alpha
_{12}}+2g_{\alpha _{1}\alpha _{2}}I_{\text{\textrm{quad}}}-2m^{2}\left(
\Delta _{2\alpha _{12}}+g_{\alpha _{12}}I_{\log }\right) \\
&=&\square _{2\rho \alpha _{12}}^{\rho }+\int \frac{\mathrm{d}^{2}k}{\left(
2\pi \right) ^{2}}\left[ \frac{g_{\alpha _{12}}k^{2}}{D_{\lambda }}+\frac{%
2k_{\alpha _{12}}}{D_{\lambda }}\right] -2m^{2}\left( \Delta _{2\alpha
_{12}}+g_{\alpha _{12}}I_{\log }\right)
\end{eqnarray}%
\begin{eqnarray}
\int \frac{\mathrm{d}^{2}k}{\left( 2\pi \right) ^{2}}\frac{2k_{\alpha _{12}}%
}{D_{\lambda }^{2}} &=&\Delta _{2\alpha _{12}}+g_{\alpha _{12}}I_{\log } \\
\int \frac{\mathrm{d}^{2}k}{\left( 2\pi \right) ^{2}}\frac{2k_{\alpha _{12}}%
}{D_{\lambda }} &=&\Delta _{1\alpha _{12}}+g_{\alpha _{1}\alpha _{2}}I_{%
\text{\textrm{quad}}}
\end{eqnarray}%
\begin{equation}
\frac{1}{2}g_{\alpha _{12}}\Delta _{1\rho }^{\rho }+\Delta _{1\alpha
_{12}}+2g_{\alpha _{1}\alpha _{2}}I_{\text{\textrm{quad}}}=\int \frac{%
\mathrm{d}^{2}k}{\left( 2\pi \right) ^{2}}\left( g_{\alpha
_{12}}k^{2}+2k_{\alpha _{12}}\right) \frac{1}{D_{\lambda }}
\end{equation}%
\begin{equation}
\square _{2\rho \alpha _{12}}^{\rho }=\int \frac{\mathrm{d}^{2}k}{\left(
2\pi \right) ^{2}}\left[ 4k^{2}k_{\alpha _{12}}-6k_{\alpha _{12}}D_{\lambda
}-g_{\alpha _{12}}k^{2}D_{\lambda }\right] \frac{1}{D_{\lambda }^{2}}.
\end{equation}%
Where the quadratic form can be made null as 
\begin{equation}
\Xi _{\alpha _{12}}^{\mathrm{quad}}=\int \frac{\mathrm{d}^{2}k}{\left( 2\pi
\right) ^{2}}\left[ 4\left( k^{2}-m^{2}\right) -4D_{\lambda }\right] \frac{%
k_{\alpha _{12}}}{D_{\lambda }^{2}}=0.
\end{equation}

\section{Bilinears reductions and the accessible values to the uniqueness
factor}

Observing the expressions%
\begin{eqnarray}
\Xi _{\nu _{23}}^{\left( a\right) } &=&2\int \frac{\mathrm{d}^{2}k}{\left(
2\pi \right) ^{2}}\left\{ \left[ \frac{8k^{2}k_{\nu _{23}}}{D_{\lambda }^{3}}%
-\frac{g_{\nu _{23}}k^{2}+6k_{\nu _{23}}}{D_{\lambda }^{2}}\right] \right. \\
&&-\left. \left[ \frac{2k_{\nu _{23}}}{D_{\lambda }^{2}}-\frac{g_{\nu _{23}}%
}{D_{\lambda }}\right] -g_{\nu _{23}}\left[ \frac{k^{2}}{D_{\lambda }^{2}}-%
\frac{1}{D_{\lambda }}\right] \right\} .  \notag
\end{eqnarray}%
If it is linear and bilinears are reduced, follow the solid resu%
\begin{equation}
\Xi _{\nu _{23}}^{\left( a\right) }=4m^{2}\int \frac{\mathrm{d}^{2}k}{\left(
2\pi \right) ^{2}}\left[ \frac{4k_{\nu _{23}}}{D_{\lambda }^{3}}-g_{\nu
_{23}}\frac{1}{D_{\lambda }^{2}}\right] =-4m^{2}\int \frac{\mathrm{d}^{2}k}{%
\left( 2\pi \right) ^{2}}\frac{\partial }{\partial k^{\nu _{3}}}\frac{k_{\nu
_{2}}}{D_{\lambda }^{2}}\equiv 0.
\end{equation}%
The last passage involves defining a surface term that appears in 4D. Here
it is finite and indisputably zero.

As the higher rank term, they appear in the violations of RAGFs and unicity
of odd amplitudes%
\begin{equation}
\Xi _{\mu _{1234}}^{\left( b\right) }=\left[ 3\Sigma _{4\rho \mu
_{1234}}^{\rho }-8\square _{3\mu _{1234}}-g_{\mu _{1234}}\Delta _{2\rho
}^{\rho }\right] ,
\end{equation}%
we will have for the first term 
\begin{equation}
3g^{\mu _{12}}\Sigma _{4\mu _{123456}}=\int \frac{\mathrm{d}^{2}k}{\left(
2\pi \right) ^{2}}\left[ \frac{144k^{2}k_{\mu _{3456}}}{D_{\lambda }^{4}}-%
\frac{8\left[ 10k_{\mu _{3456}}+k^{2}g_{(\mu _{34}}k_{\mu _{56})}\right] }{%
D_{\lambda }^{3}}\right] ,
\end{equation}%
using the formula $g^{\mu _{12}}g_{(\mu _{12}}k_{\mu _{3456})}=10k_{\mu
_{3456}}+k^{2}g_{(\mu _{34}}k_{\mu _{56})}$ and the definitions%
\begin{eqnarray}
8\square _{3\mu _{3456}} &=&\int \frac{\mathrm{d}^{2}k}{\left( 2\pi \right)
^{2}}\left[ \frac{64k_{\mu _{3456}}}{D_{\lambda }^{3}}-\frac{8g_{(\mu
_{34}}k_{\mu _{56})}}{D_{\lambda }^{2}}\right] \\
g_{(\mu _{12}}g_{\mu _{34})}\Delta _{2\rho }^{\rho } &=&g_{(\mu _{12}}g_{\mu
_{34})}\int \frac{\mathrm{d}^{2}k}{\left( 2\pi \right) ^{2}}\left[ \frac{%
2k^{2}}{D_{\lambda }^{2}}-\frac{2}{D_{\lambda }}\right] ,
\end{eqnarray}%
it is obtained the result%
\begin{eqnarray}
&&\left[ 3g^{\mu _{12}}\Sigma _{4\mu _{123456}}-8\square _{3\mu
_{3456}}-g_{(\mu _{12}}g_{\mu _{34})}\Delta _{2\rho }^{\rho }\right] \\
&=&\int \frac{\mathrm{d}^{2}k}{\left( 2\pi \right) ^{2}}\left\{ \left[ \frac{%
144k_{\mu _{3456}}}{D_{\lambda }^{3}}-\frac{8g_{(\mu _{34}}k_{\mu _{56})}}{%
D_{\lambda }^{2}}-\frac{2g_{(\mu _{12}}g_{\mu _{34})}}{D_{\lambda }}\right] 
\frac{k^{2}}{D_{\lambda }}\right.  \notag \\
&&\hspace{60pt}\left. -\left[ \frac{144k_{\mu _{3456}}}{D_{\lambda }^{3}}-%
\frac{8g_{(\mu _{34}}k_{\mu _{56})}}{D_{\lambda }^{2}}-\frac{2g_{(\mu
_{12}}g_{\mu _{34})}}{D_{\lambda }}\right] \right\} .  \notag
\end{eqnarray}

Reducing bilinears by adding and subtracting the mass makes obtaining the
identity%
\begin{equation}
\frac{k^{2}}{k^{2}-m^{2}}=1+\frac{m^{2}}{k^{2}-m^{2}}.
\end{equation}%
We reach at%
\begin{eqnarray}
&&\left[ 3g^{\mu _{12}}\Sigma _{4\mu _{123456}}-8\square _{3\mu
_{3456}}-g_{(\mu _{12}}g_{\mu _{34})}\Delta _{2\rho }^{\rho }\right] \\
&=&m^{2}\int \frac{\mathrm{d}^{2}k}{\left( 2\pi \right) ^{2}}\left\{ \frac{%
144k_{\mu _{3456}}}{D_{\lambda }^{4}}-\frac{8g_{(\mu _{34}}k_{\mu _{56})}}{%
D_{\lambda }^{3}}-\frac{2g_{(\mu _{12}}g_{\mu _{34})}}{D_{\lambda }^{2}}%
\right\} .  \notag
\end{eqnarray}%
Mass terms do not vanish identically; what remains are precisely convergent
surface terms%
\begin{eqnarray}
&&\left[ 3g^{\mu _{12}}\Sigma _{4\mu _{123456}}-8\square _{3\mu
_{3456}}-g_{(\mu _{12}}g_{\mu _{34})}\Delta _{2\rho }^{\rho }\right] \\
&=&m^{2}\int \frac{\mathrm{d}^{2}k}{\left( 2\pi \right) ^{2}}\left\{ 12\left[
\frac{12k_{\mu _{3456}}}{D_{\lambda }^{4}}-\frac{g_{(\mu _{34}}k_{\mu _{56})}%
}{D_{\lambda }^{3}}\right] +4\frac{g_{(\mu _{34}}k_{\mu _{56})}}{D_{\lambda
}^{3}}-\frac{g_{(\mu _{12}}g_{\mu _{34})}}{D_{\lambda }^{2}}-\frac{g_{(\mu
_{12}}g_{\mu _{34})}}{D_{\lambda }^{2}}\right\} ,  \notag
\end{eqnarray}%
these terms own integrands that are typical of four dimensions. Integrating
in 2D they are precisely zero%
\begin{equation}
\Delta _{3;\mu _{ij}}=-\int \frac{\mathrm{d}^{2}k}{\left( 2\pi \right) ^{2}}%
\frac{\partial }{\partial k^{\mu _{i}}}\frac{k_{\mu _{j}}}{D_{\lambda }^{2}}%
=\int \frac{\mathrm{d}^{2}k}{\left( 2\pi \right) ^{2}}\left[ \frac{4k_{\mu
_{ij}}}{D_{\lambda }^{3}}-\frac{g_{\mu _{ij}}}{D_{\lambda }^{2}}\right]
\equiv 0
\end{equation}%
\begin{equation}
\square _{4;\mu _{3456}}=-\frac{1}{2}\sum_{i=1}^{4}\int \frac{\mathrm{d}^{2}k%
}{\left( 2\pi \right) ^{2}}\frac{\partial }{\partial k^{\mu _{i}}}\frac{%
k_{\mu _{1}\cdots \hat{\mu}_{i}\cdots \mu _{4}}}{D_{\lambda }^{3}}=\int 
\frac{\mathrm{d}^{2}k}{\left( 2\pi \right) ^{2}}\left[ \frac{12k_{\mu
_{3456}}}{D_{\lambda }^{4}}-\frac{g_{(\mu _{34}}k_{\mu _{56})}}{D_{\lambda
}^{3}}\right] \equiv 0
\end{equation}%
thereby%
\begin{equation}
\left[ 3g^{\mu _{12}}\Sigma _{4\mu _{123456}}-8\square _{3\mu
_{3456}}-g_{(\mu _{12}}g_{\mu _{34})}\Delta _{2\rho }^{\rho }\right] =m^{2}%
\left[ 12\square _{4\mu _{3456}}+g_{(\mu _{34}}\Delta _{3\mu _{56})}\right]
=0
\end{equation}%
if the total derivative character of the expression is desired, we can also
write in the form%
\begin{eqnarray}
&&\left[ 3g^{\mu _{12}}\Sigma _{4\mu _{123456}}-8\square _{3\mu
_{3456}}-g_{(\mu _{12}}g_{\mu _{34})}\Delta _{2\rho }^{\rho }\right] \\
&=&m^{2}\int \frac{\mathrm{d}^{2}k}{\left( 2\pi \right) ^{2}}\left\{
-6\sum_{i=1}^{4}\frac{\partial }{\partial k^{\mu _{i}}}\frac{k_{\mu
_{1}\cdots \hat{\mu}_{i}\cdots \mu _{4}}}{D_{\lambda }^{3}}-g_{(\mu _{3}\mu
_{4}}\frac{\partial }{\partial k^{\mu _{5}}}\frac{k_{\mu _{6})}}{D_{\lambda
}^{2}}\right\} .  \notag
\end{eqnarray}

\textbf{Quadratic term in the Uniqueness factor: }We assume bilinear
reduction this term cancels identically independent from the definition of
the quadratic scalar%
\begin{equation}
U_{\alpha _{12}}^{\mathrm{quad}}=\int \frac{\mathrm{d}^{2}k}{\left( 2\pi
\right) ^{2}}\left\{ \frac{16k^{2}k_{\alpha _{12}}}{D_{\lambda }^{2}}-\frac{%
16k_{\alpha _{12}}}{D_{\lambda }}\left[ 1-m^{2}\frac{1}{D_{\lambda }}\right]
\right\}
\end{equation}%
in other words $U_{\alpha _{12}}^{\mathrm{quad}}=0$.

\addcontentsline{toc}{chapter}{Bibliography}%

\end{document}